\pdfoutput=1
\RequirePackage{lhchiggs}
\documentclass{cernyrep}
\usepackage[T1]{fontenc}
\usepackage{lmodern}
\usepackage{graphicx}
\usepackage{rotating}
\usepackage{amssymb}
\usepackage{xcolor}
\usepackage{subfigure}
\usepackage{booktabs}
\usepackage{arydshln}
\usepackage{lscape}
\usepackage{floatpag}
\floatpagestyle{plain}
\usepackage{axodraw}
\usepackage{lineno}
\usepackage{graphpap}
\usepackage{cite}

\newcommand{\kbd}[1]{\texttt{#1}\xspace}

\newcommand{\eV}{\ensuremath{\text{e\kern-0.15ex{}V}}\xspace}
\newcommand{\MeV}{\ensuremath{\text{M\eV}}\xspace}
\newcommand{\GeV}{\ensuremath{\text{G\eV}}\xspace}
\newcommand{\TeV}{\ensuremath{\text{T\eV}}\xspace}

\makeatletter
\renewcommand{\thechapter}{\@Roman\c@chapter}
\makeatother

\begin{document}

{\centering{\LARGE{\bf{Les Houches 2013: Physics at TeV Colliders\\ Standard Model Working Group Report \par }}}}

\pagenumbering{roman}

\vspace{0.7cm}
\leftline{\bf Conveners}


\noindent \emph{Higgs physics: SM issues} \\
         D. De Florian (Theory), \, 
         M. Kado (ATLAS), \,
         A. Korytov (CMS), \\
         S. Dittmaier (Electroweak Contact)
\vspace{0.1cm}

\noindent \emph{SM: Loops and Multilegs}  \\	
         N. Glover (Theory), \,
         J. Huston (ATLAS), \,
         G. Dissertori (CMS), \\
         S. Dittmaier (Electroweak Contact)
\vspace{0.1cm}

\noindent \emph{Tools and Monte Carlos} \\	
         F. Krauss (Theory), \,
         J. Butterworth (ATLAS), \,
         K. Hamilton (MC-NLO Contact), \\
         G. Soyez (Jets Contact) 
\vspace{1.0cm}

{\leftline{\bf{Abstract}}}

\vspace{0.5cm}
This Report summarizes the proceedings of the 2013 Les Houches workshop on Physics at TeV Colliders. Session 1 dealt primarily with (1) the techniques for calculating standard model multi-leg NLO and NNLO QCD and NLO EW cross sections and (2) the comparison of those cross sections with LHC data from Run 1, and projections for future measurements in Run 2. 

\vspace{0.5cm}

{\leftline{\bf{Acknowledgements}}}

\vspace{0.5cm}
We would like to thank the organizers (G.~Belanger, F.~Boudjema, P.~Gras,  D.~Guadagnoli, S. ~Gascon, J.P.~ Guillet, B. ~Herrmann, S. ~Kraml, G.H.~de Monchenault, G. Moreau, E. Pilon,  P. Slavich and D. Zerwas) and the Les Houches staff for the stimulating environment always present at Les Houches.



\newpage
{\centering{\bf{Authors\par}}}
\begin{flushleft}

J.~R.~Andersen$^{1}$, S.~Badger$^{2}$, L.~Barz\`e$^{3}$, J.~Bellm$^{4}$, F.~U.~Bernlochner$^{5}$,  A.~Buckley$^{6}$, J.~Butterworth$^{7}$,  N.~Chanon$^{8}$,  M.~Chiesa$^{9}$, A.~Cooper-Sarkar$^{10}$,  L.~Cieri$^{11}$, G.~Cullen$^{12}$, H.~van Deurzen$^{13}$, G.~Dissertori$^{8}$, S.~Dittmaier$^{14}$, D.~de Florian$^{15}$, S.~Forte$^{16}$,  R.~Frederix$^{3}$,   B.~Fuks$^{3,17}$,  J.~Gao$^{18}$, M.V.~Garzelli$^{19}$, T.~Gehrmann$^{20}$, E.~Gerwick$^{21}$, S.~Gieseke$^{22}$, D.~Gillberg$^{23}$, E.~W.~N.~Glover$^{1}$,  N.~Greiner$^{13}$,  K.~Hamilton$^{7}$, T.~Hapola$^{1}$, H.B.~Hartanto$^{24}$, G.~Heinrich$^{13}$, A.~Huss$^{14}$, J.~Huston$^{25}$, B.~J\"ager$^{26}$, M.~Kado$^{27}$, A.~Kardos$^{28}$, U.~Klein$^{29}$, F.~Krauss$^{1}$,  A.~Kruse$^{30}$, L.~L\"onnblad$^{31}$, G.~Luisoni$^{13}$, Daniel Ma\^{\i}tre$^{1}$, P.~Mastrolia$^{13,32}$, O.~Mattelaer$^{33}$, J.~Mazzitelli$^{15}$, E.~Mirabella$^{13}$, P.~Monni$^{34}$, G.~Montagna$^{9}$, M.~Moretti$^{35}$, P.~Nadolsky$^{18}$, P.~Nason$^{3}$, O.~Nicrosini$^{36}$, C.~Oleari$^{28}$, G.~Ossola$^{37,38}$, S.~Padhi$^{39}$, T.~Peraro$^{13}$, F.~Piccinini$^{36}$,  S.~Pl\"atzer$^{40}$,  S.~Prestel$^{40}$, J.~Pumplin$^{25}$, K.~Rabbertz$^{22}$, Voica Radescu$^{40}$, L.~Reina$^{41}$, C.~Reuschle$^{42}$, J.~Rojo$^{3}$, M.~Sch\"onherr$^{1}$, J.~M.~Smillie$^{43}$,  J.F.~von~Soden-Fraunhofen$^{13}$, G.~Soyez$^{44}$, R.~Thorne$^{7}$,  F.~Tramontano$^{45}$, Z.~Trocsanyi$^{19}$, D.~Wackeroth$^{46}$, J.~Winter$^{13}$, C-P.~Yuan$^{25}$, V.~Yundin~$^{13}$,  K.~Zapp$^{3}$

\end{flushleft}

\begin{itemize}

\item[$^{1}$] Institute for Particle Physics Phenomenology, Department of Physics, University of Durham, Durham DH1 3LE, UK
\item[$^{2}$] Niels Bohr Institute (NBI), University of Copenhagen, Blegdamsvej 17 DK-2100 Copenhagen, Denmark
\item[$^{3}$] PH Department, TH Unit, CERN, CH-1211, Geneva 23, Switzerland
\item[$^{4}$] Institute for Theoretical Physics, Karlsruhe Institute of Technology, D - 76128 Karlsruhe, Germany
\item[$^{5}$] University of Victoria, Victoria, British Columbia, Canada V8W 3P
\item[$^{6}$] School of Physics \& Astronomy, University of Glasgow, Glasgow, UK
\item[$^{7}$] Department of Physics and Astronomy, University College London, Gower Street, London WC1E 6BT, UK
\item[$^{8}$] Institute for Particle Physics, ETH Zurich, 8093 Zurich, Switzerland
\item[$^{9}$] Dipartimento di Fisica, Universit\`a di Pavia and INFN, Sezione di Pavia, Via A.~Bassi, 6, 27100 Pavia, Italy
\item[$^{10}$] Department of Physics, University of Oxford, Denys Wilkinson Building, Keble Road, Oxford OX1 3RH, UK
\item[$^{11}$] Dipartimento di Fisica, Universit\`a di Roma La Sapienza and INFN, Sezione di Roma, I-00185 Rome, Italy
\item[$^{12}$] DESY, Zeuthen, Platanenallee 6, D-15738 Zeuthen, Germany
\item[$^{13}$] Max-Planck-Institut f\"ur Physik, Werner-Heisenberg-Institut, F\"ohringer Ring 6, D-80805 M\"unchen, Germany
\item[$^{14}$] Physikalisches Institut, Albert-Ludwigs-Universitat Freiburg, D-79104 Freiburg, Germany
\item[$^{15}$] Departamento de F\'isica, Facultad de Ciencias Exactas y Naturales, Universidad de Buenos Aires, Pabellon I, Ciudad Universitaria (1428), Capital Federal, Argentina
\item[$^{16}$] Dipartimento di Fisica, Universit\`a degli Studi di Milano and INFN, Sezione di Milano, Via Celoria 16, I-20133 Milan, Italy
\item[$^{17}$] Institut Pluridisciplinaire Hubert Curien, Department Recherches Subatomiques, Universite de
Strasbourg, CNRS-IN2P3, 23 rue du Loess, F-67037 Strasbourg, France
\item[$^{18}$] Southern Methodist University, Dept. of Physics, Dallas, TX 75275, USA
\item[$^{19}$] MTA-DE Particle Physics Research Group, University of Debrecen, H-4010 Debrecen P.O.Box 105, Hungary
\item[$^{20}$] Institute for Theoretical Physics, University of Zurich, Winterthurerstrasse 190, CH-8057 Zurich, Switzerland
\item[$^{21}$] II. Physikalisches Institut, UniversirR Gottingen, Germany
\item[$^{22}$] Institut f\"ur Experimentelle Kernphysik, Karlsruher Institut f\"ur Technologie, D-76131 Karlsruhe, Germany
\item[$^{23}$] CERN, CH-1211 Geneva 23, Switzerland
\item[$^{24}$] Institut f\"ur Theoretische Teilchenphysik und Kosmologie, RWTH Aachen University, D-52056 Aachen, Germany
\item[$^{25}$] Department of Physics and Astronomy, Michigan State University, East Lansing, MI 48824, USA
\item[$^{26}$] PRISMA Cluster of Excellence and Institute of Physics, Johannes Gutenberg University, D-55099 Mainz, Germany
\item[$^{27}$] Laboratoire de l'Acc\'el\'erateur Lin\'eaire, CNRS/IN2P3, F-91898 Orsay CEDEX, France
\item[$^{28}$] Universit\`a di Milano-Bicocca and INFN, Sezione di Milano-Bicocca, Piazza della Scienza 3, 20126 Milan, Italy
\item[$^{29}$] University of Liverpool, Dept. of Physics Oliver Lodge Lab, Oxford St. Liverpool L69 3BX, UK
\item[$^{30}$] Univ. of Wisconsin, Dept. of Physics, High Energy Physics,2506 Sterling Hall 1150 University Ave, Madison, WI 53706, USA
\item[$^{31}$] Theoretical High Energy Physics, Department of Theoretical Physics, Lund University, S\"olvegatan 14A, SE-223 62 Lund, Sweden
\item[$^{32}$]	Dipartimento di Fisica e Astronomia, Universit\`a di Padova, and INFN Sezione di Padova, Via Marzolo 8, I-35131 Padova, Italy
\item[$^{33}$] Centre for Cosmology, Particle Physics and Phenomenology (CP3),  Universit\'e Catholique de Louvain, B-1348 Louvain-la-Neuve, Belgium
\item[$^{34}$]  Rudolf Peierls Centre for Theoretical Physics, University of Oxford, 1 Keble Road, Oxford OX1 2NP, UK
\item[$^{35}$] Dipartimento di Fisica e Science della Terra, Universita di Ferrara, and INFN Sezione di Ferrara, Via Saragat 1, 44100 Ferrara, Italy
\item[$^{36}$] INFN, Sezione di Pavia, Via A. Bassi 6, 27100 Pavia, Italy
\item[$^{37}$] Physics Department, New York City College of Technology, The City University of New York, 300 Jay Street Brooklyn, NY 11201, USA
\item[$^{38}$] The Graduate School and University Center, The City University of New York, 365 Fifth Avenue, New York, NY 10016, USA
\item[$^{39}$] Department of Physics, University of California at San Diego, La Jolla, CA 92093, USA
\item[$^{40}$] DESY, Notkestrasse 85, D-22607 Hamburg, Germany
\item[$^{41}$] Physics Department, Florida State University, Tallahassee, FL 32306-4350, USA
\item[$^{42}$] Institute for Theoretical Physics, Karlsruhe Institute of Technology, D - 76128 Karlsruhe, Germany
\item[$^{43}$] The Higgs Centre for Theoretical Physics, University of Edinburgh, Mayfield Road, Edinburgh, EH9 3JZ
\item[$^{44}$] IPhT, CEA Saclay, Orme des Merisiers, Bat 774, F-91191 Gif-sur-Yvette, cedex, France

\enlargethispage{1em}

\item[$^{45}$] Dipartimento di Fisica, Universit\`a di Napoli Federico II, and INFN, Sezione di Napoli, Complesso di Monte Sant'Angelo, via Cintia I-80126 Napoli, Italy
\item[$^{46}$] Department of Physics, SUNY at Buffalo, Buffalo, NY 14260-1500, USA
\end{itemize}


\newpage
\tableofcontents


\newpage
\pagenumbering{arabic}
\setcounter{footnote}{0}

\section{Introduction\footnote{G.~Dissertori, S.~Dittmaier, N.~Glover, J.~Huston, 
A.~Korytov, F.~Krauss}}

The Les Houches Workshop in 2011 was the first for which data from the LHC was available. In 2013, we had high statistics data for the first time, at both $7\UTeV$ and $8\UTeV$, and, most importantly, the discovery of the Higgs boson to contemplate.  The LHC data for Standard Model (SM)
processes  encompassed a very wide kinematic range, accessing transverse momenta and masses on the order of a TeV or greater. For precision understanding  at such scales, higher-order electroweak (EW) corrections need to be taken into account in addition to higher-order QCD corrections. There was  perhaps more emphasis on EW physics and corrections in the 2013 Les Houches than in previous workshops. As for the 2011 workshop, there  was no evidence of beyond-SM (BSM) processes; if such events are present in the data, they are hiding well, indicating even more the need  for precision SM calculations to indicate the presence of any deviations.

	The SM data taken at $7\UTeV$ and $8\UTeV$, with its small statistical errors and decreasing (with time) systematic errors, is useful  not only to test the theoretical predictions, but to serve as input for the global parton distribution function (PDF)  fits.  Currently, the PDF  fits are dominated by data taken at HERA and at fixed target deep-inelastic experiments, especially in the parton $x$ range from around 0.01 to 0.1. This kinematic range is especially important for  Higgs production, and in particular for the process $\Pg\Pg\rightarrow{}$Higgs. The  PDF(+$\alphas$) uncertainty is the largest theoretical uncertainty for this important process. As more (and better) data is taken at the LHC,  there is the possibility for PDF uncertainties to be reduced, especially in the TeV mass range. Until now, EW and QED corrections to  PDFs have been mostly ignored, but again, given the mass scales being considered (and the importance of processes such as  $\gamma\gamma\rightarrow \PW\PW$ at those mass scales), such considerations will become increasingly important. 

	The rapid rate of progress beyond leading-order (LO) calculations
has continued on the theoretical side, in terms of semi-automated calculations of multi-leg next-to-leading order (NLO) QCD processes, advances in  next-to-next-to-leading order (NNLO) QCD calculations, and advances in NLO EW, and mixed EW+QCD, calculations.   In the last two years, we have seen the NLO calculation of complex  multi-leg processes  such as 
for the production of 5~jets~\cite{Badger:2013yda}, $\PW+{}$5~jets~\cite{Bern:2013gka}, and Higgs~+~3~jets~\cite{Cullen:2013saa} as well as the further development of general purpose NLO codes such as 
GoSaM~\cite{Cullen:2011ac}, HELAC-NLO~\cite{Bevilacqua:2011xh}, MadLoop~\cite{Hirschi:2011pa}, OpenLoops~\cite{Cascioli:2011va}, and Recola~\cite{Actis:2012qn}. At NNLO,  there have been new results for diphoton~\cite{Catani:2011qz} and $\PZ\gamma$ production~\cite{Grazzini:2013bna}. The  complete NNLO $\Pt\bar{\Pt}$ cross section calculation is available~\cite{Baernreuther:2012ws,Czakon:2012zr,Czakon:2012pz,Czakon:2013goa}, along with  the partial (gluons only) calculation for dijet production~\cite{Ridder:2013mf,Currie:2013dwa} and for 
Higgs~+~1~jet production~\cite{Boughezal:2013uia}, with the likelihood of further  progress within the near future. The sticking point for processes with coloured final-state particles has been the infrared subtraction. Now several subtraction schemes have been  developed~\cite{Czakon:2010td,GehrmannDeRidder:2005cm,Boughezal:2011jf} and  the techniques have been applied to the processes listed above. Based on these experiences, further progress  should come more quickly. In some sense, 
part of the frontier is now NNNLO, where some approximate $\Pg\Pg\rightarrow{}$Higgs calculations have been performed~\cite{Ball:2013bra,Buehler:2013fha}, and a full calculation might be possible within the next few years~\cite{Anastasiou:2013srw,Anastasiou:2013mca,Anastasiou:2014vaa}. 

	Naively, NLO EW corrections are roughly as important as NNLO QCD corrections. However, NLO 
${\cal O}(\alpha)$ EW corrections can be  magnified by the presence of large logarithms and/or kinematical effects, such as when the hard scale(s) in the process are large compared to the W-boson mass. When all invariants for the 4-vectors of pairs of particles become large compared to the $\PW$ mass, this is referred to as the  Sudakov regime, or `Sudakov zone'.   There is a need in principle for mixed QCD+EW  calculations of order $\alphas \alpha$. Sadly, however, in most cases, the multi-scale  NNLO calculations required at this order are currently out of reach, and we have to rely on approximations as to whether for example QCD and EW  corrections are additive or multiplicative. 

	The Les Houches NLO Wish List, constructed in 2005, and added to in 2007 and 2009, was formally dis-continued in 2011, due to the  progress in NLO automation evident at that time. Instead in 2011, an NNLO Wish List was started, with a few essential processes given high  priority. These included the processes  listed above, in addition to vector-boson + jet production and double-vector-boson production. In  2013, we have gone beyond the 2011 list, and have constructed a {\em High Precision Wish List} that has to be considered as extremely ambitious,  and that will no doubt still exist several Les Houches Workshops from now. The calculation of the processes on this list will allow us to fully exploit  the data to come at the LHC at $14\UTeV$, and hopefully to understand the SM to such an extent that any evidence of new physics  encountered will be unambiguous. 

	As stressed in earlier Les Houches Workshops, it is  important not only for advanced QCD+EW calculations to be completed, but for the  calculations and their results to be available to experimenters. This has been made easier in recent years with the use of the ROOT output format  and with Rivet routines. It is now standard technology as well to allow for `on-the-fly' re-weighting of matrix element results for new PDFs,  new scales  and even new jet algorithms, and this should be incorporated, where possible, in public programs. 

	In addition to progress on fixed-order calculations, there has also been progress on resummation calculations, in particular for cross  sections with jet vetoes, or jet binning, such as used for Higgs+jets production through $\Pg\Pg$ fusion, that has lead to the restoration of the accuracy present in the related inclusive cross section predictions. 

	Another area of rapid theoretical progress has been in the combination of the fixed-order results with a parton shower, leading to particle-level predictions with improved perturbative accuracy.  The principles of {\em matching} a parton shower with NLO calculations for a particular final state are well established and have been partially or fully automated~\cite{Alioli:2010xd,Hoche:2010pf,Hirschi:2011pa,Platzer:2011bc,Hoeche:2011fd} to exploit the automated NLO codes. Alternatively, the description of multi-jet final states with parton showers can be improved by {\em merging} matrix element calculations of varying multiplicity with a parton shower. This approach has recently been refined and extended, leading to algorithms~\cite{Hoeche:2012yf,Lonnblad:2012ng} which can combine multiple NLO calculations of varying multiplicity into a single, inclusive simulation.   The MINLO method~\cite{Hamilton:2012np} for choosing the factorisation and remormalisation scales accounts for Sudakov suppression effects and enables NLO calculations to extrapolate to zero jet transverse momentum, thus offering the opportunity to match to NNLO calculations for a limited class of processes and observables~\cite{Hamilton:2012rf,Hamilton:2013fea}. Other ideas have been floated for matching NNLO parton-level calculations with a parton shower~\cite{Lonnblad:2012ix} and one hopes that in the next few years the precision of event generators for collider physics can be improved to NNLO accuracy.

	For the first time, the Les Houches Workshop coincided with a Snowmass Workshop in the US. There was a great deal of synergy between the  two workshops with regards to QCD and EW calculations and measurements. This was especially true as the focus of the two workshops were  different, with Les Houches being concerned with physics of the present-day, or near-future, and Snowmass being concerned with the physics  possible in the future with higher-energy machines. It was very informative, for example to extrapolate theoretical predictions and corrections  from $8{-}14\UTeV$ to
the higher energies being considered at Snowmass. 

	The structure of the proceedings is as follows. Later in the introduction, we discuss the new Les Houches Wish List. Then  follows in \Sref{se:ew-dict} a `dictionary' for EW corrections. 
Chapter~\ref{cha:nnlo} has contributions related to NLO automation and NNLO techniques. 
In \refC{cha:pdf} are contributions related to PDF studies, including the $\Pg\Pg\rightarrow{}$Higgs study mentioned above, and in \refC{cha:pheno} are contributions related to phenomenological studies. Finally, 
in \refC{cha:mc} are contributions related to Monte Carlo tuning and output formats. 

\subsection{The Les Houches High Precision Wish List}
\newcommand{\sigtot}{\sigma_{\mathrm{tot}}}
\newcommand{\dsig}{\rd\sigma}
\newcommand{\Pj}{\rm j}
\newcommand{\PV}{\rm V}
\newcommand{\Pc}{\rm c}

	Below we discuss the revised High Precision Wish List. As mentioned above, the list is ambitious, but not out of the question for the advances that might be expected in the next 5~years, or over the span of Run~2 at the LHC.  Historically, reducing the theoretical uncertainty by a factor of two typically takes ${\cal O}(10{-}15)$ years and one might expect that over the remaining lifetime of the LHC, a further factor of two improvement is feasible. 

For convenience, we have divided the wish list into three parts;
\begin{enumerate}
\item Final states  involving the Higgs Boson,
\item Final states  involving Jets or Heavy Quarks,
\item Final states  involving Electroweak Gauge Bosons.
\end{enumerate}
In each case, Tables~\ref{table:H}, \ref{table:T}, and \ref{table:V} show the final state, 
what is currently known, and the desired accuracy. 
Throughout, $\PV=\PZ$ or $\PW^{\pm}$. We systematically use the shorthand $\dsig$  to indicate the fully differential cross section while the accuracy is specified after the ampersand according to the following notation;  
\begin{itemize}
\item LO $\equiv$ ${\cal O}(1)$, 
\item NLO QCD $\equiv$ ${\cal O}(\alphas)$, 
\item NNLO QCD $\equiv$ ${\cal O}(\alphas^2)$, 
\item NLO EW $\equiv$ ${\cal O}(\alpha)$,
\item NNNLO QCD $\equiv$ ${\cal O}(\alphas^3)$, 
\item NNLO QCD+EW $\equiv$ ${\cal O}(\alphas\alpha)$.
\end{itemize} 
This counting of orders ${\cal O}$ is done relative to LO QCD, which is ${\cal O}(1)$
by definition, independent of the absolute power of $\alphas$ in its cross section.
Note that it is simplified in two respects: Firstly, it suppresses 
the possibility (as in $\Pt\bar\Pt$ or dijet production) that there is no uniform
scaling in $\alphas$ and $\alpha$ at tree or at the lowest loop level. In this
case EW corrections involve orders ${\cal O}(\alpha/\alphas)$ relative to LO as well.
Secondly, the counting can depend on the observable, as e.g.\ in inclusive
Higgs-boson production, where the NLO correction to the total cross section comprises
only a LO prediction for the transverse-momentum distribution for the Higgs boson.

Parametrically, $\alpha \sim \alphas^2$ so that NNLO QCD and NLO EW effects are naively of a similar size (although there may well be regions of phase space where one dominates over the other).
In this notation, $\dsig$ @ NNLO QCD + NLO EW would indicate a single code computing the fully differential cross section including both ${\cal O}(\alphas^2)$ and ${\cal O}(\alpha)$ effects.  

The following lists of LHC processes are very comprehensive, rendering it impossible
to quote all relevant higher-order calculations to each process here. We, therefore,
mostly restrict the explicitly given  theoretical references to the relevant publications of the last
2--3~years, while references to previous work can be found in the recent papers.
Whenever possible, we point to specific reviews as well.
We did also not spell out the obvious desire to have all NLO predictions matched
to QCD parton showers, as obtained for instance in the MC@NLO~\cite{Frixione:2002ik} or 
Powheg~\cite{Frixione:2007vw} approaches. Similarly, the experimental references are not intended
to be exhaustive either, but are cited to support the current or expected experimental precisions. 

Note also that we did not compile a separate wish list for decay sub-processes, but indicate
the inclusion of particle decays whenever relevant in the context of production
processes. The description of Higgs-boson decays and their accuracy have been
discussed in \Brefs{Dittmaier:2011ti,Dittmaier:2012vm,Heinemeyer:2013tqa} in great detail.
At the level of some percent the approach of describing full processes in
the factorized approach (production cross section) $\times$ (decay branching ratio) 
is limited. Instead the full resonance process, with prossible interference
effects with background, should be considered.
In the context of Higgs-boson observables, this issue is discussed in some detail in
\Brefs{Dittmaier:2012vm,Heinemeyer:2013tqa} (see also references therein);
general considerations about this issue can also be found in \refS{se:unstable}.

\begin{table}
\begin{center}
\begin{tabular}{|l|l|l|}
\hline
Process & State of the Art & Desired  
\\
\hline
$\PH    $    & $\dsig$ @ NNLO QCD (expansion in $1/\Mt$) & $\dsig$  @ NNNLO QCD  (infinite-$\Mt$ limit)              \\
             & full $m_t/\Mb$ dependence @ NLO QCD           & full $\Mt/\Mb$ dependence @ NNLO QCD   \\
             &  and @ NLO EW &   and @ NNLO QCD+EW    \\
             & NNLO+PS, in the $m_t\rightarrow \infty$ limit  &   NNLO+PS with finite top quark mass effects     \\
\hline
$\PH+\Pj$    & $\dsig$ @ NNLO QCD ($\Pg$ only) & $\dsig$ @ NNLO QCD (infinite-$\Mt$ limit)          \\
             &  and finite-quark-mass effects    &   and finite-quark-mass effects    \\
             &  @ LO QCD and LO EW  &   @ NLO QCD and NLO EW \\
\hline
$\PH+2\Pj$   & $\sigtot$(VBF) @ NNLO(DIS) QCD & $\dsig$(VBF) @ NNLO QCD + NLO EW             \\
             & $\dsig$(VBF) @ NLO EW          &                      \\
             & $\dsig(\Pg\Pg)$ @ NLO QCD  (infinite-$\Mt$ limit)      & $\dsig(\Pg\Pg)$ @ NNLO QCD  (infinite-$\Mt$ limit)              \\
             &  and finite-quark-mass effects @ LO QCD   &   and finite-quark-mass effects   \\
             &                                       &   @ NLO QCD and NLO EW\\
\hline
$\PH+\PV$    & $\dsig$ @ NNLO QCD &   with $\PH \to \Pb\bar\Pb$ @ same accuracy             \\
             & $\dsig$ @ NLO EW   &   $\dsig(\Pg\Pg)$ @ NLO QCD \\
             & $\sigtot(\Pg\Pg)$ @ NLO QCD (infinite-$\Mt$ limit)   &  with full $\Mt/\Mb$ dependence \\
\hline
$\Pt\PH$ and    & $\dsig$(stable top) @ LO QCD &  $\dsig$(top decays)  \\
$\bar\Pt\PH$    & &@ NLO QCD and NLO EW  \\
\hline
$\Pt\bar\Pt\PH$ & $\dsig$(stable tops) @ NLO QCD & $\dsig$(top decays)  \\
                &                                & @ NLO QCD and NLO EW    \\
\hline
$\Pg\Pg\to\PH\PH$ & $\dsig$ @ NLO QCD (leading $\Mt$ dependence) & $\dsig$ @ NLO QCD  \\
         & $\dsig$ @ NNLO QCD (infinite-$\Mt$ limit) & with full $\Mt/\Mb$ dependence  \\
\hline
\end{tabular}
\end{center}
\label{table:H}
\caption{Wishlist part 1 -- Higgs ($\PV=\PW,\PZ$)}
\end{table}

\subsubsection{Final states involving the Higgs Boson}

	Now that the Higgs boson has been discovered, the next key step is the detailed measurement of its properties and couplings. Already much  has been accomplished during the 2011--2012 running at the LHC, but differential measurements, for example, are still in their infancy, due to the  lack of statistics. Given its importance, a great deal of theoretical attention has already been given to calculations of the Higgs-boson production sub-processes for each of the 
production modes~\cite{Dittmaier:2011ti,Dittmaier:2012vm,Heinemeyer:2013tqa} 
including a concise summary  
of the predictions available for each channel.%
\footnote{For more references, see also \Bref{Dittmaier:2012nh}.} 
Nevertheless, as indicated in Table~\ref{table:H}, more precise calculations are needed.

\begin{itemize}
\item[$\PH$:]
The current situation is well summarized in 
Refs.~\cite{Dittmaier:2011ti,Dittmaier:2012vm,Heinemeyer:2013tqa}: we know the production cross section for the $\Pg\Pg$ fusion subprocess to NNLO QCD in the infinite-$\Mt$ limit and including finite-quark-mass effects at NLO QCD and NLO EW. 
The current experimental uncertainties associated with probing the $\Pg\Pg\to\PH$
process cross section are of the order of 20--40\%, depending on the amount of 
model-dependent assumptions.  
Theoretically, the uncertainty is of the order of 15\%, with the uncertainties   due  to PDF+$\alphas$ and higher-order corrections, as estimated through scale variations, both being on the order of 7--8\%.
The accuracy of the experimental cross section is statistically limited, with the total error expected to decrease to the order of 10\% with  $300\Ufb^{-1}$ in Run 2, running at an energy close to $14\UTeV$~\cite{Dawson:2013bba}. Thus, the desire is to improve upon our knowledge of the higher-order corrections  to $\Pg\Pg$ fusion, in parallel with improvements to the PDF+$\alphas$ uncertainty. NNNLO QCD corrections to $\Pg\Pg$ fusion in the infinite-$\Mt$ limit are already underway~\cite{Anastasiou:2013srw,Anastasiou:2013mca,Anastasiou:2014vaa} and could reduce the residual scale dependence to below 5\%.   Ultimately, one may need to know the NNLO QCD and mixed NNLO QCD+EW contributions retaining finite top-quark mass effects.  Currently, experimental physicists are using parton-shower programs which include the cross sections for Higgs production to NLO. An NNLO+PS simulation for Higgs boson production, in the infinite top mass limit, has already been presented in ~\cite{Hamilton:2012rf,Hamilton:2013fea} (which could be extended to include finite top mass effects).  

\item[$\PH+\Pj$:]	The first attempts at differential Higgs+jets measurements were made in the diphoton channel with the $8\UTeV$ data in ATLAS~\cite{TheATLAScollaboration:2013eia}. 
At $14\UTeV$, with $300\Ufb^{-1}$, there will be a very rich program of differential jet measurements, with on the order of 3000 events per experiment (in the diphoton channel) with jets above the top mass scale, thus probing inside the top-quark loop in the $\Pg\Pg$ production process. Currently, the NNLO QCD Higgs+jet cross section for the gluons-only process in the infinite-$\Mt$ limit is known~\cite{Boughezal:2013uia}, and the full cross section is expected later this year. 
The (LO) one-loop QCD and EW contributions
retaining the dependence on the masses of the particles circulating in the loop are also known, but as for the inclusive case, the finite-mass contributions to the Higgs+jet cross section at NLO QCD and NLO EW may also be needed.

\item[$\PH+2\Pj$:]	The Higgs+2jet channel is crucial in order to understand Higgs couplings, and in particular to understand the coupling to vector bosons through the vector-boson fusion (VBF) channel. VBF production itself is known to NNLO QCD in the 
double-DIS approximation
together with QCD and EW effects at NLO,
while the  $\Pg\Pg$ fusion channel is available at NLO QCD in the 
infinite-$\Mt$ limit~\cite{Campbell:2010cz,vanDeurzen:2013rv}  
and to LO QCD retaining finite-mass effects.
With $300\Ufb^{-1}$, there is the possibility of measuring 
the HWW coupling strength to the order of 5\%  which could require both the VBF and gluon fusion Higgs+2jet cross sections to NNLO QCD and finite-mass effects to NLO QCD and NLO EW accuracy~\cite{Dawson:2013bba}.


\item[$\PH+\PV$:]	Two important, but poorly known,  couplings of the Higgs boson are to the top and bottom quarks. Currently the couplings are known to  the order of 50\% or more for the bottom quark  and 100\% for the top quark.  Higgs${}\rightarrow \Pb\bar{\Pb}$ is primarily measured through the Higgs + vector-boson final state. Currently, the production cross section is known at 
NNLO QCD and at NLO EW, 
recently generalized to differential cross section including 
leptonic Z-boson and W/Z-boson decays at NNLO QCD~\cite{Ferrera:2011bk} 
and NLO EW order~\cite{Denner:2011id}, respectively.
The $\Pg\Pg\to\PH\PZ$ channel, which is known~\cite{Altenkamp:2012sx} 
at NLO QCD in the infinite-$\Mt$ limit for the total cross section, 
contributes about 14\% to the cross section
at the LHC for $14\UTeV$ and significantly increases the scale uncertainty.
The scale uncertainty is about 1\%(4\%) for $\PH\PW(\PH\PZ)$
production, the PDF uncertainty is about 3.5\%. 
The $\Pb\bar{\Pb}$ decay is currently linked to the $\PH\PV$ production process
to NLO QCD in the narrow-width approximation (NWA)~\cite{Ferrera:2013yga}, but it is desirable to
combine Higgs production and decay processes to the same order, i.e.,
NNLO in QCD and NLO in EW for the Higgs-strahlung process.
The NNLO QCD corrections to the $\PH\to\Pb\bar\Pb$ decay, in fact, are known 
fully differentially~\cite{Anastasiou:2011qx}.
With $300\Ufb^{-1}$ at a $14\UTeV$ LHC, the signal strength for the Higgs${}\rightarrow \Pb\bar{\Pb}$ final state should be measured to the 10--15\% level, shrinking to 5\% for $3000\Ufb^{-1}$~\cite{Dawson:2013bba}. 

\item[$\Pt\PH$, $\bar\Pt\PH$, $\Pt\bar\Pt\PH$:] In the presence of CP violation, the Higgs--top coupling may have both scalar and pseudoscalar components, $\kappa_t$ and $\tilde{\kappa}_t$, which are weakly bounded by the present experimental constraints.  These couplings can be probed using measurements of $\PH$ production in association with $\Pt\bar{\Pt}$, single $\Pt$, or single $\bar{\Pt}$~\cite{Ellis:2013yxa}. With $300(3000)\Ufb^{-1}$, the top-quark Yukawa coupling should be measured to approximately 15\% (5--10\%)~\cite{Dawson:2013bba}.
The $\Pt(\bar{\Pt})[\Pt\bar{\Pt}]$Higgs cross section is currently known at LO (LO) [NLO] QCD, respectively, and with stable top quarks.  In all three cases, it is necessary to know the cross section (with top decays) at NLO QCD, possibly including NLO EW effects.  
	
\item[$\PH\PH$:] The self-coupling of the Higgs boson arises from the EW symmetry breaking of the Higgs potential and measuring the triple-Higgs-boson coupling then directly probes the EW potential.  
Double-Higgs production via gluon fusion, used to measure the triple-Higgs coupling,  
is known at LO QCD with full top mass dependence,
including the leading finite-mass effects at NLO QCD~\cite{Dawson:1998py,Grigo:2013rya} and at NNLO  QCD in the infinite-$\Mt$ limit~\cite{deFlorian:2013jea}. It may be necessary to compute the full top mass dependence at NLO QCD. The production cross section for double-Higgs production is small, and the backgrounds non-negligible. Nonetheless, it is hoped that a 50\% precision on the self-coupling parameter may be possible with $3000\Ufb^{-1}$ at $14\UTeV$~\cite{Dawson:2013bba}.
Other double-Higgs production processes, such as via gluon fusion or associated production
with W/Z bosons, are mostly known to NLO QCD (excluding final states with top quarks)
and were recently discussed in \Brefs{Baglio:2012np,Frederix:2014hta}. Owing to
the strong suppression of their cross sections, their observability at the LHC is
extremely challenging.

\end{itemize}

\subsubsection{Final states involving Jets or Heavy Quarks}

\begin{table}
\begin{center}
\begin{tabular}{|l|l|l|}
\hline
Process & State of the Art & Desired  
\\
\hline
$\Pt\bar\Pt$ & $\sigtot$(stable tops) @ NNLO QCD & $\dsig$(top decays)  \\
             & $\dsig$(top decays) @ NLO QCD   & @ NNLO QCD + NLO EW  \\
             & $\dsig$(stable tops) @ NLO EW   &                       \\
\hline
$\Pt\bar\Pt+\Pj(\Pj)$ & $\dsig$(NWA top decays) @ NLO QCD & $\dsig$(NWA top decays)    \\
                 &                                   & @ NNLO QCD + NLO EW         \\
\hline
$\Pt\bar\Pt+\PZ$ &  $\dsig$(stable tops) @ NLO QCD &  $\dsig$(top decays) @ NLO QCD      \\
                 &   & + NLO EW     \\
\hline
single-top        & $\dsig$(NWA top decays)  @ NLO QCD & $\dsig$(NWA top decays)   \\
                  &                                    & @ NNLO QCD + NLO EW     \\
\hline
dijet        & $\dsig$ @ NNLO QCD ($\Pg$ only) & $\dsig$  @ NNLO QCD + NLO EW            \\
             & $\dsig$ @ NLO EW (weak) &                \\
\hline
$3\Pj$       & $\dsig$ @ NLO QCD & $\dsig$ @ NNLO QCD + NLO EW              \\
 \hline
$\ga+\Pj$ & $\dsig$ @ NLO QCD & $\dsig$ @ NNLO QCD  + NLO EW\\
          & $\dsig$ @ NLO EW &   \\
\hline
\end{tabular}
\end{center}
\label{table:T}
\caption{Wishlist part 2 -- Jets and Heavy Quarks}
\end{table}

\begin{itemize}
\item[$\Pt\bar\Pt$:]  Precision top physics is important for a number of reasons. It is by far the most massive  quark, and it is possible that new physics might have a strong coupling to top quarks; hence the need for precision predictions.  For example, a forward--backward asymmetry has been observed at the Tevatron larger than predicted by NLO QCD+EW predictions.  
The larger than expected asymmetry may be the result of new physics, due to missing
higher-order corrections, or caused by unknown problems in the experimental analysis.

        At the LHC, the dominant production mechanism for top pair production is through $\Pg\Pg$ fusion, for basically all kinematic regions. Thus, a comparison of precise top-quark measurements with similar predictions can greatly help the determination of the gluon PDF, especially at high $x$ where the current uncertainty is large. The present experimental uncertainty on the total top-quark pair cross section is  on the order of 5\% for the dilepton final state, and should improve for the lepton + jets final state to be of the same order~\cite{ATLAS:2012aa,Chatrchyan:2012bra}.  Note that a sizeable portion of that uncertainty is due to the luminosity uncertainty, emphasizing the utility of using the top pair cross section as a handle on precision normalizations.  
Currently, the total cross section is known in 
NNLO QCD~\cite{Baernreuther:2012ws,Czakon:2012zr,Czakon:2012pz,Czakon:2013goa} and to NLO EW
for stable tops (see, e.g., \Bref{Kuhn:2013zoa} and older references therein), 
but only to NLO QCD with decays of on-shell~\cite{Melnikov:2009dn} or
off-shell~\cite{Denner:2010jp,Bevilacqua:2010qb,Denner:2012yc,Cascioli:2013wga}
top quarks.
The theoretical uncertainty for predictions for the total cross section is 4\%, while differential predictions have a larger uncertainty  (currently at the NLO level). It is  expected that fully differential NNLO QCD predictions will be known in the near future, perhaps even this year. This is very important, especially for a better understanding of $\Pt\bar{\Pt}$ asymmetry measurements at the Tevatron and LHC. Note that a non-negligible contribution to the experimental uncertainty for the top-quark cross section is the extrapolation from the fiducial phase space to the full phase space. To fully exploit the physics potential, we need to be able to compare experimental fiducial cross sections to theoretical fiducial cross sections, requiring a knowledge of top-quark pair production (with decays) including NNLO QCD and NLO EW effects.

\item[$\Pt\bar\Pt+\Pj(\Pj)$:] Due to the dominant $\Pg\Pg$ sub-process production mechanism, and the large phase space for gluon emission at the LHC, most $\Pt\bar{\Pt}$ final states also contain one or more jets. 
Extending previous work on stable top quarks~\cite{Dittmaier:2007wz,Dittmaier:2008uj,Melnikov:2010iu},
the NLO QCD corrections to $\Pt\bar{\Pt}+{}$jet production 
(supplemented by parton-shower effects in \Bref{Kardos:2011qa}) were calculated 
with on-shell top decays including spin correlation effects in \Bref{Melnikov:2011qx}.
In the future, a knowledge of $\Pt\bar{\Pt}$+jet production (with decays) including NNLO QCD and NLO EW effects may be necessary.
For $\Pt\bar{\Pt}+{}$2jet production NLO QCD corrections were first worked out for
on-shell top quarks~\cite{Bevilacqua:2010ve,Bevilacqua:2011aa} and were recently
generalized to include spin-correlated top-quark decays in LO as well as
a consistent merge with a QCD parton shower~\cite{Hoeche:2014qda}. 
At least the inclusion of QCD corrections to the top-quark decays is desirable also in this case.

\item[$\Pt\bar\Pt+\PW/\PZ$:] Top pair production in association with a vector boson is an important process for comparing with $\Pt\bar{\Pt}$Higgs production, but also for measuring the coupling of the top quark with the $\PW$ or $\PZ$. 
The NLO QCD corrections in the NWA with on-shell top decays including spin correlation effects are known~\cite{Kardos:2011na,Garzelli:2011is,Campbell:2012dh,Garzelli:2012bn}
(generalizing earlier work on $\Pt\bar\Pt\PZ$ with stable top quarks~\cite{Lazopoulos:2008de}). 
It may be necessary to study the effect of hard radiation in the top decay. 

Evidence for the production of a $\Pt\bar\Pt+\PZ$ final can be found in Ref.~\cite{Chatrchyan:2013qca}.

\item[single-top:] Measurements of single-top production are important for precision top physics, and specifically the measurement of $V_{\Pt\Pb}$. The current experimental cross section precision is on the order of 10\%, and a precision of >5\% during Run 2 is likely~\cite{Agashe:2013hma}.  The single-top processes $\Pt\PW$ and $\Pt\PZ$ are more problematic. Both ATLAS~\cite{Aad:2012xca} and CMS~\cite{Chatrchyan:2012zca} have results (but less than $5\sigma$ significance) from Run~1 ($7\UTeV$) for the first channel, while the measurement of the  second channel may be possible with the $8\UTeV$ data, but with some difficulty. The $\Pt\PW$ cross section has been measured with an approximately 40\% uncertainty, dominated by statistical errors, and by the incomplete knowledge of the backgrounds. These uncertainties should decrease significantly with the $8\UTeV$ data, and certainly will with the $14\UTeV$ data.
 
Currently the single-top cross section in the various channels (including decay) is known to 
NLO in QCD~\cite{Harris:2002md,Campbell:2004ch,Campbell:2013yla} and including dominant soft-gluon and threshold effects at NNLO QCD~\cite{Kidonakis:2006bu,Kidonakis:2011wy}, with a theoretical uncertainty of the order of less than 10\%~($\Pt\PW$) and less than 5\%~($\Pt\PZ$).  A theoretical uncertainty much less than the experimental precision is desired, leading to the need for a calculation of the single-top cross section to NNLO in QCD and including NLO EW effects.

\item[dijet:] Dijet production at the LHC probes the smallest distance scales measurable, and thus is an obvious place to search for the impact of new physics. The cross section has been known to  NLO QCD for over 20 years, and considerable effort to extend the calculation to NNLO QCD has been made and the partial (gluons only) calculation is available~\cite{Ridder:2013mf,Currie:2013dwa}  with the expectation of further progress in the near future. 
The  cross section is also known at NLO EW (weak, non-photonic corrections only)~\cite{Dittmaier:2012kx}, with the weak corrections being potentially very important at high transverse momentum.  The current state of the art for global PDF fitting is NNLO QCD, with the processes currently included in the fits all known at NNLO QCD, except for inclusive jet production. Thus, a complete calculation incorporating both NNLO QCD and NLO EW effects is vital. Currently, the  experimental uncertainty for the inclusive jet cross section is between 
5\% and 10\% (in the jet transverse momentum range roughly from $200\UGeV$ to $1\UTeV$), necessitating a theoretical uncertainty below that~\cite{Aad:2011fc,Aad:2013tea,Chatrchyan:2012bja}.

\item[$3\Pj$:] The 3-jet to 2-jet ratio is a useful process to measure the running of the strong coupling constant over a wide dynamic range, in  a way that removes much of the sensitivity to the gluon distribution. Because this measurement is the ratio of two related cross  sections, many of the experimental systematic errors cancel. For example, the jet energy scale uncertainty for the ratio can be reduced to less than or equal to 1\%~\cite{Chatrchyan:2013txa}.  The measurements are currently statistically limited for jet  transverse momenta below the TeV range, but this will not be the case for the high luminosity running at $14\UTeV$. The largest uncertainties are those from the residual scale dependence at NLO QCD~\cite{Nagy:2001fj,Nagy:2003tz}, typically of the order 5\% at high $p_{\mathrm{T}}$. For this reason, it is desirable to know the cross section for 3-jet production to NNLO QCD and NLO EW, as for dijet production.

\item[$\ga+\Pj$:]  
Photon + jet production is another process that will be extremely useful for the determination of the gluon distribution, especially at high parton momentum fraction $x$. 
Due to the relatively clean measurement of the photon 4-vector, an experimental precision better than with dijet production is possible. To put the two processes on an equal theoretical basis, the calculation of photon + jet to NNLO QCD + NLO EW is desirable.

\end{itemize}

\subsubsection{Final states involving EW Gauge Bosons}

\begin{table}
\begin{center}
\begin{tabular}{|l|l|l|}
\hline
Process & State of the Art & Desired  
\\
\hline
$\PV$        & $\dsig$(lept. $\PV$ decay) @ NNLO QCD & $\dsig$(lept. $\PV$ decay) @ NNNLO QCD      \\
             & $\dsig$(lept. $\PV$ decay) @ NLO EW   & and @ NNLO QCD+EW             \\
             &                                       & NNLO+PS   \\
\hline
$\PV+\Pj(\Pj)$    & $\dsig$(lept. $\PV$ decay) @ NLO QCD & $\dsig$(lept. $\PV$ decay)  \\
             & $\dsig$(lept. $\PV$ decay) @ NLO EW  & @ NNLO QCD + NLO EW  \\
\hline
$\PV\PV'$    & $\dsig$($\PV$ decays) @ NLO QCD & $\dsig$(decaying off-shell $\PV$)  \\
             & $\dsig$(on-shell $\PV$ decays) @ NLO EW  & @ NNLO QCD + NLO EW  \\
\hline
$\Pg\Pg \to \PV\PV$    & $\dsig$($\PV$ decays) @ LO QCD & $\dsig$($\PV$ decays) @ NLO QCD              \\
\hline
$\PV\ga$    & $\dsig$($\PV$ decay) @ NLO QCD & $\dsig$($\PV$ decay) \\
             & $\dsig$(PA, $\PV$ decay) @ NLO EW  & @ NNLO QCD + NLO EW    \\
\hline
$\PV\Pb\bar\Pb$ & $\dsig$(lept. $\PV$ decay) @ NLO QCD & $\dsig$(lept. $\PV$ decay) @ NNLO QCD   \\
                & massive $\Pb$                        & + NLO EW, massless $\Pb$                          \\
\hline
$\PV\PV'\ga$ & $\dsig$($\PV$ decays) @ NLO QCD & $\dsig$($\PV$ decays)  \\
                &                              &  @ NLO QCD + NLO EW      \\
\hline
$\PV\PV'\PV''$ & $\dsig$($\PV$ decays) @ NLO QCD & $\dsig$($\PV$ decays)  \\
                &                              &  @ NLO QCD + NLO EW         \\
\hline
$\PV\PV'+\Pj$ & $\dsig$($\PV$ decays) @ NLO QCD & $\dsig$($\PV$ decays)  \\
                &                              &  @ NLO QCD + NLO EW     \\
\hline
$\PV\PV'+\Pj\Pj$ & $\dsig$($\PV$ decays) @ NLO QCD & $\dsig$($\PV$ decays)  \\
                &                              &  @ NLO QCD + NLO EW        \\
\hline
$\ga\ga$     & $\dsig$ @ NNLO QCD + NLO EW & $q_T$ resummation at NNLL matched to NNLO \\
\hline
\end{tabular}
\end{center}
\label{table:V}
\caption{Wishlist part 3 -- Electroweak Gauge Bosons ($\PV=\PW,\PZ$)}
\end{table}

\begin{itemize}
\item[$\PV$:] Vector-boson production is one of the key benchmark processes at hadron colliders. Experimentally, the cross sections can be  measured with great precision (on the order of 1--2\%, excluding luminosity uncertainties~\cite{Aad:2011dm,Chatrchyan:2014mua}),  
while the current best theoretical calculations at NNLO QCD and NLO EW have small uncertainties
(see, e.g., \Bref{Boughezal:2013cwa} and references therein).
The resulting comparisons of data to theory thus serve both as precision tests of QCD and sensitive probes of PDFs. For example, measurements of $\PW$ and $\PZ$ boson  rapidity distributions have indicated that the strange quark distribution may be larger than presented in current PDFs~\cite{Aad:2012sb}.  To take full advantage of the experimental  precision, it is necessary to know the cross section to NNNLO QCD and NNLO QCD+EW, and to implement such a cross section in a NNLO+PS format. 

\item[$\PV+\Pj(\Pj)$:] Vector boson + 1,2 jets are useful both for PDF determination (for example, the $\PZ$+jet cross section for the gluon distribution and the $\PW$+charm cross section for the strange quark PDF) as well as a well-known system in which to study the  systematics of multiple jet production, with an accessible wide kinematic range (useful for example to understand the parallel  dynamics of Higgs+jet(s) systems). 
Currently, the V+jet and V+2jet cross sections are known to NLO QCD. The NLO EW corrections 
are known for V+jet, including V~decays and off-shell effects~\cite{Denner:2009gj,Denner:2011vu,Denner:2012ts}. For Z+2jet production, the NLO EW corrections are known for on-shell 
Z~bosons~\cite{Actis:2012qn} and are in progress for the off-shell case~\cite{Denner:2013fca}. 
The resulting differential uncertainties can reach 10--20\% for high jet momenta, exceeding the experimental uncertainties~\cite{Aad:2013ysa}.  
It is desirable to know both types of cross sections at NNLO QCD + NLO EW. 

\item[$\PV\PV'$:] 
With precision measurements of double-vector-boson production ($\PV\PV'$), one has a handle on the determination of triple gauge couplings, and a possible window onto new physics. Currently, the cross sections are known to NLO QCD (with V decays) and to NLO EW (with on-shell or at least resonant V's). WZ cross sections currently have a  (non-luminosity) experimental uncertainty on the order of 10\% or less, dominated by the statistical error~\cite{Aad:2012twa,Chatrchyan:2013yaa}. The current theoretical uncertainty is on the order of 6\%. Both the experimental statistical and systematic errors will improve with more data, necessitating the need for a calculation of $\PV\PV'$ to NNLO QCD + NLO EW (with V decays). 
Recently the well-known NLO QCD corrections have been complemented by the NLO EW corrections,
first for stable W and Z~bosons~\cite{Bierweiler:2012kw,Bierweiler:2013dja,Baglio:2013toa}, 
and in the WW case also including corrections to leptonic
W-boson decays~\cite{Billoni:2013aba}.
Moreover, the EW corrections to on-shell $\PV\PV'$ production have been implemented in the Herwig
Monte Carlo generator in an approximative way~\cite{Gieseke:2014gka}.

A thorough knowledge of the VV production cross section is needed, because of measurements of triple gauge couplings and since that final state forms a background for Higgs measurements in those channels. The non-luminosity errors for the $\PV\PV$ final state are of the order of 10\% or less, with the  theoretical uncertainties approximately half that~\cite{ATLAS:2012mec,Aad:2012awa,Aad:2012twa,Chatrchyan:2012sga,Chatrchyan:2013yaa,Chatrchyan:2013oev}. 

\item[$\Pg\Pg \to \PV\PV$:]

An important piece of the VV cross section is that resulting from a $\Pg\Pg$ initial state. Formally, the $\Pg\Pg$ production sub-process is suppressed by a  factor of $\alphas^2$ with respect to the dominant $q\bar{q}$ sub-process, but still contributes 5--10\% to the cross section for typical event-selection cuts due to the large gluon flux at the LHC. 
As background to Higgs-boson studies, it can even be enhanced 
to the level of some 10\% (see, e.g., discussions in \Brefs{Dittmaier:2011ti,Dittmaier:2012vm,Heinemeyer:2013tqa} and references therein).
Currently, this subprocess is known (with lepton decays) at LO QCD. It is desirable to know the cross section to NLO. 
An approximative approach has been suggested recently in \Bref{Bonvini:2013jha}.

\item[$\PV\ga$:]

V$\gamma$ measurements serve for precision tests of the EW sector and also as a probe of new physics appearing for 
example in anomalous triple-gauge-boson couplings or in production of new vector-meson resonances decaying into V$\gamma$.
Experimental uncertainties are currently of the order of 10\% and the theoretical errors are on the order of  5--10\%~\cite{Aad:2013izg,Chatrchyan:2013nda}. 
Currently, $\PW\gamma$ production is known (with decays) at NLO QCD,
$\PZ\gamma$ production even at NNLO QCD~\cite{Grazzini:2013bna}.
The NLO EW corrections are known in the pole approximation (resonant V~bosons with decays)
in either case.
It is desirable to fully know the cross sections to NNLO QCD + NLO EW.

\item[$\PV\Pb\bar\Pb$:]

Associated Higgs-boson production (VH) with the Higgs boson decaying into a $\Pb\bar{\Pb}$ final state is a key process to measure the coupling of the Higgs boson into b-quarks. There is a significant background to this final state from $\PV\Pb\bar{\Pb}$ production. The state of the art for  calculating $\PV\Pb\bar{\Pb}$ production is at NLO QCD (including b-quark mass effects). Currently, both the experimental and theoretical systematic errors are at the order of 20\%. The experimental errors will improve with more data. Thus, it is crucial to improve the theoretical accuracy by extending the QCD calculation to NNLO (for massless b quarks). 
This includes a proper understanding of uncertainties in the approaches based on four or five
active quark flavours.

\item[$\PV\PV'\PV''$:]

Cross sections for triple-gauge-boson production are currently known to NLO QCD, but
only the NLO EW corrections to $\PW\PW\PZ$ are known in the approximation of
stable W and Z~bosons~\cite{Nhung:2013jta}.
Triple-gauge-boson production processes
serve as channels for the determination of quartic gauge boson couplings and will 
allow for a better understanding of EW symmetry breaking. 
Currently, all of these final states are very statistically limited (no measurements have been published by ATLAS or CMS for example although analyses are currently underway for the $8\UTeV$ data), but precision measurements will be possible in Run 2. Thus, it is desired for the calculation of these final states to NLO QCD + NLO EW.

\item[$\PV\PV'+\Pj(\Pj)$:]

The reasoning above applies as well to the $\PV\PV'$+j(j) final states, whose
production cross sections are currently known to NLO QCD. 
$\PV\PV'$+j production is useful as a background to Higgs-boson production and for BSM searches, 
while $\PV\PV'$+jj production contains the EW vector-boson scattering sub-process that is
particularly sensitive to EW quartic gauge-boson couplings and to details of EW symmetry breaking.
EW corrections to these process classes are yet unknown, although they are certainly
as important as QCD corrections in the vector-boson scattering channels.

\item[$\ga\ga$] Diphoton production is a very clean channel in which to carry out precision tests of perturbative QCD, as well as one of the primary decay modes used  first for the discovery of the Higgs bosons and now for detailed measurements of the same. Although sideband subtraction is a very powerful tool in which to separate resonant Higgs boson production from the non-resonant diphoton background, it is still important to gain as much understanding of the QCD mechanisms for diphoton production as possible. The diphoton cross section has been recently calculated at NNLO~\cite{Catani:2011qz} and the  NLO EW corrections are also known~\cite{Bierweiler:2013dja}. Detailed measurements of the diphoton final state system have been carried out~\cite{Aad:2012tba,Chatrchyan:2011qt}, with a considerable improvement in agreement between data and theory resulting from the calculation at NNLO. The next theoretical improvement needed is a $q_T$ resummation at NNLL matched to the NNLO calculation. If Drell-Yan and Higgs production are known in fully differential form at NNNLO, then it should be possible to extend those calculations to include all colorless final states (including
$\ga\ga$) at NNNLO. 
\end{itemize}

\def\mathswitch#1{\relax\ifmmode#1\else$#1$\fi}
\def\mathswitchr#1{\relax\ifmmode{\mathrm{#1}}\else$\mathrm{#1}$\fi}
\def\citere#1{\mbox{Ref.~\cite{#1}}}
\def\citeres#1{\mbox{Refs.~\cite{#1}}}
\newcommand{\gsim}{\;\raisebox{-.3em}{$\stackrel{\displaystyle >}{\sim}$}\;}
\newcommand{\lsim}{\;\raisebox{-.3em}{$\stackrel{\displaystyle <}{\sim}$}\;}
\def\ga{\gamma}
\def\de{\delta}
\def\be{\beta}
\def\Ga{\Gamma}
\newcommand{\llog}{{\mathrm{LL}}}
\newcommand{\fact}{{\mathrm{fact}}}

%
%
%
%
%

\noindent
\section{Dictionary for electroweak corrections%
\footnote{S.~Dittmaier}}
\label{se:ew-dict}

In the coming years, the LHC explores particle phenomena at
the TeV scale at its full centre-of-mass energy $\sqrt{s}=13{-}14\TeV$
with an increasingly higher luminosity, so that plenty of particle
processes can be investigated at the accuracy level of several percent.
This perspective, and already several measurements at
lower energies in the previous LHC run, are calling for theoretical
predictions at the level of few percent, i.e.\ electroweak (EW) radiative corrections
have to be taken into account in predictions beyond the level of crude
approximations. Since QCD radiative corrections are substantial in 
predictions for hadronic collisions, their structure and general
features are much better known in the experimental community
than their EW counterparts, which are smaller in general.

The following survey of EW higher-order issues is meant as
orientation guide for non-experts in view of EW radiative corrections.
Particular attention is paid to those EW effects that can enhance
NLO EW corrections (expected to be of the small size $\alpha/\pi$) to a size
that can compete with or even exceed the size of QCD corrections
in specific situations (e.g.\ final-state radiation, EW corrections
at high energies). Moreover, some subtle aspects 
in the calculation of EW corrections are discussed that are
crucial for consistent predictions
(e.g.\ photon--jet separation, treatment of unstable particles).

\subsection{Input-parameter scheme -- choice of $\alpha$}

A convenient choice for the input parameters of the SM is:
the electromagnetic coupling $\alpha=e^2/(4\pi)$,
the strong coupling constant $\alphas=g_{\mathrm{s}}^2/(4\pi)$,
the weak gauge-boson masses $\MW$ and $\MZ$,
the Higgs-boson mass $\MH$,
the fermion masses $m_f$, and the CKM matrix $V$.

The masses can all be defined as {\it pole masses}, defined by the locations
of the particle poles in the respective propagators, but for the quarks
it is often useful to switch to a running mass at some appropriate scale.
Properly defined observables and their predictions should be insensitive to
the perturbatively problematic light-quark masses, as discussed below in more
detail.
Recall that the Yukawa couplings do not represent free parameters, but are
fixed by the fermion masses and the other EW input parameters. Disturbing
the relation between Yukawa couplings and fermion masses, in general, 
violates EW gauge invariance and can lead to inconsistent and wrong 
predictions, especially in the calculation of EW corrections.

The definition of the CKM matrix in the presence of EW corrections, in general,
is a non-trivial task. However, apart from applications in flavour physics,
for high-energy scattering the approximations of taking all quarks other than
bottom and top massless and ignoring mixing with the third generation are appropriate.
In this approximation, the CKM matrix can only become relevant in
charged-current quark--antiquark annihilation channels, such as Drell--Yan-like 
W-boson production, and leads to global factors $|V_{ij}|^2$ to 
partonic channels with $q_i\bar q_j$ or $q_j\bar q_i$ annihilation.
Note that this statement holds at the level of NLO EW corrections as well,
because of the mass degeneracy in the first two quark generations 
where the mixing takes place.

For the boson masses $\MW$, $\MZ$, and $\MH$, real on-shell masses
are usually employed. Details about different schemes are discussed 
in Sect.~\ref{se:unstable}
in the context of the description of instability effects.
Here, we only emphasize that the weak mixing angle $\theta_{\PW}$ is not an
independent input parameter. The most common choice in EW physics follows
the on-shell prescription which defines 
$\cw\equiv\cos\theta_{\PW}\equiv\MW/\MZ$ and $\sw=\sqrt{1-\cw^2}$
via the on-shell W/Z-boson masses.
Taking $\sw$ as independent parameter in addition to $\MW$ and $\MZ$, 
e.g.\ by setting it to some ad hoc value
or to the sine of the effective weak mixing angle measured at the Z~pole,
in general breaks gauge invariance, destroys gauge cancellations, and can lead
to totally wrong results.

For the electromagnetic coupling $\alpha$ basically the choice is between three
different values: the fine-structure constant $\alpha(0)\approx1/137$
(``$\alpha(0)$-scheme''),
the effective value $\alpha(\MZ)\approx1/129$
(``$\alpha(\MZ)$-scheme''), where $\alpha(0)$ is evolved via 
renormalization-group equations
from zero-mo\-men\-tum transfer to the Z~pole, and an effective value
derived from the Fermi constant $\GF$ leading to
$\alpha_{\GF}=\sqrt{2}\GF\MW^2(1-\MW^2/\MZ^2)/\pi\approx1/132$,
defining the so-called ``$\GF$-scheme''. The various choices for $\alpha$
differ by $2{-}6\%$ and represent an important part of
the so-called {\it input-parameter scheme}.
Which scheme is the most appropriate in practice, depends on the nature
of the process under consideration. Whatever scheme will be used,
it is crucial that a common coupling factor $\alpha^n$ is used in complete
gauge-invariant subsets of diagrams, otherwise important consistency
relations (compensation of divergences, unitarity cancellations, etc.)
will be destroyed. Note that this does not necessarily mean
that there is only one value of $\alpha$, but that all $\alpha$-factors
are global factors to gauge-invariant pieces of amplitudes.

In the following we describe which input-parameter scheme is appropriate
for a contribution to a cross section (or squared matrix element) whose
LO contribution is proportional to a fixed order $\alpha_{\mathrm{s}}^m\alpha^n$, i.e.\
NLO EW contributions scale like $\alpha_{\mathrm{s}}^m\alpha^{n+1}$. 
Corrections to LO contributions scaling with
different powers $m$ or $n$ belong to disjoint gauge-invariant contributions,
which can, thus, be treated independently.
The standard QED definition of $\alpha$ employs an on-shell renormalization
condition in the Thomson limit (photon momentum transfer $Q=0$), leading to the
renormalized value $\alpha=\alpha(0)$ of the $\alpha(0)$-scheme.
Our notation and conventions used below for field-theoretical quantities and
renormalization constants follow \citere{Denner:1991kt}.
The ``best'' scheme is the one that absorbs most of the universal 
corrections resulting from EW renormalization into the corresponding lowest-order 
prediction, thereby leading to smaller corrections. 
The most prominent EW renormalization effect is the 
the running of $\alpha$, i.e.\ the question whether an EW
coupling naturally takes place at low ($Q^2=0$) or high ($|Q^2|\gsim\MW^2$)
momentum transfer $Q^2$. 
Let us first consider a process without external photons in the initial or
final state.
In the $\alpha(0)$-scheme with the on-shell charge renormalization,
each of the $n$ EW couplings leads to a relative correction $2\delta Z_e$.
The charge renormalization constant $\delta Z_e$
contains mass-singular terms like $\alpha\ln m_f$ from each light
fermion $f$ which remain uncancelled in the EW corrections.
These terms are contained in the quantity
\begin{equation}
  \Delta\alpha(\MZ) =
  \Pi^{\ga\ga}_{f\ne \Pt}(0) - \Re\{\Pi^{\ga\ga}_{f\ne \Pt}(\MZ^2)\}
\approx
\frac{\alpha(0)}{3\pi}\sum_{f\ne\mathrm{t}} N^{\mathrm{c}}_f Q_f^2
\left[\ln\biggl(\frac{\MZ^2}{m_f^2}\biggr)-\frac{5}{3}\right],
\end{equation}
with $\Pi^{\ga\ga}_{f\ne \Pt}$ denoting the photonic vacuum
polarization induced by all fermions $f$ (with charge $Q_f$)
other than the top quark (see
also \citere{Denner:1991kt}), and $N^{\mathrm{c}}_l=1$ and $N^{\mathrm{c}}_q=3$
are the colour factors for leptons and quarks, respectively.
The correction $\Delta\alpha(\MZ)\approx6\%$ quantifies the running of $\alpha$
from $Q^2=0$ to the high scale $Q^2=\MZ^2$ induced by vacuum polarization effects
of the light fermions,
\begin{equation}
\alpha(\MZ)=\alpha(0)/[1-\Delta\alpha(\MZ)],
\end{equation}
a quantity that is non-perturbative, signalled by its sensitivity to the
light-quark masses.
The numerical value for $\alpha(\MZ)$ is obtained from an experimental
analysis of
$\Pep\Pem$ annihilation into hadrons at low energies below the $\PZ$-boson 
resonance, combined with theoretical arguments using dispersion 
relations~\cite{Eidelman:1995ny}.
Eliminating $\alpha(0)$ in favour of $\alpha(\MZ)$ in the LO prediction,
i.e.\ going over to the $\alpha(\MZ)$-scheme, 
effectively subtracts the $\Delta\alpha(\MZ)$ terms from the EW corrections
and, thus, cancels all light-fermion logarithms resulting
from charge renormalization in the EW correction. This cancellation
happens at any loop order in $\alpha$, i.e.\ employing the $\alpha(\MZ)$-scheme
resums the dominant effects from the running of $\alpha$ and, 
at the same time, removes the (perturbatively unpleasant) light-quark
masses, which should have been taken from a fit to $\Pi^{\ga\ga}_{f\ne \Pt}(Q^2)$
otherwise.
For high-energy processes without external photons the $\alpha(\MZ)$-scheme
is appropriate from this point of view.

For processes with $l$ external photons ($l\le n$), however,
the relative EW correction comprises $l$ times the photonic wave-function
renormalization constant $\de Z_{AA}$, which exactly cancels the light-fermion
mass logarithms appearing in $2\de Z_e$. This statement expresses the fact that
external, i.e.\ real photons effectively couple with the scale $Q^2=0$.
Consequently, the coupling
factor $\alpha^n$ in the LO cross section should be
parametrized as $\alpha(0)^l\alpha(\MZ)^{n-l}$, and the corresponding
$\Delta\alpha(\MZ)$ terms should be subtracted 
only $(n-l)$ times from the EW correction, 
in order to absorb the
large effects from $\Delta\alpha(\MZ)$ into the LO prediction. 
For $l=n$, this scaling corresponds to the pure $\alpha(0)$-scheme,
but for $l<n$ to a mixed scheme. 

The $\GF$-scheme, finally, offers the possibility to absorb some significant
universal corrections connected with the renormalization of the
weak mixing angle into LO contributions. At NLO, the $\GF$- and $\alpha(0)$-schemes
are related according to
\begin{equation}
\alpha_{\GF}\equiv\frac{\sqrt{2}\GF\MW^2(\MZ^2-\MW^2)}{\pi\MZ^2}
=\alpha(0)\left(1+\Delta r^{(1)}\right) + {\cal O}(\alpha^3),
\label{eq:dr}
\end{equation}
where $\Delta r^{(1)}$ is the NLO EW correction to muon
decay~\cite{Sirlin:1980nh,Denner:1991kt}.
The quantity $\Delta r^{(1)}$ can be decomposed according to 
$\Delta r^{(1)}=\Delta\alpha(\MZ)-\Delta\rho^{(1)}\cw^2/\sw^2
+\Delta r_{\mathrm{rem}}$
with $\Delta\alpha(\MZ)$ from above, the universal 
(top-mass-enhanced) correction
$\Delta\rho^{(1)}=3G_\mu\Mt^2/(8\sqrt{2}\pi^2)$ 
to the $\rho$-parameter, and a very small remainder $\Delta r_{\mathrm{rem}}$.
In view of the running of $\alpha$, the $\GF$-scheme 
corresponds to an $\alpha$ at the EW scale similar to the 
$\alpha(\MZ)$-scheme, since $\Delta r^{(1)}$ contains exactly one unit
of $\Delta\alpha(\MZ)$. The $\GF$-scheme is, thus, similar to
the $\alpha(\MZ)$-scheme as far as the running of $\alpha$ is concerned,
i.e.\ it is better than the $\alpha(0)$-scheme except for the case
of external photons.
The choice between the $\GF$- or $\alpha(\MZ)$-scheme is driven by the
appearance of $\sw$ in the EW couplings. Whenever $\sw$ (or $\cw$)
is involved in an EW coupling, the corresponding EW correction 
receives a contribution from $\Delta\rho^{(1)}$ according to
$\sw^2\to\sw^2+\Delta\rho^{(1)}\cw^2$.
Using (\ref{eq:dr}), it is easy to see that the combination
$\alpha/\sw^2$, which corresponds to the SU(2) gauge coupling, 
does not receive this universal correction, since the $\Delta\rho^{(1)}$
terms from $\Delta r^{(1)}$ and the correction associated with
$\sw^2$ cancel. In other words, employing the $\GF$-scheme for 
SU(2) gauge couplings absorbs the leading correction to the $\rho$-parameter
into the LO coupling. This statement also holds at the two-loop 
level~\cite{Consoli:1989fg}.
The $\GF$-scheme is, thus, most appropriate
for describing couplings of W~bosons. For Z~bosons this scheme
absorbs at least part of the $\Delta\rho^{(1)}$ corrections because of
additional $\cw$ factors in the coupling from the weak mixing,
while the scheme is actually not appropriate for photonic couplings.
However, also here it should be kept in mind that a fixed scheme
with a global definition of couplings has to be employed within 
gauge-invariant subsets of diagrams. 
In most cases, it is advisable to use the $\GF$-scheme for couplings that
involve weak bosons, although the gauge-invariant subsets of diagrams,
in general, also contain internal photons.

The application of the various input-parameter schemes to the
neutral- and charged-current Drell--Yan processes, including also
leading EW effects beyond NLO, is discussed in
\citeres{Dittmaier:2001ay,Brensing:2007qm,Dittmaier:2009cr}.

\subsection{Electroweak corrections at high energies}

At high energies, where scattering processes involve large
scales $Q^2\gg\MW^2$, EW corrections develop large
logarithmic contributions such as 
$(\alpha/\sw^2)\ln^2(Q^2/\MW^2)$ and $(\alpha/\sw^2)\ln(Q^2/\MW^2)$ 
at NLO, and powers of those beyond NLO.
These mass-singular corrections originate from soft and/or
collinear exchange of EW gauge bosons in loop diagrams.
The corresponding logarithms grow to tens of percent in the TeV range,
i.e.\ EW corrections become very significant at high energies.

The kinematic regime in which such EW corrections are most pronounced
is characterized by the situation that all invariants
$s_{ij}=2k_i\cdot k_j$ for pairs of particles' four-momenta $k_{i,j}$
become large ($s_{ij}\gg\MW^2$) and is known as {\it Sudakov regime}.
The structure of EW corrections in this domain has
been investigated in detail 
at ${\cal O}(\alpha)$ and beyond by several groups 
(see e.g.\ \citeres{Fadin:1999bq,Kuhn:1999nn,Ciafaloni:2000df, Hori:2000tm,%
Denner:2001gw,Melles:2001dh, Beenakker:2001kf, Denner:2003wi, Jantzen:2005xi,Jantzen:2005az,%
Denner:2006jr} and references therein).
As described for example in \citeres{Denner:2001gw,Denner:2003wi,Denner:2006jr}, the
leading EW logarithmic corrections, which are enhanced by
large factors $L=\ln(s_{ij}/\MW^2)$, can be divided into an
SU(2)$\times$U(1)-symmetric part, an electromagnetic part, and a
subleading part induced by the mass difference between $\PW$ and
\PZ~bosons. The last part does not contribute to corrections
$\propto(\alpha L^2)^n$.  
The leading (Sudakov) logarithms $\propto(\alpha L^2)^n$
of electromagnetic origin cancel between
virtual and real (soft) bremsstrahlung corrections, so that
the only source of leading logarithms is,
thus, the symmetric EW 
part, which can be characterized by comprising \PW~bosons, \PZ~bosons,
and photons of a common mass $\MW$.
These leading EW Sudakov corrections can be obtained to all orders
from the respective NLO result via exponentiation. 
For the subleading EW high-energy logarithms corresponding
resummations are not fully proven, but the corrections are expected
to obey so-called infrared evolution equations, a statement that is
backed by explicit two-loop calculations.

The detailed knowledge of the tower of EW high-energy
logarithms is important for a deeper understanding of
EW dynamics, but making use of it in predictions is a subject
that deserves care:
\begin{itemize}
\item
It is certainly advisable to make use of full NLO EW corrections,
i.e.\ without applying expansions for high energies, whenever
possible for a given process. 
Non-logarithmic corrections typically amount to some percent, 
depending on the process under consideration.
A safe assessment of the
quality of logarithmic approximations usually requires the comparison
to full NLO results. 
\item
Beyond NLO, the knowledge of higher-order EW logarithms can be very useful
and exploited to improve pure NLO predictions.
However, particular care has to be taken if the EW logarithms
show large cancellations between leading and subleading terms, 
as for instance observed in the case of neutral-current fermion--antifermion
scattering~\cite{Jantzen:2005xi}.
If the full tower of logarithms of a fixed perturbative order 
is not known, it is not clear to which accuracy truncated towers
approximate the full correction. However, the known part of the tower can at least
deliver estimates for the size of missing corrections and be used in
the assessment of theoretical uncertainties.
\item
While the analytical structure of EW corrections was studied in the literature
at very detail for the Sudakov regime, there is only little knowledge 
on EW corrections beyond NLO in more general kinematical situations where 
not all invariants $s_{ij}$ are large. Note that there are many 
cross sections that are in fact not dominated by the Sudakov regime
in the high-energy limit, including all processes that are
dominated by $t$-channel diagrams. For example, 
unless specifically designed cuts are applied, reactions like
W-boson pair production via $\Pep\Pem$, $\Pp\Pp$, or $\gamma\gamma$ collisions
are dominated by the {\it Regge limit}, where the Mandelstam variable
$t$ remains small while $s$ gets large.
Moreover, it often depends on the specific observable which regime
is probed in high-energy tails of kinematical distributions.
Taking Drell--Yan processes (see, e.g., 
\citeres{Dittmaier:2001ay,Brensing:2007qm,Dittmaier:2009cr})
and dijet production~\cite{Dittmaier:2012kx}
at the LHC as examples, differential distributions in the transverse momenta
of the produced leptons or jets probe the Sudakov regime in the high-momentum tails.
On the other hand, the invariant-lepton- or jet-mass distributions of these processes
are not dominated by this regime at high scales, so that the EW 
high-energy logarithms derived in the Sudakov regime do not approximate the EW
corrections well in those observables.
\item
Since the EW high-energy logarithmic corrections
are associated with virtual soft and/or collinear weak-boson or photon
exchange, they all have counterparts in real weak-boson or photon-emission 
processes which can partially cancel  the large negative virtual corrections
(but not completely, see \citere{Ciafaloni:2000df}). 
This cancellation will not be complete, since SU(2) doublets are in general
not treated inclusively in EW corrections---a fact that is by some abuse
of language called {\it Bloch--Nordsieck violation}.%
\footnote{The Bloch--Nordsieck theorem simply does not apply to
non-Abelian gauge theories.}
To which extent the cancellation occurs depends on the experimental
capabilities to separate final states with or without weak bosons or
photons. 
Logarithmic approximations, as recently implemented in the {\sc Pythia}
shower~\cite{Christiansen:2014kba} can deliver first estimates, but 
solid predictions have to be based on complete matrix elements.
The general issue and specific examples have been discussed for example in
\citeres{Ciafaloni:2006qu,Baur:2006sn,Bell:2010gi,Chiesa:2013yma}. 
For instance, the numerical analysis~\cite{Baur:2006sn}
of neutral-current Drell--Yan production
demonstrates the effect of real
weak-boson emission in the distributions in the transverse lepton
momentum $p_{\mathrm{T},\Pl}$ and in the invariant mass
$M_{ll}$ of the lepton pair.
At the LHC, at $M_{ll}=2\TeV$
the EW corrections are reduced from about $-11\%$ to $-8\%$
by weak-boson emission. At $p_{\mathrm{T},\Pl}=1\TeV$
the corresponding reduction from about $-10\%$ to $-3\%$ is somewhat larger.
\end{itemize}

\subsection{Photonic final-state radiation off leptons}

The emission of photons collinear to incoming or
outgoing charged leptons $l$ leads
to corrections that are enhanced by large logarithms of the form
$\alpha^n\ln^n(\Ml^2/Q^2)$ with $Q$ denoting a characteristic scale of the
process. 
For lepton colliders, such as a high-energy $\Pep\Pem$ collider, 
this collinear initial-state radiation (ISR) leads to pronounced photonic
corrections, which are particularly large whenever the underlying
total or differential cross section shows strong variations
(e.g.\ near resonances or thresholds).
In the following we focus on final-state radiation (FSR) off
leptons, the situation that is also relevant at hadron colliders. 
Both for ISR and FSR, the logarithmically enhanced corrections
are universal in the sense that they do not change the nature of the cross
section of the underlying hard scattering process, but depend only
on the type and kinematics of the incoming or outgoing charged
particles. 
More precisely, ISR distorts cross sections in terms of some convolution
over the radiative energy loss in the initial state of the hard scattering,
while FSR only influences the kinematics and
acceptance of the outgoing particles.
The first-order logarithm $\alpha\ln(\Ml^2/Q^2)$ is, of course,
contained in a full (process-dependent)
NLO EW ${\cal O}(\alpha)$ correction, and likewise for
higher orders, so that $Q$ is unambiguously fixed in any completely calculated
order. In a fixed perturbative order that is not completely taken into account,
but where ISR or FSR is included in logarithmic
accuracy, the ambiguity in the scale $Q$ is part of the remaining 
theoretical uncertainty.

The universal logarithmic corrections can be evaluated 
in the so-called {\it structure-function approach}
(see \citere{Beenakker:1996kt,Arbuzov:1999cq} and references therein), 
where these logarithms are derived from
the universal factorization of the related mass singularity,
or by dedicated photonic parton showers
(see e.g.\ \citeres{Placzek:2003zg,CarloniCalame:2003ux,Golonka:2006tw}).
For FSR in a process in which a lepton $l$ with momentum $k_l$ is
produced,
the incorporation of the mass-singular logarithms takes the form of a
convolution integral over the LO cross section
$\sigma^{\mathrm{LO}}$,
\begin{equation}
  \sigma_{\llog\FSR} =
  \int \rd\sigma^{\mathrm{LO}}(k_l)
  \int^1_0 \rd z \, \Gamma_{\Pl\Pl}^{\llog}(z,Q^2) \,
  \Theta_{\cut}(z k_l) ,
\label{eq:llfsr}
\end{equation}
where $\Gamma_{\Pl\Pl}^{\llog}$ is the structure function 
describing the radiation in logarithmic accuracy.
The variable $(1-z)$ is the momentum fraction of the respective
lepton lost by collinear (single or multiple) photon emission, and
the step function $\Theta_{\cut}$ is equal to 1 if the event
passes the cuts on the rescaled lepton momentum $z k_l$ and 0
otherwise. 
The structure function is known to ${\cal O}(\alpha^5)$
in the literature, including the resummation of soft-photon effects.
At NLO, i.e.\ in ${\cal O}(\alpha)$, the structure function has the well-known
form
\newcommand{\betal}{\be_{\Pl}}
\begin{eqnarray}
\label{eq:LLll}
  \Gamma_{\Pl\Pl}^{\llog,1}(z,Q^2) &=&
  \frac{\betal}{4} \left(\frac{1+z^2}{1-z}\right)_+
\end{eqnarray}
with the variable
\begin{equation}
\betal = \frac{2\alpha(0)Q_l^2}{\pi}
\left[\ln\biggl(\frac{Q^2}{\Ml^2}\biggr)-1\right],
\end{equation}
quantifying the large logarithm, with $Q_l$ denoting the electric charge of
the lepton $l$.
Here we also indicated that $\alpha(0)$ is the appropriate 
electromagnetic coupling constant to describe these photon radiation effects.
Note that in contrast
to the parton-shower approaches to photon radiation,
the structure-function approach neglects the photon momenta transverse
to the lepton momentum. 

For FSR the {\it Kinoshita--Lee--Nauenberg} (KLN) theorem
\cite{Kinoshita:1962ur,Lee:1964is} guarantees that these logarithms cancel if
collinear lepton--photon systems are treated fully inclusively,
like in a total cross section, defined without any phase-space cuts.
Such observables are called {\it collinear safe}.
In the presence of phase-space cuts and
in differential cross sections, in general, 
mass-singular contributions survive, leading to enhanced radiation effects,
since the necessary inclusiveness for their compensation is disturbed.
At NLO, these features can be easily understood from the explicit
analytic form shown in Eq.~(\ref{eq:LLll}), where the integral
over the plus distribution vanishes when taken over the full $z$ range
(i.e.\ $\Theta_{\cut}(z k_l)=1$ for all $z$).
For differential observables the level of inclusiveness
necessary for collinear safety
can be restored by a procedure known as {\it photon recombination},
which treats collinear lepton--photon systems as
one quasi-particle. This procedure is similar to the application
of a jet algorithm in QCD.
For final-state electrons,
photon recombination automatically is involved in their
reconstruction from electromagnetic showers detected in calorimeters.
Muons, on the other hand, can be observed as {\it bare} leptons 
from their tracks in the muon chambers, but in order to reduce
large FSR corrections, observed muons are sometimes also reconstructed
as {\it dressed} leptons via photon recombination, as e.g.\
described in \citere{Aad:2011gj} for an ATLAS analysis.
Working with dressed leptons, where mass-singular FSR effects cancel,
has the advantage that the resulting cross section does not depend 
on the mass (and thus on the flavour) of the charged lepton,
i.e.\ the reconstructed lepton looks universal 
(at least electrons and muons).

We conclude this part by an appeal to experimentalists.
Obviously data should be stored for future use in such a way 
that measured total and differential cross sections
can be unambiguously confronted with new calculations in the future.
This requires that there should be at least one set of results without
any unfolding or presubtraction of FSR effects, because unfolded
results depend on details (including version dependences) 
and intrinsic uncertainties
of the tool used in the unfolding. Data should be free of such
conventions (which most often are not sufficiently well documented)
and limitations.
Making use of some standard definition of dressed leptons
(cf.~\citere{Aad:2011gj}), could be a step towards the right direction, since
photon recombination reduces unpleasantly large FSR corrections.

\subsection{QED-corrected parton distribution functions}

The inclusion of NLO EW ${\cal O}(\alpha)$ corrections to hadronic cross
sections conceptually proceeds along the same lines as the incorporation of
NLO QCD corrections, with the slight generalization that the photon
appears as parton in the hadron as well.
The ${\cal O}(\alpha)$-corrected parton cross sections
contain mass singularities 
which are due to collinear photon radiation off the
initial-state quarks or due to a collinear splitting
$\gamma\to q\bar q$ for initial-state photons.
If small quark masses are used as regulators for collinear divergences,
these singularities appear as terms of the form $\alpha\ln(m_q)$,
in dimensional regularization with strictly massless quarks they
appear as poles $\propto\alpha/\epsilon$ in $D=4-2\epsilon$ dimensions.
In complete analogy to factorization in NLO
QCD calculations, these collinear singularities are absorbed into the
quark and photon distributions.
The explicit form of the redefinition of PDFs can, e.g., be found in
\citere{Dittmaier:2009cr}.
The {\it factorization scheme} for the QED part of
the redefinition is determined by the ${\cal O}(\alpha)$ part
of the so-called {\it coefficient functions}, and the 
$\MSbar$ and DIS-like schemes are defined in complete analogy to QCD.
The $\MSbar$ scheme is motivated by formal simplicity, because it
merely rearranges the infrared-divergent terms (plus some trivial constants)
as defined in dimensional regularization.
The DIS-like scheme is defined in such a way that the DIS structure
function $F_2$ does not receive any corrections; in other words,
the radiative corrections to electron--proton DIS are implicitly
contained in the PDFs in this case.
Whatever scheme has been adopted in the extraction of PDFs from
experimental data, the same scheme has to be used when predictions
for other experiments are made using these PDFs.

In complete analogy to the pure QCD case, a factorization scale
$\mu_{\fact,\QED}$ rules up to which scale collinear QED splitting processes
in the initial state are considered to be part of the colliding hadron.
The absorption of the collinear singularities of ${\cal O}(\alpha)$
into PDFs requires the
inclusion of the corresponding ${\cal O}(\alpha)$ corrections into the
Dok\-shitzer--Gri\-bov--Lipa\-tov--Alta\-relli--Pa\-risi
(DGLAP) evolution of these distributions, which describe the
dependence of the PDFs on the factorization scales.
Note, however, that the two factorization scales
$\mu_{\fact,\QCD}$ and $\mu_{\fact,\QED}$ have to be identified, 
$\mu_{\fact}=\mu_{\fact,\QCD}=\mu_{\fact,\QED}$,
if the 
DGLAP equations are to keep their usual form as integro-differential
equation in one scale, which is the common practice.

The first set of parton distributions including QED corrections
was MRST2004QED~\cite{Martin:2004dh}, which takes into account
${\cal O}(\alpha)$ corrections to the DGLAP evolution, but
uses a theoretical model as ansatz for the photon PDF at some starting scale.
In 2013 the new set NNPDF2.3QED~\cite{Ball:2013hta} of QED-corrected PDFs
became available, which also provides error estimates for the PDFs and,
for the first time, employed data to fit the photon distribution function. 
For typical scales relevant for LHC processes, the NNPDF photon density
confirms the older MRST result for $x$-values $\gsim0.03$, but is somewhat
smaller for lower values of $x$.
Generically the impact of QED corrections on quark PDFs is below the
percent level for $x\lsim0.4$.
There are some slight shortcomings in those sets of QED-corrected
PDF sets concerning the issue of the QED factorization scheme.
The QED evolution is treated in LO only, and EW corrections are
ignored in the fit of the PDFs to data. This means that there
is some mismatch in the use of the PDFs at the level of 
non-enhanced ${\cal O}(\alpha)$ terms, which should be quite small,
and that a very conservative error estimate should treat the
difference between results obtained with various QED factorization schemes
as theoretical uncertainty.
However, as discussed in \citeres{Diener:2005me,Ball:2013hta} the
neglect of EW corrections in the PDF fit to data favours the central
value of the DIS scheme for EW corrections.
Thus, the recommendation for the NNPDF2.3QED PDF set is to use
a mixed scheme which follows the $\MSbar$ scheme for QCD and the DIS 
scheme for EW NLO corrections.

\subsection{Photon-induced processes}

The inclusion of the photon in the set of partons inside hadrons
leads to so-called photon-induced processes, i.e.\ partonic channels
with photons in the initial state, in addition to the partonic
channels of QCD. At NLO EW level, contributions from photon-induced processes
always result as crossed counterparts of photonic bremsstrahlung corrections.
For instance, quark-initiated $qq$, $q\bar q$, $\bar q\bar q$
channels always receive (real) ${\cal O}(\alpha)$ corrections from
$q\gamma$ and/or $\bar q\gamma$ scattering, where the additional $q$ or
$\bar q$ in the final state leads to an additional jet with respect to the
LO signature, similar to real NLO QCD corrections.
For specific final states with charged particles, but without net electric
charge, there is also a contribution from $\gamma\gamma$ scattering with
LO kinematics and without additional partons in the final state.
This is, for instance, the case for $\mu^+\mu^-$ or
$\PW^+\PW^-$ production.

For processes, where the diagrams of the $q\gamma$-induced corrections
have direct counterparts in the $q\Pg$ channels of the NLO QCD corrections,
the $q\gamma$ channels typically contribute at the percent level to
the hadronic cross section. This suppression originates from the 
relative coupling factor $\alpha/\alpha_{\mathrm{s}}$ and from the fact
that the photon and gluon PDFs are similar
in shape, but the gluon PDF is larger by about two orders of magnitude
for scales and $x$ values of typical LHC processes.
A similar statement holds for $\gamma\gamma$ channels whenever similar
$\Pg\gamma$ or $\Pg\Pg$ channels exist.
More significant contributions from photon-induced processes can only arise 
if the photonic channels involve diagrams without QCD counterparts,
i.e.\ diagrams where the photon couples to colour-neutral charged
particles like muons or W~bosons.
Recently discussed examples for enhanced photon-induced 
contributions are the channels
$\gamma\gamma\to\Pl^+\Pl^-$~\cite{CarloniCalame:2007cd,Dittmaier:2009cr,Boughezal:2013cwa} and
$\gamma\gamma\to\PW^+\PW^-$~\cite{Bierweiler:2012kw,Bierweiler:2013dja,Baglio:2013toa,Billoni:2013aba},
where the $\gamma\gamma$ channels comprise more than $10\%$
in certain regions of phase space.

\subsection{Photon--jet separation}

Including QCD and EW corrections in predictions for the production
of a specific final state $F$ in association with a hard jet or a hard photon
necessarily raises the issue of a proper separation of jets and photons,
since the real corrections to $F+\mathrm{jet}$ and $F+\gamma$ production
both involve final states with $F+\mathrm{jet}+\gamma$ final states.
The process of $F+\mathrm{jet}+\gamma$ production contributes to the NLO EW corrections
to $F+\mathrm{jet}$ production and to the NLO QCD corrections to 
$F+\gamma$ production at the same time.
A consistent isolation of $F+\mathrm{jet}$ from $F+\gamma$ signatures at
full (QCD+EW) NLO accuracy requires a photon--jet separation
that guarantees a proper cancellation of all infrared (soft and collinear)
singularities on the theory side and is experimentally feasible at the
particle level.

Since calorimetric information is decisive for the reconstruction of
jets and photons, it is natural to take the electromagnetic energy
fraction inside some hadronic/electromagnetic shower as criterion to
decide whether it is called a jet or a photon.
Note that this criterion, however, necessarily leads to an incomplete
cancellation of collinear singularities for photon emission
off (anti)quarks. This can be explicitly seen by inspecting the integral
(\ref{eq:llfsr}) with the structure function (\ref{eq:LLll}) 
interpreting $\Pl$ as the radiating (anti)quark. The selection criterion
cuts into the $z$-integration and destroys the cancellation of the
collinear singularity, which would require the integration to be taken over the
full $z$-range.
Two different methods have been suggested in the literature to cope
with this situation: 
one method introduces a phenomenological {\it quark-to-photon fragmentation
function} that absorbs the collinear singularity in the same way as the
PDFs absorb collinear singularities from the initial state.
Another method, known as {\it Frixione isolation},
shifts the complete collinear singularity 
to the jet side upon designing a specially adapted definition of 
the allowed hadronic energy fraction in a photonic shower:
\begin{itemize}
\item
{\it Quark-to-photon fragmentation function} \\
The quark-to-photon fragmentation
function $D_{q\rightarrow\gamma}(z_\gamma)$ describes the probability
of a quark fragmenting into a jet containing a photon carrying 
the fraction $z_\gamma$ of the total jet energy. 
Since fragmentation is a long-distance process, it cannot be
calculated entirely in perturbation theory, but
receives two types of contributions:
(a) the perturbatively calculable radiation of a photon off a quark, 
which contains a collinear divergence;
(b) the non-perturbative production of
a photon during the hadronization of the quark,
which is described by a bare non-perturbative fragmentation function
$D^{{\rm bare}}_{q\rightarrow\gamma}(z_\gamma)$.
The latter contributes to the photon-emission cross section as
\begin{equation}
\rd \sigma^{{\rm frag}}(z_{\mathrm{cut}}) =
\rd \sigma^{\LO}
\int_{z_{\mathrm{cut}}}^1 \rd z_\gamma
D^{{\rm bare}}_{q\rightarrow\gamma}(z_\gamma),
\end{equation}
where $z_{\mathrm{cut}}$ is the smallest photon energy fraction
required in the collinear photon--quark system to be identified as a photon.
The perturbative and non-perturbative (bare) contributions to the fragmentation function
are sensitive to the infrared dynamics inside the quark jet and
can a priori not be separated from each other in a unique way.
Since the infrared singularity present in the perturbative contribution
must be balanced by a divergent piece in the bare fragmentation
function, $D^{{\rm bare}}_{q\rightarrow\gamma}(z_\gamma)$
can be decomposed at a factorization scale
$\mu_{\mathrm{F}}$ into a universal divergent piece and a phenomenological 
contribution $D_{q\rightarrow\gamma}(z_\gamma,\mu_{\mathrm{F}})$, which has to
be taken from experiment,
\begin{equation}
D_{q\rightarrow\gamma}^{\mathrm{bare,DR}}(z_\gamma) =
\frac{\alpha Q_q^2}{2\pi}
\frac{1}{\epsilon}\left(
\frac{4\pi\mu^2}{\mu_{\mathrm{F}}^2}\right)^\epsilon\frac{1}{\Gamma(1-\epsilon)}
P_{ff}(1-z_\gamma) +
D_{q\rightarrow\gamma}(z_\gamma,\mu_{\mathrm{F}}),
\label{eq:massfactDR}
\end{equation}
where $Q_q$ is the electric charge of the massless quark $q$ and 
$P_{ff}(z)=(1+z^2)/(1-z)$ is the $q\to q+\gamma$ splitting function.
Here, the collinear singularity is regularized in $D=4-2\epsilon$
dimensions (with the arbitrary reference mass $\mu$) and isolated by an
$\overline{\mathrm{MS}}$-type factorization scheme; 
the respective result for
mass regularization with a small quark mass can be found in \citere{Denner:2010ia}.

In \citere{Glover:1993xc} 
a method was proposed how to measure
$D_{q\rightarrow\gamma}(z_\gamma,\mu_{\mathrm{F}})$ upon analyzing the process
$\Pe^+\Pe^-\rightarrow n\,\mathrm{jet}+\mathrm{photon}$.
The key feature of the proposed method is the democratic clustering of
both hadrons and photons into jets, while keeping track of the
photonic energy fraction in the jet, i.e.\
technically one has to deal with identified
particles in the final state that lead to infrared-non-safe observables
(with respect to collinear singularities). 
The treatment of this situation with one-cutoff phase-space slicing and
dipole subtraction is described in \citeres{Glover:1993xc} and 
\cite{Denner:2010ia}, respectively.
The first determination of $D_{q \to \gamma}(z,\mu_{0})$ was performed by the ALEPH
collaboration~\cite{Buskulic:1995au} using the ansatz
\begin{equation}
D_{q\rightarrow\gamma}^{\mathrm{ALEPH}}(z_\gamma,\mu_{\mathrm{F}})=
\frac{\alpha Q_q^2}{2\pi}   
\left[
P_{ff}(1-z_\gamma)\ln\left(\frac{\mu_{\mathrm{F}}^2}{\mu_0^2}
\frac{1}{(1-z_\gamma)^2}\right)+C
\right],
\end{equation}
with two fitting parameters, $C=-12.1$ and $\mu_0=0.22\GeV$. 
\item
{\it Frixione isolation}~\cite{Frixione:1998jh} \\
This procedure defines isolated hard photons by the requirement
that only soft partons can become collinear to the photon.
In detail, the total transverse energy $\sum_i E_{\rT,i}$ of all
partons $i$ with small distances 
$R_{i\gamma}=\sqrt{(\eta_i-\eta_\gamma)^2+(\phi_i-\phi_\gamma)^2}$
to the photon
in the pseudo-rapidity--azimuthal-angle plane is required to 
go to zero with the maximally allowed distance $R_{i\gamma}$,
i.e.\
\begin{equation}
\sum_i E_{\rT,i}\theta(\delta-R_{i\gamma}) \le {\cal X}(\delta)
\qquad \mbox{for all} \quad \delta\le\delta_0,
\label{eq:frixione}
\end{equation}
where $\delta_0$ is a measure for the size of the cone around
the photon, in which the criterion is used, and ${\cal X}(\delta)$
is an appropriate function with ${\cal X}(\delta)\to0$ for $\delta\to0$.
Specifically, the form
\begin{equation}
{\cal X}(\delta) = E_{\rT,\gamma}\epsilon_\gamma
\left(\frac{1-\cos\delta}{1-\cos\delta_0}\right)^n
\end{equation}
is suggested in Ref.~\cite{Frixione:1998jh} for a photon of transverse energy $E_{\rT,\gamma}$,
where the two parameters $\epsilon_\gamma$ and $n$ can just be taken to be $1$.
This condition excludes any hard jet activity collinear to the photon, but still
takes into account soft jet activity at a sufficiently inclusive level to
guarantee the proper cancellation of infrared singularities when calculating
NLO QCD corrections to $F+\gamma$ production with $\gamma$ being the
isolated photon.
Taking the inverse procedure to define $F+\mathrm{jet}$ production,
i.e.\ interpreting 
$F+\mathrm{jet}+\gamma$ production as photonic EW correction to
$F+\mathrm{jet}$ if condition (\ref{eq:frixione}) is not fulfilled,
formally shifts the complete contribution of the quark-to-photon fragmentation function
to $F+\mathrm{jet}$ production.
Whether this means that $F+\gamma$ production is really insensitive to 
non-perturbative corrections is not proven and sometimes under debate.
Moreover, the implementation of condition (\ref{eq:frixione}) at the 
experimental level raises issues, in particular concerning shower effects
on the hard photon kinematics and the realizability at the apex of the cone.
\end{itemize}
At the LHC, a prominent example for photon--jet separation is given by
$\PW/\PZ$ production in association with a jet or a photon, where
the NLO EW corrections to $\PW/\PZ+\mathrm{jet}$ 
production~\cite{Denner:2009gj,Denner:2011vu,Denner:2012ts} 
and the NLO QCD corrections to $\PW/\PZ+\gamma$ 
production (e.g.\ implemented in MCFM~\cite{Campbell:2011bn}) overlap.
Comparing, for instance, results on $\PW+\gamma$ production obtained with
the two separation procedures as implemented in MCFM reveals differences
at the level of $\sim2\%$ only.
Note, however, that it is not possible to make generic, process-independent
statements on such differences.

\subsection{Combination of QCD and electroweak corrections}

Naively a comparison of coupling strengths suggests that along
with NLO ${\cal O}(\alpha_{\mathrm{s}})$
and NNLO ${\cal O}(\alpha_{\mathrm{s}}^2)$ QCD corrections 
simply taking into account NLO EW corrections of 
${\cal O}(\alpha)\sim{\cal O}(\alpha_{\mathrm{s}}^2)$
would be adequate in predictions.
However, as explained in the previous sections, EW ${\cal O}(\alpha)$ corrections can be
significantly enhanced compared to the small value of $\alpha/\pi$ by large logarithms and/or
kinematical effects, raising the question about mixed QCD--EW corrections of
${\cal O}(\alpha_{\mathrm{s}}\alpha)$.
In spite of the great progress of recent years in NNLO QCD calculations,
the required multi-scale NNLO calculations needed at ${\cal O}(\alpha_{\mathrm{s}}\alpha)$
are still out of reach at present.
As long as this is the case, we have to rely on approximations---an issue that
already comes up in the combination of NLO QCD and EW corrections 
($\delta^{\NLO}_{\QCD}$ and $\delta^{\NLO}_{\EW}$)
in cross-section predictions. Schematically, the choice is between 
the two extreme variants of adding,
$1+\delta^{\NLO}=1+\delta^{\NLO}_{\QCD}+\delta^{\NLO}_{\EW}$,
or multiplying relative corrections,
$1+\delta^{\NLO}=(1+\delta^{\NLO}_{\QCD})\times(1+\delta^{\NLO}_{\EW})$,
and variants in between these two extreme cases.
The differences between different variants are of ${\cal O}(\alpha_{\mathrm{s}}\alpha)$,
and the optimal choice should minimize the remaining corrections of this order
for the most important observables.

Since long-distance effects such as soft or collinear parton emission off quarks or gluons
and collinear final-state photon radiation off leptons are known to factorize from
the actual hard scattering, the variant of factorizing QCD and EW corrections seems
to be preferable in many cases.
For the prediction of a cross section which is differential in some observable $x$ 
at the parton level, this idea of factorization can be translated
into the recipe
\begin{equation}
\frac{\rd\sigma}{\rd x} =
\frac{\rd\sigma_{\QCD}}{\rd x} \times \left[1+\delta^{\NLO}_{\EW}(x)\right]
+\frac{\rd\sigma_\gamma}{\rd x},
\end{equation}
where $\sigma_{\QCD}$ stands for the best available QCD prediction for the 
cross section, $\delta^{\NLO}_{\EW}(x)$ is the relative EW correction
differential in $x$,
and $\sigma_\gamma$ denotes the contribution from photon-induced processes.
This approach is, e.g., useful if the various contributions are calculated
independently, possibly even by independent programs.
For instance, the state-of-the-art predictions for Higgs production
via vector-boson fusion and Higgs-strahlung provided by the
LHC Higgs Cross Section Working 
Group~\cite{Dittmaier:2011ti,Dittmaier:2012vm,Heinemeyer:2013tqa}
are calculated in this way.

Similar factorization approaches can also be directly included in the
event generation within QCD-based event generators upon applying
differential reweighting factors for the EW corrections.
In simple cases where the relative EW correction $\delta^{\NLO}_{\EW}$
depends mainly on a single kinematical variable $x$,
while $\delta^{\NLO}_{\EW}$ is flat in the other relevant kinematical variables,
generated events can simply be reweighted by the factor
$1+\delta^{\NLO}_{\EW}(x)$. As discussed in \citere{Binoth:2010ra}
(Section~19), this method is quite successful for Higgs production
via vector-boson fusion, where $x$ can be identified with the
transverse momentum of the Higgs boson. Specifically,
in \citere{Binoth:2010ra} the QCD-based prediction of 
{\sc Herwig}~\cite{Corcella:2002jc} was dressed with the relative EW correction
from {\sc HAWK}~\cite{Ciccolini:2007jr,Ciccolini:2007ec}.
A similar approach has been advocated recently in
\citere{Gieseke:2014gka} where the EW 
corrections~\cite{Bierweiler:2012kw,Bierweiler:2013dja}
to on-shell diboson production (neglecting corrections to $\PW/\PZ$ decays)
were included in {\sc Herwig++}~\cite{Bahr:2008pv,Arnold:2012fq}
by reweighting factors that are double-differential in the 
partonic scattering energy and the angle of the LO kinematics.
It should be kept in mind that generally such reweighting procedures
represent some approximative compromise with advantages and weaknesses.
As in the two examples mentioned above, usually no hard photon
is generated along with the EW reweighting, but photon emission is
integrated over, i.e.\ 
in the real-photonic bremsstrahlung corrections the kinematic effects 
from hard photon emission are neglected.
While observables that are not sensitive to photonic recoil effects can
be described well via the reweighting approach
(see, e.g., \citere{Gieseke:2014gka} for examples in diboson production), the 
method is expected to fail whenever photonic recoil is important
(see, e.g., the transverse momentum of the dilepton system in $\PW$-pair
production as discussed in \citere{Billoni:2013aba}).
Before applied in practice, the applicability and reliability of the
reweighting approach should be checked by a comparison to fully
differential results.

This discussion reveals one of the crucial problems
in the combination of NLO QCD and NLO EW corrections
when following the factorization idea: The QCD and EW corrections of the
two different factors are in general defined on different phase spaces.
Only a complete calculation of ${\cal O}(\alpha_{\mathrm{s}}\alpha)$ corrections
can fully solve this problem.
Since matrix elements and phase spaces
for jet and/or photon emission factorize
in the soft and/or collinear limits of the radiated particle,
there is at least the possibility to correctly describe the leading
soft and/or collinear ${\cal O}(\alpha_{\mathrm{s}}\alpha)$ effects 
at the fully differential level.
Based on this idea, Monte Carlo generators such as
{\sc Herwig++}~\cite{Bahr:2008pv,Arnold:2012fq} and
{\sc Sherpa}~\cite{Gleisberg:2008ta} already dress QCD-based predictions 
with soft and/or collinear photon radiation (at least off final-state leptons),
in order to catch some leading ${\cal O}(\alpha_{\mathrm{s}}\alpha)$ effects
and even higher-order effects in $\alpha$.
Likewise dedicated Monte Carlo programs for Drell--Yan processes such as 
{\sc Resbos-A}~\cite{Cao:2004yy},
{\sc Horace}~\cite{Balossini:2009sa}, and
{\sc Winhac}~\cite{Placzek:2013moa}
proceed similarly.

Finally, we comment on the issue of {\it matching} fixed-order NLO calculations
of ${\cal O}(\alpha_{\mathrm{s}})$ and ${\cal O}(\alpha)$ with parton showers
that take into account multiple jet and/or photon emission in some leading
logarithmic approximation. Here the central issue is to avoid double-counting
effects at the level of one-parton and one-photon emission and to keep
full NLO accuracy after the matching.
In pure QCD, two matching prescriptions are widely in use:
{\sc MC@NLO}~\cite{Frixione:2002ik} and
{\sc Powheg}~\cite{Nason:2004rx,Frixione:2007vw}.
When matching NLO QCD+EW calculations to a pure QCD parton shower
(as e.g.\ done in \citere{Bernaciak:2012hj} for Drell--Yan processes in the
{\sc Powheg} framework),
thus, improves the NLO prediction not only in the pure QCD sector,
but also in ${\cal O}(\alpha_{\mathrm{s}}\alpha)$ by 
dressing ${\cal O}(\alpha)$ effects
with soft/collinear QCD radiation.
If the matched shower even includes photon radiation, 
where the {\sc Powheg} matching has to be generalized as described in
\citeres{Barze:2012tt,Barze:2013yca} for Drell--Yan processes,
then the effectively included ${\cal O}(\alpha_{\mathrm{s}}\alpha)$ effects
account for ${\cal O}(\alpha_{\mathrm{s}})$ effects dressed with
soft/collinear photon radiation.

As long as complete corrections of ${\cal O}(\alpha_{\mathrm{s}}\alpha)$
are missing, they are one particular source for theoretical uncertainties.
Their size can be conservatively estimated upon comparing the results obtained from
the additive and multiplicative approaches to combine NLO QCD and EW corrections.
If leading effects of ${\cal O}(\alpha_{\mathrm{s}}\alpha)$, which show
factorization properties (such as photonic FSR or EW Sudakov logarithms), 
are properly included, those effects should be excluded 
from the assessment of uncertainties at ${\cal O}(\alpha_{\mathrm{s}}\alpha)$.

\subsection{Instability effects and the treatment of $\PW/\PZ$ resonances}
\label{se:unstable}

Although external $\PW$ and $\PZ$ bosons always appear as resonances in experiments
which have to be reconstructed from their decay products, LO predictions for
$\PW/\PZ$ production processes often start from the approximation of stable
$\PW/\PZ$ bosons on their mass shell. The more realistic treatment as
resonance processes changes cross-section predictions typically 
at the level of ${\cal O}(\Ga_V/M_V)\sim3\%$ 
($M_V =$ mass, $\Ga_V$ decay width of $V=\PW,\PZ$) when the 
resonances are integrated over. For such observables off-shell effects
of the weak gauge bosons, derived from LO calculations for the resonant and
non-resonant production of the decay products,
can be counted as ${\cal O}(\alpha)$ corrections.
The remaining ${\cal O}(\alpha)$ corrections can be calculated from virtual one-loop
and real emission corrections to on-shell $\PW/\PZ$ production if the decays of the
weak gauge bosons are treated inclusively.

For an inclusion of the $\PW/\PZ$ decays in predictions, the simplest (but
somewhat crude) way is to employ the {\it narrow-width approximation} (NWA),
which treats the $\PW/\PZ$ bosons as ``stable intermediate states'', i.e.\
the full process is decomposed into on-shell $\PW/\PZ$ production and
on-shell $\PW/\PZ$ decays. This decomposition results from the limit
$\Ga_V\to0$ in the squared matrix element of the full resonance process,
where the squared propagator factor (momentum transfer $k$) behaves like
\begin{equation}
\frac{1}{|k^2-M_V^2+\ri M_V\Ga_V|^2} 
\; \asymp{\Ga_V\to0} \; \frac{\pi}{M_V\Ga_V} \, \de(k^2-M_V^2).
\end{equation}
The $1/\Ga_V$ factor on the r.h.s.\ is part of the well-known {\it branching ratio}
which emerges after the inclusive integration over the $V$ decay phase space.
If cuts are imposed on the decay products, or if distributions in kinematical
variables of those are considered, in general effects of spin correlations 
between $\PW/\PZ$ production and decay (or between various decaying gauge bosons)
appear. 
The naive NWA, which employs unpolarized production cross sections and
decay widths, can be easily improved to include these correlations
by properly combining production and decay parts for definit polarization states.
Note that ${\cal O}(\alpha)$ corrections to a cross section in NWA do not
only consist of corrections to the $V$-production cross section and to the
relevant branching ratio, but also comprise the off-shell effects of ${\cal O}(\Ga_V/M_V)$
mentioned above.
Furthermore, it has been shown~\cite{Kauer:2007zc}
that naive estimates often underestimate the
theoretical uncertainties of the NWA;
in scenarios with cascade decays, which are quite common in non-standard models,
the NWA can even fail completely, as demonstrated in \citere{Berdine:2007uv}.

A detailed description of a resonance process, keeping the full differential
information of the kinematics of the decay products,
has to be based on complete
matrix elements for the full process, including both resonant and
non-resonant diagrams.
Note that in standard perturbation theory, particle propagators do not include
decay widths. The decay widths rather appear in the 
propagator denominators only after a Dyson summation of self-energy diagrams
(or at least the imaginary parts thereof near resonances). 
Depending on the details of the field-theoretical definition of the unstable
particle's mass, different results are obtained for the resonant propagator.
Writing generically 
\begin{equation}
P_V(k) = \frac{1}{k^2-M_V^2+\ri M_V\Ga_V(k^2)} 
\label{eq:PropV}
\end{equation}
for the $V$-propagator factor, two frequently used versions are:
\begin{itemize}
\item
{\it Fixed width} (FW): $\Ga_V(k^2)=\Ga_V=\mathrm{const.}$ \\
In this parametrization, the complex squared mass $M_V^2-\ri M_V\Ga_V$ plays the
role of the complex location of the pole in the propagator, i.e.\ it is
a field-theoretically sound, gauge-invariant quantity resulting from an
all-order definition 
(see \citere{Gambino:1999ai,Grassi:2001bz} and older references therein). 
The real and imaginary parts of this complex
quantity define the {\it pole mass} and its associated {\it pole width}.
\item
{\it Running width} (RW): $\Ga_V(k^2)=\Ga_V\times k^2/M_V^2\times\theta(k^2).$ \\
This behaviour results from the so-called {\it on-shell definition} (OS) of the
gauge-boson masses (see, e.g., \citere{Denner:1991kt}), $M_V^{\mathrm{OS}}$,
which are tied to the zeroes of the real parts of the respective self-energies.
This definition has the drawback that it is gauge dependent at the two-loop
level and beyond. The values for the $\PZ$- and $\PW$-boson masses have,
however, been determined within this scheme at LEP and the Tevatron.
\end{itemize}
For reasons of theoretical consistency, the pole definition of mass and width
are clearly preferable. Fortunately, there is a simple translation from one
scheme to the other~\cite{Bardin:1988xt,Beenakker:1996kn}:
\begin{equation}
M_V = \frac{M_V}{\sqrt{1+\Ga_V^2/M_V^2}}\Bigg|_{\mathrm{OS}},
\qquad
\Ga_V = \frac{\Ga_V}{\sqrt{1+\Ga_V^2/M_V^2}}\Bigg|_{\mathrm{OS}},
\label{eq:mwdef}
\end{equation}
so that $M_{\PZ,\mathrm{OS}}-\MZ\approx 34\MeV$ and
$M_{\PW,\mathrm{OS}}-\MW\approx 27\MeV$.
For precision EW physics, in particular for a precision $\MW$ measurement,
it is important to be consistent in the use of a scheme and the respective input.

The FW and the RW prescriptions
represent different parametrizations of resonances,
but none of them represents a consistent scheme to calculate full cross sections
for resonance processes, since both in general lead to gauge-dependent results.
This is due to the fact that using Eq.~(\ref{eq:PropV}) for propagator factors
necessarily mixes different orders in perturbation theory in practice,
since the width term in the propagator results from partial all-order resummations, 
but the cross-section calculation stops at some finite loop level.
While the FW usually at least leads to acceptable LO results, 
the $k^2$-dependence of the width term in the RW propagator can enhance
gauge-breaking terms already in LO predictions to a level that completely
destroys predictions. Various examples are, e.g., discussed in
\citeres{Beenakker:1996kn,Denner:1999gp,Dittmaier:2002ap}.
Proper cross-section predictions should be based on consistent,
gauge-invariant schemes, or at least be validated by comparing to
such results.

In a nutshell, we briefly describe the schemes that are most
frequently used in LHC physics to deliver
gauge-invariant predictions at LO and NLO:
\begin{itemize}
\item
{\it Factorization schemes} \\
Different variants of factorizing resonant structures from amplitudes have
been suggested and used in the literature, but they all share the idea
to separate a simple resonance factor from complete (gauge-invariant) 
amplitudes or from gauge-invariant subsets of diagrams.
In \citere{Dittmaier:2001ay}, for instance, the virtual electroweak
correction to Drell--Yan-like W~production was factorized from the
resonant LO amplitude, so that the relative correction factor did not
involve resonance factors anymore.
At first sight, such schemes seem to be very simple, 
but for more complicated processes it is highly non-trivial to guarantee 
the aimed precision (e.g.\ NLO) everywhere in phase space and to
match virtual and real corrections.
Simply modifying LO cross sections with fudge factors containing
decay widths for resonances in general introduces spurios ${\cal O}(\alpha)$
terms destroying NLO EW accuracy.
A consistent implementation of weak NLO corrections
to $\PZ$-boson production in a factorization scheme 
is, e.g., described in \citere{Dittmaier:2009cr}.
\item
{\it Pole scheme} \\
The pole scheme exploits the fact that both the location of the
$V$~propagator pole and its residue in amplitudes are gauge-independent
quantities.
The idea~\cite{Stuart:1991xk,Aeppli:1993rs}
is, thus, to first isolate the residue for the
considered resonance and subsequently to introduce a finite decay width
only in the gauge-independent resonant part. If done carefully this
procedure respects gauge invariance and can be used to make uniform
predictions in resonant and non-resonant phase-space regions.
There is some freedom in the actual implementation of the scheme,
because the resonant part of an amplitude is not uniquely determined by
the propagator structure alone, but depends on a specific phase-space
parameterization and in most cases also on the separation of
polarization-dependent parts (spinors, polarization vectors).
Taking the pole prescription literally, the scattering amplitude
on the resonance pole (i.e.\ with $k^2$ being complex) involves
matrix elements with complex kinematical variables, which is a subtle
issue~\cite{Passarino:2010qk}. This {\it complex pole scheme} was applied to
Higgs production via gluon fusion at the LHC in \citere{Goria:2011wa},
with particular emphasis on a heavy, i.e.\ very broad, Higgs boson.
For narrow resonances such as the weak gauge bosons
$V=\PW,\PZ$, the complex kinematics can be avoided by suitable
expansions in $\Gamma_V/M_V$, as e.g.\ done in \citere{Dittmaier:2009cr}
for single-$\PZ$ production.

The pole scheme offers a well-defined separation between resonant
and non-resonant parts of a cross section, i.e.\ in some sense
a definition of {\it signal} and {\it background}
for the production of a resonance. This scheme is, thus, ideal
for a parametrization of a resonance by so-called
{\it pseudo-observables}, such as total and partial decay widths,
peak cross sections, effective couplings, etc.
Moreover,
the consistent separation of signal and background contributions
is an ideal starting point to further improve the description of the
signal by higher-order corrections, a point that is particularly important
if the signal dominates over the background and, thus, deserves higher precision.

\item
{\it Pole approximation} \\
A pole approximation---in contrast to a full pole-scheme 
calculation---results
from a resonant amplitude defined in the pole scheme upon
neglecting non-resonant parts.
Consequently, NLO calculations in pole approximation deliver NLO accuracy
only in the neighborhood of resonances, but not in off-shell tails.
Note that NLO corrections in pole approximations do not only involve
independent corrections to the production and to the decay of resonances
(called {\it factorizable} corrections),
but also comprise soft-photon or soft-gluon exchange between production
and decay subprocesses (called {\it non-factorizable} corrections).
Different variants of 
{\it double-pole approximations} have, for instance, been employed
to describe W-pair production with decays in $\Pep\Pem$ annihilation
at LEP2 (reviewed, e.g., in \citere{Grunewald:2000ju}).
Recently, this concept has been applied to W-pair production at the
LHC in \citere{Billoni:2013aba}.
\item
{\it Effective field theories} \\
Approaches~\cite{Beneke:2003xh,Beneke:2004km,Hoang:2004tg} 
via effective field theories (EFT)
deliver a field-theoretically elegant way to carry out pole expansions owing to their
formulation via Lagrangians and effective actions.
Like pole approximations, their validity is restricted to the resonance region,
but they offer the combination with further expansions, e.g.\ around thresholds,
and suggest better possibilities to carry out dedicated resummations.
The EFT shows limitations or complications if detailed
information on differential properties of observables is needed, since different
degrees of freedom are integrated out.
For $\Pep\Pem$ collisions the EFT approach was, for instance,
used to evaluate NLO EW corrections to the W-pair production cross section
near the $\PW\PW$ threshold~\cite{Beneke:2007zg}, 
including leading EW higher-order effects beyond NLO.
Currently, the formalism is applied to the NLO QCD corrections to
top-quark pair production at the LHC, with first results on the
quark--antiquark annihilation channel~\cite{Falgari:2013gwa}.
\item
{\it Complex-mass scheme} \\
This scheme was introduced in \citere{Denner:1999gp} for
LO calculations and generalized to NLO in
\citere{Denner:2005fg}. In this approach the squared
W- and Z-boson masses are consistently identified with 
the complex values $M_V^2-\ri M_V\Ga_V$ $(V=\PW,\PZ)$,
not only in the $V$-propagators, but also in the couplings,
which therefore become complex and involve 
a complex weak mixing angle via 
$\cw^2=1-\sw^2=(\MW^2-\ri\MW\Ga_{\PW})/(\MZ^2-\ri\MZ\Ga_{\PZ})$.
Otherwise the usual perturbative calculus with Feynman
rules and counterterms works without modification.
All relations following from gauge invariance are respected,
because the gauge-boson masses are modified only by an analytic continuation. 
NLO calculations deliver uniform predictions with NLO accuracy everywhere in phase 
space, i.e.\ both in resonant and non-resonant regions.
Spurious terms spoiling unitarity are unproblematic, since they are of (N)NLO in
an (N)LO calculation (i.e.\ of higher order)
without any unnatural amplification, because
unitarity cancellations, which are ruled by gauge invariance, are respected.

The complex-mass scheme, first used for four-fermion production in $\Pep\Pem$ 
physics~\cite{Denner:2005fg},
is frequently used in NLO calculations for weak gauge-boson or Higgs-boson
processes at the LHC, for example
for the off-shell Higgs-boson decays
$\PH\to\PW\PW/\PZ\PZ\to4\,$fermions at NLO accuracy in the
Monte Carlo program {\sc Prophecy4f}~\cite{Bredenstein:2006rh,Bredenstein:2006ha} and
in the previously mentioned 
NLO EW calculations for single-$\PZ$~\cite{Dittmaier:2009cr} and
$\PW/\PZ+\mathrm{jet}$~\cite{Denner:2009gj,Denner:2011vu,Denner:2012ts}
at the LHC.
As discussed in the context of the (two-loop) calculation
of NLO EW corrections to Higgs production via gluon 
fusion~\cite{Actis:2008ug,Actis:2008uh},
the complex-mass scheme delivers also a sound description of particle
thresholds in loop amplitudes (which can be viewed as resonances inside loops).
\end{itemize}
Finally, we want to reemphasize that
this survey is just meant to assist non-experts in the field in finding
their bearings upon putting some keywords into context. 
It is by no means comprehensive or complete, i.e.\ several issues 
concerning unstable particles are left
untouched here, such as signal--background interferences or many
technical details and subtleties.
There is still much work to be done until all relevant resonance
processes at the LHC are theoretically described in a satisfactory 
way---a statement that is also backed by the recent report of the
Higgs CAT~\cite{Passarino:2013bha}.

\subsection*{Acknowledgments}

The author (S.D.) is grateful to A.~Denner and A.~M\"uck for valuable
discussions and comments on this contribution.




\newpage

\chapter{NLO automation and (N)NLO techniques}
\label{cha:nnlo}
\def\eq#1{{(\ref{#1})}}
\def\LO{{\rm LO}}
\def\NLO{{\rm NLO}}

\section{The first use case for BLHA2 extensions: {\sc NJet} plus Herwig++/Matchbox\footnote{S. Badger, S. Pl\"atzer, V. Yundin}}

We present first results from interfacing NJet to
Herwig++/Matchbox. This combination has provided the first working
example for extensions of the BLHA interface regarding tree-level, as
well as colour and spin correlated matrix elements on top of virtual
corrections.

\subsection{Introduction}

We explore the possibility of a more general interface between next-to-leading order~(NLO)
Monte-Carlo event generators and automated matrix element generators. Apart from specialized
implementations, currently the BLHA~\cite{Binoth:2010xt} has been used to provide loop level
matrix elements. The purpose of this study is to use the updated BLHA2 accord~\cite{Alioli:2013nda} to provide
more general matrix elements to the Monte-Carlo program for use in the real radiation contributions as NLO.

Our starting point is the differential cross section to NLO accuracy in QCD in hadron-hadron collisions
written using the Catani-Seymour subtraction scheme~\cite{Catani:1996vz}:
\begin{align}
  d\sigma_n&(\mu_R, \mu_F; H_1(p_1)+H_2(p_2) \to X) = \nonumber\\&
  \sum_{k=0}^{K} d\sigma_n^{{\rm N}^k\LO}(\mu_R, \mu_F; H_1(p_1)+H_2(p_2) \to X) + \mathcal{O}(\alpha_s^{n-1+K}) \\
  d\sigma_n^{\LO}&(\mu_R, \mu_F; H_1(p_1)+H_2(p_2) \to X) = \nonumber\\&
  \sum_{i,j} f_{i/H_1}(x_1,\mu_F) f_{j/H_2}(x_2,\mu_F) d\widehat{\sigma}_n^{B}(\mu_R, \mu_F; i(x_1p_1)+j(x_2p_2)\to X) dx_1dx_2 \\
  d\sigma_n^{\NLO}&(\mu_R, \mu_F; H_1(p_1)+H_2(p_2) \to X) = \nonumber\\&
  \sum_{i,j} f_{i/H_1}(x_1,\mu_F) f_{j/H_2}(x_2,\mu_F) d\widehat{\sigma}_n^{V}(\mu_R, \mu_F; i(x_1p_1)+j(x_2p_2)\to X) dx_1dx_2 \nonumber\\+&
  \sum_{i,j} f_{i/H_1}(x_1/z,\mu_F) f_{j/H_2}(x_2,\mu_F) d\widehat{\sigma}_n^{I}(\mu_R, \mu_F, z; i(x_1p_1)+j(x_2p_2)\to X) dx_1dx_2dz \nonumber\\+&
  \sum_{i,j} f_{i/H_1}(x_1,\mu_F) f_{j/H_2}(x_2/z,\mu_F) d\widehat{\sigma}_n^{I}(\mu_R, \mu_F, z; i(x_1p_1)+j(x_2p_2)\to X) dx_1dx_2dz \nonumber\\+&
  \sum_{i,j} f_{i/H_1}(x_1,\mu_F) f_{j/H_2}(x_2,\mu_F) d\widehat{\sigma}_n^{RS}(\mu_R, \mu_F; i(x_1p_1)+j(x_2p_2)\to X) dx_1dx_2
\end{align}
where $f_{i/H}(x, \mu_F)$ is the parton distribution function giving
the probability of finding a parton $i$ with momentum fraction $x$
inside the hadron $H$. The partonic cross sections can then be written
in terms of matrix elements and dipole subtraction terms as follows:
\begin{align}
  d\widehat{\sigma}_n^{B} &= c(\alpha_s,\mu_R)\, d{\rm PS}_n
        \sum_{\lambda_h=\pm} B_n(\{\lambda_h\}) \\
  d\widehat{\sigma}_n^{V} &= c(\alpha_s,\mu_R)\alpha_s(\mu_R)\,d{\rm PS}_n
        \sum_{\lambda_h=\pm} V_n(\mu_R,\{\lambda_h\}) \\
  d\widehat{\sigma}_n^{I} &= c(\alpha_s,\mu_R)\alpha_s(\mu_R)\,d{\rm PS}_n \sum_{k,l}
        \sum_{\lambda_h=\pm} \tilde{B}^{kl}_n(\{\lambda_h\}) \mathcal{D}^{kl}(z) \\
  d\widehat{\sigma}_n^{RS} &= c(\alpha_s,\mu_R)\alpha_s(\mu_R)\, d{\rm PS}_{n+1}\left\{
  \sum_{\lambda_h=\pm} B_{n+1}(\{\lambda_h\}) - \sum_{k,l} S_{n+1}^{kl}(\{\lambda_h\})\right\}
  \label{eq:RSdef}
\end{align}
where
\begin{align}
  B_n(\{\lambda_h\}) &=
  \sum_{i,j} \left(A_i^{(0)}(p_1^{\lambda_1},\ldots,p_n^{\lambda_n})\right)^\dagger C^{(0)}_{ij} A_j^{(0)}(p_1^{\lambda_1},\ldots,p_n^{\lambda_n})
  \label{eq:born}\\
  V_n(\mu_R, \{\lambda_h\}) &=
  \sum_{i,j} \left(A_i^{(0)}(p_1^{\lambda_1},\ldots,p_n^{\lambda_n})\right)^\dagger C^{(1)}_{ij} A_j^{(1)}(\mu_R, p_1^{\lambda_1},\ldots,p_n^{\lambda_n})
  \label{eq:virt}\\
  \tilde{B}^{kl}_n(\{\lambda_h\}) &=
  \sum_{i,j} \left(A_i^{(0)}(p_1^{\lambda_1},\ldots,p_n^{\lambda_n})\right)^\dagger \tilde{C}^{(0),kl}_{ij} A_j^{(0)}(p_1^{\lambda_1},\ldots,p_n^{\lambda_n})
  \label{eq:borncc}\\
  S_{n+1}(\{\lambda_h\}) &= \nonumber\\
  \Big| \sum_{\lambda_P=\pm} \mathcal{A}_n&(p_1^{\lambda_1},\ldots,
  p_{k-1}^{\lambda_{k-1}}, p_{k+l}^{\lambda_{P}}, p_{k+1}^{\lambda_{k+1}},\ldots,
  p_{l-1}^{\lambda_{l-1}}, p_{l+1}^{\lambda_{l+1}},\ldots,
  p_{n+1}^{\lambda_{n+1}}) \mathcal{S}^{kl\to P}(-p_{k+l}^{-\lambda_P}; p_k^{\lambda_k}, p_l^{\lambda_l})\Big|^2
  \label{eq:sub}
\end{align}
$\mathcal{D}^{kl}$ are the integrated dipoles and $\mathcal{S}$ represents a
polarized splitting function.  The colour sums over the indices $i$ and $j$ run
over the independent set of primitive amplitudes defined together with the
colour matrices $C^{(L)}_{ij}$ and colour correlated matrices
$\tilde{C}^{(L),lm}_{ij}$ whose entries are polynomials in $N_c$.

The subtraction terms must be expanded before they can be written in terms of
helicity and colour correlated tree-level matrix elements. There are clearly
four terms in the expansion though only two are independent due to parity
symmetry.  These are labelled using $\lambda_P=\pm$ in the amplitude and
$\lambda_P^\dagger=\pm$ in the conjugate amplitude.  The first of these,
$(\lambda_P,\lambda_P^\dagger)=(+,-)$ , is the same colour correlated born
amplitudes appearing in the integrated subtraction terms while the second
involves the off-diagonal elements $(\lambda_P,\lambda_P^\dagger)=(+,+)$.
This term is the spin-correlated born amplitude can be written as a usual
colour correlated born together with flipping the helicity of the $P^{\rm th}$
particle in the conjugate amplitude:
\begin{align}
  &B^{\text{flip}(P), kl}_n(\lambda_1,\dots,\lambda_n)
  = \nonumber\\&
  \sum_{i,j}
  \left(A_i^{(0)}(p_1^{\lambda_1},\ldots,p_P^{-\lambda_P},\ldots,p_n^{\lambda_n})\right)^\dagger
    \tilde{C}^{(0),kl}_{ij}
  A_j^{(0)}(p_1^{\lambda_1},\ldots,p_P^{\lambda_P},\ldots,p_n^{\lambda_n})
  \label{eq:bornsc}
\end{align}
Once collected in terms of helicity summed objects, the real-subtraction terms are,
\begin{align}
  \sum_{\lambda_h=\pm} S_{n+1}^{kl}(\{\lambda_h\}) &=
      D^{kl\to p}(p_k, p_l) \sum_{\lambda_h'=\pm} B^{kl}_n(\{\lambda_h'\}) \nonumber\\&
    + \left(D^{\text{flip}(P), kl\to P}(p_k, p_l) \sum_{\lambda_h'=\pm} B^{\text{flip}(P), kl}_n(\{\lambda_h'\}) + (c.c)\right),
\end{align}
where $\{\lambda_h'\}$ is are the $n$-point helicity configurations. Note that we have used parity symmetry to write,
\begin{align}
  \sum_{\lambda_h=\pm} \tilde{B}^{kl}_n(\{\lambda_{h=1,n}\}) = 2 \sum_{\lambda_h=\pm} \tilde{B}^{kl}_n(\{+,\lambda_{h=2,n}\}).
\end{align}
While the first term proportional to the dipole $D^{kl\to P}$ is phase independent, the spin correlated
term depends on a complex phase which cancels with the phase in $D^{\text{flip}(P), kl\to P}$.
When passing spin correlated amplitudes via the BLHA interface the necessary phase information
is extracted using the polarization vector $\varepsilon_+^{\mu}(p,q)$ provided via the function,
\begin{verbatim}
PolVec(double* p, double* q, double* eps)
\end{verbatim}
For the current set-up we pass the helicity summed objects between {\sc
NJet}~\cite{Badger:2012pg} and the Herwig++/Matchbox~\cite{Bahr:2008pv}.  The
BLHA2 keyword \verb|AmplitudeType| takes options defined as follows
\begin{center}
\begin{tabular}[h]{cl}
  {\tt AmplitudeType} & function \\
  \hline
  {\tt loop} & $\sum\limits_{} c(1,\mu_R) V_{\mu_R, n}(h)$ \\
  {\tt tree} & $\sum\limits_{} c(1,\mu_R) B_{n}(h)$ \\
  {\tt cctree} & $\sum\limits_{} c(1,\mu_R) B^{kl}_{n}(h)$ \\
  {\tt sctree} & $\sum\limits_{} c(1,\mu_R) B^{\text{flip}(P),kl}_{n}(h, m)$ \\
  \hline
\end{tabular}
\end{center}
The colour correlated tree amplitudes form a real symmetric $n\times n$ matrix
which is accessed using the \verb|OLP_EvalSubProcess2| function to return a
$n(n-1)/2$ array of type \verb|double| where $B^{kl}$ is the $k+l(l-1)/2$-th
element.  The spin correlated amplitudes form a complex $n\times n$ matrix
which is accessed using the \verb|OLP_EvalSubProcess2| function to return a
$2n^2$ array of type \verb|double| where $\Re(B^{\text{flip}(P),kl})$ is the $(2k+2nl)$-th element
and $\Im(B^{\text{flip}(P),kl})$ is the $(2k+2nl+1)$-th element.

\subsection{Proof of Concept and Validation: Multi-jet Production}

In this section we describe the application of the BLHA2 interface to
calculate jet production at NLO QCD. We have compared all tree-level
and one-loop amplitudes required for two and three jet production, as
well as selected colour and spin correlated matrix elements to
NLOJet++~\cite{Nagy:2003tz} (via a dedicated
interface~\cite{Kotanski:2011}) and found full agreement. Let us
stress that the comparison of the one loop contributions, {\it i.e.}
one loop matrix elements and integrated subtraction dipoles, also
provided another non-trivial check of the framework, since the
conventions of the finite one loop pieces differ between NLOJet++ and
the BLHA2 interface. On top of this, there are different
regularization schemes used, however both of these complications did
not complicate the implementation as Matchbox provides a transparent
way of communicating conventions on the one loop contributions to the
integrated dipoles. To present few more details on how the interface
is working, we show an example BLHA2 order file produced by Herwig++
for the $d\bar{d}\to g g$ channel, and the contract file produced by
{\tt njet.py} in figure~\ref{fig:ordercontract}. These files are
automatically processed; Herwig++ is run as with any other builtin
process by a simple addition to the Matchbox input file,
\begin{small}
\begin{verbatim}
library HwMatchboxNJet.so
cd /Herwig/MatrixElements/Matchbox/Amplitudes
create Herwig::NJetAmplitude NJet
insert /Herwig/MatrixElements/Matchbox/PPFactory:Amplitudes 0 NJet
\end{verbatim}
\end{small}
and the process of interest, say three-jet production, is enabled by
\begin{small}
\begin{verbatim}
set /Herwig/MatrixElements/Matchbox/PPFactory:OrderInAlphaS 3
do /Herwig/MatrixElements/Matchbox/PPFactory:Process p p j j j
\end{verbatim}
\end{small}
 Note that the coupling power is always taken with respect to the
 leading order process.

\begin{figure}
\begin{center}
\framebox(235,300){%
    \parbox{225\unitlength}{\tt\scriptsize
\# OLP order file created by Herwig++/Matchbox\\
InterfaceVersion          BLHA2\\
Model                     SM\\
CorrectionType            QCD\\
IRregularisation          CDR\\
Extra HelAvgInitial       no\\
Extra ColAvgInitial       no\\
Extra MCSymmetrizeFinal   no\\
NJetReturnAccuracy        yes\\
NJetRenormalize           yes\\
NJetNf                    2\\
AlphasPower               2\\
AlphaPower                0\\
AmplitudeType             loop\\
1 -1 $->$ 21 21\\
AmplitudeType             cctree\\
1 -1 $->$ 21 21\\
AmplitudeType             sctree\\
1 -1 $->$ 21 21\\
AlphasPower               3\\
AmplitudeType             tree\\
1 -1 $->$ 21 21 21
}}\parbox{5\unitlength}{\ }\framebox(235,300){%
    \parbox{225\unitlength}{\tt\scriptsize
\# OLP order file created by Herwig++/Matchbox\\
\# Generated by njet.py, do not edit by hand.\\
\# Signed by NJet 364882640.\\
\# 1021 1 1e-05 0.01 2 3 1 1 3 2\\
InterfaceVersion BLHA2 $|$ OK\\
Model SM $|$ OK\\
CorrectionType QCD $|$ OK\\
IRregularisation CDR $|$ OK\\
Extra HelAvgInitial no $|$ OK\\
Extra ColAvgInitial no $|$ OK\\
Extra MCSymmetrizeFinal no $|$ OK\\
NJetReturnAccuracy yes $|$ OK\\
NJetRenormalize yes $|$ OK\\
NJetNf 2 $|$ OK\\
AlphasPower 2 $|$ OK\\
AlphaPower 0 $|$ OK\\
AmplitudeType loop $|$ OK\\
1 -1 $->$ 21 21 $|$ 1 1 \# 2 1 0 1 (-1 -2 3 4)\\
AmplitudeType cctree $|$ OK
1 -1 $->$ 21 21 $|$ 1 2 \# 2 1 0 3 (-1 -2 3 4)\\
AmplitudeType sctree $|$ OK\\
1 -1 $->$ 21 21 $|$ 1 3 \# 2 1 0 5 (-1 -2 3 4)\\
AlphasPower 3 $|$ OK\\
AmplitudeType tree $|$ OK\\
1 -1 $->$ 21 21 21 $|$ 1 4 \# 6 1 0 2 (-1 -2 3 4 5)\\
}}
\caption{\label{fig:ordercontract}Example order and contract files of the BLHA2 interface.}
\end{center}
\end{figure}

\begin{figure}
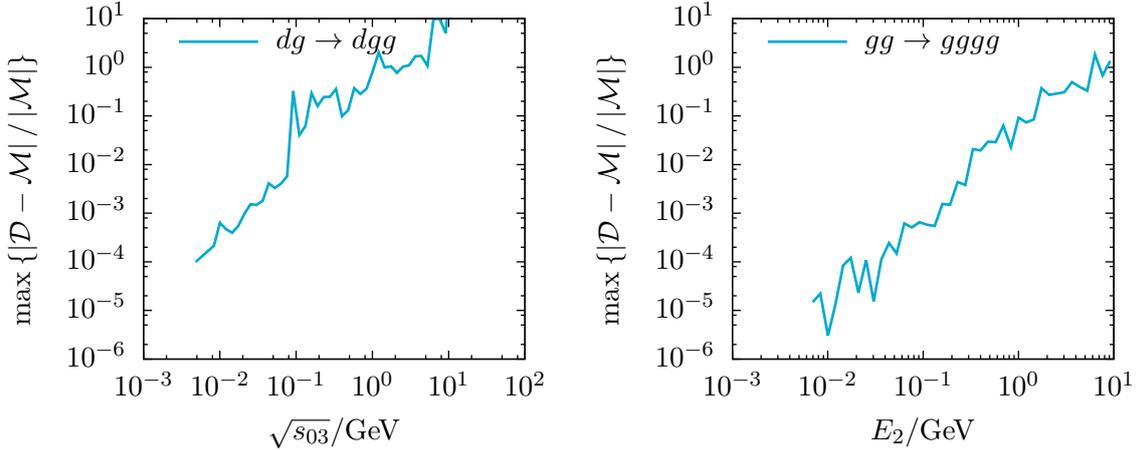

\begin{center}
\input{njet-herwig/dgdgg-0-3-plot}
\input{njet-herwig/gggggg-2-2-plot}
\end{center}
\caption{\label{figures:sub}Validation of the subtraction for a
  collinear (left) and a soft limit (right). We show the maximum
  relative deviation of the subtraction terms to the real emission
  matrix element, as a function of $s_{03}=2 p_0\cdot p_3$ for
  configurations which are collinear but not soft, and as a function
  of the gluon energy $E_2$ for purely soft configurations. The
  incoming partons are labelled $0,1$ and outgoing partons are
  numbered from $2$ onwards.}
\end{figure}
\begin{figure}
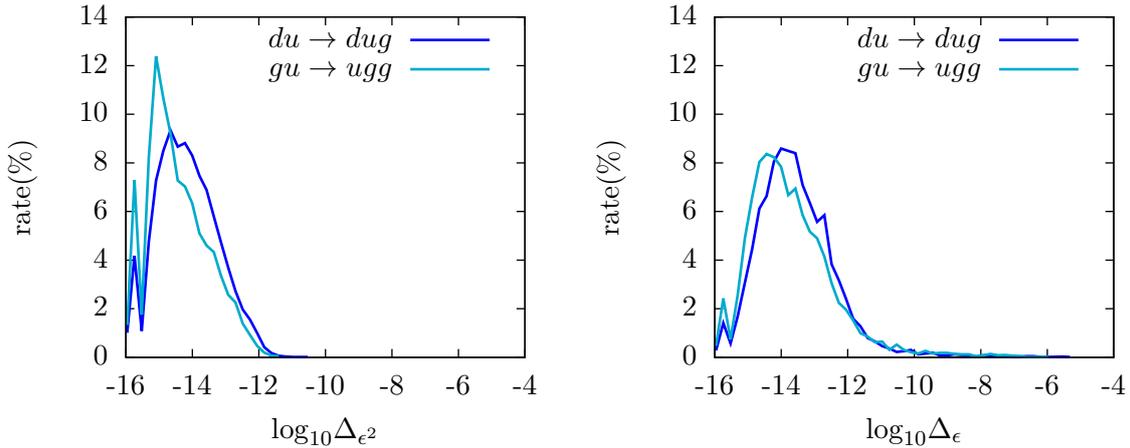

\begin{center}
\input{njet-herwig/pole2-plot}
\input{njet-herwig/pole-plot}
\end{center}
\caption{\label{figures:poles}Validation of the integrated subtraction
  terms. We show the distribution of events as a function of the
  relative difference of the $\epsilon^2$ and $\epsilon$ pole
  coefficients, $\Delta_{\epsilon^2}$ and $\Delta_{\epsilon}$
  respectively.}
\end{figure}
Besides the low-level check of matrix elements on a point-by-point
basis, Matchbox offers a variety of checks to enable a full NLO cross
section calculation, including the cancellation of $\epsilon$-poles
between integrated subtraction terms and one-loop amplitudes, as well
as tests of the singularity cancellation in the unresolved limit of
the real emission matrix element. In figures~\ref{figures:sub} and
\ref{figures:poles} we present few examples of this kind, showing full
functionality of the setup.

\subsection{Outlook \& Conclusions}

We have implemented a tested the BLHA2 interface as a method to pass
all loop-level and tree-level-type matrix elements relevant for all
components of the NLO calculation.  The new features of the BLHA2
allow this to be done in a relatively clean way allowing us to make
maximum use of the matrix elements computed by automated codes. The
code used to produce the matrix elements is publicly available from
\url{https://bitbucket.org/njet/njet/}. The Herwig++ counterpart of
the interface will be included in the next 2.7.x series release of
Herwig++, see \url{http://herwig.hepforge.org}, and the present work
has formed the basis for interfacing Herwig++ to other OLP codes as
well \cite{Bellm:2014}.

Helicity sampling has been shown to be advantageous in a number of cases. From
the formalism we used here it is clear that the born, virtual and integrated
subtraction contributions can have the sum over helicity trivially moved into
the MC program at the cost of passing one extra integer defining the helicity
configuration.  The real+subtraction terms on the other hand are
not so trivial to decompose since the sum of dipoles does not commute
with the sum over helicities. It would be interesting to apply the
techniques developed in reference \cite{Czakon:2009ss} for future
applications of the BLHA interface. We would also like to point out
further work in extending the BLHA2 standard to a level where actual
amplitudes can be interfaced, this being very beneficial to enable
recent developments on improved parton showering along the lines of
\cite{Platzer:2012qg,Platzer:2013fha}. We will of course also explore
phenomenological studies on multi-jet production within a matched or
merged simulation setup.

\subsection*{Acknowledgements}
We would like to thank the Les Houches workshop for hospitality
offered during which some of the work contained herein was
performed. SP acknowledges partial support from the Helmholtz Alliance
`Physics at the Terascale'.
VY acknowledges support by the Alexander von Humboldt Foundation,
in the framework of the Sofja Kovaleskaja Award 2010, endowed
by the German Federal Ministry of Education and Research.

\clearpage
\def\eq#1{{(\ref{#1})}}
\def\LO{{\rm LO}}
\def\NLO{{\rm NLO}}
\newcommand{\gosam}{{\textsc{Go\-Sam}{}}}
\newcommand{\herwig}{{\textsc{Herwig}++}}
\newcommand{\matchbox}{{\textsc{Matchbox}}}
\newcommand{\herwigmatchbox}{{\textsc{Herwig}++/\textsc{Matchbox}}}
\newcommand{\fortran}{{\textsc{Fortran90}}}

\section{GoSam plus Herwig++/Matchbox\footnote{J.~Bellm, S.~Gieseke, N.~Greiner, G.~Heinrich, S.~Pl\"atzer, C.~Reuschle and 
J.F.~von~Soden-Fraunhofen}}

We describe how \gosam~and \herwig~are linked via the new version of
the Binoth-Les Houches-interface (BLHA2), emphasizing new features of
the interface.  We also give a phenomenological example and comment on
prospects to match NLO calculations with a parton shower and
hadronisation in a fully automated way.

\subsection{Introduction}

Next-to-leading order (NLO) calculations matched to a Monte Carlo
program which can provide not only the phase space integration, but
also a parton shower and subsequent hadronisation, should be the new
standard for LHC data analysis.

To this aim, automated tools providing the various ingredients of an
NLO calculation at parton level are indispensable.  These tools
comprise a part providing the one-loop virtual corrections, and parts
providing the ingredients which do not involve loops, i.e. the real
radiation matrix element, the subtraction terms for the infrared
singularities, and the Born matrix element. The full NLO code
therefore can be divided into the categories ``one-loop provider"
(OLP) and ``Monte Carlo program" (MC).

The past years have seen enormous progress in the development of
programs for automated one-loop calculations for multi-particle final
states~\cite{Berger:2008sj,Hahn:2010zi,Campbell:2010ff,Arnold:2011wj,Bevilacqua:2011xh,Hirschi:2011pa,Cullen:2011ac,Cascioli:2011va,
  Actis:2012qn,Badger:2012pg}.  On the Monte Carlo side, a lot of
progress has been achieved in matching NLO real radiation matrix
elements with a parton shower, see
e.g.~\cite{Frixione:2007vw,Alioli:2010xd,Hoeche:2011fd,Platzer:2011bc,Platzer:2012bs,Hoeche:2012yf}.

The various one-loop matrix element generators and Monte Carlo
programs each have different focuses and strengths.  Therefore, in
order to achieve maximal flexibility in connecting various OLP's with
different MC's, it is desirable to have a standard interface allowing
to link these two parts smoothly.  Such an interface has been
initiated at the Les Houches 2009 workshop~\cite{Binoth:2010xt}, known
as the ``Binoth-Les Houches-Accord" (BLHA) and has been updated in
2013~\cite{Alioli:2013nda}.  The latter update will be called BLHA2 in
the following.

Here we present the combination of the one-loop program
\gosam~\cite{Cullen:2011ac} with the Monte Carlo program
\herwig~\cite{Bellm:2013lba,Bahr:2008pv} based on the BLHA2 interface,
with particular emphasis on the new features of the interface.

\subsection{Linking \gosam~and Herwig++/Matchbox via the BLHA2 interface}

Various function calls, as defined according to BLHA2, have been
implemented on both sides in order to enable the communication between
OLP and MC.

\subsubsection{Herwig++/Matchbox}
\label{sec:herwig}

On the side of \herwig~the interface has been realized within the
\matchbox~module~\cite{Platzer:2011bc}, pro\-vi\-ding functionality to
perform hard process generation at the level of NLO QCD accurcay and
easing the setup of run time interfaces to external codes for hard
process generation.  \herwigmatchbox~provides thereby the phase space
generation and integration, as well as the organization of the matrix
elements and of the appropriate real radiation subtraction terms for
the infrared singular regions. The class {\tt \small GoSamAmplitude} has
been implemented in the \matchbox~module, to be included through the
shared object file {\tt \small HwMatchboxGoSam.so}, which itself is globally
operated through the class {\tt \small MatchboxOLPME}.

To configure \herwig~accordingly a prefix with the
installation path ({\tt \small DIR}) to \gosam~has to be specified prior to
the installation: {\tt \small ./configure --with-gosam=DIR}.

The user only needs to enable the interface from the
\herwigmatchbox~process input files, using for example the following syntax:
{\small 
\begin{verbatim}
library HwMatchboxGoSam.so
cd /Herwig/MatrixElements/Matchbox/Amplitudes
create Herwig::GoSamAmplitude Gosam
set Gosam:GoSamSetup /desiredpath/gosam_setup.in
insert /Herwig/MatrixElements/Matchbox/PPFactory:Amplitudes 0 Gosam
\end{verbatim}
}
where a {\tt \small GoSamAmplitude} object {\tt \small Gosam} will be
created. This amplitude can be used to calculate a specific process, which can be 
enabled from the \herwigmatchbox~process input files, as is usual for a \herwigmatchbox~calculation, 
using for example the following syntax:
{\small 
\begin{verbatim}
do PPFactory:Process p p e+ e- j
\end{verbatim}
}
upon which the calculation will be performed fully automatically. This includes Catani-Seymour 
dipole subtraction for NLO calculations as well as matching procedures if required from the input 
file, etc. From a user point of view this means that, except for creating and inserting the {\tt \small GoSamAmplitude} 
object, nothing changes.

Internally, according to BLHA, the process setup is communicated
between OLP and MC by exchanging an order file and a contract file. On
the side of \herwigmatchbox~this is attended to by the functions
{\tt \small GoSamAmplitude::fillOrderFile()} and
{\tt \small GoSamAmplitude::checkOLPContract()}, where an intermediate
process-ID map is used to translate the numbering of the process ID's
between \gosam~and \herwigmatchbox. The contract is ``signed" by
calling the appropriate \gosam-OLP python script within the function
{\tt \small GoSamAmplitude::signOLP()}, whereupon the \gosam-OLP python script reads the 
\gosam~configuration input file {\tt \small /desiredpath/gosam\_setup.in}\footnote{Various \gosam~specific 
options can be set within the \gosam~configuration input file.}. 
Subsequently the process setup is started and the
corresponding \fortran~matrix elements are created by \gosam,
organized into the various subprocesses and helicity configurations
thereof. 
The function {\tt \small GoSamAmplitude::buildGoSam()} then calls
the appropriate {\tt \small make} routine, provided by \gosam, whereupon the
compilation of the various \fortran~matrix elements is performed.

Since \herwigmatchbox~does momentarily not possess the means to create
its own tree-level matrix elements, \gosam~is also used to provide
the various tree-level matrix elements (usual Born matrix elements as
well as spin- and colour-correlated Born matrix elements). The various
subprocesses, specified in the order/contract file, are therefore
grouped into various amplitude types ({\tt \small Tree}, {\tt \small ccTree},
{\tt \small scTree} and {\tt \small Loop}) as well as the different powers of
$\alpha_s$, as specified according to BLHA2.

The function {\tt \small GoSamAmplitude::startOLP()} loads the object file
{\tt \small libgolem\_olp.so} from the \fortran~matrix element
compilation. It also calls the \gosam~\fortran~function
{\tt \small OLP\_Start()}, as well as the \gosam~\fortran~function
{\tt \small OLP\_SetParameter()} to set the initial parameters of the
calculation. If required, it also calls for the
\gosam~\fortran~function {\tt \small OLP\_PrintParameter()}, which prints the
parameters of the calculation into an output file. From the
\herwigmatchbox~process input files this feature can be enabled as
follows (see example above):
{\small
\begin{verbatim}
set Gosam:PrintParameter On
\end{verbatim}
}
\noindent The runtime evaluation for each phase-space point is performed by the functions 
{\small
\begin{verbatim}
GoSamAmplitude::evalSubProcess(),
GoSamAmplitude::evalColourCorrelator(),
GoSamAmplitude::evalSpinColourCorrelator(),
\end{verbatim}
}
depending on the various amplitude types. Within those functions the
\gosam~\fortran~function {\tt \small OLP\_Eval SubProcess2()} is called, where
the type of ouput depends on the amplitude type in question.

To acquire the spin-correlated Born matrix elements the exchange of
the polarization vectors of the associated gluons is required. For a
specific spin-correlation the corresponding polarization vector is
communicated through the function
{\tt \small GoSamAmplitude::plusPolarization()} and the
\gosam~\fortran~function {\tt \small OLP\_Polvec()}, as specified according
to BLHA2. The spin-colour-correlated Born matrix element can be
written as
\begin{multline}
\langle{\cal M}_\mu|{\mathbf C}_{ij} C^{\mu\nu}|{\cal M}_\nu\rangle =
\frac{1}{Q^2} \left[
\langle{\cal M}|{\mathbf C}_{ij}|{\cal M}\rangle
\left(-C Q^2 + |\epsilon_+\cdot q|^2\right)\right.\\ +\left.
2{\rm Re}\left((\epsilon_+\cdot q)^2
\begin{cases}
\langle{\cal M}_-|{\mathbf C}_{ij}|{\cal M}_+\rangle & \text{outgoing }g\\
\langle{\cal M}_+|{\mathbf C}_{ij}| {\cal M}_-\rangle & \text{incoming }g\\
\end{cases}
\right)\right]
\end{multline}
where ${\cal M}_\pm$ refers to the amplitude with positive/negative
gluon helicity and a sum over helicities is understood if ${\cal M}$
carries no subscript. The polarization vector which is communicated
through the function {\tt \small OLP\_Polvec()} is hereby $\epsilon_+$
\cite{Alioli:2013nda} and the standard is that of outgoing gluons,
i.e. $\langle{\cal M}_-|{\mathbf C}_{ij}|{\cal M}_+\rangle$ for the
spin-correlated result. In the case of incoming gluons the
polarization vectors are simply complex conjugated within the function
{\tt \small GoSamAmplitude::plusPolarization()}, which yields the correct
result since only the real part is of interest here.

The calls to the \fortran~functions of \gosam~are realized by external
function calls in the class {\tt \small GoSamAmplitude}:
{\small
\begin{verbatim}
extern "C" void OLP_Start(const char*, int*);
extern "C" void OLP_Polvec(double*,double*,double*);
extern "C" void OLP_SetParameter(char*,double*,double*,int*); 
extern "C" void OLP_PrintParameter(char*); 
extern "C" void OLP_EvalSubProcess2(int*,double*,double*,double*,double*);
\end{verbatim}
}
The class {\tt \small GoSamAmplitude} also provides an automated way to
hand over the correct electroweak parameters according to the chosen
electroweak scheme in the \herwig~process input files, via the
function {\tt \small OLP\_SetParameter()}. \gosam~automatically chooses the
appropriate scheme upon recognizing the corresponding electroweak
parameters. Note also that, regarding regularization schemes,  conventional 
dimensional regularization (cdr) as well as dimensional reduction (dred) are supported 
and yield the same results.

\subsubsection{\gosam}
\begin{figure}[h]
\begin{center}
\framebox(200,300){%
    \parbox{180\unitlength}{\tt\scriptsize
\# OLP order file created by \\
\# Herwig++/Matchbox for GoSam\\

InterfaceVersion         BLHA2\\
MatrixElementSquareType  CHsummed\\
CorrectionType           QCD\\
IRregularisation         CDR\\

AlphasPower              1\\
AmplitudeType ccTree\\
1 -1 $->$ -11 11 21 \\
...\\
21 3 $->$ -11 3 11 \\

AmplitudeType scTree\\
1 -1 $->$ -11 11 21\\ 
...\\ 
21 3 $->$ -11 3 11 \\

AmplitudeType Loop\\
1 -1 $->$ -11 11 21\\ 
...\\
21 3 $->$ -11 3 11 \\
\\
AlphasPower              2\\
AmplitudeType Tree\\
1 1 $->$ -11 1 1 11\\ 
...\\
21 21 $->$ -11 -3 3 11 }}\parbox{5\unitlength}{\ }\framebox(200,300){%
    \parbox{180\unitlength}{\tt\scriptsize
\# vim: syntax=olp\\
\#@OLP GOSAM 2.0.beta\\
\#@IgnoreUnknown False\\
\#@IgnoreCase False\\
\#@SyntaxExtensions \\
InterfaceVersion BLHA2 $|$ OK\\
MatrixElementSquareType CHsummed $|$ OK\\
CorrectionType QCD $|$ OK\\
IRregularisation CDR $|$ OK\\
AlphasPower 1 $|$ OK\\
AmplitudeType ccTree $|$ OK\\
1 -1 $->$ -11 11 21 $|$ 1 131\\
...\\
21 3 $->$ -11 3 11 $|$ 1 70\\
AmplitudeType scTree | OK\\
1 -1 $->$ -11 11 21 $|$ 1 145\\
...\\
21 3 $->$ -11 3 11 $|$ 1 71\\
AmplitudeType Loop $|$ OK\\
1 -1 $->$ -11 11 21 $|$ 1 137\\
...\\
21 3 $->$ -11 3 11 $|$ 1 63\\
AlphasPower 2 $|$ OK\\
AmplitudeType Tree $|$ OK\\
1 1 $->$ -11 1 1 11 $|$ 1 42\\
...\\
21 21 $->$ -11 -3 3 11 $|$ 1 106\\}}
\caption{Automatically created order and contract file for Z+jet. The various subprocesses
  are given a process ID by the OLP.}
\label{figure:ordercontract}
\end{center}
\end{figure}  
On the \gosam~side, dedicated OLP modules are provided which host the
interface.  It should be emphasized again that the user does not
need to have a detailed knowledge of these modules, but only interacts
via the user interface of the Monte Carlo program, in the present case
the \herwig~process input files.

The function {\tt \small  OLP\_Start(char* fname, int* ierr)}, which was
already in place with the first version of the BLHA interface, takes
the name of the contract file as first argument and initializes the
OLP setup.  The function {\tt \small OLP\_Info(char olp\_name,char
  olp\_version,char message)} prints the \gosam~ version, svn revision
number and references which should be cited when using the program.

An important new feature of BLHA2 is the function 
{\tt \small OLP\_SetParameter()}, 
which allows to set parameters like $\alpha_s$, $\alpha$, masses,
widths etc.  These parameters are set in the \herwig~process input
files.  Note that this function can also be used to exchange dynamic
parameters like $\alpha_s(\mu^2)$ at runtime.

At runtime, the OLP returns the values for the virtual amplitude to the 
MC via the function {\tt \small OLP\_EvalSub Process2()}. 
As compared to the original function {\tt \small OLP\_EvalSubProcess()}, the
new function {\tt \small OLP\_Eval SubProcess2()} does not contain the passing
of coupling constants anymore, as their values are now passed
separately by using {\tt \small OLP\_SetParameter()}\footnote{The function
  {\tt \small OLP\_EvalSubProcess()}, using BLHA1 conventions, is also still
  in operation in \gosam.}.  It also has a new argument appended which
serves to assess the accuracy of the result returned by the OLP.  The
arguments of {\small \tt OLP\_EvalSubProcess2()} are described in
detail in ref.~\cite{Alioli:2013nda}.  Here we point out that the
amplitude types {\tt \small scTree} and {\tt \small ccTree}, denoting spin- and
colour-correlated Born matrix elements, can also be provided by
\gosam, as described already in section \ref{sec:herwig}.  This allows
\herwigmatchbox~to construct the real radiation subtraction terms for
the infrared singular regions.  To this end \gosam~also provides the
function {\tt \small OLP\_Polvec()} as described in section \ref{sec:herwig}.

At this point we would like to stress another useful feature of
\gosam, which is the ability to read in model files in the {\tt \small UFO}
(Universal Feynrules Output)~\cite{Degrande:2011ua} format and which
is particularly useful for calculations beyond the Standard Model.
The {\tt \small UFO} format also provides human readable name attributes for
the model parameters, as well as the SLHA
identifiers~\cite{Skands:2003cj}, which are processed by the
subroutine {\tt \small read\_slha\_file}.

\subsection{Example Z+jet at NLO}

\begin{figure}[t]
\begin{center}
\parbox{8cm}{\includegraphics[width=0.5\textwidth]{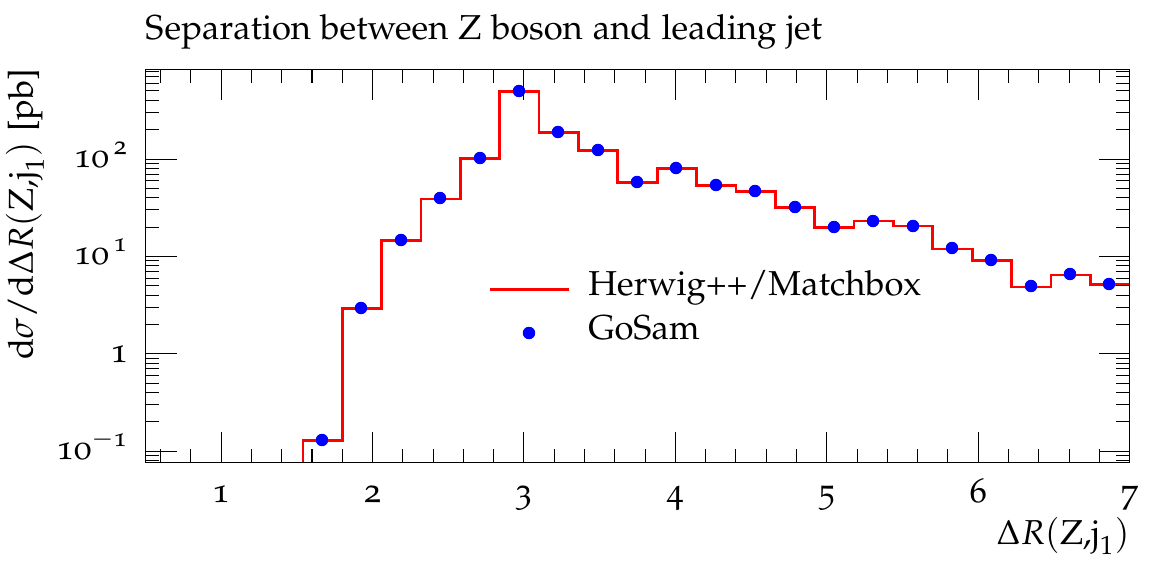}}\parbox{8cm}{\input{gosam-herwig/subtr/uubare+e-gg-1-4-plot}}
\caption{Left: The fixed order parton level R-separation between the Z
  boson and the leading jet in Z+jet production at NLO. Right: Testing
  the numerical stability in the collinear and soft region shows the
  expected behaviour. As an example we show the sum over all
  Catani-Seymour-dipoles\,\cite{Catani:1996vz} divided by the real
  contribution, at phase-space points were the $\bar{u}g_{1}$
  invariant mass is smaller than 10 GeV. Here $s_{14}$ denotes the invariant mass 
  between the $\bar{u}$ and the first gluon in the list.}
\label{fig:nlo}
\end{center}
\end{figure}  

\begin{figure}[t]
\begin{center}
\parbox{8cm}{\includegraphics[width=0.5\textwidth]{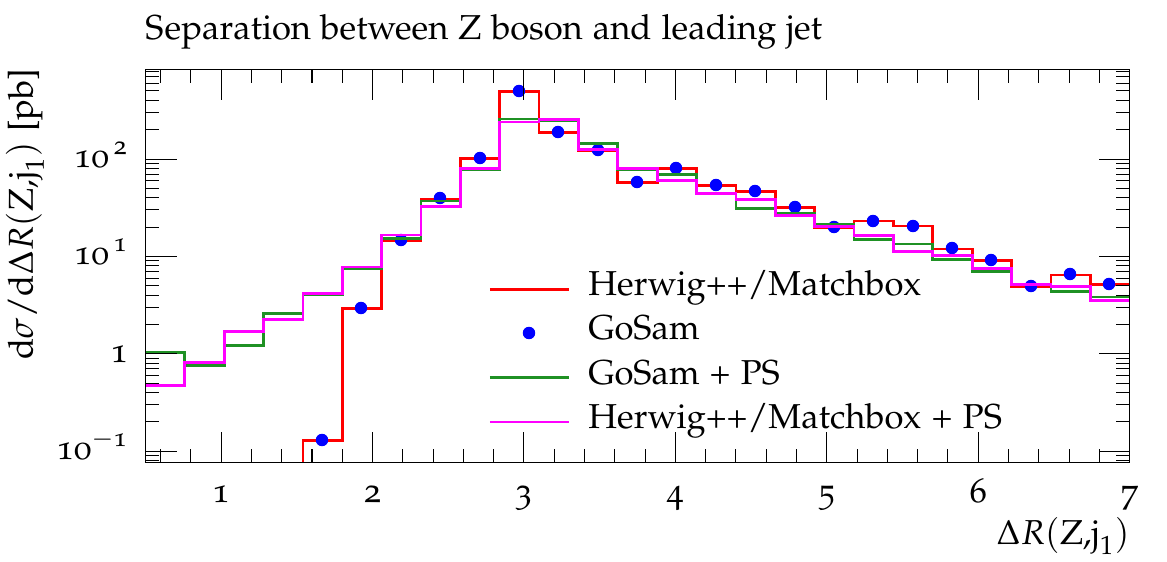}}\parbox{8cm}{\includegraphics[width=0.50\textwidth]{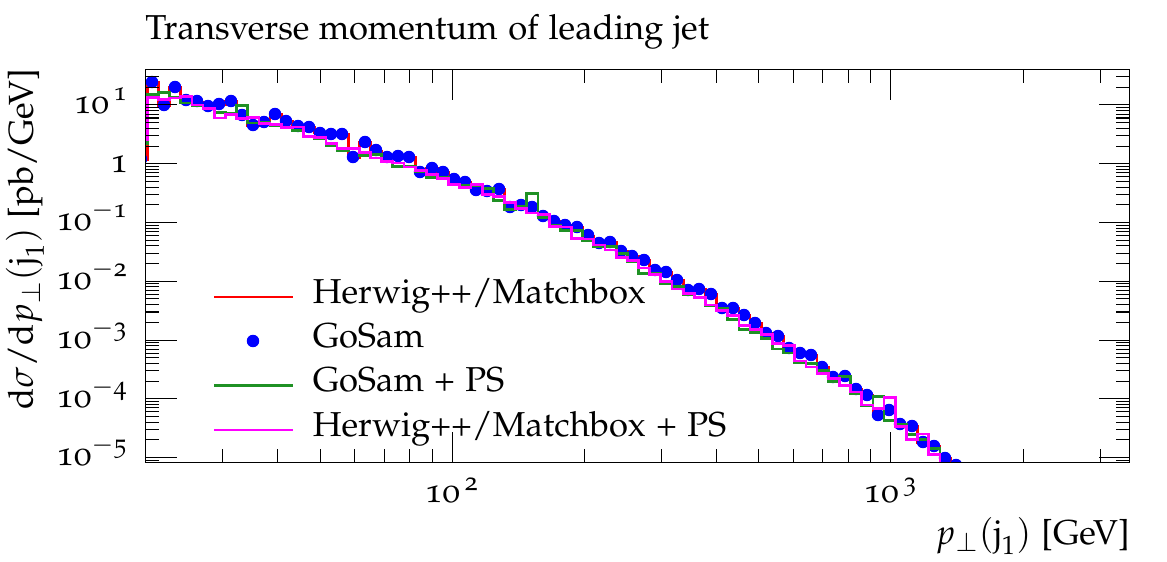}}
\caption{Left: The R-separation between the Z-boson and the leading
  jet in Z+jet production. Right: The transverse momentum of the
  leading jet. Both with and without parton shower matching. The --
  for this process -- inclusive observable $p_\perp$ of the first jet
  is hardly affected by the parton shower. The red and pink lines show
  the results with the \herwigmatchbox~built-in matrix elements, the
  blue dots and green lines show the results of \herwigmatchbox~with
  the \gosam~ matrix elements.}
\label{fig:PS}
\end{center}
\end{figure}  

As an example we use the built-in process $pp\to Z/\gamma^*+$\,jet
$\to e^+e^-+$\,jet\, of \herwigmatchbox~and compare it against the
same automatically generated \gosam~process.  The corresponding, yet
shortened, order and contract files are shown in
fig.~\ref{figure:ordercontract}. Here we want to stress again that the whole 
compuation proceeds fully automatically only upon a few commands in the 
\herwig~input files, which includes the automatic generation of the order and 
the contract file. To get agreement between the
\herwigmatchbox~built-in and the \gosam{} matrix elements, we need to make the
same assumptions on both sides: While in the \herwigmatchbox~built-in matrix
elements the diagrams with photon- and Z-radiation from bottom- and
top-triangles are not implemented, because they are numerically
negligible, \gosam~produces these by default. For the comparison,
these contributions were eliminated by using the appropriate filter
options in the \gosam~configuration input file.

To calculate the cross sections and distributions shown in
fig.~\ref{fig:nlo} and fig.~\ref{fig:PS} we use the minimal set of
cuts

\begin{displaymath}
 66\, \text{GeV} < m_{e^+e^-} <116\, \text{GeV} , \hspace{5mm} p_{\text{T},\text{j}}>20\, \text{GeV}\ ,\hspace{5mm}  |\eta_\text{j}|<5\ ,
\end{displaymath}
and the dynamical scale
\begin{displaymath}
 \mu_r=\mu_F=m_{e^+e^-}\ .
\end{displaymath}

The scale choice may not be ideal from a phenomenological point of
view, but it shows that dynamical scales work fine within the
interface. The shower matched distributions are obtained using the
dipole shower outlined in \cite{Platzer:2009jq} and a modified version
of the MC@NLO-type matching described in \cite{Platzer:2011bc}, which
accounts for subleading colour contributions. The matching can be enabled, 
as is usual for \matchbox~processes, with a single command line:
\begin{verbatim}
set PPFactory:ShowerApproximation DipoleMatching
\end{verbatim}

\subsection{Conclusions \& Outlook}

The combination of one-loop providers (OLP) and Monte Carlo programs
(MC) using a standardized interface has been proven to be a successful
strategy for automated NLO calculations.  We have presented
phenomenological results obtained through the automatized combination
of the Monte Carlo program \herwigmatchbox~with the one-loop program
\gosam.  The link between the two codes is based on the new version of
the Binoth-Les-Houches interface (BLHA2), which has been worked out at
the Les Houches 2013 workshop, including proof-of-concept interfaces
to other OLPs, \cite{Badger:2014}. In particular, the new feature to
pass spin- and colour-correlated tree level matrix elements from the
one-loop provider to the Monte Carlo program allows the
\herwigmatchbox~ module to provide the full NLO real radiation part.
As a first application we calculated the NLO QCD corrections to Z+jet
production at the LHC.  We validated the results with the
\gosam~matrix elements against a corresponding built-in implementation
in \herwigmatchbox.

In combination with the \herwig~shower, the full chain from process
definition to the production of showered event samples at NLO accuracy
is thus provided in a fully automated way.  This opens the door for a
multitude of phenomenological applications.  As both \gosam{} and
\herwig{} are also able to import model files in {\tt \small UFO} format, and
offer various features which are important for calculations involving
unstable particles, the applications are not limited to QCD, but also
well suited for electroweak corrections and BSM calculations.

\subsection*{Acknowledgements}
This work was supported in part by the Marie Curie Initial Training
Network \emph{MCnetITN}, the German Federal Ministry of Education and
Research (BMBF), the Helmholtz Alliance ``Physics at the Terascale'' and
the Karlsruhe School of Elementary Particle and Astroparticle Physics
(KSETA).

\clearpage

\chapter{Parton distribution functions}
\label{cha:pdf}
\def\gsim{\mathrel{\rlap{\lower4pt\hbox{\hskip1pt$\sim$}}
    \raise1pt\hbox{$>$}}}         
\def\lsim{\mathrel{\rlap{\lower4pt\hbox{\hskip1pt$\sim$}}
    \raise1pt\hbox{$<$}}}         

\section{PDF dependence of the Higgs production cross section in gluon fusion from HERA data%
\footnote{A.~Cooper-Sarkar, S.~Forte, J.~Gao, J.~Huston, P.~Nadolsky,
J.~Pumplin, V.~Radescu, J.~Rojo, R.~Thorne and C.-P.~Yuan}}
%



We document a detailed comparison of 
theoretical settings and fitting methodologies employed 
in the determination of parton distribution functions (PDFs) of
the proton, focusing on the factors driving the PDF uncertainty
in Higgs boson production in the gluon fusion channel.
In order to understand the moderate discrepancies among predictions
obtained using CT10, HERAPDF1.5, MSTW08 and NNPDF2.3  PDF ensembles,
we first complete a benchmarking comparison of computations
of NNLO neutral current DIS cross sections used 
in the fitting codes of these groups. Then, 
we compare results of the fits performed 
by following the methodologies of the four groups to a reduced 
common set of DIS data from HERA experiments.
We conclude that the predictions for $gg\to H$ observables,
obtained when the PDFs are constrained using only the HERA DIS data,
are in fairly good agreement, within the PDF uncertainty. 



\subsection{Introduction\label{sec:intro}}

Following the discovery of the Higgs boson at the LHC,
it is crucial to measure all of its properties precisely
in order to identify the underlying mechanism of
electroweak symmetry breaking.
 PDF uncertainty is one of the
limiting factors in the accuracy with which one can extract the Higgs
boson couplings from experimental data. Its control 
is needed for precise Higgs characterization~\cite{Heinemeyer:2013tqa}.

It has been shown that recent
NNLO PDFs from various PDF analysis groups are in generally good agreement.
However, moderate
differences due to the gluon PDFs are observed around the Higgs mass region,
at $\approx 125\, {\rm GeV}$. This leads to discrepancies at the level
of 1 to 2 standard deviations~\cite{Ball:2012wy} in the predictions for
the Higgs production cross sections through gluon fusion,
which is the dominant production channel at the LHC.
In part because the origin of this spread in the 
results is not understood, the PDF4LHC
convention~\cite{Botje:2011sn}, adopted by the Higgs working group~\cite{Dittmaier:2011ti,Heinemeyer:2013tqa}, recommends a conservative estimate for the 
combined PDF+$\alpha_s$ uncertainty based on the envelope 
of results obtained using different PDF sets. 
If this sizable uncertainty is of theoretical origin, it 
would remain irreducible unless its source is identified. 
Whereas the problem may be resolved after including more
experimental data, this is by no means guaranteed, and thus it is
important to investigate the origin of the
differences among the various PDF sets and, if possible, reduce or
even eliminate it. 

There have been in the past several
related PDF benchmarking studies, some in the context of other
Les Houches workshops.
The evolution of PDFs was benchmarked in ref.~\cite{Giele:2002hx}, where
the HOPPET~\cite{Salam:2008qg} and
PEGASUS~\cite{Vogt:2004ns} codes were compared, including the effect of
scale variations.
This exercise was
extended within the context of the HERA-LHC workshop to NNLO evolution and polarized
PDF evolution~\cite{Dittmar:2005ed}.
Differences in general-mass heavy-quark schemes for DIS charm structure
functions, $F_{2,c}$, were studied in detail in ref.~\cite{Binoth:2010ra},
where good agreement between the ACOT, FONLL and TR schemes was found up to
different treatments of powerlike corrections.
Finally, other recent benchmarking exercises have focused on the comparison
of PDFs and their predictions for
LHC cross sections~\cite{Ball:2012wy,Demartin:2010er,
Watt:2011kp,Alekhin:2011sk,Alekhin:2010dd,Alekhin:2012ig},
and the aforementioned differences in the gluon PDFs were already noted.

The main goal of this contribution is to begin a benchmarking exercise
aimed at advancing the control of PDF dependence of 
the $gg\to H$ cross section. We will restrict ourselves to the comparison
of the PDFs based on the general-mass schemes for heavy-quark
scattering contributions, as  it has been noted in a number of studies
~\cite{CooperSarkar:2007ny,Thorne:2012az,Ball:2013gsa,Thorne:2014toa}
that the PDFs (and $\alpha_s$) can differ rather significantly if a
fixed-flavor number (FFN) scheme is used.\footnote{Studies of
the gluon PDF and predictions for the Higgs and other benchmark LHC cross
sections in the FFN scheme are available 
in \cite{Alekhin:2010dd,Alekhin:2012ig,Alekhin:2013nua}.}
The recent NNLO sets utilizing the general-mass schemes include
CT10~\cite{Gao:2013xoa}, HERAPDF1.5~\cite{CooperSarkar:2011aa}, MSTW2008~\cite{Martin:2009iq} and NNPDF2.3~\cite{Ball:2012cx}.
Except for HERAPDF1.5,
which is based on inclusive HERA data only, three other PDF
sets include information from a variety of data sets from fixed-target and
collider DIS, as well as from hadron-hadron scattering processes, 
including, in some cases, the LHC data.
Detailed comparisons of these NNLO PDFs and the respective LHC predictions 
can be found in a recent benchmarking study~\cite{Ball:2012wy}.
In all these fits, the backbone of the fit is provided by
the combined HERA-1 inclusive DIS data~\cite{Aaron:2009aa}, which also
plays an important role in constraining the gluon PDFs in the
Higgs mass region~\cite{Dulat:2013kqa}.
Recently combined charm quark production measurements at HERA 
have also been published~\cite{Abramowicz:1900rp}, 
which introduce an additional constraint on the gluon PDF.

Given the relevance of the HERA data for pinning down the gluon PDF,
and the fact that this data set underwent careful
examination of all the experimental uncertainties, we have chosen
to initiate this benchmarking exercise by first studying the fits 
that include exclusively the HERA-1 data.
In this highly controlled environment,  we can compare the output
from the different fitting codes and try to isolate
possible sources of discrepancy related to, for example,
implementation of reduced cross section calculations,
alternative parametrizations, kinematical
cuts used in the fits, and the different definitions of $\chi^2$ used in
determining the fit quality. We can then study how these differences
translate into the predictions for Higgs cross sections 
in the gluon fusion channel.

As a prerequisite for this benchmarking exercise, we compared
the NNLO QCD predictions for the neutral-current DIS cross sections 
from four different codes, but always using the same PDF set. 
This is an important starting point:
we show that, when all groups use common PDFs and settings, they find 
good agreement among their NC DIS cross sections, 
with (small) remaining differences arising from known sources 
such as the selection of the heavy-quark scheme. 

This contribution is organized as follows. In Section 2
we briefly recall the status of the PDF dependence of the
NNLO Higgs cross section in gluon fusion. In Section 3
we present the results of the benchmarking 
of the NNLO neutral current DIS structure functions from different
groups using a common PDF set.
Section 4 contains the main results of this study,
where fits to HERA only data are performed, and the impact of various
theoretical and methodological settings is assessed.
Finally in Section 5 we summarize our main results and outline
the continuation of this benchmarking exercise beyond the HERA-only fits.

\subsection{The Higgs cross section via 
gluon fusion obtained with recent NNLO PDF sets\label{sec:global}}

\begin{figure}[h]
\centering
 \includegraphics[width=0.48\textwidth]{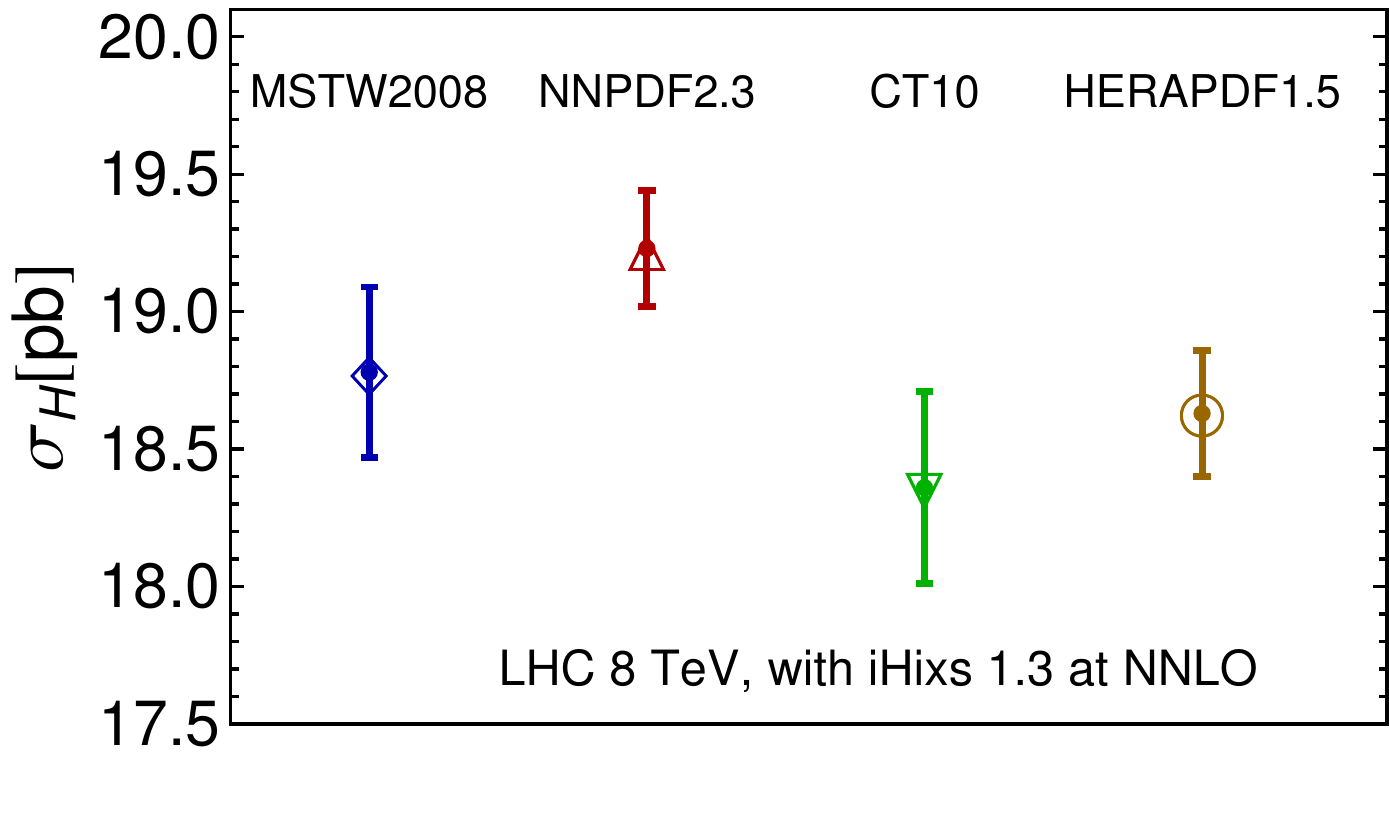}
\hspace{0.2in}
 \includegraphics[width=0.48\textwidth]{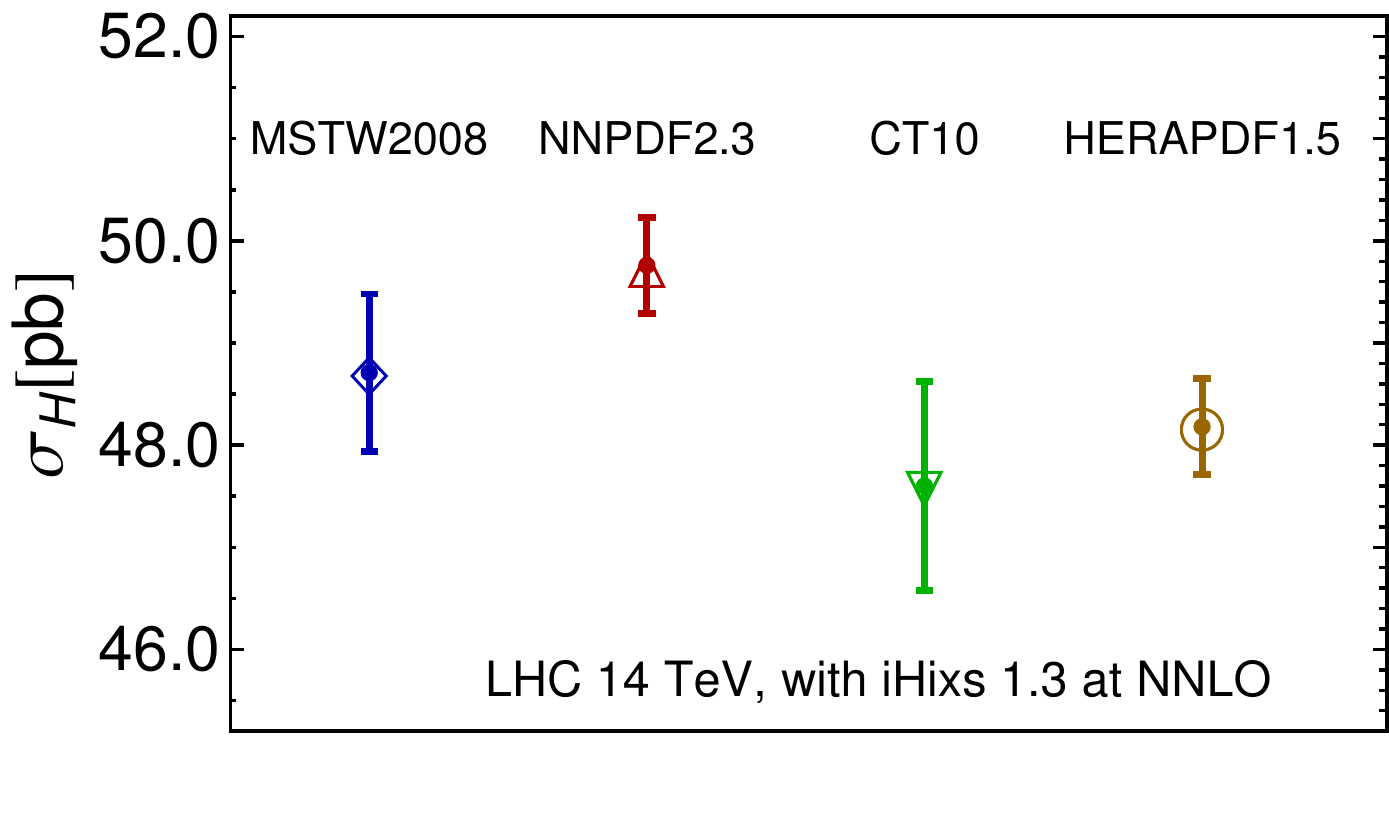}
\vspace{-1ex}
 \caption{\label{fig:globalinc}
Predictions for inclusive cross sections of the SM Higgs boson production
through gluon fusion based on recently published NNLO PDFs.}
\end{figure}

We first  review the settings of the four recent NNLO PDF determinations
that will be studied in this paper.
Table~\ref{tab:global1} summarizes the heavy-quark scheme, the scale at
which the PDFs are parametrized, the charm-quark mass, 
and the kinematic cuts on DIS data.
 Note that, in addition to these differences in settings,
the four PDF analyses also differ in the data sets that they include.
We compare the corresponding predictions for the Higgs cross section
in the gluon fusion channel
in Table~\ref{tab:global2} and Fig.~\ref{fig:globalinc}.
These predictions use versions of the PDF sets with a common
$\alpha_s(M_Z)$ value of 0.118 
and are calculated with iHixs1.3~\cite{Anastasiou:2011pi} at NNLO in QCD,
with both Higgs boson mass and QCD scales set to 125 GeV 
in the heavy top-quark limit.
We show  the PDF uncertainties at the 68\% confidence level (c.l.).
For the LHC at 8 TeV, the two extreme
results, CT10
and NNPDF2.3, differ by about 4$\%$, or $2\sigma$ if their uncertainties are
added in quadrature.
This spread is the origin of the significant  PDF+$\alpha_s$
uncertainty in the $gg\rightarrow H$ channel~\cite{Heinemeyer:2013tqa}.
A similar results holds for the LHC at 14 TeV. 

\begin{table}[h]
\centering
\begin{tabular}{|c|c|c|c|c|c|c|}
\hline
PDFs set & Ref. & Initial & Heavy-quark& $m_c$ & $Q^2$ cut & $W^2$ cut \tabularnewline
 &  & scale [GeV] & scheme  & mass [GeV] & $\ge$ ${\rm GeV}^2$ & $\ge$ ${\rm GeV}^2$ \tabularnewline
\hline
\hline
CT10 & \cite{Gao:2013xoa} & 1.3 & S-ACOT-$\chi$~\cite{Guzzi:2011ew} & 1.3 & 4 & 12.25 \tabularnewline
\hline
HERAPDF1.5 & \cite{CooperSarkar:2011aa} &1.378 & default TR' & 1.4 & 3.5 & N.A.${}^*$
\tabularnewline
\hline
MSTW'08 & \cite{Martin:2009iq} & 1.0 & default TR'~\cite{Thorne:2006qt} & 1.4 & 2.0 & 15.0 \tabularnewline
\hline
NNPDF2.3 ($N_f=5$) & \cite{Ball:2012cx} & 1.414 & FONLL-C~\cite{Forte:2010ta} & 1.414 & 3.0 & 12.5  \tabularnewline
\hline
\end{tabular}
{\protect{\begin{flushleft}\small ${}^*$All HERA data have $W^2 > 350$ GeV$^2$.\end{flushleft}}}
\caption{Basic inputs of the recently published NNLO PDFs based on a GM-VFN scheme.\label{tab:global1}}

\end{table}

\begin{table}[h]
\centering
\begin{tabular}{|c|c|c|c|c|}
\hline
$\sigma_H$ [pb] & CT10& MSTW'08 & NNPDF2.3  & HERAPDF 1.5 \tabularnewline
\hline
\hline
LHC 8 TeV &       18.36$\pm 0.35$ &      18.78$\pm 0.31$ &  19.23$\pm 0.21$ &         18.63$\pm 0.23$  \tabularnewline
\hline
LHC 14 TeV &        47.60$\pm 1.02$ &    48.71$\pm 0.77$ &   49.76$\pm 0.47$ &        48.18$\pm 0.47$ \tabularnewline
\hline
\end{tabular}
\caption{Predictions for inclusive cross sections of SM Higgs boson production
through gluon fusion utilizing recently published NNLO PDFs.
    \label{tab:global2}}
\end{table}

\begin{figure}[h]
\centering
 \includegraphics[width=0.55\textwidth]{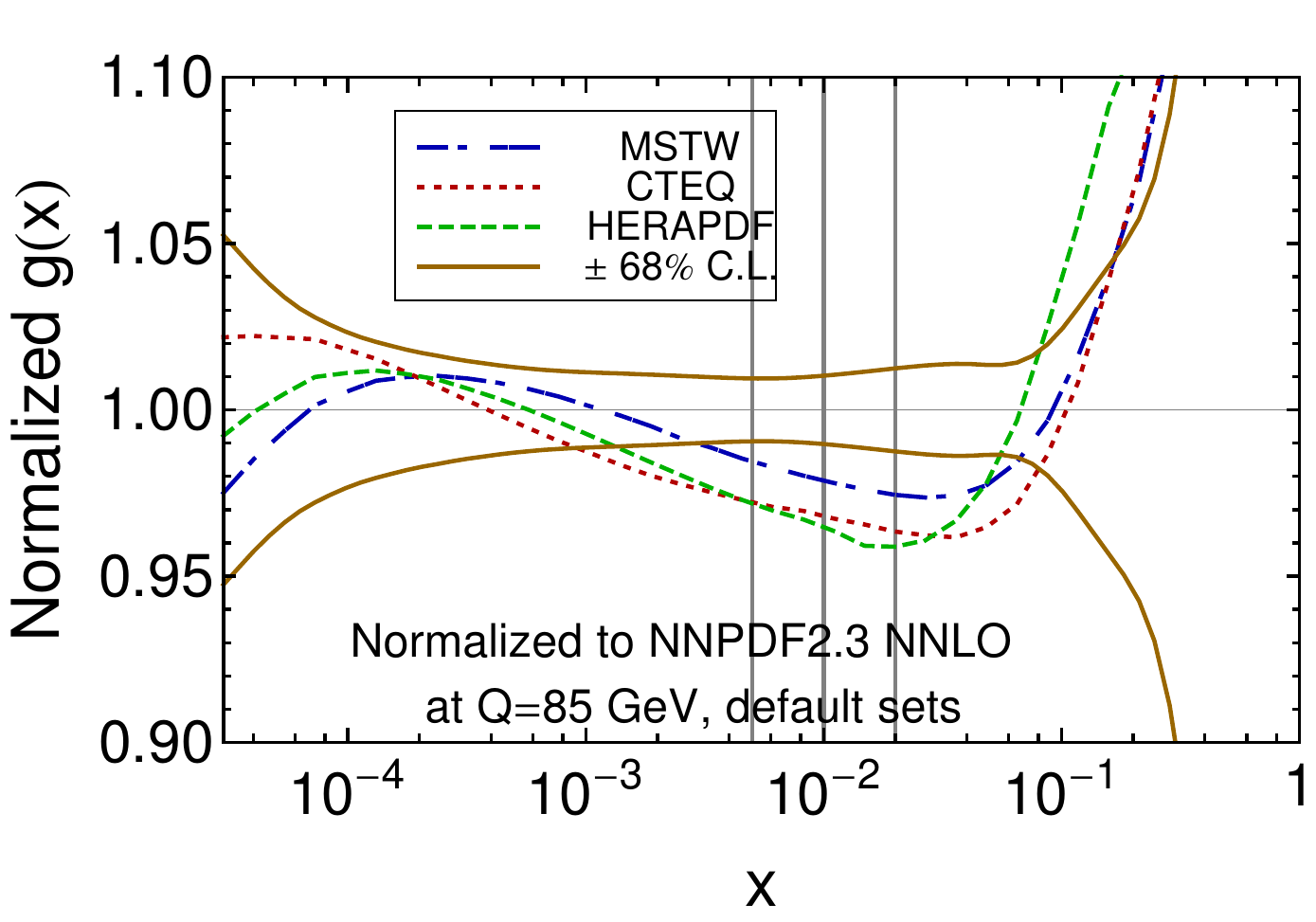}
\vspace{-1ex}
 \caption{\label{fig:globalglu} Comparison of central gluon PDFs from
 recently published NNLO PDF sets at a common scale $Q=85\ {\rm GeV}$,
 normalized to the NNPDF2.3 gluon PDF and superimposed on the NNPDF2.3 68\% c.l. PDF uncertainty.}
\end{figure}

\begin{figure}[h]
\centering
 \includegraphics[width=0.48\textwidth]{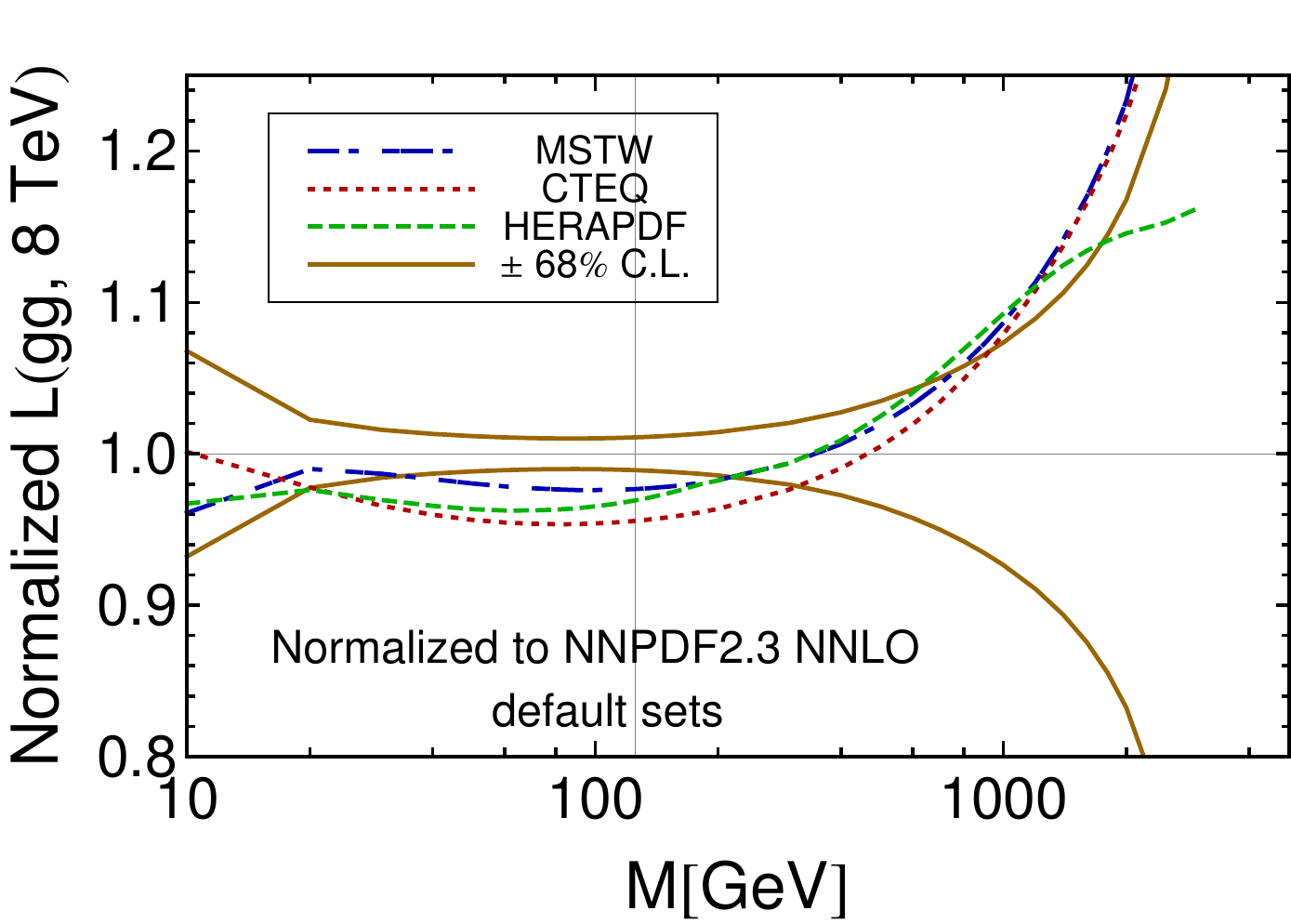}\hspace{0.2in}
 \includegraphics[width=0.48\textwidth]{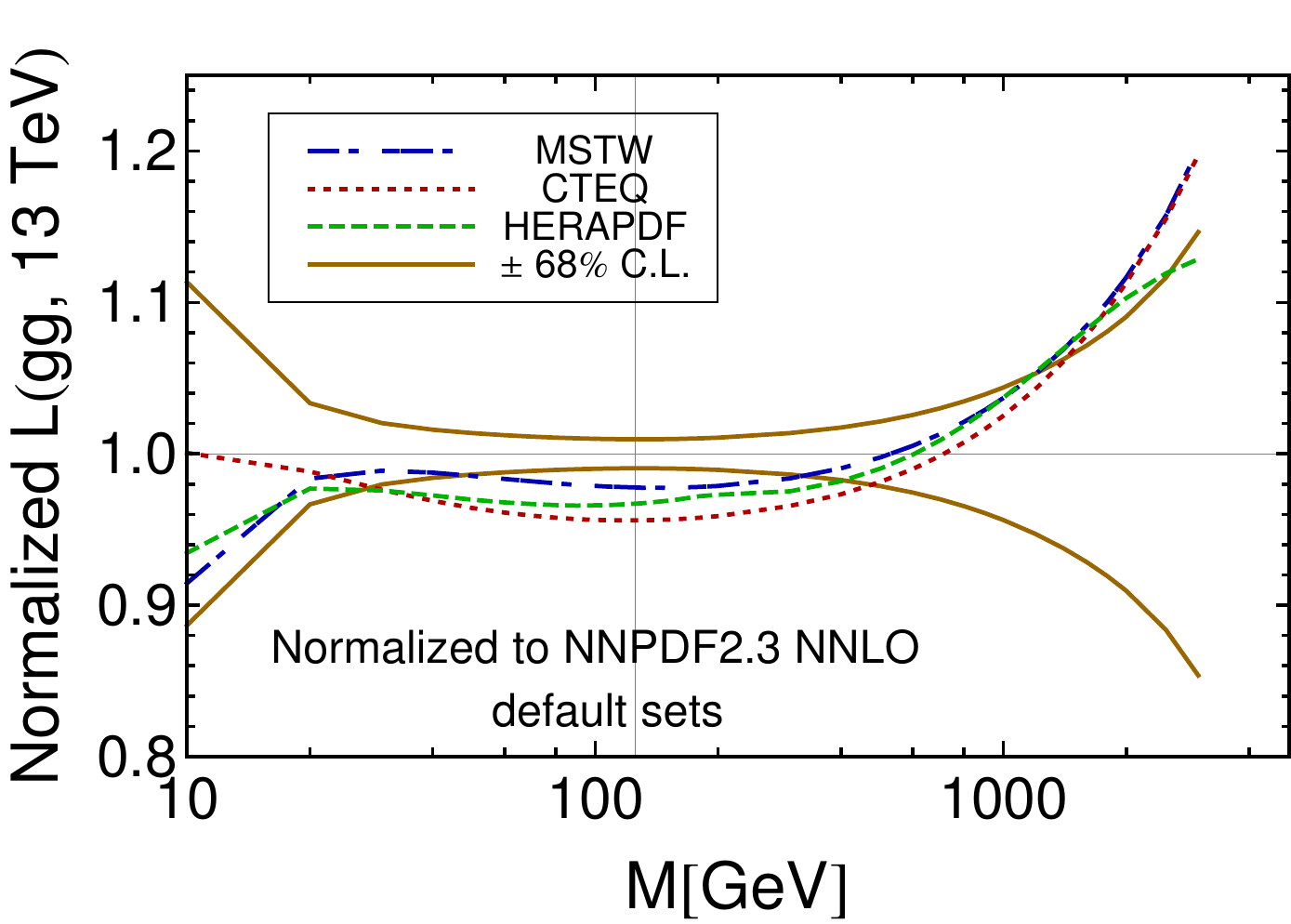}
\vspace{-1ex}
 \caption{\label{fig:globallum} Comparison of the gluon-gluon parton luminosity as
 a function of invariant mass at the LHC 8 and 13 TeV from recently published
 NNLO PDFs, normalized to the NNPDF 2.3 central prediction. The factorization scale
 is set to the invariant mass. The vertical line corresponds to the 
mass of the SM Higgs boson. }
\end{figure}

These discrepancies of the Higgs cross sections
can be traced back to the differences in the normalization
and shape of the corresponding gluon PDFs,
as shown in Fig.~\ref{fig:globalglu}.
In this plot all gluon PDFs are normalized to the gluon PDF of the 
first, or central, set of the NNPDF2.3 ensemble, and only 
the PDF uncertainty of the NNPDF2.3 ensemble is shown in the figure 
for simplicity.  
The three vertical lines indicate
the $x$ values of $0.005$, $0.01$, and $0.02$, covering typical $x$ values for Higgs production in the
central rapidity region, $x= m_H/\sqrt s \sim 0.016\, (0.009)$ for LHC 8 (14) TeV.
We can see that at $Q=85\,{\rm GeV}$, and for $x$ from 0.005 to 0.1,
the gluon PDF from CT10 lies below that of NNPDF2.3. MSTW lie in between,
while HERAPDF is close to CT10 at $x < 0.02$ and larger than CT10 at $x>0.02$. 
Fig.~\ref{fig:globallum} shows a comparison of
the gluon-gluon parton luminosity as a function of the invariant mass of
the final state. The factorization scale is set to the invariant
mass. The vertical line indicates the mass of the SM Higgs boson at 125 GeV. 
For higher
invariant masses ($> 400 $ GeV), all predictions are within 68\%
c.l. uncertainties of NNPDF2.3 until very high masses, ($> 1000 $ GeV).

\begin{figure}[h]
\centering
 \includegraphics[width=0.48\textwidth]{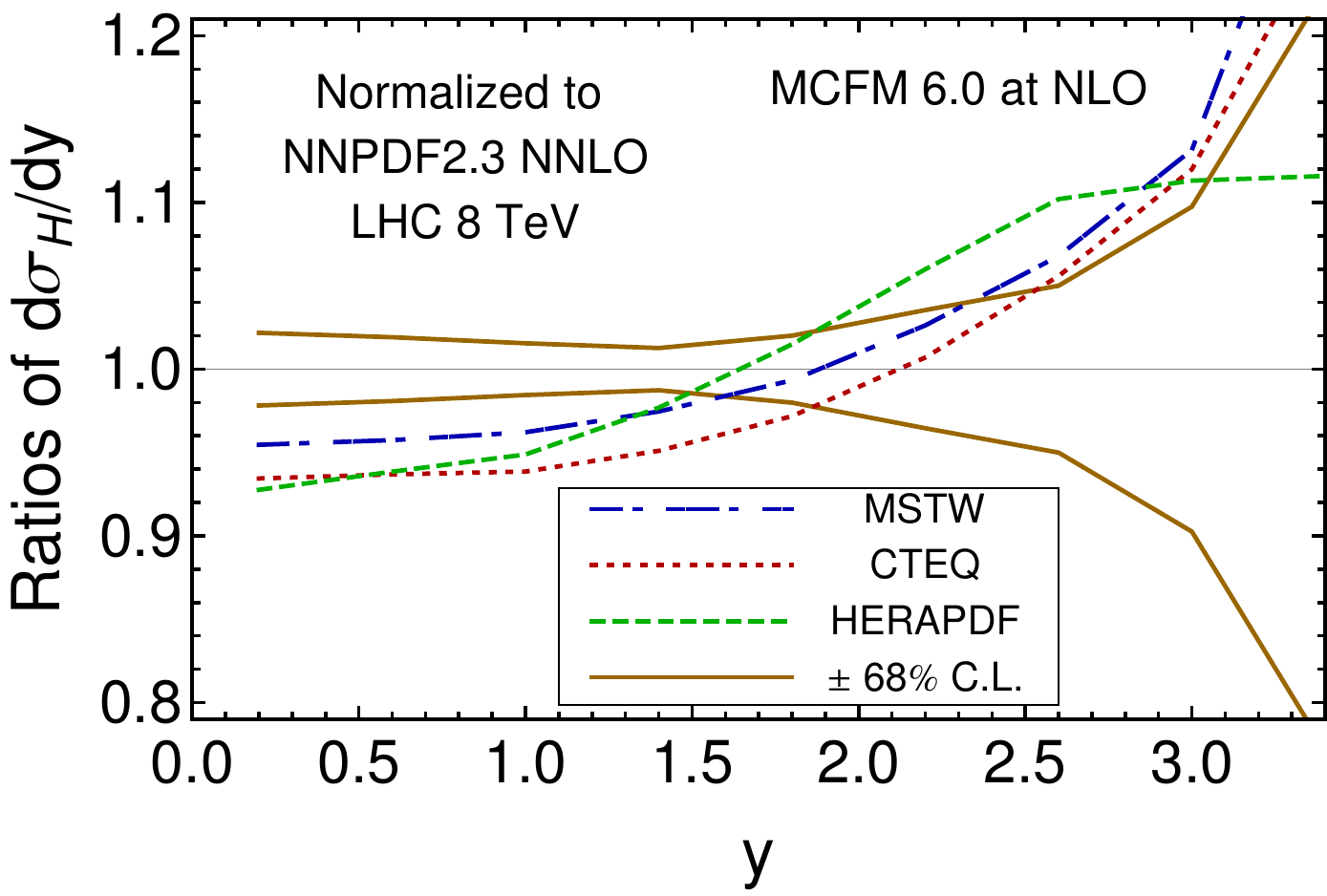}\hspace{0.2in}
 \includegraphics[width=0.48\textwidth]{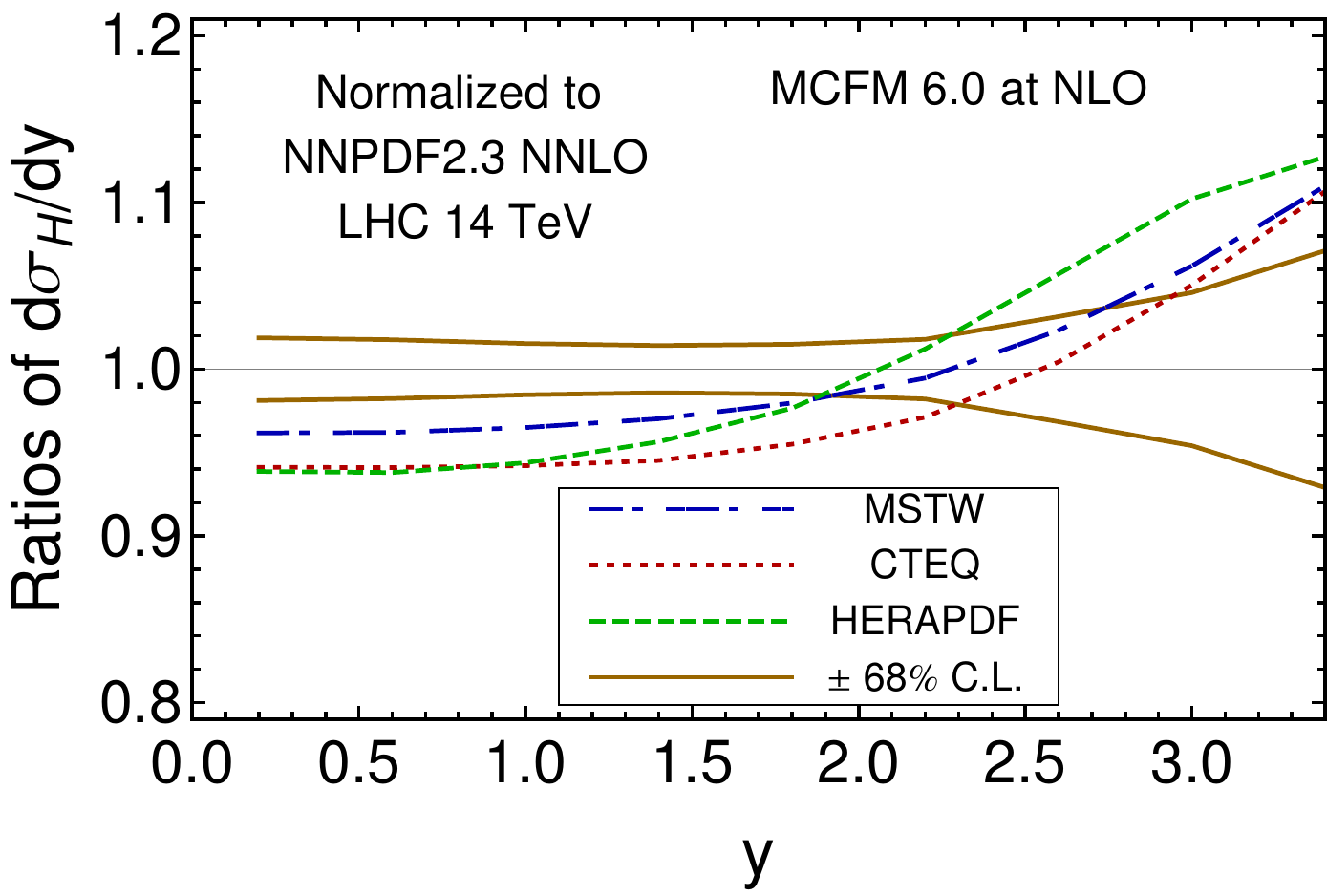}
\vspace{-1ex}
 \caption{\label{fig:globalrap} Comparison of the rapidity distribution
 of the SM Higgs boson in production through gluon fusion from recently published
 NNLO PDFs, normalized to the NNPDF 2.3 central prediction. The matrix
 elements of hard scattering are calculated at NLO.}
\end{figure}

To illustrate the sensitivity of the Higgs cross section to the gluon PDF,
we also plot the rapidity distribution of the Higgs boson from gluon fusion in
Fig.~\ref{fig:globalrap}. The predictions are calculated at NLO
using MCFM 6.0~\cite{Campbell:2010ff}, 
with both the Higgs mass and QCD scales set at 125 GeV, 
and knowing that the NNLO/NLO $K$-factors cancel well in these ratios.
We see that
in the central production region ($|y|<1$), which contributes
 the bulk of the total cross section,
the difference between CT10 and NNPDF2.3 can reach 6\%, i.e., it is larger 
than in the inclusive rate.
At higher rapidities the predictions of all groups overshoot that
of NNPDF2.3, reflecting the behavior of their large-$x$ gluon
PDFs in Fig.~\ref{fig:globalglu}.

In Section 4 we will investigate the origin of these differences
in the gluon PDFs in a simpler scenario than that of the global fit, namely,
by comparing the PDF fits of the HERA-1 DIS data set alone.
However, before doing that, we want to make sure that the differences
among the theoretical predictions for NNLO DIS neutral current
cross sections in different fitting codes are under control, and
with this aim we discuss now a new benchmarking of DIS structure
functions.

\subsection{Benchmark comparisons 
of NNLO neutral current DIS cross sections\label{sec:str}}

As in the previous benchmarking studies, in this section 
we adopt the Les Houches toy PDFs and settings ~\cite{Binoth:2010ra}.
The corresponding {\tt LHAPDF} grid file was generated 
with the {\tt APFEL} program~\cite{Bertone:2013vaa}
by evolving the Les Houches toy PDFs from the starting scale 
$Q_0^2=$ 2 GeV$^2$.
We set the bottom and top masses to infinity, and
use as a boundary condition $\alpha_s(\mu)=0.35$ at a scale $\mu$ 
approaching $Q_0$ from below.
Theoretical predictions are made at the $(x,Q^2)$ values of the
neutral current positron  HERA-1 combined data set~\cite{Aaron:2009aa}
for $Q^2 \ge 3.5$ GeV$^2$.

%
Modern GM-VFN schemes, like FONLL-C~\cite{Forte:2010ta} used by NNPDF, TR~\cite{Thorne:2012az} used by MSTW and HERAPDF,
and S-ACOT-$\chi$~\cite{Guzzi:2011ew} used by CTEQ, are expected to be
very similar at NNLO, and we will
demonstrate the convergence here.

For the comparisons of this section,
 the MSTW and HERAPDF results will both use the optimal TR heavy-quark
scheme~\cite{Thorne:2012az}, instead of the standard scheme that is used
in the published NNLO PDF sets.
As noted in~\cite{Thorne:2012az}, the change in the TR scheme leads to only an
extremely small change in the MSTW fit quality 
and the resulting PDFs at NNLO, i.e. small fractions of a percent
in the Higgs cross section at the LHC.
The NNPDF2.3 results for structure functions
will be obtained using the {\tt APFEL} code, instead of using the
{\tt FastKernel} NNPDF fitting code, because of its ease of use (the
predictions of the two codes agree
at the per mille level anyway.).
The HERAPDF results on structure functions are obtained with
the {\tt HERAFitter} code~\cite{herafitter, Botje:2010ay, Aaron:2009aa, Aaron:2009kv}.

In the following, we will compare the total (with subscript ``$tot$'')
and charm-quark contributions (``$c$'') to inclusive $F_2$, $F_3$,
and $F_L$ structure functions, which are defined as in~\cite{Aaron:2009aa}
and include all scattering contributions coupled to the 
photon and Z boson currents. 
The light quark contributions (``$lt$'') are defined by
subtracting the charm-quark contributions from the total structure
functions. This means that $F_c$ is defined as the structure function
that one would get if only the electromagnetic 
charge of the charm quark was nonzero~\cite{Forte:2010ta,Guzzi:2011ew},
and then $F_{lt}$ is implicitly defined by
\begin{equation}
F_{tot}\equiv F_{lt}+F_c=\sum_{l=1}^{N_l}F_l+F_c,
\end{equation}
where $F_l$ is the structure function obtained when only the charge
of the $l$-th quark is nonzero.
With this definition, at  NNLO $F_l$ receives a contribution from the
process with charm in the final state, such as
 $\gamma^*u\rightarrow u c\bar c$, as well as from
diagrams containing the charm quark in virtual loops. Note that there
is a cancellation of terms with large logs of $Q^2/m_c^2$ between
these two contributions.  

The NNLO results for $F_{2,lt(c)}$, $F_{L,lt(c)}$,
and the reduced cross sections obtained
for a representative subset of the HERA-1 data points, using the four
different codes, are collected  in
Tables~\ref{tab:DISbenchct}-\ref{tab:DISbenchnn}.
These tables should be useful for future comparisons, for instance for
validation of  structure function codes.
In the following we compare and discuss these results, both at NLO and at NNLO.
Comparisons at LO are not meaningful because of the different perturbative
ordering between different codes.

\begin{table}
\centering
\begin{tabular}{|c|c|c|c|c|c|c|c|}
\hline
 $Q^2$ [${\rm GeV^2}$] & $x$ & $y$ & $F_{2,lt}$ & $F_{2,c}$ & $F_{L,lt}$ & $F_{L,c}$ & $\sigma^{+}_{r,NC}$ \tabularnewline
\hline
\hline
2.7 &   0.0000309   &   0.86398 &   0.7768  &   0.1411  &   0.1267  &   0.00844 &   0.8189  \tabularnewline
2.7 &   0.00013 &   0.20523 &   0.6275  &   0.09204 &   0.1173  &   0.00592 &   0.7163  \tabularnewline
2.7 &   0.02    &   0.00133 &   0.3871  &   0.00723 &   0.0849  &   0.00074 &   0.3943  \tabularnewline
4.5 &   0.0000618   &   0.71999 &   0.964   &   0.193   &   0.1866  &   0.01879 &   1.0583  \tabularnewline
4.5 &   0.00032 &   0.13896 &   0.7187  &   0.1136  &   0.1547  &   0.01191 &   0.8305  \tabularnewline
4.5 &   0.013   &   0.00342 &   0.4317  &   0.02011 &   0.0898  &   0.00259 &   0.4518  \tabularnewline
10. &   0.00013 &   0.85281 &   1.192   &   0.31011 &   0.2535  &   0.04618 &   1.2886  \tabularnewline
10. &   0.0008  &   0.12352 &   0.8064  &   0.16391 &   0.1758  &   0.0263  &   0.9685  \tabularnewline
10. &   0.02    &   0.00494 &   0.4614  &   0.03152 &   0.0766  &   0.00564 &   0.4929  \tabularnewline
120.    &   0.0016  &   0.74111 &   1.1038  &   0.41841 &   0.179   &   0.08637 &   1.3852  \tabularnewline
120.    &   0.008   &   0.14822 &   0.6915  &   0.18639 &   0.0905  &   0.03846 &   0.8761  \tabularnewline
120.    &   0.18    &   0.00659 &   0.3312  &   0.00591 &   0.0128  &   0.00095 &   0.3371  \tabularnewline
650.    &   0.0085  &   0.84779 &   0.768   &   0.24564 &   0.0791  &   0.04104 &   0.9216  \tabularnewline
650.    &   0.032   &   0.20072 &   0.5249  &   0.09203 &   0.0364  &   0.01444 &   0.6122  \tabularnewline
650.    &   0.25    &   0.02569 &   0.252   &   0.00306 &   0.0059  &   0.00032 &   0.2545  \tabularnewline
2000.   &   0.032   &   0.6929  &   0.5502  &   0.10329 &   0.0328  &   0.01347 &   0.5973  \tabularnewline
2000.   &   0.13    &   0.15202 &   0.3634  &   0.0164  &   0.011   &   0.00172 &   0.3698  \tabularnewline
2000.   &   0.25    &   0.07905 &   0.244   &   0.00327 &   0.0049  &   0.00027 &   0.2433  \tabularnewline
8000.   &   0.13    &   0.68224 &   0.3948  &   0.01837 &   0.01    &   0.00149 &   0.284   \tabularnewline
8000.   &   0.18    &   0.43917 &   0.3307  &   0.00903 &   0.0069  &   0.00066 &   0.2649  \tabularnewline
8000.   &   0.25    &   0.31621 &   0.254   &   0.00356 &   0.0043  &   0.00022 &   0.2148  \tabularnewline
\hline
\end{tabular}
\caption{Predictions for NC DIS structure functions and reduced cross sections from
HERA-1~\cite{Aaron:2009aa} at NNLO, obtained with the CT10 code~\cite{Gao:2013xoa} and Les Houches toy PDFs \cite{Binoth:2010ra}.
    \label{tab:DISbenchct}}
\end{table}
\begin{table}
\centering

\begin{tabular}{|c|c|c|c|c|c|c|c|}
\hline
 $Q^2$ [${\rm GeV^2}$] & $x$ & $y$ & $F_{2,lt}$ & $F_{2,c}$ & $F_{L,lt}$ & $F_{L,c}$ & $\sigma^{+}_{r,NC}$ \tabularnewline
\hline
\hline
2.7 &   0.0000309   &   0.86398 &   0.7821  &   0.17992 &   0.261   &   0.01067 &   0.7628  \tabularnewline
2.7 &   0.00013 &   0.20523 &   0.6318  &   0.10519 &   0.1711  &   0.00669 &   0.7324  \tabularnewline
2.7 &   0.02    &   0.00133 &   0.3893  &   0.01013 &   0.0829  &   0.00073 &   0.3995  \tabularnewline
4.5 &   0.0000618   &   0.71999 &   0.972   &   0.2175  &   0.2507  &   0.01961 &   1.0596  \tabularnewline
4.5 &   0.00032 &   0.13896 &   0.7249  &   0.11996 &   0.1685  &   0.01122 &   0.8429  \tabularnewline
4.5 &   0.013   &   0.00342 &   0.435   &   0.02257 &   0.0885  &   0.00244 &   0.4575  \tabularnewline
10. &   0.00013 &   0.85281 &   1.2026  &   0.32346 &   0.2759  &   0.04743 &   1.2958  \tabularnewline
10. &   0.0008  &   0.12352 &   0.8135  &   0.16653 &   0.1771  &   0.02543 &   0.9783  \tabularnewline
10. &   0.02    &   0.00494 &   0.465   &   0.03274 &   0.0797  &   0.00588 &   0.4978  \tabularnewline
120.    &   0.0016  &   0.74111 &   1.107   &   0.40842 &   0.1824  &   0.08717 &   1.3761  \tabularnewline
120.    &   0.008   &   0.14822 &   0.693   &   0.18279 &   0.0935  &   0.03914 &   0.8739  \tabularnewline
120.    &   0.18    &   0.00659 &   0.3317  &   0.00594 &   0.0149  &   0.00113 &   0.3376  \tabularnewline
650.    &   0.0085  &   0.84779 &   0.7681  &   0.2396  &   0.0817  &   0.04192 &   0.9133  \tabularnewline
650.    &   0.032   &   0.20072 &   0.5249  &   0.09021 &   0.0385  &   0.01505 &   0.6103  \tabularnewline
650.    &   0.25    &   0.02569 &   0.252   &   0.00305 &   0.0068  &   0.00022 &   0.2546  \tabularnewline
2000.   &   0.032   &   0.6929  &   0.5489  &   0.10207 &   0.0345  &   0.01416 &   0.5939  \tabularnewline
2000.   &   0.13    &   0.15202 &   0.3632  &   0.0163  &   0.0121  &   0.00189 &   0.3695  \tabularnewline
2000.   &   0.25    &   0.07905 &   0.2439  &   0.00326 &   0.0055  &   0.00019 &   0.2432  \tabularnewline
8000.   &   0.13    &   0.68224 &   0.3937  &   0.01893 &   0.011   &   0.0017  &   0.2837  \tabularnewline
8000.   &   0.18    &   0.43917 &   0.3301  &   0.00928 &   0.0077  &   0.00076 &   0.2649  \tabularnewline
8000.   &   0.25    &   0.31621 &   0.2537  &   0.00365 &   0.0048  &   0.00017 &   0.2148  \tabularnewline
\hline
\end{tabular}

\caption{Same as Table~\ref{tab:DISbenchct}, using the \tt HERAFitter \rm code~\cite{herafitter, Botje:2010ay, Aaron:2009kv, Aaron:2009aa}.
    \label{tab:DISbenchhe}}
\end{table}

\begin{table}

\centering

\begin{tabular}{|c|c|c|c|c|c|c|c|}
\hline
 $Q^2$ [${\rm GeV^2}$] & $x$ & $y$ & $F_{2,lt}$ & $F_{2,c}$ & $F_{L,lt}$ & $F_{L,c}$ & $\sigma^{+}_{r,NC}$ \tabularnewline
\hline
\hline
2.7 &   0.0000309   &   0.86398 &   0.7809  &   0.17478 &   0.2611  &   0.01104 &   0.7604  \tabularnewline
2.7 &   0.00013 &   0.20523 &   0.6316  &   0.10218 &   0.1713  &   0.00678 &   0.7292  \tabularnewline
2.7 &   0.02    &   0.00133 &   0.3895  &   0.01016 &   0.0835  &   0.00074 &   0.3997  \tabularnewline
4.5 &   0.0000618   &   0.71999 &   0.9713  &   0.2174  &   0.2488  &   0.02008 &   1.0623  \tabularnewline
4.5 &   0.00032 &   0.13896 &   0.7249  &   0.12013 &   0.1678  &   0.0113  &   0.8431  \tabularnewline
4.5 &   0.013   &   0.00342 &   0.4351  &   0.02258 &   0.0888  &   0.00246 &   0.4577  \tabularnewline
10. &   0.00013 &   0.85281 &   1.2024  &   0.32374 &   0.2735  &   0.04786 &   1.2991  \tabularnewline
10. &   0.0008  &   0.12352 &   0.8134  &   0.16673 &   0.1762  &   0.02555 &   0.9784  \tabularnewline
10. &   0.02    &   0.00494 &   0.465   &   0.03268 &   0.0797  &   0.00592 &   0.4977  \tabularnewline
120.    &   0.0016  &   0.74111 &   1.1067  &   0.40833 &   0.181   &   0.08729 &   1.3784  \tabularnewline
120.    &   0.008   &   0.14822 &   0.6929  &   0.18255 &   0.0929  &   0.03919 &   0.8736  \tabularnewline
120.    &   0.18    &   0.00659 &   0.3317  &   0.0059  &   0.0147  &   0.00113 &   0.3376  \tabularnewline
650.    &   0.0085  &   0.84779 &   0.7669  &   0.23899 &   0.0811  &   0.04189 &   0.9126  \tabularnewline
650.    &   0.032   &   0.20072 &   0.5243  &   0.08985 &   0.0382  &   0.01503 &   0.6093  \tabularnewline
650.    &   0.25    &   0.02569 &   0.2519  &   0.00302 &   0.0067  &   0.00037 &   0.2545  \tabularnewline
2000.   &   0.032   &   0.6929  &   0.5462  &   0.1013  &   0.0341  &   0.01409 &   0.5896  \tabularnewline
2000.   &   0.13    &   0.15202 &   0.3624  &   0.01615 &   0.0119  &   0.00188 &   0.3682  \tabularnewline
2000.   &   0.25    &   0.07905 &   0.2437  &   0.00322 &   0.0054  &   0.00031 &   0.2429  \tabularnewline
8000.   &   0.13    &   0.68224 &   0.3893  &   0.0186  &   0.0107  &   0.00167 &   0.2756  \tabularnewline
8000.   &   0.18    &   0.43917 &   0.3277  &   0.00914 &   0.0075  &   0.00075 &   0.2607  \tabularnewline
8000.   &   0.25    &   0.31621 &   0.2525  &   0.0036  &   0.0047  &   0.00026 &   0.2129  \tabularnewline
\hline
\end{tabular}

\caption{Same as Table~\ref{tab:DISbenchct}, using the MSTW08 code~\cite{Martin:2009iq}.
    \label{tab:DISbenchms}}
\end{table}

\begin{table}

\centering

\begin{tabular}{|c|c|c|c|c|c|c|c|}
\hline
 $Q^2$ [${\rm GeV^2}$] & $x$ & $y$ & $F_{2,lt}$ & $F_{2,c}$ & $F_{L,lt}$ & $F_{L,c}$ & $\sigma^{+}_{r,NC}$ \tabularnewline
\hline
\hline
2.7 &   0.0000309   &   0.86398 &   0.7809  &   0.13392 &   0.1035  &   0.0089  &   0.8324  \tabularnewline
2.7 &   0.00013 &   0.20523 &   0.6309  &   0.08936 &   0.0981  &   0.00628 &   0.7176  \tabularnewline
2.7 &   0.02    &   0.00133 &   0.3884  &   0.0069  &   0.0736  &   0.00073 &   0.3953  \tabularnewline
4.5 &   0.0000618   &   0.71999 &   0.9629  &   0.19426 &   0.1706  &   0.01982 &   1.0656  \tabularnewline
4.5 &   0.00032 &   0.13896 &   0.7181  &   0.1138  &   0.1425  &   0.01266 &   0.8302  \tabularnewline
4.5 &   0.013   &   0.00342 &   0.431   &   0.01932 &   0.0828  &   0.00293 &   0.4503  \tabularnewline
10. &   0.00013 &   0.85281 &   1.1897  &   0.31837 &   0.247   &   0.05032 &   1.2962  \tabularnewline
10. &   0.0008  &   0.12352 &   0.8046  &   0.16446 &   0.1714  &   0.02859 &   0.9673  \tabularnewline
10. &   0.02    &   0.00494 &   0.4601  &   0.03125 &   0.0742  &   0.00642 &   0.4913  \tabularnewline
120.    &   0.0016  &   0.74111 &   1.1083  &   0.40632 &   0.1794  &   0.08744 &   1.3769  \tabularnewline
120.    &   0.008   &   0.14822 &   0.6939  &   0.18171 &   0.0907  &   0.03899 &   0.8738  \tabularnewline
120.    &   0.18    &   0.00659 &   0.3319  &   0.00614 &   0.013   &   0.00098 &   0.338   \tabularnewline
650.    &   0.0085  &   0.84779 &   0.771   &   0.23832 &   0.0795  &   0.0412  &   0.9169  \tabularnewline
650.    &   0.032   &   0.20072 &   0.5268  &   0.08994 &   0.0367  &   0.01452 &   0.612   \tabularnewline
650.    &   0.25    &   0.02569 &   0.2524  &   0.00307 &   0.006   &   0.00032 &   0.255   \tabularnewline
2000.   &   0.032   &   0.6929  &   0.5519  &   0.10081 &   0.0332  &   0.01354 &   0.5965  \tabularnewline
2000.   &   0.13    &   0.15202 &   0.3643  &   0.01615 &   0.0112  &   0.00174 &   0.3704  \tabularnewline
2000.   &   0.25    &   0.07905 &   0.2443  &   0.00324 &   0.005   &   0.00027 &   0.2436  \tabularnewline
8000.   &   0.13    &   0.68224 &   0.3953  &   0.01807 &   0.0104  &   0.00152 &   0.2847  \tabularnewline
8000.   &   0.18    &   0.43917 &   0.3311  &   0.0089  &   0.0072  &   0.00067 &   0.2655  \tabularnewline
8000.   &   0.25    &   0.31621 &   0.2541  &   0.00352 &   0.0044  &   0.00023 &   0.2151  \tabularnewline
\hline
\end{tabular}
\caption{Same as Table~\ref{tab:DISbenchct}, using the {\tt APFEL} code~\cite{Bertone:2013vaa} with the FONLL-C scheme.
These results agree at the per mille level with those of the $N$-space {\tt FastKernel}
code adopted in the NNPDF fits.
    \label{tab:DISbenchnn}}
\end{table}

\begin{figure}[b]
\centering
\small
 \includegraphics[width=0.48\textwidth]{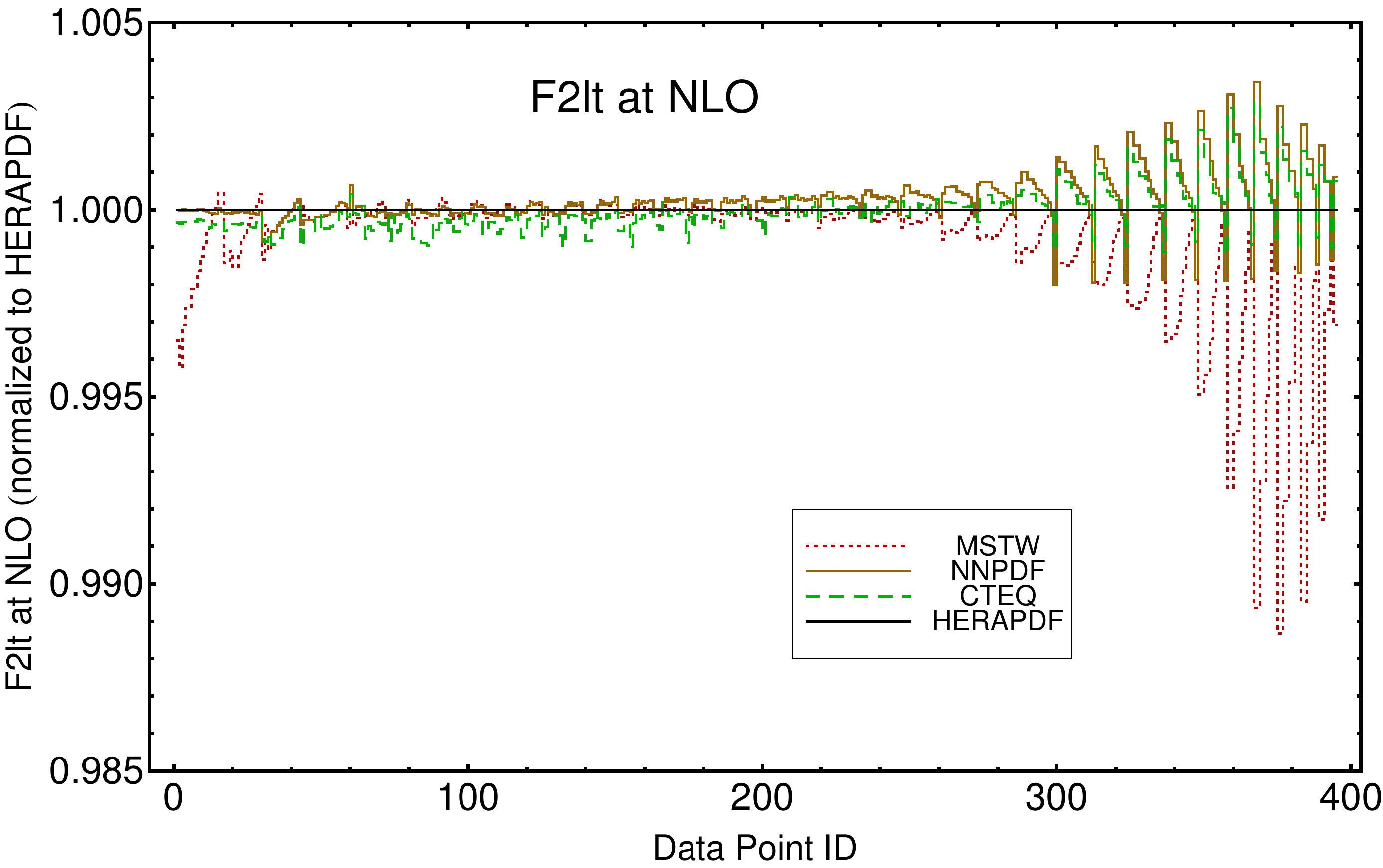}
 \includegraphics[width=0.48\textwidth]{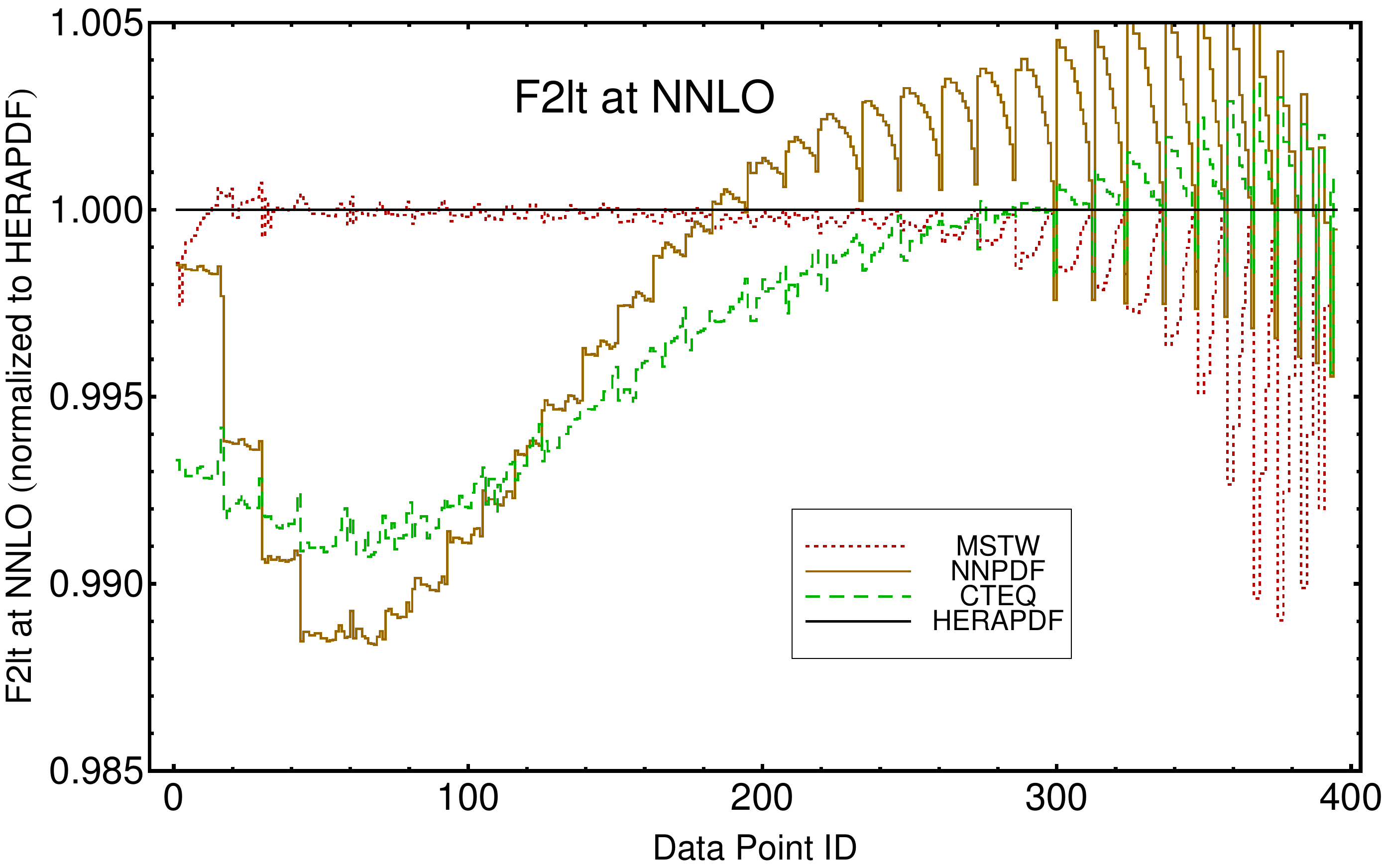}
\vspace{-1ex}
 \caption{\label{fig:STRf2lt} Structure function $F_{2,lt}(x, Q)$
 at NLO and NNLO from all groups, normalized to the results from the
 HERAPDF1.5  code. }
\end{figure}

In Fig.~\ref{fig:STRf2lt} we plot 
the light structure function $F_{2,lt}$ at NLO and NNLO,
normalized to the HERAPDF results.
The data point indexing is ordered in increasing $Q^2$
(starting from $Q^2=$2.7 ${\rm GeV^2}$), then increasing $x$.
 We find
 excellent agreement at NLO among all groups, with  small fluctuations
at the per mille level at low
$Q$ due to numerical precision, and  moderate oscillations of at most $1.5\%$
at high $Q$ from
different treatment of electroweak corrections and $Z$ or $\gamma/Z$ interference terms.
Note that  at large $Q^2$
experimental uncertainties  are typically $>10\%$, so these
differences have no impact on the fit results.

At NNLO we observe some systematic shifts of 1\% between MSTW/HERAPDF and
CTEQ/NNPDF especially around the data point 50, which corresponds to
$Q^2\sim6$~GeV$^2$. However, we have checked that if
we use a 3-flavor scheme we find very good agreement, within 0.2\%,
for $F_{2,lt}$ at NNLO in that $Q$ region. Therefore, we attribute
these shifts to the aforementioned contribution
from the processes with  $\gamma^*q\rightarrow q c\bar c$
contributions, or charm quark loops, 
which first appear at NNLO and are known to be implemented differently 
in the codes that were compared. As with other flavor scheme
differences, there should be convergence at higher orders. 

\begin{figure}[p]
\centering
\small
 \includegraphics[width=0.48\textwidth]{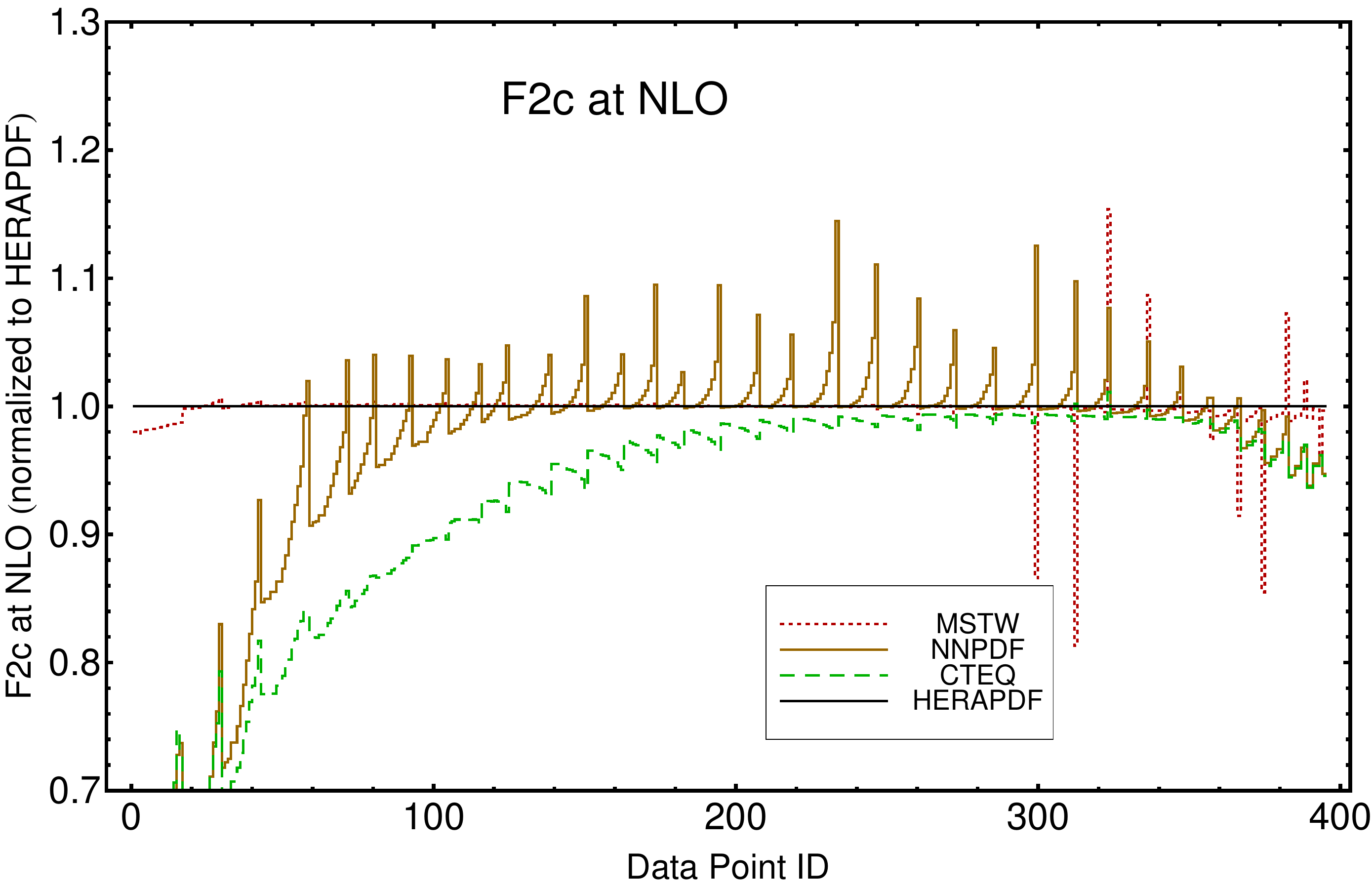}
 \includegraphics[width=0.48\textwidth]{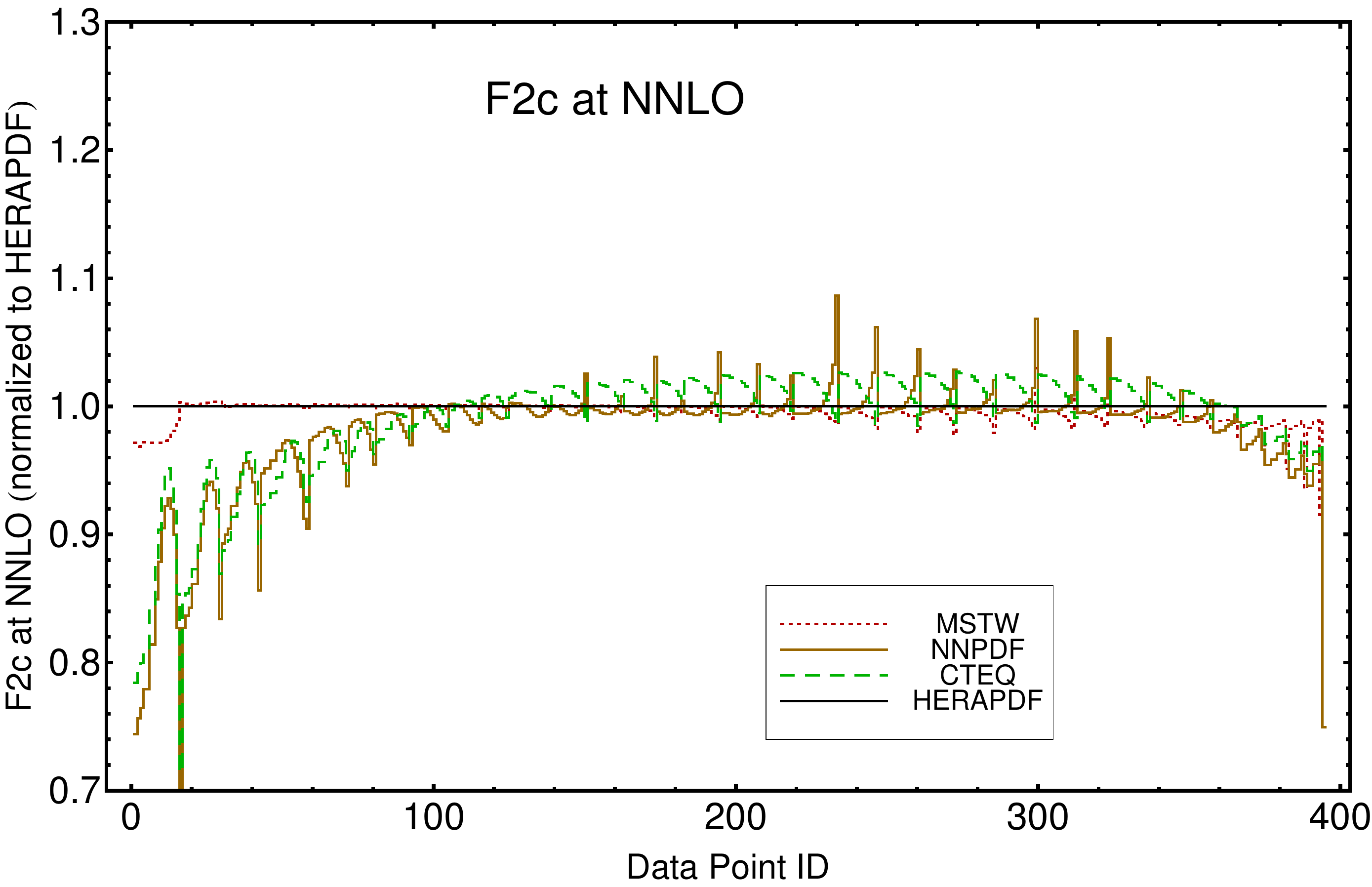}
\vspace{-1ex}
 \caption{\label{fig:STRf2c} Same as Fig.~\ref{fig:STRf2lt}, for the
 structure function $F_{2,c}$. }
\end{figure}

Turning now to the heavy-quark structure function,
Fig.~\ref{fig:STRf2c} shows
the results for $F_{2,c}$. MSTW and HERAPDF agree as they should, 
since they both use the optimal TR' scheme. We also observe 
close similarity between CTEQ and NNPDF, which are based 
on the numerically close schemes S-ACOT-$\chi$, FONLL-A (at NLO) and
FONLL-C (at NNLO). In the low-$Q$ region, CTEQ and NNPDF
are in very good agreement, and both are smaller than MSTW/HERAPDF.
In the intermediate-$Q$ region, all groups agree within a couple of percent.
It has been further verified that, if
one uses the NNPDF or CTEQ scheme definition in the MSTW code at NNLO,
even better agreement is found between the groups.
The ``spike'' structure for NNPDF
is due to the use of a different kinematical variable  
in the GM-VFN scheme definition, i.e. $x$ is used in the FONLL 
scheme~\cite{Forte:2010ta}, while $\chi = x(1+4m_c^2/Q^2)$ 
is used in the TR'~\cite{Thorne:2006qt} and
S-ACOT-$\chi$~\cite{Guzzi:2011ew} schemes.  
This effect is mainly seen at the highest $x$ point in each $Q^2$ bin,
where $F_2^c$ is a small contribution to the total.

The overall trends in $F_{2,c}$ are consistent with the results 
obtained in the context of the 2010 Les Houches 
heavy-quark benchmarks~\cite{Binoth:2010ra} (see also the follow-up
comparisons with S-ACOT-$\chi$ at NNLO in \cite{Guzzi:2011ew}). 
The differences are reduced
when going from NLO and NNLO, in agreement with the general
expectation that, at each order in
$\alpha_s$, the GM-VFN heavy-quark schemes are equivalent
up to contributions proportional to $m_c^2/Q^2$ at one higher order. 
Apart from the residual differences in their heavy-quark schemes, 
the CTEQ and NNPDF codes evaluate
${\cal O}(\alpha_s^2)$ heavy-quark scattering contributions by
interpolation of integral tables \cite{Riemersma:1994hv}
and direct integration \cite{Forte:2010ta}, respectively. This
contributes to some of the minor differences seen 
between the CTEQ and NNPDF $F_{2,c}$ values at NNLO.

\begin{figure}[p]
\centering
\small
 \includegraphics[width=0.48\textwidth]{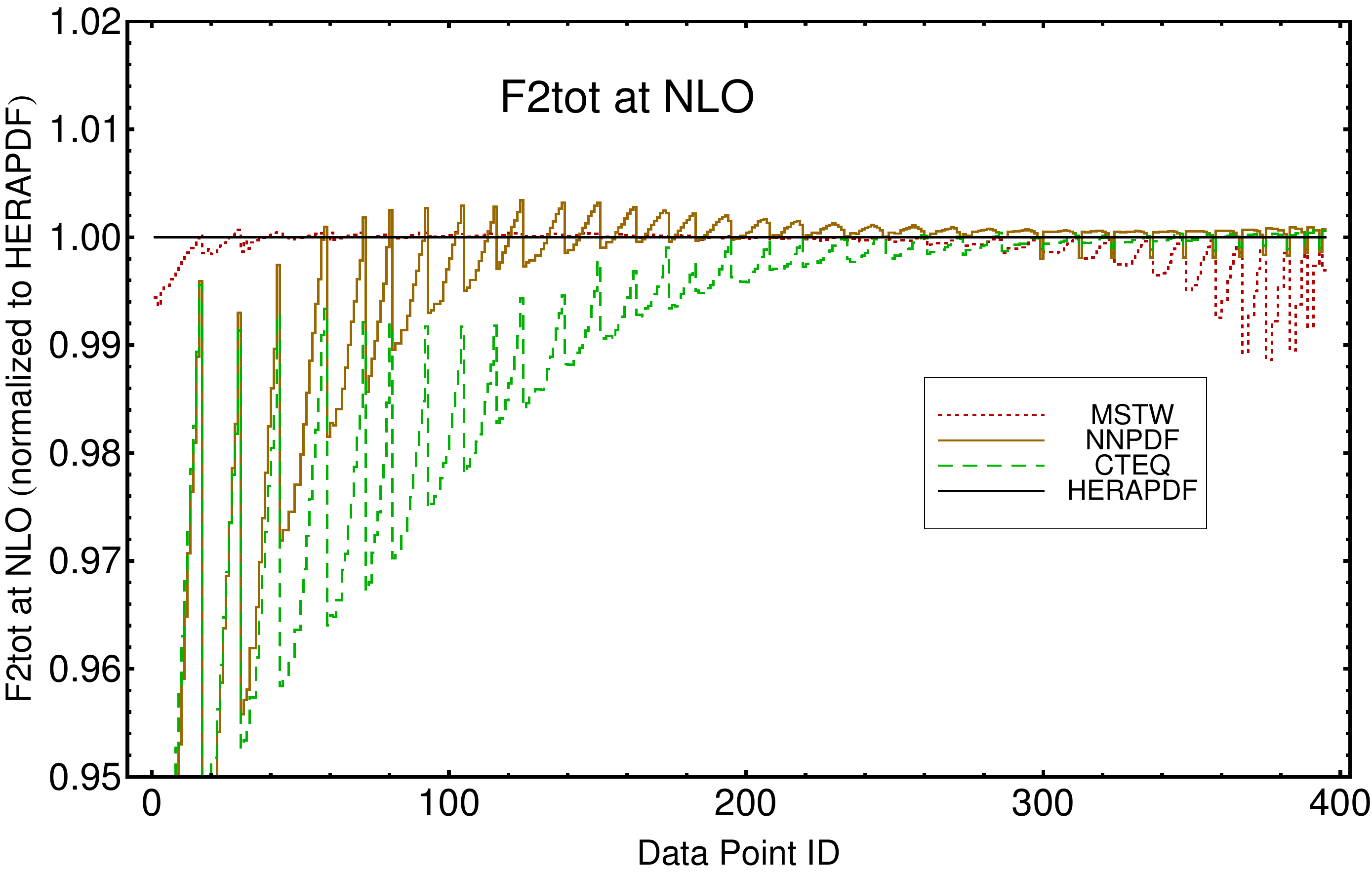}
 \includegraphics[width=0.48\textwidth]{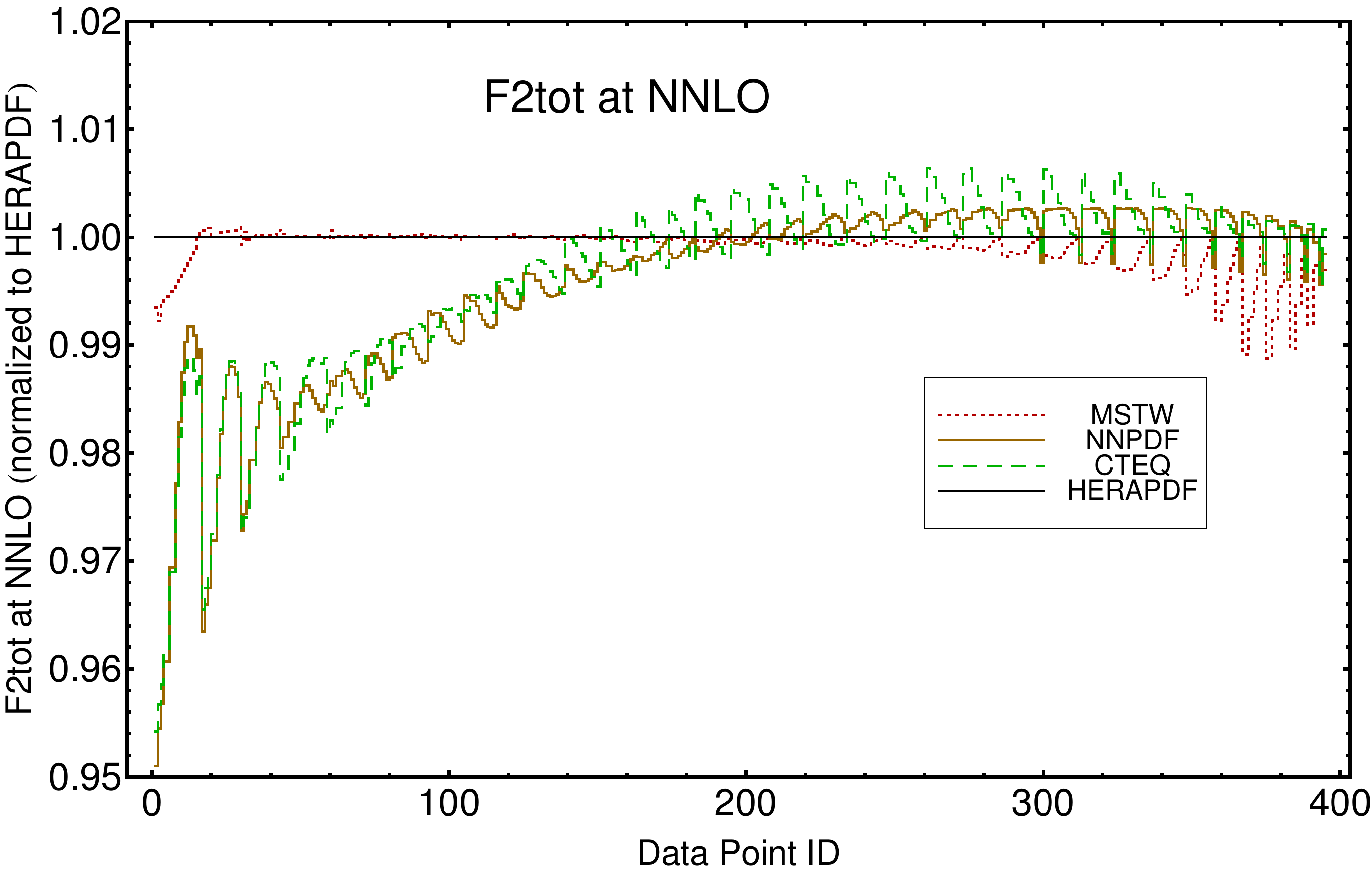}
\vspace{-1ex}
 \caption{\label{fig:STRf2tot} Same as Fig.~\ref{fig:STRf2lt}, for the
 structure function $F_{2,tot}$. }
\end{figure}

Fig.~\ref{fig:STRf2tot} presents comparisons of $F_{2,tot}$, which is impacted
by all effects seen in Figs.~\ref{fig:STRf2lt} and \ref{fig:STRf2c}.
Again the differences decrease as we go from NLO to NNLO.

\begin{figure}[p]
\centering
\small
 \includegraphics[width=0.48\textwidth]{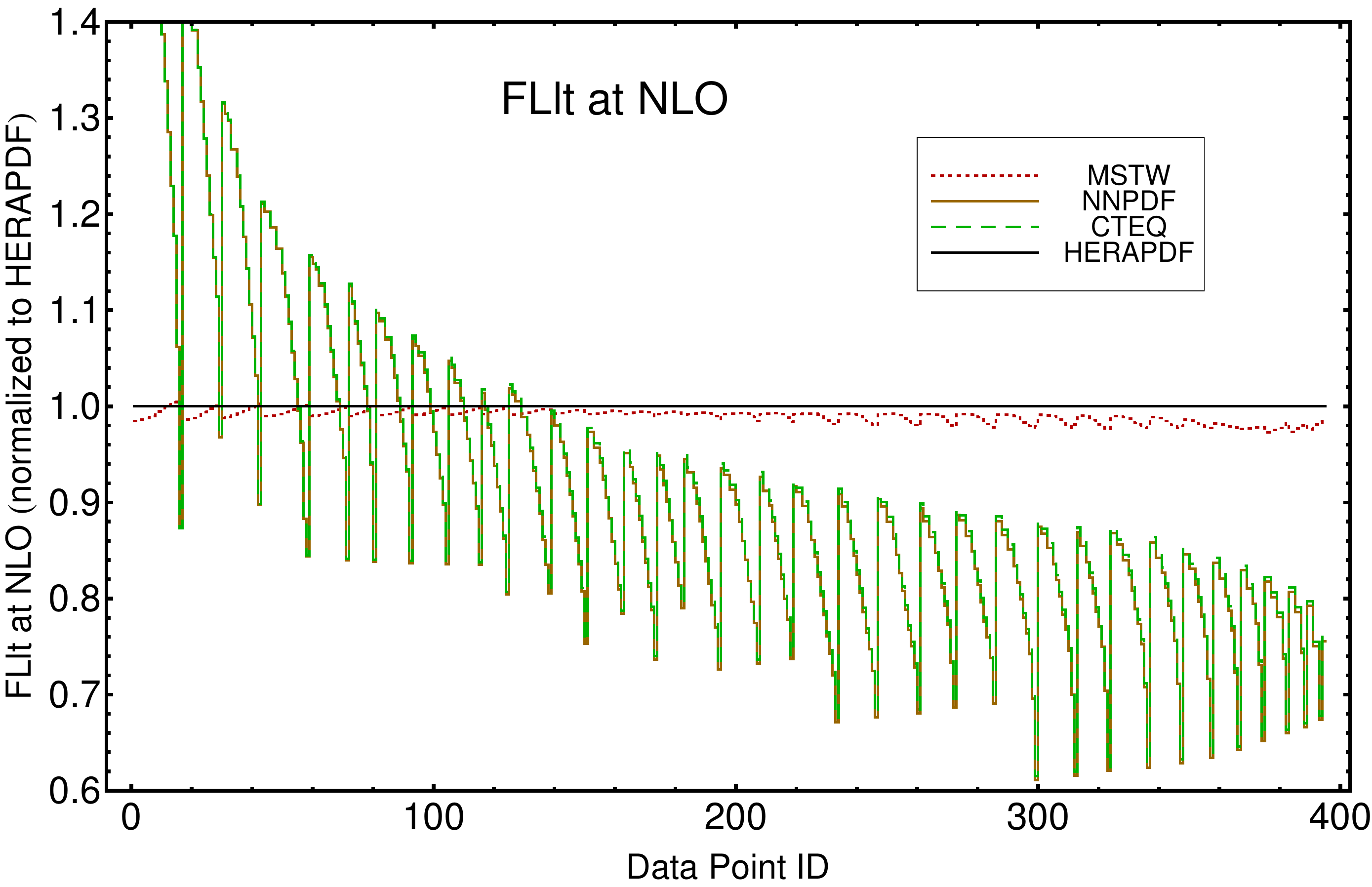}
 \includegraphics[width=0.48\textwidth]{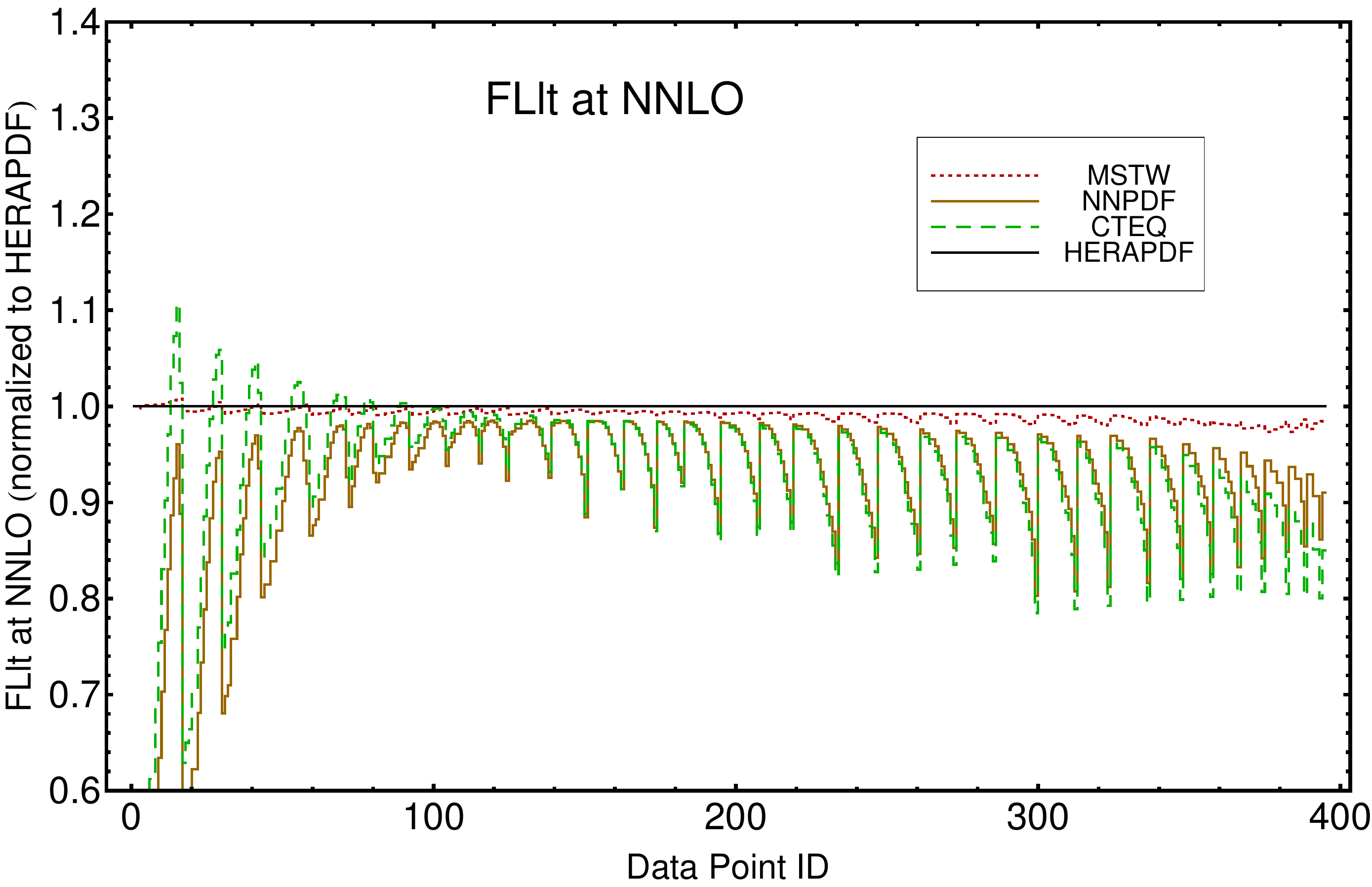}
\vspace{-1ex}
 \caption{\label{fig:STRfllt} Structure function $F_{L,lt}(x, Q)$
 at nominal NLO and NNLO from all groups, normalized to the results from the
 HERAPDF1.5  code. The counting of perturbative orders for $F_{L}(x, Q)$
differs in the CTEQ/NNPDF and HERAPDF/MSTW conventions, as discussed in the main text.}
\end{figure}

\begin{figure}[p]
\centering
 \includegraphics[width=0.48\textwidth]{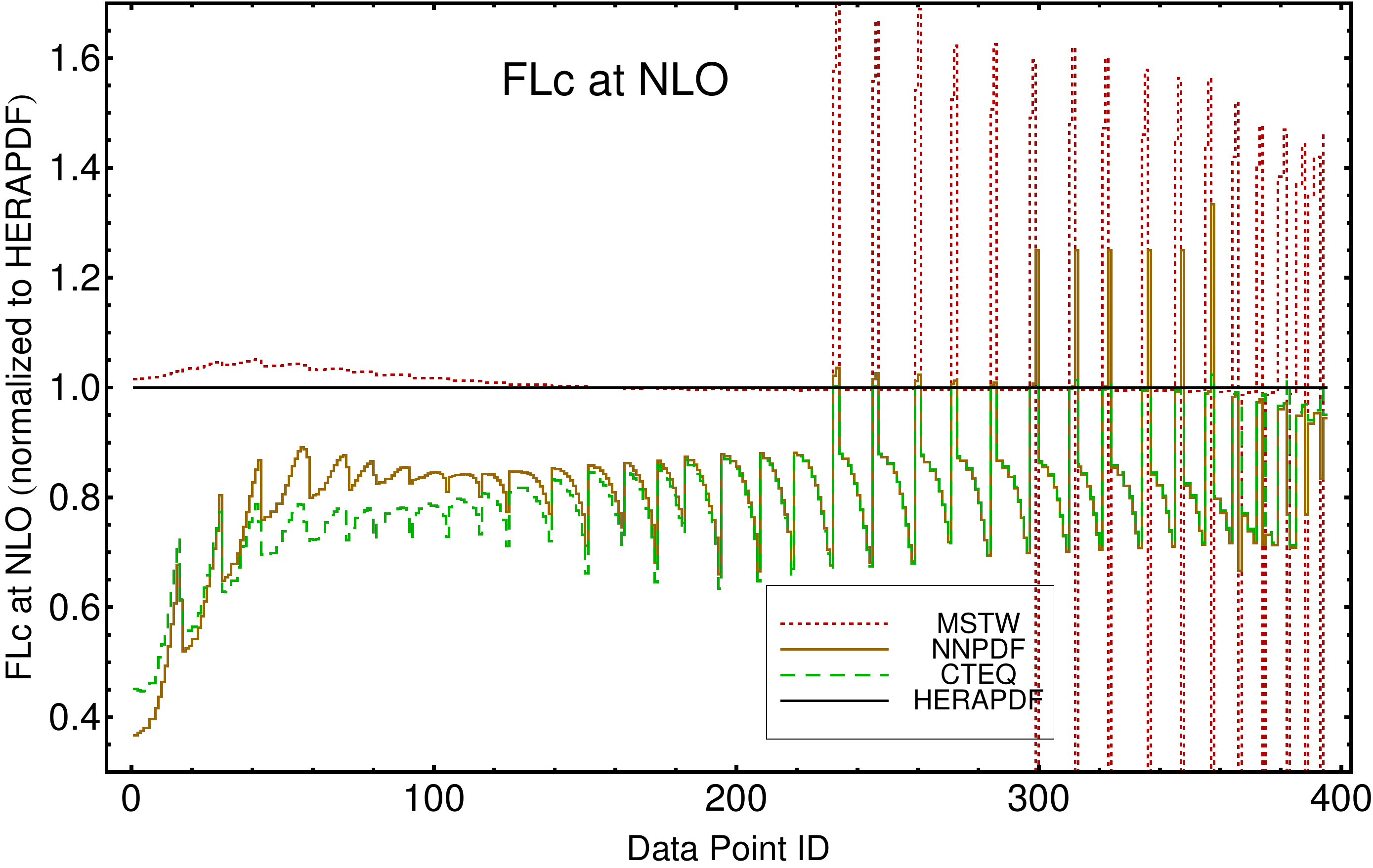}
 \includegraphics[width=0.48\textwidth]{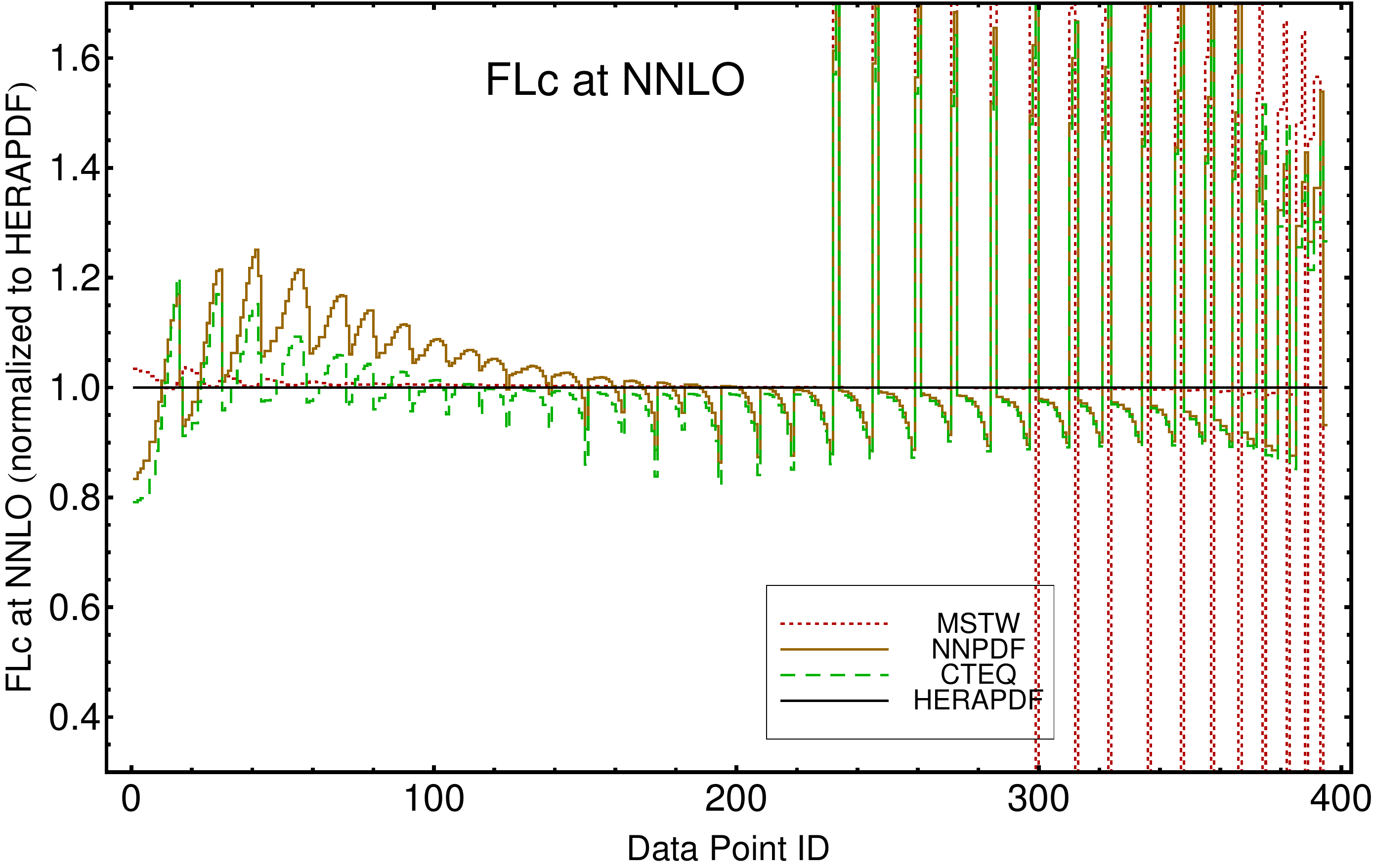}
\vspace{-1ex}
 \caption{\label{fig:STRflc} Same as Fig.~\ref{fig:STRfllt}, for the
 structure function $F_{L,c}$. }
\end{figure}

\begin{figure}
\centering
 \includegraphics[width=0.48\textwidth]{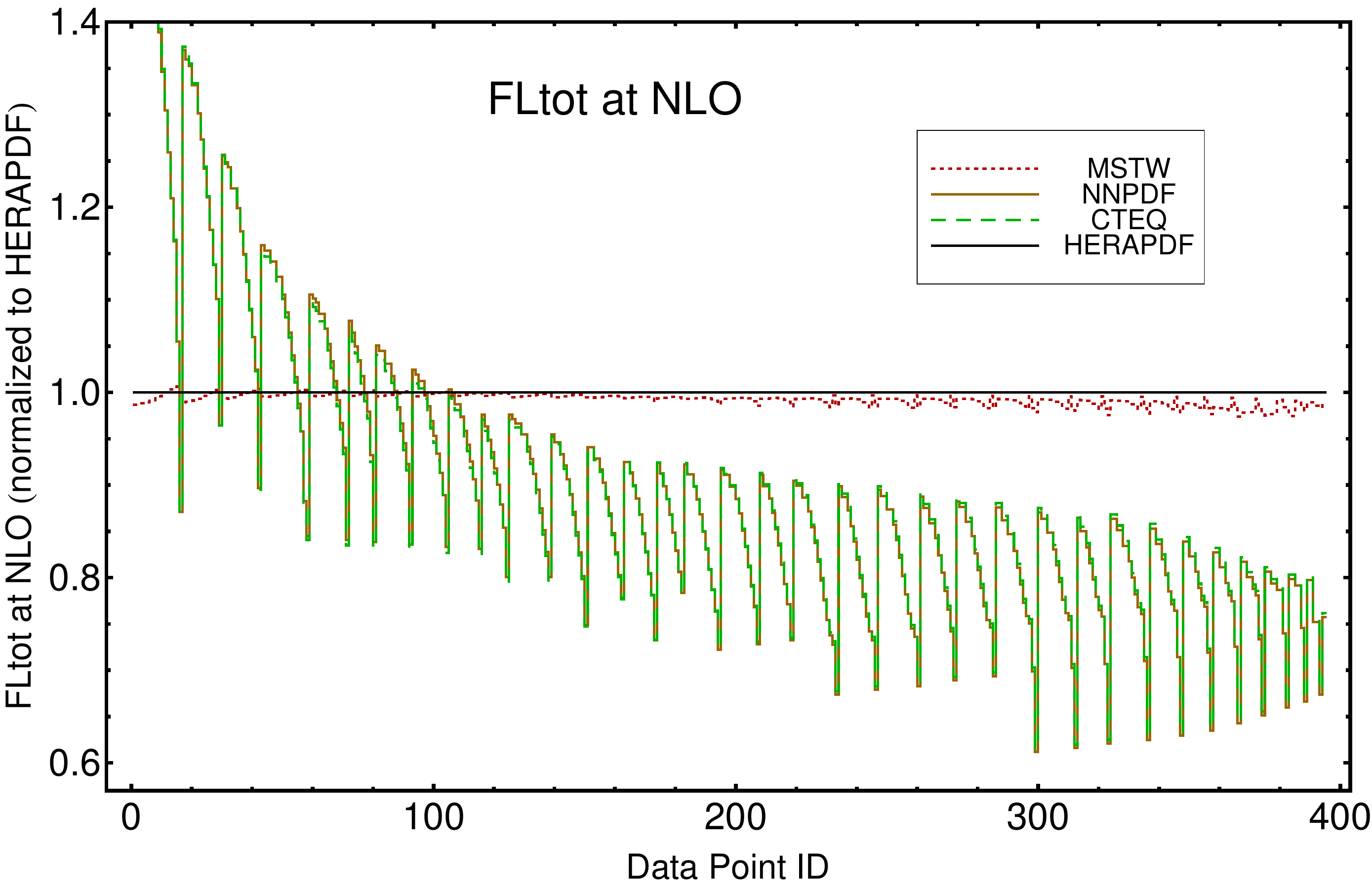}
 \includegraphics[width=0.48\textwidth]{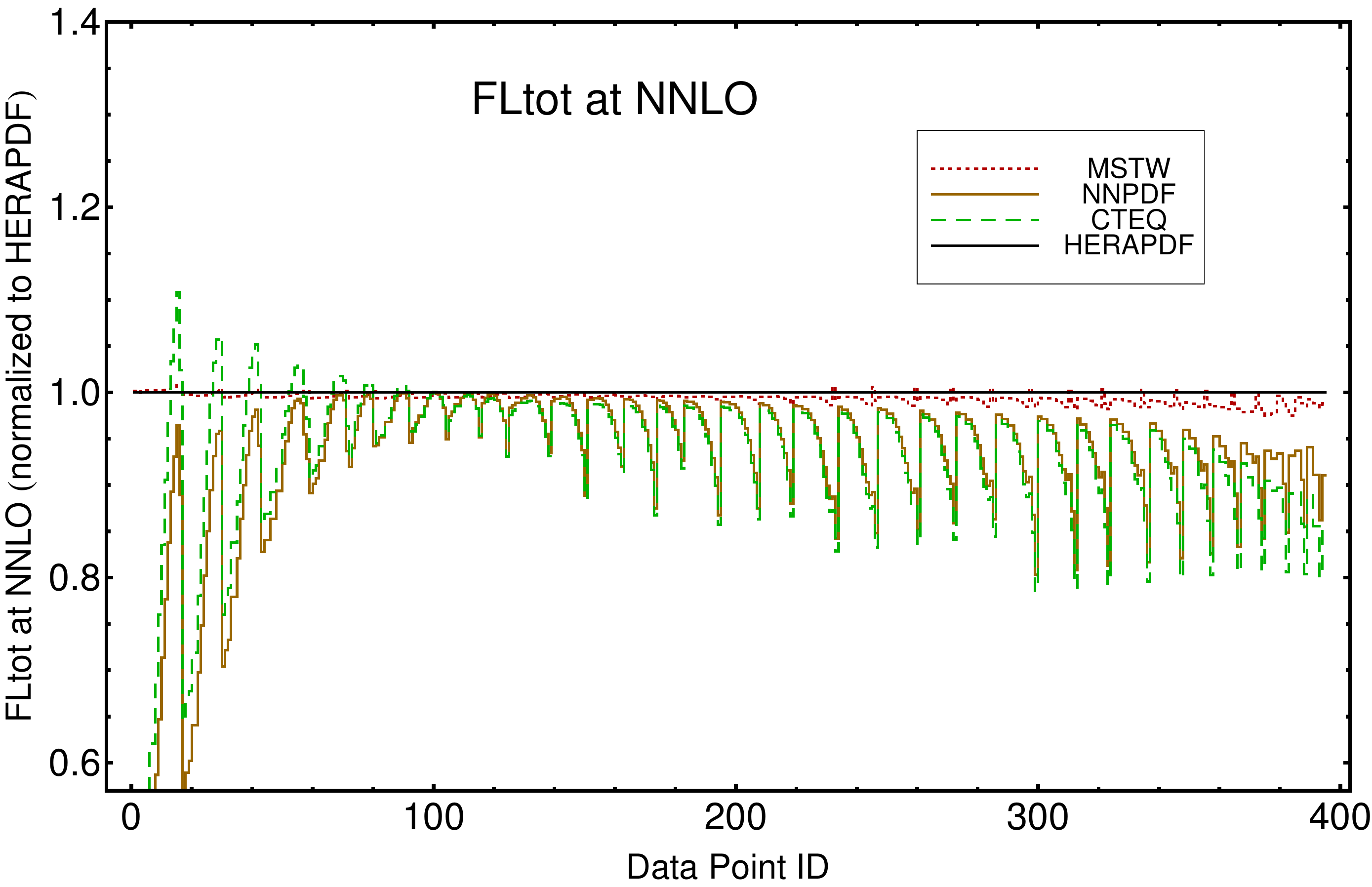}
\vspace{-1ex}
 \caption{\label{fig:STRfltot} Same as Fig.~\ref{fig:STRfllt}, for the
 structure function $F_{L,tot}$. }
\end{figure}

We now turn to the other structure functions.
Figs.~\ref{fig:STRfllt}-\ref{fig:STRfltot} show results for
$F_{L}$. Here different counting of
perturbative orders adopted by the groups 
results in large discrepancies 
between the 
MSTW/HERAPDF and CTEQ/NNPDF $F_L$ values of nominally the same orders, 
e.g., at LO $F_L$ vanishes
for CTEQ/NNPDF, while it is of ${\mathcal O}(\alpha_s)$ for MSTW/HERAPDF.
We have checked explicitly that, once we use the ${\mathcal O}(\alpha_s)$
(${\mathcal O}(\alpha_s^2)$) Wilson coefficients for $F_L$ with LO (NLO) toy 
PDFs both in CTEQ and MSTW codes, the respective CTEQ and MSTW values agree 
at the 1-2\% level. We also find very good agreement on $F_3$ among all groups.

\begin{figure}
\centering
 \includegraphics[width=0.48\textwidth]{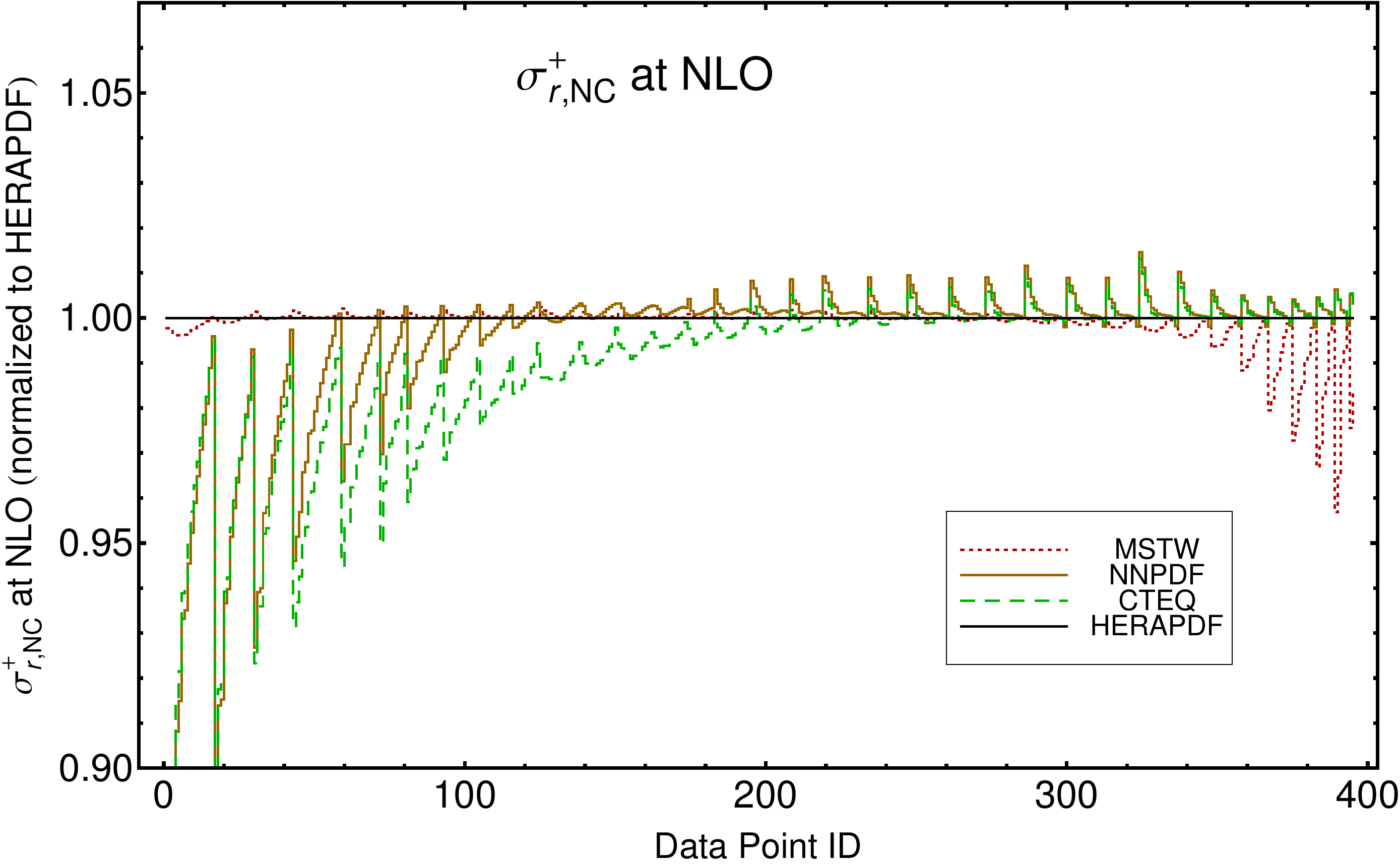}
 \includegraphics[width=0.48\textwidth]{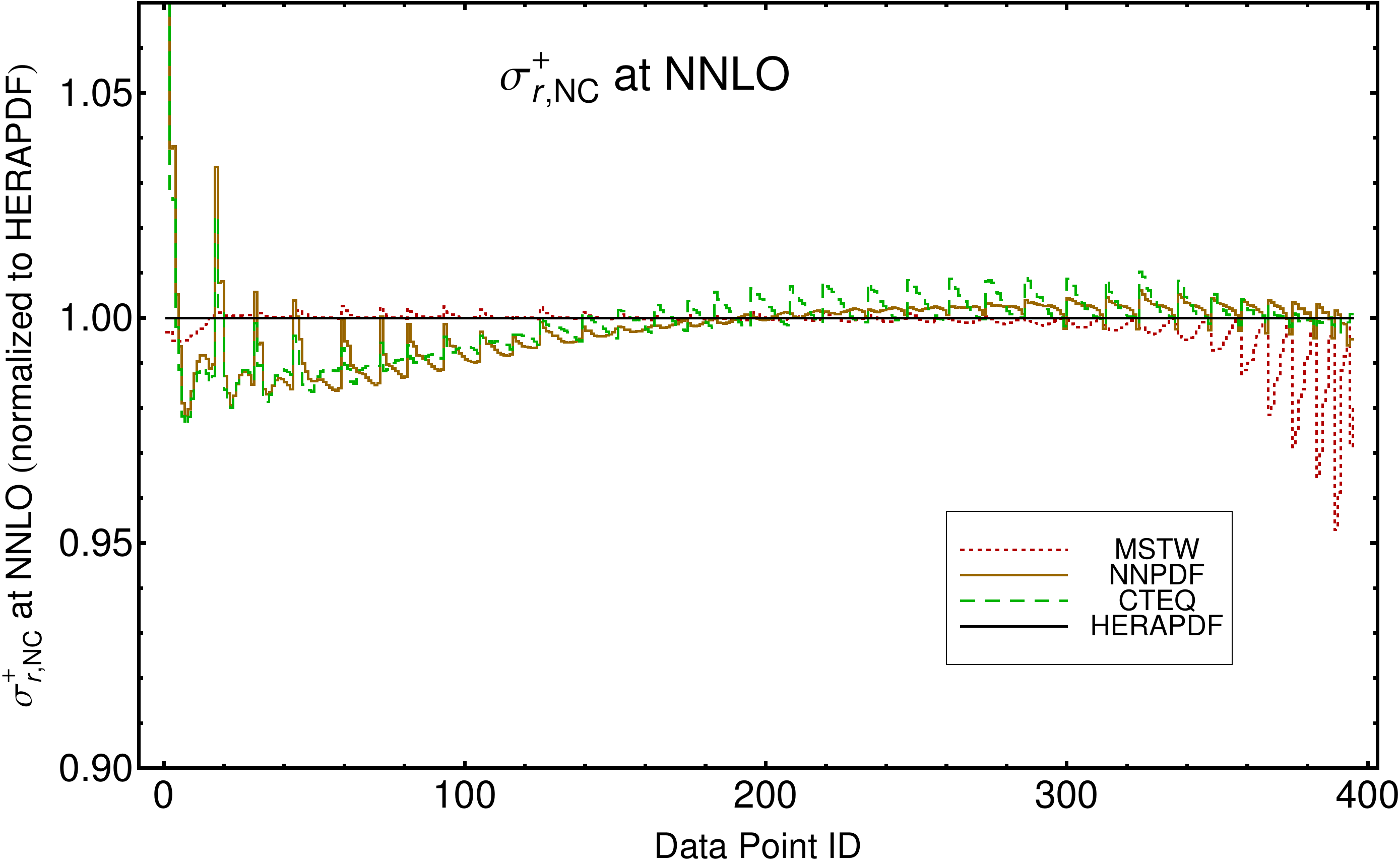}
\vspace{-1ex}
 \caption{\label{fig:STRred} Same as Fig.~\ref{fig:STRf2lt}, for
 the HERA reduced cross section, $\sigma^{+}_{r,NC}$. }
\end{figure}

Finally, in Fig.~\ref{fig:STRred}
we show a comparison of the HERA reduced cross sections constructed from
$F_2$, $F_L$, and $F_3$~\cite{Aaron:2009aa}, which can be
directly compared to the data. Again we see
excellent agreement between all groups, and significant convergence from NLO to NNLO.
Except for the extremely high
$Q$ region, where there exist some differences in the implementations
of electroweak corrections (and where data 
points have very large errors), all curves agree
within 1\% in the 
moderate-$Q$ region,
and 2\% in the low-$Q$ region. The MSTW and HERAPDF predictions are
essentially identical, as they should be (other than the differences due to
electroweak corrections or
numerical precision). CTEQ and NNPDF are quite close, too, even at low $Q$,
due to the similarities of the S-ACOT-$\chi$ and FONLL-A (FONLL-C) 
heavy-quark schemes at NLO (NNLO).

We conclude from this benchmark exercise
that the differences in theoretical predictions
of neutral current DIS cross sections between CT, HERAPDF,
MSTW and NNPDF are well understood as mostly arising from
the choice of the respective heavy-quark
schemes, non-identical counting of perturbative orders, 
electroweak contributions, and that these theoretical 
differences are small at NNLO 
compared to the experimental margin of error.
Therefore, we
expect that the differences in heavy-quark schemes should not have a large
impact on the fits using only HERA-1 data or on the Higgs cross sections
predicted from them at the NNLO.
This will be investigated in the next section.
%

Before closing this section, we note
that in the following fits, NNPDF uses the
truncated solution for the DGLAP evolution equations and the
$\alpha_s$ renormalization-group evolution.
Modifications resulting from this choice (essentially from a 
different treatment of higher-order perturbative corrections 
as compared to the exact solutions
used in the CT, MSTW and HERAPDF codes) should be quite small at NNLO.
We have verified that it is certainly negligible for the
running of $\alpha_s$. Second, larger differences (of order 10\%) are
observed between the group's predictions for charged-current DIS cross
sections, which are still evaluated at NLO. As the experimental
uncertainties on the charged-current DIS cross sections are still rather
significant, the current differences between the codes 
in the charged-current sector are not anticipated to be of the same
consequence for the gluon PDF as those in the neutral-current DIS cross
sections. 

\subsection{Benchmark PDF fits to HERA-1 data\label{sec:bench}}

\subsubsection{Setup\label{sec:set}}

We will first describe the setup used for the fits to the reduced data set 
based on the combined HERA-1 DIS data only. At this stage we performed
the baseline fit based on the default settings, as well as a
series of alternative fits designed to assess dependence on the input  
assumptions. The kinematical ranges of the data sets included in the fits,
and the number of points in each measurement, are summarized in
Table~\ref{tab:herakin}. The baseline fit includes only the combined 
inclusive  data with $Q^2>5 {\rm GeV}^2$ 
to minimize the differences due to the heavy-quark 
schemes. Some alternative fits also include the combined charm quark production
data and vary the lower $Q^2$ cut or other input parameters. 
We assume the strong coupling constant
to be $\alpha_s(M_Z)=0.118$ and set the pole mass of the
charm quark to a common value of $1.4\,{\rm GeV}$
in the baseline fit and to the default values of each group 
in the variant fits. NNLO QCD theory is used throughout. 

In the HERA-1-only fits, the data do not constrain
PDFs in the extremely large $x$ region or
some of the flavor combinations, e.g.,
$\bar u-\bar d$, $(s+\bar s)/(\bar u+\bar d)$, $s-\bar s$. 
The remaining degrees of freedom are associated
mostly with $g$, $u_v$, $d_v$, and $\bar u+\bar d$ PDFs.
Therefore, CTEQ fits use a smaller number of free PDF parameters than in their
global fits, as well as updated parametrization forms for some flavors
based on Chebyshev polynomials~\cite{Pumplin:2009bb,Martin:2012da}. 
The MSTW HERA1-only fit is done using an updated
parametrization based on Chebyshev polynomials for some flavors, 
which has been shown
\cite{Martin:2012da} to lead to a change in the Higgs cross section at
the LHC of only a small fraction of a percent compared to the
published MSTW'08 parametrizations. The HERAPDF baseline fit uses
the same number of free parameters as in HERAPDF1.5, which is smaller
than in CTEQ and MSTW global fits. Similarly, the baseline neural-network
parametrizations of NNPDF are the same as in their global fits and
produce very large uncertainty bands for their unconstrained PDF
combinations. The dependence on the PDF parametrization form is
further discussed in Sec.~\ref{sec:var}.

All PDF analyses utilize the figure-of-merit $\chi^2$ to find the best
fit and estimate the PDF uncertainties. In the presence of correlated
systematics, $\chi^2$ can be written as a function of 
the PDF parameters ${a}$ and nuisance parameters ${\lambda}$,
\begin{equation}
\chi^2({a},{\lambda})=\sum_{k=1}^{N_{pt}}{1\over s_k^2+\sigma_k^2}\left(D_k-T_k-\sum_{\alpha =1}^{N_{\lambda}}
\beta_{k,\alpha}\lambda_{\alpha}\right)^2+\sum_{\alpha=1}^{N_{\lambda}}\lambda_{\alpha}^2,
\label{chi2}
\end{equation}
where $N_{pt}$ and $N_{\lambda}$ are 
the total numbers of data points and correlated
systematic errors, respectively~\cite{Pumplin:2002vw,Pumplin:2001ct}. 
$T_k$ are the theoretical predictions, and $D_k$ are the
central values of experimental measurement. 
$s_k$, $\sigma_k$, and $\beta_{k,\alpha}$
are the statistical, uncorrelated systematic, and correlated
systematic errors, including luminosity error. The available experimental errors are published in the form of
fractional errors. Depending on how they are normalized, there exist
several definitions of the $\chi^2$
function~\cite{Ball:2012wy,Aaron:2009aa,Gao:2013xoa}, 
some of which are reviewed in Table~\ref{tab:chidef}. 
The table entries list the variables that
multiply the corresponding fractional errors.

The usage of
$\chi^2$ with the $d_1$ definition in the fit would lead
to the D'Agostini bias from the luminosity error~\cite{D'Agostini:1993uj}.
The published HERAPDF set adopts the $d_2$ definition.
MSTW and NNPDF use the $d_3$ definition for their published
sets\footnote{NNPDF actually
use the $T^{(0)}$ method~\cite{Ball:2009qv}, in which the
theoretical values multiplying the correlated systematic errors 
are held constant in a series of iterations. 
MSTW2008 apply a quartic penalty to the normalization uncertainty.}.
The CT10 PDF sets
use the $d_4$ definition by default. The justification for $d_{2,3,4}$ requires one 
to draw a clear distinction between the additive and multiplicative
experimental errors, which has not been done in the combined HERA-1
publications. Instead, in the published combined HERA-1 data set, the
uncertainty associated with the $\chi^2$ definition in the combination 
procedure has been counted as one of the correlated systematics. In
the HERA1-only baseline fits, all four groups used the $d_4$
definition. The effects of using the alternative $\chi^2$ definitions
is discussed in Sec.~\ref{sec:var}.

\begin{table}

\centering

\begin{tabular}{|c|c|c|c|c|c|}
\hline
 & NC ($e^+p$) & NC ($e^-p$) & CC ($e^+p$) & CC ($e^-p$)  & Charm \tabularnewline
\hline
\hline
$Q^2$ range [${\rm GeV}^2$] & 0.045-30000  & 90-30000 & 300-15000 & 300-30000  & 2.5-1000 \tabularnewline
\hline
$x$ range & 0.621$\times10^{-6}$-0.65 & 0.0013-0.65 & 0.008-0.4 & 0.013-0.4  & 3$\times 10^{-5}$-0.05  \tabularnewline
\hline
$N_{pt}$, $Q^2\ge 2\,{\rm GeV}^2$  & 408 & 145 & 34 &34    & 52 \tabularnewline
\hline
$N_{pt}$, $Q^2\ge 3.5\,{\rm GeV}^2$ & 379 &145 & 34 &34   & 47 \tabularnewline
\hline
$N_{pt}$, $Q^2\ge 5\,{\rm GeV}^2$ & 353 &145 & 34 & 34  & 47 \tabularnewline
\hline
\end{tabular}

\caption{Kinematic ranges and number of points in the HERA-1-only fits.
    \label{tab:herakin}}
\end{table}

\begin{table}

\centering

\begin{tabular}{|c|c|c|c|c|}
\hline
 & $s_k$ (sta.) & $\sigma_k$ (unc. sys.) & $\beta_{k,NL}$ (cor., not lum.) & $\beta_{k,L}$ (lum.)\tabularnewline
\hline
\hline
 $d_1$(exp.)& $D_k$ & $D_k$ & $D_k$ & $D_k$ \tabularnewline
\hline
 $d_2$(HERAPDF)& $\sqrt{D_k\cdot T_k}$ & $T_k$ & $T_k$ & $T_k$ \tabularnewline
\hline
 $d_3$(NNPDF/MSTW)& $D_k$ & $D_k$ & $D_k$ & $T_k$ \tabularnewline
\hline
 $d_4$(CTEQ)& $D_k$ & $D_k$ & $T_k$ & $T_k$ \tabularnewline
\hline
\end{tabular}

\caption{Normalization factors in the definitions of experimental
  errors adopted by the PDF analysis groups.
    \label{tab:chidef}}
\end{table}

\subsubsection{Baseline fits\label{sec:base}}

In their baseline fits, each group provides a best-fit candidate set, which 
yields the lowest $\chi^2$ for the HERA-1 inclusive DIS data. 
The NNPDF group additionally provides an estimate of the PDF
uncertainty, serving as a measure of the agreement 
between the best fits of the different
groups. CTEQ and MSTW have both examined the Higgs 
cross section variation using the
Lagrange Multiplier method \cite{Stump:2001gu}, 
each obtaining a 68\% c.l. uncertainty of about
0.4 pb on $\sigma_H$ at 8 TeV (corresponding to $\Delta \chi^2=1$), 
which is a bit bigger than 
the standard HERAPDF1.5 uncertainty of $\approx$0.25 pb.

\begin{table}[tb]
\centering
\begin{tabular}{|c|c|c|c|c|}
\hline
566 data points & CTEQ& MSTW & NNPDF  & HERAPDF \tabularnewline
\hline
\hline
$\chi^2$ & 521.8& 514.8  & 548.5    & 535.0          \tabularnewline
\hline
lum. shift &  -0.19&    0.27  &   0.16   &  0.18 \tabularnewline
\hline
max. shift &    1.64&    1.51  &   1.82      &     1.81 \tabularnewline
\hline
$\sigma_H$[pb], $8\,{\rm TeV}$  &    17.86  &    18.25  &   18.60$\pm 1.10$     &     18.82 \tabularnewline
\hline
$\sigma_H$[pb], $14\,{\rm TeV}$ &    46.37  &    47.38  &   48.76$\pm 2.26$     &     48.78 \tabularnewline
\hline
\end{tabular}

\caption{Outputs of the best baseline fits, including values of 
$\chi^2$ as defined in Eq.~(\ref{chi2}), 
luminosity and maximal systematic shifts, and 
Higgs cross sections.
    \label{tab:basefit}}
\end{table}

\begin{figure}[p]
\centering
 \includegraphics[width=0.48\textwidth]{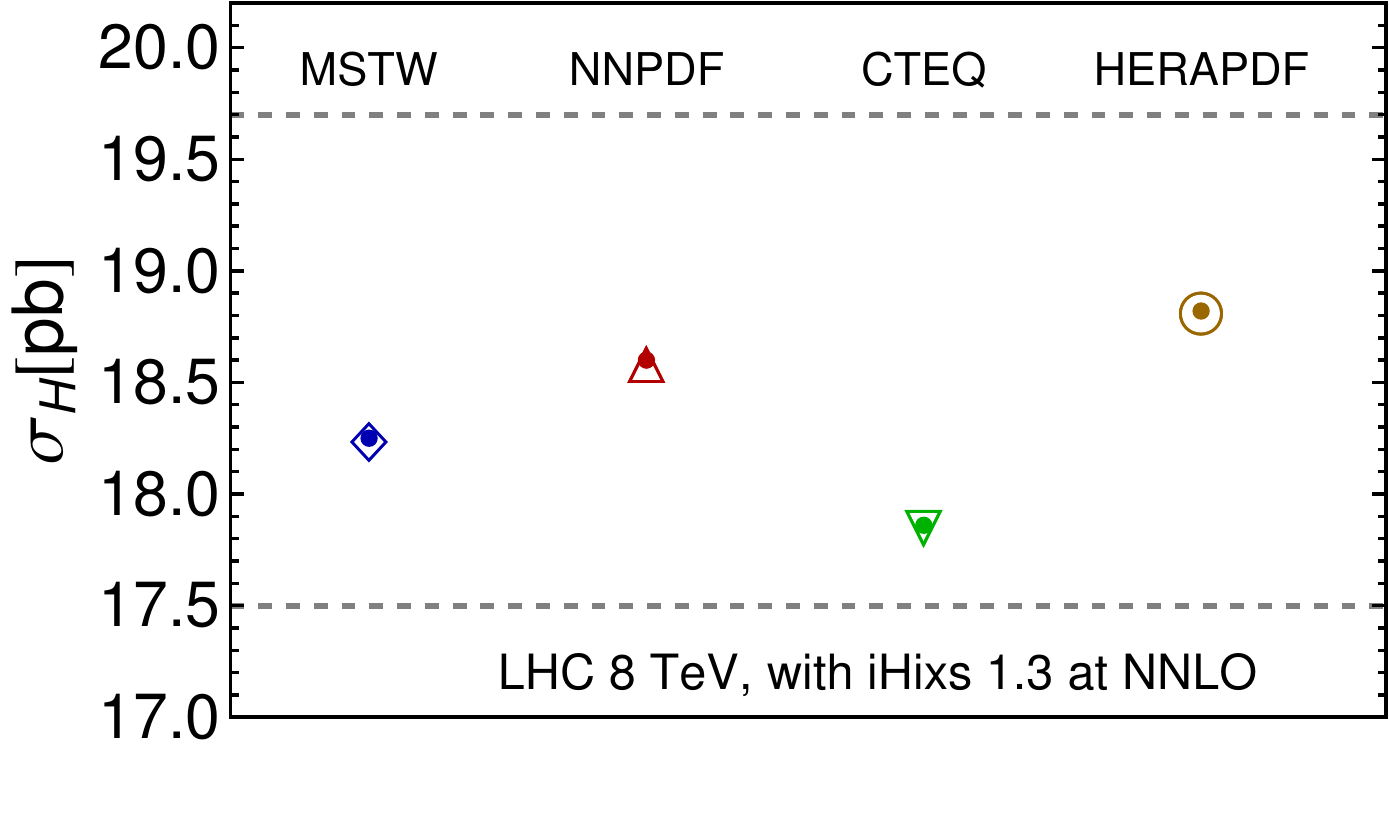}\hspace{0.2in}
 \includegraphics[width=0.48\textwidth]{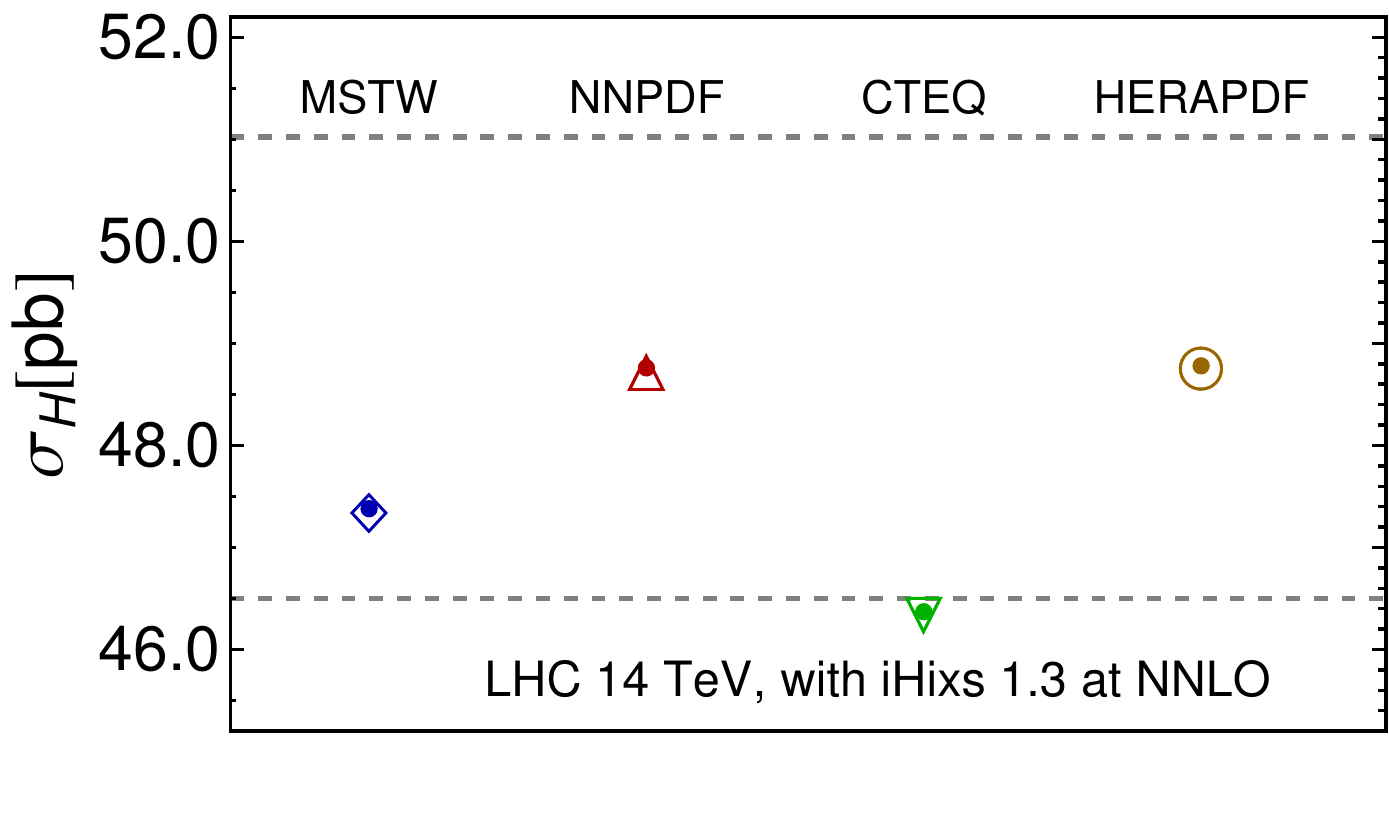}
\vspace{-1ex}
 \caption{\label{fig:baseinc}
Predictions for inclusive cross sections of the SM Higgs boson
production through gluon fusion based on the HERA-1-only NNLO PDFs.
The settings are the same as in Fig.~\ref{fig:globalinc}. The dashed
lines are the edges of the $1\sigma$ uncertainty band determined using
 the NNPDF set (and including PDF uncertainties only).}
\end{figure}

\begin{figure}[p]
\centering
 \includegraphics[width=0.55\textwidth]{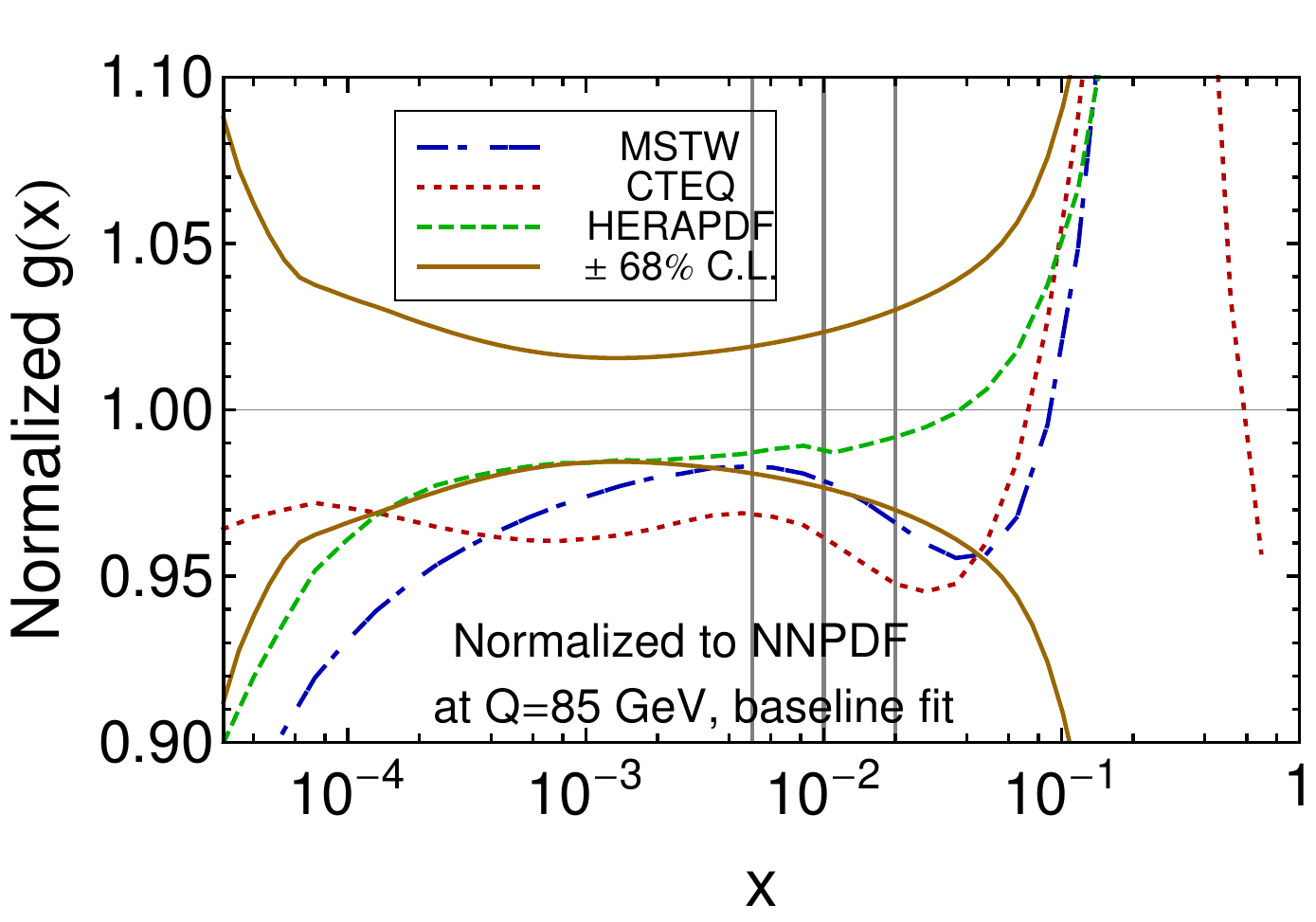}
\vspace{-1ex}
 \caption{\label{fig:baseglu} Comparison of the gluon PDFs from
 the baseline HERA-1 NNLO fits at a common scale $Q=85\ {\rm GeV}$,
 normalized to the NNPDF central prediction.}
\end{figure}

\begin{figure}[p]
\centering
 \includegraphics[width=0.48\textwidth]{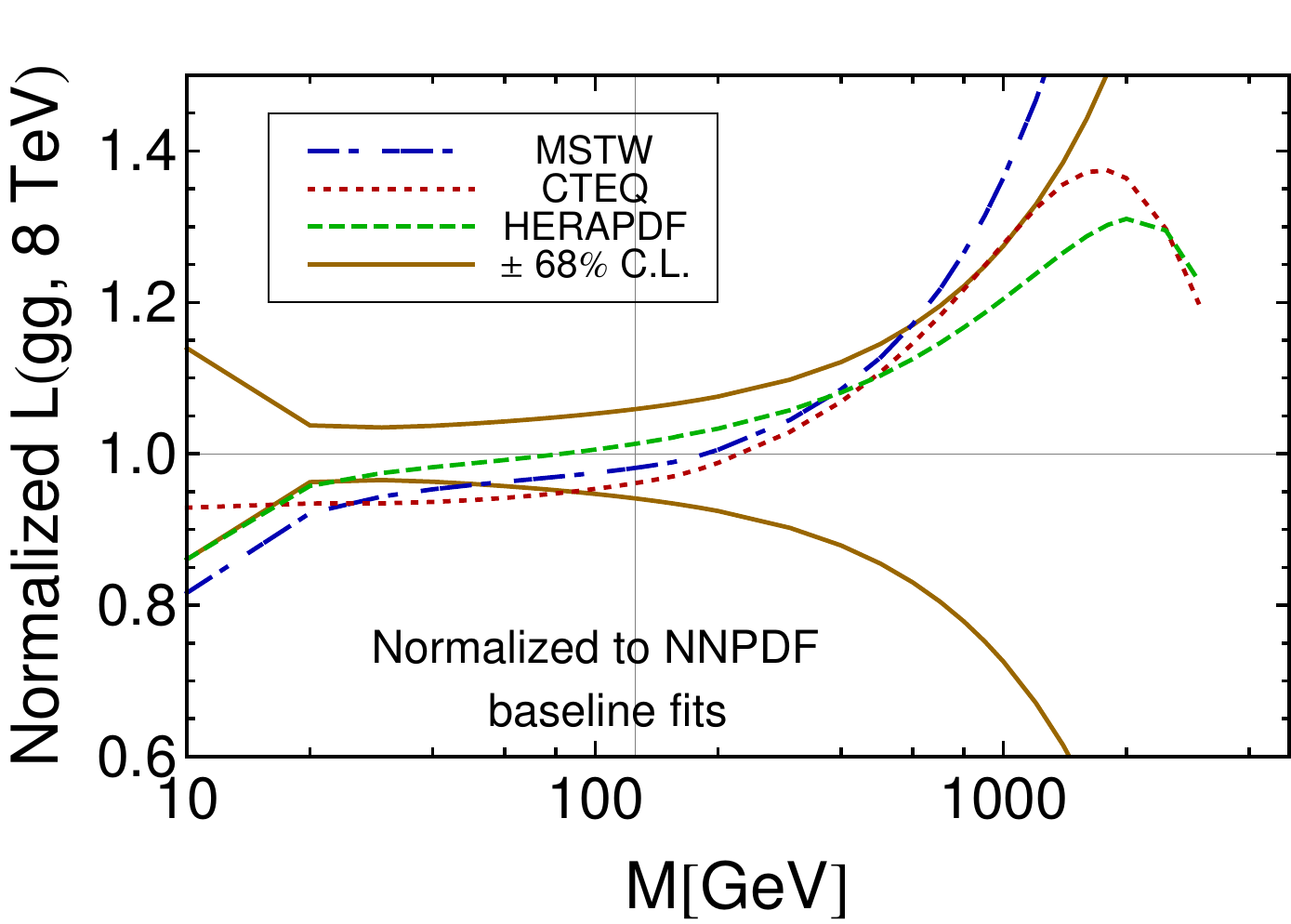}\hspace{0.2in}
 \includegraphics[width=0.48\textwidth]{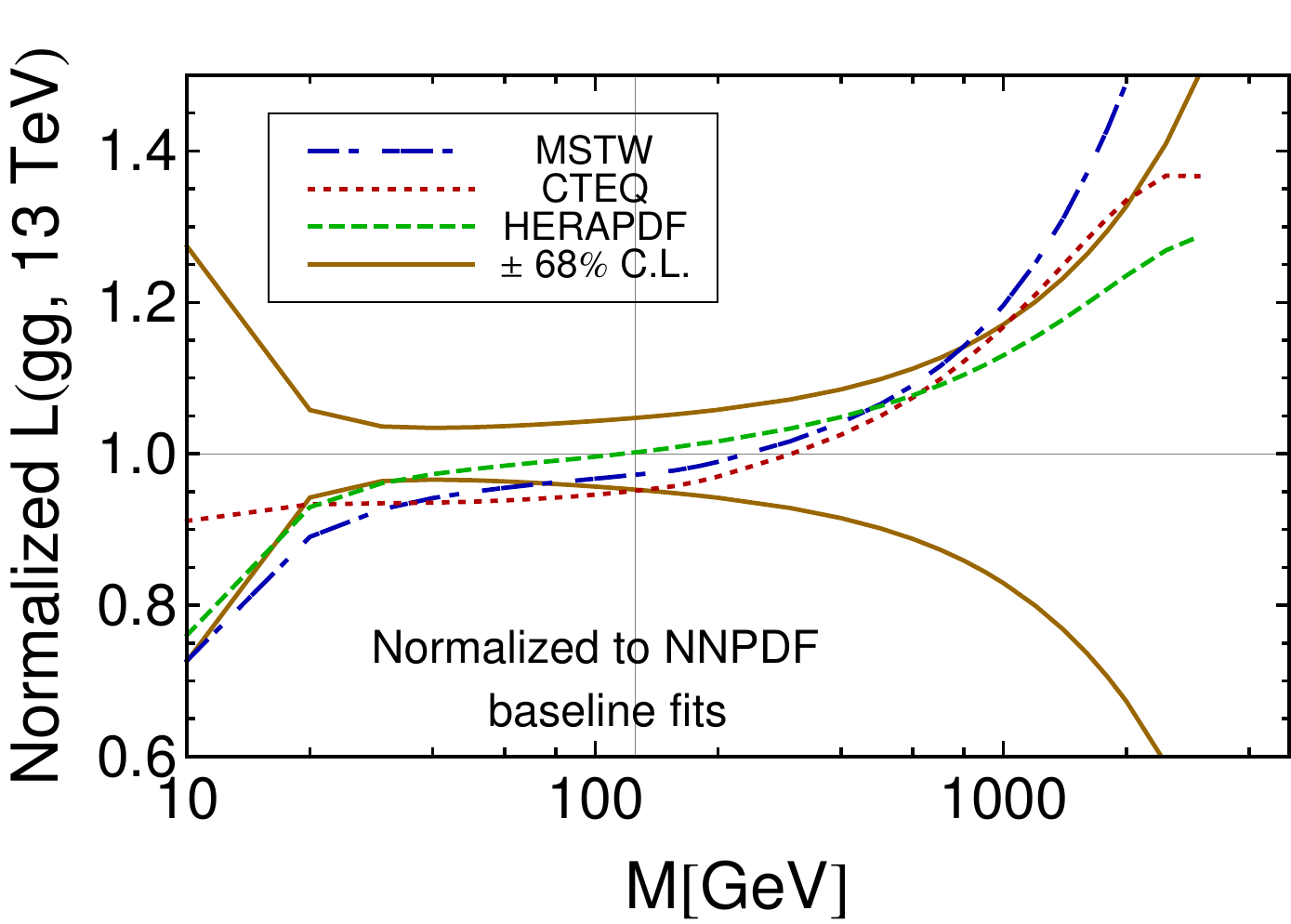}
\vspace{-1ex}
 \caption{\label{fig:baselum} Comparison of the gluon-gluon parton luminosity as
 a function of invariant mass at the LHC 8 and 13 TeV from the
 HERA-1-only NNLO fits, 
normalized to the NNPDF central prediction. The factorization scale
 is set to the invariant mass. The vertical line indicates mass of the SM Higgs boson.}
\end{figure}

\begin{figure}[tb]
\centering

 \includegraphics[width=0.48\textwidth]{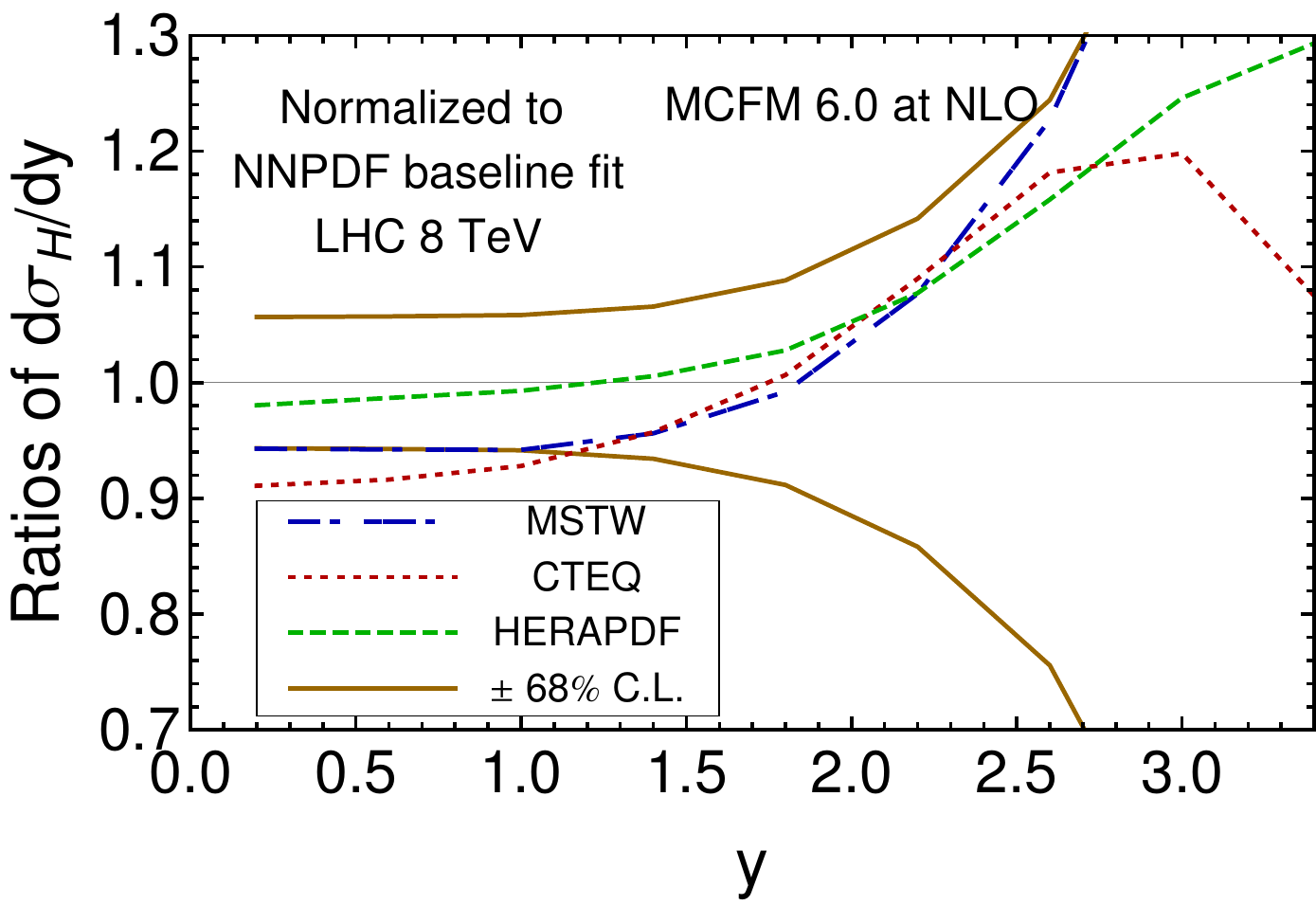}\hspace{0.2in}
 \includegraphics[width=0.48\textwidth]{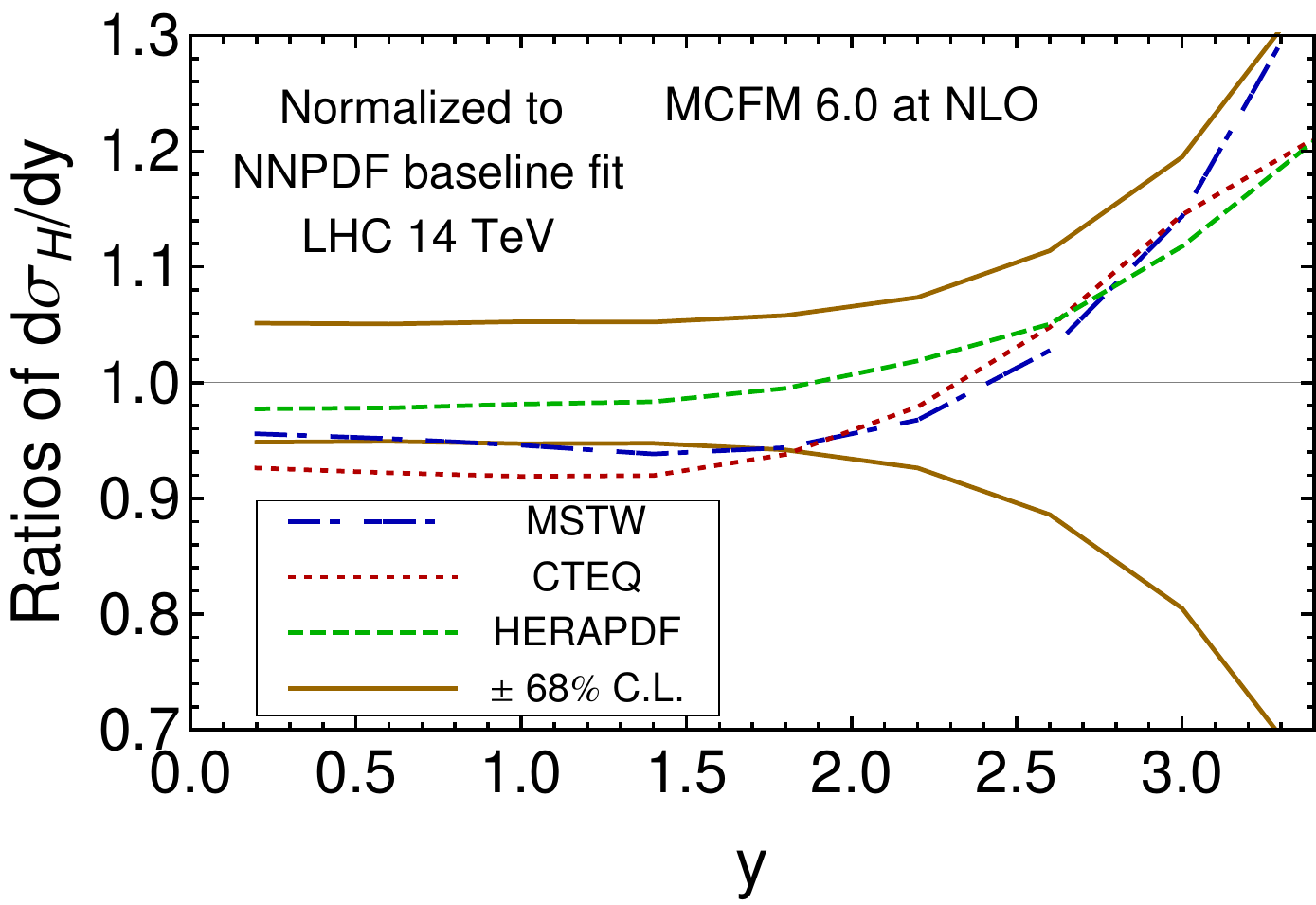}
\vspace{-1ex}
 \caption{\label{fig:baserap} Comparison of the rapidity distribution
 of the SM Higgs boson in production through gluon fusion for baseline
 HERA-1-only NNLO PDFs, normalized to the NNPDF central prediction. The
 settings are the same as in Fig.~\ref{fig:globallum}.}
\end{figure}

In Table~\ref{tab:basefit} the first three rows show the quality of
the best fits from 
the four groups, including total $\chi^2$, luminosity shift and
maximum systematic 
shift required. We see that all these best fits describe the HERA-1
inclusive data 
pretty well, with $\chi^2/d.o.f.$ being less than unity. The
systematic shifts are less than 2$\sigma$ and show similarities among
the different groups. The Higgs production cross sections 
through gluon fusion at the LHC for 8 and 14 TeV are also included in
Table~\ref{tab:basefit} and illustrated in 
Fig.~\ref{fig:baseinc}. 
Comparing with Table~\ref{tab:global2} we find that, for all groups
except HERAPDF, the cross sections 
are lowered by a similar amount compared to the public sets: by 
0.6 pb at 8 TeV and 
1.2 pb at 14 TeV. The cross sections
from HERAPDF are increased a little compared to HERAPDF1.5 
due to the removal of the preliminary HERA-2 data included in
HERAPDF1.5.  The trends
of relative differences between the groups are very similar to those for
 the public sets. The spread of the Higgs cross
sections is still at the level of $4-5$~\%. The PDF uncertainties in
the NNPDF column are larger than those predicted by NNPDF2.3, 
which has many additional data sets included. All predictions are now
within the $1\sigma$ error bands of NNPDF. 

Fig.~\ref{fig:baseglu} shows a plot for the gluon PDFs that is
analogous to Fig.~\ref{fig:globalglu}. The hierarchy of CT, MSTW, and
NNPDF predictions at the 
values of $x$ relevant for Higgs production remains similar to that of
Fig.~\ref{fig:globalglu}. The HERAPDF curve moved above both CT and MSTW curves 
at these $x$, in contrast to Fig.~\ref{fig:globalglu}, where it is below CT. 
At large $x$ the relative gluon shapes are very similar to those in the
global fits, with the central gluon PDF from NNPDF being lower than
those of the other three groups in the $x\gsim 0.2$
region, but still within the large gluon uncertainty from all of
them. (The HERA DIS data set does not impose a strong constraint on
$g(x,Q)$ at high $x$.) However, for
$x\lsim 10^{-3}$ the shapes and ordering of gluon PDFs change, with the
MSTW and CTEQ gluons staying below the NNPDF one in the central Higgs
production region. Fig.~\ref{fig:baselum}
presents plots for the gluon-gluon parton luminosity similar to
Fig.~\ref{fig:globallum}. All predictions are located inside, or close to the PDF
uncertainties predicted by NNPDF for invariant masses above 100 GeV.

Fig.~\ref{fig:baserap} is a counterpart of Fig.~\ref{fig:globalrap}
and shows Higgs rapidity distributions obtained with the HERA-1 baseline PDFs. 
The spread of the distribution in the central region is
a little larger than the spread of the inclusive cross sections.  It
generally amounts to
$\sim1.5 \sigma$ of the uncertainties predicted by NNPDF.
While all curves moved somewhat in Fig. 15  as compared  to Fig. 4, 
the HERAPDF curve underwent the most pronounced 
change between the two sets of figures,
mostly as a consequence of the removal of the HERA-2 data set from their fit.
The ratio to NNPDF is  flatter in the rapidity 
distribution for the HERA-1-only fit.

\subsubsection{Dependence of the fits on input assumptions\label{sec:var}}

In addition to the fits with the baseline settings, we carried out
various exploratory fits to understand sensitivity to the input parameters.

\begin{figure}[p]
\centering

 \includegraphics[width=0.48\textwidth]{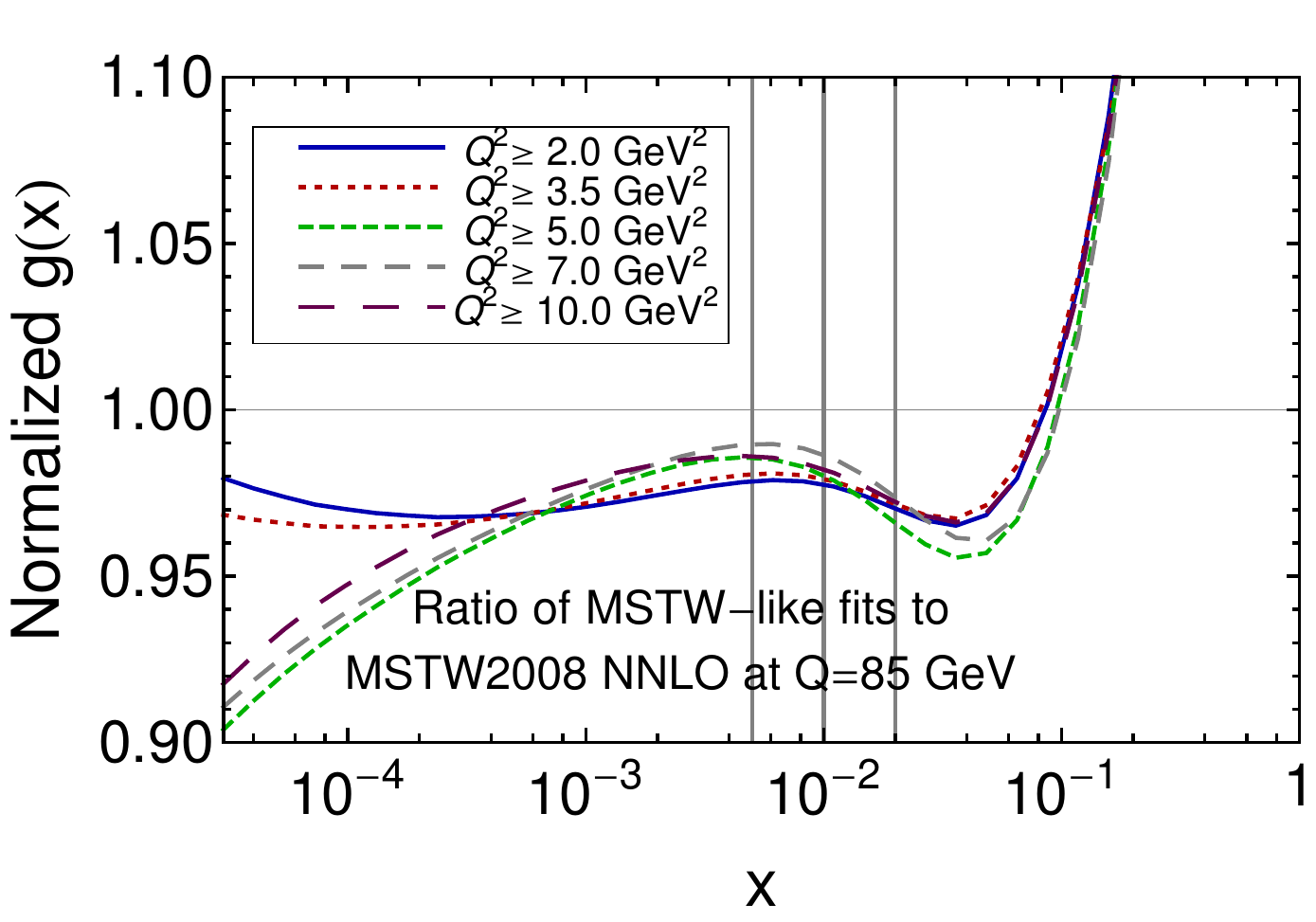}\hspace{0.2in}
 \includegraphics[width=0.48\textwidth]{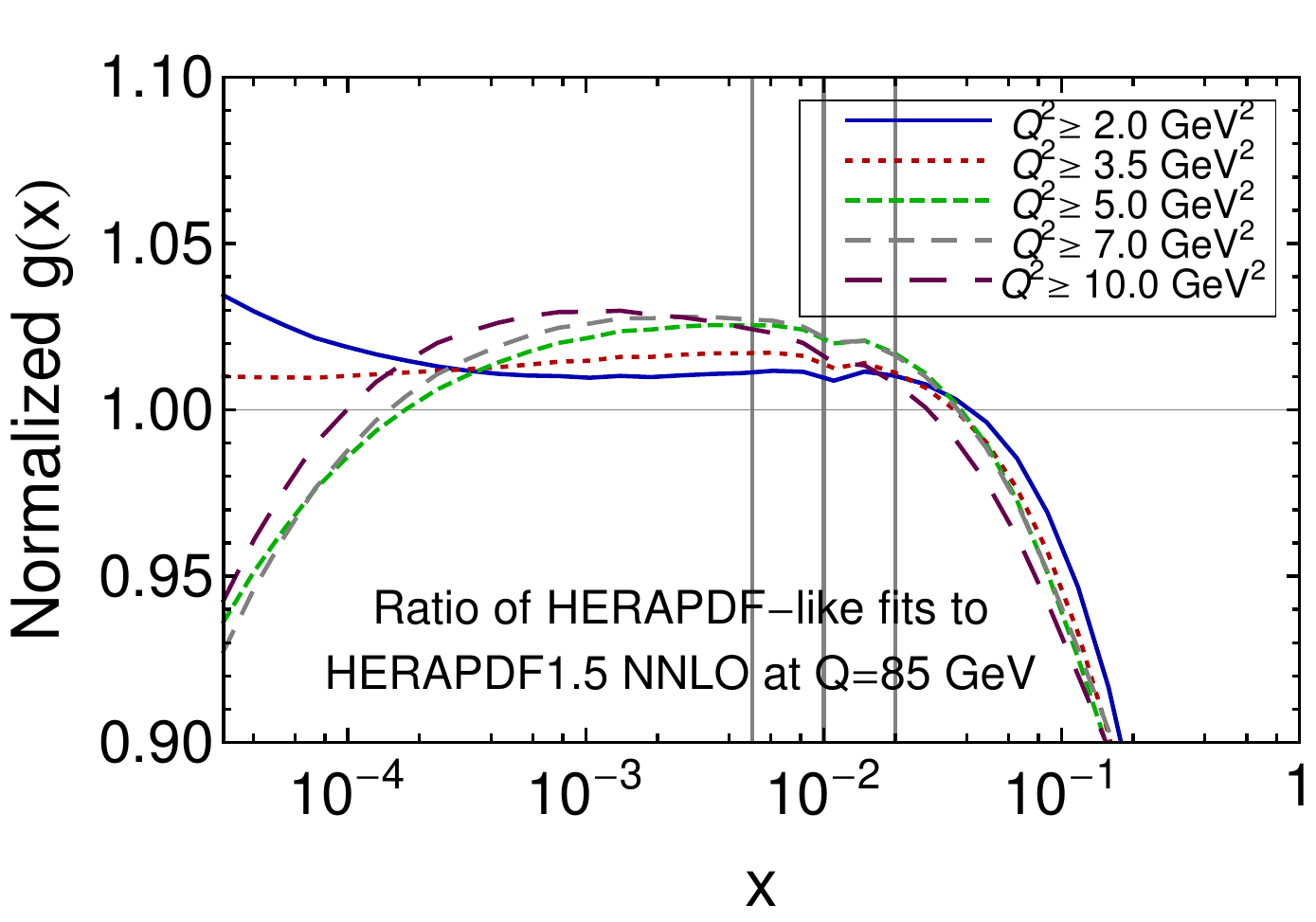}
\vspace{-1ex}
 \caption{\label{fig:msq2} Dependence of the gluon PDF on the $Q^2$
   cut in HERA-1 inclusive data selection, from MSTW and HERAPDF fits.}
\end{figure}

\begin{figure}[p]
\centering

 \includegraphics[width=0.48\textwidth]{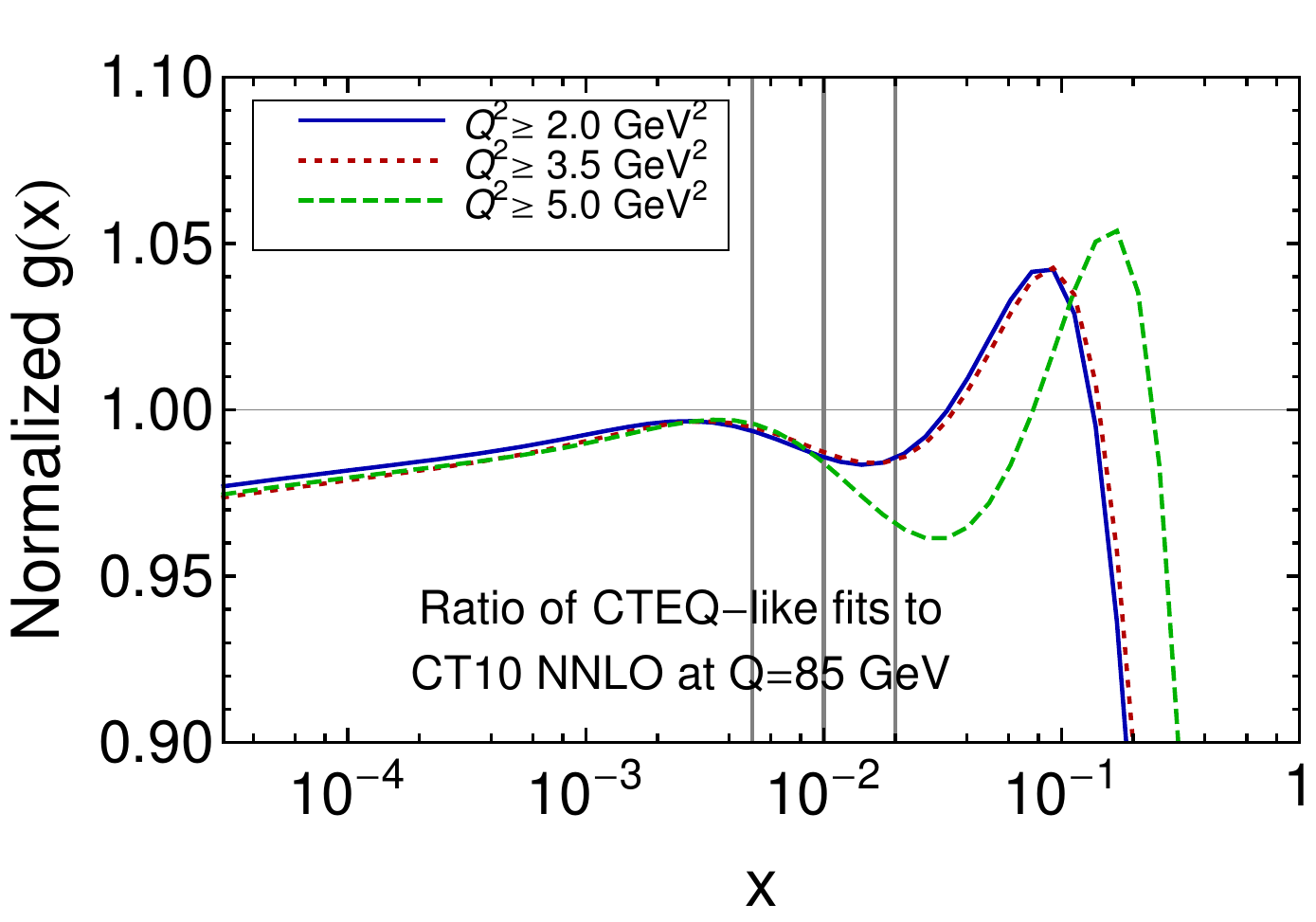}\hspace{0.2in}
 \includegraphics[width=0.48\textwidth]{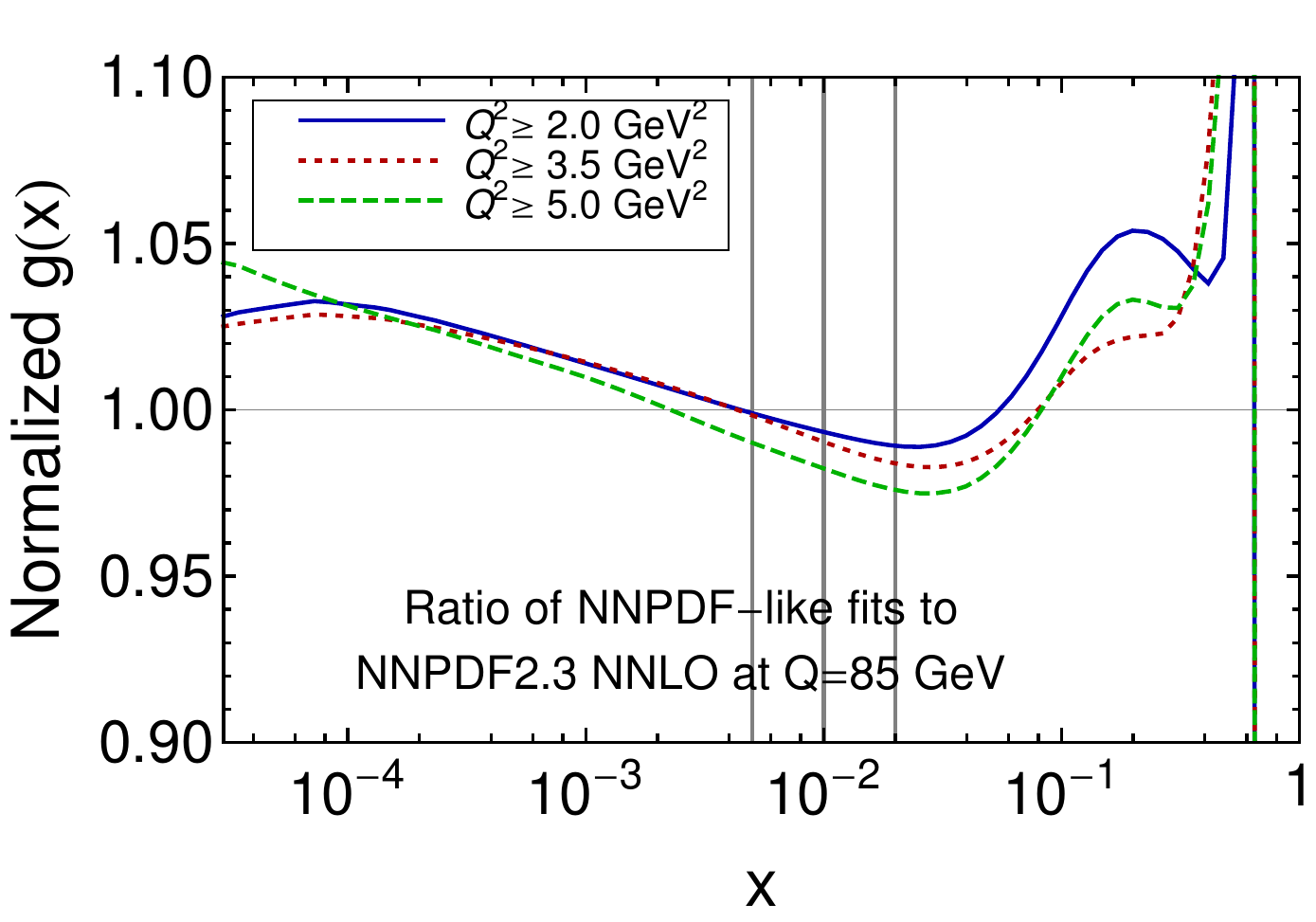}
\vspace{-1ex}
 \caption{\label{fig:nnq2} Dependence of the gluon PDF on the $Q^2$
   cut in HERA-1 inclusive data selection, for CTEQ and NNPDF fits.}
\end{figure}

{\bf Choice of the $Q^2$ cut.}
The internal consistency of the HERA-1 inclusive DIS data set was examined 
by varying the low $Q^2$ cut for data selection.
Fig.~\ref{fig:msq2} shows the best fits of MSTW and HERAPDF with $Q^2$ cuts
of $2$, $3.5$, $5$, $7.5$, and $10\,{\rm GeV^2}$. Fig.~\ref{fig:nnq2} gives similar plots for CT10 and NNPDF fits with $Q^2$ cuts of $2$, $3.5$, and $5\,{\rm GeV}$. The
gluon PDFs can vary by up to 2\% around the central region of Higgs production
depending on the chosen $Q^2$ cut. However, the inclusive $gg\rightarrow H$
cross sections are
less sensitive to the $Q^2$ cut due to compensation from the gluon PDFs across
the whole range of $x$ which is accessed.
For example, Table~\ref{tab:xsecq2} shows the relative variations of the inclusive cross sections.
Cross sections from MSTW, CTEQ, and HERAPDF fits are rather stable, with variations
of about 1\%. The NNPDF cross sections show a larger variation of about 2\%.

\begin{table}[p]

\centering

\begin{tabular}{|c|c|c|c|c|}
\hline
 & CTEQ& MSTW & NNPDF  & HERAPDF \tabularnewline
\hline
\hline
Range of low $Q^2$ cuts [${\rm GeV}^2$] & 2.0-5.0& 2.0-10.0& 2.0-5.0 & 2.0-10.0 \tabularnewline
\hline
$\delta\sigma_H/\sigma_H $ [\%], LHC 8 TeV &    1.3& 0.9   & 2.6       &  1.1         \tabularnewline
\hline
$\delta\sigma_H/\sigma_H $ [\%], LHC 14 TeV & 1.2 &  0.8   & 2.2           & 1.4       \tabularnewline
\hline
\end{tabular}

\caption{Relative variations in predicted Higgs cross sections when changing 
the $Q^2$ cut in selection of the HERA-1 inclusive DIS data.
    \label{tab:xsecq2}}
\end{table}

\begin{figure}
\centering

 \includegraphics[width=0.55\textwidth]{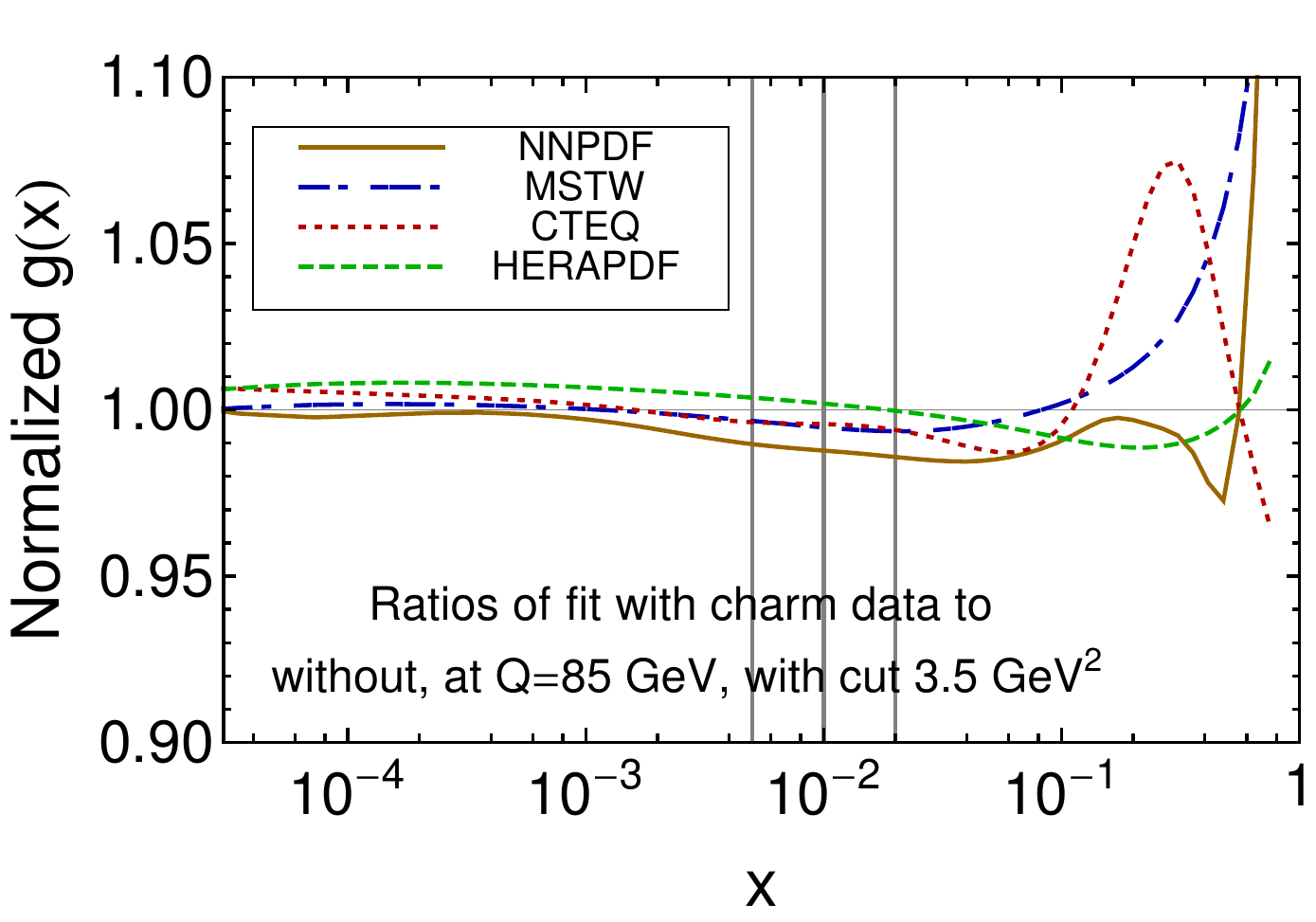}
\vspace{-1ex}
 \caption{\label{fig:charmdep} Changes of the gluon PDF when including the
 HERA-1 charm quark production data in addition to the inclusive DIS data.}
\end{figure}

{\bf Inclusion of charm production.}
Another group of fits included 
the combined HERA-1 charm quark production data in addition to
the inclusive data. Fig.~\ref{fig:charmdep} shows the ratios of the gluon PDFs from fits with and without
the charm data for different groups. In the comparison we choose a $Q^2$ cut of
3.5 ${\rm GeV^2}$ and the charm quark mass of the default value of each group as
in Table~\ref{tab:global1}. The charm quark production data generally
prefer slightly smaller gluon
PDFs around the Higgs mass region, especially in the NNPDF PDFs,
which show a reduction
of about 1.5\%. The relative changes of the predictions for Higgs cross sections
when including the charm data are shown in Table~\ref{tab:charm}.
There is little variation, the largest change being a reduction of
about $2 \%$ for the NNPDF fits.

\begin{table}

\centering

\begin{tabular}{|c|c|c|c|c|}
\hline
 & CTEQ& MSTW & NNPDF  & HERAPDF \tabularnewline
\hline
\hline
$\delta\sigma_H/\sigma_H $ [\%], LHC 8 TeV &    -0.7  & -0.4   & -2.0     &  -0.2         \tabularnewline
\hline
$\delta\sigma_H/\sigma_H $ [\%], LHC 14 TeV&  -0.5  &  -0.4   & -1.9         & 0.2       \tabularnewline
\hline
\end{tabular}

\caption{
    \label{tab:charm} Relative changes in the predicted Higgs cross
    sections upon including the HERA-1 charm production data 
in addition to the inclusive DIS data.}
\end{table}

{\bf Parametrization dependence} was studied by repeating the fits
using alternative PDF parametrization forms and input scales $Q_0$. 

The requirement of the positivity imposed on the gluon PDF 
or gluon-mediated cross sections
alters the span of allowed gluon PDF shapes at $x<10^{-3}$ and also
may affect very large $x$ via the momentum sum rule. 
All published PDF sets except for
CTEQ allow a negative gluon PDF at very small $x$ and $Q$. 
In the alternative series, HERAPDF replaced the second gluon term,
introduced to allow small $x$ negativity, by a positive definite form with an
additional polynomial. In this case the input scale was set to 1.9 GeV$^2$.
Similarly MSTW have removed the second term in the original 
parametrization and included one more Chebyshev
polynomial while simultaneously changing the initial scale to $1.3\,\rm{GeV}$,
the same as for CTEQ, but lower than for HERAPDF.
They checked that changing the input scale while maintaining the
same parametrization form resulted in a fit quality and PDFs
essentially identical to those using the default input scale of 1 GeV.

The  NNPDF fits use the same very general parametrization for the
HERA-only fit as in their global fit, hence little parametrization
dependence is expected. In  the public global fit,
NNPDF does not require a positive definite gluon (though positivity of
observables such as $F_L$ is imposed). NNPDF also examined how
their PDFs would change if positivity on their gluon PDFs is imposed.

\begin{figure}[H]
\centering
\includegraphics[width=0.48\textwidth]{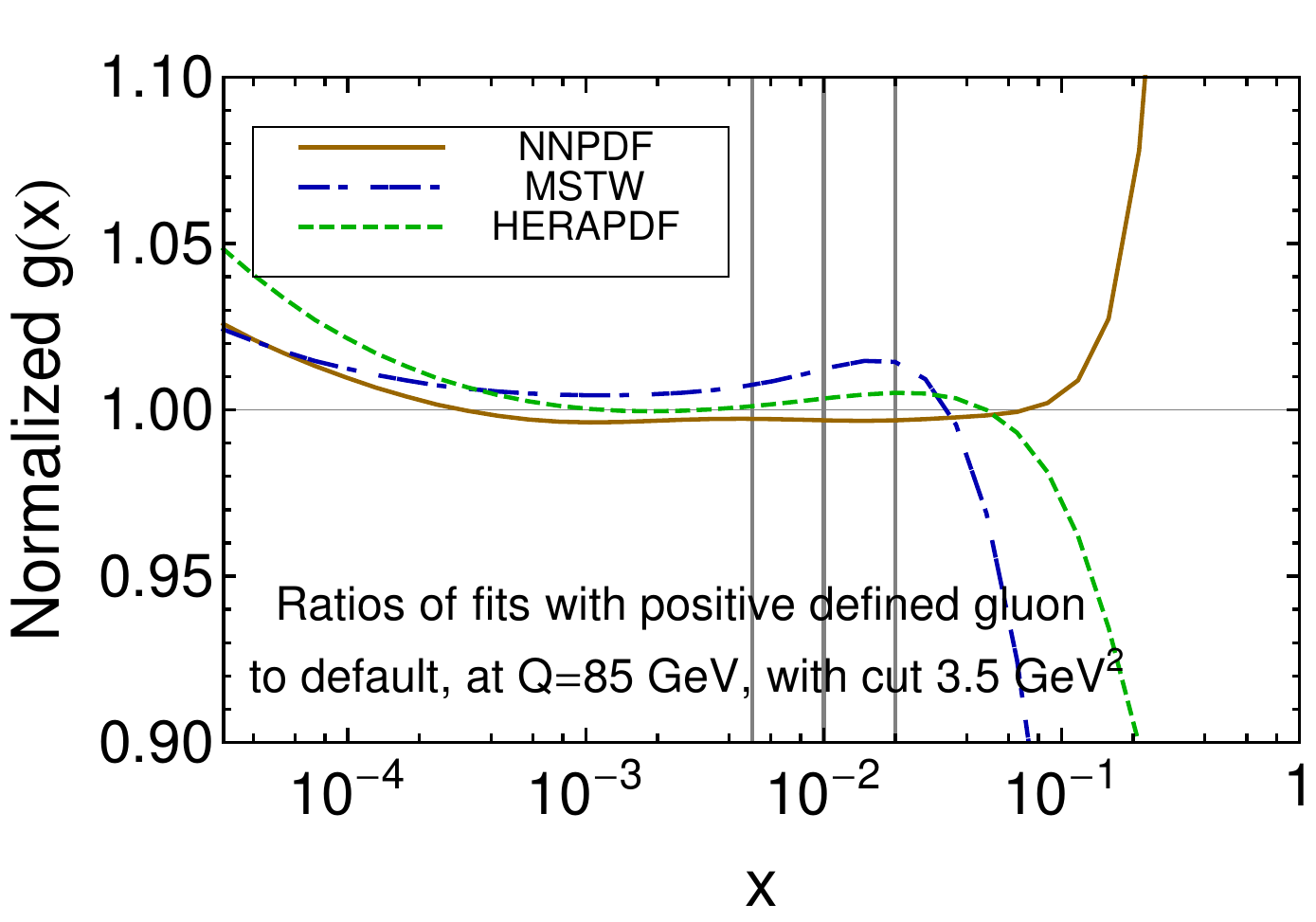}
 \includegraphics[width=0.48\textwidth]{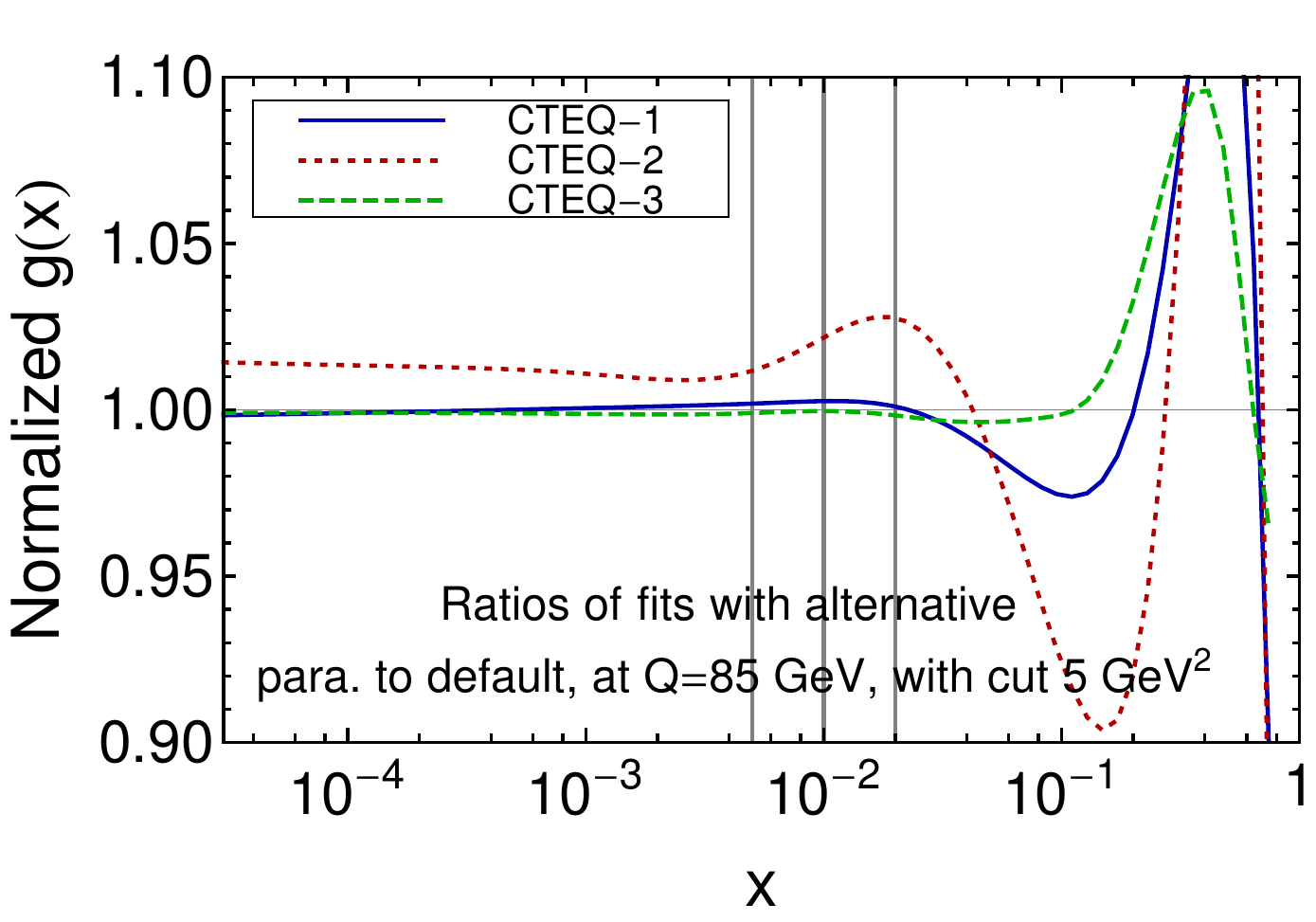}\\
(a) \hspace{2in} (b)
\vspace{-1ex}
\caption{\label{fig:pardep} Dependence of the gluon PDF on the choice of
parametrization forms of the PDFs for (a) MSTW, NNPDF, and HERAPDF fits
with HERA-1 inclusive and charm quark production data, and (b) CTEQ fits
with only HERA-1 inclusive data.}
\end{figure}

Fig.~\ref{fig:pardep}a
shows the ratios of gluon PDFs with positive-definite and
default parametrizations, from HERAPDF, MSTW, and NNPDF. In the case of
the HERAPDF or NNPDF fits, the gluon PDFs are only affected by the
positivity requirement at small
$x\sim 10^{-4}$ and large $x\sim 0.2$, and the Higgs cross sections
are reduced or increased by about half percent. The gluon PDF in the MSTW
fit could change by 10\% at $x\sim 0.1$ and induce a
reduction of the Higgs cross sections by 4\%. However, the MSTW fit with
positive gluon has a $\chi^2$ larger by 15 units and thus is not favored
by the HERA-1 data.

\begin{figure}[H]
\centering
 \includegraphics[width=0.48\textwidth]{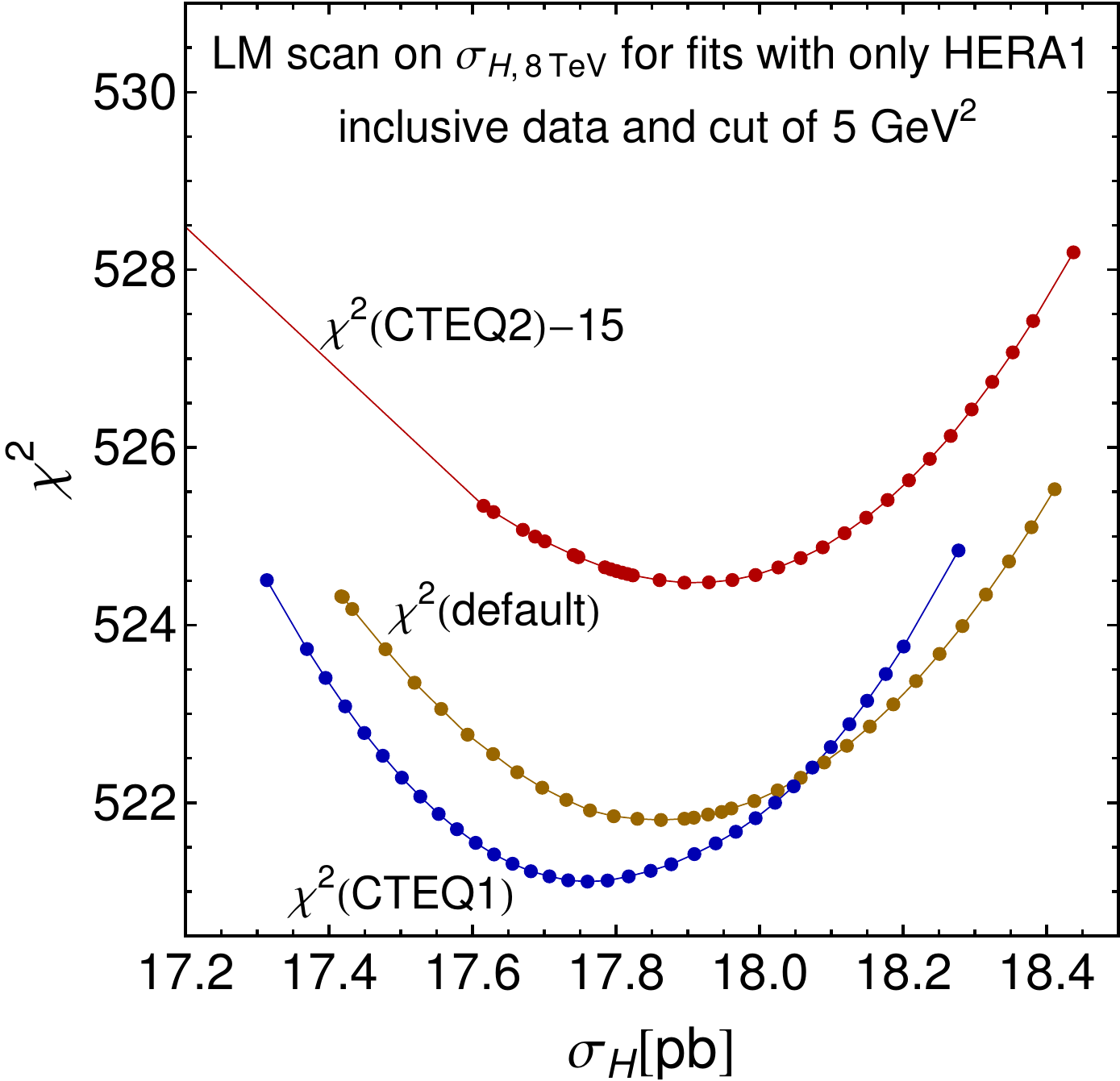}
\vspace{-1ex}
\caption{\label{fig:ctscan} LM scan of the Higgs cross section at the
LHC 8 TeV for CTEQ fits with different parametrization forms
and only HERA-1 inclusive data.}
\end{figure}

The baseline parametrization of HERA-only fits by CTEQ uses 
a positive-definite gluon and Chebyshev polynomials for some flavors. In
addition, we explored three alternative parametrizations: CTEQ-1, a
more flexible parametrization with Chebyshev polynomials; CTEQ-2,
equivalent to the CT10 NNLO parametrization with fewer free
parameters; and CTEQ-3, which is the baseline parametrization in which one
Chebyshev polynomial is replaced by a term proportional to $x^p\,
(1-x)^{64}$ to allow a negative gluon. The best-fit $\chi^2$ values are about
521 for the parametrizations with Chebyshev polynomials (baseline,
CTEQ-1, and CTEQ3) and 539 (somewhat worse) for CTEQ-2. 
The shapes of the gluon PDF for the alternative parametrizations,
compared to the alternative parametrization, are illustrated 
in Fig.~\ref{fig:pardep}b. We see some variation in the shape of
$g(x,Q)$, especially for $x \sim 0.1$, but the total Higgs production
cross sections remain rather stable. The $\chi^2$ vs. $\sigma_H$
dependence (for the LHC 8 TeV) in a Lagrange multiplier (LM) scan 
for the baseline, CTEQ-1, and CTEQ-2 parametrizations 
is illustrated in  Fig.~\ref{fig:ctscan}. 
All three scans show consistent results for the values of the best-fit cross
sections and overall dependence of $\chi^2$ on $\sigma_H$. For the CTEQ-3
form (allowing for a negative gluon), the convergence of the fit is
worse for some values of $\sigma_H$ in the LM scan. Nevertheless, the
best-fit value of $\sigma_H$ for CTEQ-3 is 17.9 pb, which is close to the
baseline result.

\begin{figure}[tb]
\centering
 \includegraphics[width=0.55\textwidth]{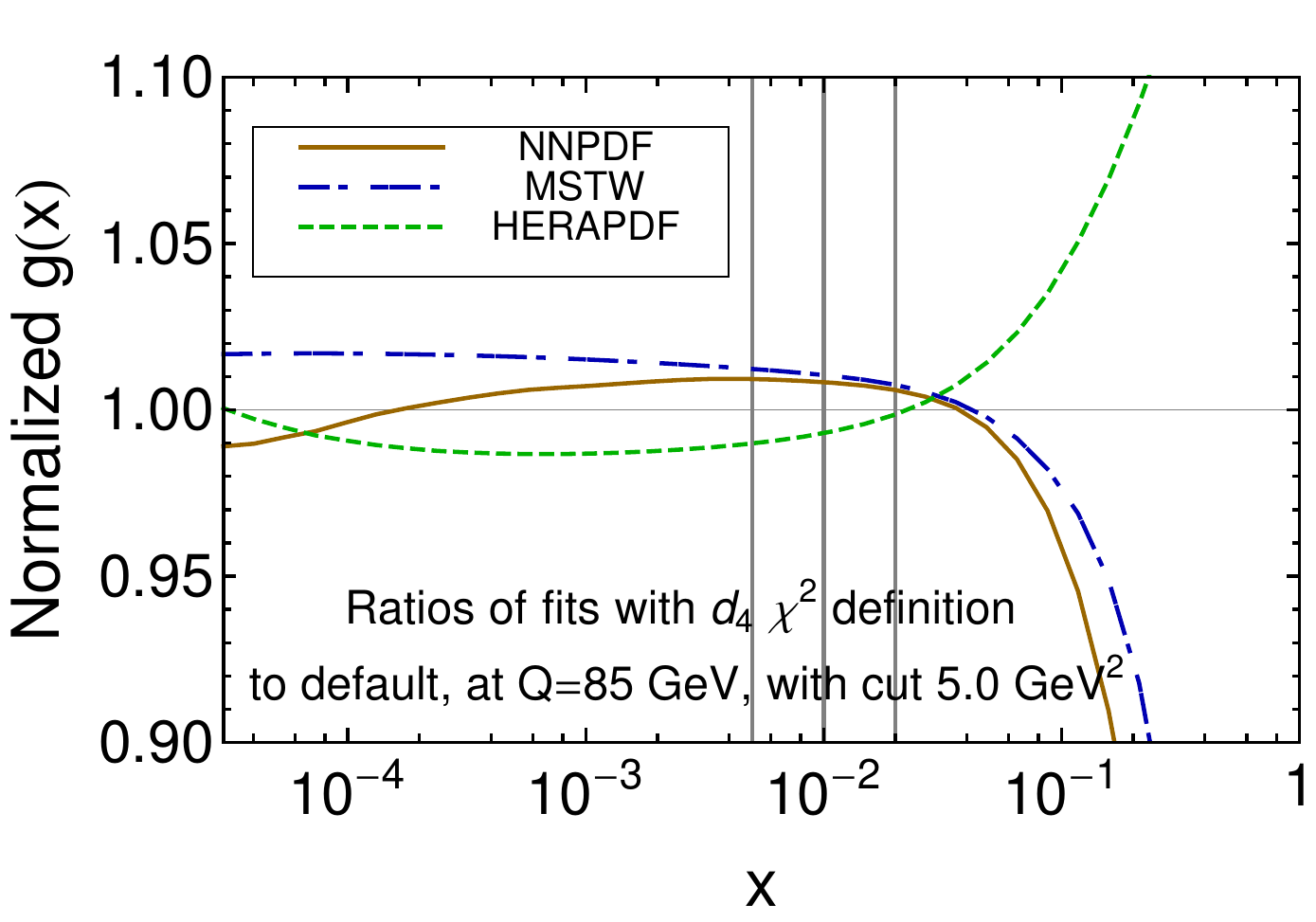}
\vspace{-1ex}

 \caption{\label{fig:chidep} Dependence of the gluon PDF on $\chi^2$ definition used,
 for MSTW, NNPDF, and HERAPDF fits with only HERA-1 inclusive data.}
\end{figure}

{\bf Definition of $\chi^2$.}
As mentioned in Sec.~\ref{sec:set}, several definitions of $\chi^2$ can be implemented
in the fits with only HERA-1 data. The dependence on the $\chi^2$
definition may be viewed
as an additional systematic uncertainty. In Fig.~\ref{fig:chidep} we plot ratios of the gluon PDFs
from the fit with the $d_4$ definition of the $\chi^2$ to the gluon PDFs from the fit with the default
$\chi^2$ definition of HERA, MSTW, and NNPDF groups. CTEQ fits always
use the $d_4$ definition and are not included in the figure. Other settings, besides the $\chi^2$ definition, are the same in this
comparison. We can see that the $\chi^2$ definition can have some impact on the shape of the
gluon PDF, and thus on the rapidity distribution of the Higgs boson. For example, when we change
from the $d_3$ definition to the $d_4$ definition the gluon PDF is enhanced in the small $x$ region ($< 0.01$)
and reduced for $x> 0.1$. An  opposite trend is observed
when changing from the $d_2$ definition
to $d_4$.  However, the inclusive cross
sections are still stable, with changes less than 1\%. This is shown in 
Table~\ref{tab:xsecchi2}.

\begin{table}[tb]
\centering

\begin{tabular}{|c|c|c|c|}
\hline
 & MSTW & NNPDF  & HERAPDF \tabularnewline
\hline \hline $\delta \sigma_H/\sigma_H$[\%], LHC 8 TeV& -0.4 & -1.3
& 1.2 \tabularnewline \hline $\delta \sigma_H/\sigma_H$[\%], LHC 14
TeV& 0.4 & -0.6 & 0.2 \tabularnewline \hline
\end{tabular}

\caption{Relative changes in the predicted Higgs cross sections when 
the default $\chi^2$ definition of each group is replaced by $d_4$
definition. \label{tab:xsecchi2}}
\end{table}

{\bf Treatment of heavy quarks} and other theoretical choices, such
as the values of factorization and renormalization scales,
may affect the gluon PDFs. Dependence on parameters 
of the heavy-quark schemes and mass of the charm quark 
has been studied before
\cite{Martin:2010db,Binoth:2010ra,Abramowicz:1900rp,Gao:2013wwa}, and
relatively small impact on the gluon PDF at a fixed $\alpha_s$ has
been observed. In the context of the current study, we varied some
settings of the heavy-quark schemes and observed the changes in
the gluon PDFs. In CTEQ fits, the pole charm mass and
the $\lambda$ parameter introduced by the slow rescaling convention in the
S-ACOT-$\chi$ scheme \cite{Nadolsky:2009ge,Guzzi:2011ew}
has been varied to gauge the associated theoretical uncertainty due to
the heavy-quark contributions. [The baseline CTEQ 
fit assumes $\lambda=0$ and $m_c=1.4 \mbox{ GeV}$.] The dependence on $m_c$ and $\lambda$ in the CTEQ HERA-1 only fit is compatible with the behavior 
seen in the global fits \cite{Gao:2013wwa}, 
with the Higgs cross sections changing within 2-3\% under variation of $m_c$ and $\lambda$ before appreciable increase in $\chi^2$ is reached. 
In the MSTW study the gluon and Higgs cross sections have been obtained from fits which use the  
FONLL-C scheme used by NNPDF at NNLO, 
and HERAPDF have performed fits using the full ACOT scheme. 
In all cases, variations were small, 
though not totally negligible.  Overall, this suggests that
higher-order effects  
beyond NNLO in HERA DIS cross sections compare to a fraction 
of the current experimental uncertainty. 

\subsection{Summary and outlook\label{sec:con}}

This work is motivated by the need to understand, and possibly reduce,
the discrepancies between different PDF sets related to the predictions for
the cross sections for inclusive Higgs boson production via gluon fusion 
at the LHC, in order to improve the prospects of Higgs boson characterization.
In this contribution
we have performed a first step in this direction, by comparing in detail
the outcomes of the PDF fits by CT, HERAPDF, MSTW and NNPDF
groups to the well-understood combined HERA-1 data set.
As an important spin-off, we have carried out a benchmark comparison of
NNLO neutral-current DIS structure functions, which have shown reasonable
agreement between the four fitting codes 
when using the same toy PDF set. We have acquired 
good understanding of the remaining differences arising
from the use of diverse  GM-VFN
schemes and counting of perturbative orders,
and observed a distinctly improving agreement when going from NLO to NNLO.

We have then completed a series of benchmark fits at NNLO using
 only HERA-1 data and studied the predictions of the resulting
PDFs for the gluon fusion Higgs cross sections.
In general the HERA-1 only fits prefer slightly smaller
Higgs cross sections than in the global fits, though they
are fully consistent within the larger uncertainties of
the former. 
Predictions from different groups for the Higgs cross sections
show a spread at a similar level as in the published NNLO PDF sets, 
and with mostly similar hierarchy. 
However, since in the HERA-1 only fits the PDF uncertainties
are substantially larger than in the global fits, we find
that the predictions from all groups agree well within 
the PDF uncertainty in this case.
Various systematic effects on the Higgs cross section predictions
have also been studied, leading to effects at the level of a percent or two,
 well within the PDF uncertainties.

Therefore, the main conclusion of this work is that, in the
HERA-1-only fits, the predictions of all four fitting groups
are consistent within the PDF uncertainties.
There are however some interesting trends, for instance the hierarchy between
NNPDF, MSTW and CT seems to be maintained both in the global and
HERA1-only fits. This could be related to different methodological
choices associated with the selection of the gluon parametrization 
or alike issues, some of which were investigated in Sec.~\ref{sec:var}.

The next natural step will be to continue this
exercise by adding additional experimental data sets into the
PDF fits sequentially.
This will require both benchmarking the theoretical
predictions used by each group for the different observables and ensuring
that all groups use exactly the same data points, uncertainties and
definition of the systematic uncertainties and covariance matrix.
The data that could be added to the HERA-1 data include
fixed-target neutral current DIS data, charged current neutrino DIS data, vector boson and inclusive
jet production data, and future HERA-2 combined data,
until a data set similar to that
used in the three global fits is reproduced.\footnote{
A first exercise in this direction by the NNPDF Collaboration is
presented as a separate contribution to these proceedings.}
Another possible way forward would be to compare the impact of
new LHC data that is sensitive to the gluon PDF
in the region relevant for Higgs production,
like direct photon production~\cite{d'Enterria:2012yj,Carminati:2012mm},
vector boson production in association with jets~\cite{Malik:2013kba},
top quark pair production~\cite{Czakon:2013tha}
and inclusive jet production~\cite{Watt:2013oha}.
\
We hope that, by building on the results and strategy presented in
this contribution, this future work will contribute to reduction 
of PDF uncertainties at the NNLO level of accuracy that is called for
by the LHC physics program.

\subsection*{Acknowledgments}
This work at MSU and SMU 
was supported by the U.S. DOE Early Career Research Award
DE-SC0003870, U.S. National Science Foundation under Grant No. PHY-0855561, 
and by Lightner-Sams Foundation. 
The work of RST is supported partly by the London Centre 
for Terauniverse Studies (LCTS), using funding from the European
Research Council via the Advanced Investigator Grant 267352 and is
also funded by the Science and Technology Research Council (STFC). 
The work of S.F. is supported by an Italian PRIN 2010 grant and by a European
EIBURS grant. This research was supported in part by the European
Commission through the ``HiggsTools'' Initial Training Network
PITN-GA-2012-316704. 



\section{Dataset sensitivity of the $gg\to H$ cross-section in the NNPDF analysis\footnote{S.~Forte and J.~Rojo}}

\subsection{Introduction}

Parton distributions provide a dominant source of uncertainty on all
Higgs production modes at the LHC: combined PDF+$\alpha_s$
uncertainties are around $\sim 8\%$ for gluon fusion, the 
production mode with largest cross section.
In a particularly unfortunate situation, 
for this process, and for the Higgs mass of $m_H=125$ GeV,  
differences between the three PDF
sets that enter the PDF4LHC recommendation~\cite{Botje:2011sn} adopted by the
Higgs Cross Section Working Group predictions~\cite{Dittmaier:2011ti},
namely
 CT10~\cite{Gao:2013xoa},  MSTW~\cite{Martin:2009iq} and
NNPDF2.3~\cite{Ball:2012cx}, are such that their gluon-gluon
luminosities differ more than for any other mass value~\cite{Ball:2012wy}. The
reason for this discrepancy could be methodological: this is
investigated in a companion contribution to these proceedings, where
methodological differences between CTEQ, HERAPDF, NNPDF and MSTW are
studied in the controlled setting of fitting to a common and
consistent dataset, the combined HERA-I data.

 However, the reason could also be that there are
tensions, i.e. minor inconsistencies, between data included in a
global fit. Results could then differ either because of small
differences in the dataset adopted, or because different fits
respond differently to data inconsistencies. If this is the case, the
discrepancies would eventually be resolved as more and more data
become available, specifically from the LHC, but in practice this
might take a very long time. It is therefore important to understand
whether this might be the case.

With this motivation, in this  contribution we explore, in the framework
of the NNPDF2.3 global analysis, how the Higgs cross section in the
gluon fusion channel is affected by the choice of fitted dataset.
First of all we present a correlation study to quantify which of the experiments
in the global fit affect more the gluon PDF in the region of $x$ relevant for
Higgs production.
Then we present the results of a wide variety of NNPDF2.3-like fits with
different datasets, and study how the Higgs production cross-section changes
as these different choices of  fitted dataset are adopted.

As mentioned above, the present study is based on the NNPDF2.3 framework~\cite{Ball:2012cx},
but now varying the fitted dataset.
In all the fits that are presented in this contribution, the value of
$\alpha_s(M_Z)=0.119$ is adopted.
This value is  consistent with the current PDG average~\cite{Beringer:1900zz}
and also  with a direct determination from the NNPDF2.1NNLO fit~\cite{Ball:2011uy,Ball:2011us,Lionetti:2011pw}.
Theoretical predictions for the inclusive $gg\to H$ cross-section at
NNLO were
computed with the {\tt iHixs} code, {\tt v1.3.3}~\cite{Anastasiou:2012hx}.
A similar, more detailed study was presented in the framework of the CT10 global
fits in~\cite{Dulat:2013kqa}, and an earlier related study in the
MSTW08 framework 
can be found in~\cite{Thorne:2011kq}.

\subsection{Gluon sensitivity to individual  datasets}

We first study the sensitivity of the gluon PDF to the
 different data included
in the global fit, with the motivation of quantifying which 
datasets  mostly
constrain $g(x,m_H)$ in the region relevant to the $gg\to H$ cross-section, namely 
$x\in \left[ 0.005, 0.05\right]$.
Note that the total Higgs production cross-section in the gluon fusion channel
is dominated by the threshold region, which means that even beyond
leading order the result is mostly controlled by 
the gluon luminosity at $\tau=\frac{m^2_H}{s}$,
i.e. the gluon PDF at  $x=\sqrt\frac{m^2_H}{s}\approx 0.01$.

This sensitivity of the gluon on each individual dataset
 can be quantified by  computing the correlation~\cite{Demartin:2010er,Alekhin:2011sk} 
between the gluon PDF $g(x,m_H)$ and the contribution to the
 $\chi^2$ from the given experiment. This is given by
\begin{equation}\label{corrdef}
\rho(x) = \frac{\left\langle g^{(k)}(x,m_H)\chi^{2(k)} \right\rangle_{\rm rep}-\left\langle g^{(k)}(x,m_H)
\right\rangle_{\rm rep} \left\langle \chi^{2(k)} \right\rangle_{\rm rep}}{\sigma_{g(x,m_H)}\sigma_{\chi^2}}
\end{equation}
where averages are computed over the sample of $k=1,\ldots,N_{\rm rep}$ replicas.
We normalize the correlation to the number of datapoints, i.e. such
that $\rho=N_{\rm dat}$ means complete correlation, 
thereby accounting 
for the fact that datasets with a larger number of points will carry
more weight in the global fit. In practice this is done by multiplying
by $N_{\rm dat}$ the correlation coefficient defined according to
Eq.~\ref{corrdef}.

\begin{figure}[h]
\begin{center}
\includegraphics[width=0.43\textwidth]{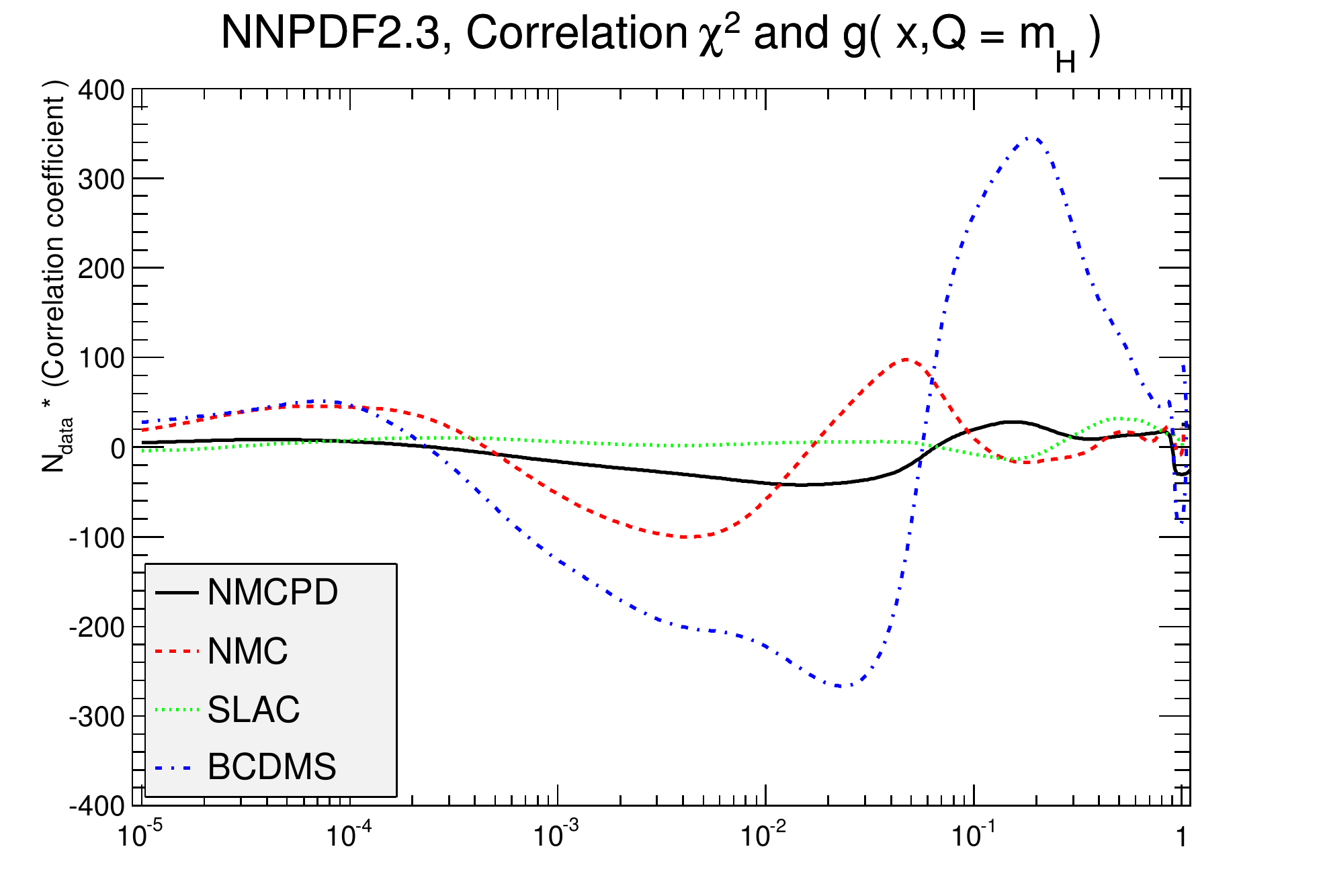}
\includegraphics[width=0.43\textwidth]{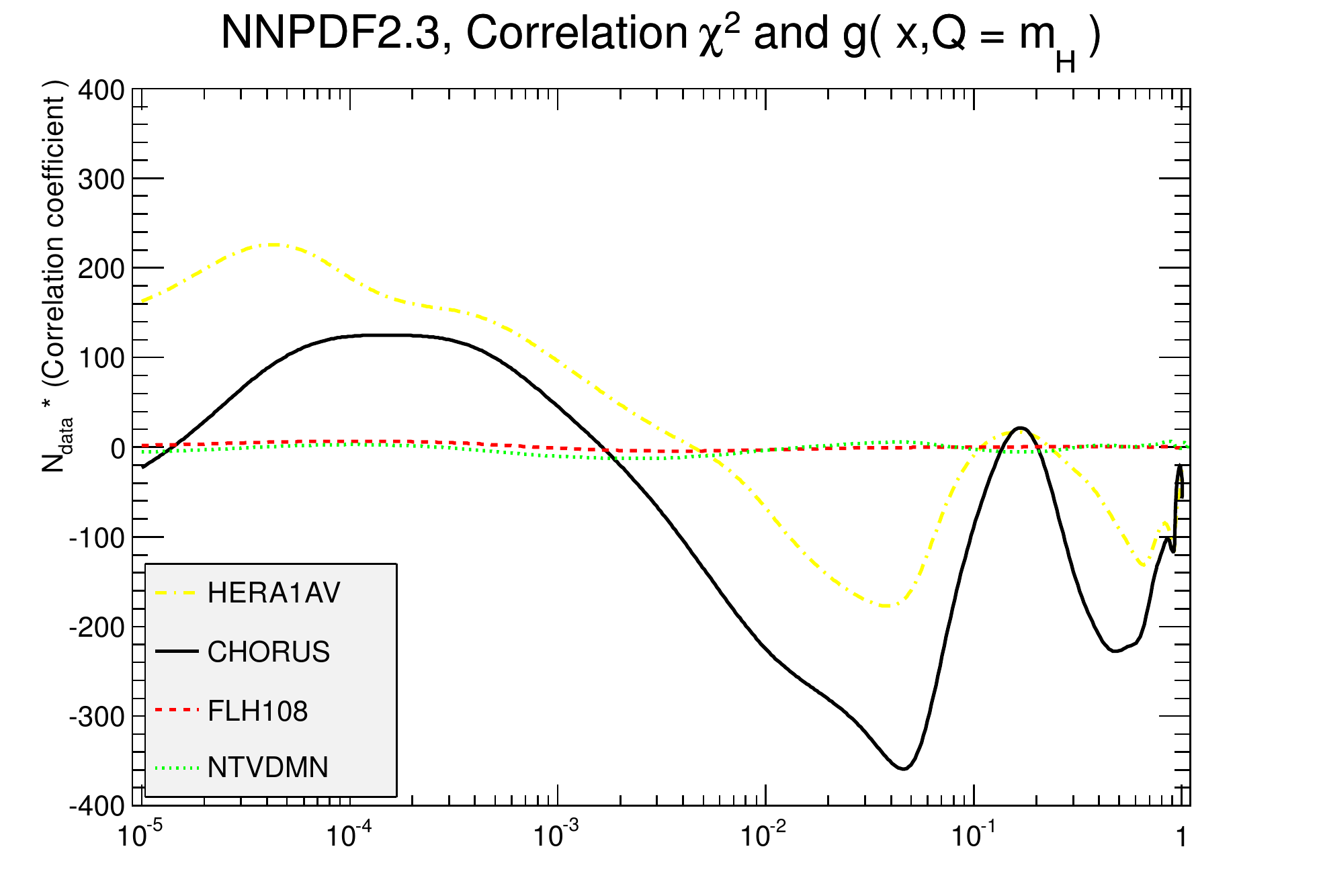}
\includegraphics[width=0.43\textwidth]{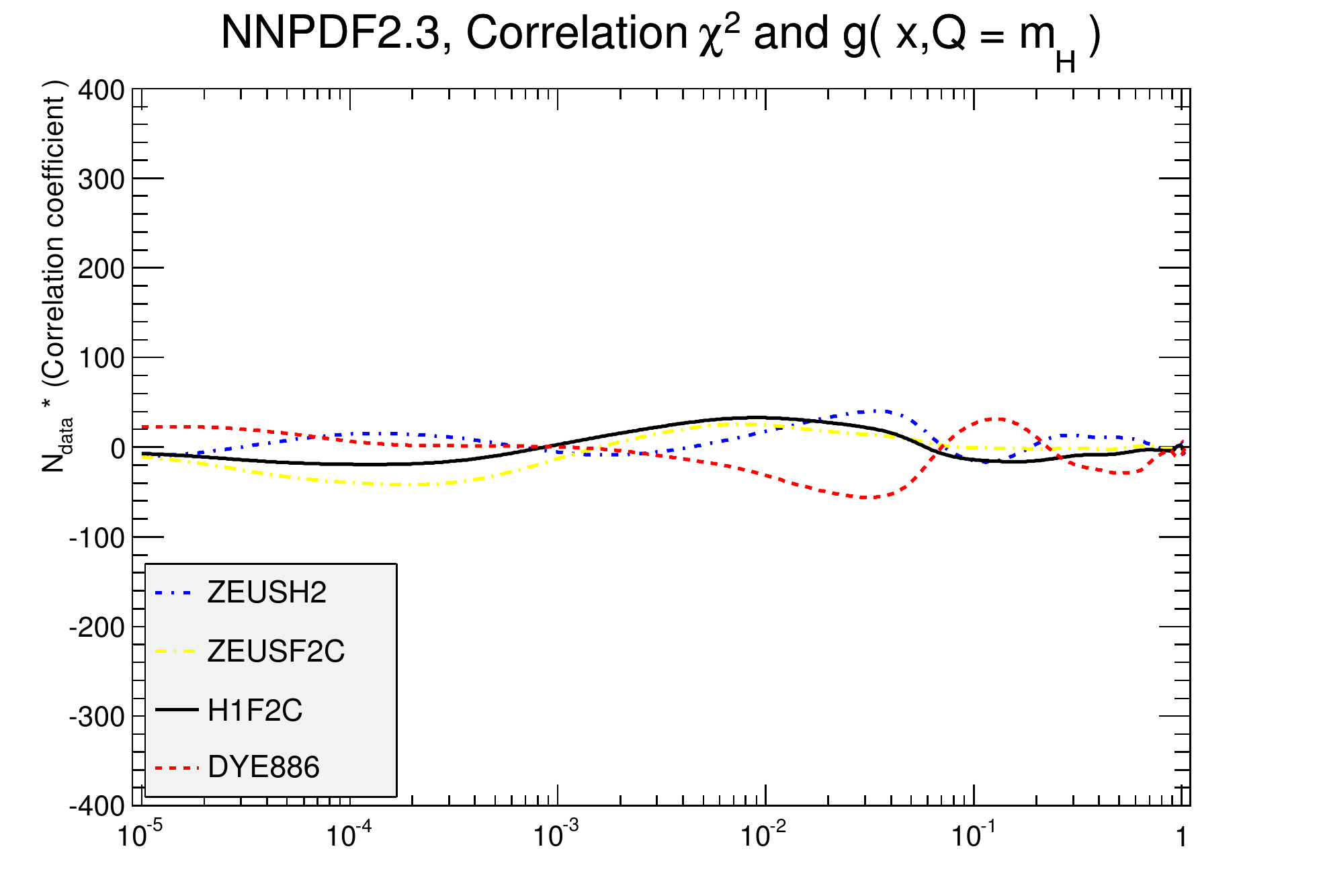}
\includegraphics[width=0.43\textwidth]{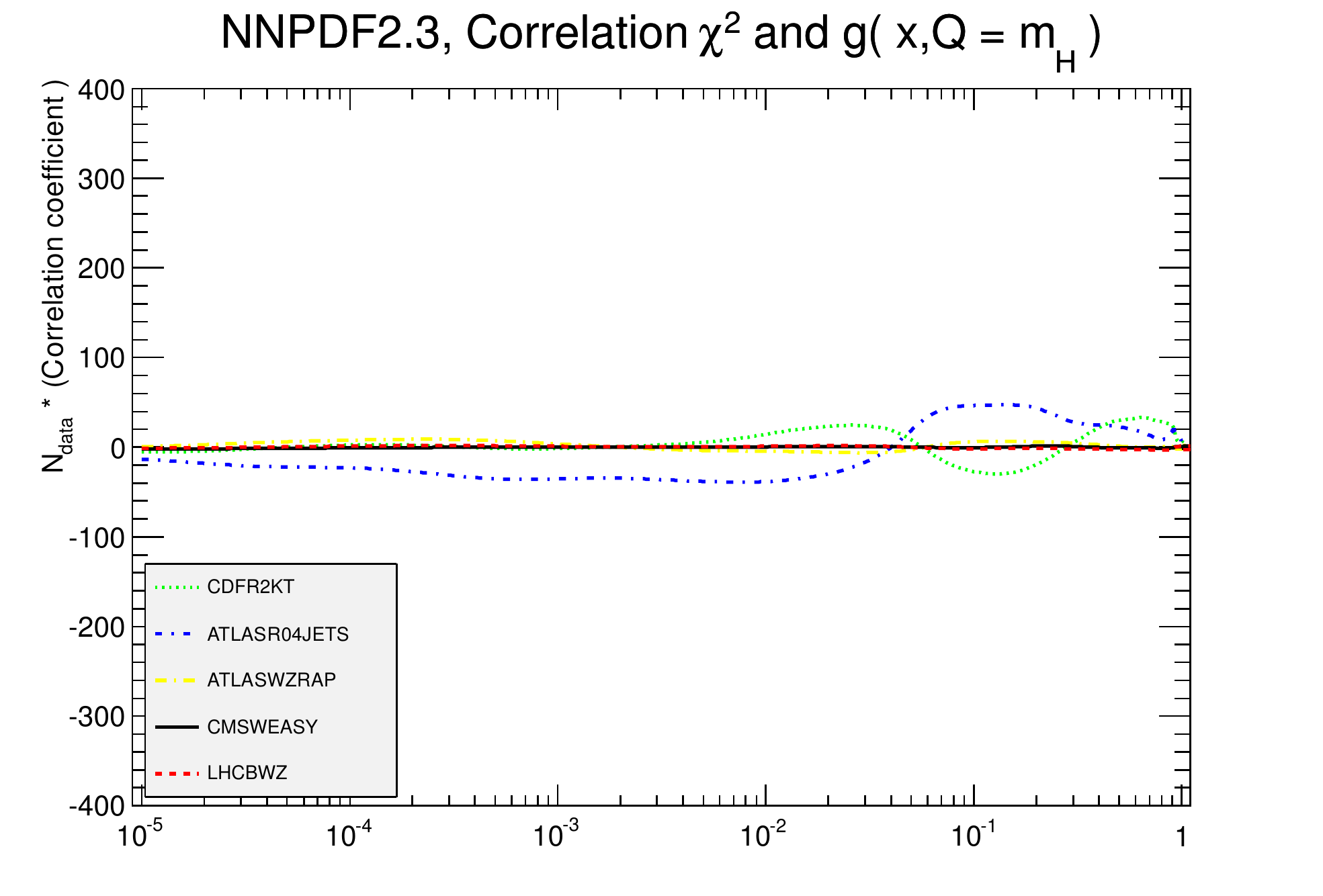}
 \caption{The correlation coefficient (normalized to $N_{\rm dat}$) 
between the gluon PDF at $Q=m_H$
and the $\chi^2$ for a representative subset of the
experiments included
in the NNPDF2.3 NNLO fit, as a function of
$x$. 
}
\label{fig:correlation}
\end{center}
\end{figure}

In Fig.~\ref{fig:correlation} we show the correlation
between the gluon PDF $g(x,Q=m_H)$
and the $\chi^2$ for different datasets in the NNPDF2.3 fit.
From these correlation profiles, we see that in the region
relevant for the Higgs cross-section in the gluon fusion channel,
the experiments which carry highest  weight in the global fit are some 
of the DIS fixed target data, in particular BCDMS and CHORUS,
as well as the inclusive HERA-I data.
Other experiments have a smaller weight in the global fit $\chi^2$
because of the smaller number of points, like ATLAS jets or HERA charm data,
but they could still have a non-negligible impact as 
they may affect PDFs in regions for which experimental information is
scarce. 

Having established which data have a potential impact on the results
of the fit, we now study in detail how the NNPDF2.3 predictions for
the Higgs cross section are affected when some of these datasets
are added to or removed from the global fit.
These results however are indirect, since they do not allow us to quantify how the Higgs
$gg\to H$ cross-section (with the associated PDF uncertainties) vary when only
a subset of the data is included, as compared to the reference value from
the NNPDF2.3 global fit.
To address this issue, in the next section we study the 
dataset dependence of $\sigma(gg\to H)$ in the NNPDF2.3 fit

\subsection{Dataset dependence of $\sigma(gg\to H)$ in the NNPDF2.3 fit}

We have performed a large number of variants of the NNPDF2.3NNLO fit
keeping all the settings unchanged and varying only the fitted dataset.
For each of these fits, we have then computed the $gg\to H$ cross-section
with {\tt iHixs}, and compared the results with the reference NNPDF2.3 values,
as well as with those of CT10 and MSTW08 NNLO fits.
In all cases we use $\alpha_s(M_Z)=0.119$

\begin{figure}[h]
\begin{center}
\includegraphics[width=0.75\textwidth]{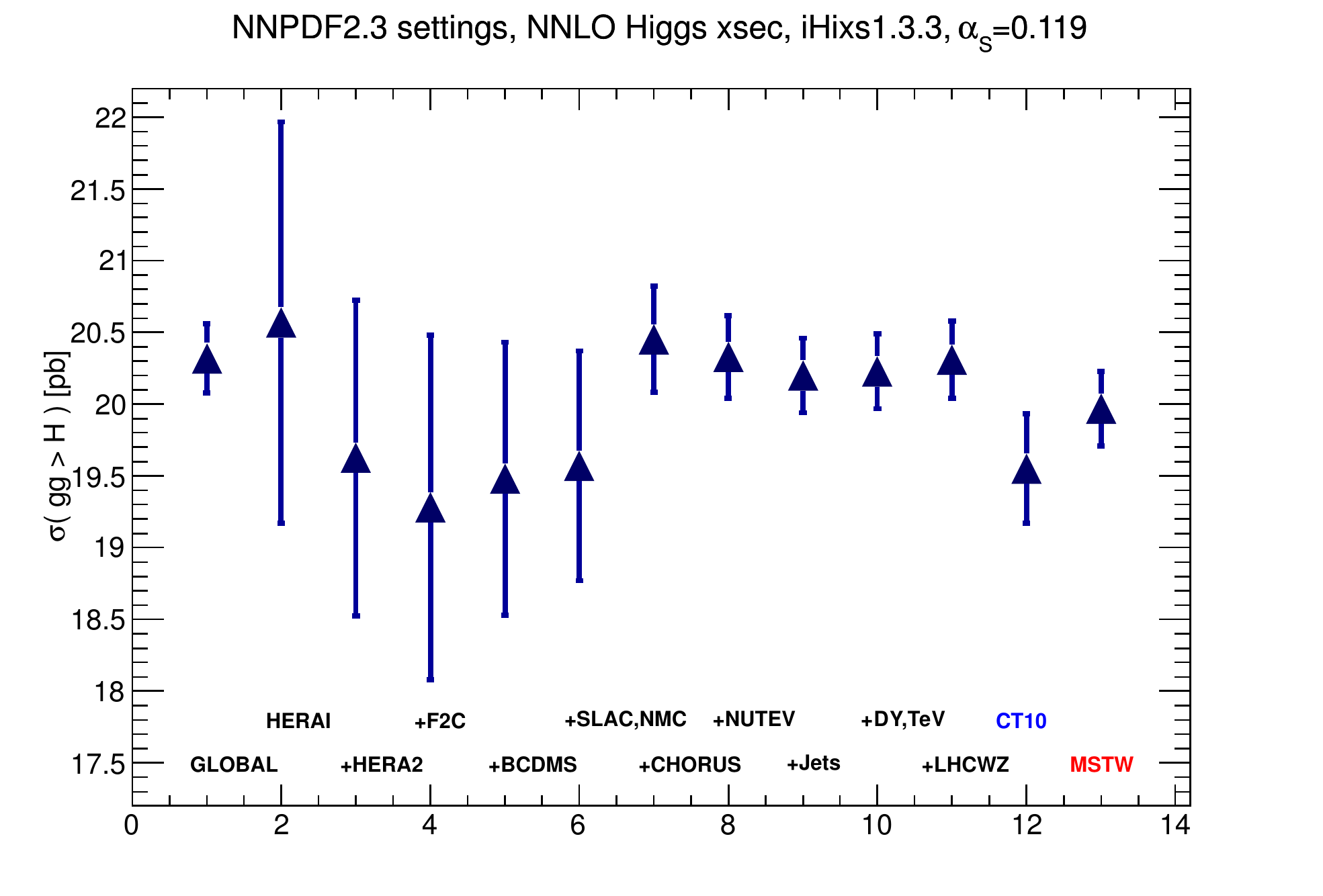}
 \caption{NNLO predictions for the Higgs production cross-section
in the gluon fusion channel for NNPDF2.3-like fits with
different datasets. Starting from the HERA-I dataset (second point
from the right), we sequentially add all datasets until the
global fit result (first point) is reproduced. For reference,
we also show the corresponding values from CT10 and MSTW, with
the same value of $\alpha_s(M_Z)=0.119$.
}
\label{fig:higgs1}
\end{center}
\end{figure}

In our first two exercises, we start with a fit to HERA data only, and
then we build up the dataset of the global fit by adding experiments
or groups of experiments one after the other.
In the first case, shown in Fig.~\ref{fig:higgs1}, we start with
experiments which should be subject to relatively smaller systematics,
namely deep-inelastic scattering (DIS), starting with the very clean
combined HERA-I data (which  have minimal systematics),
then adding combined HERA $F_2^c$ data, then HERA-2 ZEUS data, and
then fixed target DIS data, first with charged lepton beams and then
with neutrino beams. We then add first, jet data (Tevatron and LHC)
which are the hadron collider data which should mostly impact the
gluon (though at rather larger $x$), 
and finally other (Drell-Yan, $W$ and $Z$ production) 
hadron collider data, which  
we do not expect to have a big impact, first Tevatron
and then LHC. The reader is referred to Ref.~\cite{Ball:2012cx} and
references therein for a detailed list and discussion of the various
datasets (including number of datapoints, kinematic cuts, and references).
 For reference,
we also show the corresponding values from CT10 and MSTW, with
the same value of $\alpha_s(M_Z)=0.119$.

When only HERA data is used the PDF uncertainties are large,
around 7\%, because only indirect constraints on the gluon PDF for $x\sim 0.01$
are provided.
Still, it is remarkable that even with a single dataset, PDF uncertainties
are already below the 10\% range. Furthermore, no  manifest
inconsistencies appear: each time a dataset is added, the
cross-section always moves within the 1-sigma band of the previous result, 
and PDF uncertainties are reduced.
Despite their overall consistency, it is clear that different data pull
in different directions. In particular, 
HERA-II and HERA $F_2^c$ data seem to pull the cross section towards
smaller values; charged-lepton fixed-target DIS data then have little
impact; however, neutrino DIS data 
increase the cross-section
and reduce the PDF uncertainties substantially. Hadron collider data
finally  have a minor effect.

This exercise raises several questions: are the charm data also
pulling the cross-section down, or only the HERA-2 data? Are neutrino
data pulling 
the cross section up because of the lack of hadron collider data, or
would they even if the hadron collider data were not included? Are
hadron collider data having little effect only because they were added
in the very end?

\begin{figure}[h]
\begin{center}
\includegraphics[width=0.75\textwidth]{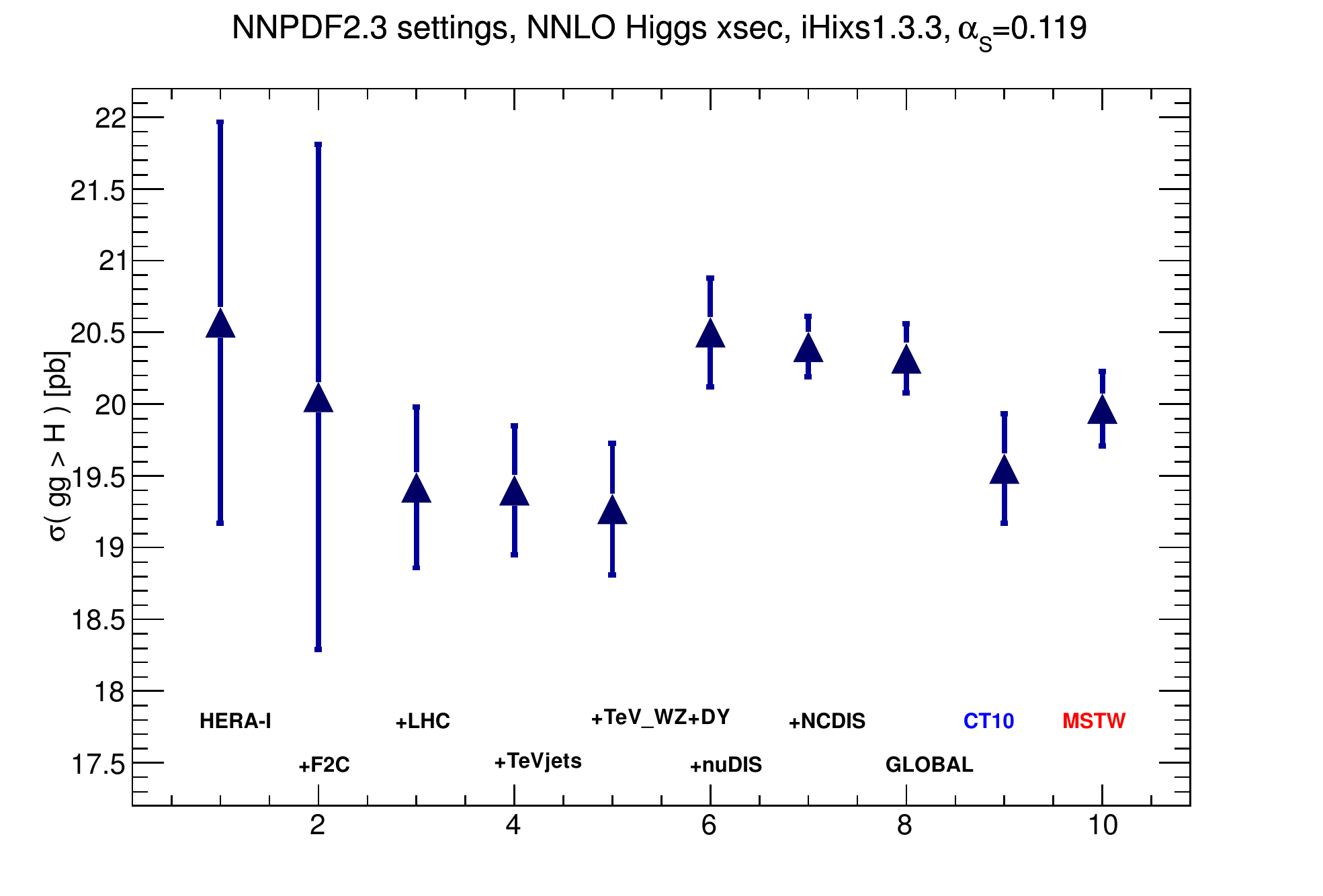}
 \caption{Same as  Fig.~\ref{fig:higgs1}, but
changing the order in which  datasets are added.
}
\label{fig:higgs2}
\end{center}
\end{figure}

In order to try to answer some of these questions, we repeat the
exercise but now changing the order in which datasets are added. To
answer the first question, we add the charm data directly to the
HERA-I data. To answer the third question, we then add the LHC data
immediately after the HERA-I and charm data. To answer the second
question, we only add neutrino data after all  hadron collider
data have been also added. Results are shown in Fig.~\ref{fig:higgs2}.

A tentative conclusion would be that first, the charm data do pull the
cross section down, without affecting much the uncertainty. Second,
LHC data reduce the PDF errors substantially and also lead to
smaller cross-sections (always within 1-sigma). And third,
neutrino DIS data increase
the cross-section by a factor around 2-sigma (this time without affecting much
PDF uncertainties). Again, all other data have a minor  effect. This
exercise shows that a ``collider only'' fit leads to a smaller Higgs cross
section, something that was already noticed in
Ref.~\cite{Ball:2012cx}. Such a collider-only fit is not subject to
uncertainties related to the treatment of lower-scale data, nuclear
targets, and neutrino beams, and thus in principle more reliable. It
is important to understand, however, that PDF uncertainties for the 
 collider-only fit are significantly larger, and results are
 consistent at a 90\% confidence level with those of the global
 fit. Therefore, we cannot conclude that there is any clear evidence
 of inconsistency in the global fit, which remains thus the most
 reliable result, even though there is some indication that  LHC data
 and neutrino DIS data do pull the fit in opposite directions.

\begin{figure}[h]
\begin{center}
\includegraphics[width=0.75\textwidth]{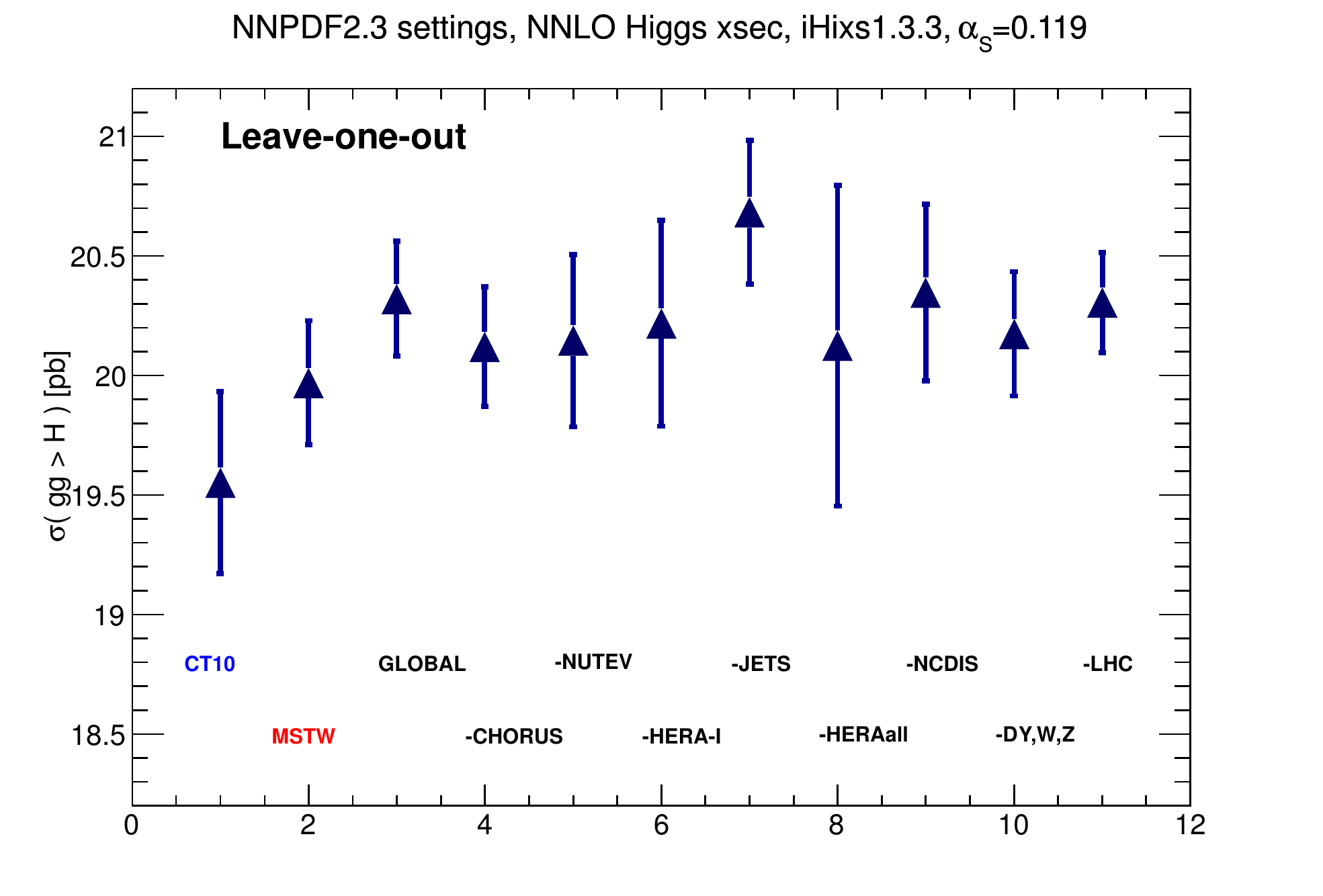}
 \caption{Same as  Fig.~\ref{fig:higgs1}, but
with one dataset at a time removed from the global fit.
}
\label{fig:higgs3}
\end{center}
\end{figure}

In order to try to elucidate this, we have
performed an exercise in which, with the goal of understanding the
impact of each individual data set,  we remove
from the global fit only a single dataset at a time.
Results are shown in Fig.~\ref{fig:higgs3}.
Again, the overall consistency of the global fit
is  quite remarkable, since no experiment leads to a pull larger that 
one--sigma, and typically rather less.
Results seem consistent with those of the previous exercise: removing
neutrino data the cross-section goes down, which confirms that
neutrino data tend to increase the cross-section. Removing jet data,
the cross-section goes up, which confirms that hadron collider
data  seem to pull the cross section in the opposite direction as
neutrino data, and suggests that it is jet data which have this
effect. The remaining datasets seem to have a more moderate impact either
in terms of absolute shift of the cross section (such as LHC data) or
in terms of their pull, i.e. the amount the cross-section moves in
units of the standard deviation (such as HERA data).

Whereas these studies start providing some indications, more detailed
investigations would be needed in order to reach a clear-cut
conclusion. These include studying multiple correlations: for example,
it was shown in Ref.~\cite{Ball:2011us} that the impact of a
particular set of DIS data on $\alpha_s$ depends on whether jet data
are or are not included in the fit: namely, it may turn out that
effects (including possible inconsistencies) are only revealed when a
pair of data (or more) are simultaneously considered (and included or
removed). Also, it might be worth 
studying more detailed statistical indicators, such as those which can
be defined when including new data through Bayesian
reweighting~\cite{Ball:2010gb,Ball:2011gg}.

\subsection*{Conclusion}
In summary, we have seen that there is no clear sign of inconsistency
for any of the  data included in the NNPDF2.3 global
fit. However, there does seem to be some indication of possible
tension between data, with neutrino DIS data pulling the Higgs cross
section up, and the prediction from a collider-only fit leading to a
rather lower cross-section than that of the global fit, with a rather
larger uncertainty and compatible with it at a 90\% confidence level. 
More detailed studies within the NNPDF framework, 
including investigations of multiple
correlations, could shed further light on these issues, but it would
be especially interesting if it were possible to compare the behavior
of different global fits upon the inclusion of individual datasets.

Of course, future measurements of LHC processes are going to provide
significant further constraints, and are likely to  eventually resolve all
discrepancies. 
These include
isolated photon production~\cite{d'Enterria:2012yj},
the total cross-sections and differential
distributions in 
top quark pair production~\cite{Czakon:2013tha},  the transverse
momentum distribution of electroweak vector bosons~\cite{Malik:2013kba},
inclusive jet and dijet producton data~\cite{Chatrchyan:2012bja} as
well as ratios of the above cross-sections between
13 and 8 TeV~\cite{Mangano:2012mh}.

\subsection*{Acknowledgments}
We would like to acknowledge useful discussions with R.~D.~Ball, J.~Huston,
P.~Nadolsky, J.~Pumplin, and R.~Thorne.
We thank the organizers of the Les Houches "Physics at TeV colliders 2013" workshop
were part of this work was performed.
The work of S.~F. is supported by an Italian PRIN 2010 and by a European
EIBURS grant

\clearpage

\section{New access to PDF data via LHAPDF6\footnote{A.~Buckley}}

I present an overview of problematic technical issues in the LHAPDF system, and
of a complete rewrite of LHAPDF which addresses them, plus prospects for future
development of systems for PDF data access in HEP.

\subsection{History and evolution of LHAPDF}

LHAPDF v5 and earlier themselves arose out of a Les Houches
meeting\,\cite{Whalley:2005nh}, as the need for a scalable system to replace
PDFLIB became pressing. The problem with PDFLIB was that the data for
interpolating each PDF was stored in the library, and as PDF fitting became
something of an industry (particularly with the CTEQ and MRST collaborations
producing many sets), this model was no longer viable due to the explosion in
the size of the compiled library.

LHAPDF was originally intended to address this problem by instead storing only
the parameters of each parton density fit at a fixed low scale and then using
standard DGLAP evolution in $Q$ via QCDNUM\,\cite{Botje:2010ay} to evolve this, build
a dynamic interpolation grid, and thereafter work as before.

However, by the mid-2000s and version 4 of LHAPDF, this model had also broken
down. Each PDF parameterisation required custom code to be included in the
LHAPDF library, and the bundled QCDNUM within LHAPDF had itself become
significantly outdated: upgrading it was not an option due to the need for
consistent behaviour between LHAPDF versions. Additionally PDF fitting groups,
concerned that the built-in QCDNUM evolution would not precisely match that used
by their own fitting code, had universally chosen to supply full interpolation
grid files rather than evolution starting conditions, and LHAPDF had acquired a
large collection of set-specific interpolation routines to read and use these
files.  At the time of writing, of the many actively used PDFs in use for LHC
simulation and phenomenology only the CTEQ6L1\,\cite{Pumplin:2002vw} PDF uses QCDNUM
evolution; all others are interpolation-based.

At the same time as these trends back to interpolation-based PDF provision, user
demand resulted in new features for simultaneous use of several PDFs -- a
so-called ``multiset'' mode. The implementation of this simply involves
multiplying the amount of allocated interpolation space by a factor of
\kbd{NMXSET}, which defaults to a value of 3. Hence in a default LHAPDF\,5
build, it is possible to switch rapidly between up to three PDF sets (although
only one of these ``slots'' can be active at any time).

\subsection{Memory/speed and correctness problems with LHAPDF5}

Aside from the maintenance issue of requiring new interpolation code in LHAPDF
in order to use a new PDF set (an issue ameliorated some time ago by addition of
special PDF names for various orders of generic spline interpolation), the major
problems with LHAPDF v5 relate to the technical implementation of the various
interpolation routines and the multiset mode.

Both these issues are rooted in Fortran's static memory allocation. As usual,
the interpolation routines for various PDFs operate on large arrays of floating
point data declared as common blocks. However, in practice they are not used
commonly, but rather each PDF ``wrapper'' code operates on its own array. As the
collection of supported PDF sets has become larger, the memory requirements of
LHAPDF have continually grown, and with version 5.9.1 (the final version in the
v5 series) more than 2\,GB is declared as necessary to use it at all. In
practice this is not a ``real'' issue -- operating systems are not so foolish as
to allocate memory to store uninitialized data for PDFs which are not being used
-- but on memory managed systems such as the LHC Grid where 2\,GB is the maximum
level permitted per core/process, it is enough to block any LHAPDF\,5 jobs from
running.

A workaround solution was provided some time ago for this problem: a so-called
``low memory mode'' in which LHAPDF\,5 could be built, reducing the static
memory footprint within acceptable limits. This has the effect of only providing
interpolation array space for one member in each PDF set, which is usually
sufficient for event generation but unworkable for uncertainty studies in which
each event must be re-evaluated or reweighted to each member in the PDF error
set: unless the user structures their code runs in a rather unnatural way,
constant re-initialisation of the single PDF slots from the data file slows
operations to a crawl. For this reason, PDF reweighting studies for the LHC have
had to be done using special, often private, user builds of LHAPDF rather than
the standard versions distributed by the LHC experiments. There is danger here
for error and inconsistency.

Notably, low-memory mode is incompatible with the recent trend to compute (in
particular NLO) PDF reweightings within an MC generator run rather than post
hoc, since every event involves upwards of 40 slow PDF re-initializations which
dominate the generation time: large scale experiment MC production with this
feature enabled in aMC@NLO\,\cite{Hirschi:2011pa},
Sherpa\,\cite{Gleisberg:2008ta}, or POWHEG\,\cite{Alioli:2010xd} requires an
improved LHAPDF.

Further options exist for selective disabling of LHAPDF support for particular
PDF families, as an alternative way to reduce the memory footprint. However,
since this highly restricts the parton density fits which can be used, it has
not found much favour.

The last set of problems with LHAPDF\,5 relate, concerningly, to the
\emph{correctness} of the output. For example different generations of PDF fit
families share the same interpolation code, although they may have different
ranges of validity in $x$--$Q$ phase space, and wrong ranges are sometimes
reported. The reporting of $\Lambda_\text{QCD}$ metadata has notoriously been
dysfunctional for a long time and cannot be fixed, with the result that shower
MC generators which attempt to acquire an appropriate \alphas definition from
the PDF will fail -- indeed PYTHIA\,6's many tunes depend on this behaviour! And
since the multiset mode is only implemented as a multiplying factor on the size
and indexing offsets, reported values of metadata such as \alphas and $x$,$Q$
boundaries does not match the currently active PDF slot, but rather the last set
to be initialized.

All of these problems stem from the combination of the static nature of Fortran
memory handling and from the way that evolving user demands on LHAPDF forced
retro-fitting of features such as grid interpolation and multiset mode on to a
system not designed to incorporate them. These combine with other ``emergent''
features such as the lack of any versioned connection between the PDF data files
and the library, the menagerie of interpolation grid data formats (each group
has its own, and sometimes several), and the need for code additions to use new
sets, to make LHAPDF\,5 difficult both to use and to maintain.

\subsection{LHAPDF\,6: a new library design and programmatic interface}

LHAPDF v6 is a completely new PDF library system, written specifically to
address all of the above problems, both technical and maintenance/manpower
related. As so many of the problems fundamentally stem from Fortran's memory
handling (at least prior to Fortran\;90) and the bulk of new experimental and
event generator code is written in C++, we have also chosen to write the new
LHAPDF\,6 in object oriented C++.

The central code/design object in LHAPDF\,6 is the \kbd{PDF}, an interface class
representing parton density functions for a variety of parton
flavours.\footnote{The nomenclature around ``PDFs'', ``PDF sets'' etc. is
  historically rather sloppy: we use the LHAPDF tradition of referring to a
  single multi-flavour fit as a ``PDF'' or ''member'', and several such PDFs as
  a ``set''; single-flavour parton densities do not exist as named entities in
  the code.}  An extra object, \kbd{PDFSet} is provided purely for (significant)
convenience in accessing PDF set metadata and all the members in the set,
e.g. for making systematic variations within a set.

As LHAPDF\,5 included special case treatments for PDFs including a photon or
gluino constituent, which were rather ``hacky'', LHAPDF\,6 allows completely
general flavours, identified using the standard PDG Monte Carlo~ID
code\,\cite{Beringer:1900zz} scheme. An alias of 0 for 21 = gluon is also supported, for
backward compatibility and the convenience of being able to access all QCD
partons with a for-loop from -6 to 6. A hypothetical \kbd{PDF} might hence
declare,for example, that it only contains explicit up, down and gluino
flavours, and all other parton flavours would return $xf(x,Q^2) = 0$ everywhere
(this PDF would not be expected to describe data particularly well, but there is
no \emph{technical} obstacle to its construction!) $xf(x,Q^2)$ values may be
retrieved from the library either for a single flavour at a time, for all
flavours simultaneously as a \kbd{int} $\to$ \kbd{double} \kbd{std::map}, or for
the standard QCD partons as a \kbd{std::vector} of \kbd{double}s. The latter
method can also be used to fill a pre-existing \kbd{vector} for improved speed.

A key feature in the LHAPDF\,6 design is a powerful ``cascading metadata''
system, whereby any information (integer, floating point, string, or lists of
them) can be attached to a PDF, a PDF set, or the global configuration of the
LHAPDF system via a string-valued lookup key. If, for example, $\alphas(M_Z)$ is
not defined on a \kbd{PDF}, the system will automatically fall back to looking
in its containing set and then the LHAPDF configuration for a value before
throwing an error. This information is set by default in the
PDF/set/configuration data files using the standard YAML\,\cite{yamlweb}
syntax. The access to this information is via the general \kbd{Info} class, from
which \kbd{PDFSet} and \kbd{PDFInfo} are specialised. The info system is used to
compactly implement a multitude of metadata properties, from quark and $Z$
masses to the PDG ID of the parent particle (to allow for identifiable nuclear
PDFs), and the error treatment, confidence level, etc. to allow automated error
computation. All metadata set from file may also be explicitly overridden in the
user code. PDF sets can contain metadata keys expressing the version of LHAPDF
required to correctly process it, and an integer data version key to allow for
tracking of bugfixes to the data file: these provide a robust way to track PDF
data versions, and also to declare that a PDF is unvalidated (with no
\kbd{DataVersion} key, or a declared negative version).

Following the strong trend towards interpolated PDFs, whereby PDF groups can
obtain arbitrarily faithful representations of their fits by increasing the
sampling density, and away from internally QCD-evolved PDFs, the only internally
provided PDF type in LHAPDF\,6 is an interpolation on a rectangular grid in
$x$--$Q^2$ space. The \kbd{PDF} class, however, is abstract in the sense that
any means of PDF evolution may be implemented according to the interface that it
defines. The built-in grid interpolation is provided in the internal
\kbd{GridPDF} class and associated helper structures for handling the grid and
flavour decomposition. In fact, each PDF may contain many distinct grids in
$Q^2$, in order to allow for parton density discontinuities (or discontinuous
gradients) across quark mass thresholds: however these subgrids, and the $x$,
$Q^2$ sampling points (interpolation ``knots'') within them must be the same for
all flavours in the PDF. The interpolation is currently performed using cubic
splines in $\log Q$--$\log x$ space, but as the interpolator algorithm is
configurable via a metadata key there is the possibility of evolving better
interpolators in a controlled way without changing previous PDF behaviours.
Internally, PDF querying is natively done via $Q^2$ rather than $Q$, since event
generator shower evolution naturally occurs in a squared energy (or $p_\perp$)
variable and it is advisable to minimise expensive calls of \kbd{sqrt}. For this
log-based interpolation measure, however, the logarithms of (squared) knot
positions are also pre-computed in the interpolator construction to avoid
excessive \kbd{log} calls. Custom extrapolators are also possible, but at
present only ``freeze'' and ``throw error'' extrapolation handlers are provided
for handling PDFs outside their fitted range.

The interpolation PDF data files use a single uniform plain text format. This
starts with a YAML-format header section (to provide member-specific metadata
overrides), then blocks of numbers for each $x$--$Q^2$ subgrid, each started by
a header declaring the knot positions and encoded parton flavours. As opposed to
LHAPDF\,5, where each PDF set was encoded in a single text data file, the
LHAPDF\,6 format is that each set is a \emph{directory}, which contains one
``.info'' file of set-wide metadata, plus one ``.dat'' file for each PDF member
in the set. This permits much faster lookup of set-level metadata, and random
access to single members in the set. If desired, the member content of a set
could be stripped down to save space in special applications. It is possible,
using the \kbd{zlibc} library\,\cite{zlibcweb}, to zip the data files in the
directory for space-saving and faster data reading. The data parser has
internally been optimised for fast reading of ASCII numerical data.  As there is
no set-specific or family-specific handling code in the library, new PDFs may be
made and privately tested (or even released separately from LHAPDF) without
needing a new version of the LHAPDF library. The only update required for a new
official set is an updated version of the \kbd{pdfsets.index} file, which
provides a lookup to PDF names and member numbers from the global LHAPDF ID
code. The latest version of this file, along with the PDF data itself compressed
as one tarball per set, is available for download from the LHAPDF website\,\cite{lhapdfweb}.

Since \alphas evolution is key to correct PDF evolution and usage, LHAPDF\,6
contains implementations of \alphas running via three methods: an analytic
approximation, a cubic spline interpolation in $\log Q$, and numerical solution
of the ODE -- the latter used to dynamically populate an interpolation grid
which is used thereafter for performance reasons. These evolvers are specified
(cf. grid interpolators and extrapolators) via metadata keys on the PDF member
or set, which can be programmatically overridden. Flavour thresholds/masses,
orders of QCD running, and fixed points/$\Lambda_\text{QCD}$ are all correctly
handled in the analytic and ODE solvers, and subgrids are available in the
interpolation.

The most distinctive change in LHAPDF\,6 is how the user manages the memory
associated with \kbd{PDF} objects, namely they are now fully responsible for
it. A user may create as many or as few \kbd{PDF}s at runtime as they wish --
there is neither a necessity to create a whole set at a time, nor any need to
re-initialise objects, nor a limitation to \kbd{NMXSETS} concurrent PDF
sets. The flip-side to this flexibility is that the user is also responsible for
cleaning up the memory use afterwards, either with manual calls to \kbd{delete}
or by use of e.g. smart pointers. PDFs or info objects (or interpolators,
extrapolators, \alphas objects\dots) are created by the \kbd{new} operator in
helper functions, via string-valued identifiers or the global LHAPDF~ID code,
which is still in use and will continue to be allocated for submitted
PDFs. Unlike in LHAPDF\,5, the different PDF objects exist independently of each
other and can be used concurrently, e.g. in multi-threaded programming, although
as always with concurrent programming care must be taken to avoid accidentally
sharing memory between the threads.

Finally, since uptake of LHAPDF\,6 is realistically contingent on the mass of
pre-existing code continuing to work (while providing a more friendly and
powerful alternative interface) legacy interfaces have been provided to the
Fortran LHAPDF and PDFLIB interfaces, and to the LHAPDF\,5 C++ interface. C++
preprocessor hooks have been provided to allow smooth migration to v6 based on
the version of LHAPDF found on the system at compile time. These legacy
interfaces have been successfully tested with a variety of MC generator codes.

\subsection{Migration and validation programme}

A major task, at least as substantial as writing the new library, has been the
migration of PDFs from the multitude of version 5 formats to the new format and
interpolator, and then validating their faithfulness to the originals. This has
been done in several steps, starting with a Python script which used the
LHAPDF\,5 interface (with some extensions) to extract the grid knots and dump
the PDF data at the original knot points into the new format. This script has
undergone extensive iteration, as support was added for subgrids,
member-specific metadata, etc. etc., and to allow more automation of the
conversion process for hundreds of PDFs. The choice was made to only convert the
most recent PDF sets in each family unless there were specific requests for
earlier ones: this collection is over 200 PDF sets, and at a conversion rate of
approximately one second per member it takes several hours to re-dump the full
collection.

After dumping, the set directories are zipped into tarballs and uploaded to the
LHAPDF website. PDFs which have been approved by their original authors are
available in the ``main'' download area, while those yet to be blessed are
contained in a subsidiary ``unvalidated'' staging area. Validated PDFs are also
made available on CERN AFS, at
\path{/afs/cern.ch/sw/lcg/external/lhapdfsets/current}.

\begin{figure}[t]
  \centering
  \includegraphics[width=0.48\textwidth]{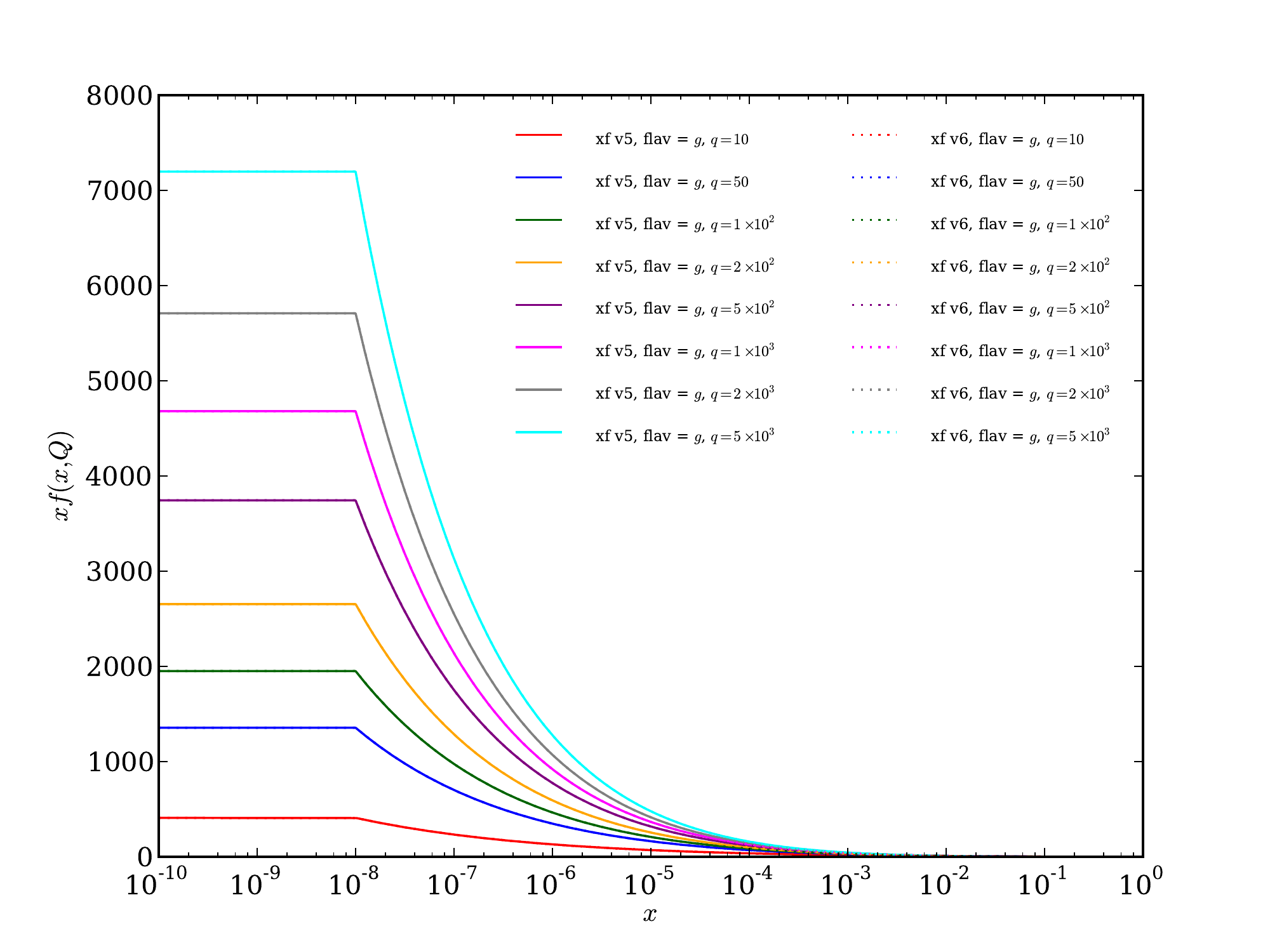}
  \includegraphics[width=0.48\textwidth]{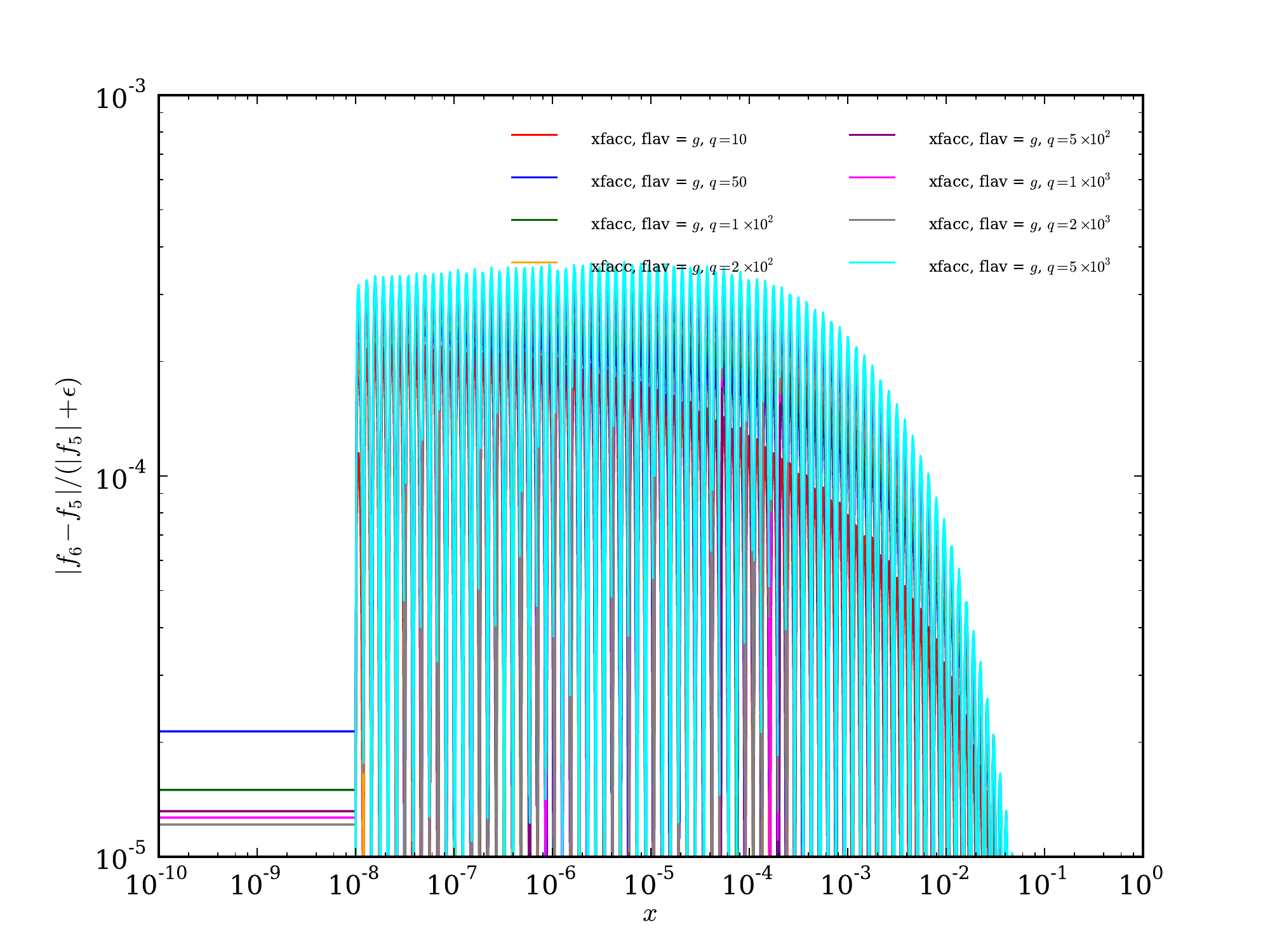}
  \caption{Example comparison plots for the validation of the CT10nlo central
    gluon PDF, showing the raw PDF behaviour as a function of $x$, and the
    corresponding v5 vs. v6 regularised accuracy metric. The differences between
    v5 and v6 cannot be seen in the left-hand plot, since as seen on the right
    the fractional differences are everywhere below one part in 1000.}
  \label{fig:ct10cmp}
\end{figure}
To become validated, a comparison system has been developed which uses the same
C++ code to dump PDF $xf$ values in scans across $x$ and $Q$ (as well as \alphas
values in $Q$) at a wide range of values in each variable, from both LHAPDF\,5
and 6. The corresponding data files from each version are then compared to each
other using a difference metric which corresponds to the fractional deviation of
the v6 value from the original v5 one in regions where the $xf$ value is large,
but which suppresses differences as the PDFs go to zero, to minimise false
alarms. An ad hoc difference tolerance of $10^{-3}$ was chosen on consultation
with PDF authors as a level to which no-one would object, despite differences in
opinion on e.g. preferred interpolation schemes. This level, as illustrated in
Fig.~\ref{fig:ct10cmp} for the CT10nlo central PDF member validation, has been
achieved almost everywhere for the majority of PDFs. Where differences do occur
(e.g. on MSTW set flavour thresholds before they were explicitly added to the
migration script) they are typically restricted to single points on the extremes
of the PDF phase space, or as transient spikes on flavour thresholds: these
cases can then be individually examined to determine if there is a genuine
problem, which so far has not been the case. This validation and plotting
process is much slower than PDF conversion, and generates huge volumes of plots
(several $x$ and $Q$ scans of each flavour in each member of each set) and hence
we have so far used only the central members of each set for validation. At the
time of writing, all NNPDF and MRST/MSTW sets have been approved by their
authors, as have several heavily-used CTEQ sets. Approval of the remainder of
CTEQ sets, and the HERA, ATLAS, and ABM sets, is pending.

\subsection{Status, limitations and prospects}

After a lengthy testing period and useful user feedback, the first official
LHAPDF\,6 version was released in August 2013. Since then several further
developments and bugfixes (as well as increased PDF set support) has occurred
and at the time of writing version 6.0.5 may be downloaded, along with data
files, from \url{http://lhapdf.hepforge.org}. An automatic data
downloader/manager script is provided with the library to help with PDF data
management. Author approval of the remaining unvalidated sets is hoped to happen
soon.

At present the scope of LHAPDF\,6 is intentionally more LHC-focused than
LHAPDF\,5: we are mindful of the small amount of development manpower and the
interests of those involved. Accordingly, no QCD evolution is planned for the
library although we will consider proposals and code patches, and the class
interface is available for anyone to plug their own implementations into, should
they so wish. Similarly, there is at present no plan to include automatic
nuclear corrections or PDFs of the resolved virtual photon: we believe that the
former is better done explicitly using one of several nuclear corrections tools,
and the latter is not presently of interest at the LHC (although EW constituents
of nucleons are fully supported) and can be implemented for fixed values of
virtuality and impact parameter using the current interface. Double-parton PDFs
and generalised PDFs for CCFM QCD evolution are also considered as special cases
and are not directly supported by the LHAPDF\,6 interface (after consultation
with PDF authors.)

The library is currently considered in a production-ready state, and several
generators and LHC experiments have updated their code interfaces to make use of
it. The performance has been found to be (of course) vastly better than
LHAPDF\,5 in terms of memory usage, by explicit design. It has also been
pleasing to find that design decisions such as the split member data files lead
to a measurably faster start-up time, and that the possibility of evolving
single flavours at a time (as opposed to LHAPDF\,5, which always evolved all
flavours) can produce significant speed-ups (of order a factor of 3) in MC event
generation and event reweighting. We have scheduled several developments to get
even better performance, such as caching interpolation weights between calls to
a PDF interpolator, and optimising the interpolation code to make use of
vectorized instructions on modern CPUs. Use of GPGPUs has been suggested, but we
advise that anyone wanting that functionality contact the LHAPDF developers with
a plan for their implementing at least a proof of concept version. Zipped access
to PDF data files is already possible by slightly tricky use of \kbd{zlibc}, but
this could also become a built-in library feature if it can be implemented
without incurring unacceptable dependency and build difficulties.

Some physics improvements have also been suggested, such as use of a $\log
(1-x)$ measure for high-$x$ interpolation, a low-$Q$ extrapolation scheme based
on computed anomalous scaling dimensions, and (optional) distinction between
quark masses and flavour thresholds in \alphas evolution.

We are pleased that a year after the project started, LHAPDF\,6 is a mature,
performant and powerful library for future PDF access at Run\,2 of the LHC and
beyond, with an active and increasing user base. Our thanks to all who helped us
to get here!

\subsection*{Acknowledgements}

AB thanks the Les Houches workshop once again for the hospitality and
enjoyable/productive working environment in the Chamonix valley, and
acknowledges support from a Royal Society University Research Fellowship, a CERN
Scientific Associateship, an IPPP Associateship, and the University of Glasgow's
Leadership Fellow programme.

\clearpage

\clearpage

\chapter{Phenomenological studies}
\label{cha:pheno}

\section{Inclusive jet cross section at the LHC
  and the impact of weak radiative corrections\footnote{%
    S.~Dittmaier, A.~Huss and K.~Rabbertz}}

\subsection{Introduction}
\label{sec:incJet:intro}

Quantum chromodynamics (QCD) describes the strong interaction between
quarks and gluons, which are confined within bound states and cannot
be observed as free particles. Instead, in high-energetic hadron
collisions, collimated streams of particles, known as ``jets'', are produced
that convey essential information about the hard scattering process
between the hadrons' constituents. The investigation of these jets,
technically defined through jet algorithms, allows to compare
predictions from perturbation theory with measurements and to look for
new phenomena at the highest accessible transverse momenta. In
particular, the energy scale dependence of the strong coupling can be
determined, and the parton distribution functions (PDFs), notably the
gluon PDF, can be constrained.

The high quality and precision of the measurements performed by the
experiments at the Large Hadron Collider (LHC) at CERN demand a
similar level of accuracy in the theoretical predictions. The QCD
corrections to jet production are known up to next-to-leading order
(NLO) in perturbation
theory~\cite{Ellis:1990ek,Ellis:1992en,Giele:1994gf}, and a
substantial effort is currently put into the computation of the
corrections at next-to-next-to-leading order (NNLO), where the purely
gluonic channel, retaining the full dependence on the number of
colours, was completed recently~\cite{Ridder:2013mf,Currie:2013dwa}.

Probing the high-energy regime beyond \Unit{1}{\UTeV} of jet transverse
momentum, electroweak (EW) corrections might become large because of
the appearance of Sudakov-type and other high-energy logarithmic terms
in the calculations. In particular, the leading term is given by
$\alpha_\mathrm{w}\ln^2(Q^2/M_\PW^2)$, where $Q$ denotes the typical scale of
the hard-scattering reaction, $M_\PW$ is the $\PW$-boson mass, and
$\alpha_\mathrm{w}=\alpha/\sw^2=e^2/(4\pi\sw^2)$ is derived from the $SU(2)$
gauge coupling $e/\sw$ with $\sw$ being the sine of the weak mixing
angle $\theta_\rw$. The potential impact of these EW corrections has
to be evaluated in order to accompany the NNLO QCD prediction.

This article is organized as follows: In Section~\ref{sec:incJet:theory}, the
purely weak radiative corrections of $\mathcal{O}(\alphas^2\alpha)$ to the
inclusive jet production are presented, where the restriction to the
weak corrections, denoted by $\alphas^2\alpha_\mathrm{w}$, is motivated by the
aforementioned logarithmic enhancements. In the following
section~\ref{sec:incJet:cms} the measurement is introduced, to which the
theory predictions are compared in
Section~\ref{sec:incJet:comp}. Section~\ref{sec:incJet:summary} summarizes the
results and gives a short outlook.

\subsection{Weak radiative corrections and tree-level electroweak
  contributions}
\label{sec:incJet:theory}

The weak corrections form a well-defined gauge-invariant subset of the
full EW ones and can be supplemented by the photonic contributions at
a later time. In case of the photonic corrections to jet production 
the high-energy
logarithmic terms correspond to the well-known infrared singularities,
which cancel against bremsstrahlung corrections and therefore do not
lead to logarithmically enhanced contributions. The results shown in
the following are based on the calculation of the respective
corrections to dijet production presented in
Ref.~\cite{Dittmaier:2012kx}, which includes a detailed discussion of
the numerical results presented in the form of distributions in the
dijet invariant mass and the transverse momenta of the leading and
sub-leading jets. Corrections at this order have been previously
calculated for the single-jet inclusive cross section in
Ref.~\cite{Moretti:2006ea}, and preliminary results to dijet
production were shown in Ref.~\cite{Scharf:2009sp}.

\begin{figure}[b]
  \centering {\footnotesize(a)}\quad $\left\lvert
    \vcenter{\hbox{\includegraphics[scale=0.8,trim=0em 1.3em 0em
      1.3em]{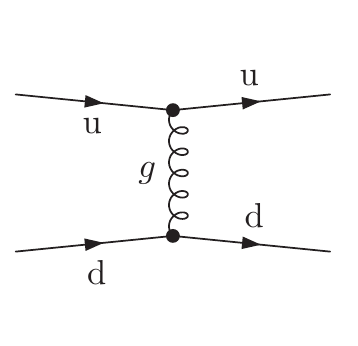}}} \right\rvert^2$ \qquad\quad
  {\footnotesize(b)}\quad $\left\lvert
    \vcenter{\hbox{\includegraphics[scale=0.8,trim=0em 1.3em 0em
      1.3em]{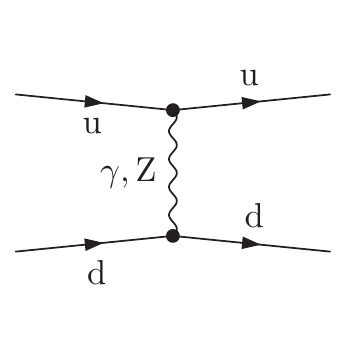}}} +
    \vcenter{\hbox{\includegraphics[scale=0.8,trim=0em 1.3em 0em
      1.3em]{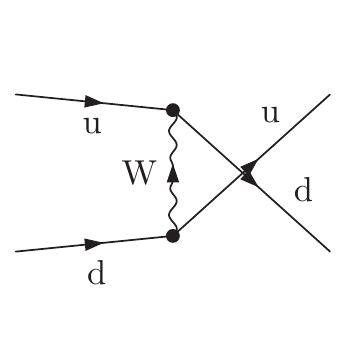}}} \right\rvert^2$
  \\[1em]
  {\footnotesize(c)}\quad $2\,\Re\,\left\lbrace
    \vcenter{\hbox{\includegraphics[scale=0.8,trim=0em 1.3em 0em
      1.3em]{incjet_main/incJet/TreeQCD}}} \times \left(
      \vcenter{\hbox{\includegraphics[scale=0.8,trim=0em 1.3em 0em
        1.3em]{incjet_main/incJet/TreeW}}} \right)^* \right\rbrace$
  \caption{The tree-level contributions to the process
    $\PQu+\PQd\to\PQu+\PQd$ of the orders (a)~$\alphas^2$, (b)~$\alpha^2$,
    and (c)~$\alphas\alpha$.}
  \label{fig:incJet:tree}
\end{figure}

The EW interaction does not only affect jet cross sections through
radiative corrections, but in case of the four-quark subprocesses the
production of jets can also occur via the exchange of EW gauge bosons
already at leading order (LO).
As a consequence, the Born cross section receives, in addition to the
pure QCD contribution of $\mathcal{O}(\alphas^2)$, further contributions of
$\mathcal{O}(\alphas\alpha)$ and $\mathcal{O}(\alpha^2)$ from the interference
between QCD and EW diagrams and the squares of the EW amplitudes.
The different contributions that need to be considered
at tree level are illustrated exemplarily for the case of the partonic
subprocess $\PQu+\PQd\to\PQu+\PQd$ in Fig.~\ref{fig:incJet:tree}. The
interference term of $\mathcal{O}(\alphas\alpha)$ only contributes to the
product between $t$- and $u$-channel diagrams due to the colour
structure. Note that all EW gauge bosons are accounted for in the LO
prediction, i.e.\ the photonic contributions are included.

\begin{figure}
  \centering
  \begin{align*}
    \text{\footnotesize(a)}\quad \left\lbrace
      \vcenter{\hbox{\includegraphics[scale=0.8,trim=0em 1.3em 0em
        1.3em]{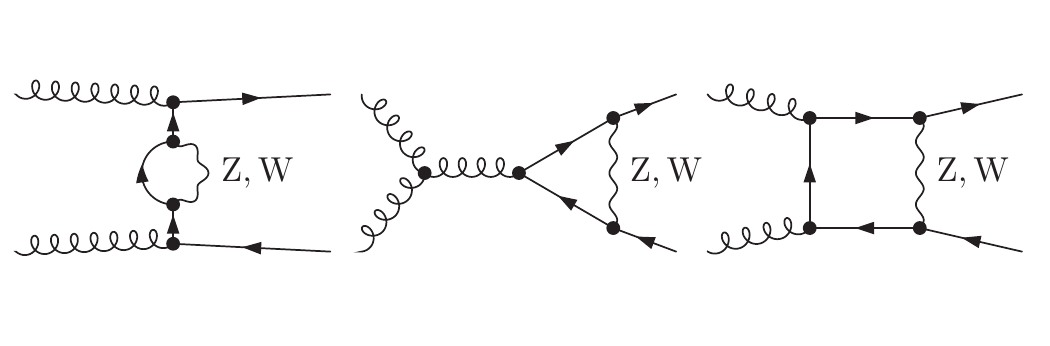}}} \ldots\right\rbrace\times
    &\left\lbrace \vcenter{\hbox{\includegraphics[scale=0.8,trim=0em 1.3em
        0em 1.3em]{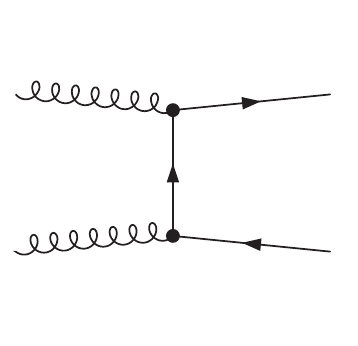}}} \ldots\right\rbrace^*
    \\[0.8em]
    \text{\footnotesize(b)}\quad \left\lbrace
      \vcenter{\hbox{\includegraphics[scale=0.8,trim=0em 1.3em 0em
        1.3em]{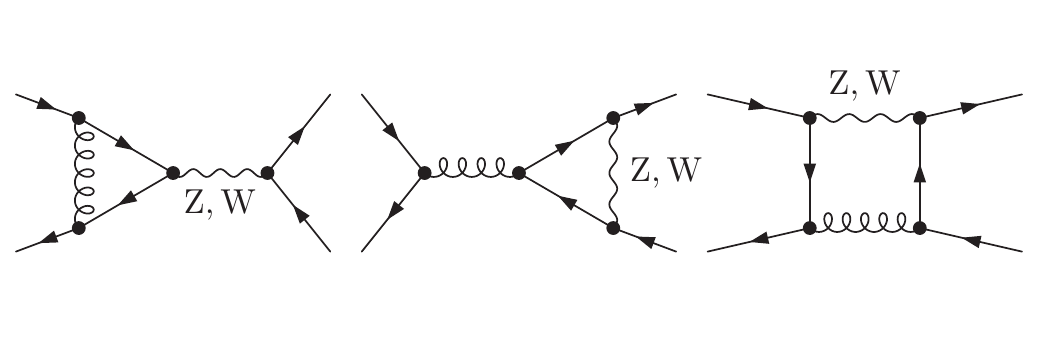}}} \ldots\right\rbrace\times
    &\left\lbrace \vcenter{\hbox{\includegraphics[scale=0.8,trim=0em 1.3em
        0em 1.3em]{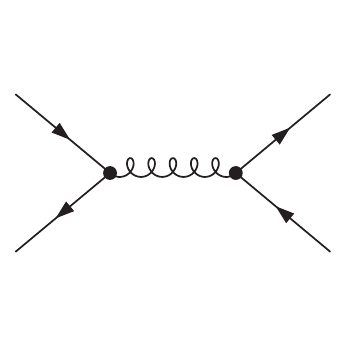}}} \ldots\right\rbrace^*
    \\[0.8em]
    \text{\footnotesize(c)}\quad \left\lbrace
      \vcenter{\hbox{\includegraphics[scale=0.8,trim=0em 1.3em 0em
        1.3em]{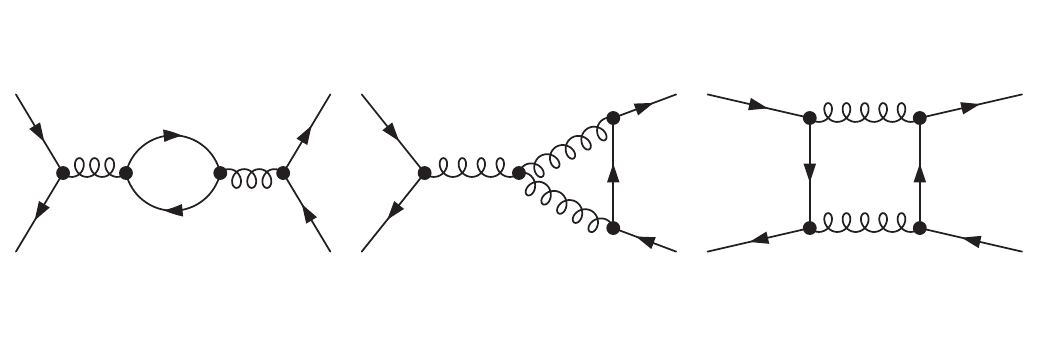}}} \ldots\right\rbrace\times
    &\left\lbrace \vcenter{\hbox{\includegraphics[scale=0.8,trim=0em 1.3em
        0em 1.3em]{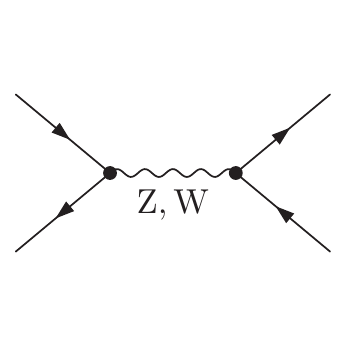}}} \ldots\right\rbrace^*
  \end{align*}
  \caption{The virtual corrections of $\mathcal{O}(\alphas^2\alpha_\mathrm{w})$
    illustrated in terms of typical interferences for the (a)
    two-gluon--two-quark and (b), (c) four-quark subprocess.}
  \label{fig:incJet:virt}
\end{figure}

As motivated in the introduction, the NLO calculation is restricted to
the purely weak corrections defined by the order $\alphas^2\alpha_\mathrm{w}$.
A selection of interference contributions that enter the calculation of
the virtual corrections is shown in Fig.~\ref{fig:incJet:virt}. The purely
gluonic channel does not receive corrections through this order, and
the corrections to the subprocesses with two gluons and two quarks
correspond to the weak $\mathcal{O}(\alpha_\mathrm{w})$ corrections to the LO
$\mathcal{O}(\alphas^2)$ cross section which are shown in
Fig.~\ref{fig:incJet:virt}(a)\@. The fact that diagrams of both,
$\mathcal{O}(\alphas)$ and $\mathcal{O}(\alpha_\mathrm{w})$, occur at LO in case of the
four-quark process leads to two different ways to obtain corrections
of the order $\alphas^2\alpha_\mathrm{w}$. Firstly, the one-loop diagrams of
$\mathcal{O}(\alphas\alpha_\mathrm{w})$ interfered with the LO QCD diagram, similarly
to the corrections to the two-gluon--two-quark subprocess, which is
illustrated in Fig.~\ref{fig:incJet:virt}(b)\@. Secondly, the one-loop NLO
QCD diagrams of $\mathcal{O}(\alphas^2)$ interfered with the LO weak
diagrams as shown in Fig.~\ref{fig:incJet:virt}(c)\@. Note, however, that a
strict separation of the weak and QCD corrections is not possible,
which is reflected by the appearance of diagrams such as the third
loop diagram in Fig.~\ref{fig:incJet:virt}(b)\@. This box diagram can be
obtained in two ways: by considering the weak corrections to the
tree-level diagram shown on the right of Fig.~\ref{fig:incJet:virt}(b), and
by attaching a virtual gluon to the tree-level diagram in
Fig.~\ref{fig:incJet:virt}(c)\@. This signals that all contributions of the
defining order $\alphas^2\alpha_\mathrm{w}$ have to be taken consistently into
account and, furthermore, the corresponding real-emission corrections
need to be included in the case of the four-quark processes to ensure
the proper cancellation of infrared divergences. Further details on
the calculation of the corrections can be found in
Ref.~\cite{Dittmaier:2012kx}.

\subsection{Measurement of the inclusive jet cross section with the CMS
  detector}
\label{sec:incJet:cms}

The inclusive jet cross section that is compared to theory predictions
has been measured with the CMS detector~\cite{Chatrchyan:2008aa} in
2011 from proton--proton collision data at $\sqrt{s}=\Unit{7}{\UTeV}$. The
recorded data correspond to \Unit{5.0}{\Ufb$^{-1}$} of integrated luminosity and
reach up to \Unit{2}{\UTeV} in jet transverse momentum $p_{\mathrm{T}}$ and 2.5 in absolute
jet rapidity $\lvert y\rvert$. The measurement is performed double-differentially
with five equally sized bins of $\Delta\lvert y\rvert = 0.5$ up to $\lvert y\rvert =
2.5$ and a binning in jet $p_{\mathrm{T}}$ that follows the jet $p_{\mathrm{T}}$ resolution of
the central detector. The minimal $p_{\mathrm{T}}$ imposed on any jet is \Unit{114}{\UGeV}
while the upper reach in $p_{\mathrm{T}}$ is limited by the available amount of
data and decreases with $\lvert y\rvert$.

Employing the {\textsc{FastJet}} package~\cite{Cacciari:2011ma} jets are
reconstructed using the collinear- and infrared-safe anti-$k_\mathrm{T}$
clustering algorithm~\cite{Cacciari:2008gp} with a jet size parameter
of $R=0.7$. The measured cross sections, corrected for detector
effects, are published in~Ref.~\cite{Chatrchyan:2012bja} and are
available via the {\textsc{HepData}}
project~\cite{Whalley:1989mt,Buckley:2010jn} including correlations
between systematic and statistical experimental uncertainties.
Non-perturbative QCD corrections, see Sect.~\ref{sec:incJet:comp}, that were
used in the CMS reference are provided in the {\textsc{HepData}} files as well.

The measurement is subject to four categories of experimental
uncertainties: the jet energy scale, the luminosity, the unfolding,
and a residual uncertainty. The jet energy scale as the dominant
source of systematic uncertainty affects the jet cross section to
about 5--25\% of uncertainty. The luminosity, which is known to a
precision of 2.2\%, corresponds to a global normalization uncertainty.
The unfolding corrects for detector effects and is related to the jet
energy resolution and the modelling of the detector response in
detailed Monte Carlo simulations. It leads to an uncertainty on the
cross section of the order of 3--4\%. Diverse residual effects are
summarized into an uncorrelated uncertainty of 1\%.

\subsection{Comparison between theory and measurement}
\label{sec:incJet:comp}

\begin{figure}
  \centering
  \includegraphics[scale=0.8]{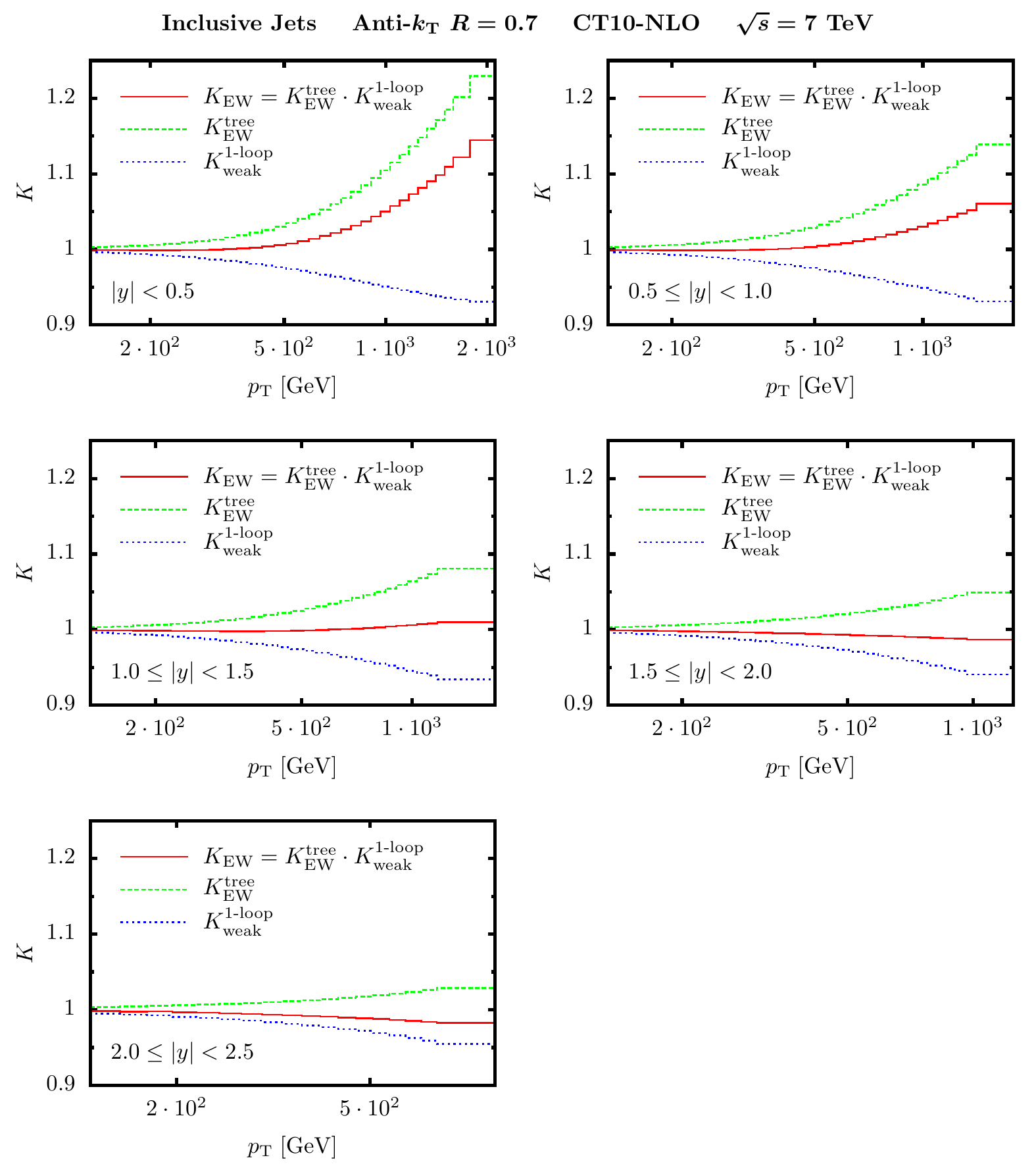}
  \caption{The EW correction factors to the
    transverse-momentum distribution for inclusive jet production for
    each rapidity bin, divided into the weak radiative corrections of
    $\mathcal{O}(\alphas^2\alpha_\mathrm{w})$ (dotted blue), the tree-level
    contributions of $\mathcal{O}(\alphas\alpha,\;\alpha^2)$ (dashed
    green), and the combination of both contributions (solid red).}
  \label{fig:incJet:ew}
\end{figure}

\begin{table}
  \centering
  \scalebox{0.6}{\begin{tabular}{|c|c| |c|c| |c|c| |c|c| |c|c|}
 \multicolumn{2}{|c||}{$\lvert y \rvert<0.5$} 
		   & \multicolumn{2}{ c||}{$0.5\leq\lvert y \rvert<1.0$} 
		   & \multicolumn{2}{ c||}{$1.0\leq\lvert y \rvert<1.5$} 
		   & \multicolumn{2}{ c||}{$1.5\leq\lvert y \rvert<2.0$} 
		   & \multicolumn{2}{ c| }{$2.0\leq\lvert y \rvert<2.5$} \\[.6em] 
 $p_\mathrm{T} \; [\mathrm{GeV}]$ & $ \left.\begin{subarray}{l}K^\text{tree}_\text{EW}\\K^\text{1-loop}_\text{weak}\end{subarray}\right\} K_\mathrm{EW}$ 
		   & $p_\mathrm{T} \; [\mathrm{GeV}]$ & $ \left.\begin{subarray}{l}K^\text{tree}_\text{EW}\\K^\text{1-loop}_\text{weak}\end{subarray}\right\} K_\mathrm{EW}$ 
		   & $p_\mathrm{T} \; [\mathrm{GeV}]$ & $ \left.\begin{subarray}{l}K^\text{tree}_\text{EW}\\K^\text{1-loop}_\text{weak}\end{subarray}\right\} K_\mathrm{EW}$ 
		   & $p_\mathrm{T} \; [\mathrm{GeV}]$ & $ \left.\begin{subarray}{l}K^\text{tree}_\text{EW}\\K^\text{1-loop}_\text{weak}\end{subarray}\right\} K_\mathrm{EW}$ 
		   & $p_\mathrm{T} \; [\mathrm{GeV}]$ & $ \left.\begin{subarray}{l}K^\text{tree}_\text{EW}\\K^\text{1-loop}_\text{weak}\end{subarray}\right\} K_\mathrm{EW}$ \\[.5em] \hline 
 $114 - 133$  & $ \left.\begin{subarray}{l}1.002\\0.997\end{subarray}\right\} 0.999 $  &  $114 - 133$  & $ \left.\begin{subarray}{l}1.002\\0.997\end{subarray}\right\} 0.999 $  &  $114 - 133$  & $ \left.\begin{subarray}{l}1.003\\0.997\end{subarray}\right\} 0.999 $  &  $114 - 133$  & $ \left.\begin{subarray}{l}1.003\\0.996\end{subarray}\right\} 0.999 $  &  $114 - 133$  & $ \left.\begin{subarray}{l}1.003\\0.996\end{subarray}\right\} 0.999 $ \\ 
 $133 - 153$  & $ \left.\begin{subarray}{l}1.003\\0.996\end{subarray}\right\} 0.999 $  &  $133 - 153$  & $ \left.\begin{subarray}{l}1.003\\0.996\end{subarray}\right\} 0.999 $  &  $133 - 153$  & $ \left.\begin{subarray}{l}1.003\\0.996\end{subarray}\right\} 0.999 $  &  $133 - 153$  & $ \left.\begin{subarray}{l}1.003\\0.995\end{subarray}\right\} 0.999 $  &  $133 - 153$  & $ \left.\begin{subarray}{l}1.004\\0.995\end{subarray}\right\} 0.998 $ \\ 
 $153 - 174$  & $ \left.\begin{subarray}{l}1.004\\0.995\end{subarray}\right\} 0.999 $  &  $153 - 174$  & $ \left.\begin{subarray}{l}1.004\\0.995\end{subarray}\right\} 0.999 $  &  $153 - 174$  & $ \left.\begin{subarray}{l}1.004\\0.995\end{subarray}\right\} 0.999 $  &  $153 - 174$  & $ \left.\begin{subarray}{l}1.004\\0.994\end{subarray}\right\} 0.999 $  &  $153 - 174$  & $ \left.\begin{subarray}{l}1.004\\0.993\end{subarray}\right\} 0.998 $ \\ 
 $174 - 196$  & $ \left.\begin{subarray}{l}1.005\\0.994\end{subarray}\right\} 0.999 $  &  $174 - 196$  & $ \left.\begin{subarray}{l}1.005\\0.994\end{subarray}\right\} 0.999 $  &  $174 - 196$  & $ \left.\begin{subarray}{l}1.005\\0.993\end{subarray}\right\} 0.999 $  &  $174 - 196$  & $ \left.\begin{subarray}{l}1.005\\0.993\end{subarray}\right\} 0.998 $  &  $174 - 196$  & $ \left.\begin{subarray}{l}1.005\\0.993\end{subarray}\right\} 0.998 $ \\ 
 $196 - 220$  & $ \left.\begin{subarray}{l}1.006\\0.993\end{subarray}\right\} 0.999 $  &  $196 - 220$  & $ \left.\begin{subarray}{l}1.006\\0.993\end{subarray}\right\} 0.999 $  &  $196 - 220$  & $ \left.\begin{subarray}{l}1.006\\0.992\end{subarray}\right\} 0.998 $  &  $196 - 220$  & $ \left.\begin{subarray}{l}1.006\\0.991\end{subarray}\right\} 0.998 $  &  $196 - 220$  & $ \left.\begin{subarray}{l}1.006\\0.991\end{subarray}\right\} 0.997 $ \\ 
 $220 - 245$  & $ \left.\begin{subarray}{l}1.007\\0.991\end{subarray}\right\} 0.999 $  &  $220 - 245$  & $ \left.\begin{subarray}{l}1.008\\0.991\end{subarray}\right\} 0.999 $  &  $220 - 245$  & $ \left.\begin{subarray}{l}1.008\\0.991\end{subarray}\right\} 0.998 $  &  $220 - 245$  & $ \left.\begin{subarray}{l}1.008\\0.990\end{subarray}\right\} 0.997 $  &  $220 - 245$  & $ \left.\begin{subarray}{l}1.007\\0.989\end{subarray}\right\} 0.996 $ \\ 
 $245 - 272$  & $ \left.\begin{subarray}{l}1.009\\0.990\end{subarray}\right\} 0.999 $  &  $245 - 272$  & $ \left.\begin{subarray}{l}1.009\\0.990\end{subarray}\right\} 0.999 $  &  $245 - 272$  & $ \left.\begin{subarray}{l}1.009\\0.989\end{subarray}\right\} 0.998 $  &  $245 - 272$  & $ \left.\begin{subarray}{l}1.009\\0.988\end{subarray}\right\} 0.997 $  &  $245 - 272$  & $ \left.\begin{subarray}{l}1.008\\0.987\end{subarray}\right\} 0.995 $ \\ 
 $272 - 300$  & $ \left.\begin{subarray}{l}1.011\\0.988\end{subarray}\right\} 0.999 $  &  $272 - 300$  & $ \left.\begin{subarray}{l}1.011\\0.988\end{subarray}\right\} 0.999 $  &  $272 - 300$  & $ \left.\begin{subarray}{l}1.011\\0.987\end{subarray}\right\} 0.998 $  &  $272 - 300$  & $ \left.\begin{subarray}{l}1.010\\0.986\end{subarray}\right\} 0.996 $  &  $272 - 300$  & $ \left.\begin{subarray}{l}1.009\\0.986\end{subarray}\right\} 0.994 $ \\ 
 $300 - 330$  & $ \left.\begin{subarray}{l}1.013\\0.987\end{subarray}\right\} 1.000 $  &  $300 - 330$  & $ \left.\begin{subarray}{l}1.013\\0.986\end{subarray}\right\} 0.999 $  &  $300 - 330$  & $ \left.\begin{subarray}{l}1.013\\0.985\end{subarray}\right\} 0.998 $  &  $300 - 330$  & $ \left.\begin{subarray}{l}1.012\\0.984\end{subarray}\right\} 0.996 $  &  $300 - 330$  & $ \left.\begin{subarray}{l}1.010\\0.984\end{subarray}\right\} 0.994 $ \\ 
 $330 - 362$  & $ \left.\begin{subarray}{l}1.016\\0.985\end{subarray}\right\} 1.000 $  &  $330 - 362$  & $ \left.\begin{subarray}{l}1.015\\0.984\end{subarray}\right\} 1.000 $  &  $330 - 362$  & $ \left.\begin{subarray}{l}1.015\\0.983\end{subarray}\right\} 0.998 $  &  $330 - 362$  & $ \left.\begin{subarray}{l}1.013\\0.982\end{subarray}\right\} 0.995 $  &  $330 - 362$  & $ \left.\begin{subarray}{l}1.011\\0.982\end{subarray}\right\} 0.993 $ \\ 
 $362 - 395$  & $ \left.\begin{subarray}{l}1.019\\0.983\end{subarray}\right\} 1.001 $  &  $362 - 395$  & $ \left.\begin{subarray}{l}1.018\\0.982\end{subarray}\right\} 1.000 $  &  $362 - 395$  & $ \left.\begin{subarray}{l}1.017\\0.981\end{subarray}\right\} 0.998 $  &  $362 - 395$  & $ \left.\begin{subarray}{l}1.015\\0.980\end{subarray}\right\} 0.995 $  &  $362 - 395$  & $ \left.\begin{subarray}{l}1.013\\0.979\end{subarray}\right\} 0.992 $ \\ 
 $395 - 430$  & $ \left.\begin{subarray}{l}1.022\\0.981\end{subarray}\right\} 1.002 $  &  $395 - 430$  & $ \left.\begin{subarray}{l}1.021\\0.980\end{subarray}\right\} 1.001 $  &  $395 - 430$  & $ \left.\begin{subarray}{l}1.019\\0.979\end{subarray}\right\} 0.998 $  &  $395 - 430$  & $ \left.\begin{subarray}{l}1.016\\0.978\end{subarray}\right\} 0.994 $  &  $395 - 430$  & $ \left.\begin{subarray}{l}1.014\\0.977\end{subarray}\right\} 0.991 $ \\ 
 $430 - 468$  & $ \left.\begin{subarray}{l}1.026\\0.979\end{subarray}\right\} 1.004 $  &  $430 - 468$  & $ \left.\begin{subarray}{l}1.025\\0.978\end{subarray}\right\} 1.002 $  &  $430 - 468$  & $ \left.\begin{subarray}{l}1.022\\0.977\end{subarray}\right\} 0.998 $  &  $430 - 468$  & $ \left.\begin{subarray}{l}1.018\\0.976\end{subarray}\right\} 0.994 $  &  $430 - 468$  & $ \left.\begin{subarray}{l}1.016\\0.974\end{subarray}\right\} 0.990 $ \\ 
 $468 - 507$  & $ \left.\begin{subarray}{l}1.030\\0.976\end{subarray}\right\} 1.006 $  &  $468 - 507$  & $ \left.\begin{subarray}{l}1.028\\0.976\end{subarray}\right\} 1.003 $  &  $468 - 507$  & $ \left.\begin{subarray}{l}1.025\\0.974\end{subarray}\right\} 0.998 $  &  $468 - 507$  & $ \left.\begin{subarray}{l}1.020\\0.973\end{subarray}\right\} 0.993 $  &  $468 - 507$  & $ \left.\begin{subarray}{l}1.017\\0.972\end{subarray}\right\} 0.989 $ \\ 
 $507 - 548$  & $ \left.\begin{subarray}{l}1.035\\0.974\end{subarray}\right\} 1.008 $  &  $507 - 548$  & $ \left.\begin{subarray}{l}1.032\\0.973\end{subarray}\right\} 1.005 $  &  $507 - 548$  & $ \left.\begin{subarray}{l}1.028\\0.972\end{subarray}\right\} 0.999 $  &  $507 - 548$  & $ \left.\begin{subarray}{l}1.023\\0.971\end{subarray}\right\} 0.993 $  &  $507 - 548$  & $ \left.\begin{subarray}{l}1.019\\0.969\end{subarray}\right\} 0.988 $ \\ 
 $548 - 592$  & $ \left.\begin{subarray}{l}1.040\\0.972\end{subarray}\right\} 1.011 $  &  $548 - 592$  & $ \left.\begin{subarray}{l}1.037\\0.971\end{subarray}\right\} 1.007 $  &  $548 - 592$  & $ \left.\begin{subarray}{l}1.031\\0.969\end{subarray}\right\} 0.999 $  &  $548 - 592$  & $ \left.\begin{subarray}{l}1.025\\0.968\end{subarray}\right\} 0.992 $  &  $548 - 592$  & $ \left.\begin{subarray}{l}1.021\\0.966\end{subarray}\right\} 0.987 $ \\ 
 $592 - 638$  & $ \left.\begin{subarray}{l}1.046\\0.969\end{subarray}\right\} 1.014 $  &  $592 - 638$  & $ \left.\begin{subarray}{l}1.042\\0.968\end{subarray}\right\} 1.009 $  &  $592 - 638$  & $ \left.\begin{subarray}{l}1.034\\0.967\end{subarray}\right\} 1.000 $  &  $592 - 638$  & $ \left.\begin{subarray}{l}1.027\\0.965\end{subarray}\right\} 0.991 $  &  $592 - 638$  & $ \left.\begin{subarray}{l}1.023\\0.963\end{subarray}\right\} 0.985 $ \\ 
 $638 - 686$  & $ \left.\begin{subarray}{l}1.053\\0.966\end{subarray}\right\} 1.018 $  &  $638 - 686$  & $ \left.\begin{subarray}{l}1.047\\0.965\end{subarray}\right\} 1.011 $  &  $638 - 686$  & $ \left.\begin{subarray}{l}1.038\\0.964\end{subarray}\right\} 1.001 $  &  $638 - 686$  & $ \left.\begin{subarray}{l}1.030\\0.962\end{subarray}\right\} 0.991 $  &  $638 - 686$  & $ \left.\begin{subarray}{l}1.026\\0.959\end{subarray}\right\} 0.984 $ \\ 
 $686 - 737$  & $ \left.\begin{subarray}{l}1.060\\0.964\end{subarray}\right\} 1.022 $  &  $686 - 737$  & $ \left.\begin{subarray}{l}1.053\\0.963\end{subarray}\right\} 1.013 $  &  $686 - 737$  & $ \left.\begin{subarray}{l}1.042\\0.961\end{subarray}\right\} 1.001 $  &  $686 - 737$  & $ \left.\begin{subarray}{l}1.033\\0.959\end{subarray}\right\} 0.990 $  &  $686 - 905$  & $ \left.\begin{subarray}{l}1.029\\0.955\end{subarray}\right\} 0.983 $ \\ 
 $737 - 790$  & $ \left.\begin{subarray}{l}1.068\\0.961\end{subarray}\right\} 1.026 $  &  $737 - 790$  & $ \left.\begin{subarray}{l}1.059\\0.960\end{subarray}\right\} 1.017 $  &  $737 - 790$  & $ \left.\begin{subarray}{l}1.046\\0.958\end{subarray}\right\} 1.002 $  &  $737 - 790$  & $ \left.\begin{subarray}{l}1.036\\0.956\end{subarray}\right\} 0.990 $ \\ 
 $790 - 846$  & $ \left.\begin{subarray}{l}1.076\\0.959\end{subarray}\right\} 1.032 $  &  $790 - 846$  & $ \left.\begin{subarray}{l}1.065\\0.957\end{subarray}\right\} 1.020 $  &  $790 - 846$  & $ \left.\begin{subarray}{l}1.050\\0.955\end{subarray}\right\} 1.003 $  &  $790 - 846$  & $ \left.\begin{subarray}{l}1.039\\0.952\end{subarray}\right\} 0.989 $ \\ 
 $846 - 905$  & $ \left.\begin{subarray}{l}1.085\\0.956\end{subarray}\right\} 1.037 $  &  $846 - 905$  & $ \left.\begin{subarray}{l}1.072\\0.954\end{subarray}\right\} 1.023 $  &  $846 - 905$  & $ \left.\begin{subarray}{l}1.055\\0.952\end{subarray}\right\} 1.004 $  &  $846 - 905$  & $ \left.\begin{subarray}{l}1.042\\0.949\end{subarray}\right\} 0.989 $ \\ 
 $905 - 967$  & $ \left.\begin{subarray}{l}1.094\\0.954\end{subarray}\right\} 1.044 $  &  $905 - 967$  & $ \left.\begin{subarray}{l}1.079\\0.952\end{subarray}\right\} 1.026 $  &  $905 - 967$  & $ \left.\begin{subarray}{l}1.059\\0.949\end{subarray}\right\} 1.005 $  &  $905 - 967$  & $ \left.\begin{subarray}{l}1.045\\0.945\end{subarray}\right\} 0.988 $ \\ 
 $967 - 1032$  & $ \left.\begin{subarray}{l}1.104\\0.951\end{subarray}\right\} 1.050 $  &  $967 - 1032$  & $ \left.\begin{subarray}{l}1.086\\0.949\end{subarray}\right\} 1.030 $  &  $967 - 1032$  & $ \left.\begin{subarray}{l}1.064\\0.946\end{subarray}\right\} 1.006 $  &  $967 - 1248$  & $ \left.\begin{subarray}{l}1.049\\0.941\end{subarray}\right\} 0.987 $ \\ 
 $1032 - 1101$  & $ \left.\begin{subarray}{l}1.115\\0.949\end{subarray}\right\} 1.058 $  &  $1032 - 1101$  & $ \left.\begin{subarray}{l}1.093\\0.946\end{subarray}\right\} 1.034 $  &  $1032 - 1101$  & $ \left.\begin{subarray}{l}1.068\\0.942\end{subarray}\right\} 1.007 $ \\ 
 $1101 - 1172$  & $ \left.\begin{subarray}{l}1.126\\0.946\end{subarray}\right\} 1.065 $  &  $1101 - 1172$  & $ \left.\begin{subarray}{l}1.101\\0.943\end{subarray}\right\} 1.038 $  &  $1101 - 1172$  & $ \left.\begin{subarray}{l}1.074\\0.939\end{subarray}\right\} 1.008 $ \\ 
 $1172 - 1248$  & $ \left.\begin{subarray}{l}1.137\\0.944\end{subarray}\right\} 1.073 $  &  $1172 - 1248$  & $ \left.\begin{subarray}{l}1.109\\0.941\end{subarray}\right\} 1.043 $  &  $1172 - 1684$  & $ \left.\begin{subarray}{l}1.081\\0.934\end{subarray}\right\} 1.010 $ \\ 
 $1248 - 1327$  & $ \left.\begin{subarray}{l}1.148\\0.942\end{subarray}\right\} 1.081 $  &  $1248 - 1327$  & $ \left.\begin{subarray}{l}1.117\\0.938\end{subarray}\right\} 1.047 $ \\ 
 $1327 - 1410$  & $ \left.\begin{subarray}{l}1.160\\0.940\end{subarray}\right\} 1.090 $  &  $1327 - 1410$  & $ \left.\begin{subarray}{l}1.125\\0.935\end{subarray}\right\} 1.052 $ \\ 
 $1410 - 1497$  & $ \left.\begin{subarray}{l}1.171\\0.938\end{subarray}\right\} 1.098 $  &  $1410 - 1784$  & $ \left.\begin{subarray}{l}1.139\\0.931\end{subarray}\right\} 1.060 $ \\ 
 $1497 - 1588$  & $ \left.\begin{subarray}{l}1.185\\0.936\end{subarray}\right\} 1.109 $ \\ 
 $1588 - 1784$  & $ \left.\begin{subarray}{l}1.202\\0.934\end{subarray}\right\} 1.122 $ \\ 
 $1784 - 2116$  & $ \left.\begin{subarray}{l}1.229\\0.931\end{subarray}\right\} 1.144 $ \\ 
\end{tabular}}
  \caption{The numerical values of the correction factors for each
    transverse-momentum $p_{\mathrm{T}}$ and rapidity $\lvert y\rvert$ bin,
    c.f.\ Fig.~\ref{fig:incJet:ew}.}
  \label{tab:incJet:ew}
\end{table}

The EW corrections studied here consist of the two contributions
introduced in Sect.~\ref{sec:incJet:theory}. They have been evaluated with
the renormalization and factorization scales, $\mu_\mathrm{R}$ and $\mu_\mathrm{F}$, set to
the maximum jet $p_{\mathrm{T}}$, $p_{\mathrm{T,max}}$. The results are presented in
Fig.~\ref{fig:incJet:ew} as a function of the transverse momentum $p_{\mathrm{T}}$ for the
five rapidity bins in terms of correction factors $K$:

\begin{description}
\item[\boldmath The weak radiative corrections of
  $\mathcal{O}(\alphas^2\alpha_\mathrm{w})$] denoted by $K^\text{1-loop}_\text{weak}$
  are negligible for small transverse momenta and become increasingly
  negative towards higher scales. This corresponds to the typical
  behaviour known from the weak high-energy logarithms, which become
  large in the so-called ``Sudakov regime'', where all 
  scalar invariants of different momenta in the 
  hard scattering are required to be simultaneously much larger than
  the weak gauge-boson masses. For instance, as discussed in more
  detail in Ref.~\cite{Dittmaier:2012kx}, the tails of the dijet
  invariant-mass distributions do not probe this Sudakov regime, but
  are dominated by the Regge (forward) region. As a consequence, the
  weak corrections to the dijet invariant-mass distribution turn out
  to be much smaller than in the case of the inclusive jet $p_{\mathrm{T}}$
  considered here, where the Sudakov regime is probed at high $p_{\mathrm{T}}$.
\item[\boldmath The tree-level contributions of
  $\mathcal{O}(\alphas\alpha,\;\alpha^2)$] with the associated correction
  factor $K^\text{tree}_\text{EW}$ solely appear in the four-quark
  subprocesses and are negligible at small transverse momenta, where
  the cross section is dominated by gluon-induced channels. Towards
  higher transverse momenta, the PDFs are probed at larger values of
  the momentum fraction $x$, so that the relative size of the
  quark--quark luminosity becomes more and more relevant compared to
  the gluon-induced channels. Hence, the tree-level contributions
  become large at high $p_{\mathrm{T}}$ and are positive at the LHC.
\end{description}

It is observed that these two contributions are of the same generic
size, however, with opposite sign, leading to significant
cancellations in the combined correction factor Eq.~\eqref{eq:incJet:K_EW}.%
\footnote{At the Tevatron proton--anti-proton collider, the tree-level
  contribution is dominated by the quark--anti-quark induced
  channels. They turn out to be negative at high $p_{\mathrm{T}}$, so that they
  further enhance the negative weak corrections.}  In the central
region ($\lvert y\rvert < 1.0$) the tree-level contributions exceed the weak
radiative ones, leading to a net correction that is positive and
increases towards higher jet $p_{\mathrm{T}}$ up to $15\%$ for $p_{\mathrm{T}} = \Unit{2}{\UTeV}$ and
$\lvert y\rvert < 0.5$. In the more forward region with $2.0 \leq \lvert y\rvert < 2.5$,
on the other hand, the weak corrections slightly surpass the
tree-level contributions leading to negative corrections that amount
to $-2\%$ for transverse momenta of $p_{\mathrm{T}} \approx \Unit{800}{\UGeV}$. The
numerical values of the respective correction factors are given in
Table~\ref{tab:incJet:ew} separately for each bin in $p_{\mathrm{T}}$ and $\lvert y\rvert$.

\begin{figure}
  \centering
  \begin{tabular}{cc}
    \multicolumn{2}{c}{\makebox[\textwidth]{\fontfamily{phv}\fontseries{b}\fontshape{n}\selectfont%
        \footnotesize\boldmath\hfill{}Inclusive Jets\hfill{}anti-$k_\mathrm{T}$$R=0.7$\hfill{}CT10-NLO\hfill{}%
        $\sqrt{s} = \Unit{7}{\UTeV}$\unboldmath\hfill}} \\
    \includegraphics[width=0.42\textwidth]{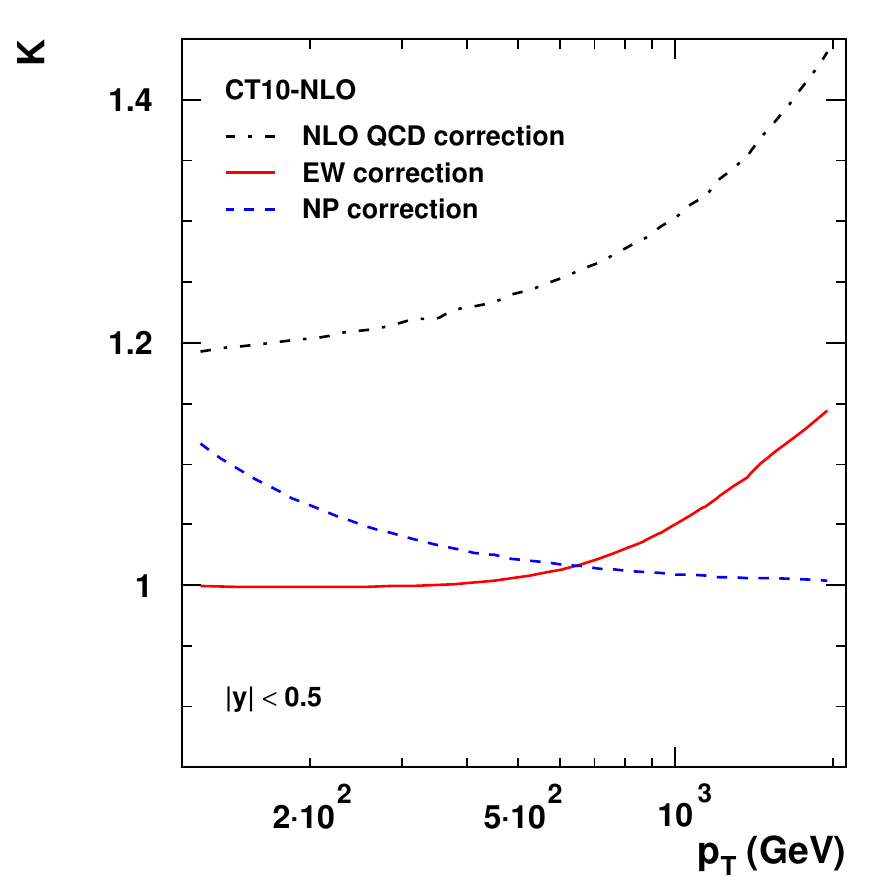} &
    \includegraphics[width=0.42\textwidth]{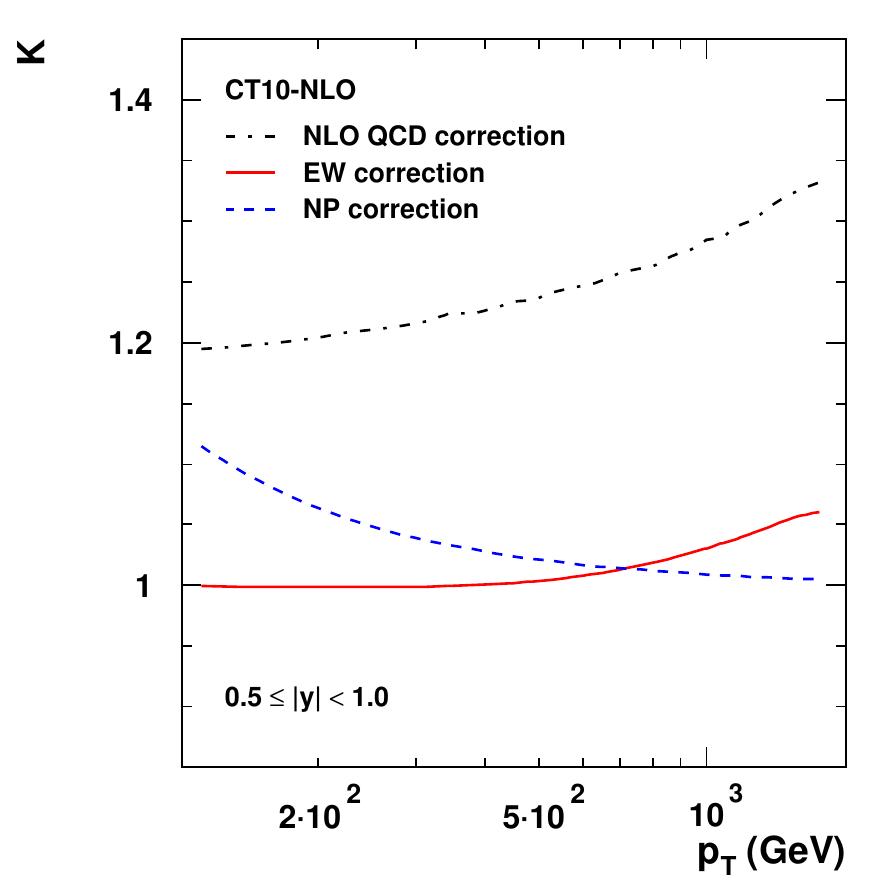} \\
    \includegraphics[width=0.42\textwidth]{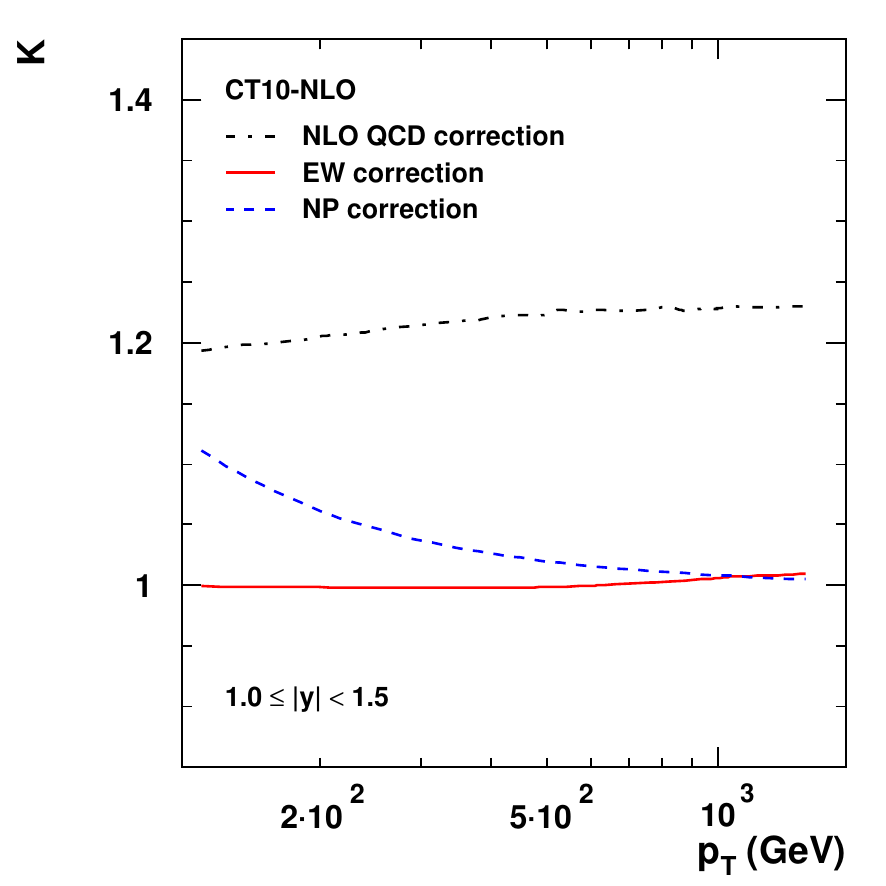} &
    \includegraphics[width=0.42\textwidth]{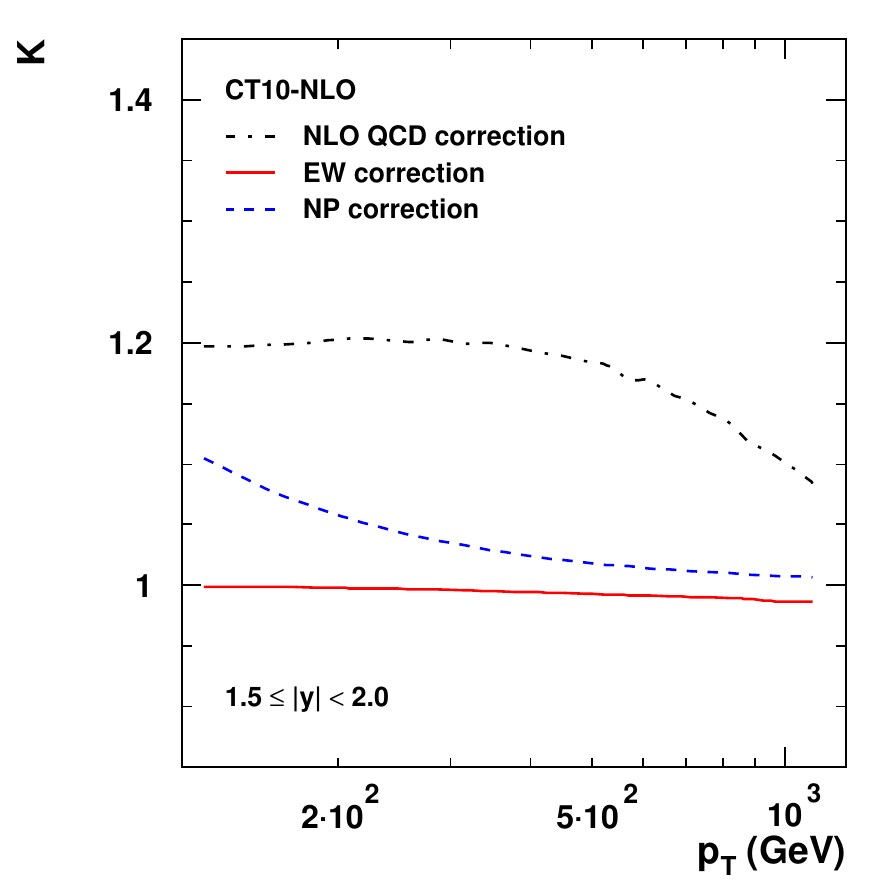} \\
    \includegraphics[width=0.42\textwidth]{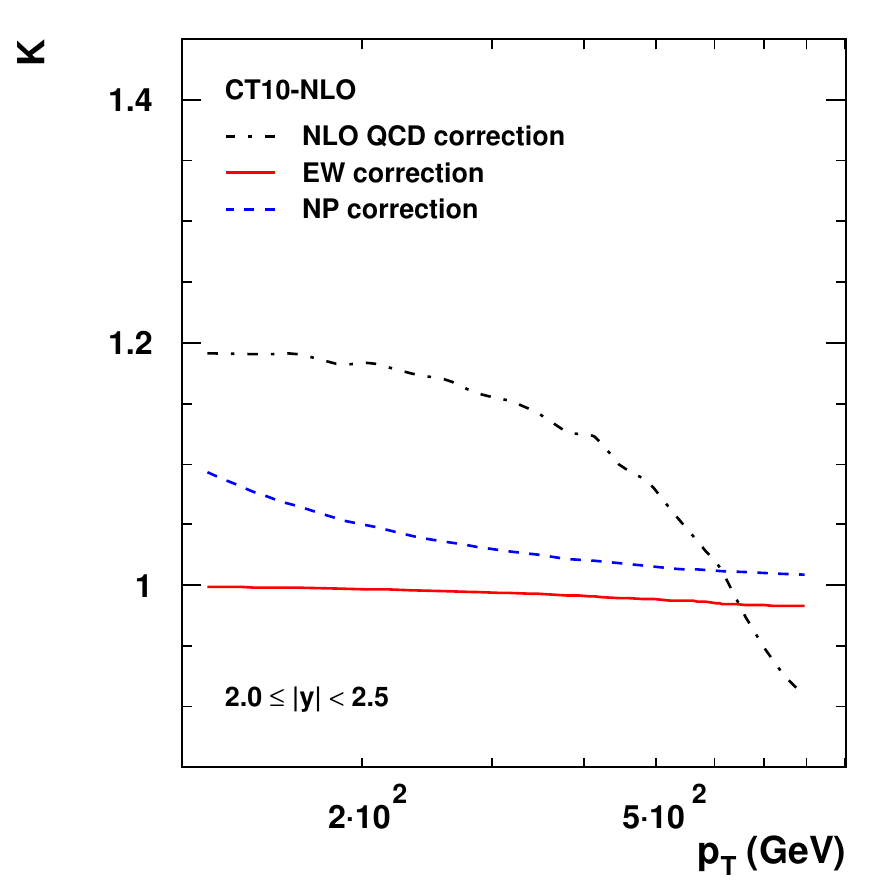} &
  \end{tabular}
  \caption{The three correction factors that enter the theory--data
    comparison as given by the NLO QCD corrections (``NLO QCD'' dash-dotted
    black), the EW corrections (``EW'' solid red), and the non-perturbative 
    effects (``NP'' dashed blue).}
  \label{fig:incJet:corr}
\end{figure}

The different corrections that enter the comparison between theory and
measurement are shown in Fig.~\ref{fig:incJet:corr} and comprise the pure QCD
corrections at NLO, the EW corrections discussed above, and
non-perturbative corrections.

The NLO QCD predictions are obtained using
{\textsc{NLOJet++}}~\cite{Nagy:2001fj,Nagy:2003tz} within the framework of
{\textsc{fastNLO}}~\cite{Britzger:2012bs} and represent the dominant
corrections, which are typically of the order of $20\%$. They have
been derived for renormalization and factorization scales of $\mu_\mathrm{R} =
\mu_\mathrm{F} = p_{\mathrm{T,jet}}$ for each jet, which differs from the scale choice
$\mu_\mathrm{R} = \mu_\mathrm{F} = p_{\mathrm{T,max}}$ in the EW corrections.
However, differences in the QCD corrections induced by the two scale 
settings are negligible at high $p_{\mathrm{T}}$ and amount to at most $+4\%$ at 
lowest jet $p_{\mathrm{T}}$.

The NP corrections account for the underlying event, typically
described in terms of multiple-parton interactions, and the
hadronization process. Both cannot be described within perturbative QCD,
but are modelled by Monte Carlo (MC) event generators, which introduce
additional free parameters that have to be tuned to measurements. 
In Ref.~\cite{Chatrchyan:2012bja} the CMS Collaboration compared the two
MC event generators {\textsc{pythia6}}~\cite{Sjostrand:2006za} and
{\textsc{herwig++}}~\cite{Bahr:2008pv}, which offer different models and
implementations of the NP effects. The NP correction is defined as the
relative difference between the relevant cross sections predicted with and without
multiple-parton interactions and hadronization. The central value of
this correction is estimated as the average of the predictions by
{\textsc{pythia6}} and {\textsc{herwig++}} and the uncertainty as $\pm$ half of their
difference. The exact numbers are available from the {\textsc{HepData}} record
of the CMS data and amount to about $+10\%$ at low $p_{\mathrm{T}}$ decreasing
steadily towards higher jet $p_{\mathrm{T}}$'s. Beyond about \Unit{800}{\UGeV} they become
negligible.

\begin{figure}
  \centering
  \includegraphics[scale=0.75,trim=0 0 17
  0]{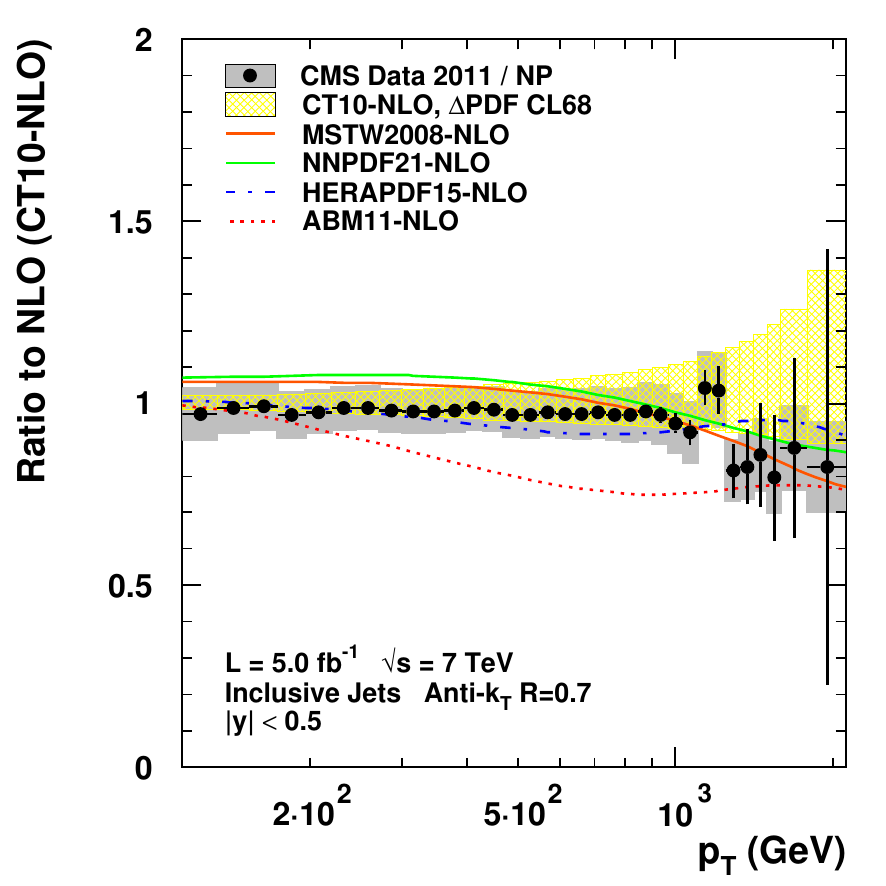}
  \includegraphics[scale=0.75,trim=52 0 0
  0,clip]{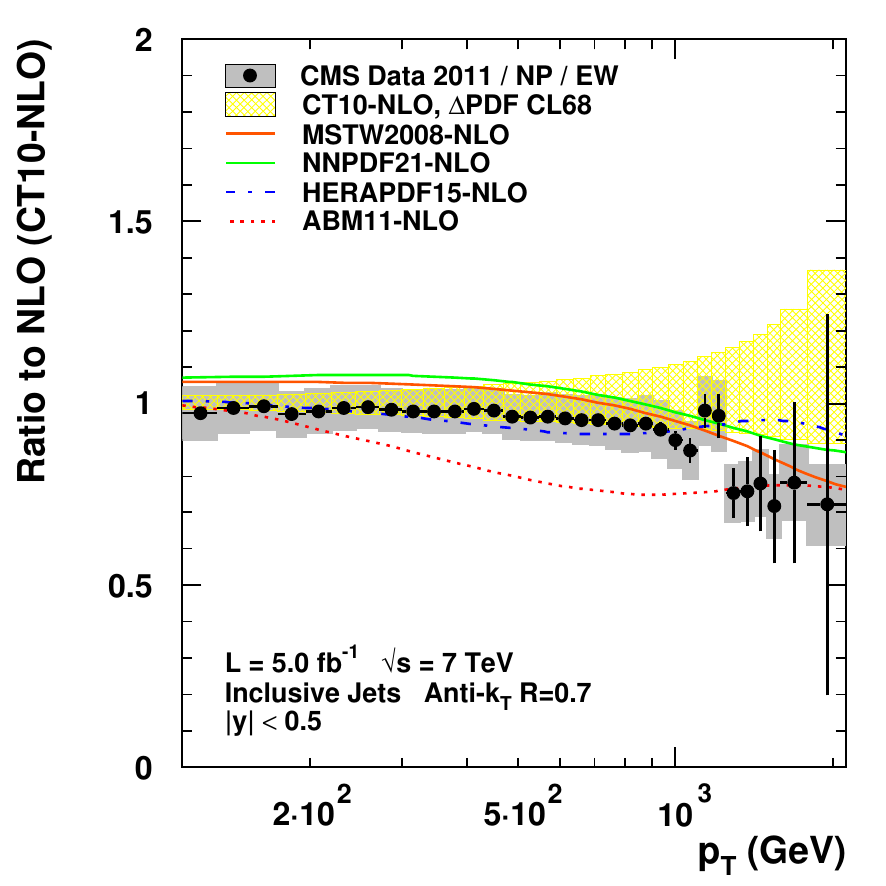}
  \includegraphics[scale=0.75,trim=0 0 17
  0]{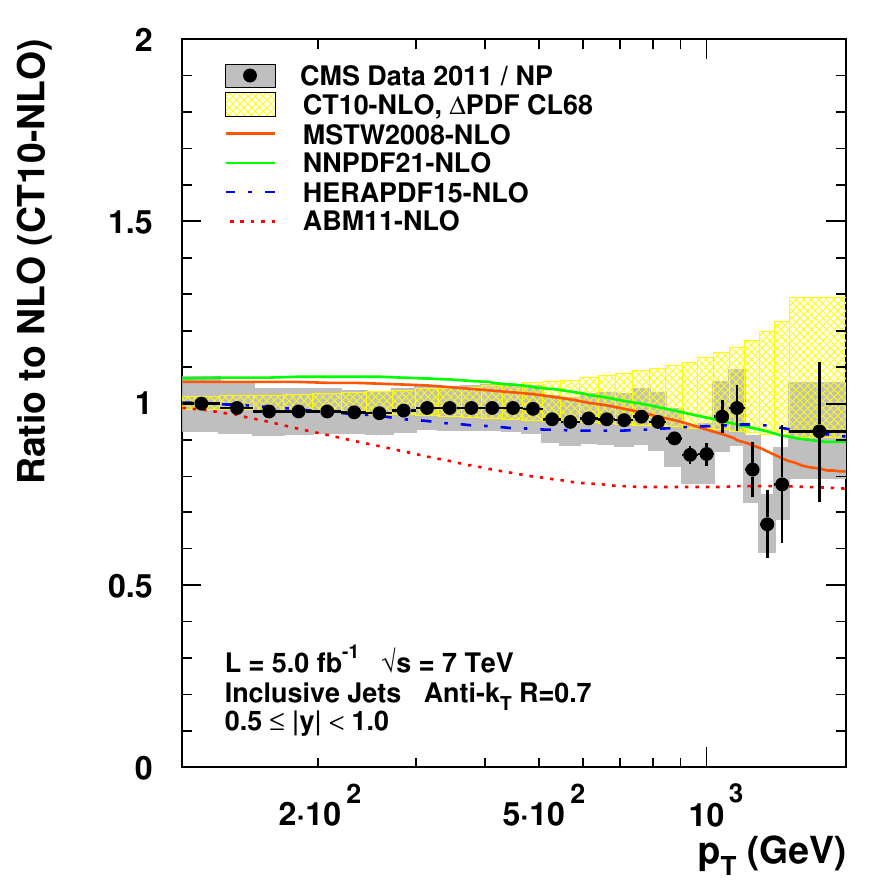}
  \includegraphics[scale=0.75,trim=52 0 0
  0,clip]{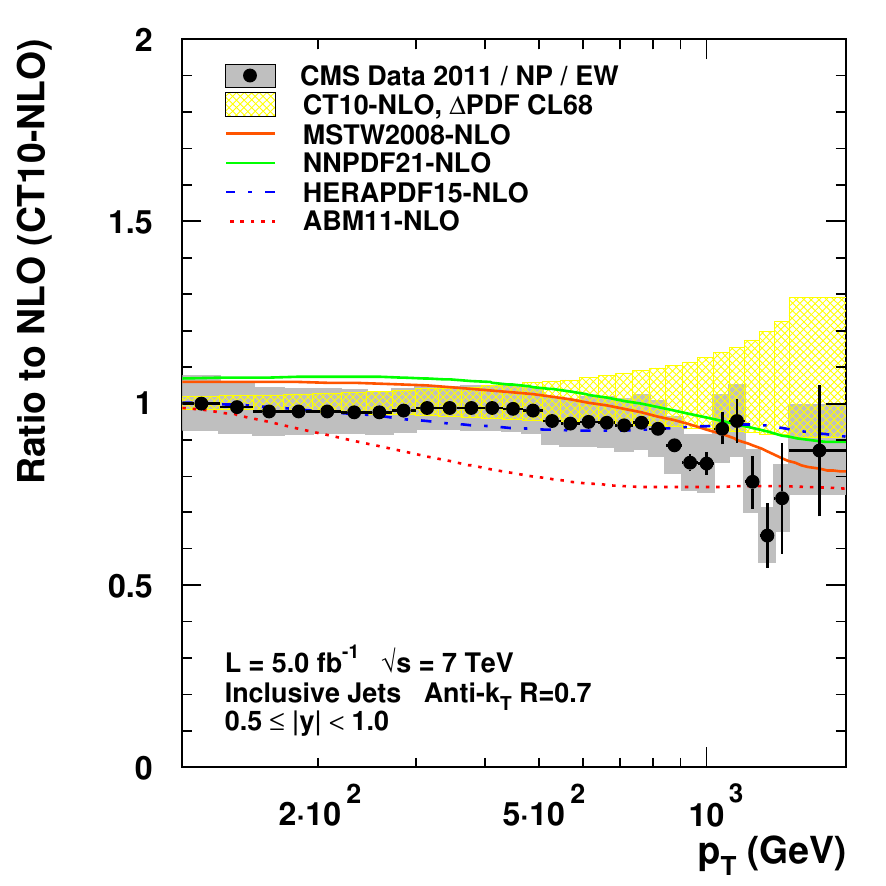}
  \includegraphics[scale=0.75,trim=0 0 17
  0]{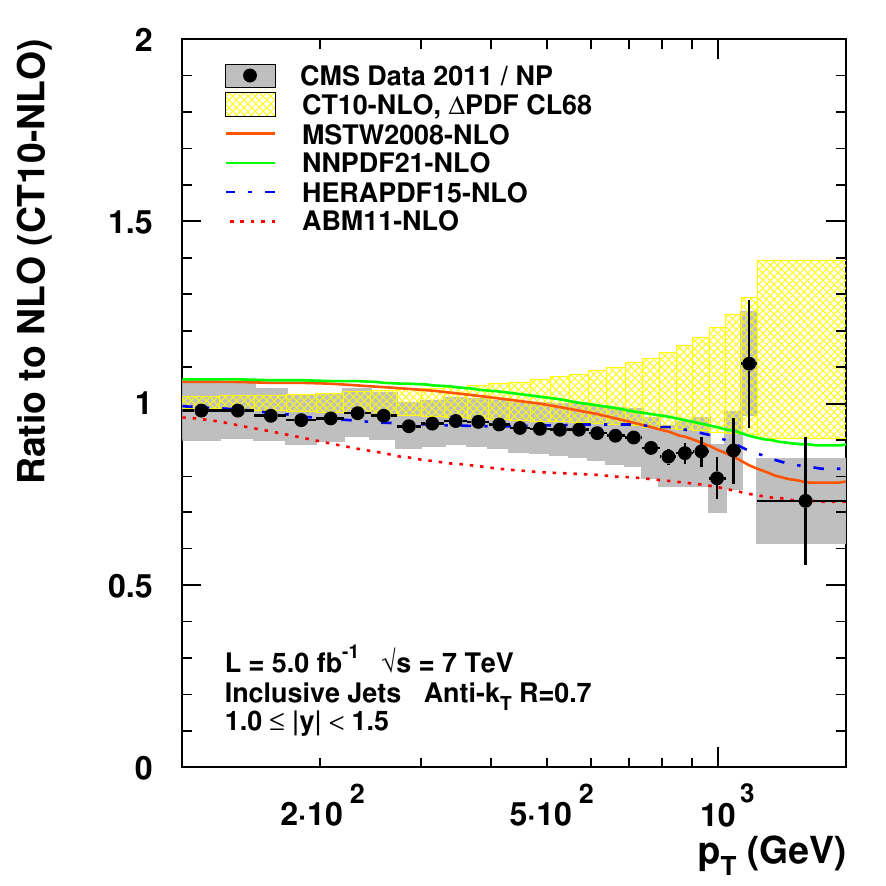}
  \includegraphics[scale=0.75,trim=52 0 0
  0,clip]{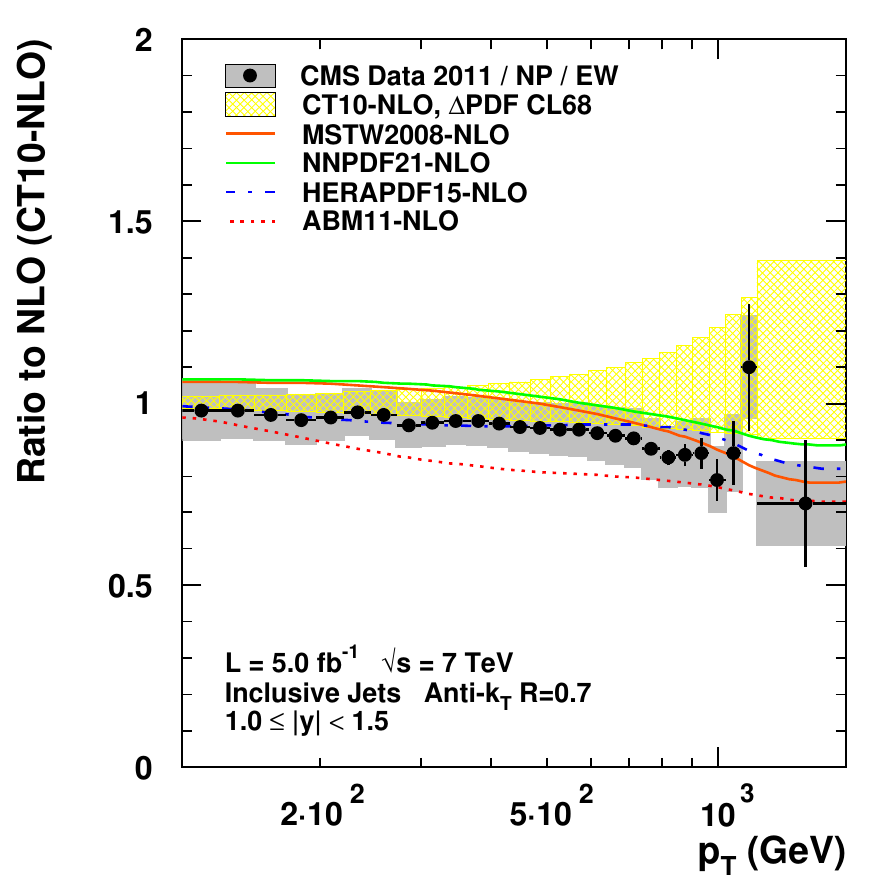}
  \caption{The theory--data comparison for the rapidity bins $\lvert y\rvert <
    0.5$, $0.5 \leq \lvert y\rvert < 1.0$, and $1.0 \leq \lvert y\rvert < 1.5$,
    illustrated in terms of ratios with respect to the NLO QCD
    prediction obtained with the \textsc{CT10-NLO} PDF set. On the
    left, only NP correction factors are taken into account in this
    ratio, and on the right also the EW corrections.}
  \label{fig:incJet:data123}
\end{figure}

\begin{figure}
  \centering
  \includegraphics[scale=0.75,trim=0 0 17
  0]{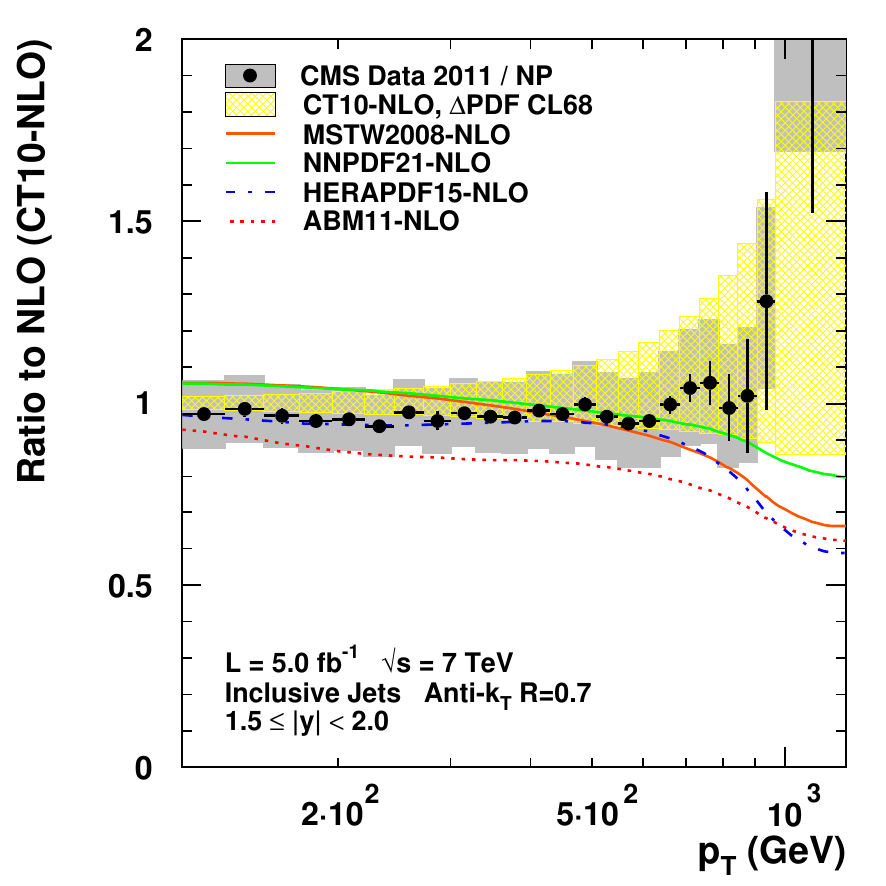}
  \includegraphics[scale=0.75,trim=52 0 0
  0,clip]{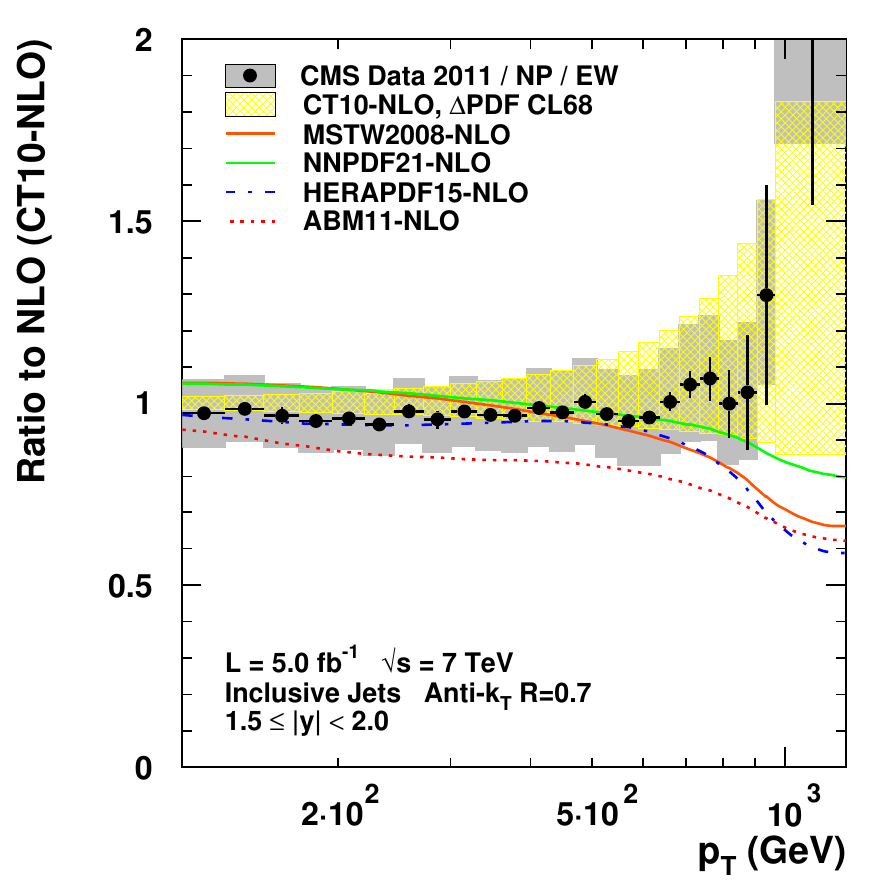}
  \includegraphics[scale=0.75,trim=0 0 17
  0]{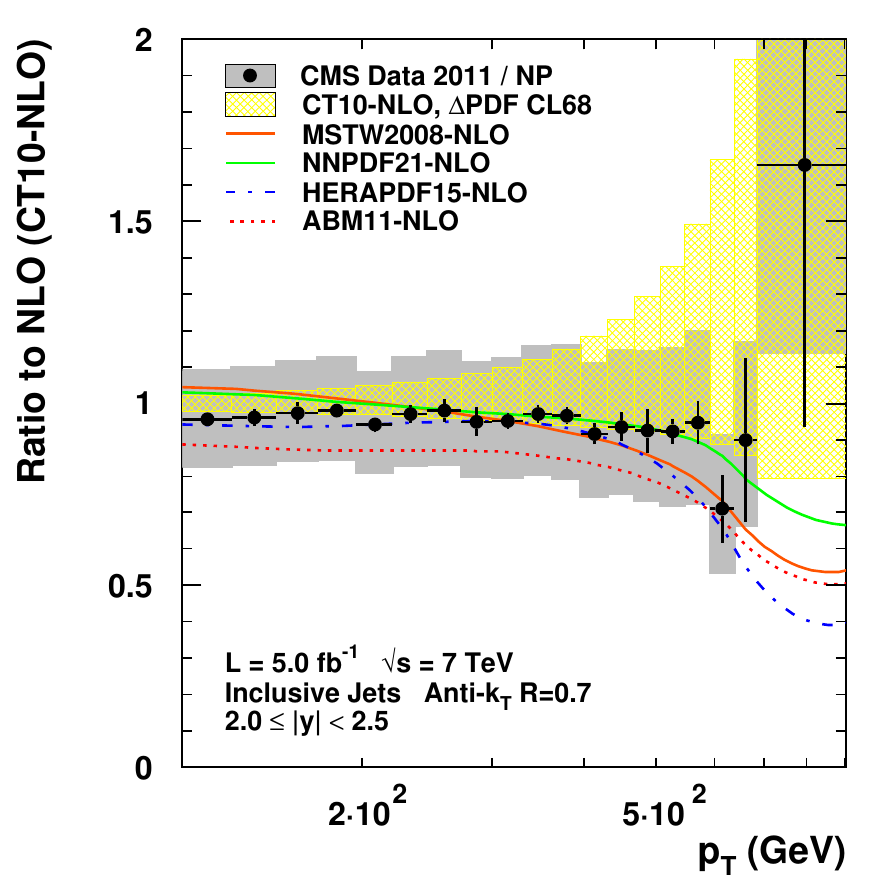}
  \includegraphics[scale=0.75,trim=52 0 0
  0,clip]{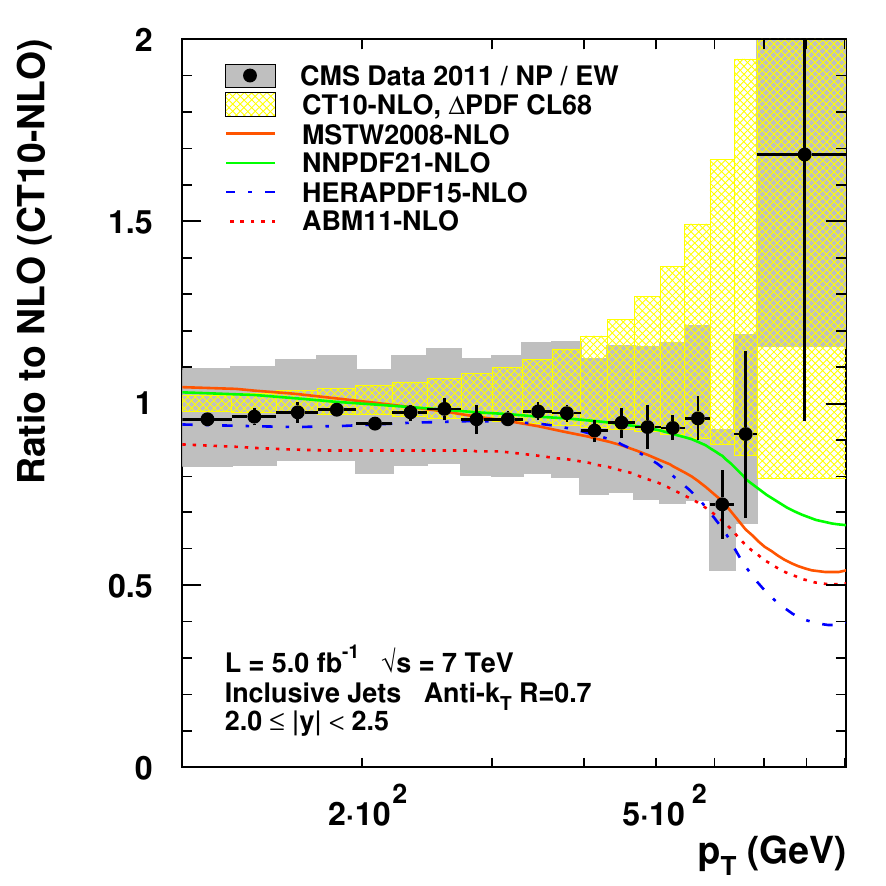}
  \caption{The theory--data comparison for the rapidity bins $1.5 \leq
    \lvert y\rvert < 2.0$, and $2.0 \leq \lvert y\rvert < 2.5$, illustrated in terms of
    ratios with respect to the NLO QCD prediction obtained with the
    \textsc{CT10-NLO} PDF set. On the left, only NP correction factors
    are taken into account in this ratio, and on the right also the EW
    corrections.}
  \label{fig:incJet:data45}
\end{figure}

The total correction factor to be applied to a leading-order QCD
prediction is then given by
\begin{equation}
  \label{eq:incJet:K_tot}
  K_\text{tot} = K_\text{QCD} \cdot K_\text{NP} \cdot K_\text{EW}\,,
\end{equation}
where
\begin{equation}
  \label{eq:incJet:K_EW}
  K_\text{EW} =
  K^\text{1-loop}_\text{weak} \cdot K^\text{tree}_\text{EW} \equiv
  (1+\delta^\text{1-loop}_\text{weak}) \cdot (1+\delta^\text{tree}_\text{EW}) \,.
\end{equation}
More details on the definition of the relative correction factors
$\delta^\text{1-loop}_\text{weak}$ and $\delta^\text{tree}_\text{EW}$
can be found in Ref.~\cite{Dittmaier:2012kx}, where an alternative
approach was employed in combining the two contributions, namely
$K_\text{EW}=(1+\delta^\text{1-loop}_\text{weak}+\delta^\text{tree}_\text{EW})$,
which differs, however, only through higher-order terms.
Figures~\ref{fig:incJet:data123} and~\ref{fig:incJet:data45} summarize the
comparison between theory and experiment separately for each rapidity
bin in terms of ratios with respect to the NLO QCD prediction obtained
with the \textsc{CT10-NLO} PDF set~\cite{Lai:2010vv}. In order to
examine the impact of the choice of different PDF sets, the plots
further include the ratios of the NLO QCD predictions derived with
various other PDF
sets~\cite{Martin:2009iq,Ball:2011mu,Aaron:2009aa,Alekhin:2012ig}.%
\footnote{Since photonic NLO corrections, which require a PDF
  redefinition via the factorization of photonic collinear
  singularities, are not yet included, $\mathcal{O}(\alpha)$-corrected PDF
  sets such as \textsc{NNPDF2.3QED} should not be used in combination
  with these EW corrections.} %
The dependence on the PDF set almost perfectly cancels in the relative
correction factor of the weak radiative corrections
$K^\text{1-loop}_\text{weak}$.  The tree-level part
$K^\text{tree}_\text{EW}$, on the other hand, exhibits a small
dependence on the choice for the PDFs, which is expected since these
contributions are sensitive to the relative size of the quark PDFs to
the gluon PDF\@.  For each rapidity bin the left plot displays the
data points that have been divided by the NP correction factor,
whereas in the right plot they are further divided by the EW
corrections. Differences are visible in particular at central rapidity
with $\lvert y\rvert < 1.0$. Because of the large statistical uncertainty of
the measurement in the relevant region no significant impact on fits
of PDFs and/or the strong coupling constant are expected.  This will
be different, once larger amounts of data become available at 8 or
even \Unit{13}{\UTeV} centre-of-mass energy.

\subsection{Summary and Outlook}
\label{sec:incJet:summary}

Electroweak corrections to the inclusive jet cross section have been
derived double-differentially in the jet transverse momentum $p_{\mathrm{T}}$ and
absolute rapidity $\lvert y\rvert$. They comprise corrections of the order
$\alphas^2\alpha_\mathrm{w}$ as well as contributions of
$\mathcal{O}(\alphas\alpha,\;\alpha^2)$ to the LO QCD prediction. This
calculation complements the previously published results of
Ref.~\cite{Dittmaier:2012kx}, where the dijet production process was
studied in detail. The Sudakov-type logarithms present in the weak
corrections become sizeable in the high-$p_{\mathrm{T}}$ tail of the
transverse-momentum distribution. They are negative throughout and can
amount to a $-7\%$ correction for $\lvert y\rvert < 0.5$ and $p_{\mathrm{T}} \approx
\Unit{2}{\UTeV}$ at the LHC for \Unit{7}{\UTeV} centre-of-mass energy. The tree-level EW
contributions are of the same generic size and positive at the LHC,
which induces large cancellations between the two, so that the
impact of the EW corrections to the inclusive jet cross section turns
out to be of moderate size.

Accounting for the sizeable QCD corrections at NLO and estimating
non-perturbative effects, which are small at high $p_{\mathrm{T}}$, the presented
theory predictions have been compared to data published by the CMS
experiment at the LHC for a centre-of-mass energy of
\Unit{7}{\UTeV}~\cite{Chatrchyan:2012bja}. For the given amount of \Unit{5}{\Ufb$^{-1}$} of
integrated luminosity the statistical uncertainties of the measurement
are too large at jet transverse momenta beyond \Unit{1}{\UTeV} to detect a
significant impact of the EW corrections. However, it will be most
interesting to repeat this comparison with data collected at higher
centre-of-mass energies and with more data. At \Unit{8}{\UTeV} for example, the
LHC delivered roughly four times the luminosity of the \Unit{7}{\UTeV} running
period.

Furthermore, the two contributions that enter the EW corrections
behave differently with respect to an increase of the collider
energy~\cite{Dittmaier:2012kx}, which can strongly affect the
accidental cancellations observed here. For a fixed value of the jet
transverse momentum, the scales appearing in the large logarithms that
are responsible for the bulk of the weak corrections will remain
mostly unchanged. This in turn means that for a given value of the
transverse momentum, the weak radiative corrections will not be
affected much by the collider energy. The tree-level contributions, on
the other hand, are strongly dependent on the parton luminosities, in
particular, the $q$ and $\bar{q}$ PDFs. For fixed values of the
transverse momentum, an increase in the collider energy corresponds to
probing smaller values of the momentum fractions $x$ in the
PDFs. Therefore, the tree-level contributions will decrease for higher
collision energies and the impact of the weak one-loop corrections are
expected to become more strongly pronounced.

\clearpage

\section{NNLO QCD and NLO EW Drell Yan background predictions for new gauge boson searches\footnote{U. Klein}}

\subsection{Introduction}

New gauge boson searches belong to the flagship measurements at the high energy frontier. In particular, searches for Z' and W' bosons are 'golden channels' due to their clean final state signatures and rather high rates. However, they require a precise modelling of the large neutral current (NC) and charged current (CC) Drell Yan backgrounds, which have  the same final state signatures.
For this kind of searches, 
very high invariant masses could and will be explored with the available
LHC c.m.s. energies of 8~TeV in Run~I and of anticipated 14~TeV in Run~II, respectively.
High invariant masses challenge the understanding of a kinematic region where NNLO  QCD corrections ${\cal{O}}(\alpha_s^2)$ and NLO electroweak corrections $\cal{O}(\alpha)$ become of similar size  (c.f. discussion at this workshop). In the following a few observations and practical, novel methods  related to the modelling of NNLO QCD and NLO EW high mass Drell Yan cross sections are summarised. Part of the methodologies have been developed in the context of a recent Z' search analysis~\cite{ATLAS-CONF-2013-017}. 

\subsection{Framework and tools}

Drell Yan production in proton-proton collisions  can be calculated with high precision and  over a wide kinematic range 
 up to next-to-next-to-leading-order (NNLO) in the strong coupling constant  using recent versions of the programs FEWZ~\cite{Gavin:2010az,Gavin:2012sy,Li:2012wn} and DYNNLO~\cite{Catani:2007vq,Catani:2009sm}. Both programs calculate vector boson production and decays with
full spin correlations and finite width effects. They are used widely by LHC experimentalist since they allow the application of important kinematic 
phase space requirements. In addition, powerful and fast programs like ZWPROD~\cite{Hamberg:1991} and VRAP~\cite{vrap} are used for total cross section calculations and estimates of e.g. factorisation and renormalisation scale uncertainties as well as to cross check the predictions of the various programs for high invariant masses. 
\begin{figure}[htbp]
\begin{center}
\includegraphics[trim = 0mm 93mm 0mm 90mm, clip, width=0.8\textwidth]{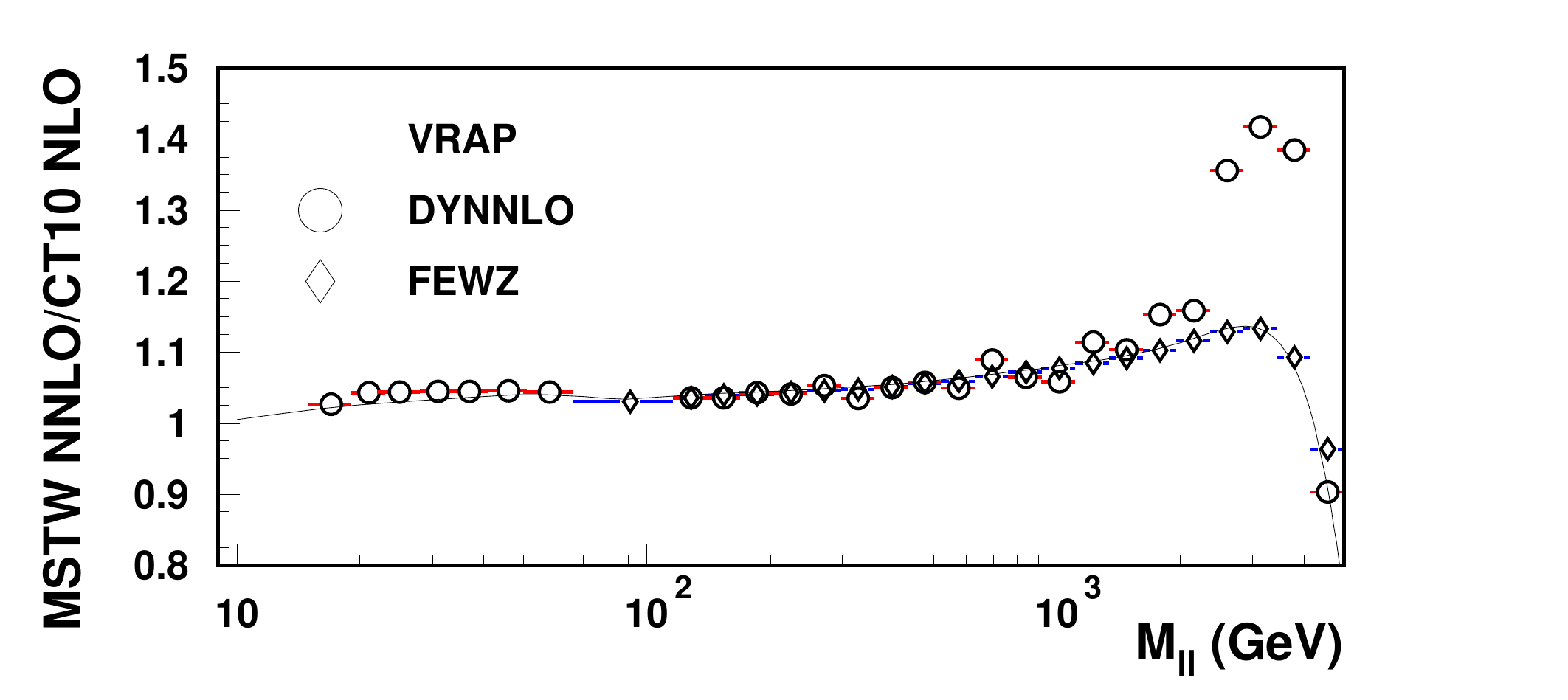}
\caption{\label{ewqcd:fig1} NNLO-to-NLO QCD k-factors for total NC Drell Yan cross sections at $\sqrt{s}=8$~TeV using VRAP 0.9, FEWZ 3.1.b2 and DYNNLO 1.3. See text and Ref.~\cite{lhapdfweb} for the details of the PDFs.}
\end{center}
\end{figure}
Such a cross check revealed that the total NNLO QCD cross section predictions by  FEWZ ~3.1.b2 and VRAP~0.9 are in excellent agreement  for invariant dilepton masses from 10 to 5000~GeV,  whereas the DYNNLO~1.3 version appears not to be suitable for off-resonant cross section calculations.
This observation is illustrated in Fig.~\ref{ewqcd:fig1} for NC DY channel using an LHC c.m.s. energy of $\sqrt{s}=8$~TeV. The calculation of the QCD NNLO-to-NLO k-factors shown uses the proton parton distribution functions MSTW2008nnlo at NNLO QCD and CT10 at NLO QCD (a similar observation was obtained for CC DY but is not shown here)~\cite{Grazzini:2013_private}.
 An updated version of DYNNLO~1.4 is now available which shows for total LO QCD NC DY cross sections in the mass range 60 to 5000~GeV an excellent agreement~\cite{Grazzini:2013_private}, however, more detailed cross checks are worthwhile also for CC DY production and specific phase space cuts also for the resonant region where the experimental precision is very high (less than a few percent). For example, the fiducial resonant W cross section predictions varied within 1\%  between FEWZ and DYNNLO (using the same SM and electroweak parameters for both programs)  as reported by ATLAS~\cite{Aad:2011dm} already in 2012.

\begin{figure}[htbp]
\begin{center}
\includegraphics[trim = 0mm 0mm 0mm 90mm, clip, width=0.495\textwidth]{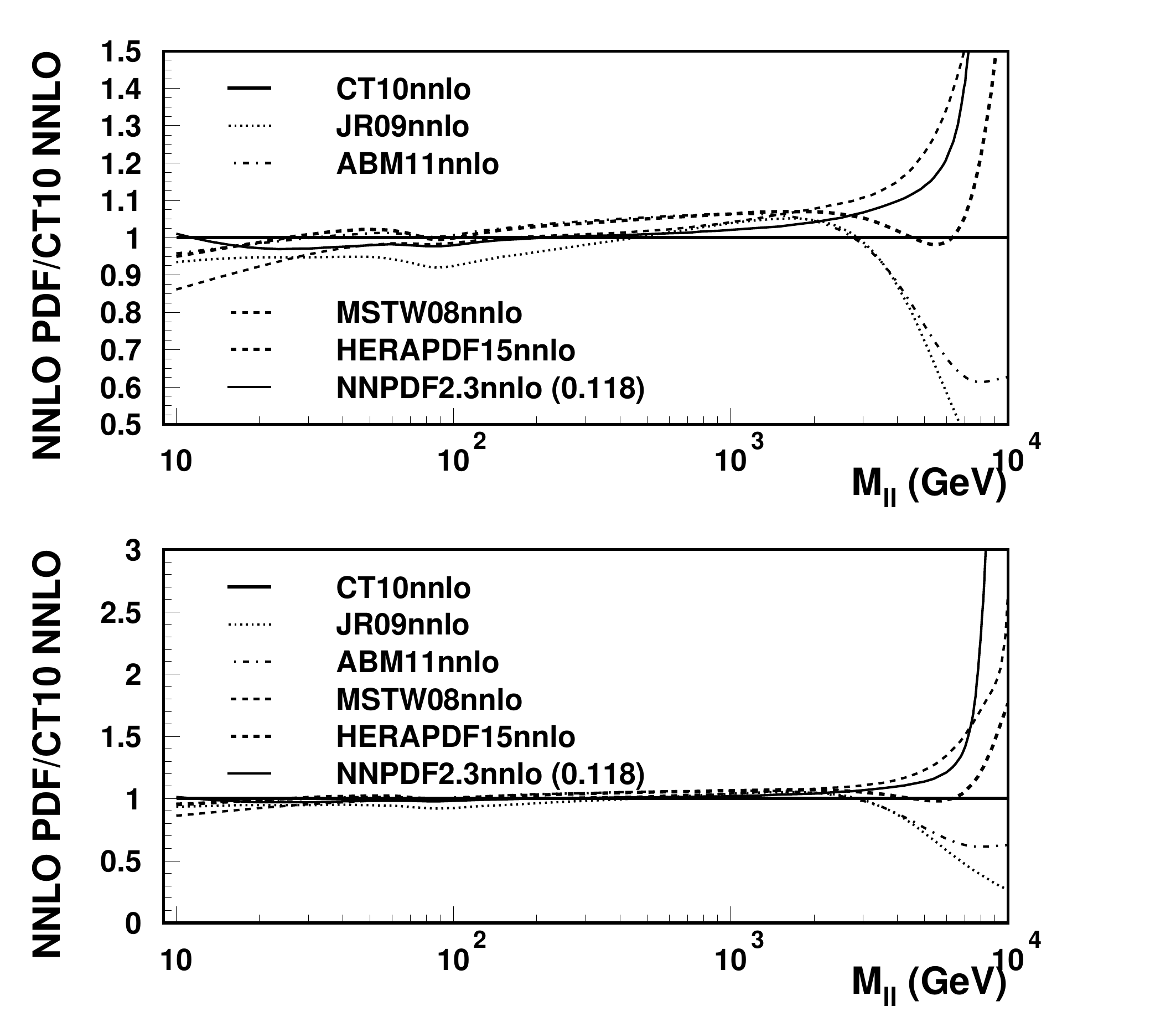}
\includegraphics[trim = 0mm 0mm 0mm 90mm, clip, width=0.495\textwidth]{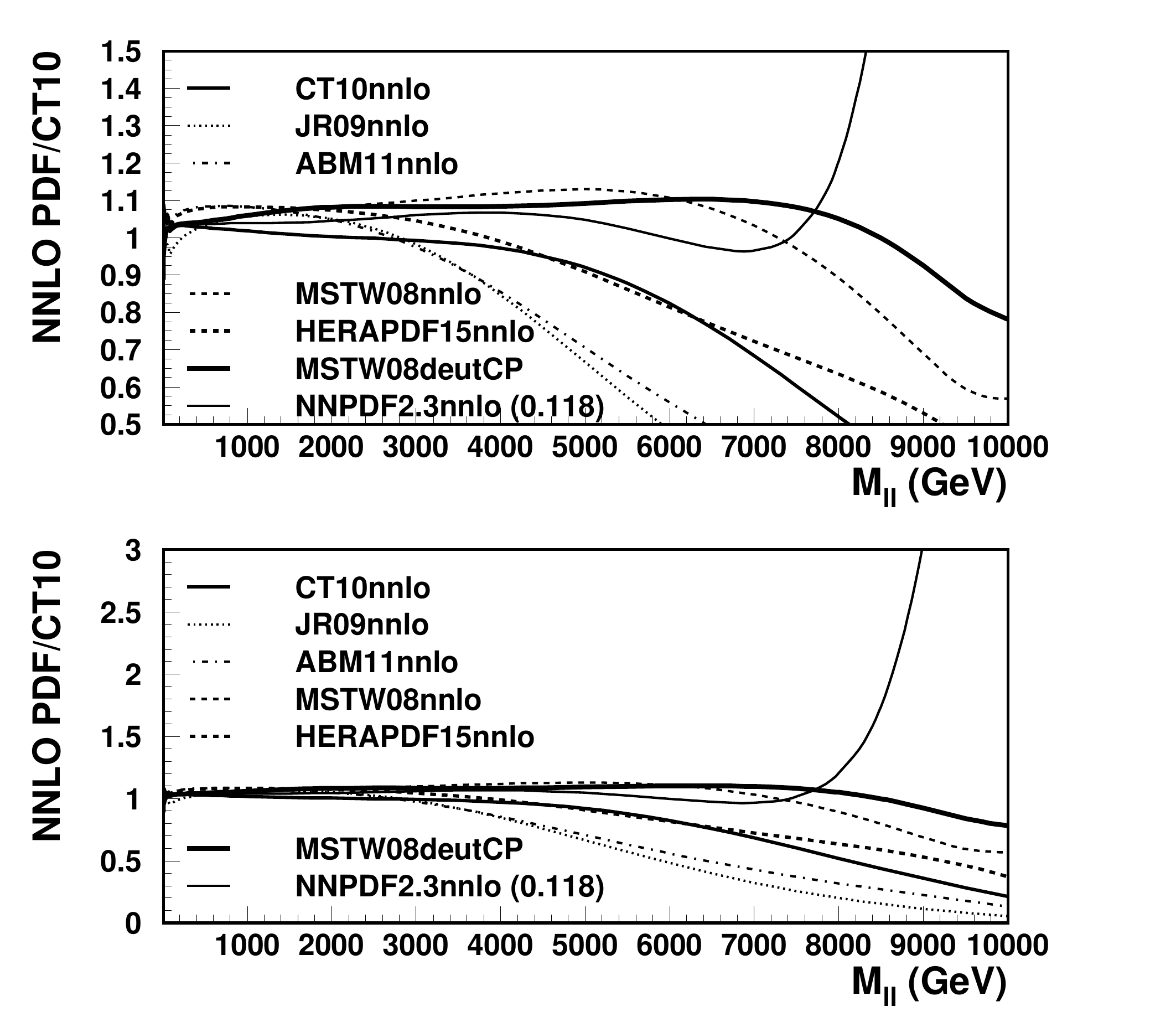}
\caption{\label{ewqcd:qcd14} Invariant mass dependence of the total NC Drell Yan production cross section predictions for modern NNLO PDFs w.r.t. the CT10nnlo PDF (left plots) and the QCD NNLO-to-NLO QCD k-factors w.r.t. to CT10 NLO QCD (right plots). The lower and upper  panels show the same quantities but with different  $y$-axis ranges from 0 to 3  and 0.5 to 1.5), respectively. The LHC energy is  $\sqrt{s}=14$ TeV. 
 Calculations are based on VRAP 0.9 and the NNLO (NLO) PDFs~\cite{lhapdfweb} as indicated in the legend.}
\end{center}
\end{figure}

The QCD NNLO-to-NLO k-factor for LHC Run~II at  $\sqrt{s}=14$ TeV are illustrated for NC Drell Yan production in Fig.~\ref{ewqcd:qcd14} (right). The QCD k-factors show a wide spread for invariant masses larger than about 4~TeV reflecting the lack of knowledge in the proton parton distribution functions at high Bjorken-$x$ values.
The corresponding spread in the predictions at NNLO QCD using modern PDFs with respect to the CT10nnlo PDF is shown in Fig.~\ref{ewqcd:qcd14} (left).

While the QCD k-factors are rather insensitive to the choice of the SM model inputs and the electroweak parameter scheme, care has to be taken for the absolute cross section predictions including those NLO electroweak corrections which are not addressed already in the unfolding of the experimental data. 
It is well know from the literature, see e.g.~\cite{Dittmaier:2009cr} for a brief introduction into the most commonly used electroweak schemes, that the $G_\mu$ electroweak scheme is well suited for Drell Yan production.

\begin{table}[htb]
  \begin{center}
\begin{tabular}{lrlr}
\hline
\hline
    $M_Z$                       & 91.1876 GeV                        & $V_{ud}$      & 0.97427  \\
    $\Gamma_Z$                  & 2.4949 GeV                            & $V_{us}$      & 0.22534  \\
    $\Gamma(Z \rightarrow ll)$  & 0.084 GeV                             & $V_{ub}$      & 0.00351  \\
    $M_W$                       & 80.385 GeV                        & $V_{cd}$      & 0.22520  \\
    $\Gamma_W$                  & 2.0906 GeV                            & $V_{cs}$      & 0.97344  \\
    $\Gamma(W \rightarrow l\nu)$ & 0.22727 GeV                          & $V_{cb}$      & 0.0412   \\
    $M_H$                       & 125     GeV                        & $V_{td}$      & 0.00867  \\
    $m_t$                       & 173.5   GeV                        & $V_{ts}$      & 0.0404   \\
    $G_F$                       & $1.1663787 \times 10^{-5}$ GeV$^{-2}$ & $V_{tb}$      & 0.999146 \\
    $\sin^2\theta_W$             & 0.22289722252391828                   &               &       \\
    $\alpha_G$                  & $7.56239563669733848 \times 10^{-3}$  &               &       \\
    $vec_{up}$                  & 0.40560740660288463                   &               &       \\
    $vec_{dn}$                  & -0.70280370330144226                  &               &       \\
    $vec_{le}$                  & -0.10841110990432690                  &               &       \\

    \hline
    \hline
    \end{tabular}
    \caption{Electroweak input parameters in the $G_\mu$ scheme for the NC and CC Drell Yan cross section calculations.}
    \label{ewqcd:tabEWpar}
      \end{center}
\end{table}
In the calculations presented here, the electroweak scheme is set to
the $G_\mu$ scheme according to the details outlined in
~\cite{Dittmaier:2009cr} and calculated by SANC~\cite{Bardin:2012jk}.
A summary of the values is given in Tab.~\ref{ewqcd:tabEWpar}.

\begin{figure}[htbp]
\begin{center}
\includegraphics[trim = 0mm 93mm 0mm 90mm, clip, width=0.8\textwidth]{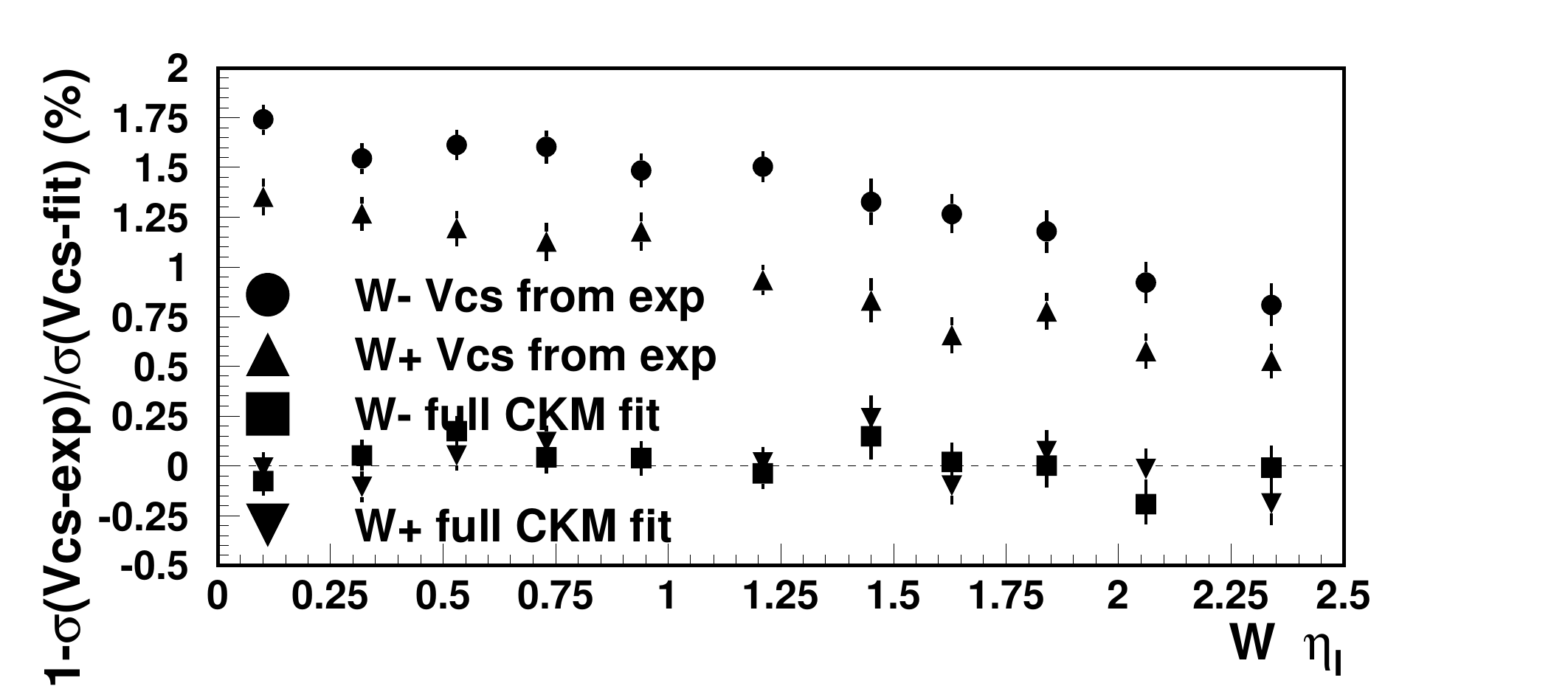}
\caption{\label{ewqcd:vcs} Effect of the choice of  ${\tt V_{cs}}$ on the fiducial pseudo-rapidity $\eta_\ell$ distributions of resonant W cross sections at $\sqrt{s}=7$~TeV. Calculations are performed with  DYNNLO 1.3 NLO QCD and the CT10 NLO PDF. }
\end{center}
\end{figure}

The CKM values are  taken from electroweak fits based on all precision observables as reported in~\cite{PDG}. Here the values of the CKM fit~\cite{PDG} are used. 
The pseudo-rapidity $\eta_\ell$ distribution of resonant single W$^\pm$ production is  sensitive to the choice of the value of ${\tt V_{cs}}$. Fig.\ref{ewqcd:vcs} illustrates 
the effect  of changing the fitted ${\tt V_{cs}}$ to the currently best experimentally known value of ${\tt V_{cs}}=1.006\pm0.023$~\cite{PDG} on the fiducial pseudo-rapidity $\eta_\ell$ distributions of resonant single W$^+$ and  W$^-$ production.
Since all other CKM matrix elements relevant for CC Drell Yan production are determined with high precision, the use of either the fitted or the measured values  (except the above described ${\tt V_{cs}}$ features) give the same results otherwise.

The free Standard Model input parameters are chosen  from the 2012 PDG~\cite{PDG}, except the partial leptonic decay width, $\Gamma(W \rightarrow l\nu)$.
The PDG value of  $\Gamma(W \rightarrow l\nu)=226.36$~MeV includes already higher order QCD and more importantly higher order electroweak effects and the use of this value would be then inconsistent with the set-up of higher order electroweak programs like e.g. SANC~\cite{Bardin:2012jk,Bondarenko:2013nu}. 
To get a consistent evaluation of higher order electroweak effects (except QED FSR) also on the  W decay kinematics, thus the LO  partial width $\Gamma(W \rightarrow l\nu)=227.27$~MeV has to be used in the QCD and EW calculations.

\subsection{Combining NNLO QCD with NLO EW corrections}

In Drell Yan production, the by far dominant part of the higher order electroweak corrections is the final state QED radiation (QED FSR) from the final state leptons.
The LHC Drell Yan Monte Carlo simulations often use PHOTOS~\cite{Golonka:2006photos} as an "afterburner"  for the modelling of QED FSR, and the experimental data are usually unfolded for QED FSR effects. PHOTOS and SANC QED FSR agree quantitatively very well~\cite{Arbuzov:2012dx}. The SANC program can be thus used to calculate the electroweak corrections excluding QED FSR for both NC and CC Drell Yan.
Those missing  HO EW terms include contributions from initial state photon radiation (ISR QED), 
electroweak loop corrections and initial and final state photon interferences. 
The recent FEWZ 3.1 versions allow for the NC Drell Yan channel to select the $G_\mu$ electroweak scheme and and the simultaneous calculation of NLO electroweak corrections~\cite{Li:2012wn} .
To enable the calculation of NLO EW effects except QED FSR and to match thus QCD and EW predictions to experimental data the following EW control flags had been introduced into FEWZ~3.1.a3 (and maintained in the updated versions), where FEWZ reads the control flag ({\tt EWflag}) and uses it as a binary number~\cite{Li:2012_private}:
\begin{itemize}
{\tt 
\item EWflag = 0 = (0000)\_2 means nothing off (Weak, ISR$*$FSR, ISR, FSR all on)
\item EWflag = 1 = (0001)\_2 means FSR off
\item EWflag = 2 = (0010)\_2 means ISR off
\item EWflag = 4 = (0100)\_2 means ISR$*$FSR off
\item EWflag = 8 = (1000)\_2 means Weak off
\item EWflag = 7 = (0111)\_2 means ISR$*$FSR, ISR and FSR all off (only Weak is on)
}
\end{itemize}
\begin{figure}[htbp]
\begin{center}
\includegraphics[trim = 0mm 0mm 0mm 90mm, clip, width=0.75\textwidth]{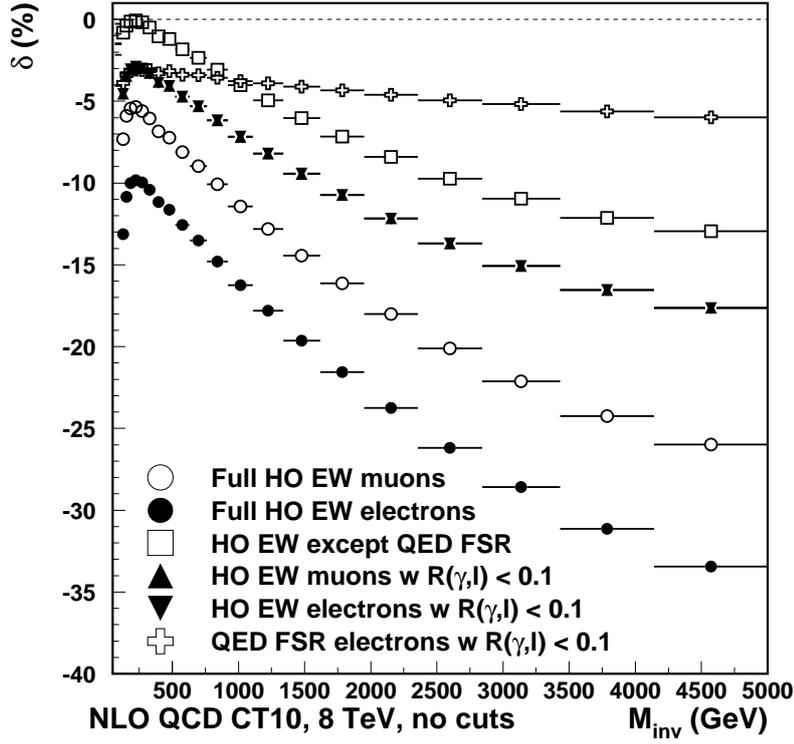}
\caption{\label{ewqcd:ewZ} 
Invariant mass dependence of  various higher order EW contributions ($\delta$ in \%)  for NC Drell Yan production at $\sqrt{s}=8$~TeV. Shown are the full NLO EW corrections for electrons (full circles) and muons (open circles), the full NLO EW corrections using a recombination of the FSR photon and lepton with in a cone of 0.1 (w R$(\gamma,l)<0.1$) for electrons (full triangles top down) and muons (full triangles top up), the NLO EW corrections except QED FSR (open squares), the QED FSR corrections using a recombination of the FSR photon and electron (open crosses) with in a cone of 0.1 (w R$(\gamma,l)<0.1$).
 Calculations are based on FEWZ 3.1.b2 NLO QCD and NLO EW using the CT10 NLO PDF.}
\end{center}
\end{figure}
Various NLO QCD and NLO EW calculations have been performed for NC Drell Yan over a wide invariant mass range, where each mass bin has been calculated separately to avoid biases in the calculations, using the CT10 NLO PDF and the {\tt EWflag} options as described above in FEWZ 3.1.b2. The HO EW corrections show a strong invariant mass dependence as illustrated in Fig.~\ref{ewqcd:ewZ} for the NC Drell Yan production at $\sqrt{s}=8$~TeV and the expected large QED FSR contributions.  
The significant reduction of the QED FSR contributions  using a recombination of the FSR photon and lepton with in a cone of 0.1 (w R$(\gamma,l)<0.1$), usually called "dressed leptons" by experimenters,  is shown in Fig.~\ref{ewqcd:ewZ} as well. The ATLAS strategy of dressing leptons has been widely discussed within the LPCC EW working group initiated by Ref.~\cite{Belloni_Klein_2012}. It is important to note that any lepton dressing can be performed on Monte Carlo generator level only,
i.e. experimental measurements can be unfolded to various levels of QED FSR corrections based on the QED FSR code and EW parameter settings as implemented in the used Monte Carlos. 

FEWZ LO QCD based HO EW correction calculations, except QED FSR (FEWZ {\tt EWflag=1}), and corresponding SANC calculations, performed  in the $G_\mu$ scheme, agree very well for  NC Drell Yan process. 
Hence in the following the FEWZ results are used to demonstrate  an improved understanding of the application of HO EW corrections to (N)NLO QCD predictions relevant for an estimate of an EW (except QED FSR) systematic uncertainty for very high invariant masses. The method developed here has been  transferred successfully to CC Drell Yan production where no combined NNLO QCD and NLO EW code is available yet, and the results of various external programs need to be combined in a consistent way.

Various methodologies of combining HO QCD and HO EW effects for DY production have been discussed over the past decade, see e.g. also this and other Les Houches workshop contributions and references therein. From the viewpoint of an experimenter performing new gauge boson searches and SM DY measurements, the practical question needed an answer  how mass dependent EW systematic uncertainties could be estimated when further mixed ${\cal{O}}(\alpha \alpha_s)$ contributions are not calculated yet. 

Using fixed order LO QCD calculations for a given EW parameter scheme as the baseline, $\sigma_{LO\_QCD}$, there are mainly two methodologies which are called here in short "factorised approach", Eq.~\ref{ewqcd:eqf}, and  "additive approach", Eq.~\ref{ewqcd:eqa} to construct a combined NNLO QCD and NLO EW DY cross section, $\sigma_{NNLO\_QCD+NLO\_EW}$. In the context of the here discussed  new gauge boson searches, "EW" refers here always to the HO EW corrections except QED FSR (also sometimes called "missing HO EW corrections", $\delta_{miss}$, since those corrections are missing in the experimental Monte Carlos while QED FSR is included already).\\
Factorised approach:{
\begin{eqnarray}\label{ewqcd:eqf}
\sigma_{NNLO\_QCD+NLO\_EW}  & = & k_{QCD} \times k_{EW} \times \sigma_{LO\_QCD} \\
& k_{QCD } = & \frac {\sigma_{NNLO\_QCD} }{\sigma_{LO\_QCD}} \\
& k_{EW} = & \frac {\sigma_{NLO\_EW,LO\_QCD} }{\sigma_{LO\_QCD}}. 
\end{eqnarray}
}
Additive approach:{
\begin{eqnarray}\label{ewqcd:eqa}
\sigma_{NNLO\_QCD+NLO\_EW} & = & \sigma_{NNLO\_QCD} + \Delta \sigma_{LO\_QCD+NLO\_EW} \\
	& = & \sigma_{NNLO\_QCD} \big ( 1 + \frac{\Delta \sigma_{LO\_QCD+NLO\_EW} }{\sigma_{NNLO\_QCD} } \big ). 
\end{eqnarray}
}
The factorised approach, Eq.~\ref{ewqcd:eqf}, assumes that the HO EW corrections are the same for all orders of QCD, and thus can be determined based on LO QCD and then transferred to any higher order in QCD. 
The additive approach, Eq.~\ref{ewqcd:eqa}, is e.g. followed in the NC DY FEWZ code~\cite{Li:2012wn} for the combination of HO QCD and EW corrections. This approach assumes that the EW except QED FSR corrections  are mainly additive in their nature and the same term needs to be added for all orders of QCD, thus  the relative fraction of HO EW corrections for each order of QCD is changing. 
\begin{figure}[hbtp]
\begin{center}
\includegraphics[trim = 0mm 0mm 0mm 0mm, clip, width=0.82\textwidth]{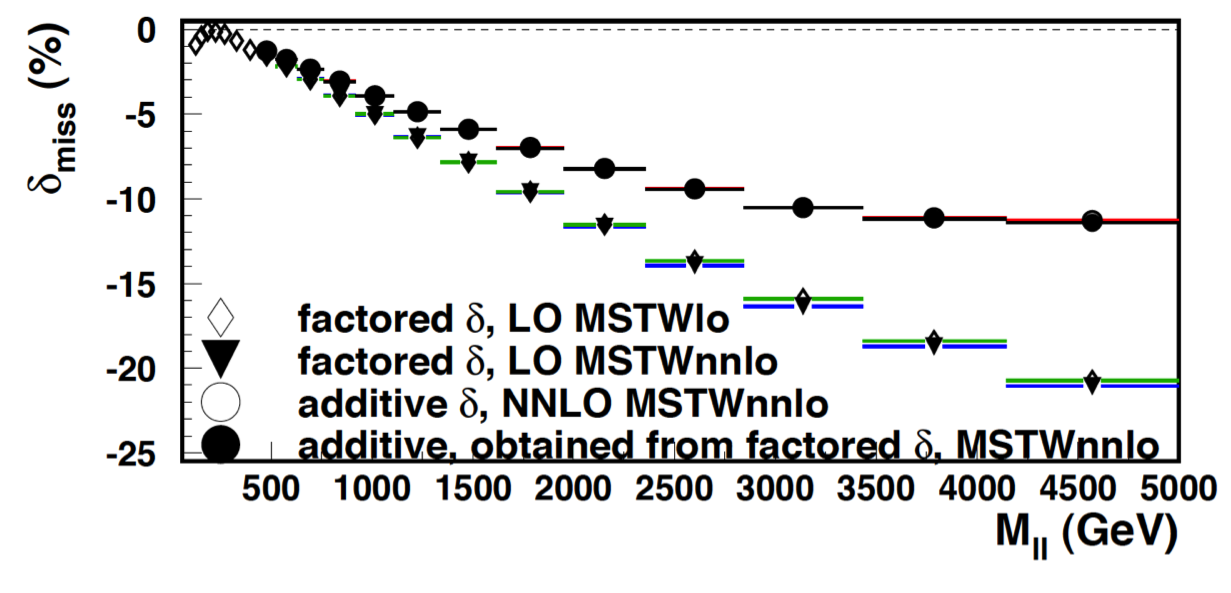}
\caption{\label{ewqcd:ewqcd} HO EW except QED FSR ($\delta_{miss}$ in \%) corrections for NC Drell Yan production. The  results of the factorised  approach based on LO QCD  and the MST2008lo LO PDF (open diamonds) and MSTW2008nnlo NNLO PDF (full triangles on top)  are shown. Using the MSTW2008nnlo PNNLO PDF, the results of a full, one-step NNLO QCD and NLO EW (open circles) and the additive approach according to Eq.~\ref{ewqcd:eqa} (full circles) are shown. Calculations are based on FEWZ 3.1.b2 and $\sqrt{s}=8$~TeV.}
\end{center}
\end{figure}
Here a careful analysis has been performed to understand the transfer of the additive EW term to all orders of QCD. It has been found that consistent results can only be achieved if the term 
\begin{equation}\label{ewqcd:eqdelta}
\Delta \sigma_{LO\_QCD+NLO\_EW} = \sigma_{LO\_QCD+NLO\_EW} - \sigma_{LO\_QCD} 
\end{equation}
is calculated at LO QCD with exactly the same PDF as the wanted highest order QCD result. For example, if a NNLO QCD prediction using the MSTW2008nnlo PDF is aimed for, then the additive EW corrections term can be calculated with high precision based on LO QCD but using the MSTW2008nnlo PDF. Similarly e.g. for a NLO QCD prediction using the CT10 PDF, this would require the HO EW calculation done based on LO QCD using CT10. The results of this study are shown in Fig.~\ref{ewqcd:ewqcd} where the mass dependence of the HO EW except QED FSR corrections are shown for the factorised and the additive approach, and the excellent agreement of the resulting NNLO QCD and NLO EW except QED FSR predictions using either directly FEWZ 3.1.b2 or the strategy according to Eq.~\ref{ewqcd:eqa}. The formalism described above is robust and gives a prescription also applicable for CC DY processes where the NNLO QCD and NLO EW calculation   rely on different external programs for the QCD and the EW part.
Also indicated in Fig.~\ref{ewqcd:ewqcd} is the expected weak PDF dependence of the HO EW corrections.
For higher invariant masses the discrepancies between the additive and factorised approaches are significant, and the results of the two approaches are used to get an estimate of the mass dependent systematic uncertainty for the HO EW except QED FSR  corrections for NC and CC Drell Yan processes.
Since roughly speaking, the factorised approach maximises potential mixed ${\cal{O}}(\alpha \alpha_s)$ contributions but  with its sign is unknown,  the following strategy has been suggested for each invariant mass bin:
The central value of the combined NNLO QCD and NLO EW except QED FSR prediction is taken from the additive approach while the difference to the factorised approach results is taken as a double sided systematic uncertainty estimate.  The resulting mean HO EW except QED FSR corrections with those symmetric uncertainties are shown in Fig.\ref{ewqcd:ew14} (lower plot) for the expected LHC Run II at $\sqrt{s}=14$~TeV. Also illustrated in Fig.~\ref{ewqcd:ew14} (upper plot) is the rather weak energy dependence for the HO EW except QED FSR corrections.
\begin{figure}[htbp]
\begin{center}
\includegraphics[trim = 0mm 0mm 0mm 90mm, clip, width=0.8\textwidth]{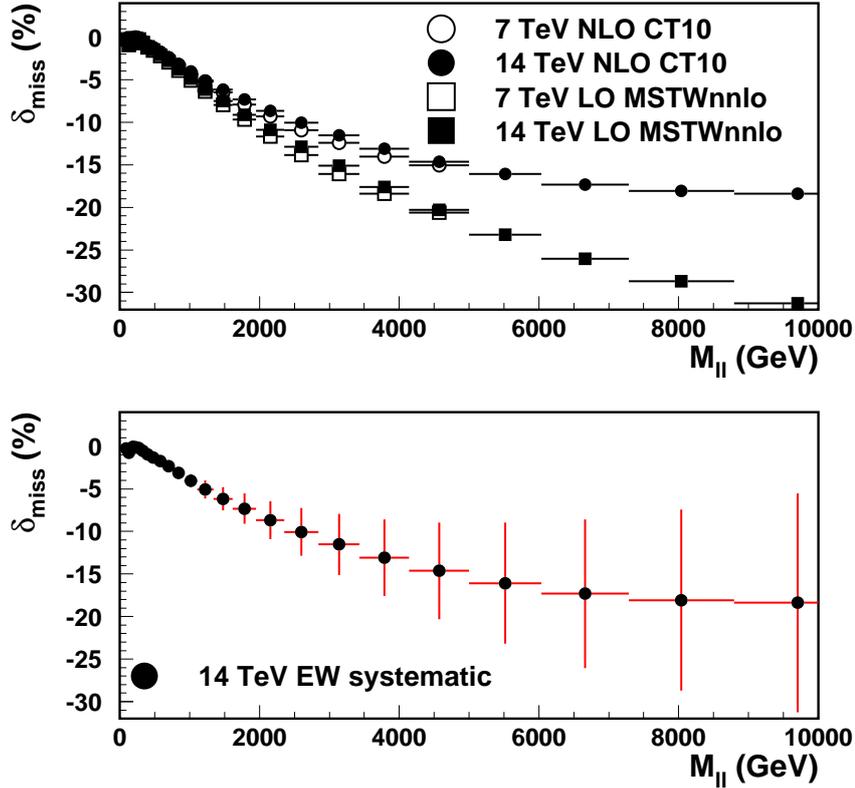}
\caption{\label{ewqcd:ew14} HO EW except QED FSR ($\delta_{miss}$ in \%) corrections for NC Drell Yan production. The  results of the factorised (additive) approach based on LO QCD (NLO QCD) for $\sqrt{s}=7$~TeV are shown in the upper plot with empty squares (empty circles) and for 14 TeV with full squares (full circles). An estimate of the EW except QED FSR systematic uncertainty, see text, is shown in the lower plots for $\sqrt{s}=14$~TeV.
All calculations are based on FEWZ 3.1.b2  and the PDFs as indicated in the legend.}
\end{center}
\end{figure}

\subsection{Matching Monte Carlo programs and external calculations}

\begin{figure}[htp]
\begin{center}
\includegraphics[trim = 0mm 90mm 0mm 90mm, clip, width=0.86\textwidth]{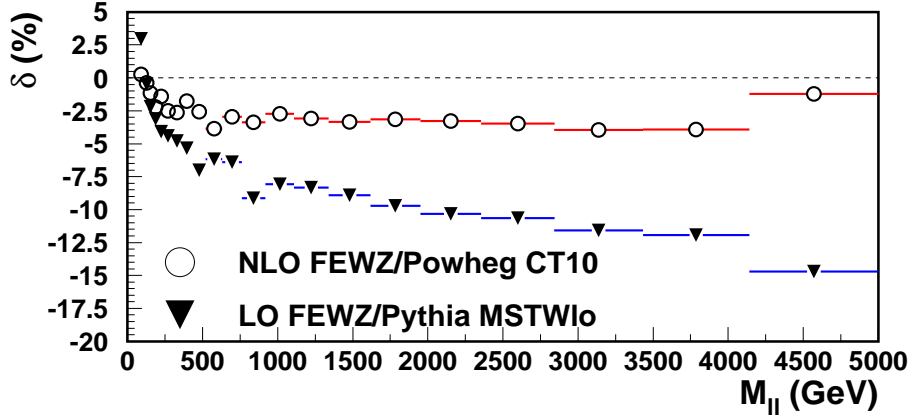}
\caption{\label{ewqcd:match} Deviation  ($\delta$ in \%)  between external NLO (open circles ) and LO (full triangles on top) QCD predictions using FEWZ 3.1.b2 with the well-defined $G_{\mu}$ scheme and currently used NLO (POWHEG) and LO (PYTHIA) Monte Carlo generators for NC Drell Yan production at $\sqrt{s}=8$~TeV and over a wide kinematic range in the invariant mass.}
\end{center}
\end{figure}

The  methodologies discussed in the previous chapter work reliably for total cross sections and well defined variables like the invariant mass or the pseudo-rapidity of the final state leptons, but fail e.g for  variables like the $p_T(\ell)$ or  $p_T(Z,W)$ which rely on a modelling of soft gluon resummation and parton shower effects usually taken care of in Monte Carlo programs.
While the LHC Monte Carlo programs are powerful tools for the description of complex QCD processes, they are not very well defined w.r.t. the electroweak part and the EW parameter scheme used.
Here, a study has been performed to understand the matching of a well-defined external QCD calculation using the $G_{\mu}$ scheme with parameters according to Tab.~\ref{ewqcd:tabEWpar} and currently PDG guided EW parameter settings as used in LHC Monte Carlo generations. Fig.~\ref{ewqcd:match}
shows the deviations between external NLO (LO) QCD predictions using FEWZ 3.1.b2 with the well-defined $G_{\mu}$ scheme and currently used NLO (POWHEG) and LO (PYTHIA) Monte Carlo generators for NC Drell Yan production at $\sqrt{s}=8$~TeV. The observations in Fig.~\ref{ewqcd:match} illustrates that depending on the Monte Carlo generation used an additional "matching" could be needed to normalise the Monte Carlo to the best theoretical available cross section prediction. It also is apparent from Fig.~\ref{ewqcd:match} that this could be an issue relevant for high invariant masses specifically.

For high mass new gauge boson searches thus the following novel methodology has been developed to 
transfer the best external NNLO QCD and NLO EW except QED FSR knowledge to LHC Drell Yan Monte Carlo samples interfaced to PHOTOS for QED FSR (and using SANC for the HO EW except QED FSR corrections in a consistent way; other QED FSR models may require other HO EW except QED FSR calculations for high invariant masses to maintain overall consistency of the full HO EW corrections) :
\begin{eqnarray}\label{ewqcd:eqmatch}
\sigma_{NNLO\_QCD+NLO\_EW} & = & k_{\rm fit} \times \sigma_{MC} \\
& k_{\rm fit} = & \frac {\sigma_{NNLO\_QCD+NLO\_EW} }{\sigma_{MC}}. \label{ewqcd:eqmatch1}
\end{eqnarray}
The factor $k_{\rm fit}$  is obtained by a mass-dependent fit of the ratios of total cross sections per invariant mass bin as described in in Eq.~\ref{ewqcd:eqmatch1}. Thus this factor automatically corrects for any possible mismatches in the EW parameter scheme and "forces" the mass-dependent Monte Carlo Drell Yan cross section to the "best" NNLO QCD and NLO EW knowledge. Further systematic uncertainties due to the knowledge of the PDFs and the EW except QED FSR corrections can be then assigned.
The fit function of 
$k_{\rm fit}(M_{inv})$ and its uncertainties can be conveniently applied as weights to each Monte Carlo event and the  "best" NNLO QCD and NLO EW knowledge could be propagated to more complex variables with kinematic cuts. 

\subsection{Photon-induced background contributions}
 
 The subject of photon-induced dilepton  contributions has been discussed in the theory community since a decade (c.f. other contributions to this workshop). Recently this subject has been revived from the experimenter's point of view~\cite{Klein_2012} showing the potentially large size of such contributions for off-peak (low and high mass) NC Drell Yan production at the LHC. Furthermore, indications are found in a recent Standard Model high mass Drell Yan analysis performed by ATLAS that LHC data may include such backgrounds~\cite{Aad:2013iua}. 
 In Ref.~\cite{Aad:2013iua} also the radiation of real (on-shell) W and Z bosons has been quantified for the experimental conditions following the procedure outlined in Ref.\cite {Baur:2006sn}. Using the same technique, the contributions due to real W and Z boson radiation have been also calculated for the 2012 Z' analysis~\cite{ATLAS-CONF-2013-017}, but those background contributions are much smaller than the photon-induced background contributions and hence not discussed here further.
 \begin{figure}[htbp]
\begin{center}
\includegraphics[trim = 0mm 93mm 0mm 90mm, clip, width=0.75\textwidth]{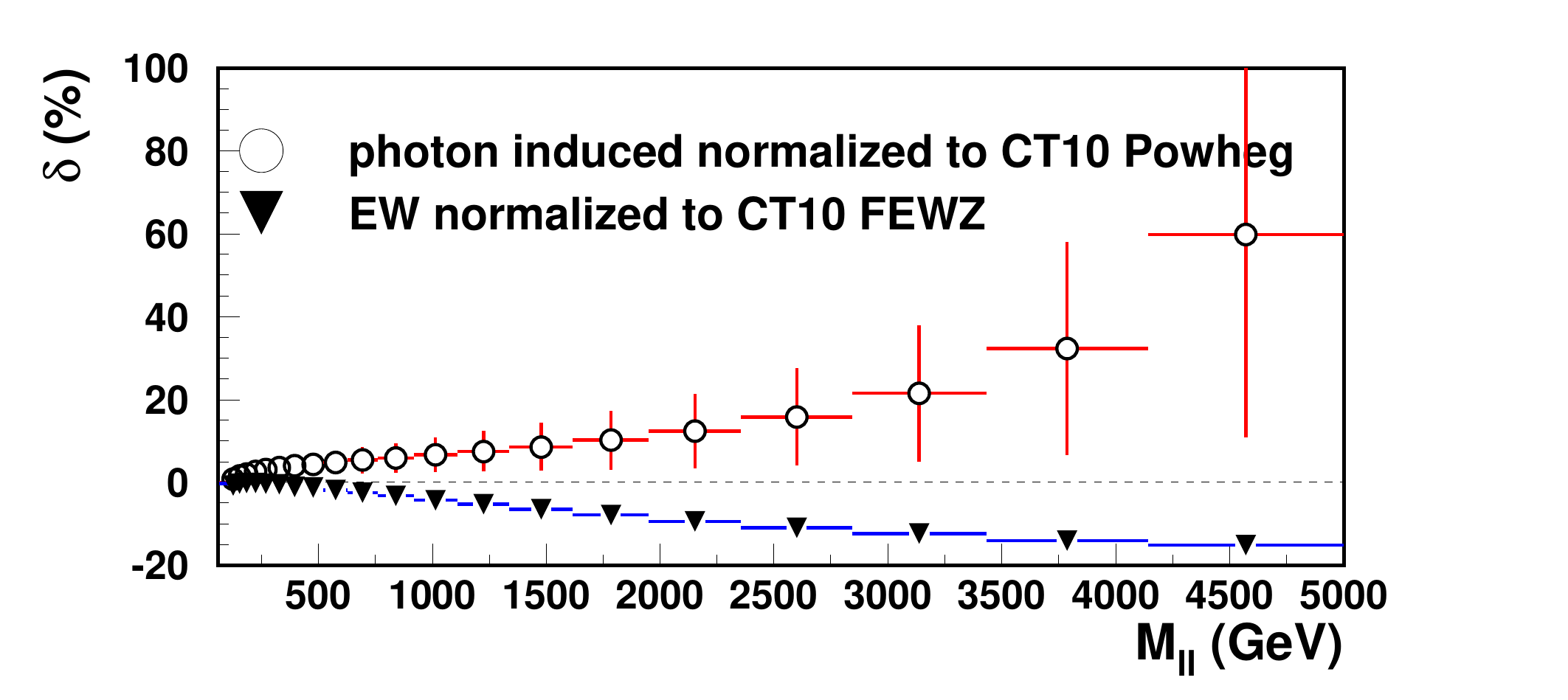}
\includegraphics[trim = 0mm 93mm 0mm 90mm, clip, width=0.75\textwidth]{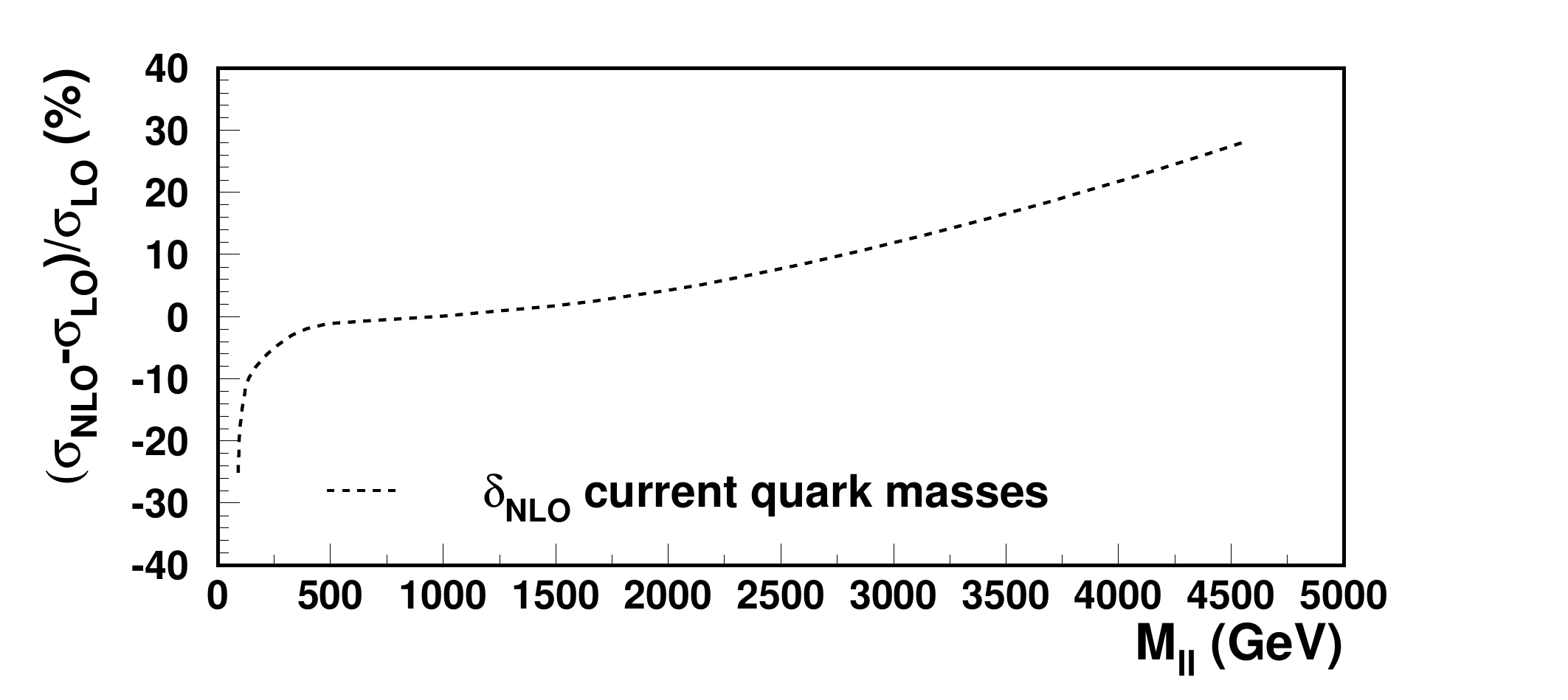}
\caption{\label{ewqcd:pinlo} Photon induced background contributions calculated with $p_{T,\ell}>25$~GeV and $\eta_{\ell}<2.5$  MRST2004QED photon PDF and $\sqrt{s}=8$~TeV over a wide invariant mass range. Shown are the leading order photon induced dilepton contribution using the mean of the two available MRST2004qed predictions as the central values and the deviations between the two predictions and the central value as an uncertainty estimate (upper plot). The LO photon induced contribution is displayed as a fraction of a NLO POWHEG NC DY prediction. Overlaid is also the HO EW except QED FSR correction including its systematic uncertainty estimate as described in the text.  Next-to-leading order photon induced dilepton production contribution w.r.t. to the dominant LO photon-induced cross sections over a wide invariant mass range (lower plot). Calculations are performed with MCSANC~\cite{Bondarenko:2013nu}. using the nominal MRST2004qed PDF.}
\end{center}
\end{figure}
 Since our knowledge of photon-induced processes is very poor,
 only rough estimates are possible using the MRST2004QED PDF~\cite{Martin:2004dh}. Here a modified version of the MRST2004qed PDF is used which offers two parameterisations~\cite{Martin:2004dh_private}, i.e. reflecting the sensitivity to the choice of the quark mass, either current quark masses as used for the nominal parameterisation or constituent quark masses as used for an alternative parameterisation. 

First estimates of the size of photon-induced contributions have been performed at parton level only due to the lack of full Monte Carlo simulations at that time. Since dilepton production $\gamma\gamma \to \ell\ell$ is an additional background which can be significant at high invariant masses, it is important that the fiducial phase space requirements are applied. 
The general  strategy is then that this background contribution is quantified separately and added to the theory prediction including its model uncertainty in data-theory comparisons, see e.g.  Ref.~\cite{Aad:2013iua}. This strategy allows for a re-analysis of the data in the future if  the knowledge about this background may have been improved.

The photon induced background contributions for kinematic requirements of  $p_{T,\ell}>25$~GeV and $\eta_{\ell}<2.5$  are shown in Fig.~\ref{ewqcd:pinlo} (upper plot) for a LHC c.m.s. energy of $\sqrt{s}=8$~TeV over a wide invariant mass range.  The potential contributions could sizeable and may reach up to 50\%  - 100 \% of the fiducial NC DY cross section at invariant masses of 4 to 5~TeV. Such background contributions may then also significantly reduced expected HO EW effects (also shown in Fig.~\ref{ewqcd:pinlo} upper plot). NLO, quark-photon initiated dilepton production contributions are expected to be negative at lower invariant masses but may add even more backgrounds at higher masses, see Fig.~\ref{ewqcd:pinlo} (lower plot). The current procedure to estimate an uncertainty due this background relies on a simple mean between the two available MRST2004qed predictions at LO to avoid single-sided systematic uncertainties. It was also lively discussed at this workshop that LHC experiments should aim for a measurement of this background employing specific phase space regions.

\subsection{NNLO QCD scale uncertainties}
\begin{figure}[htbp]
\begin{center}
\includegraphics[trim = 0mm 93mm 0mm 90mm, clip, width=0.95\textwidth]{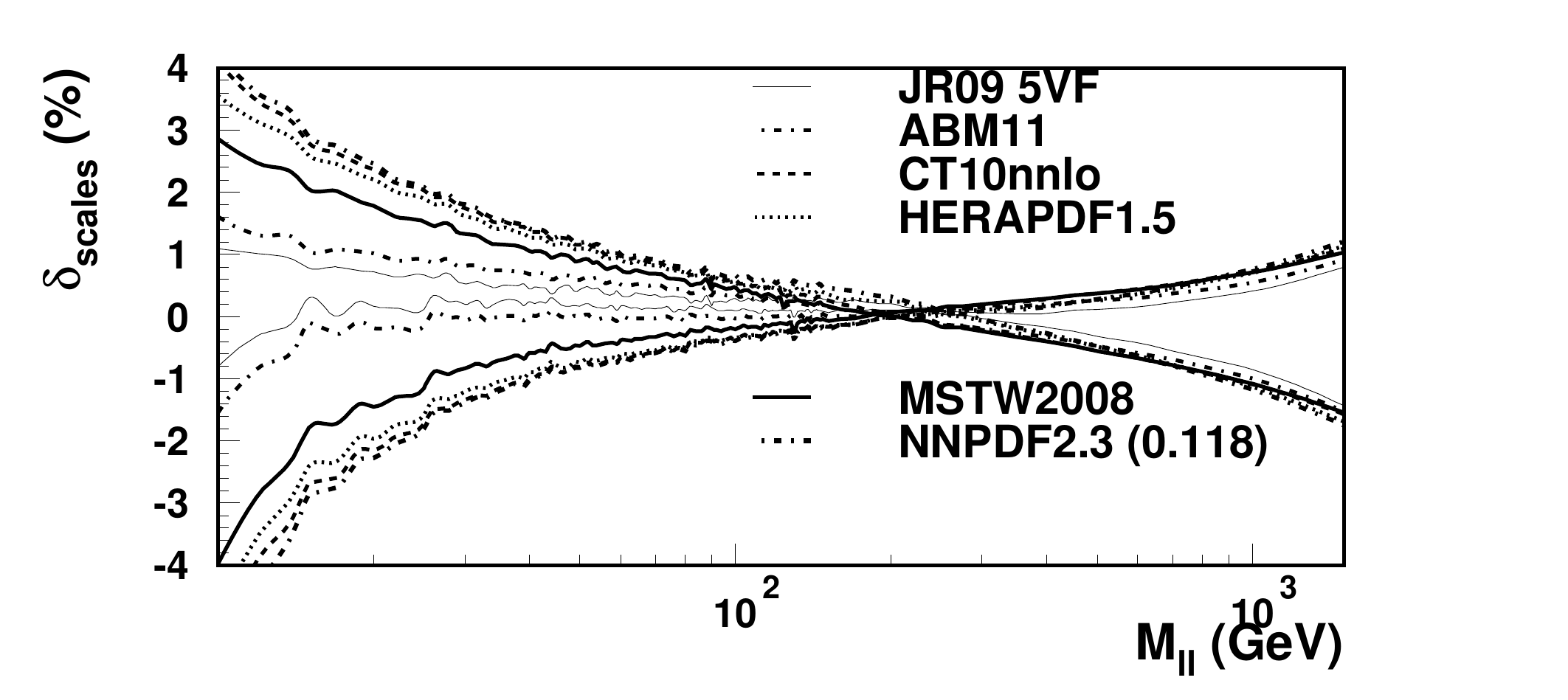}
\caption{\label{ewqcd:ncscale} Effect on the total NC Drell Yan NNLO QCD cross section predictions due to the  change of the renormalisation and factorisation scales simultaneously by a factor 2 or 1/2 for an invariant mass range of 10 to 2000 GeV. See Ref.~\cite{lhapdfweb} for the details of the NNLO QCD PDFs; the NNPDF calculations is obtained for $\alpha_s=0.118$. Calculations are based on VRAP 0.9 and $\sqrt{s}=7$~TeV.}
\end{center}
\end{figure}
\begin{figure}[htbp]
\begin{center}
\includegraphics[trim = 0mm 0mm 0mm 90mm, clip, width=0.95\textwidth]{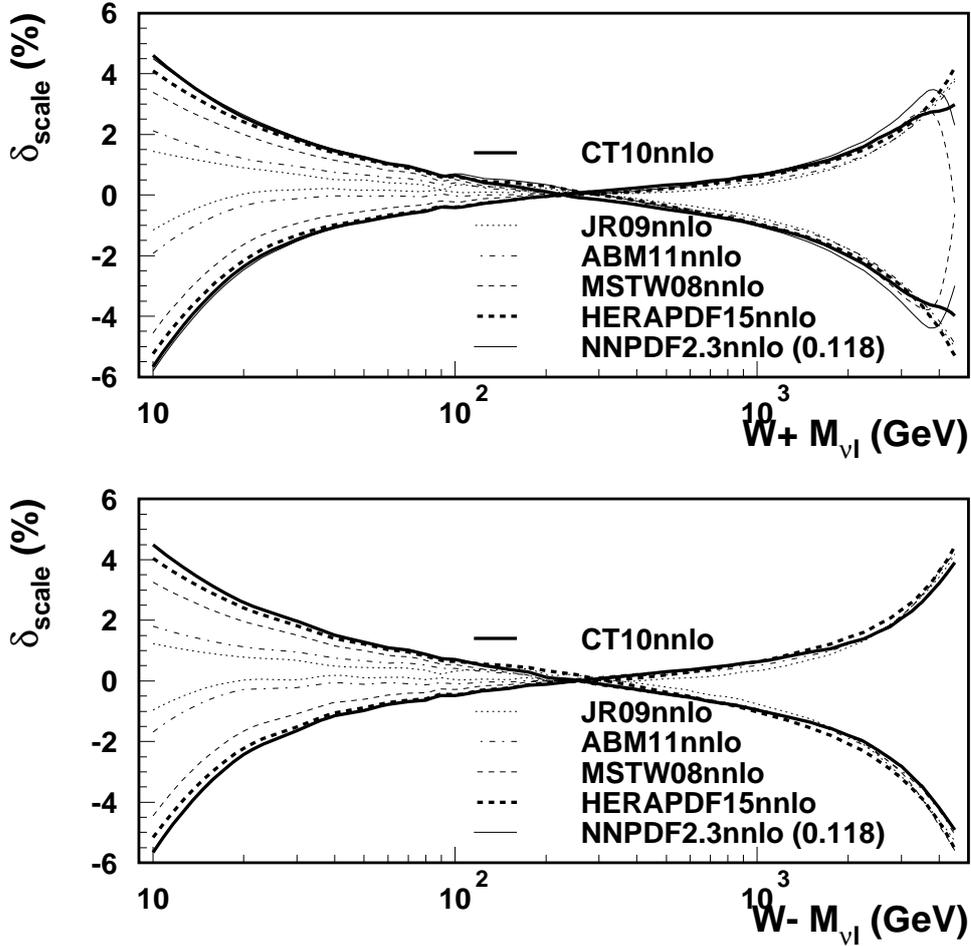}
\caption{\label{ewqcd:ccscale} Effect on the total CC Drell Yan NNLO QCD cross section predictions due to the  change of the renormalisation and factorisation scales simultaneously by a factor 2 or 1/2 for an invariant mass range of 10 to 5000 GeV. See Ref.~\cite{lhapdfweb} for the details of the NNLO QCD PDFs; the NNPDF calculations is obtained for $\alpha_s=0.118$. Calculations are based on VRAP 0.9 and $\sqrt{s}=8$~TeV.}
\end{center}
\end{figure}

  The nominal renormalisation and factorisation scales for
the calculation of DY cross sections are set to either directly to the invariant mass in VRAP or to the mean of an invariant mass bin in FEWZ (DYNNLO). 
The variation of the renormalisation and factorisation scales simultaneously by a factor of 2 or 1/2 can be conveniently calculated in NNLO QCD for the total Drell Yan production cross section using the program VRAP. 
Fig.\ref{ewqcd:ncscale} and Fig.\ref{ewqcd:ccscale} show the resulting effects for NC and CC Drell Yan, respectively, for all modern NNLO parton distribution functions. The mass-dependencies of the scale variations are observed to very similar for NC and CC Drell Yan, and also to  show very similar and significant  dependencies on the PDF choice. For low and high invariant masses the scale uncertainties are rising, even at NNLO QCD, and deserve further detailed studies.

%

\subsection*{Acknowledgements}

The electroweak and photon-induced background studies of this work have been done together with Dimitry
Bardin, Serge Bondarenko, and 
Lidia Kalinovskaya with the support of the grant CERN-RFBR-Scientific-Cooperation 12-02-91526-CERN\_a, and I am very thankful for this exceptional and fruitful collaboration. I also like to thank  Lance Dixon, Frank Petriello, Ye Li, Seth Quackenbush and Massimiliano Grazzini for many useful discussions over the past years. 
Part of this work were supported by my PPSC20 grant "Precision modelling of hadron structure at highest energies". Thanks to my ATLAS and LPCC WG colleagues for the co-operation.
Last but not least it is my pleasure  to thank warmly the Les Houches workshop 
for the hospitality and perfect organisation. 


\section{Electroweak Sudakov corrections to $Z/\gamma + $~jets at the LHC%
\footnote{M. Chiesa, L. Barz\`e, G. Montagna, M. Moretti, O. Nicrosini, 
   F. Piccinini, F. Tramontano}}

\subsection{Introduction}

Important searches for new phenomena beyond the Standard Model 
(SM) at the proton-proton ($pp$) collider LHC are based on the 
analysis of events with jets and missing transverse 
momentum ($\rlap\slash{\!p_T}$). 
Typical examples of such studies are searches for squarks and gluinos in 
all-hadronic reactions containing high-$p_T$ jets, missing transverse 
momentum 
and no electrons or muons, as predicted in many supersymmetric extensions 
of the SM. 
These final states can appear in a number of R-parity conserving models where 
squarks and gluinos can be produced in pairs and subsequently decay 
to standard 
strongly interacting particles plus neutralinos that escape 
detection, thus giving rise to a large amount of 
$\rlap\slash{\!p_T}$. Typically, the event selections adopted at 
run I of the LHC, at center of mass energies of 7 and 8~TeV, require 
the leading jet $p_T$ larger than 130~GeV or 
the single jets $p_T$'s larger than 50~GeV. 
Moreover, the signal region is defined by $m_{\rm eff} > 1000 $~GeV, 
where $m_{\rm eff} = \sum_i  {p_T}_i   + \rlap\slash{\!\!E_T}$, 
or $H_T > 500$~GeV and $|\,\, \rlap\slash{\!\!\vec{H}_T} | > 200$~GeV, 
where $H_T = \sum_i {p_T}_i$ and \,  
$\rlap\slash{\!\!\vec{H}_T} = 
- \sum_i \vec{p_t}_i$~\cite{Aad:2011ib,Collaboration:2011ida,cms:2012mfa}. 

The main SM backgrounds to the above mentioned signal(s) are given by the 
production of weak bosons accompanied by jets ($W/Z + n$~jets), 
pure QCD multiple jet events and $t \bar{t}$ production. Among these 
processes only $Z + n$~jets (in particular with $Z \to \nu \bar\nu$) 
constitutes an irreducible background, particularly relevant for final states 
with 2 and 3 jets. Because new physics signals could manifest themselves 
as a mild deviation with respect to  the large SM background, precise 
theoretical predictions for the processes under consideration are needed. 
In order to reduce the theoretical systematic uncertainties, 
the experimental procedure for the irreducible background determination 
relies on data driven methods. For instance, a measurement of 
the cross section for $Z(\to \nu {\bar \nu}) + n$~jets could be done 
through the measured cross section for $Z(\to l^+ l^-) + n$~jets times 
the ratio $\frac{{\cal B}(Z \to \nu \bar \nu)}{{\cal B}(Z \to l^+ l^-)}$, 
which is known with high precision from LEP1 data. 
With the exception of the ratio of 
branching ratios, the method is free of theoretical systematics. However, 
due to the low production rate of $Z(\to l^+ l^-) + n$~jets, 
in particular in the signal regions, this method results to be 
affected by large 
statistical uncertainty. Other possible choices of reference processes 
are $W + n $~jets and $\gamma + n $~jets. In both cases the statistics is 
not a limitation and the required theoretical input is the ratio 
\begin{equation}
R^n_V = \left[ \frac{d\sigma(Z(\to \nu \bar \nu) + n\, 
            {\rm jets})}{dX} \right] / 
        \left[\frac{d\sigma(V + n\, 
            {\rm jets})}{dX}\right] \, ,
\label{eq:ratioth}
\end{equation}
where $X$ is the observable under consideration 
and $V = W$, $\gamma$. Additional sources 
of uncertainties, specific for each channel, are the contamination of 
other processes, such as $t \bar t$ events (for the case of $W + $~jets) 
and photon isolation (for the case of $\gamma + $~jets). Recent theoretical 
work has been devoted to the study of the theoretical uncertainties related 
to the ratio of Eq.~(\ref{eq:ratioth}). 
In particular, in Ref.~\cite{Malik:2013kba} a study of the cancellation 
in $R^{n}_W$ of systematic theoretical uncertainties originating 
from higher-order QCD corrections, including scale variations 
and choice of PDF's, has been presented. 
A first detailed analysis of the impact of higher-order QCD corrections, 
PDF's and scale choice to $R^{1,2,3}_\gamma$ 
has been shown in Ref.~\cite{Ask:2011xf}. 
More recently, the level of theoretical uncertainty induced 
by QCD higher-order corrections  
in the knowledge of $R^2_\gamma$ and $R^3_\gamma$, relying on the comparison 
of full NLO QCD calculations with parton shower simulations matched to 
LO matrix elements, has been discussed in Refs.~\cite{Bern:2011pa} 
and \cite{Bern:2012vx}, respectively. All these studies point out 
that many theoretical systematics related to pQCD and PDF's 
largely cancel in the ratio and the corresponding theoretical uncertainty 
in the ratio $R^i_\gamma$ can be safely estimated to be within 10\%. 

What is not expected to cancel 
in the ratio is the contribution of higher-order electroweak (EW) 
corrections, which are different for the processes 
$Z(\to \nu {\bar \nu}) + n$~jets and 
$\gamma + n$~jets. Moreover these effects can be enhanced 
by Sudakov logarithms. 
For $R^1_\gamma$ the higher-order EW corrections have been calculated 
in Refs.~\cite{Kuhn:2005gv,Maina:2004rb,Stirling:2012ak}, where it is shown 
that for a vector boson transverse momentum 
of 2~TeV they are of the order of 20\%. 
In view of the forthcoming run II of the LHC, it is therefore necessary 
to quantitatively estimate the effects of EW corrections to $R^n_\gamma$, 
in particular for $n \geq 2$.

In this contribution we present first numerical results on the 
impact of EW Sudakov corrections to $R^2_\gamma$ and $R^3_\gamma$, 
for two realistic event selections at the LHC, at a center of mass energy of 
14~TeV. These results have been obtained by means of the algorithmic 
implementation of EW Sudakov corrections discussed in 
Ref.~\cite{Chiesa:2013yma}, which we describe below. 

\subsection{Implementation of EW Sudakov corrections in ALPGEN}
For energy scales well above the EW scale, EW radiative corrections are 
dominated by double and single logarithmic contributions 
(DL and SL, respectively) whose argument involves the ratio of the energy 
scale to the mass of the weak bosons. 
These logs are generated by diagrams in which 
virtual and real gauge bosons are radiated by external leg particles, 
and correspond to the soft and collinear singularities appearing in QED 
and QCD, i.e. when massless 
gauge bosons are involved. At variance with this latter case, the weak bosons 
masses put a physical cutoff on these ``singularities", so that virtual 
and real weak bosons corrections can be considered separately. Moreover, 
as the radiation of real weak bosons is in principle detectable, for those 
event selections where one does not include real weak bosons radiation, 
the physical effect of (negative) virtual 
corrections is  singled out, and can amount to tens of per cent. Since these 
corrections 
originate from the infrared structure of the EW theory, they are  
``process independent" in the sense that they depend only on the external 
on-shell 
legs\cite{Ciafaloni:1998xg,Beccaria:1999fk,Fadin:1999bq,
Ciafaloni:2000df,Ciafaloni:2000rp,Denner:2000jv,Denner:2001gw,
Ciafaloni:2001vt,Beenakker:2001kf,Denner:2003wi,
Kuhn:2004em,Kuhn:2005az,Kuhn:2005gv,Kuhn:2007qc,Accomando:2006hq}. 
As shown by Denner and Pozzorini in Refs.~\cite{Denner:2000jv,Denner:2001gw}, 
DL corrections can be accounted for by factorizing a proper correction 
which depends on flavour and kinematics of all possible pairs of electroweak 
charged external legs. SL corrections can be accounted for by factorizing 
an appropriate radiator function associated with each individual external 
leg. 
Notice that our implementation includes correctly all single logarithmic 
terms of ${\cal O}(\alpha^2 \alpha_s^n)$ of both UV and infrared origin, 
as detailed in Ref.~\cite{Pozzorini:2001rs}. 
The above algorithm has been implemented 
in {\tt ALPGEN v2.14}~\cite{Mangano:2002ea}, 
where all the contributing 
tree-level amplitudes are automatically provided. Since the matrix elements 
in 
{\tt ALPGEN} are calculated within the unitary gauge, for the time being we 
do not implement the corrections for the amplitudes involving longitudinal 
vector bosons, which, according to Refs.~\cite{Denner:2000jv,Denner:2001gw}, 
are calculated by means of the Goldstone Boson Equivalence Theorem. 
For the processes $Z + 2$~jets and $Z + 3$~jets, this approximation affects 
part of the ${\cal O}(\alpha^3)$ and ${\cal O}(\alpha^3 \alpha_s)$ 
contributions, respectively, 
and we checked that in view of 
our target precision of few percent it can be 
accepted~\footnote{The logarithms of photonic origin have not been 
considered in this realization since they can be treated separately 
together with their real counterpart 
for the processes under investigation. At any rate, these gauge invariant 
contributions (at the leading order $\alpha_s^{njets} \alpha$) 
for sufficiently inclusive experimental setup give rise to  rather 
moderate corrections. }. 

\subsection{Numerical results}
\begin{figure}[h]
\includegraphics[scale=0.5]{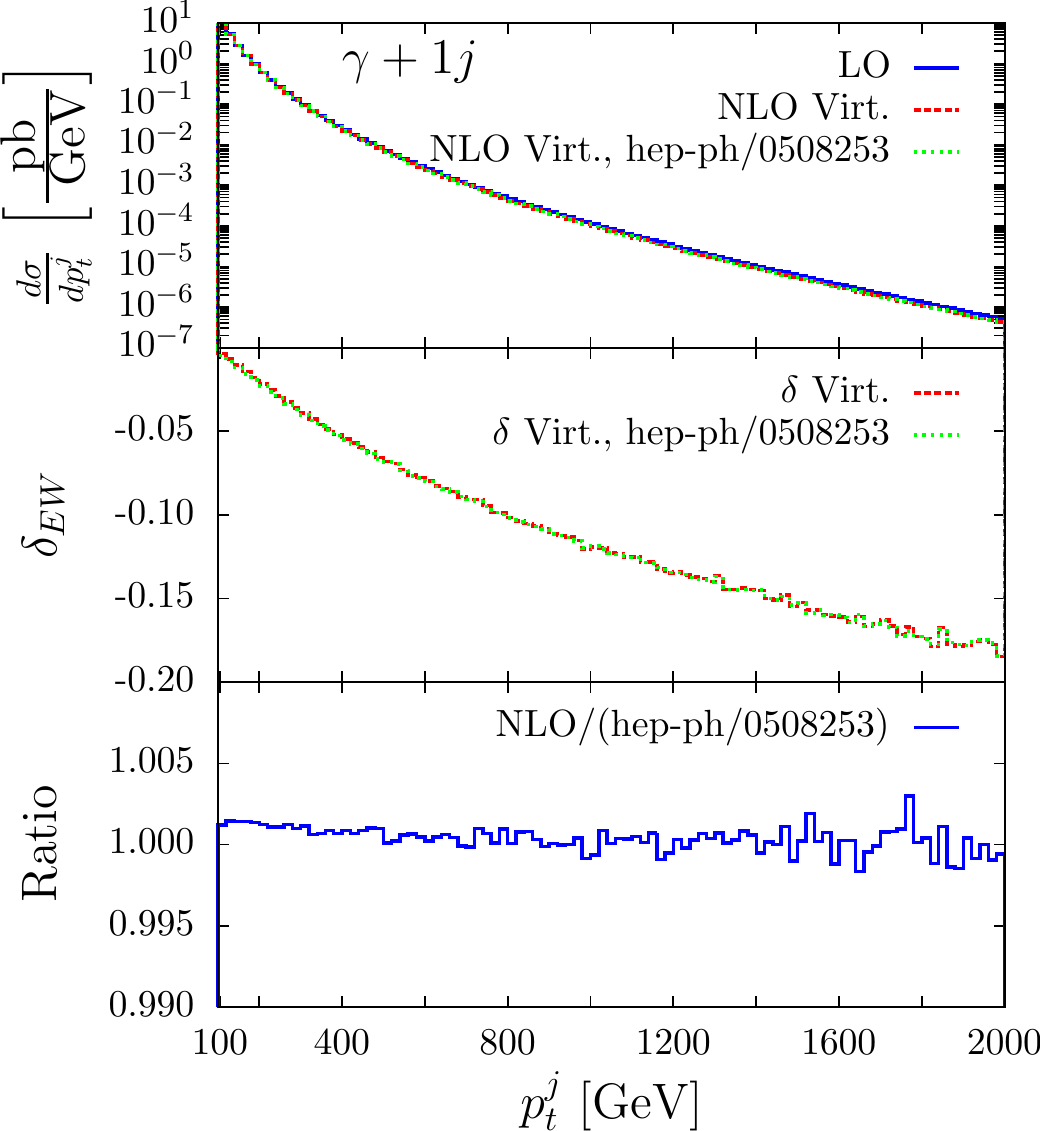}
\includegraphics[scale=0.5]{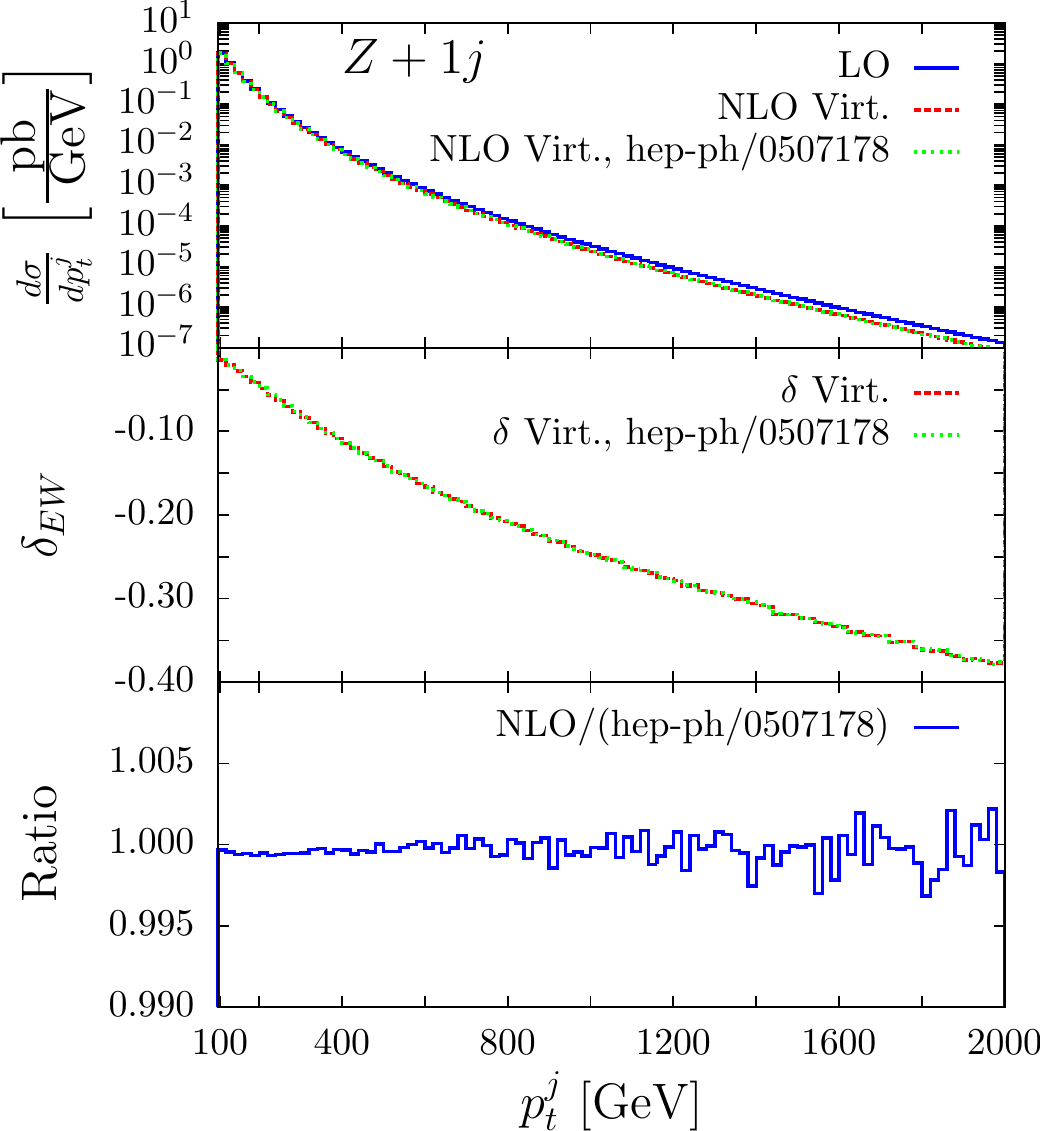} 
\includegraphics[scale=0.5]{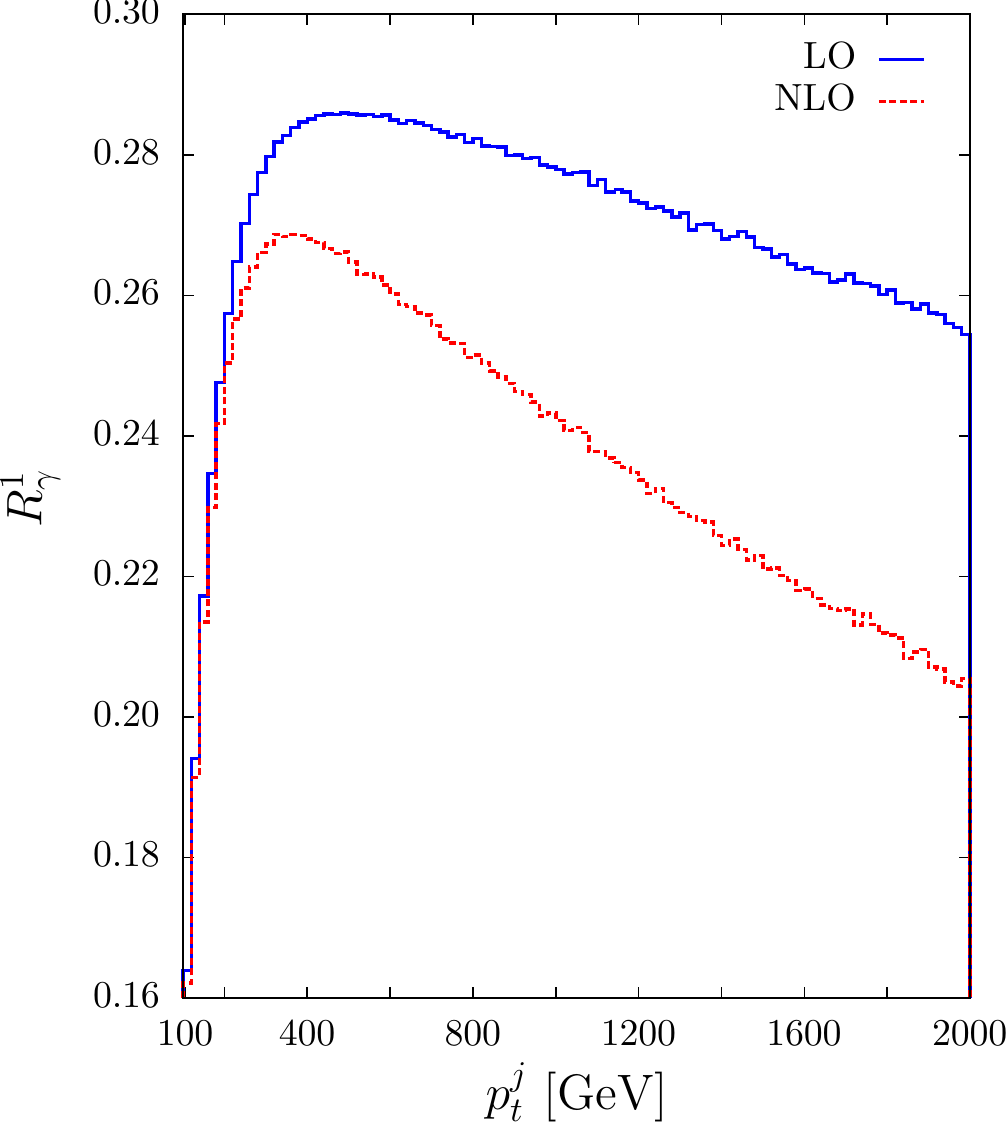} 
\caption{\label{fig:R_1} $p^{j}_{t}$ distribution at $\sqrt{s} = 14$~TeV 
for $\gamma+1$~jet and 
$Z(\to \nu {\bar \nu}) + 1$~jet and 
their ratio $R^1_\gamma$ at the LO (solid blue lines) and at 
approximated NLO (dashed red lines). Green dotted lines are the results of 
Refs.~\cite{Kuhn:2004em,Kuhn:2005az,Kuhn:2005gv} 
when the difference between the $W$
and the $Z$ masses is not neglected. The ratio of our predictions 
with the ones of Refs.~\cite{Kuhn:2004em,Kuhn:2005az,Kuhn:2005gv} is shown in 
the lower panels of the first two plots for $\gamma+1$~jet 
and $Z(\to \nu {\bar \nu})+1$~jet, respectively.}
\end{figure}

Our numerical results have been obtained by using the code {\tt ALPGEN v2.14} 
with default input parameters/PDF set. The algorithm has been validated with 
results available in the literature. 
In Fig.~\ref{fig:R_1} we show our predictions for the jet $p_{t}$ 
distribution 
for $\gamma + 1$~jet, $Z(\to \nu {\bar \nu}) + 1$~jet and 
for the ratio $R^1_\gamma$ at $\sqrt{s} = 14$~TeV in the region
\begin{equation}
p^{j}_{t}>100 \, {\rm GeV}, \quad |y(j)|< 2.5, \quad |y(V)|< 2.5 \; \; 
(V= \gamma,Z)
\label{eq:1jcut}
\end{equation} 
and we compare our results with the ones obtained from 
Refs.~\cite{Kuhn:2004em,Kuhn:2005az,Kuhn:2005gv} by including also the 
terms proportional to $\log(\frac{M_W}{M_Z})$: the level of agreement we 
find is well within the percent level.
Moreover, we estimated the corrections to 
$p_T^Z$ and to the leading jet $p_T$ distributions in the large tails 
for the process $Z + 2 $~jets with only one fermionic current, 
as discussed in Ref.~{\cite{Actis:2012qn}}, finding good agreement.  
Finally, for $Z + 2 $~jets we cross-checked our results with the 
automatic package 
{\tt GOSAM v1.0}~\cite{Cullen:2011ac}, with the event selection adopted in 
the present study. Since the electroweak renormalization is not yet available 
in the present version of {\tt GOSAM}, we subtracted the logarithmic terms 
due to the renormalization counterterms from the formulae of 
Refs.~\cite{Denner:2000jv,Denner:2001gw} and tested the asymptotic behaviour 
of all relevant distributions.   
Since the relative weight 
of two-quark and four-quark subprocesses is about 75\% and 25\% 
for total cross sections, while 
for the observables under consideration and in the high tails 
is about 50\% each at the LO, respectively, 
we performed this analysis for 
different subprocesses involving both one and two fermionic currents  
such as $q \bar q \to Z g g$, $q \bar q \to Z q'' \bar q''$, 
$q q \to Z q q$ and $q q' \to Z q q'$ (with $q$ and $q'$ belonging 
to the same isodoublet).
For all the above cases we found that the shape of the 
distributions 
predicted by the two calculations is in good agreement.

\begin{figure}[h]
\includegraphics[scale=0.5]{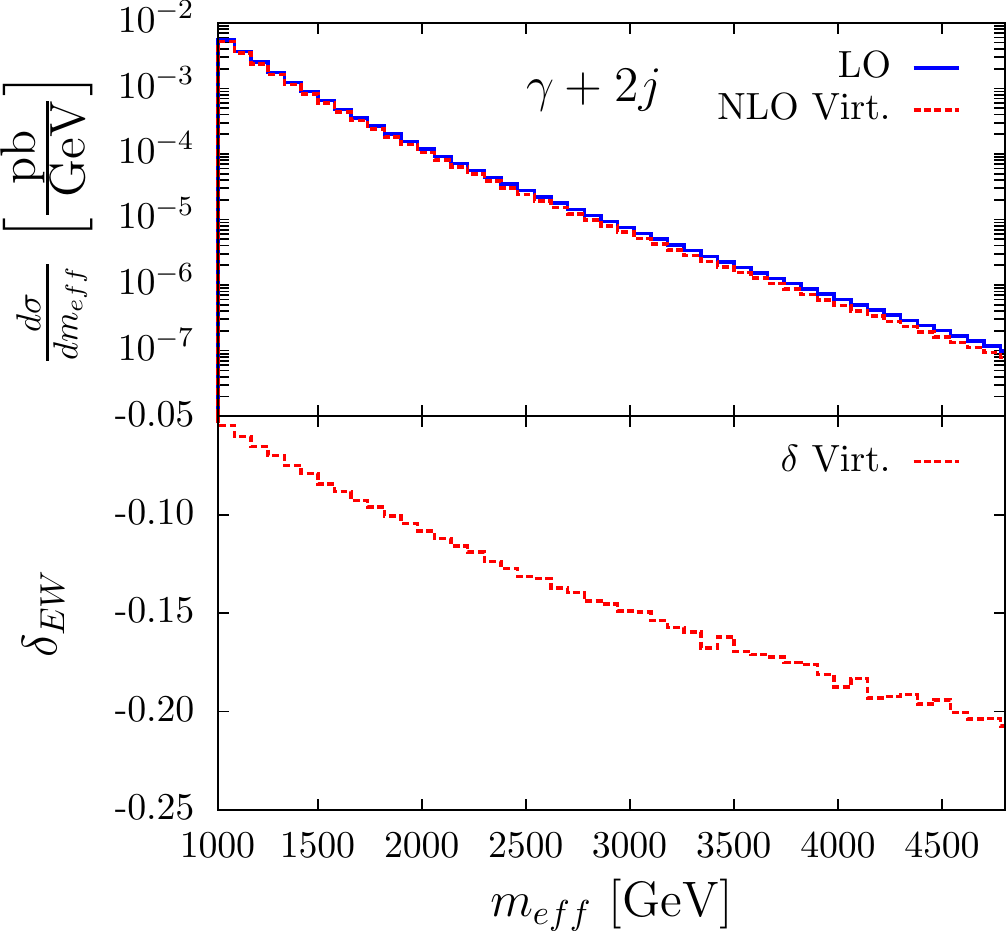}
\includegraphics[scale=0.5]{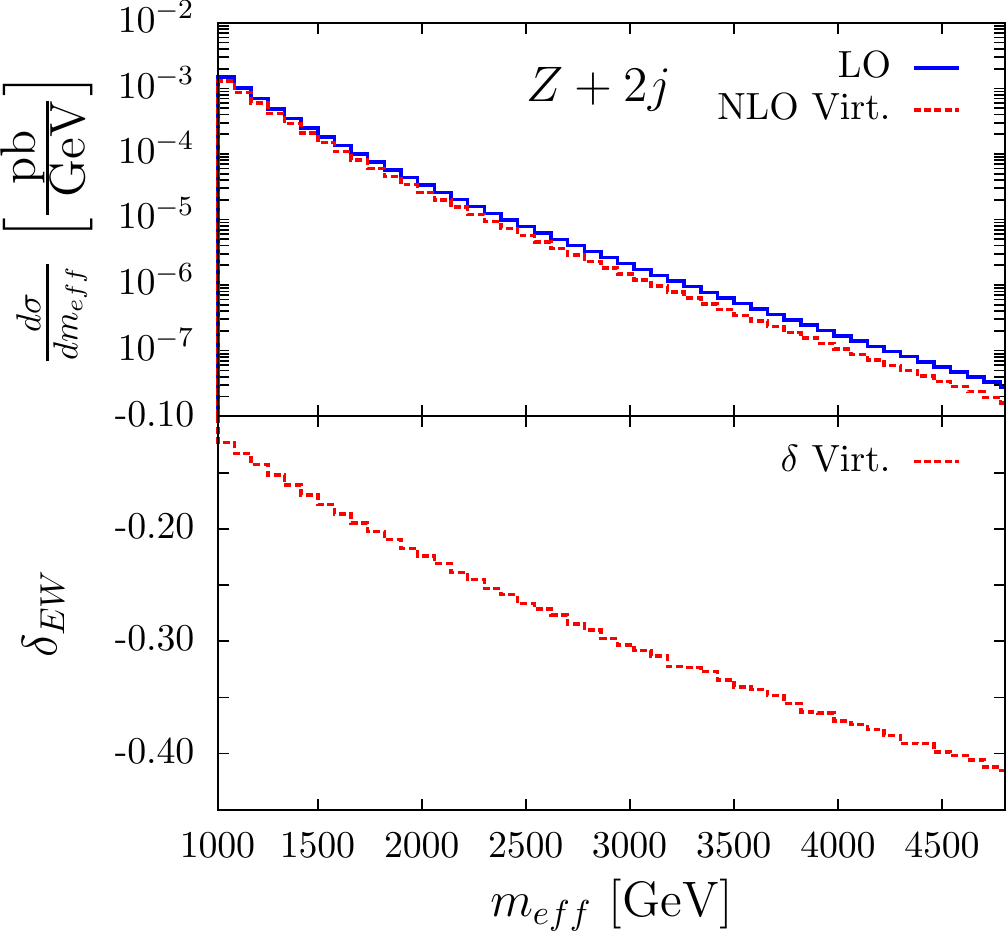} 
\includegraphics[scale=0.5]{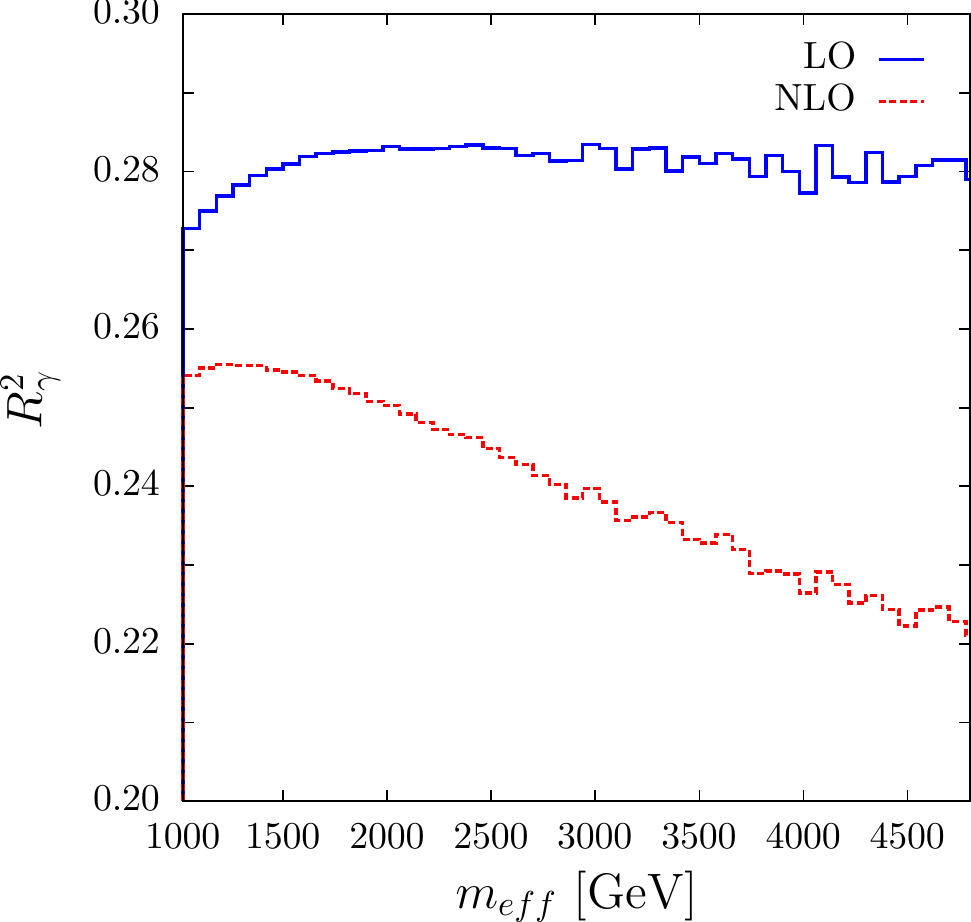}
\caption{\label{fig:R_2} $m_{\rm eff}$ distribution at $\sqrt{s} = 14$~TeV 
for $\gamma + 2$~jets and $Z(\to \nu {\bar \nu})+2$~jets and their 
ratio $R^2_\gamma$ 
under the ATLAS event selection (same notation of Fig.~\ref{fig:R_1}).}
\end{figure}

\begin{figure}[h]
\includegraphics[scale=0.5]{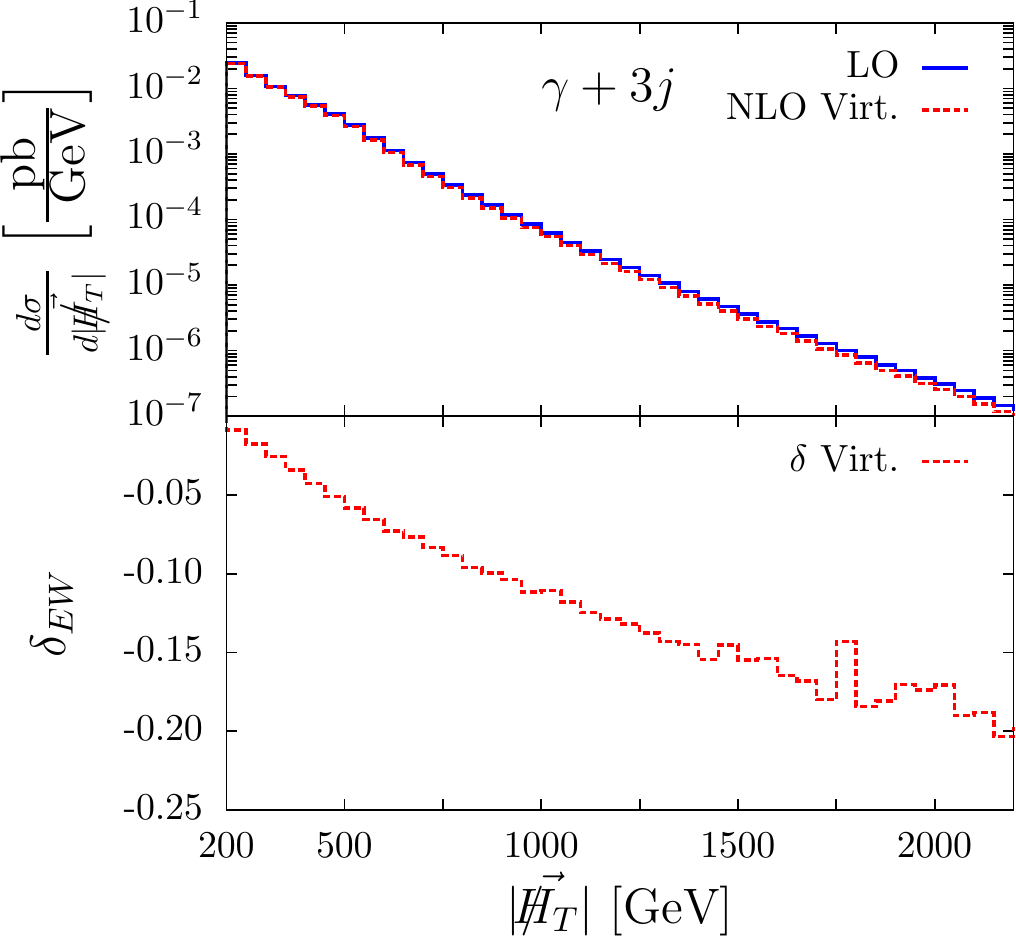}
\includegraphics[scale=0.5]{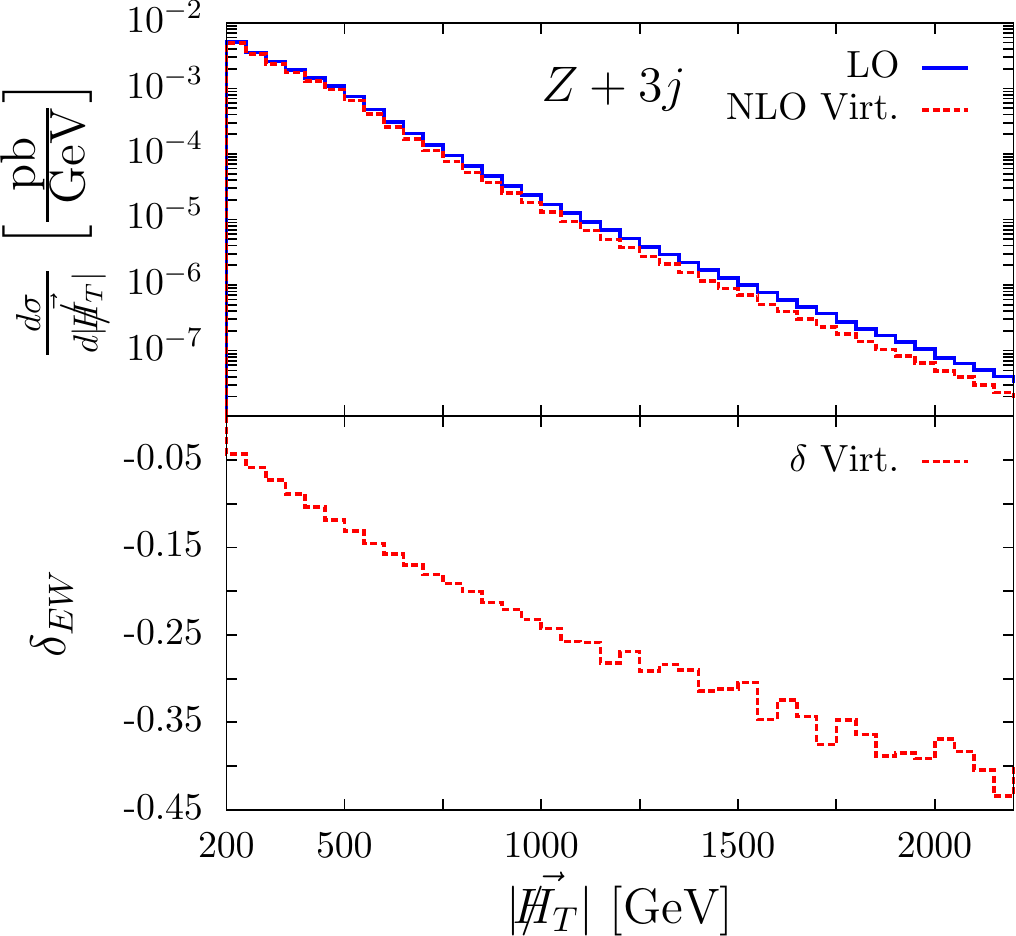}
\includegraphics[scale=0.5]{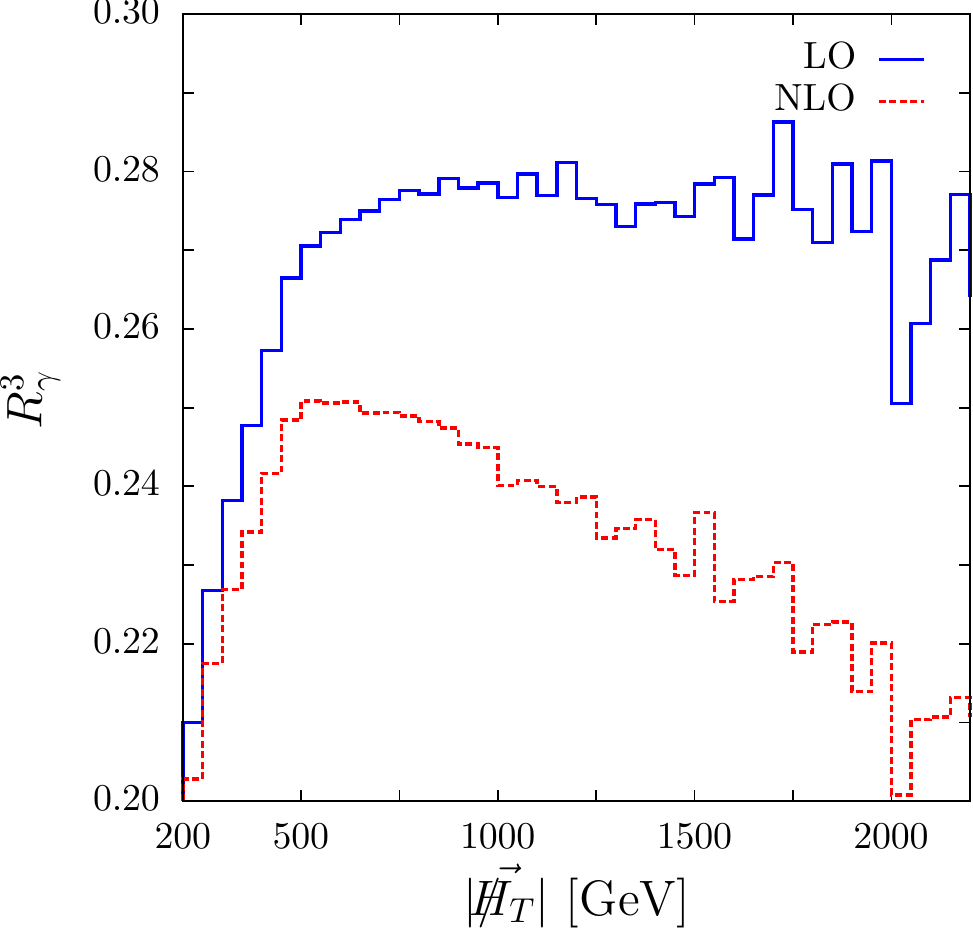}
\caption{\label{fig:R_3} $|\,\, \rlap\slash{\!\!\vec{H}_T} |$ distribution 
at $\sqrt{s} = 14$~TeV 
for $\gamma + 3$~jets and $Z(\to \nu {\bar \nu})+3$~jets and 
their ratio $R^3_\gamma$ 
under the CMS event selection (same notation of Fig.~\ref{fig:R_1}).}
\end{figure}

We present our results on $R^{2,3}_\gamma$ for two sets of cuts that  
mimic the real experimental event selections used by ATLAS and CMS 
in run I analysis. 
For the $Z(\to \nu {\bar \nu}) + 2$~jets and $\gamma + 2$~jets 
final states we consider 
the observable/cuts adopted by ATLAS, namely 
\begin{eqnarray}
&& m_{\rm eff} > 1~{\rm TeV} \qquad \, \, \, \, \, 
\rlap\slash{\!\!E_T}/m_{\rm eff} > 0.3  \nonumber\\
&& p_T^{j_1} > 130~{\rm GeV} \qquad \, \, 
p_T^{j_2} > 40~{\rm GeV} \quad \, \, |\eta_{j}| < 2.8 \nonumber \\
&& \Delta\phi ({\vec p}_T^j,\rlap\slash{\!\vec{p}_T}) > 0.4 \quad 
\Delta R_{(j_1, j_2)} > 0.4  \, 
\label{eq:atlascut}
\end{eqnarray}
where $j_1$ ($j_2$) are the leading (next-to-leading) $p_T$ jets, the 
missing momentum $\, \rlap\slash{\!\vec{p}_T}$ is defined as 
minus the sum of the jet transverse momenta 
and $\rlap\slash{\!\!E_T}$ is the magnitude of 
$\, \rlap\slash{\!\vec{p}_T}$. 
For the $Z(\to \nu {\bar \nu}) + 3$~jets and $\gamma + 3$~jets 
final states we consider the observables/cuts used by CMS, namely
\begin{eqnarray}
&& H_T > 500~{\rm GeV} \qquad \, \, \, \, \, \, 
|\,\, \rlap\slash{\!\!\vec{H}_T} | > 200~{\rm GeV}  \nonumber\\
&& p_T^j > 50~{\rm GeV} \qquad  \, \, |\eta_j| < 2.5 
\quad \Delta R_{(j_i, j_k)} > 0.5 \nonumber\\
&& \Delta\phi ({\vec p}_T^{j_1, j_2},\, \rlap\slash{\!\!\vec{H}_T}  ) > 0.5 
\qquad  \Delta\phi (\vec{p}_T^{j_3},\, \rlap\slash{\!\!\vec{H}_T}) > 0.3 
\label{eq:cmscut}
\end{eqnarray}
Fig.~\ref{fig:R_2} shows the $m_{\rm eff}$ distribution 
for $\gamma + 2$~jets and $Z(\to \nu {\bar \nu}) + 2$~jets, with the 
ATLAS cuts of Eq.~(\ref{eq:atlascut}). 
In the tails of the distributions the effect 
of EW Sudakov corrections is of the order of $-20$\% 
and $-40$\% for the processes 
$\gamma +2$~jets and $Z(\to \nu {\bar \nu})+2$~jets, respectively. 
As a result, the ratio 
$R^2_\gamma$ of the two distributions, which is almost flat at the LO, 
decreases at the NLO of about $8$\% for $m_{\rm eff} \simeq 1$~TeV 
up to $20$\% for $m_{\rm eff} \simeq 4.5$~TeV.
Fig.~\ref{fig:R_3} is the analogous of Fig.~\ref{fig:R_2} 
for the observable $|\,\, \rlap\slash{\!\!\vec{H}_T} |$ 
under CMS conditions of 
Eq.~(\ref{eq:cmscut}) for the processes $Z(\to \nu {\bar \nu}) + 3$~jets and 
$\gamma +3$~jets at $\sqrt{s} = 14$~TeV. 
As in the two jets case, EW Sudakov corrections are larger 
for $Z(\to \nu {\bar \nu})+3$~jets (where they are of the order of $-45$\% 
for $|\,\, \rlap\slash{\!\!\vec{H}_T} | \simeq 2$~TeV), 
while for $\gamma +3$~jets 
the size of the corrections is of the order of $-20$\% in the high 
$|\,\, \rlap\slash{\!\!\vec{H}_T} |$ tail.
Due to the different impact of EW corrections to $\gamma +3j$ 
and $Z(\to \nu {\bar \nu})+3j$, 
the shape of their ratio $R^3_\gamma$ changes at the NLO w.r.t. LO 
and decreases of about 
$20$\% for $|\,\, \rlap\slash{\!\!\vec{H}_T} | \simeq 2$~TeV.

To summarize, the theoretical accuracy on the quantity $R^n_\gamma$ 
is relevant in the calibration of the process 
$Z(\to \nu {\bar \nu}) + n$~jets, an important irreducible background 
to New Physics searches at the LHC. 
Contrary to higher-order QCD corrections, which largely cancel in the 
ratio, as shown in the literature, EW NLO corrections are sizeable, 
in particular in the phase space regions interesting for 
New Physics searches. 
We computed the NLO EW Sudakov corrections 
to $Z(\to \nu {\bar \nu}) + n$~jets, 
$\gamma + n$~jets and to the ratio $R^n_\gamma 
= \frac{d\sigma(Z(\to \nu {\bar \nu})+n\, 
    {\rm jets})}{d\sigma(\gamma + n \, {\rm jets})}$ 
(where $n = 1,2,3$) and we showed our results for the LHC 
at $\sqrt{s} = 14$~TeV. 
In the typical event selections used for New Physics searches based on the 
signature\,\, $\rlap\slash{\!\!E_T} +n$~jets, 
electroweak Sudakov corrections to 
$Z(\to \nu {\bar \nu}) + n$~jets and $\gamma + n$~jets in the tails of 
the distributions 
considered are of the order of $-40$\% and $-20$\%, respectively. Since the 
impact of EW corrections to $\gamma + n$~jets and 
$Z(\to \nu {\bar \nu}) + n$~jets is different, 
the shape of the ratio $R^n_\gamma$ changes at NLO accuracy w.r.t. LO, 
decreasing of about $20$\% (slightly depending on the observable considered) 
in the regions where the corrections are larger. Thus EW Sudakov 
corections are important in view of a target uncertainty on the ratio 
$R^n_\gamma$ of the order of 10\%. 
In this contribution we do not address the issue of real weak corrections, 
which is left to a future study. While the results presented in 
Ref.~\cite{Chiesa:2013yma} suggest a non-negligible contribution to 
the signatures $Z(\to \nu {\bar \nu}) + 2$~jets and 
$Z(\to \nu {\bar \nu}) + 3$~jets, preliminary investigations 
of Ref.~\cite{Bern:2012vx} on $R^3_\gamma$ show modest effects, 
at the \% level at most. 

\noindent
{\bf Acknowledgments}\\
\noindent
This work was supported in part by the Research Executive Agency (REA) of the 
European Union under the Grant Agreement number 
PITN-GA-2010-264564 (LHCPhenoNet), and by the Italian Ministry of University 
and Research under the PRIN project 2010YJ2NYW.
The work of L.B. is supported by the ERC grant 291377, ``LHCtheory - 
Theoretical predictions and analyses of LHC physics: advancing the precision 
frontier". F.P. and M.M. would like to thank the CERN PH-TH Department 
for partial support and hospitality during several stages of the work. 
F.P. and G.M. thank the organizers of the LH13 Workshop for their kind 
invitation. 

\clearpage

\def\eq#1{{(\ref{#1})}}

\section{Jet rates from recursion relations\footnote{E.~Gerwick, S.~Pl\"atzer}}

We present an improved recursive construction for analytic 
jet rates of arbitrary flavor at next-to-double-leading-logarithmic 
accuracy.  There are significant computational improvements in 
this formulation of the recursive algorithm as the explicit sum 
over integer partitions is avoided at each step in the recusion.  
We discuss this property and comment on future improvements.

\subsection{Introduction}
There has been recent interest in analytic all-orders jet rates
\cite{Platzer:2009jq,Gerwick:2013haa,Gerwick:2012fw} for the purpose of 
a meaningful analytic comparison to parton shower Monte Carlo 
\cite{Schumann:2007mg,Platzer:2011bc,Sjostrand:2007gs,Gleisberg:2008ta,Bahr:2008pv}.  
For example, the distribution of shower paths in the parton shower 
arriving at a particular phase space point is only known 
\emph{a postori}, after generating a potentially large amount of 
data.  As hadron collider physics turns to more QCD intensive 
observables, one would like to know this probability distribution 
\emph{a priori} \cite{Bauer:2007ad,Bauer:2006mk}. 
In this note we present progress on enumerating the spitting histories in 
observable, one would like to know this probability distribution 
\emph{a priori} \cite{Bauer:2007ad,Bauer:2006mk}. 
In this note we present progress on enumerating the splitting histories in 
a final state parton cascade based on an analytically solvable recursive 
evolution equation.
\subsection{Recursive Formula}

We begin by setting forth our notation.  In general, 
we will be dealing with analytic expressions for 
$n$-jet rates containing leading and sub-leading 
logarithms.
To this end, we define the $\mathcal{O}(\alpha_s^n)$ 
NDLL accurate contribution to jet production 
\begin{equation}
\Gamma_{n}(\mu^2, q^2) \;\; = \;\; 
a_s^{n}
\left[\sum_k 
\sum_{i=0}^{n-1} 
\sum_{l=0}^{n}
\left( c^{(l,n)}_{ik}
 \log \left(\frac{q^2}{\mu^2} \right)^{2n}  
+ 
\tilde{c}^{(l,n)}_{ik} 
\log \left(\frac{q^2}{\mu^2} \right)^{2n-1}   
\right) \right] .
\label{efers}
\end{equation}
The sum on $l$ is over emitted partons in the event classified 
as either real or virtual, while the sum over $i$ is over the 
number of primary emissions in a particular splitting history.  The 
sum on $k$ is over all splitting histories of the same order 
in $l$ and $i$ which becomes non-trivial starting 
only at $\mathcal{O}(\alpha_s^4)$ with respect to the core 
process.  Also we have defined $a_s = \alpha_s /\pi$ and 
for concreteness $q$ is the initiating scale and $\mu$ the 
IR cut-off associated with the resolution scale for jets.  Below 
this scale there is complete real virtual cancellation.  

Relating the expansion in \eq{efers} to the terms of the 
all-orders expression for the $m$-jet fraction we have
\begin{equation}
f_{m} \;\; = \;\; 
\sum_{n=m}^{\infty} \Gamma_n
^{(l= m )}
\label{rectorate}
\end{equation}
where the superscript indicates that we pick out only the 
terms with $l=m$ in the first sum of \eq{efers}.  The virtue of the $3$ sums in 
\eq{efers} is that all coefficients stand in one-to-one relation 
with a diagrammatic splitting history ($c$ at LL and $\tilde{c}$ at 
NDLL).  Note that the definition of $c^{(l,n)}_{ik}$ differs 
from that in Ref.~\cite{Gerwick:2013haa} by including the 
color factor.  In effect, \eq{rectorate} tells us that by 
computing $\Gamma_n$ we may always recover the rates. 

The recursive formula derived in Ref.~\cite{Gerwick:2013haa} 
for the LL coefficients can now be rewritten as a recursion relation 
for $\gamma$ itself.  We found there 
\begin{equation}
\Gamma_{n} = \left(a_s L^2 \right) \sum_{l=0}^{n-1}  
b^{p}_{n-l} (-1)^{n-l}
\left( \Gamma_{n-1} d^{(n)}  \; + \;
\sum_{p(n)} \frac{1}{S} \prod_{\sigma_i=\{ \sigma_1,\cdots\, \sigma_{r}\} } \; 
\Gamma^{(\sigma_i)}_{\sigma_i-1} \right).
\label{masform_eg}
\end{equation}
Here we define $p$ as as the degree of the stripped 
splitting history expressed in terms of primordial coefficients 
(discussed in more detail in \cite{Gerwick:2013haa}).  Also 
we define the binomial coefficient $b^{i}_{j} \equiv i!/(j!(i-j)!)$.  
The factor $S$ is the over-all permutation symmetry factor 
of a particular splitting history.  The proof of \eq{masform_eg} was 
sketched in Ref. \cite{Gerwick:2013haa} from the generating 
functional in the coherent branching formalism.
A few additional comments are in order regarding the terms 
of \eq{masform_eg}.
\begin{itemize}
\item  The first sum on $l$ counts the number of real gluons at a given order in $n$.  Thus 
the $l=n$ term corresponds to the coefficient where all emissions are resolved.
\item  The sum inside brackets is over the integer partitions $p(n)$ of $n$, while the 
product is over the corresponding coefficient of each element of the partition.  In 
practice, as $p(n) \sim \exp(\sqrt{n})$ this becomes the most computationally 
intensive step at high parton multiplicity.
\item  The number of previous components $i$ needed for the 
$\Gamma_{n}$ recursion, and thus necessarily stored in 
memory, also grows with $n$, as the degree of the splitting 
history must be tracked.
\item  We find an expression for the exclusive rate $\sigma_{n}$ 
inserting all $\Gamma_i$ with $i\le n$ in \eq{rectorate}.
\end{itemize}
\medskip

We now provide a different recursive formulation for the real emission
components of $\Gamma_n$ which provides significant improvements, both
by including sub-leading logarithmic pieces and boosting numerical
properties at large $n$.  This constitutes a crucial step towards
realistic analytic constructions of shower histories.

For the resolved components it is sufficient to sum over the $c$ and $\tilde{c}$ 
coefficients in \eq{efers}, as these are treated on equal footing.  We now define
\begin{equation}
\Gamma_{n}(\mu^2, q^2) \;\; = \;\; 
a_s^n
\left[ C^{\text{LL}}_{nk} 
 \log \left(\frac{q^2}{\mu^2} \right)^{2n}
+ C^{\text{NLL}}_{nk} 
 \log \left(\frac{q^2}{\mu^2} \right)^{2n-1}
 \right] ,
\label{inters}
\end{equation}
for the emission of $n-1$ partons from an initial configuration 
k.  The $k$ index is neglected in the definition of \eq{efers} 
for simplicity, but is easily restored.  Here, we keep it explicit.
According to \eq{inters}, we claim accuracy on 
the integrated multi-parton splitting function to the level 
of $\alpha^n L^{2n-1}$, which amounts to 
next-to-double-leading-logarithmic (NDLA) accuracy.

Each coefficient $C^{\text{LL}}_n$ and $C^{\text{NLL}}_n$ 
is degenerate on a potentially large number of splitting 
histories.  
Using the 	modified coherent branching formalism, the 
sub-leading logarithms in \eq{inters} are propagated 
through to the $n$-jet rates.  A simple evolution equation 
reads
\begin{multline}
\Gamma_{nk}(\mu^2,Q^2) = \\
\frac{\alpha_s}{2\pi} \sum_{m=0}^{n-1}\sum_{i,j} \int_{4\mu^2}^{Q^2} \frac{{\rm d}q^2}{q^2}
\int_{\mu/q}^{1-\mu/q} {\rm d} z \,P_{k\to ij}(z) \Gamma_i^{(m)}(\mu^2,z^2 q^2) \Gamma_j^{(n-m-1)}(\mu^2,(1-z)^2 q^2) \ ,
\label{evaol}
\end{multline}
with $\Gamma_k^{(0)}= 1$, effectively enforcing the $0$-th order 
expanded Sudakov boundary condition.  The sum on $n$ is now 
over partitions of partons of length $2$ while the sum over $i,j$ is 
over flavors (q/g) and is a purely NDLA effect.  
The diagrammatic form of \eq{evaol} is
\begin{alignat}{5}
\hbox{\includegraphics[width=0.55\textwidth]{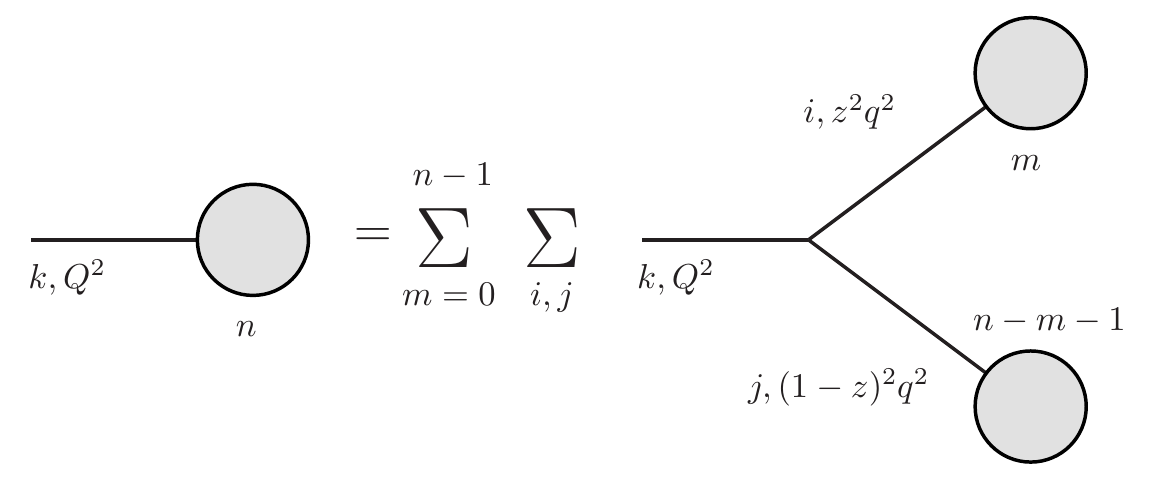}}
\label{gen_proof}
\end{alignat}
Now we insert the general expression \eq{inters} into the previous 
evolution equation \eq{evaol} and expand the respective coefficients.  
We find 
\begin{alignat}{5}
\Gamma_{nk}(\mu^2,Q^2)  =  & \left(  \frac{\alpha_s}{2\pi}\right)^n \sum_{m=0}^{n-1} \sum_{i,j} \int_{4\mu^2}^{Q^2} \frac{{\rm d}q^2}{q^2} \times \notag \\
&\left[  C_{i,m}^{\text{LL}} C_{i,n-m-1}^{\text{LL}} \sum_{\alpha = 0}^{2m} \sum_{\beta = 0}^{2(n-m)-2}
b^{2m}_{\alpha}\ b^{2(n-m)-2}_\beta\ 
L^{2n-2-\alpha-\beta}\ \gamma_{k\to ij}^{\alpha,\beta} \right. + \notag \\
& C_{i,m}^{\text{NLL}} C_{i,n-m-1}^{\text{LL}} \sum_{\alpha = 0}^{2m-1} \sum_{\beta = 0}^{2(n-m)-2}
b^{2m-1}_{\alpha}\ b^{2(n-m)-2}_\beta\ 
L^{2n-3-\alpha-\beta}\ \gamma_{k\to ij}^{\alpha,\beta} +  \notag \\
& C_{i,m}^{\text{LL}} C_{i,n-m-1}^{\text{NLL}} \sum_{\alpha = 0}^{2m} \sum_{\beta = 0}^{2(n-m)-3}
b^{2m}_{\alpha}\ b^{2(n-m)-3}_\beta\ 
L^{2n-3-\alpha-\beta}\ \gamma_{k\to ij}^{\alpha,\beta} + \notag \\
& \left. C_{i,m}^{\text{NLL}} C_{i,n-m-1}^{\text{NLL}} \sum_{\alpha = 0}^{2m-1} \sum_{\beta = 0}^{2(n-m)-3}b^{2m-1}_{\alpha}\ b^{2(n-m)-3}_\beta\ 
L^{2n-4-\alpha-\beta}\ \gamma_{k\to ij}^{\alpha,\beta} \right] \ ,
\end{alignat}
where
\begin{equation}
\gamma_{k\to ij}^{\alpha,\beta} \equiv \int_{\mu/q}^{1-\mu/q} {\rm d}z P_{k\to ij}(z) \ln^\alpha z^2 \ln^\beta (1-z)^2.
\label{gamdef}
\end{equation}
As is already evident from \eq{inters}, the splitting 
function $P(z)$, with singular pieces in both $z \to 0$ and 
$(1-z)\to 0$, picks out the logarithmically enhanced 
coefficients of the corresponding $\Gamma$.  
To this end, we decompose the splitting function as
\begin{equation}
P_{k\to ij}(z) = \frac{P_{k\to ij}^{(-)}}{1-z} + \frac{P_{k\to ij}^{(+)}}{z} + P_{k\to ij}^{(0)} (z) \ ,
\end{equation}
where $P_{k\to ij}^{(0)} (z)$ is regular at both $z\to 0$, $z\to 1$ and we 
let $\langle P_{k\to ij}^{(0)} \rangle$ denote its average over $z\in(0,1)$.  
This allows us to extract the logarithmic pieces from \eq{gamdef}.  The 
relevant contributions are
\begin{alignat}{5}
\gamma_{k\to ij}^{\alpha,\beta} \; = \; P_{k\to ij}^{(-)} \frac{(-1)^\beta}{2(\beta+1)}L^{\beta+1}\delta_{\alpha,0} +
P_{k\to ij}^{(+)} \frac{(-1)^\alpha}{2(\alpha+1)}L^{\alpha+1}\delta_{\beta,0} + 
\langle P_{k\to ij}^{(0)} \rangle \delta_{\alpha,0}\delta_{\beta,0} + {\cal O}(\mu) \ .
\end{alignat}
The leading and next-to-leading logarithmic coefficients, defined in \eq{inters},  
are given by sums of coefficients from lower $n$, with pre-factors assigned by 
the integrated splitting functions.  We find
\begin{alignat}{5}
&C_{k,n}^{\text{LL}} = \frac{1}{4n} \sum_{m=0}^{n-1}\sum_{i,j}C_{i,m}^{\text{LL}}C_{j,n-m-1}^{\text{LL}}
\left( P_{k\to ij}^{(-)} \xi_{2n-2m-2} + P_{k\to ij}^{(+)} \xi_{2m}  \right) \notag \\
& C_{k,n}^{\text{NLL}} = \frac{1}{2(2n-1)} \times  \sum_{m=0}^{n-1}\sum_{i,j}\left[
P_{k\to ij}^{(-)}\left(C_{i,m}^{\text{NLL}} C_{j,n-m-1}^{\text{LL}}\xi_{2n-2m-2} + 
C_{i,m}^{\text{LL}} C_{j,n-m-1}^{\text{NLL}}\xi_{2n-2m-3} \right) \right. + \notag \\
& \qquad \quad \; \left. P_{k\to ij}^{(+)}\left(C_{i,m}^{\text{NLL}} C_{j,n-m-1}^{\text{LL}}\xi_{2m-1} + 
C_{i,m}^{\text{LL}} C_{j,n-m-1}^{\text{NLL}}\xi_{2m} \right) +
2 \langle P_{k\to ij}^{(0)} \rangle C_{i,m}^{\text{LL}} C_{j,n-m-1}^{\text{LL}} \right]\, .
\label{sec_cons}
\end{alignat}
with initial condition $C_{k,0}^{\text{LL}}=1$, $C_{k,0}^{\text{NLL}}=0$.  
We define here
\begin{equation}
\xi_n=\sum_{k=0}^n\frac{(-1)^k}{k+1} b^n_k = \begin{cases}\frac{1}{n+1} & n\ge 0 \\ 0 & \text{else}\end{cases} \ .
\end{equation}

\subsection{Discussion}

\underline{Partitions of integers:}

We clarify here the relation between the two recursive 
constructions, \eq{masform_eg} and \eq{sec_cons}.  In \eq{sec_cons} 
there is no sum over integer partitions of the number of partons $n$.  
This dramatically reduces the (potentially prohibitive at high multiplicity) 
computational expense.  It is useful to inductively analyze precisely how 
the second construction \eq{sec_cons} aviods this complication.

The coefficients $C_{k,n}^{\text{LL}}$ contain information on all splitting 
histories up to $n$, including the partitions (and sub-partitions) of $n$.  
Now consider the partition $\{\sigma_1 ,\sigma_2, \cdots\,  \sigma_q\}$ of 
length $q$ of $n$ so that~$\sigma_1 + \sigma_2 + \cdots \, \sigma_q =n$.  
We take the term included in $\Gamma_n$  which is proportional 
to $c_{\sigma_1}c_{\sigma_2} \cdots c_{\sigma_q}$.  Now we will display 
how all partitions of length $q$ of $n+1$ are generated. 

Consider the partition $\{\sigma_2, \cdots\,  \sigma_q\}$ of 
length $q-1$ in $\Gamma_{n-\sigma_1}$ which we label 
$\Gamma_{n-\sigma_1}^{q-1}$.  The $z\to 0$ region of the 
integral
\begin{equation}
P(z) \, \Gamma_{\sigma_1} (\mu^2,z^2 q^2)\, \Gamma_{n-\sigma_1}^{q} (\mu^2,(1-z)^2 q^2) 
\end{equation}
gives the contribution corresponding to the partition 
$\{\sigma_1+1,\sigma_2, \cdots\,  \sigma_q\}$.  Similarly, the term 
$\Gamma_{\sigma_2}\Gamma_{n-\sigma_2}$ generates 
$\{\sigma_1,\sigma_2+1, \cdots\,  \sigma_q\}$.  It is clear that 
this procedure exhausts all partitions of $n+1$ of length $q$, 
which are now contained in $\Gamma_{n+1}$.  
The only partition not includes here is of length $n+1$, which 
clearly originates from the non-singular regions of $P(z)\Gamma_{0}\Gamma_{n}$.  
\smallskip

\noindent 
\underline{Systematic improvements:}

As presented the recursive contruction only includes the resolved 
coefficienst which are leading order in $\alpha_s$.  Including the 
virtual coefficients along with the NDLL terms, one should have 
enough information to include the leading effects of the running 
coupling which formally start at one power higher in the coupling 
together with sub-leading logarithms.  At this stage a direct comparison 
with parton shower Monte Carlo is possible.

\noindent 
\underline{Validation:} 

We have compared results from the recursion
relation presented here (at LL), with the recursion relation discussed
previously. The NDLC calculation has been verified for the
coefficients quoted in \cite{Webber:2010vz}
  
\noindent 
\underline{Outlook:} 

Both the original recursion proposed in \cite{Gerwick:2013haa}, and
the new approach with the inclusion of NDLC terms are available as
symbolic Mathematica code, while we have also developed a highly
efficient numerical implementation in C++. Work is ongoing on
including the unresolved components in the new approach, as well.  We
foresee providing the results in both in the analytic and numeric
tools which will in turn enable us to perform more realistic cross
checks of the jet rate coefficients against parton shower
implementations.

\subsection*{Conclusions}

Our prescription for computing analytic jet rate coefficients is relevant 
a more complete understanding of the probability distributions of 
exclusive final states.  These are typically only accessed via parton 
shower Monte Carlo, although there are recent attempts to put these 
on more solid analytic footing 
\cite{Platzer:2009jq,Bauer:2007ad,Bauer:2006mk}.

\subsection*{Acknowledgements}
We would like to thank the Les Houches workshop for hospitality
offered during which some of the work contained herein was performed,
and the stimulating environment to foster collaboration on comparing
and developing two different approaches to the same problem. SP
acknowledges partial support from the Helmholtz Alliance `Physics at
the Terascale'.

=======
\section{Study of the average number of hard jets\footnote{J.R.~Andersen, D.~Ma\^{\i}tre}}

\subsection{Introduction}
Jet vetos are an important ingredient in many analysis, in particular for
analysis involving a Higgs boson produced through vector boson fusion. It is
necessary to understand the impact of the jet vetos on theoretical
predictions and their uncertainties. For this purpose it is useful to test
the description offered by different tools in a process offering more
available and precise data. In this contribution we investigate the extent to
which pure perturbative tools (no underlying event, no hadronisation) can
describe dijet events when vetos on further jets are imposed.

We compare theoretical predictions obtained using HEJ \cite{Andersen:2009nu,Andersen:2009he,Andersen:2011hs} and NLO predictions obtained by BlackHat+Sherpa \cite{BH,pureQCD,sherpa1,sherpa2,amegicI,amegicII}. We base the analysis on the observables measured by the Atlas collaboration described in \cite{Aad:2011jz}.

\begin{figure}
\includegraphics[scale=0.5]{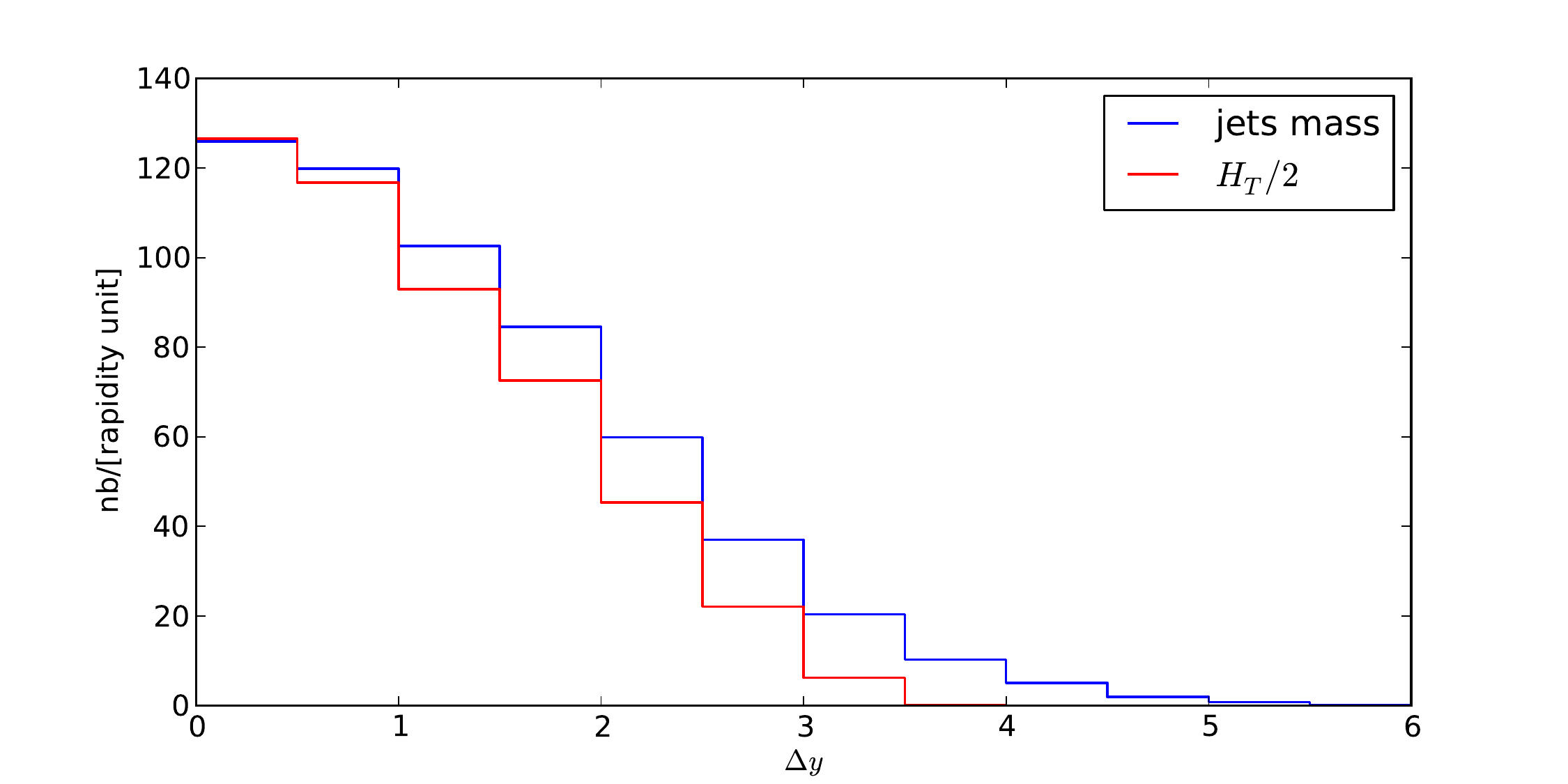}
\caption{Distribution of inclusive 2-jet events with no jets in the ragion of
  rapidity between the two jets of highest transverse momentum, for two
  choices of factorisation and renormalisation scale.}
\label{fig:scales}
\end{figure}

For the NLO predictions we used the invariant mass of all jets as the scale
and not the usual $H_T/2$ variable since in configurations with two widely
separated jets with moderate transverse momenta (the type of configuration
dominant for most observables in this study) $H_T$ is not a good scale to
represent the scale of the process, as it is much lower than the partonic
center of mass energy. The fact that this scale is too low in these
configurations is evidenced by the fact that distributions turn negative for
the corresponding observables, as can be seen in Figure~\ref{fig:scales},
illustrating the cross section for dijet production with no further jets in
the region of rapidity between the two hardest jets, as obtained with two
different choices for the scale in the calculation of dijets@NLO.

There are different ways of defining the gap fraction at NLO given the ambiguity in the inclusion of higher order terms. In this contribution we will follow two approaches, equivalent up to higher order terms, trying to identify whether one agrees with the data better than the ther.

\subsection{Gap fractions for a fixed $p_t$-threshold} 

In this section we use $Q_0$ to denote the minimum transverse momentum for
the jets. The gap fraction is given by:
\begin{equation}
g=\frac{\sigma_{Y/pt}(Q_0)}{\sigma_{tot}},
\end{equation}
where $\sigma_{Y/pt}(Q_0)$ is the cross section with the constraint that
there are no additional jets with $p_T$ above $Q_0$ between (in rapidity) the
two leading $p_T$ jets ($pt$) or the forward/backward jets ($Y$). We use the
notation $g=i$ to specify how many jets are in the region of rapidity between
the tagging dijets (the condition for this to happen depend on the context,
e.g. whether the tagging dijet system is defined by the most energetic jets or the most forward and backward). $j=i$ is used to specify the number of jets. In both cases the equal sign correspond to an exclusive final state, for example $j=2$ means exactly two jets, while inequalities mean inclusive final state $g\geq 1$ means one or more jets in the gap.  

We investigate two different approximations for the gap fraction, which are
both equivalent up to $\alpha_s^3$. These are
\begin{eqnarray}\label{gapNLO}
g&=&\frac{\sigma_{g=0}}{\sigma_{tot}}
=1-\frac{\sigma_{g>=1}}{\sigma_{tot}}\nonumber\\
&=&1-\frac{\sigma^{\rm{NLO},j\geq 3}_{g\geq 1}}{\sigma^{\rm{NLO},j\leq 2}}
=1-\frac{\sigma^{\rm{NLO},j>=3}-\sigma^{\rm{NLO},j=3}_{g=0}-\sigma^{LO,j=4}_{g=0} }{\sigma^{\rm{NLO},j\leq 2}}
\end{eqnarray}
and 
\begin{eqnarray}\label{gapExSum}
g=\frac{\sigma_{g=0}}{\sigma_{tot}}
=\frac{\sigma^{\rm{NLO},j=2}_{g=0}+\sigma^{\rm{NLO},j=3}_{g=0}+\sigma^{\rm{NLO},j\geq 4}_{g=0}}{\sigma^{\rm{NLO},j=2}+\sigma^{\rm{NLO},j=3}+\sigma^{\rm{NLO},j\geq 4}}
\end{eqnarray}
In the case where the jets defining the dijet system are the forward/backward ones, the formulae simplify to
\begin{eqnarray}
g_Y&=&\frac{\sigma_{g=0}}{\sigma_{tot}}
=\frac{\sigma^{\rm{NLO},j=2}}{\sigma^{\rm{NLO},j\geq 2}}
\end{eqnarray}
and
\begin{eqnarray}
g_Y=\frac{\sigma_{g=0}}{\sigma_{tot}}
=\frac{\sigma^{\rm{NLO},j=2}}{\sigma^{\rm{NLO},j=2}+\sigma^{\rm{NLO},j=3}+\sigma^{\rm{NLO},j\leq 4}}.
\end{eqnarray}

\begin{figure}
\includegraphics[scale=0.45]{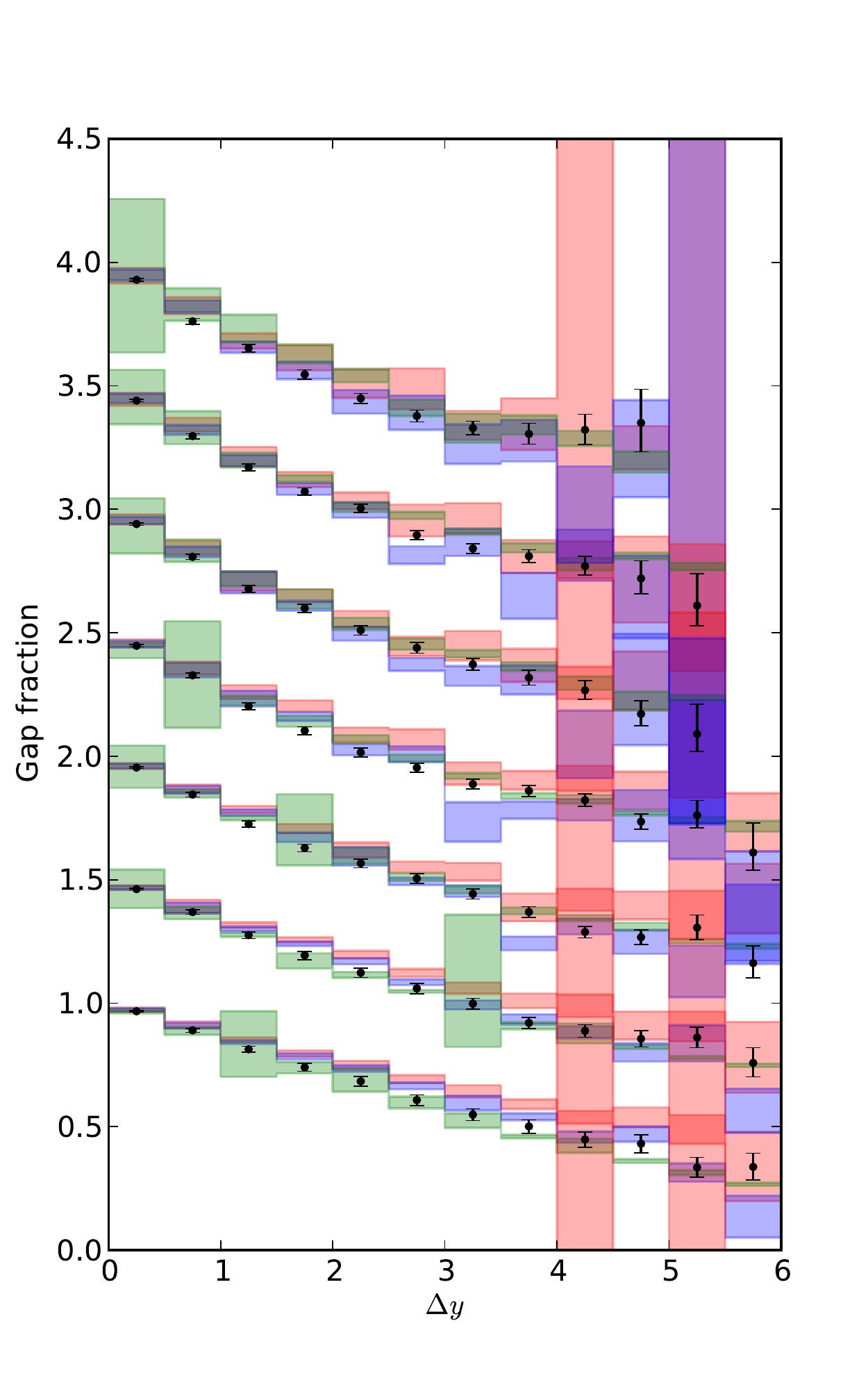}
\includegraphics[scale=0.45]{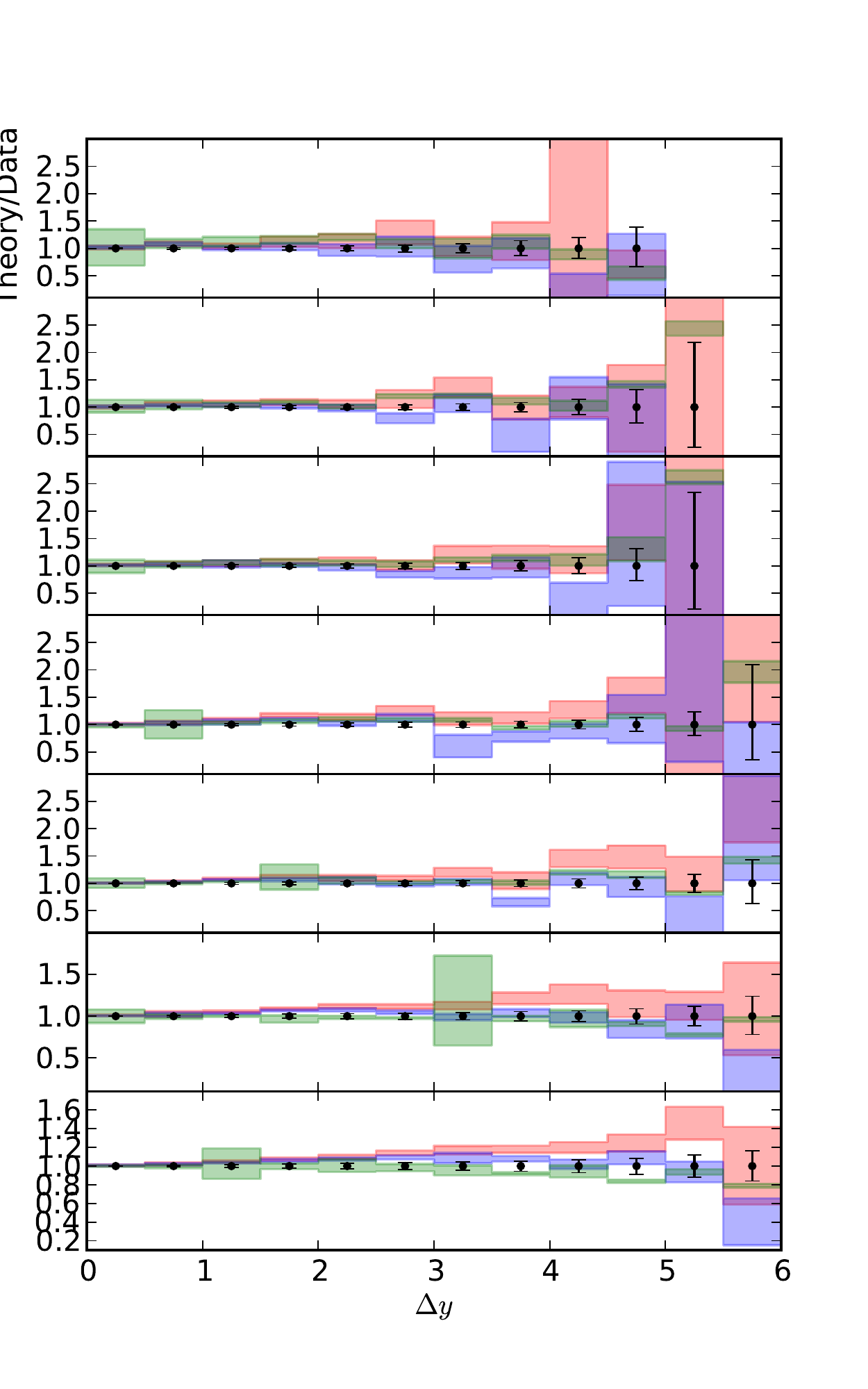}
\caption{Gap fraction as a function of $\Delta y$ for various slices of $\bar p _T$. The jets defining  $\bar p _T$ and $\Delta y$ are the two jets with the largest $p_T$.}
\label{fig:3a}
\end{figure}

Figure \ref{fig:3a} shows the gap fraction as a function of the the rapidity difference between the two highest transverse momentum jet. The curves corresponds to different slices of the average transverse momentum of the jets
\[\bar p_T=\frac{1}{2}\left(p_T^{\rm j_1}+p_T^{\rm j_2}\right).\]
From the top to the bottom the slices are:
\begin{eqnarray}\label{ptSlices}
240\,{\rm GeV}< &\bar p_T& < 270\, {\rm GeV}\nonumber\\
210\,{\rm GeV}< &\bar p_T& < 240\, {\rm GeV}\nonumber\\
180\,{\rm GeV}< &\bar p_T& < 210\, {\rm GeV}\nonumber\\
150\,{\rm GeV}< &\bar p_T& < 180\, {\rm GeV}\nonumber\\
120\,{\rm GeV}< &\bar p_T& < 150\, {\rm GeV}\nonumber\\
90\,{\rm GeV}< &\bar p_T& < 120\, {\rm GeV}\nonumber\\
70\,{\rm GeV}< &\bar p_T& < 90\, {\rm GeV}
\end{eqnarray}
The green curve corresponds to the HEJ prediction, while the blue and red
curves correspond to the NLO predictions of formulae (\ref{gapNLO}) and
(\ref{gapExSum}), respectively. The band represent the statistical Monte
Carlo intgration only. For this observable, the NLO-based prediction fares
marginally better compared to ATLAS data\cite{Aad:2011jz} when using
Eq.~\eqref{gapExSum} rather than Eq.~\eqref{gapNLO}.

\begin{figure}
\includegraphics[scale=0.45]{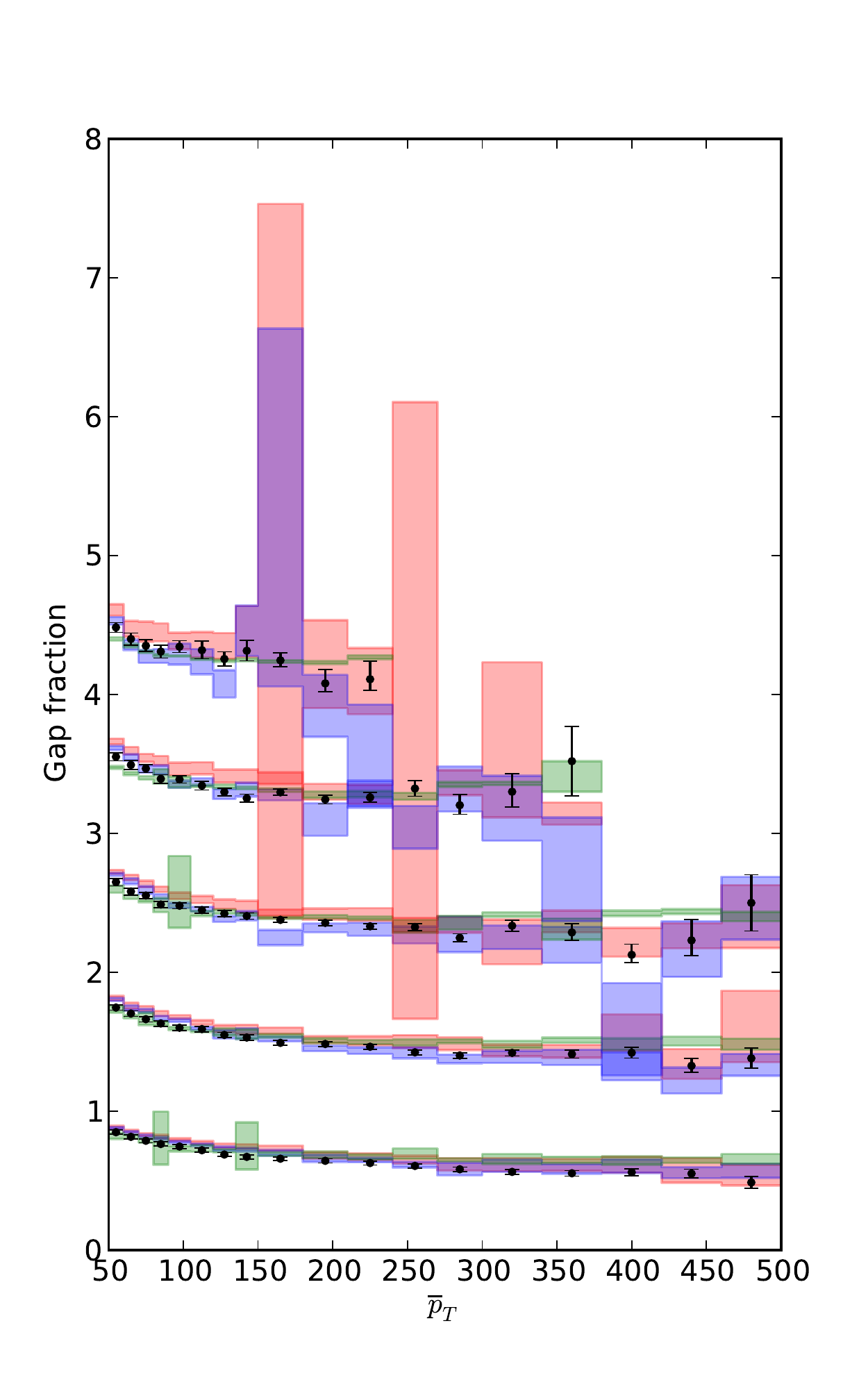}
\includegraphics[scale=0.45]{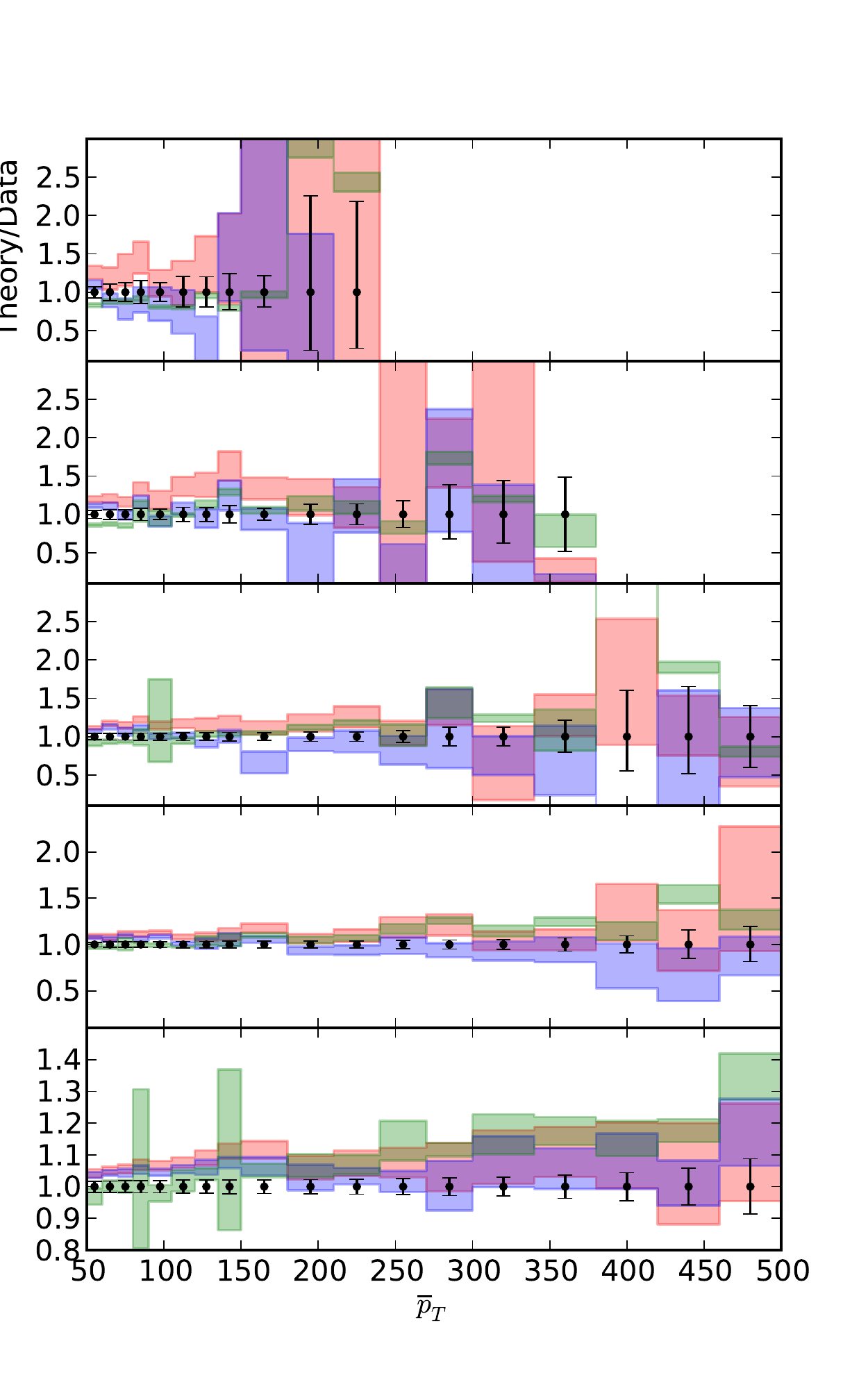}
\caption{Gap fraction as a function of $\bar p _T$ for various slices of $\Delta y$. The jets defining  $\bar p _T$ and $\Delta y$ are the two jets with the largest $p_T$.}
\label{fig:4a}
\end{figure}

In Figure \ref{fig:4a} we present the gap fraction as a function of the average transverse momentum of the highest transverse momentum jets for different slices of rapidity difference between them. The slices are give, from top to bottom by
\begin{eqnarray}\label{deltaySlices}
5<&\Delta y&<6\nonumber\\
4<&\Delta y&<5\nonumber\\
3<&\Delta y&<4\nonumber\\
2<&\Delta y&<3\nonumber\\
1<&\Delta y&<2\,.
\end{eqnarray} 
The green curve corresponds to the HEJ prediction, while the blue and red
curves correspond to the NLO predictions of formulae (\ref{gapNLO}) and
(\ref{gapExSum}), respectively. The band represent the statistical Monte
Carlo intgration only. For this observable the NLO-based predictions are
again better using the gap formula (\ref{gapExSum})
rather than that of  (\ref{gapNLO}); the former overshooting the data less
than the latter. The difference between the two formulae becomes larger as
the rapidity separation between the jets increases, which is fully understood
from the increased of yet higher order terms for increasing rapidity spans.


\subsection{Mean number of jets}
\begin{figure}
\includegraphics[scale=0.45]{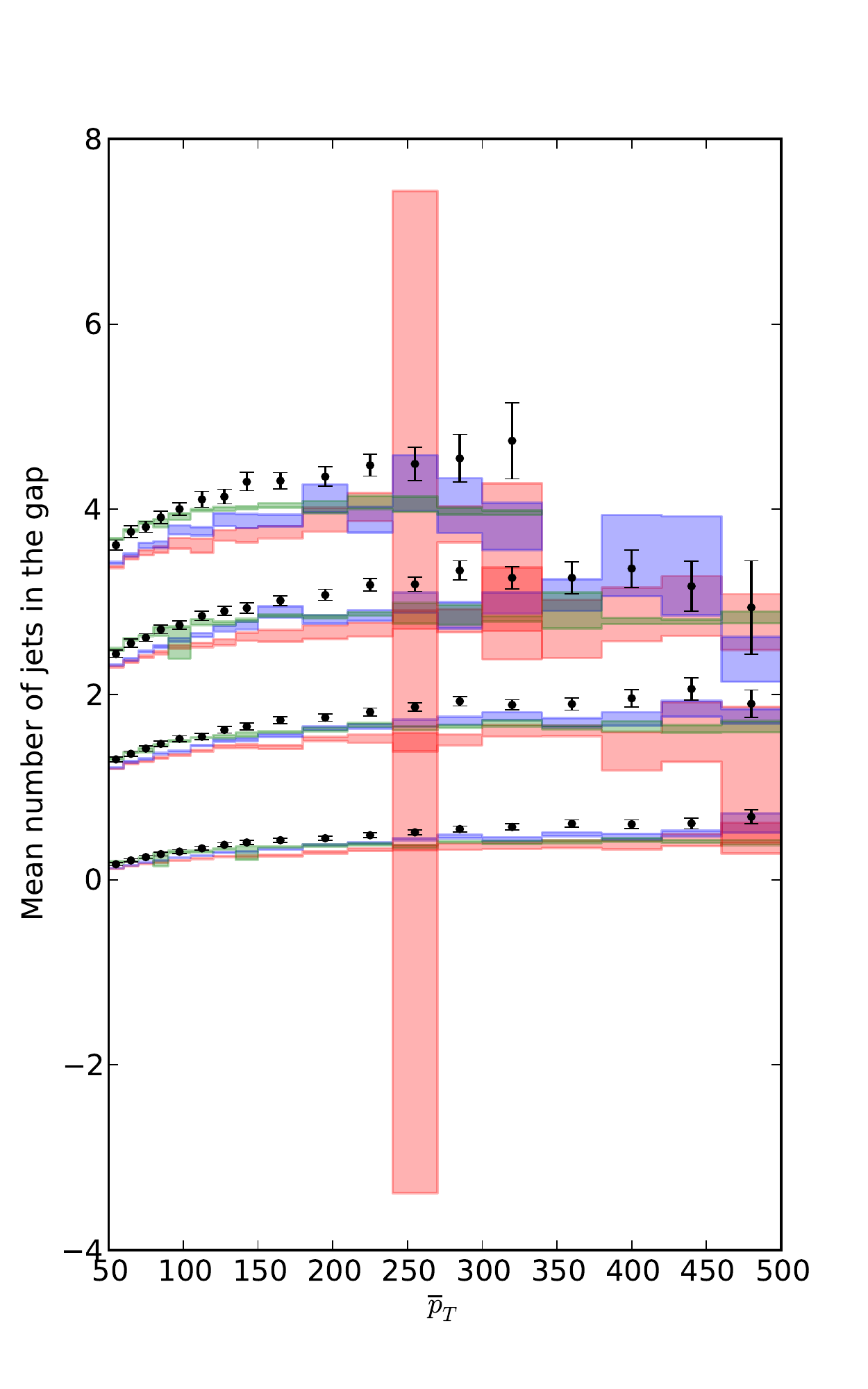}
\includegraphics[scale=0.45]{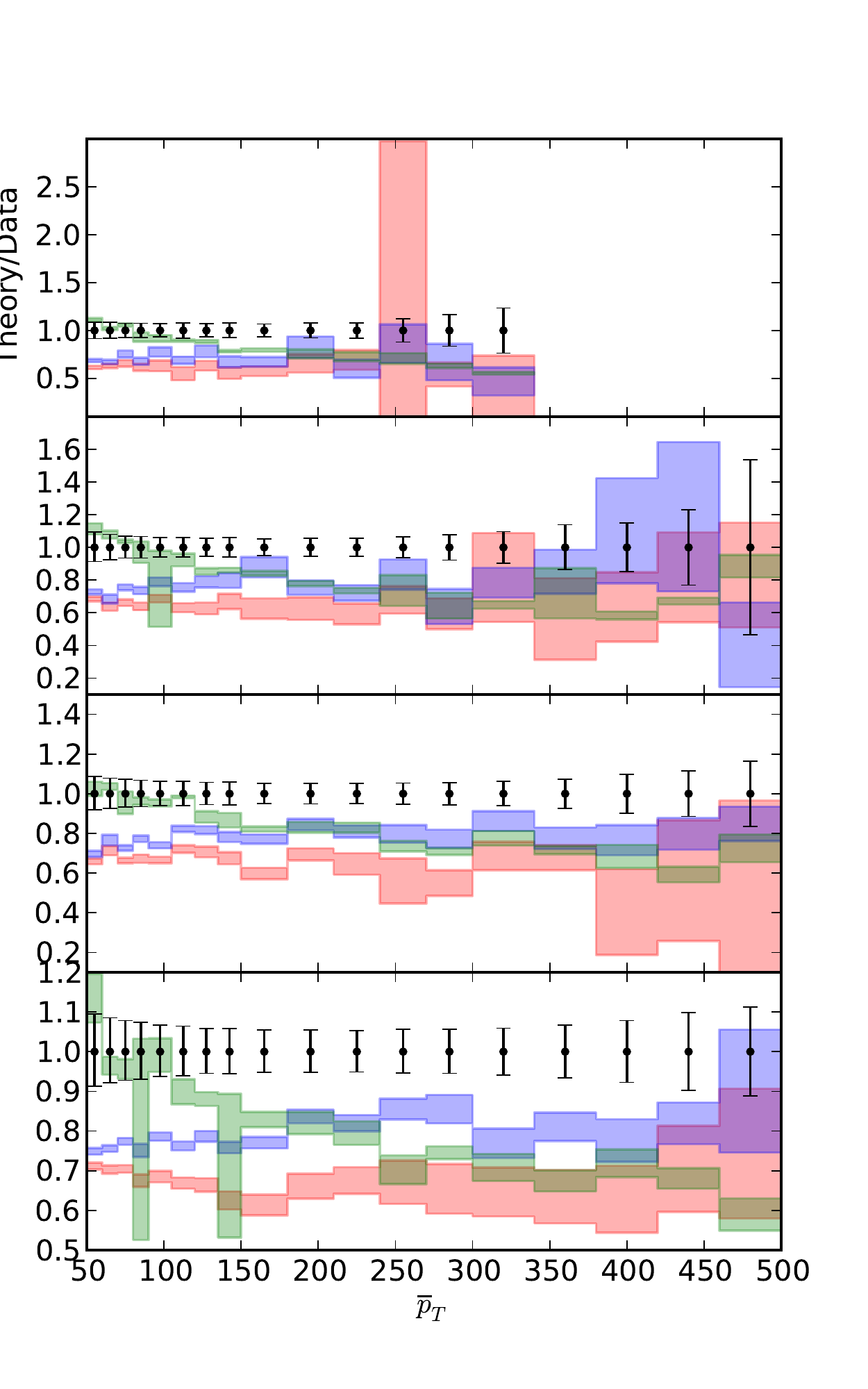}
\caption{Mean number of jets in the gap as a function of $\bar p _T$ for various slices of $\Delta y$. The jets defining  $\bar p _T$ and $\Delta y$ are the two jets with the largest $p_T$. }
\label{fig:6a}
\end{figure}

The mean number of jets in the relevant region of rapidity is obviously defined as 
\begin{eqnarray}
<N_{gap}>=\frac{\sigma_{g=1}+2\sigma_{g=2}+3\sigma_{g=3}+\dots}{\sigma_{tot}}.
\end{eqnarray}
Based on NLO-calculations, we can construct equivalent predictions for the
average number of jets as
\begin{eqnarray}\label{ngagExSum}
<N_{gap}>=\frac{1(\sigma_{g=1}^{j=3}+\sigma_{g=1}^{j>=4})+2\sigma_{g=2}^{j>=4}+3\sigma_{g=3}^{j=5}}{\sigma^{\rm{NLO},j=2}+\sigma^{\rm{NLO},j=3}+\sigma^{\rm{NLO},j\leq 4}}
\end{eqnarray}
and
\begin{eqnarray}\label{ngapNLO}
<N_{gap}>=\frac{1(\sigma_{g=1}^{j=3}+\sigma_{g=1}^{j=4,LO})+2\left(\sigma_{g=2}^{j=4,LO}\right)}{\sigma^{j\leq 2,\rm{NLO}}}
\end{eqnarray}
In the case where the jets defining the dijet system are the forward/backward ones, the formulae simplify to
\begin{eqnarray}
<N_{gap}>=\frac{1\sigma_{g=1}^{j=3}+2\sigma_{g=2}^{j=4}+3\sigma_{g=3}^{j=5}}{\sigma^{\rm{NLO},j=2}+\sigma^{\rm{NLO},j=3}+\sigma^{\rm{NLO},j=4}}
\end{eqnarray}
and
\begin{eqnarray}
<N_{gap}>=\frac{1(\sigma_{g=1}^{j=3})+2\left(\sigma_{g=2}^{j=4,LO}\right)}{\sigma^{\rm{NLO},j\leq 2}}
\end{eqnarray}
In Figure~\ref{fig:6a} we show the average number of jets in the gap as a function of the average transverse momentum for the last four $\Delta y$ slices in Eq.~\ref{deltaySlices}. Both the HEJ and the NLO prediction undershoot the data with an increasing difference at large average transverse momentum. The ratio looks similar for the all slices for the HEJ prediction while the disagreement gets worse for increasing rapidity difference for the NLO predictions.   

\subsection{Conclusions}
In this contribution we have compared predictions for gap probabilities and
related observables from HEJ and BlackHat+Sherpa with experimental data from
the ATLAS collaboration. Our analysis shows that both HEJ and NLO can provide
a very good (and fully perturbative) description of e.g.~the gap fraction in
the full range of rapidities and all slices of $\bar p _T$
(Fig.~\ref{fig:3a}). 

However, neither HEJ nor the NLO combination provide a perfect description of
the data when it comes to e.g.~the average number of soft (20GeV) jets in
events with two much harder jets (Figure~\ref{fig:6a}. This is perhaps not
surprising, since this is the region where the systematic resummation
provided by a parton shower is necessary.

Furthermore, there is no clear preference for one or the other NLO
'stategies' to define the gap fraction. 
In future work we hope to understand to which extent perturbative methods can describe the gap fraction and extend the study to all distributions in \cite{Aad:2011jz}. The statistical samples used for this analysis leaves us with quite large uncertainties, for further studies more statistics will be needed.

\clearpage

\section[Uncertainties in ${pp\to h + 2}$ jets production through 
         gluon fusion\footnote{%
  J.R.~Andersen, 
  F.U.~Bernlochner, 
  D.~Gillberg, 
  K.~Hamilton, 
  T.~Hapola, 
  J.~Huston, 
  A.~Kruse, 
  L.~L\"onnblad,
  P.~Nason, 
  S.~Prestel, 
  M.~Sch\"onherr,
  J.M.~Smillie, 
  K.~Zapp}]{Uncertainties in $\mathbf{pp\to h + 2}$ jets production through 
         gluon fusion\footnote{%
  J.R.~Andersen, 
  F.U.~Bernlochner, 
  D.~Gillberg, 
  K.~Hamilton, 
  T.~Hapola, 
  J.~Huston, 
  A.~Kruse, 
  L.~L\"onnblad,
  P.~Nason, 
  S.~Prestel, 
  M.~Sch\"onherr,
  J.M.~Smillie, 
  K.~Zapp}}
\subsection{INTRODUCTION}
\label{hdijet:sec:intro}
The process of Higgs boson production in association with dijets is of
interest for several reasons. The interference between the two production
channels of Weak-boson fusion (WBF) and gluon fusion (GF) is
insignificant \cite{Andersen:2007mp,Bredenstein:2008tm,Dixon:2009uk}, so it is
justified to discuss the processes separately. This study is concerned with
the description of the gluon fusion contribution to Higgs boson production in
association with dijets. 

The gluon fusion channel is of interest not just as a background to the
direct study through WBF of the couplings between the weak bosons and the
Higgs boson, but also because the structure of the heavy-quark-mediated
coupling of the Higgs boson to gluons can be studied beyond the question of
normalisation. The structure of the higgs-couplings means that for a
$CP$-specific coupling the cross section displays a striking dependence on
the azimuthal angle between rapidity-separated jets. In contrast, the
weak-boson-fusion component is relatively
flat\cite{Plehn:2001nj,Klamke:2007cu}.

In this region of phase space of large rapidity spans of the jets, the
component of gluon fusion and weak-boson fusion exhibit very different
patterns of radiative corrections. This is caused by a dominant colour octet
exchange between the dijets in the case of gluon-fusion, whereas the
weak-boson fusion component has no colour exchange between the
jets \cite{Dokshitzer:1987nc,Dokshitzer:1991he}. For the gluon fusion sample,
this enhances the contribution from jet multiplicities beyond the two
required, and necessitates a generalisation of the Born-level jet analysis to
stabilise the extraction of the azimuthal
dependence \cite{Andersen:2010zx}. Furthermore, the enhanced contribution from
further hard radiation severely tests the current limits of the perturbative
description.

With a satisfactory description of each component, the very different
radiation patterns in WBF and GF can be utilised to study each contribution
separately (for extraction and verification of the $CP$-structure of the
couplings).

The purpose of the current study is to investigate the variation in the
current description of the gluon-fusion contribution, in particular in the
testing region of the cuts employed to enhance the weak-boson fusion
contribution to Higgs boson production in association with dijets.


\subsection{GENERATORS}
\label{hdijet:sec:gens}

In this section all generators employed in this study, including their 
settings and methods of determining the uncertainties, are briefly reviewed.

\subsubsection{aMC@NLO}
\label{hdijet:sec:amcatnlo}

Events were generated using {\sc MadGraph}5\_aMC@NLO 2.0~\cite{Alwall:2011uj} 
for exclusive jet multiplicities ($pp\to h+0j$, $pp\to h+1j$, $pp\to h+2j$), 
each at NLO accuracy.  The virtual corrections were taken from {\sc Mcfm} 
\cite{Campbell:2006xx,Campbell:2010cz}. 
Jet multiplicities are then merged using the {\sc FxFx} merging prescription 
\cite{Frederix:2012ps}.  Events are then showered with 
{\sc Herwig}\scalebox{0.8}{++}~\cite{Bahr:2008pv}, where events 
are matched to the hard process.  To avoid double counting, the jet 
multiplicity after showering must match the jet multiplicity of the 
hard process for which NLO accuracy is desired.  This means that only 
events with 0 jets after showering are kept if they were generated with 
0 partons in the matrix element.  Because NLO is only desired for Higgs 
in association with up to two jets, events which have more than two jets 
after showering are kept if the hard process was generated with two jets.  
In addition, jets after showering must be matched with the hard partons 
of the generated event by imposing a $\Delta R=1$ requirement between jets 
before and after showering. All matched jets after showering must have 
$p_\perp$ larger than the merging scale.

Events were produced using the CT10 parton distribution function 
\cite{Lai:2010vv} throughout. The renormalisation and factorisation 
scales, $\mu_R$ and $\mu_F$, are set following the MiNLO procedure 
\cite{Hamilton:2012np}. The merging scale was chosen to be 30 GeV 
for the nominal sample. Scale variations were found by varying 
$\mu_R$ and $\mu_F$ up and down by factors of two.
%

\subsubsection{\textsc{Hej}}
\label{hdijet:sec:hej}

High Energy Jets ({\sc Hej}) describes hard, wide angle (high energy-)
emissions to all orders and to all multiplicities. The predictions are based
on events generated according to an all-order resummation, merged with
high-multiplicity full tree-level matrix-elements. The explicit resummation
is built on an approximation to the n-parton hard scattering matrix element
\cite{Andersen:2009nu,Andersen:2009he,Andersen:2011hs} which becomes exact in
the limit of wide-angle emissions, ensuring leading logarithmic accuracy for
both real and virtual corrections. These logarithmic terms are important when
the partonic invariant mass is large compared to the typical transverse
momentum in the event. This is precisely the situation which arises in
typical ``VBF'' cuts, including those used in this study.  Matching to the
full tree level accuracy for up to three jets is obtained by supplementing
the resummation with a merging
procedure~\cite{Andersen:2008ue,Andersen:2008gc}.

The implementation of this framework in a fully-flexible Monte Carlo event
generator is available at \texttt{http://hej.web.cern.ch}, and produces
exclusive samples for events with at least two jets.  The predictions include
resummation also for events with up to two un-ordered emissions,
i.e.~contributions from the first sub-leading configurations~\cite{Andersen:2014xx}.

The factorisation and renormalisation scales can be chosen arbitrarily, just
as in a standard fixed-order calculation. Here, we have chosen to evaluate
two powers of the strong coupling at a scale given by the Higgs mass, and for
the central predictions the remaining scales are evaluated at $\mu_R=H_T/2$. Thus,
for the $n$-jet tree-level evaluation, 
\begin{equation}
  \alpha_s^{n+2}=\alpha^2_s(m_h)\cdot \alpha^n_s(\mu_R).
\end{equation}
The scale variation bands shown in the plots here correspond to varying
$\mu_F$ and $\mu_R$ independently by a factor of two in either direction, but
discarding evaluations where any ratio is bigger than two. The CT10nlo
\cite{Lai:2010vv,Gao:2013xoa} parton distribution functions were used in the
predictions.


\subsubsection{\textsc{PowhegBox}}
\label{hdijet:sec:powhegbox}

Samples were generated using the latest version of \textsc{PowhegBox} 
(SVN revision 2550). The produced Les Houches files were
passed through \textsc{Pythia}~8 \cite{Sjostrand:2007gs} (v8.1.6) 
for parton showering (ISR+FSR). This was done for default 
\textsc{Powheg} \texttt{ggF}, \texttt{MiNLO-HJ} and \texttt{MiNLO-HJJ} 
for each of the scale 
variation settings. CT10 \cite{Lai:2010vv} was used both for the 
matrix element and hardest emission calculation in {\sc PowhegBox} 
and parton showering in \textsc{Pythia}~8 \cite{ATLAS:2011krm,ATLAS:2012uec}. 
The total cross section of these samples are summarised in 
Tab.\ \ref{hdijet:sec:powhegbox}.

\begin{table}[t]
  \begin{center}
    \begin{tabular}{l|ccccccc}\hline\hline
      $(\mu_R{}/m_h,\mu_F{}/m_h)$ & (0.5,0.5) & (0.5,1) & (1,0.5) & (1,1)
      & (1,2) & (2,1) & (2,2) \\\hline
      {\sc Powheg} \texttt{ggF} &
      15.83 &
      15.90 &
      13.20 &
      13.29 &
      13.34 &
      11.27 &
      11.31\\
      \texttt{MiNLO-HJ} &
      16.58 &
      16.68 &
      12.32&
      12.58&
      12.74&
      10.76&
      10.90\\
      \texttt{MiNLO-HJJ} &
      22.36 &
      23.58 &
      16.89 &
      17.74 &
      18.19 &
      13.14 &
      13.35 
      \\\hline\hline
    \end{tabular}
  \end{center}
  \caption{
            Inclusive $pp\to h$ cross sections in pb for \textsc{Powheg} 
            \texttt{ggF}, \texttt{MiNLO-HJ} and \texttt{MiNLO-HJJ} 
            for various scale choices.
            \label{hdijet:tab:powhegbox_xs}
          }
\end{table}

The {\sc Pythia}~8 default parton showering interface to the 
{\sc PowhegBox} was used that picks up the \texttt{SCALUP} parameter 
from the LesHouches file and restricts parton shower emissions to 
this scale (aka ``wimpy showers'').  Parton showering/resummation 
uncertainties were evaluated by producing separate samples  
with the \texttt{SCALUP} parameter varied by factors of 0.75 and 1.5.
A cross check was performed replacing the \texttt{SCALUP} restriction 
by the ``power shower''+``$p_{\rm T}$ veto'' approach\footnote
{
  See \texttt{Pythia8/examples/main31}.
}. The difference between the two approaches is (well) within the 
\texttt{SCALUP} variation.

Additional results switching off ISR and/or FSR and including 
hadronisation and/or MPI are provided at 
\texttt{https://dgillber.web.cern.ch/dgillber/Rivet/Powheg\_ggF.html}.

For the study presented in Sec.\ \ref{hdijet:sec:results} the 
\texttt{MiNLO-HJJ} setup was used throughout as it gives the highest formal 
accuracy for most observables under scrutiny. Please note: for 
the sake of consistency in describing correllations inbetween the 
different observable classes the \texttt{MiNLO-HJJ} setup was also 
used for zero- and one-jet inclusive observables where its formal 
accuracy falls below the \texttt{MiNLO-HJ} setup.

\subsubsection{\textsc{Pythia}~8}
\label{hdijet:sec:pythiaeight}

The {\sc unlops} method \cite{Lonnblad:2012ix} allows to incorporate 
multiple next-to-leading order calculations into a parton shower 
event generator. The implementation of this NLO merging scheme 
in \textsc{Pythia}~8 \cite{Sjostrand:2007gs} uses external next-to-leading 
order matrix element generators to produce events distributed 
according to inclusive NLO cross sections. Higher-order terms 
(of the {\sc umeps} merging scheme \cite{Lonnblad:2012ng}) are supplied 
by reweighting additional tree-level input samples. In the {\sc unlops} 
approach, the merged prediction for an observable is

\begin{eqnarray}
\label{hdijet:eq:unlops-full-inclusive}
\langle \mathcal{O} \rangle
 &=&
 \sum_{m=0}^{M-1}~ \int d\phi_0 \int\!\cdots\!\int 
     \mathcal{O}(S_{+mj})
     ~\Bigg\{
   \overline{\textnormal{B}}_{m}
 + \left[\vphantom{\sum}\widehat{\textnormal{B}}_{m}\right]_{-m,m+1}
\nonumber\\
&&\quad~
 - \sum_{i=m+1}^{M} \int_s\overline{\textnormal{B}}_{i\rightarrow m}
 - \sum_{i=m+1}^{M} \left[\vphantom{\sum}\int_s\widehat{\textnormal{B}}
                             _{i\rightarrow m}\right]_{-i,i+1}
 - \sum_{i=M+1}^{N} \int_s\widehat{\textnormal{B}}_{i\rightarrow m}
~\Bigg\}\nonumber\\
&&{}+
   \int d\phi_0 \int\!\cdots\!\int 
   \mathcal{O}(S_{+Mj})\Bigg\{~
   \overline{\textnormal{B}}_{M}
 + \left[\vphantom{\sum}\widehat{\textnormal{B}}_{M}\right]_{-M,M+1}
 - \sum_{i=M+1}^{N} \int_s\widehat{\textnormal{B}}_{i\rightarrow M}~
   \Bigg\}\nonumber\\
&&{}+
  \sum_{n=M+1}^N
  \int d\phi_0 \int\!\cdots\!\int 
  \mathcal{O}(S_{+nj})~\left\{ \widehat{\textnormal{B}}_{n}
                               - \sum_{i=n+1}^{N} \int_s
                                   \widehat{\textnormal{B}}
                                     _{i\rightarrow n} ~\right\}~.
\end{eqnarray}
The notation is defined in \cite{Lonnblad:2012ix}. For this study, 
$M=2$ and $N=3$. The {\sc PowhegBox} generator \cite{Alioli:2010xd} is 
used to produce the inclusive NLO samples $\overline{\textnormal{B}}_{0}$ 
\cite{Alioli:2008tz}, $\overline{\textnormal{B}}_{1}$ and 
$\overline{\textnormal{B}}_{2}$ \cite{Campbell:2012am}, while additional 
tree-level inputs are taken from {\sc MadGraph/MadEvent} 
\cite{Alwall:2011uj,Maltoni:2002qb}. The merging scale criterion 
(see Appendix A.1 of \cite{Lonnblad:2012ix}) is used to remove infrared 
divergences due to additional jets at Born level. This cut is applied 
internally in {\sc Pythia}~8, so that no user interference with e.g.\ 
the {\sc PowhegBox} source code becomes necessary. All input samples are 
computed with fixed scales $\mu_F$ and $\mu_R$. Terms of 
$\mathcal{O}\left(\alpha_s^{n+2}\right)$ and higher due to running 
scales are included, in a controlled way, by reweighting tree-level 
samples with factors to produce the CKKW-L 
\cite{Lonnblad:2001iq,Lavesson:2005xu,Lonnblad:2011xx} scale setting 
prescription.

This NLO merging scheme replaces the approximate 
$\alpha_s^{n+1}$-terms of the {\sc umeps} tree-level merging method with 
the desired full fixed-order expressions, but otherwise leaves 
higher-order corrections untouched. Particularly, no higher-order 
terms are introduced by reweighting virtual corrections. Note 
further that the method keeps the inclusive cross section fixed to 
the zero-jet NLO cross section, without producing spurious logarithmic 
artifacts of the merging procedure, and without introducing 
additional finite terms.

To produce results, we generate fixed-order input events with 
{\sc PowhegBox} and {\sc MadEvent} at $E_\text{CM} = 8$ TeV, using 
CT10nlo \cite{Lai:2010vv,Gao:2013xoa} parton distributions, massless 
b-quarks, vanishing coupling between b-quarks and Higgs-boson and a 
stable Higgs-boson of mass $m_h = 125$ GeV. We define the merging 
scale as the minimal relative shower $p_{\perp\text{evol}}$ and use 
a merging scale value of $\rho_\text{ms} = 15$ GeV. The samples are 
merged within {\sc Pythia}~8.180. {\sc Pythia}~8 is run with Tune 
AU2-CTEQ6L1 \cite{ATLAS:2012uec}, which uses CTEQ6L1 
\cite{Pumplin:2002vw} parton distributions. Masses and widths are, 
naturally, set equal in {\sc PowhegBox}, {\sc MadEvent} and 
{\sc Pythia}~8. We assess uncertainties by producing two sets of 
variations:
\begin{itemize}
  \item Parton shower variation: Use {\sc PowhegBox} and {\sc MadEvent} inputs
	generated with $\mu_F=\mu_R=m_h$, and choose the {\sc Pythia}~8
	starting scale $\mu_Q$ for adding emissions to zero-parton
	events as $\mu_{Q,\text{low}}= \tfrac{1}{2}\,m_h$,
	$\mu_{Q,\text{central}}=m_h$ or $\mu_{Q,\text{high}}=2\,m_h$.
	Construct an envelope from the results of these variations.
	Only the starting scale for adding emissions to zero-parton
	events can be varied, as the starting scale for showering
	$n\!\geq\!1$-parton events is locked by the CKKW-L prescription.
  \item Matrix element variation: Vary the renormalisation and
	factorisation scales in {\sc PowhegBox} and {\sc MadEvent}
	as $\mu_{F,\text{low}} = \mu_{R,\text{low}} = \tfrac{1}{2}\,m_h$,
	$\mu_{F,\text{central}} = \mu_{R,\text{central}} = m_h$
	or $\mu_{F,\text{high}} = \mu_{R,\text{high}}= 2\,m_h$, keep
	the {\sc Pythia}~8 starting scale $\mu_Q$ for adding
	emissions to zero-parton events fixed at $\mu_Q=m_h$.
	Construct an envelope from the results of the variations.
\end{itemize}
We do not attempt a more complete uncertainty study including variation 
of renormalisation and factorisation scales in the parton shower, mainly 
because we have not decided on a strategy of how to deal with such 
uncertainties, and other fixed-order uncertainties, in the context of 
parton shower resummation and tuning. However, the envelope of the 
uncertainty bands presented here should, for most cases, yield a 
reasonably conservative estimate. Also, for this perturbative-physics 
driven study, we do not address the (potentially important) 
uncertainties due to non-perturbative modelling in the event generator.

\subsubsection{\textsc{Sherpa}}
\label{hdijet:sec:sherpa}

The {\sc Sherpa} event generator implements the {\sc meps@nlo} 
\cite{Hoeche:2012yf,Gehrmann:2012yg} technique to merge multiple 
{\sc nlops} matched \cite{Hoeche:2011fd} samples of successive jet 
multiplicities into an inclusive calculation that retains both 
the NLO accuracy for every individual jet multiplicity and the 
parton shower's resummation properties in resumming emission 
scale hierarchies wrt.\ inclusive process. This implementation 
has already been successfully applied to various background 
processes to Higgs production channels \cite{Cascioli:2013gfa,
  Cascioli:2013era}. Also Higgs production through gluon fusion 
has been studied in detail with this approach recently 
\cite{Hoeche:2014lxa}.

For the contribution to this comparative study the subprocesses 
$pp\to h+0$ jets, $pp\to h+1$ jet and $pp\to h+2$ jets are 
calculated at next-to-leading order and $pp\to h+3$ jets at 
leading order accuracy. The one-loop matrix elements are taken 
from \cite{Dawson:1990zj,Djouadi:1991tka}, \cite{deFlorian:1999zd,
  Ravindran:2002dc} and \cite{Campbell:2006xx,Campbell:2010cz}, 
respectively. The LO $pp\to h+3$ jets process, as it is merged on 
top of the NLO $pp\to h+2$ jets process by the \textsc{menlops} 
mehtod \cite{Gehrmann:2012yg}, receives a local $K$-factor in 
dependence of the $pp\to h+2$ jets composition in terms of 
$\mathbb{S}$- and $\mathbb{H}$-events. For the central prediction, 
the individual jet multiplicities were separated by 
$Q_\text{cut}=20$ GeV. The renormalisation and factorisation 
scales were set according to the CKKW prescription, i.e.\ 
$\alpha_s^{2+n}(\mu_R)=\alpha_s(m_h^2)\,\alpha_s(t_1)\cdots\alpha_s(t_n)$ 
and $\mu_F=m_h$ for the core process. The parton shower starting scale 
$\mu_Q$ was set to $m_h$. Throughout, the CT10nlo \cite{Lai:2010vv,
  Gao:2013xoa} PDF set was used.

The uncertainties on the central prediction were estimated by 
varying $\mu_R$ and $\mu_F$ independently by the customary factor 
of two and the parton shower starting scale by a factor $\sqrt{2}$. 
The merging scale was varied in the range $\{15,20,30\}$ GeV. 
Further, the intrisic parton shower uncertainties were estimated 
by varying the evolution variable and the recoil scheme according 
to \cite{Hoeche:2014lxa}. Non-perturbative uncertainties which may 
very well play a role in analyses looking into hadronic activity 
within jets widely separated in rapidity 
\cite{Hoche:2012wh,AlcarazMaestre:2012vp} are not assessed in this study.

\subsection{RESULTS}
\label{hdijet:sec:results}

This study aims at quantifying the gluon fusion contribution 
and its perturbative uncertainty to typical VBF Higgs plus two 
jets event selections at the LHC at a centre-of-mass energy of 8 TeV. 
To summarise Sec.\ \ref{hdijet:sec:gens}, 
the following parameters were then varied to estimate the 
respective generator's uncertainty:\vspace*{2mm}\\\vspace*{2mm}
\begin{tabular}{ll}
 \textsc{Hej}:& $\mu_R,\mu_F\in [\tfrac{1}{2},2]$ separately \\
 a\textsc{Mc@Nlo}:& $\mu_R,\mu_F\in [\tfrac{1}{2},2]$ \\
 \textsc{PowhegBox}: & $\mu_R,\mu_F\in [\tfrac{1}{2},2]$ separately, $\texttt{SCALEUP}\in[0.75,1.5]$ \\
 \textsc{Pythia}~8: & $\mu_R,\mu_F\in [\tfrac{1}{2},2]$, $\mu_Q\in[\tfrac{1}{\sqrt{2}},\sqrt{2}]$ \\
 \textsc{Sherpa}: & $\mu_R,\mu_F\in [\tfrac{1}{2},2]$ separately, $\mu_Q\in[\tfrac{1}{\sqrt{2}},\sqrt{2}]$, $Q_\text{cut}\in[15,20,30]$
\end{tabular}\\
A common {\sc Rivet} \cite{Buckley:2010ar,RivetAnalysis} 
analysis implementing the observables for the event selections 
described below has been used.
This analysis is inclusive with respect to the Higgs, i.e.\ there 
are no phase space cuts on the Higgs and all decay channels are 
included. 
Concerning the presence of additional jets the analysis defines 
several stages of cuts. 

\begin{figure}[t!]
  \includegraphics[width=0.47\textwidth]{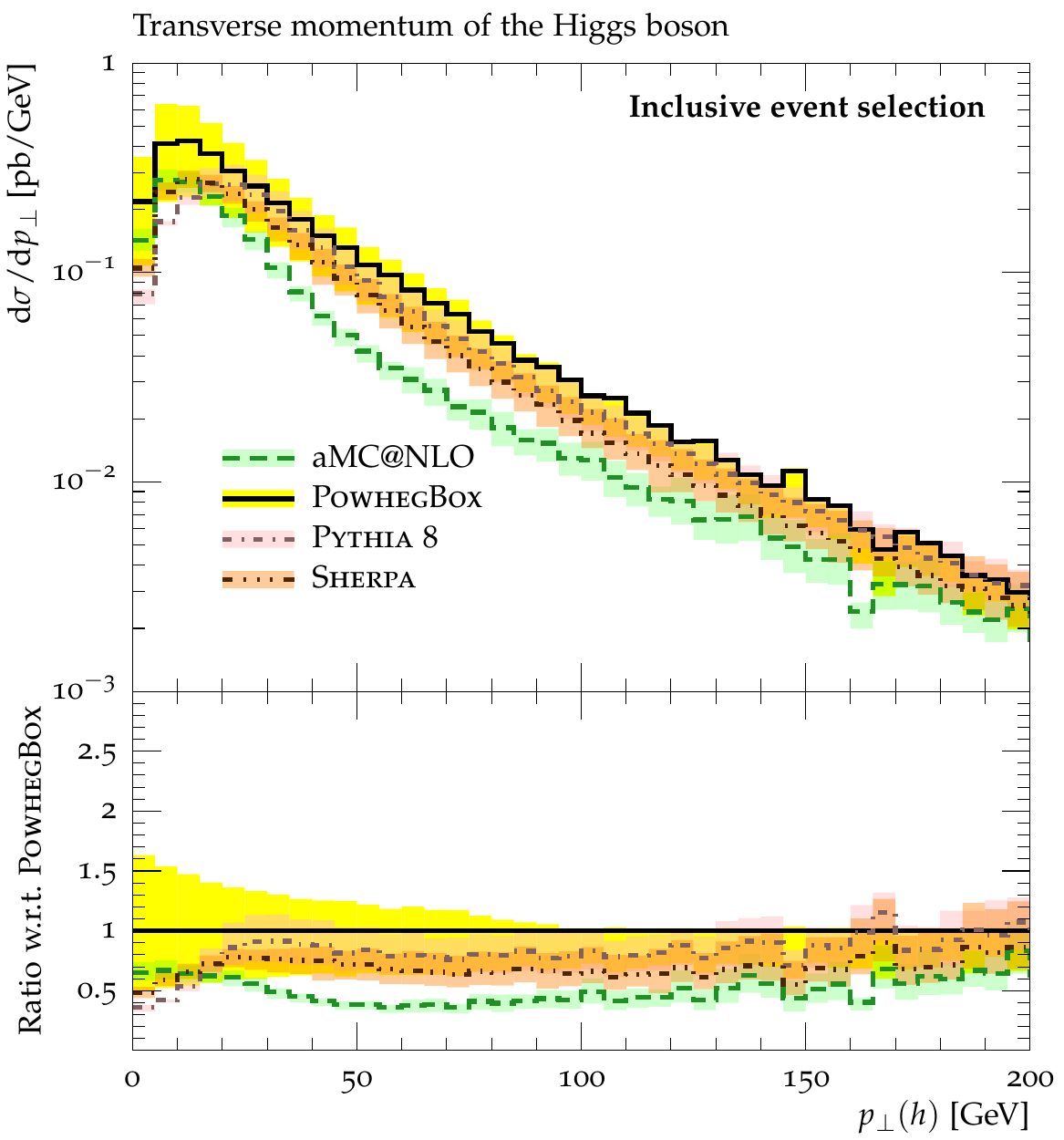}\hfill
  \includegraphics[width=0.47\textwidth]{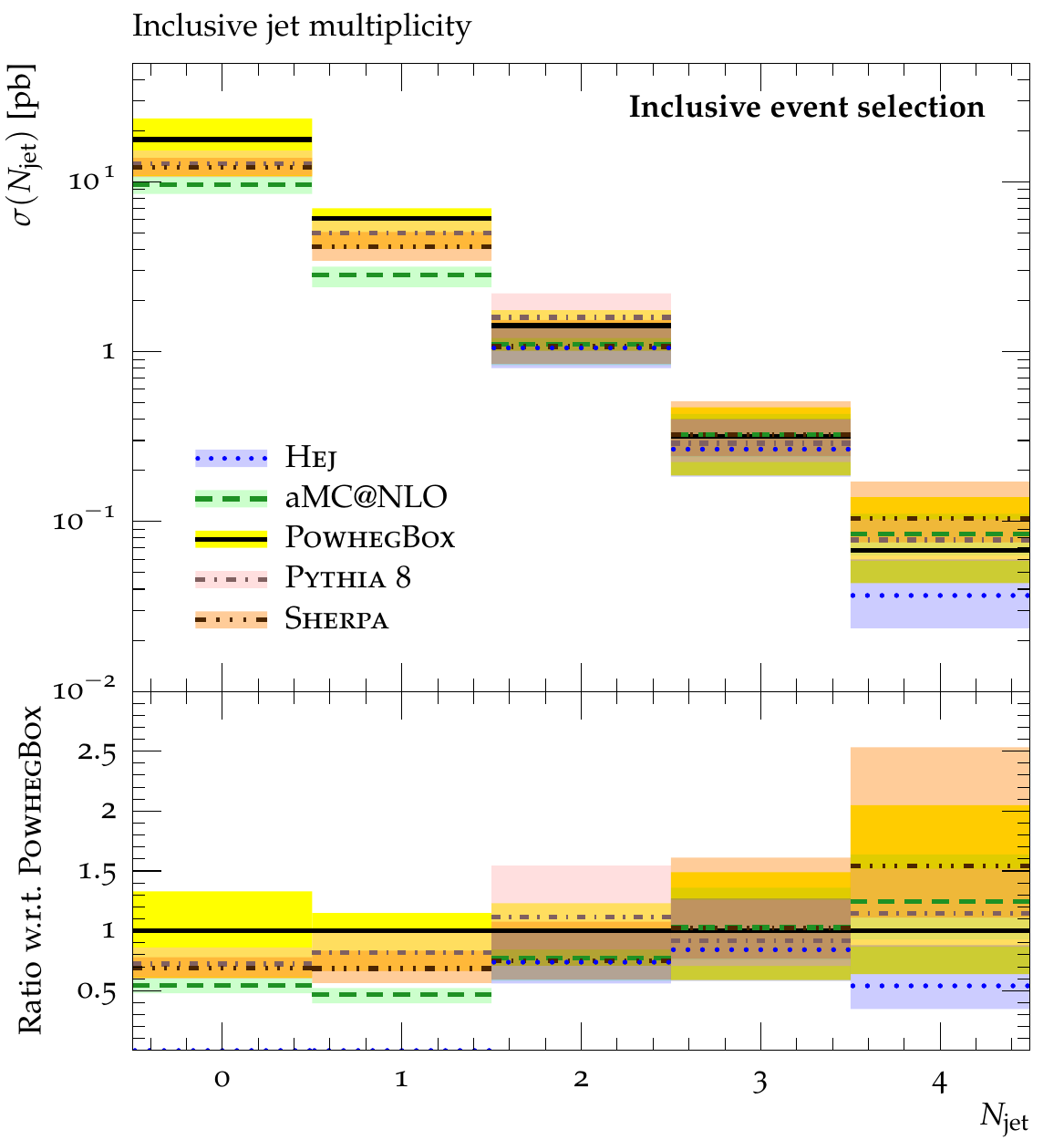}
  \caption{
            Transverse momentum of the Higgs boson (left) and 
            inclusive jet multiplicity (right) as predicted by 
            the different generators. 
	    The individual sources of uncertainties used to generate 
	    the respective bands are described in Sec.\ \ref{hdijet:sec:gens}.
	    Note that for \textsc{PowhegBox} the \texttt{MiNLO-HJJ} 
	    prediction has been used throughout, despite its lowered 
	    accuracy for inclusive and one-jet observables, for 
	    consistencies sake.
            \label{hdijet:fig:hpt_njet}
          }
\end{figure}

\smallskip

First we study the transverse momentum of the Higgs boson for inclusive 
Higgs production as obtained in the shower based approaches to study 
the overall consistency of their fixed-order and resummation properties.
The results are shown in Fig.\ \ref{hdijet:fig:hpt_njet} on the left hand 
side. \textsc{PowhegBox}, \textsc{Pythia}\,8 and \textsc{Sherpa} 
are compatible within uncertainties, where 
\textsc{PowhegBox} reports somewhat larger central values and 
uncertainties due to the usage of the \texttt{MiNLO-HJJ}\footnote
{
  As this study focusses on $pp\to h+2$ jets topologies \texttt{MiNLO-HJJ} 
  is used throughout despite its reduced uncertainty for more inclusive 
  observables for consistencies sake.
}, cf.\ 
Tab.\ \ref{hdijet:tab:powhegbox_xs}. aMC@NLO has a similar description 
of the Sudakov region but its predictions fall below those of 
the other generators for larger transverse momenta. \textsc{Hej} 
generates at least two jets and thus is not included in the 
comparison for this observable. 

The following observables all depend on the presence of at least one jet. 
Jets are reconstructed using the anti-kt algorithm provided 
by \textsc{FastJet}~\cite{Cacciari:2011ma} with radius parameter 
$R=0.4$ within the pseudo-rapidity range $|\eta| < 5$ and are 
required to have a minimal transverse momentum of 
$p_\perp > 30$ GeV. 
The predictions for the number of jets, shown in Fig.\ 
\ref{hdijet:fig:hpt_njet} on the right hand side, are largely 
consistent, only aMC@NLO has somewhat less cross section 
in the zero- and one-jet inclusive bins. The inclusive 
dijet cross section is consistent within uncertainties 
for all generators. From here on, as a minimum of two jets 
accompanying the Higgs are always required, the \textsc{Hej} 
prediction is included and it is found to be in good agreement 
with the shower based approaches. 
For the higher inclusive jet multiplicities the spread and also 
the uncertainties are significantly larger. This is expected, 
as the theoretical accuracy decreases there for all generators. 
This is especially true for the inclusive four jet rate 
since -- with the exception of \textsc{Hej} -- the fourth jet 
comes from the parton shower only. 

\begin{figure}[t]
  \includegraphics[width=0.47\textwidth]{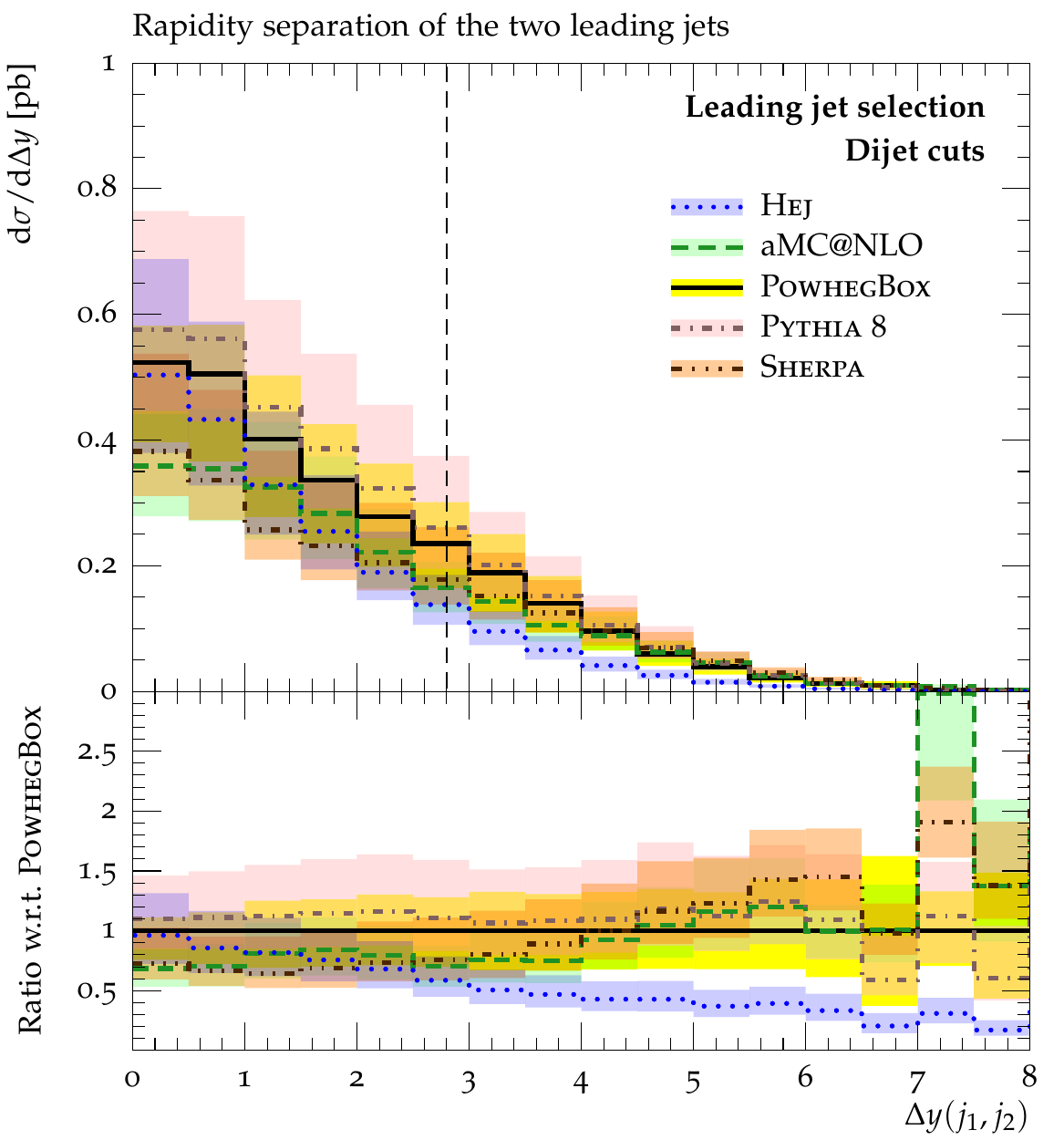}\hfill
  \includegraphics[width=0.47\textwidth]{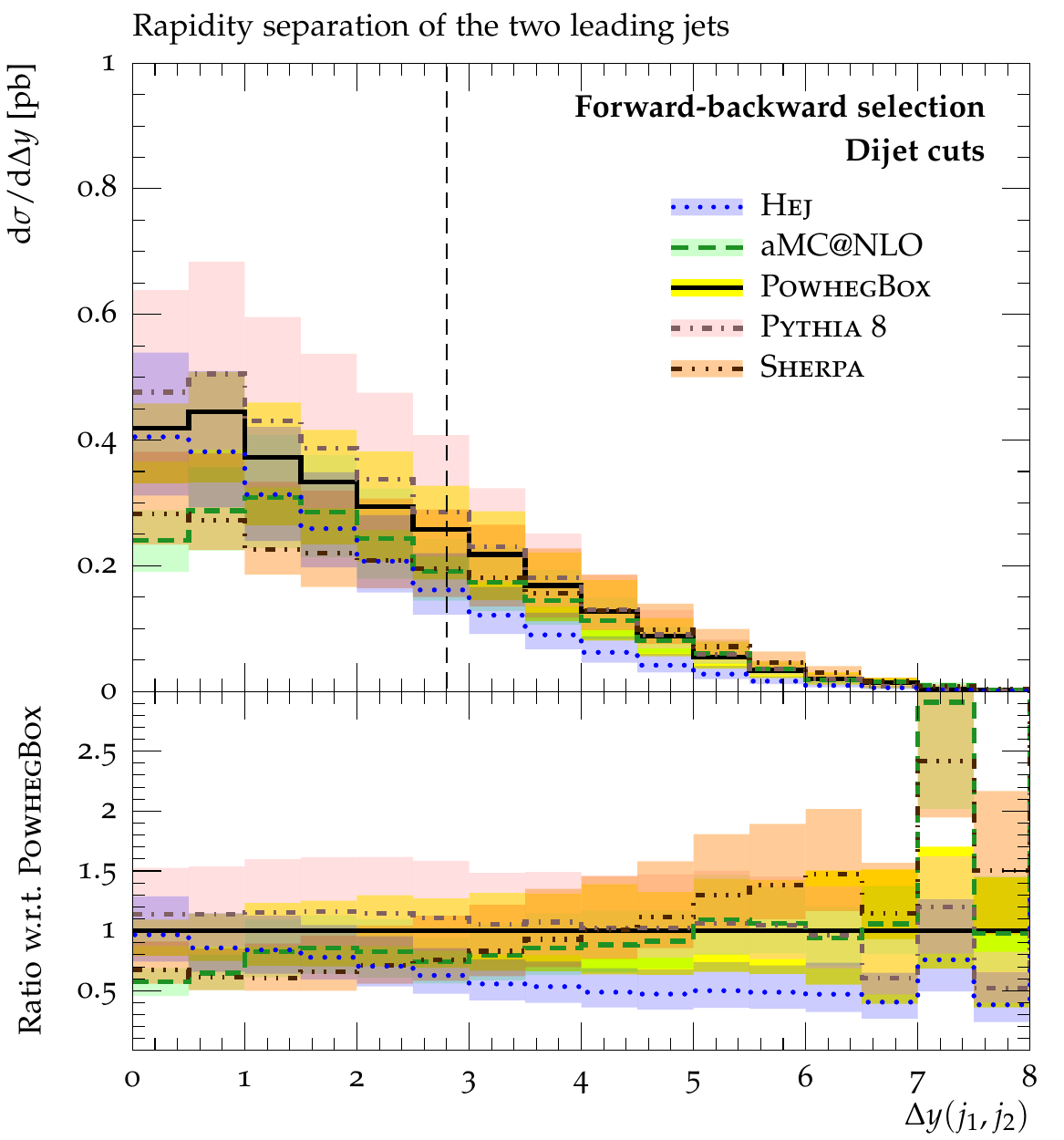}
  \caption{
            Rapidity difference of both tagging jets in the leading jet 
            selection (left) and the forward-backward selection (right)
            as predicted by the different generators. 
	    The individual sources of uncertainties used to generate 
	    the respective bands are described in Sec.\ \ref{hdijet:sec:gens}.
	    The location of the VBF cut is indicated.
            \label{hdijet:fig:dyjj_AB}
          }
\end{figure}

\begin{figure}[t!]
  \includegraphics[width=0.47\textwidth]{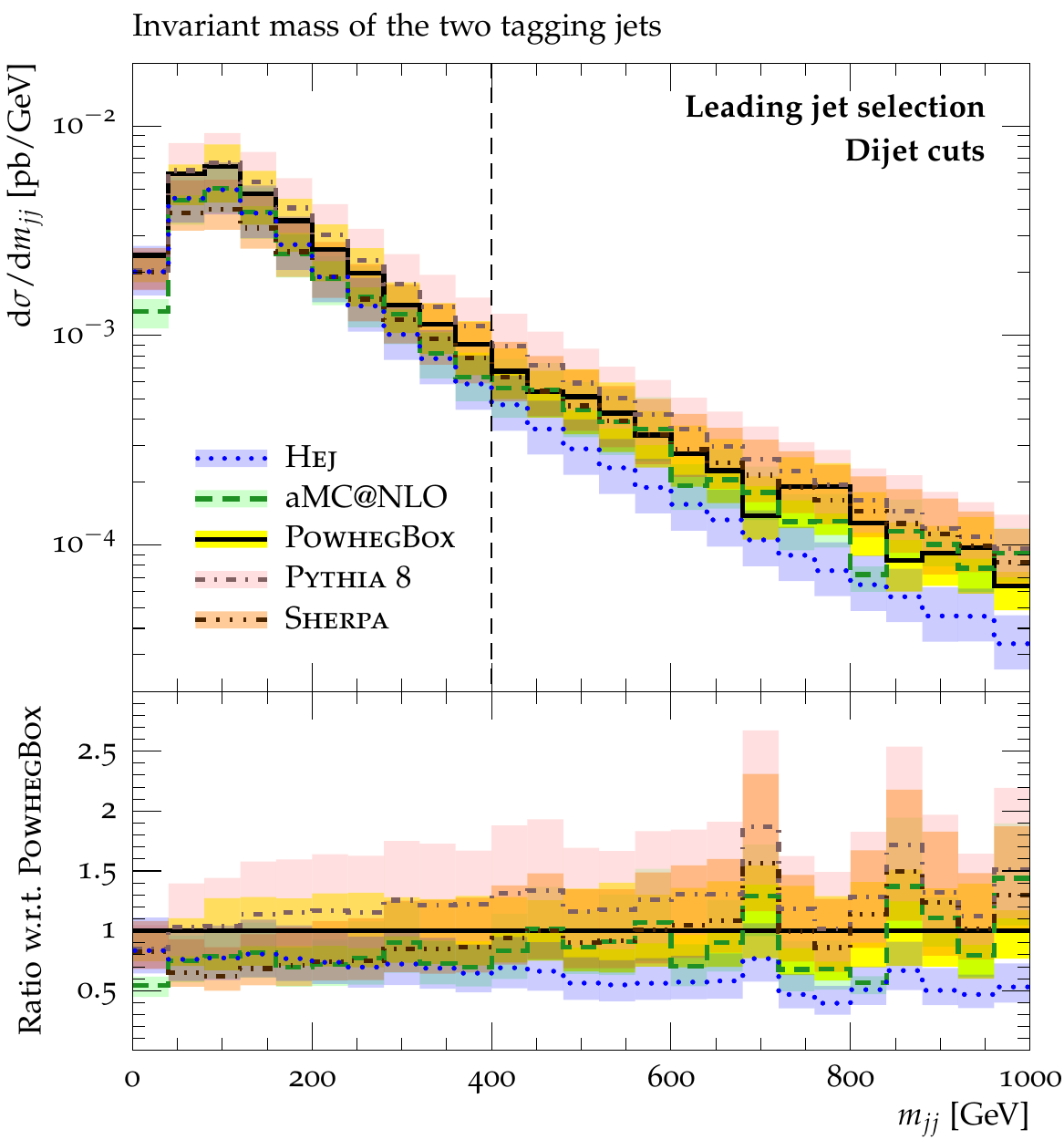}\hfill
  \includegraphics[width=0.47\textwidth]{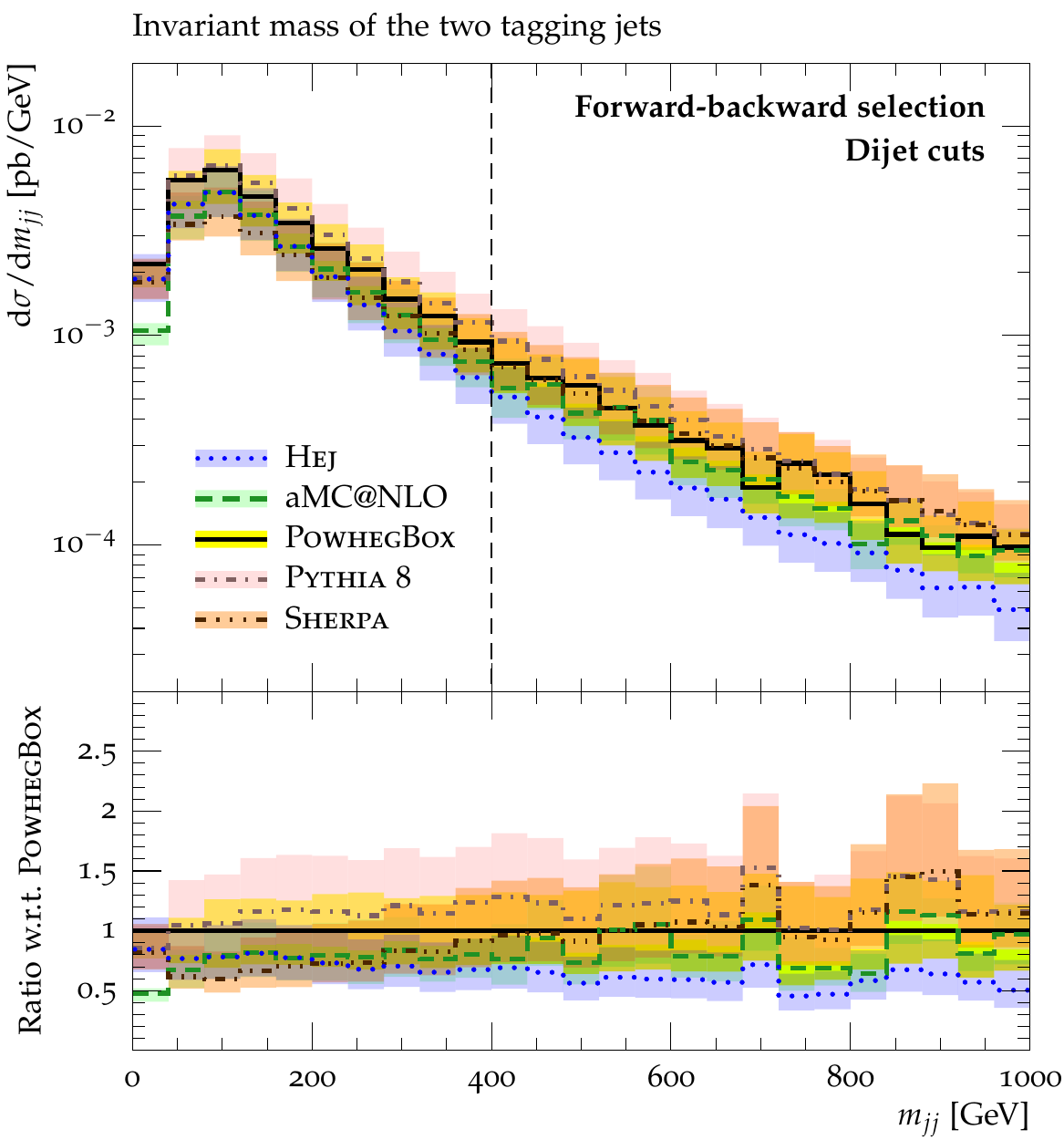}
  \caption{
            Invariant mass of both tagging jets in the leading jet 
            selection (left) and the forward-backward selection (right)
            as predicted by the different generators. 
	    The individual sources of uncertainties used to generate 
	    the respective bands are described in Sec.\ \ref{hdijet:sec:gens}.
	    The location of the VBF cut is indicated.
            \label{hdijet:fig:mjj_AB}
          }
\end{figure}

For the Higgs+dijet selection two 
alternatives are considered: In the \emph{leading jet selection} 
the two hardest jets in the event constitute the tagging jets, while 
in the \emph{forward-backward selection} the most forward and 
the most backward (in rapidity) jet are defined as tagging jets. 
For events with exactly two jets the two selections are obviously 
identical, for events with more than two jets they can differ. 
This general dijet selection defines the second level of cuts. 
The third level is a VBF-like selection in which the two tagging 
jets are required to have a rapidity separation 
$\Delta y(j_1,j_2)>2.8$ and a large invariant mass 
$m_{jj}>400$ GeV. The invariant mass and rapidity 
separation of the tagging jets are shown in Figs.\ 
\ref{hdijet:fig:dyjj_AB} and \ref{hdijet:fig:mjj_AB}, 
respectively, for both dijet selections before the VBF cuts. 
As expected, the rapidity separation between the two tagging jets, 
Fig.\ \ref{hdijet:fig:dyjj_AB}, is larger in the forward-backward 
than in the leading jet selection. In this observable differences 
between the generators become visible: While \textsc{PowhegBox} 
and \textsc{Pythia}\,8 are consistent in shape but differ slightly 
in normalisation, \textsc{Sherpa} and aMC@NLO  predict significantly 
more cross section at large $\Delta y$ but also differ in normalisation. 
A possible origin of such differences within the parton shower based 
event generators is the behaviour of their respective parton shower, 
which affects distributions both explicitly by radiating additional 
partons and (for some approaches) through the computation of 
Sudakov form factors. The slightly higher cross section in the 
\textsc{Pythia}~8 prediction can, for example, be traced to the 
high value of $\alpha_s$ in Tune AU2-CTEQ6L1. \textsc{Hej}, on 
the contrary, falls off faster 
at large $\Delta y$. This can be understood from the 
systematic resummation of virtual correction in $\log \frac{s}{t}$ 
which become important at large $\Delta y$. In the leading jet 
selection there is the additional effect that when jets are produced 
at large rapidity differences the void in between them tends to be 
filled with additional hard jets unordered in $p_\perp$. 
These differences are, however, partially covered by the uncertainties. 

The invariant mass distribution is almost identical in the two 
dijet selections. The predictions are all consistent in shape 
but differ somewhat in normalisation with \textsc{Pythia}~8 at the 
upper and \textsc{Hej} on the lower end of the normalisation spectrum. 
Noteworthy is that \textsc{Sherpa}, and to some extent also 
\textsc{Pythia}~8, predicts more cross 
section at larger $m_{jj}$ than the other generators, while 
\textsc{Hej} predicts slightly less. Here a discussion of how the 
aMC@NLO, \textsc{PowhegBox}, \textsc{Pythia}~8 and \textsc{Sherpa} 
treat configurations with at least three jets which make up a large 
fraction of the events in the high invariant mass as well as the large 
$\Delta y$ regions, seems expedient. Treating three jet configurations 
as a separate contribution, as is the case in \textsc{Pythia}~8 and 
\textsc{Sherpa}, leads to assigning them a renormalisation scale 
defined through $\alpha_s^{2+3}(\mu_R)=\alpha_s^2(m_h)\,\alpha_s(t_1)\,
\alpha_s(t_2)\,\alpha_s(t_3)$ for ordered jet emissions in the CKKW 
approach ($t_i$ being 
the reconstructed emission/clustering scales of the respective approach) 
whereas treating three jet configurations as real emission corrections 
to the underlying two jet configuration would result in assigning them a 
coupling $\alpha_s^{2+3}(\mu_R)$ where the renormalisation scale is defined 
through that underlying two jet configuration 
as $\alpha_s^{2+2}(\mu_R)=\alpha_s^2(m_h)\,\alpha_s(t_1)\,\alpha_s(t_2)$. 
Though the difference is of higher order, the latter definition clearly 
leads to higher values of $\mu_R$ in most cases, and therefore lower 
cross sections. However, note 
that this simple picture is significantly muddied and partly remedied 
by the treatment of the real emission configuration in \textsc{Powheg}-like 
approaches and the interplay of $\mathbb{S}$- and $\mathbb{H}$-events in 
MC@NLO-like calculations. In case of \textsc{Hej}, the slight depletion of 
the cross section at large $m_{jj}$ is also closely related to that at large 
$\Delta y$: instigated by the filling of large rapidity intervals through 
emissions unordered in transverse momentum leads to a potentially 
different selection of tagging jets.
In conclusion, however, despite all differences in the respective 
calculations, the resulting predictions are largely covered by the 
uncertainties. Still, the differences seen for $\Delta y$ and 
$m_{jj}$ have a non-negligible effect on the efficiency of the 
VBF cuts, as seen in the following.

\begin{figure}[t!]
  \includegraphics[width=0.47\textwidth]{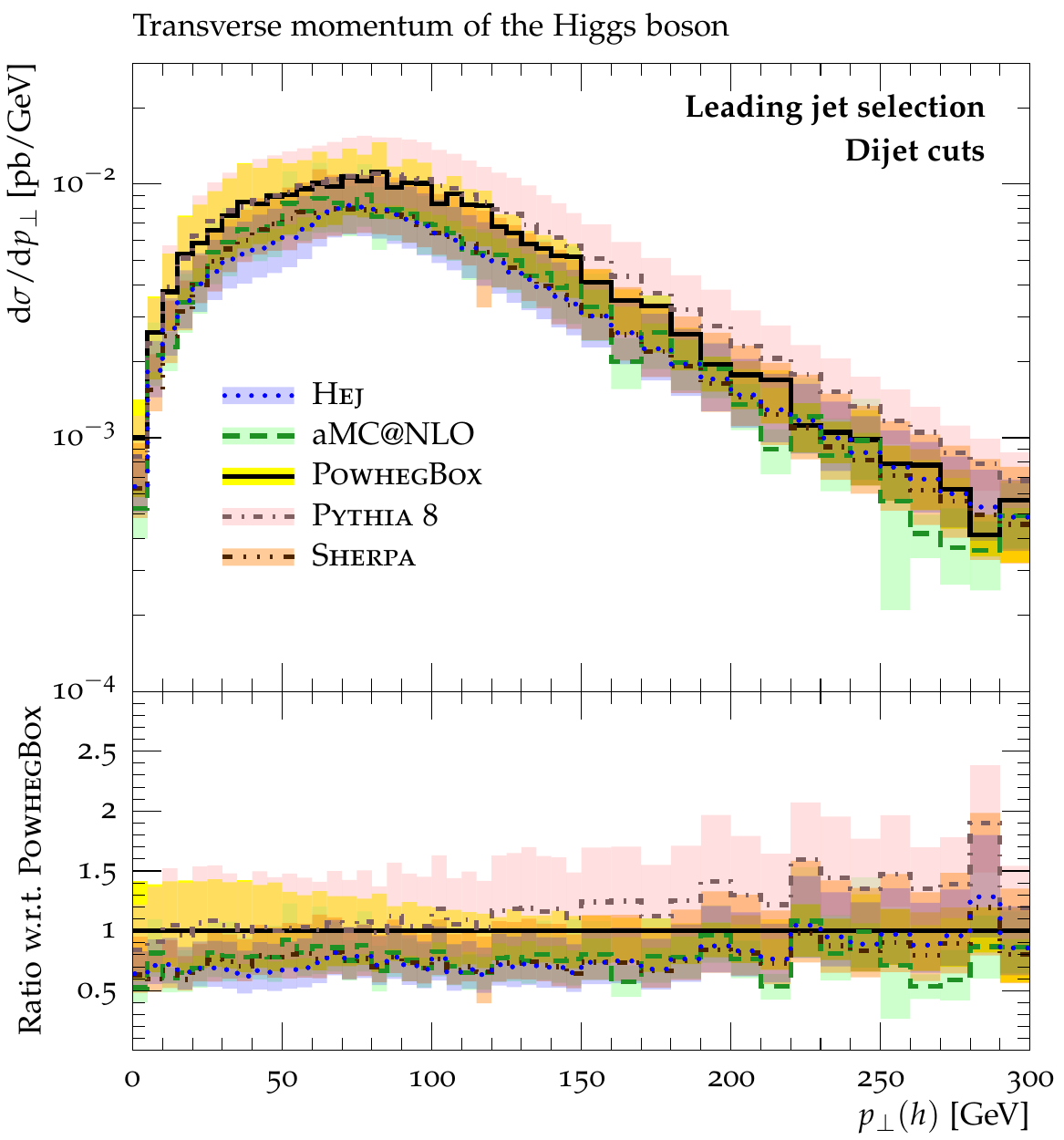}\hfill
  \includegraphics[width=0.47\textwidth]{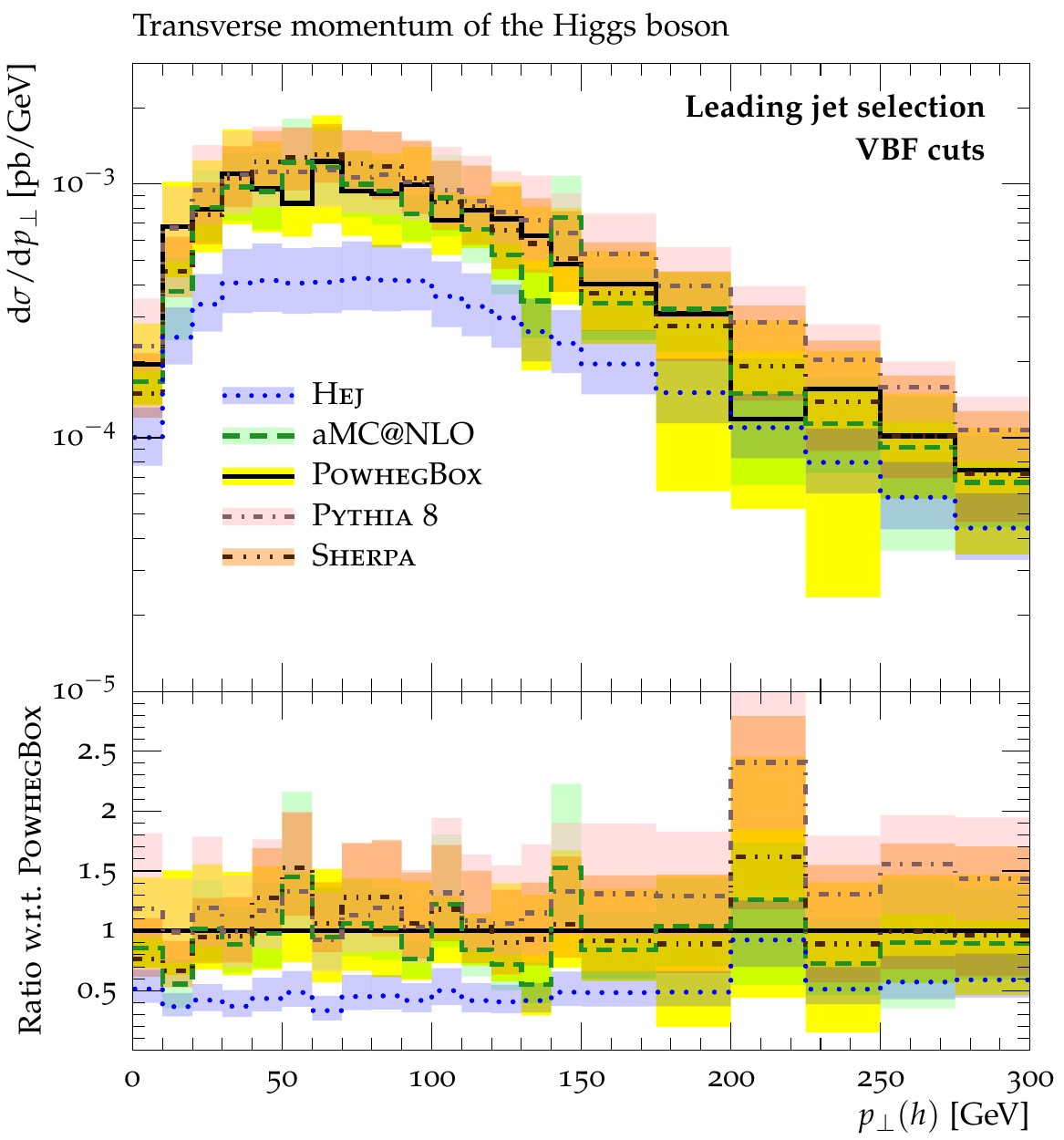}
  \caption{
            Transverse momentum of the Higgs boson produced in association 
            with two jets before (left) and after (right) the application 
            of the VBF selection cuts in the leading jet selection
            as predicted by the different generators. 
	    The individual sources of uncertainties used to generate 
	    the respective bands are described in Sec.\ \ref{hdijet:sec:gens}.
            \label{hdijet:fig:hpT_DV}
          }
\end{figure}

The Higgs transverse momentum in the leading jet selection before and 
after VBF cuts is displayed in Fig.\ \ref{hdijet:fig:hpT_DV}. It 
comes out to be very similar in between the different generators, again 
except for small differences in the normalisation. Only the predictions 
of \textsc{PowhegBox} seem to have a slight tilt as compared to the other 
generators before and after VBF cuts. Noteworthy is also that while the 
cross sections after VBF cuts are largely in agreement among aMC@NLO, 
\textsc{PowhegBox}, \textsc{Pythia}~8 and \textsc{Sherpa} the cross 
section predicted by \textsc{Hej} now drops outside the uncertainties of 
the parton shower based approaches. This is a result of a trend already 
seen on Fig.\ \ref{hdijet:fig:dyjj_AB}.

\begin{figure}[t!]
  \includegraphics[width=0.47\textwidth]{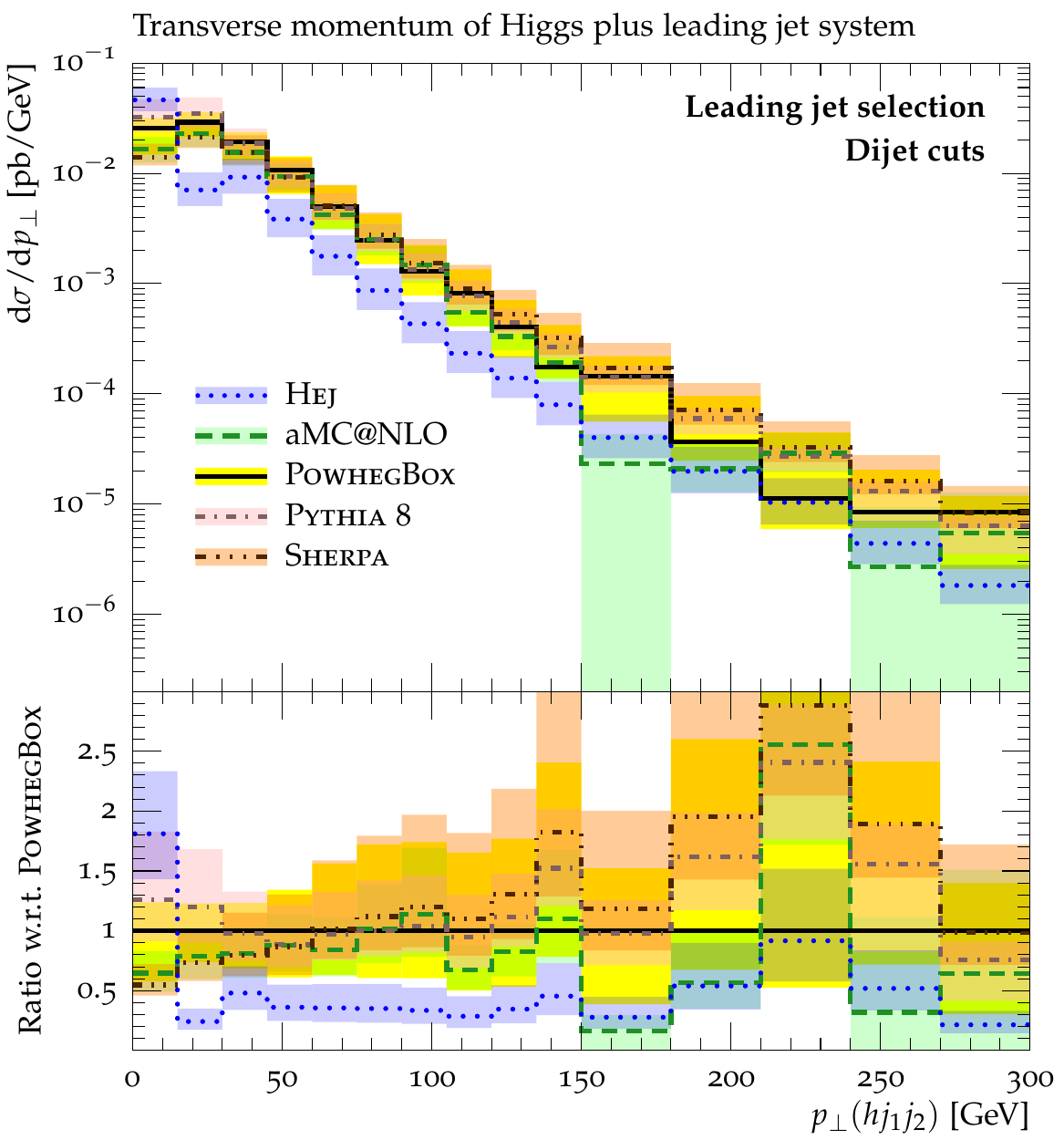}\hfill
  \includegraphics[width=0.47\textwidth]{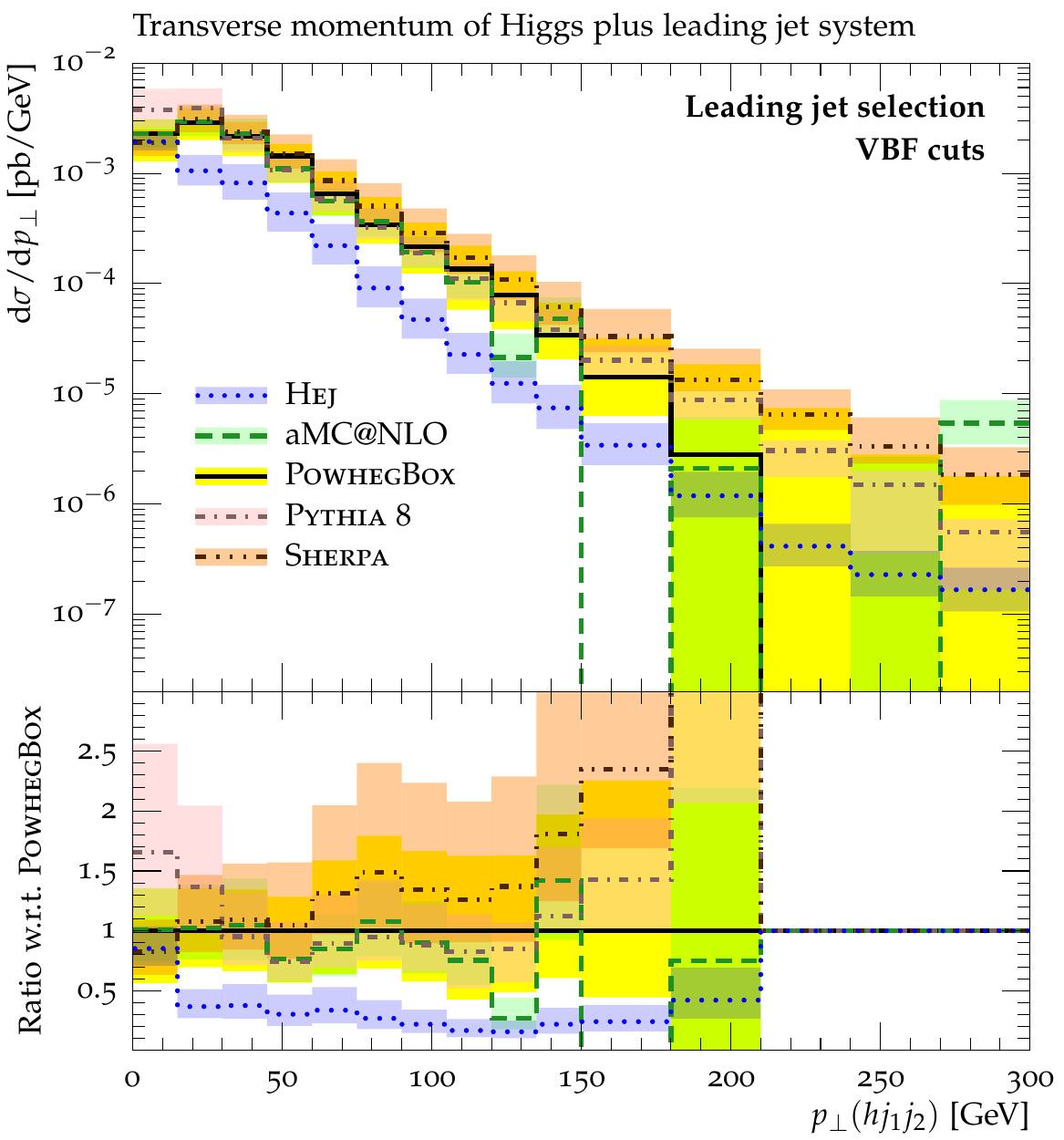}
  \caption{
            Transverse momentum of the Higgs boson plus tagging jets 
            system before (left) and after (right) the application 
            of the VBF selection cuts in the leading jet selection 
            as predicted by the different generators. 
	    The individual sources of uncertainties used to generate 
	    the respective bands are described in Sec.\ \ref{hdijet:sec:gens}.
            \label{hdijet:fig:hjjpT_DV}
          }
\end{figure}

Fig.\ \ref{hdijet:fig:hjjpT_DV} shows the transverse momentum 
distribution of the system consisting of the Higgs boson and the 
two tagging jets in the leading jet selection, again before and 
after VBF cuts. The results are largely compatible within the 
uncertainties, with \textsc{Pythia}~8 and \textsc{Sherpa} 
predicting a harder spectrum than aMC@NLO, \textsc{PowhegBox} or 
\textsc{Hej}. Again, as this distribution is dominated by three jet 
topologies, this can be understood by the same reasoning as for Fig.\ 
\ref{hdijet:fig:mjj_AB}. \textsc{Hej} does not exhibit the typical Sudakov 
shape with a maximum around 25 GeV as it is not matched to a parton 
shower and, therefore, does not contain DGLAP resummation. 

\begin{figure}[t!]
  \includegraphics[width=0.47\textwidth]{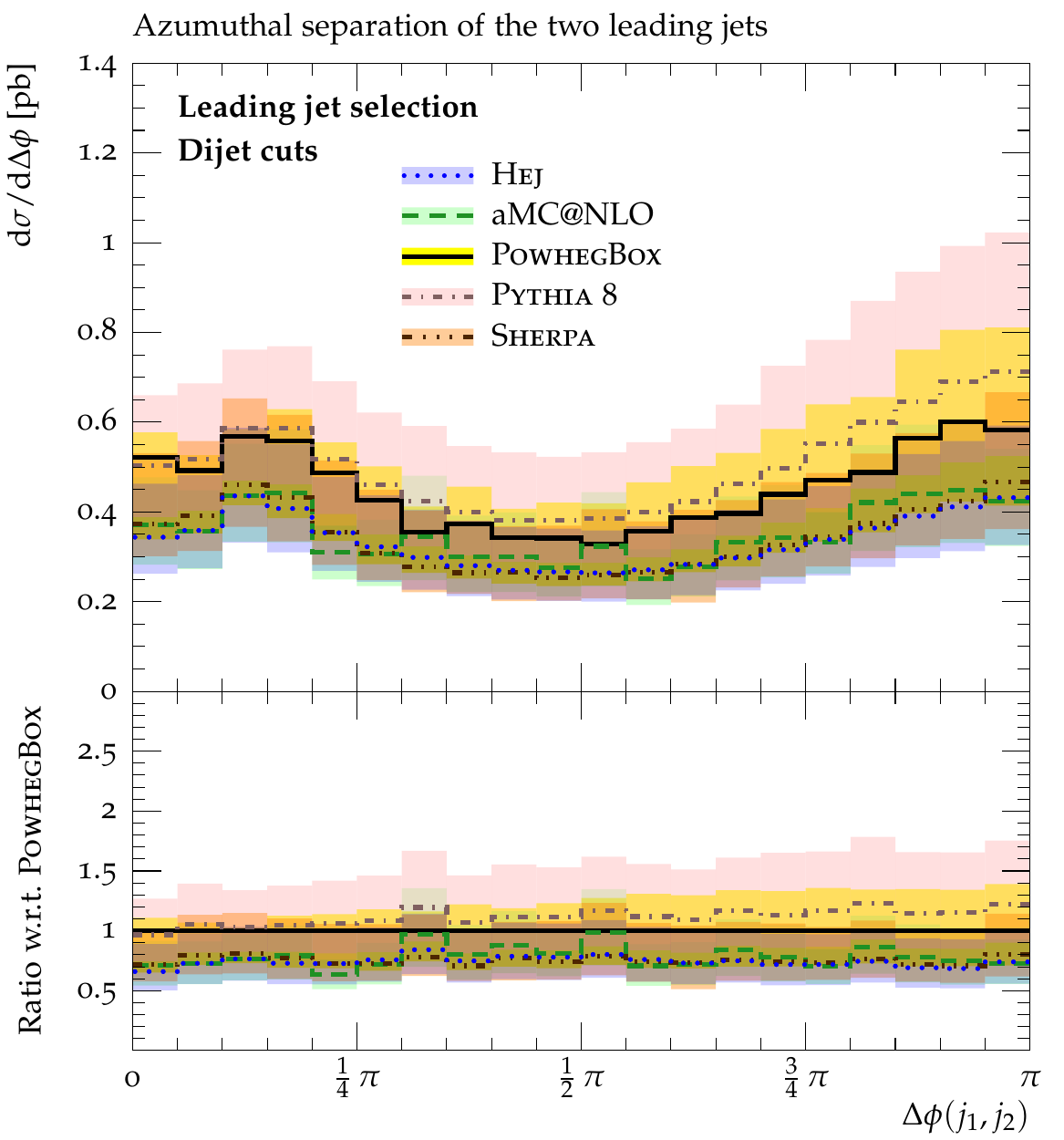}\hfill
  \includegraphics[width=0.47\textwidth]{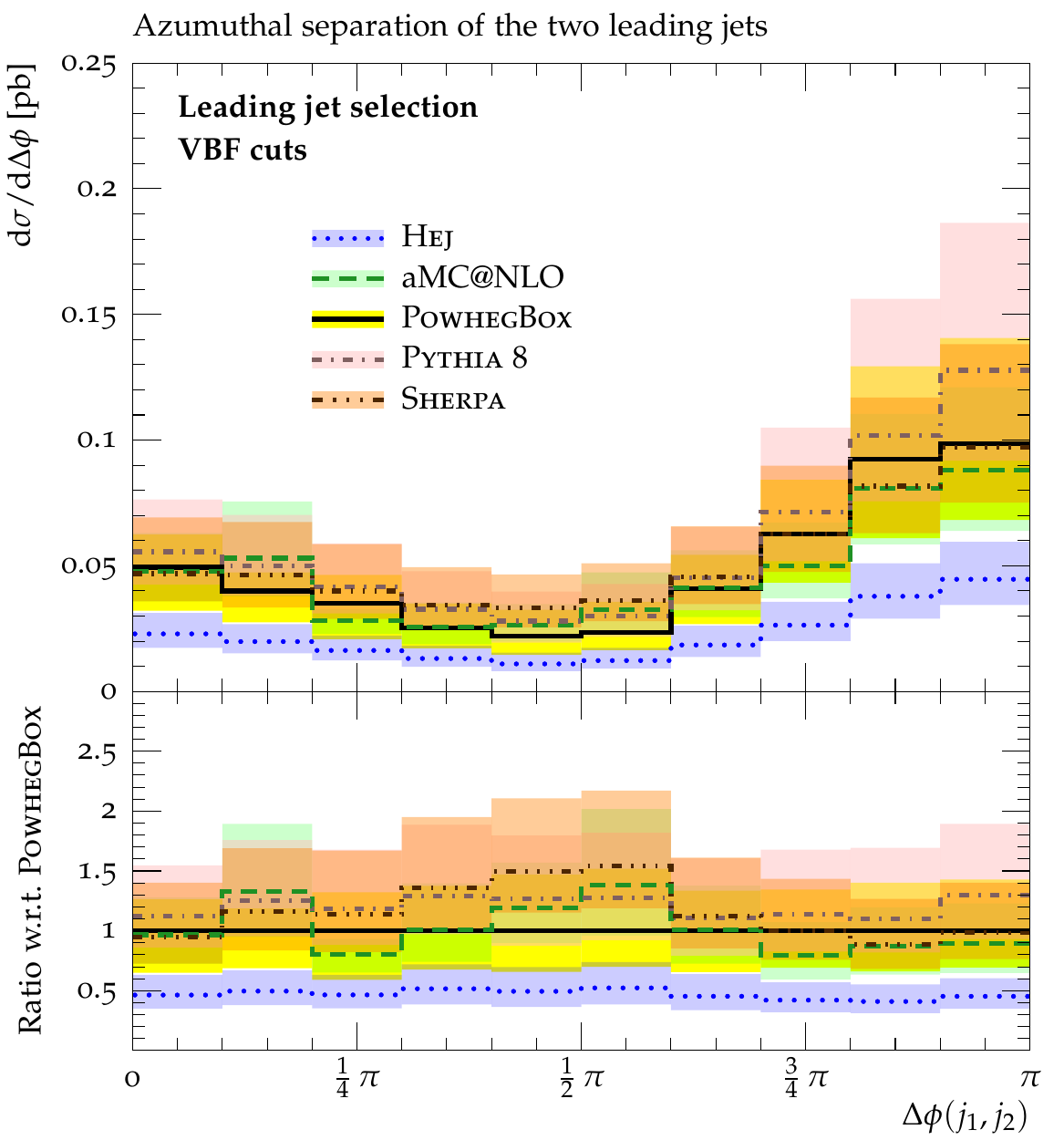}
  \caption{
            Azimuthal separation of the tagging jets before (left) and 
            after (right) the application of the VBF selection cuts in 
            the leading jet selection 
            as predicted by the different generators. 
	    The individual sources of uncertainties used to generate 
	    the respective bands are described in Sec.\ \ref{hdijet:sec:gens}.
            \label{hdijet:fig:jjdphi_DV}
          }
\end{figure}

In the case of the azimuthal separation between the tagging jets, 
shown again for the leading jet selection before and after VBF 
cuts in Fig.\ \ref{hdijet:fig:jjdphi_DV}, the shapes are very 
similar before the VBF cuts. But after VBF cuts \textsc{Sherpa}, 
and to a lesser extent aMC@NLO, predict slightly more cross section 
in the region $\Delta \phi(j_1,j_2)\sim \pi/2$, 
whereas \textsc{Pythia}~8 predicts slightly more events in the back-to-back 
region than the other generators. The \textsc{Pythia}~8 result can 
to some extend again be attributed to the usage of CTEQ6L1 PDFs 
and its rather high $\alpha_s$-value in Tune AU2-CTEQ6L1 used for 
its parton showering.

\begin{figure}[t!]
  \includegraphics[width=0.47\textwidth]{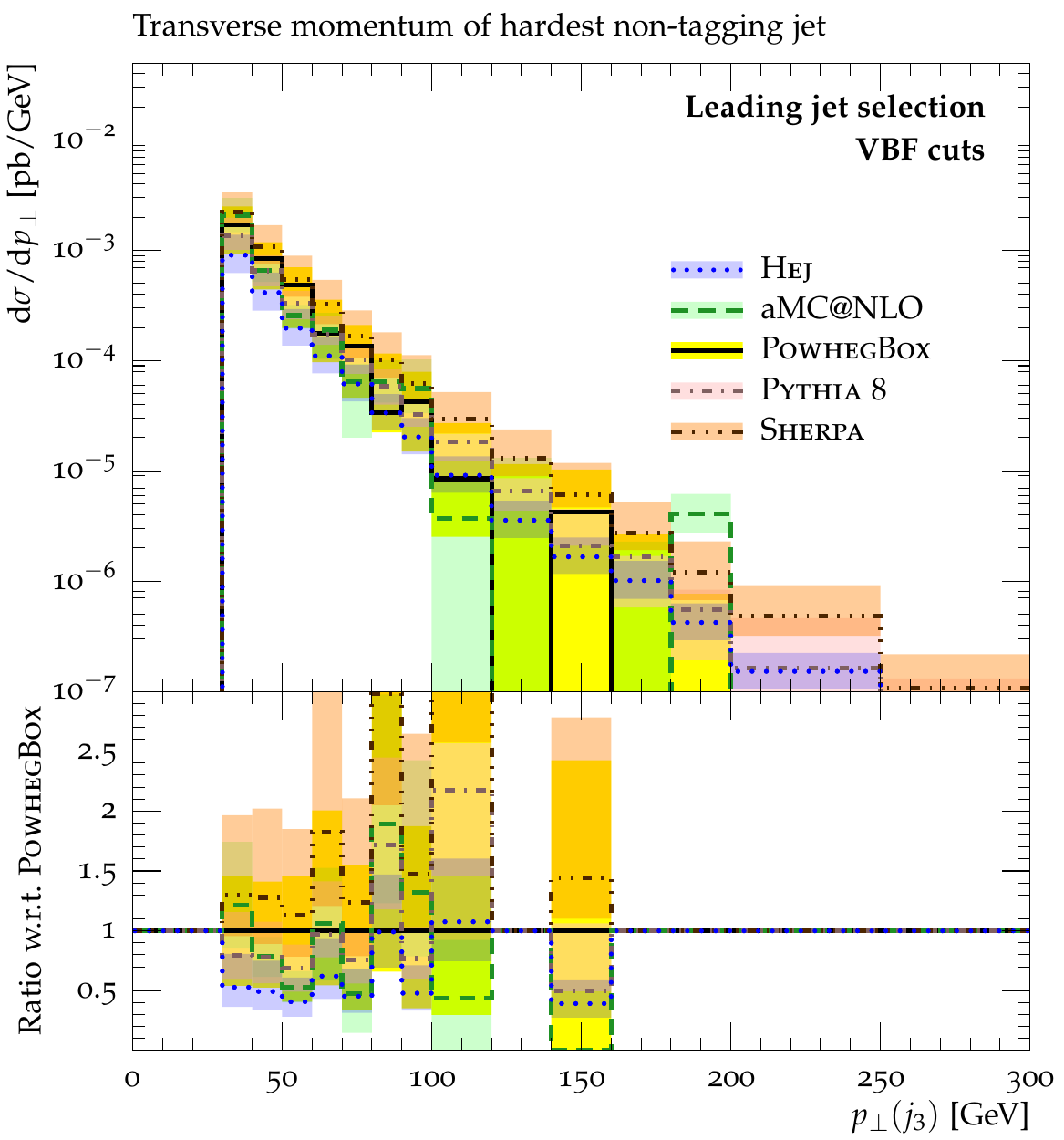}\hfill
  \includegraphics[width=0.47\textwidth]{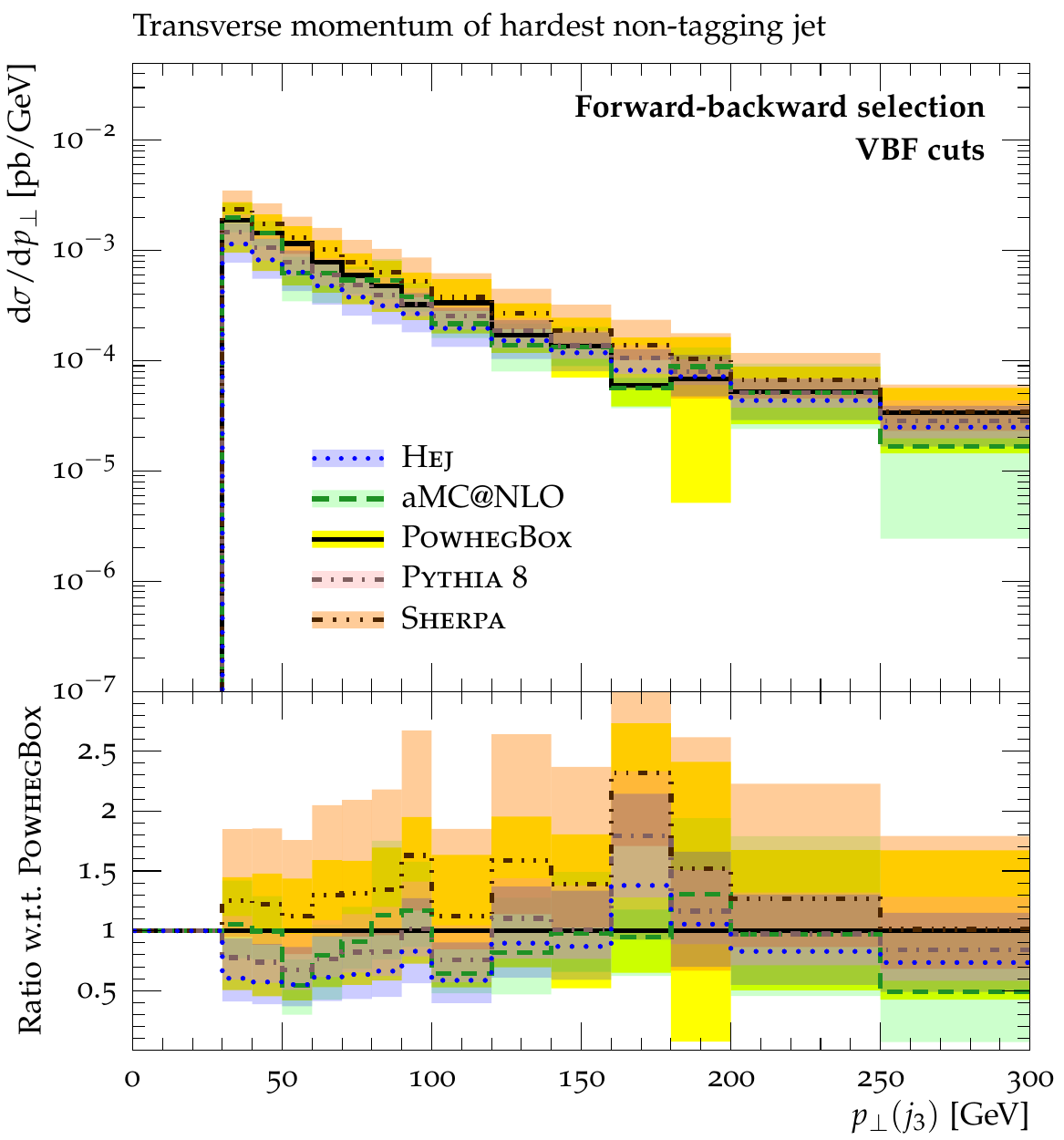}
  \caption{
            Transverse momentum of the hardest non-tagging jet in the 
            leading jet selection (left) and the forward-backward 
            selection (right) after VBF cuts 
            as predicted by the different generators. 
	    The individual sources of uncertainties used to generate 
	    the respective bands are described in Sec.\ \ref{hdijet:sec:gens}.
            \label{hdijet:fig:j3pT_AB}
          }
\end{figure}

Fig.\ \ref{hdijet:fig:j3pT_AB} displays the transverse 
momentum of the hardest non-tagging jets after VBF cuts for 
both the leading jet and the 
forward-backward selection. Naturally, in the forward-backward 
selection, not necessarily taking the two hardest jets as tagging 
jets, the $p_\perp$ of this third jet is much higher than in the 
leading jet selection. All generators predict roughly similar 
shapes for both distributions, but at very different rates. 
\textsc{Sherpa} predicts the largest cross-section for such events 
containing at least three jets, followed by aMC@NLO and 
\textsc{PowhegBox}, while \textsc{Pythia}~8 and \textsc{Hej} 
predict the lowest rates. 

\begin{table}[t]
  \begin{center}
    \begin{tabular}{|l|c|c|c|c|c|}
      \multicolumn{6}{ c }{Cross sections in the leading jet selection} \\
      \hline
      \multirow{2}{*}{Generator}
	& \multirow{2}{*}{$\sigma_\text{dijet}$ [pb]}
	& $\sigma_\text{dijet}$ [pb]
	& \multirow{2}{*}{$\sigma_\text{VBF}$ [pb]}
	& $\sigma_\text{VBF}$ [pb]
	& $\sigma_\text{VBF}$ [pb]\\
      & & $\!y_{j_\text{bw}}<y_h<y_{j_\text{fw}}\!$
      & & $\!y_{j_\text{bw}}<y_h<y_{j_\text{fw}}\!$
	& $\!y_{j_\text{bw}}<y_{j_3}<y_{j_\text{fw}}\!$ \\
      \hline
      {\sc Hej} \vphantom{$\int\limits_a^b$}
	& $1.053_{-0.253}^{+0.374}$  
	& $0.337_{-0.075}^{+0.107}$  
	& $0.070_{-0.017}^{+0.025}$  
	& $0.057_{-0.013}^{+0.019}$  
	& $0.0177_{-0.0055}^{+0.0092}$  \\
      \hline
      aMC@NLO \vphantom{$\int\limits_a^b$}
	& $1.106_{-0.272}^{+0.316}$  
	& $0.440_{-0.109}^{+0.125}$  
	& $0.149_{-0.039}^{+0.051}$  
	& $0.131_{-0.035}^{+0.044}$  
	& $0.0338_{-0.0088}^{+0.0114}$  \\
      \hline
      {\sc PowhegBox} \vphantom{$\int\limits_a^b$}\hspace*{-2pt}
	& $1.426_{-0.415}^{+0.328}$  
	& $0.583_{-0.181}^{+0.178}$  
	& $0.154_{-0.048}^{+0.050}$  
	& $0.135_{-0.042}^{+0.043}$  
	& $0.0345_{-0.0167}^{+0.0165}$  \\
      \hline
      {\sc Pythia} 8 \vphantom{$\int\limits_a^b$}
	& $1.590_{-0.385}^{+0.612}$  
	& $0.650_{-0.162}^{+0.262}$  
	& $0.183_{-0.049}^{+0.083}$  
	& $0.160_{-0.043}^{+0.072}$  
	& $0.0277_{-0.0071}^{+0.0110}$  \\
      \hline
      {\sc Sherpa} \vphantom{$\int\limits_a^b$}
	& $1.073_{-0.225}^{+0.462}$  
	& $0.422_{-0.082}^{+0.185}$  
	& $0.165_{-0.039}^{+0.071}$  
	& $0.138_{-0.033}^{+0.061}$  
	& $0.0461_{-0.0126}^{+0.0265}$  \\
      \hline
    \end{tabular}
  \end{center}
  \caption{
	    Cross section for $pp\to h+2$ jet production within the dijet 
	    and VBF cuts as predicted by the different generators 
	    in the \emph{leading jet} selection. Additionally the cross 
	    sections with the additional requirement that the Higgs 
	    boson is produced in between both tagging jets or that 
	    a third jet is present and between both tagging jets are 
	    given. $y_{j_\text{fw}}$ and $y_{j_\text{bw}}$ are the 
	    rapidities of the forward and the backward tagging jet. 
	    The individual sources of uncertainties used to generate 
	    the respective bands are described in Sec.~\ref{hdijet:sec:gens}.
            \label{hdijet:tab:xs_A}
          }
\end{table}

\begin{table}[t!]
  \begin{center}
    \begin{tabular}{|l|c|c|c|c|c|}
      \multicolumn{6}{ c }{Cross sections in the forward-backward selection} \\
      \hline
      \multirow{2}{*}{Generator}
	& \multirow{2}{*}{$\sigma_\text{dijet}$ [pb]}
	& $\sigma_\text{dijet}$ [pb]
	& \multirow{2}{*}{$\sigma_\text{VBF}$ [pb]}
	& $\sigma_\text{VBF}$ [pb]
	& $\sigma_\text{VBF}$ [pb]\\
      & & $\!y_{j_\text{bw}}<y_h<y_{j_\text{fw}}\!$
      & & $\!y_{j_\text{bw}}<y_h<y_{j_\text{fw}}\!$
	& $\!y_{j_\text{bw}}<y_{j_3}<y_{j_\text{fw}}\!$ \\
      \hline
      {\sc Hej} \vphantom{$\int\limits_a^b$}
	& $1.053_{-0.253}^{+0.374}$  
	& $0.384_{-0.089}^{+0.130}$  
	& $0.103_{-0.028}^{+0.044}$  
	& $0.086_{-0.022}^{+0.035}$  
	& $0.0585_{-0.0190}^{+0.0323}$  \\
      \hline
      aMC@NLO \vphantom{$\int\limits_a^b$}
	& $1.106_{-0.272}^{+0.316}$  
	& $0.512_{-0.127}^{+0.147}$  
	& $0.183_{-0.047}^{+0.058}$  
	& $0.163_{-0.041}^{+0.050}$  
	& $0.0796_{-0.0198}^{+0.0237}$  \\
      \hline
      {\sc PowhegBox} \vphantom{$\int\limits_a^b$}\hspace*{-2pt}
	& $1.426_{-0.415}^{+0.328}$  
	& $0.658_{-0.214}^{+0.199}$  
	& $0.197_{-0.068}^{+0.068}$  
	& $0.177_{-0.061}^{+0.060}$  
	& $0.0878_{-0.0394}^{+0.0472}$  \\
      \hline
      {\sc Pythia} 8 \vphantom{$\int\limits_a^b$}
	& $1.590_{-0.385}^{+0.612}$  
	& $0.716_{-0.175}^{+0.282}$  
	& $0.220_{-0.055}^{+0.093}$  
	& $0.195_{-0.049}^{+0.082}$  
	& $0.0726_{-0.0173}^{+0.0288}$  \\
      \hline
      {\sc Sherpa} \vphantom{$\int\limits_a^b$}
	& $1.073_{-0.225}^{+0.462}$  
	& $0.499_{-0.099}^{+0.229}$  
	& $0.218_{-0.052}^{+0.102}$  
	& $0.189_{-0.045}^{+0.091}$  
	& $0.1129_{-0.0296}^{+0.0656}$  \\
      \hline
    \end{tabular}
  \end{center}
  \caption{
	    Cross section for $pp\to h+2$ jet production within the dijet 
	    and VBF cuts as predicted by the different generators 
	    in the \emph{forward-backward} selection. Additionally the cross 
	    sections with the additional requirement that the Higgs 
	    boson is produced in between both tagging jets or that 
	    a third jet is present and between both tagging jets are 
	    given. $y_{j_\text{fw}}$ and $y_{j_\text{bw}}$ are the 
	    rapidities of the forward and the backward tagging jet. 
	    The individual sources of uncertainties used to generate 
	    the respective bands are described in Sec.~\ref{hdijet:sec:gens}.
            \label{hdijet:tab:xs_B}
          }
\end{table}

Finally, the cross sections at the different selection levels are 
summarised in Tabs.~\ref{hdijet:tab:xs_A} and \ref{hdijet:tab:xs_B}. 
While good agreement is found for inclusive $pp\to h+2$ jets production, 
as already seen in the distributions, the cross sections in \textsc{Hej} 
are generally smaller than those predicted by the other generators once 
VBF cuts are applied. This can be understood by the difference in the 
description of additional hard radiation from an underlying configuration. 
While all generators largely agree for the observables before 
VBF cuts within their respective uncertainties, both \textsc{PowhegBox} and 
\textsc{Pythia}~8 are significantly higher than either aMC@NLO, 
\textsc{Sherpa} or \textsc{Hej}. The differences to 
\textsc{Hej} are larger for the additional requirement of having the 
Higgs boson produced within the tagging jets (in rapidity).

After VBF cuts the picture is slightly changed, emphasising their different 
efficiencies across the generators. The effect of the cuts on predictions 
of \textsc{Hej} is larger by a factor of 1.5-2 as compared to the parton 
shower based generators. This is again understood from the effects seen 
in Fig.\ \ref{hdijet:fig:dyjj_AB}. The cross sections predicted by the 
parton shower based approaches 
are in somewhat better agreement than before this set of additional cuts, 
though their respective uncertainties have increased in some cases, 
especially in the forward-backward selection. The probability to also find 
the Higgs boson in between both tagging jets now is substantially increased. 
This nicely illustrates the effect of the VBF cuts shaping the $pp\to h+2$ 
jets events into a VBF-like topology with a large rapidity span that is very 
likely to contain the Higgs. The probability to find another jet in between 
the tagging jets again shows larger variability among the generators, as was 
already observed in Fig.\ \ref{hdijet:fig:j3pT_AB}. Here in both selections 
\textsc{Sherpa} predicts the largest rates whereas those of \textsc{Hej} 
and \textsc{Pythia}~8 are the smallest.

It is interesting to observe that as, by construction, the rapidity span 
is larger in the forward-backward selection than in the dijet selection, 
not only is the cross section after VBF cuts enlarged, but also have both 
the probability to find the Higgs and to find an additional jet in between 
the two tagging jets increased.  In particular the latter is interesting 
in the context of using jet vetoes to suppress the gluon fusion contribution 
to VBF-like signatures. This possibility is not studied here, as it also 
requires careful treatment of the underlying event, which can also produce 
additional jets.

The full set of observables studied can be found at \\
\texttt{http://phystev.in2p3.fr/wiki/2013:groups:sm:higgs:hdijetsresults}.

\subsection{CONCLUSIONS}
\label{hdijet:sec:conclusions}

The aim of this study was to quantify the gluon fusion contribution 
to $h+2$ jets with a VBF-like signature and its uncertainties. The 
strategy was to obtain state-of-the-art NLO or multi-jet merged NLO 
predictions from the event generators \textsc{Hej}, aMC@NLO, 
\textsc{PowhegBox}, \textsc{Pythia}\,8 and \textsc{Sherpa} 
including perturbative uncertainties and do a detailed comparison 
of both rates and distributions.
The events were analysed with a common, generic Higgs analysis. 
Distributions are plotted at different levels of cuts going from 
inclusive Higgs via Higgs+dijet to a VBF-like selection. 

The gluon fusion contribution to Higgs events with a VBF-like 
signature turns out to be of the order of 0.1-0.2 pb and is 
thus non-negligible. This is especially relevant as observables 
such as $\Delta\phi(j_1,j_2)$ exhibit different shapes in Higgs 
production through gluon fusion as opposed to production through 
weak-boson fusion, diluting characteristic patterns used to measure 
Higgs properties. The predictions by the NLO event generators 
are in good agreement with one-another within their respective 
uncertainty estimates. In all cases (except for aMC@NLO)
all sources of perturbative uncertainties, stemming from the 
fixed-order cross-section input, the matched parton shower 
resummation (in the form of its starting scale), and, where 
applicable, also the unphysical multijet merging parameter, 
were accounted for. Additional uncertainties originating in the 
parton shower splitting functions (evolution variables, choice of 
the scales of the strong coupling, recoil schemes) are current 
topics of investigation.

The differences in cross section are reflected in the normalisation 
of the distributions. The shapes are largely consistent. 
\textsc{Pythia}\,8 and \textsc{Sherpa} have been observed to have 
a tendency to predict slightly harder transverse momentum spectra, 
but these differences are mostly covered by the reported uncertainties. 

Although it is reassuring that the predictions by different 
generators are generally consistent, the uncertainties remain 
rather large. Taking the perturbative uncertainties reported 
by the generators and the spread of the predictions seriously, 
one arrives at an uncertainty estimate on the cross section that 
is roughly a factor 2. One possible avenue for the reduction of 
these uncertainties is the experimental study of the VBF 
phase space on various signal processes, e.g.\ \cite{Aad:2014dta},
to constrain the models. It is further important to note that 
this study only investigated perturbative uncertainties. 
Noneperturbative uncertainties stemming from the modelling of the 
parton-to-hadron fragmentation and especially multiple parton 
interaction, however, may also play an important role in analyses 
with gap fractions \cite{Hoche:2012wh}.
\subsubsection*{Acknowledgements}

We would like to thank the organisers of Les Houches 2013 for 
repeatedly hosting such a fantastic and
fruitful workshop. We 
would also like to thank all participants for a lively and 
stimulating atmosphere
and lots of inspiring discussions.

KH gratefully acknowledges support from Durham University 
IPPP through their associateship program.
TH's and MS's work was supported by the Research Executive Agency (REA) 
of the European Union under the Grant Agreement number 
PITN-GA-2010-264564 (LHCPhenoNet). MS further acknowledges support by 
PITN-GA-2012-315877 (MCnet). JMS is funded by a Royal Society University 
Research Fellowship.

\section{Higgs boson plus di- and tri-jet production at NLO in QCD%
\footnote{G.~Cullen, H.~van Deurzen, N.~Greiner, J.~Huston, G.~Luisoni,
P.~Mastrolia, E.~Mirabella, G.~Ossola, T.~Peraro, F.~Tramontano, J.~Winter and V.~Yundin}}

\subsection{Introduction}

Among the different production mechanisms of a Higgs boson ($H$) within the
Standard Model, Gluon-Gluon Fusion (GGF) via a virtual top-quark loop
is the one with the largest cross section at the Large Hadron Collider
(LHC). Although direct measurements of the Higgs boson properties in
this channel, without applying any vetoes on additional jets, are difficult
due to the large QCD background, precise theoretical predictions for
the associate production of a Higgs boson and jets in GGF are important
for several aspects. On the one hand the possibility to reliably estimate
the theoretical uncertainty when a jet veto is applied heavily relies
on the knowledge of inclusive and exclusive cross sections for Higgs
boson production and additional jets, on the other hand the production
of a Higgs boson together two jets in GGF is one of the main
irreducible backgrounds to the production of a Higgs boson in
Vector Boson Fusion (VBF), which allows to directly probe the coupling
of the Higgs boson to other electroweak bosons.

The leading order (LO) contribution to the production of a Higgs boson
in association with two jets ($H$+2-jets) and three jets ($H$+3-jets),
retaining the full top-mass ($m_t$) dependence, have been computed
respectively in Refs.~\cite{DelDuca:2001eu,DelDuca:2001fn}, and
Ref.~\cite{Campanario:2013mga}. These calculations showed that large
top-mass approximation ($m_{t} \to \infty$) is valid whenever the mass
of the Higgs particle and the $p_T$ of the jets are not much larger
than the mass of the top quark. In the results presented here we adopt
this approximation and introduce a set of effective vertices, which
directly couple the Higgs particle to two, three and four
gluons~\cite{Djouadi:1991tka,Dawson:1990zj}. Feynman rules for these vertices can be found e.g. in
Ref.~\cite{vanDeurzen:2013aaa}. The next-to-leading order (NLO)
corrections for $H$+2-jets in GGF at LHC were first computed in
Ref.~\cite{Campbell:2006xx,Campbell:2010cz} using amplitudes computed
in Refs.~\cite{DelDuca:2004wt,
  Dixon:2004za,Badger:2004ty,Ellis:2005qe,Ellis:2005zh,Berger:2006sh,Badger:2006us,Badger:2007si,Glover:2008ffa,Badger:2009hw,Dixon:2009uk,Badger:2009vh}. These
amplitudes have been also used to obtain matched NLO plus shower
predictions~\cite{Campbell:2012am,Hoeche:2014lxa}. More recently they
have been recomputed using for the first time an automated tool for
the evaluation of both tree-level and loop amplitudes in
Ref.~\cite{vanDeurzen:2013aaa}. A similar setup was also used to
compute the first NLO results for $H$+3-jets in
GGF~\cite{Cullen:2013saa}.

In these proceedings we present some new results for $H$+3-jets at NLO
to which we applied a set of ATLAS-like cuts, and compare it with
predictions for $H$+2-jets at NLO.  In Section~\ref{h3jsec:calc_setup}
the computational setup used to obtain the various predictions is
presented. Compared to the original setup used in
Ref.~\cite{Cullen:2013saa}, here the newest developments regarding the
evaluation of loop amplitudes with integrand reduction techniques
based on Laurent expansion were used~\cite{Mastrolia:2012bu}. In
Section~\ref{h3jsec:results} we present our results and discuss
them. They include predictions for the total cross sections and for
several relevant distributions, together with scale uncertainty bands.

\subsection{Calculational details}\label{h3jsec:calc_setup}
All parton-level predictions, which we present in these proceedings,
have been produced by combined generator packages.  The $H$+2-jets
samples were obtained with the \textsc{GoSam}+\textsc{Sherpa} package
while those describing $H$+3-jets production required the combination
of contributions from \textsc{GoSam}+\textsc{Sherpa} and the
\textsc{MadGraph/Dipole/Event} framework.  More specifically, the
$H$+3-jets samples were built from the Born plus virtual (BV) terms as
provided by \textsc{GoSam+Sherpa} and the dipole-subtracted real
emission (RS) contribution as well as the integrated subtraction
terms (I) as produced by \textsc{MadGraph/Dipole/Event}.  We checked
the consistency of our hybrid MC integration for $H$+3-jets on
$H$+2-jets, verifying that the full cross section at NLO agrees with
the corresponding result for the integration of all contributions
(BVIRS) obtained with \textsc{GoSam}+\textsc{Sherpa} alone.  Moreover, for
$H$+3-jets we found excellent agreement between \textsc{MadGraph} and
\textsc{Sherpa} for the LO cross section.

\subsubsection{Virtual amplitudes}
The virtual 1-loop amplitudes are generated and computed using the
\textsc{GoSam} framework~\cite{Cullen:2011ac}, which combines the algebraic
generation of $d$-dimensional integrands via Feynman
diagrams~\cite{Nogueira:1991ex, Vermaseren:2000nd, Reiter:2009ts,
  Cullen:2010jv}, with the numerical evaluation based upon
integrand-reduction~\cite{Ossola:2006us, Ossola:2007bb, Ellis:2007br,
  Ossola:2008xq,Mastrolia:2008jb, Mastrolia:2012bu, Mastrolia:2012an},
as implemented in \textsc{Samurai} \cite{Mastrolia:2010nb,Mastrolia:2012bu,Mastrolia:2012du,vanDeurzen:2013pja}
and \textsc{Ninja} \cite{Peraro:2014cba,Mastrolia:2012bu,vanDeurzen:2013saa,Peraro:2013oja}
and tensor integral calculus, as implemented in \textsc{Golem95} \cite{Cullen:2011kv,Guillet:2013msa}.

The virtual amplitudes for the production of Higgs plus two- and
three-jets presented in
Ref.\cite{vanDeurzen:2013aaa} and Ref.\cite{Cullen:2013saa}
respectively, were obtained by using the \textsc{Samurai} reduction library.
For the results presented here, instead, \textsc{GoSam} was used to generate the
input for the new library \textsc{Ninja} \cite{Peraro:2014cba}, which computes the loop amplitude using a novel technique
based on the interplay of the reduction at the integrand level and the
Laurent expansion of the numerator \cite{Mastrolia:2012bu}, and
proved to be in general faster and more stable compared to the
standard integrand reduction method \cite{vanDeurzen:2013saa}. In fact, by employing \textsc{Ninja}, the evaluation time per
phase-space point reduced by a factor of 50\%, and the fraction of unstable point dropped below the permille level.
The scalar loop integrals are computed using \textsc{OneLOop}~\cite{vanHameren:2010cp}.

In order to deal with the complexity of the amplitudes of the processes computed here, several improvements were introduced into \textsc{GoSam}. They will become public with the release of the new version of the program, whose final development is currently in progress. Among the improvements, there is the possibility to exploit the optimized manipulation of polynomial expressions available in \textsc{FORM}~4.0~\cite{Kuipers:2012rf}.

The one-loop amplitudes for Higgs production in GGF containing the effective $ggH$ coupling lead to integrands that may exhibit
numerators with rank larger than the number of the denominators. More details about the treatment of these so-called higher-rank integrals can be found in Refs.~\cite{vanDeurzen:2013aaa,Cullen:2013saa,vanDeurzen:2013pja,Guillet:2013msa}.

\subsubsection{GoSam+Sherpa specifics}
Within \textsc{Sherpa}~\cite{Gleisberg:2008ta} the tree amplitudes for the Born contributions and real corrections were obtained using the automatic amplitude generator \textsc{Amegic}~\cite{Krauss:2001iv}, which also automatically generates the subtraction terms and their integrated counterpart using an implementation~\cite{Gleisberg:2007md} of the Catani-Seymour dipoles subtraction~\cite{Catani:1996vz}. The integration of the virtual amplitudes was also performed using \textsc{Sherpa}. The amplitudes generated with \textsc{GoSam} were interfaced to \textsc{Sherpa} via the Binoth-Les-Houches Accord (BLHA)~\cite{Binoth:2010xt,Alioli:2013nda}, which sets the standards for the communication between a Monte Carlo program and a One Loop Program, and allows to generate and link the two codes fully automatically. The integration of the $H$+3-jets amplitudes is carried out by generating $\mathcal{O}(10^6)$ events, sampled on a MC grid trained on the Born matrix element, and weighted with the sum of the Born and the virtual amplitudes.

\subsubsection{MadGraph/Dipole/Event specifics}
In the \textsc{MadGraph} setup the tree-level amplitudes are obtained using
\textsc{MadGraph}~4~\cite{Stelzer:1994ta,Alwall:2007st} and the subtraction
terms are provided by
\textsc{MadDipole}~\cite{Frederix:2008hu,Frederix:2010cj}, which also
implements Catani-Seymour dipole subtraction. The numerical
integration is done using \textsc{MadEvent}~\cite{Maltoni:2002qb}. We
verified the independence of our result under the variation of the so
called $\alpha$-parameter that fixes the amount of subtractions around
the divergences of the real corrections.

\subsubsection{Parameter settings and scale variations}\label{hj3nlo_sec_params}

The various $H$+jets calculations are performed for proton-proton collisions at
$8\mathrm{\:TeV}$ center-of-mass energy using the five-flavour scheme,
massless $b$ quarks and a vanishing Yukawa $bbH$ coupling. We only
consider the effective coupling of gluons to the Higgs boson evaluated
in the infinite top-mass limit. The Higgs boson is stable with a mass
of $m_H=126\mathrm{\:GeV}$. To guarantee consistent setups, we use the
\texttt{cteq6l1} ($\alpha_\mathrm{s}(M_Z)=0.1298$) and
\texttt{CT10nlo} ($\alpha_\mathrm{s}(M_Z)=0.118$) parton density
functions (PDFs) to produce our main results at leading and next-to
leading orders in QCD, respectively.

Scale uncertainties were determined following the common procedure
applied in fixed-order calculations. Accordingly, we vary the
renormalization and factorization scales, $\mu_\mathrm{R}$ and
$\mu_\mathrm{F}$, by factors of $2.0$ and $0.5$.

Everywhere, but in the effective coupling of the Higgs to the gluons,
the renormalization and factorization scales are set to
\begin{equation}
\mu_{F}=\mu_{R}=\frac{\hat{H}_T}{2}=\frac{1}{2}\left(\sqrt{m_{H}^{2}+p_{T,H}^{2}}+\sum_{i}|p_{T,i}|\right)\,,
\end{equation}
where the sum runs over the final state jets. The strong coupling
is therefore evaluated at different scales according to
$\alpha_s^2\rightarrow\alpha_s^2(m_H)\alpha_s^2(\hat{H}_T/2)$
for $H$+2-jets, and
$\alpha_s^5\rightarrow\alpha_s^2(m_H)\alpha_s^3(\hat{H}_T/2)$
for $H$+3-jets.

\subsubsection{Kinematic requirements}\label{hj3nlo_sec_cuts}

In a 2-jets analysis, there are several ways of defining the two
tagging jets accompanying the production of the Higgs boson. For this
study, we apply a leading jet selection where the two highest $p_T$
jets are specified to be the tagged ones. The third jet therefore is
defined as the hardest untagged jet of each event. Jets are
constructed utilizing the anti-$k_T$ jet finding algorithm implemented in
{\sc FastJet}~\cite{Cacciari:2005hq,Cacciari:2008gp,Cacciari:2011ma} and a
separation of $R=0.4$, $p^\mathrm{(jet)}_t>30\mathrm{\:GeV}$ and
$|\eta^\mathrm{(jet)}|<4.4$. Of course, our inclusive $H$+jets samples
fulfill the tagging jet requirements by definition; in the 3-jets
case, we provide predictions where even the first untagged jet is
described at NLO accuracy.

\subsection{Next-to-leading order results}\label{h3jsec:results}

We divide our result section into two parts: first, we will discuss
the cross sections, which we obtain in the different jet multiplicity
bins and at different orders of $\alpha_\mathrm{s}$. Then we present a
selection of transverse momentum and rapidity observables to better
quantify the effects of higher-order corrections and scale variations
on the $H$+jets differential final states.

\subsubsection{Inclusive cross sections and multiplicity ratios}

\begin{table}[t!]
  \centering\small
  \begin{tabular}{cp{10mm}lrlp{10mm}c}
    \toprule\\[-8pt]
    Sample     &&\multicolumn{3}{l}{Cross sections for Higgs boson plus}\\[4pt]
    $K$-factor && $\ge2$ jets & $f_3$ & $\ge3$ jets
    && $r_{3/2}$\\[4pt]
    \midrule\\[-10pt]
    & LO &\\[0pt] 
    \midrule\\[-7pt]
    $H$+2-jet2  \scriptsize(LO PDFs) && $1.23~^{+37\%}_{-24\%}$ \\[8pt]
    $H$+3-jets \scriptsize(LO PDFs) && $(0.381)$ & $1.0$ & $0.381~^{+53\%}_{-32\%}$
                                    && $0.310~^{0.347}_{0.278}$ \\[8pt]
    \hdashline\\[-5pt] 
    $H$+2-jets  \scriptsize(NLO PDFs)&& $0.970~^{+33\%}_{-23\%}$ \\[8pt]
    $H$+3-jets \scriptsize(NLO PDFs)&& $(0.286)$ & $1.0$ & $0.286~^{+50\%}_{-31\%}$
                                    && $0.295~^{0.332}_{0.265}$ \\[5pt]
    \midrule\\[-10pt]
    & NLO &\\[0pt] 
    \midrule\\[-7pt]
    $H$+2-jets   && $1.590~^{-4\%}_{-7\%}$ & $0.182$ &
                   $0.289~^{+49\%}_{-31\%}$ \\[8pt]
    $H$+3-jets  && $(0.485)$ & $1.0$ &
                   $0.485~^{-3\%}_{-13\%}$ &&
                   $0.305~^{0.307}_{0.284}$ \\[20pt]
    $K_2$, $K_3$ \scriptsize(LO PDFs for LO)
                && $1.29~~~^{0.911}_{1.59}$ && $1.27~~~^{0.806}_{1.63}$ \\[5pt]
    $K_2$, $K_3$ \scriptsize(NLO PDFs for LO)
                && $1.64~~~^{1.19}_{1.98}$ && $1.70~~~^{1.10}_{2.13}$ \\[5pt]
    \bottomrule
  \end{tabular}
  \caption{\label{hj3nlo_tab_xsecs}
    Cross sections in pb for the various parton-level Higgs boson plus
    jet samples used in this study. The upper and lower parts of the
    table show the LO and NLO results, respectively, together with
    their uncertainties (in percent) from varying scales by factors of
    two, up (subscript position) and down (superscript position).
    NLO-to-LO $K$-factors, $K_n$, for both the inclusive 2-jets
    ($n=2$) and 3-jets ($n=3$) bin, the cross section ratio $r_{3/2}$
    and $m$-jet fractions, $f_m$, are given in addition.}
\end{table}

Table~\ref{hj3nlo_tab_xsecs} lists the inclusive jet cross sections
$\sigma_n$ of all our parton-level calculations using the parameters
and constraints as given in Secs.~\ref{hj3nlo_sec_params} and
\ref{hj3nlo_sec_cuts}. Three jet multiplicity bins, $n=2,3,4$, are
relevant to this study and given by requiring at least $n$ jets. We employ
two of these to compare the results of the $H$ plus two-jet and
three-jet calculations with each other. Except for the down-scale
variation ($\mu=\mu_\mathrm{R,F}/2$), we find the NLO rate effect to
be substantial as indicated by the large $K$-factors. Interestingly,
in all cases, they are rather similar in size for both jet bins $n=2$
and $n=3$. While the LO cross sections show a strong dependence on
joint renormalization and factorization scale variations, the NLO
cross section uncertainties due to these $\mu_\mathrm{R,F}$ variations
are considerably reduced to approximately $15\%$.

Another interesting indicator that we consider at LO and NLO is the
three-to-two jet cross section ratio, generally defined as
$r_{(n+1)/n}=\sigma_{n+1}/\sigma_{n}$. Owing to the similar $K$-factors,
it remains fairly stable at $\sim30\%$, roughly a factor 3 times
higher than a sole $\alpha_\mathrm{s}$ effect. Just for completeness,
we have also listed the inclusive three-jet fraction, $f_3$,
belonging to each sample.

\subsubsection{Differential observables}

The set of observables we want to present here includes Higgs boson
and single-jet transverse momenta, $p_{T,x}$, as well as their
respective rapidities, $y_x$ ($x\in\{H,j_{1,2,3}\}$). As an example of
a multi-object observable, we also discuss the $p_T$ distribution of
the $Hj_1j_2$ system.

\begin{figure}[t!]
  \centering
  \includegraphics[width=0.33\textwidth]{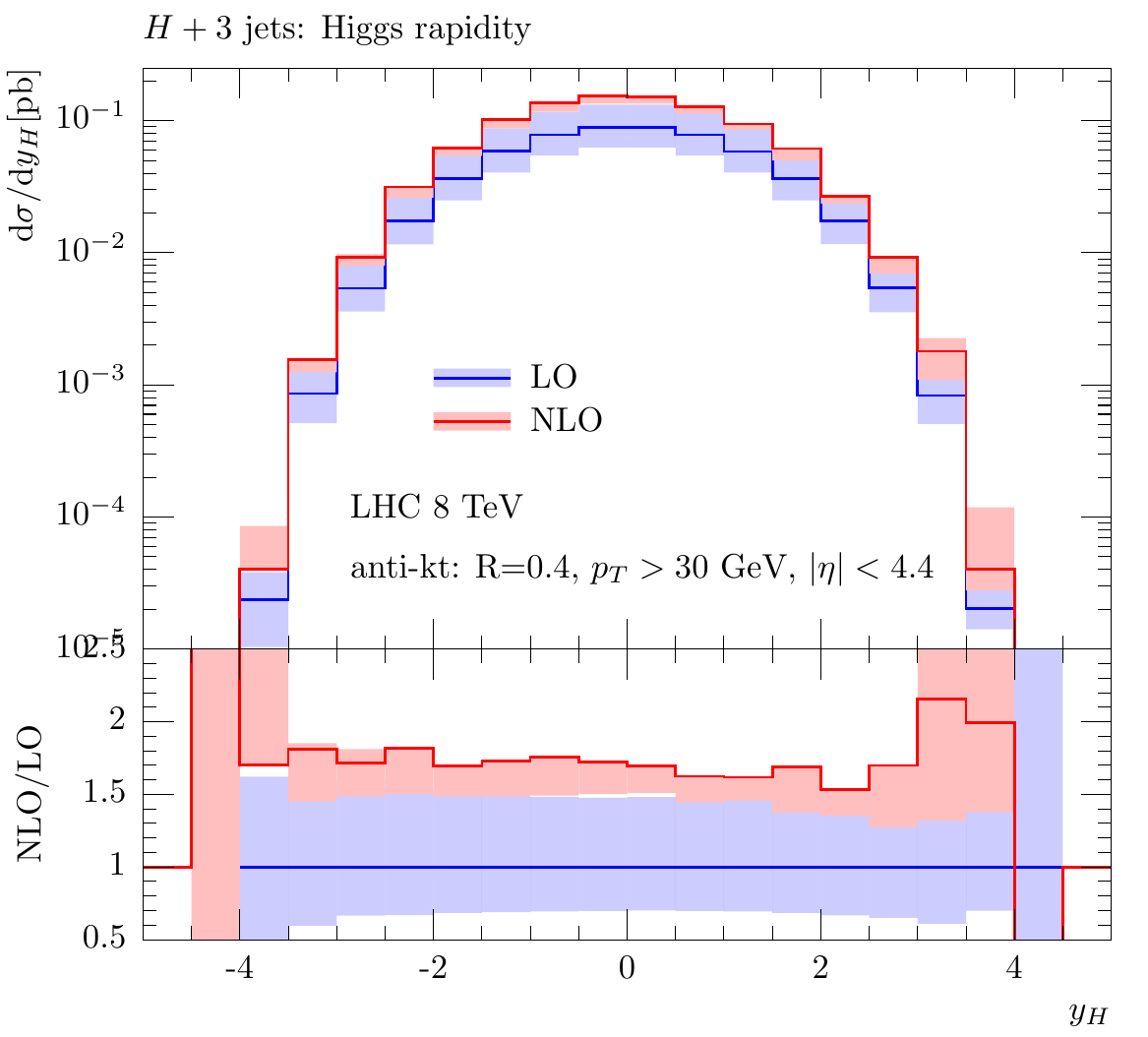}\hskip0mm
  \includegraphics[width=0.33\textwidth]{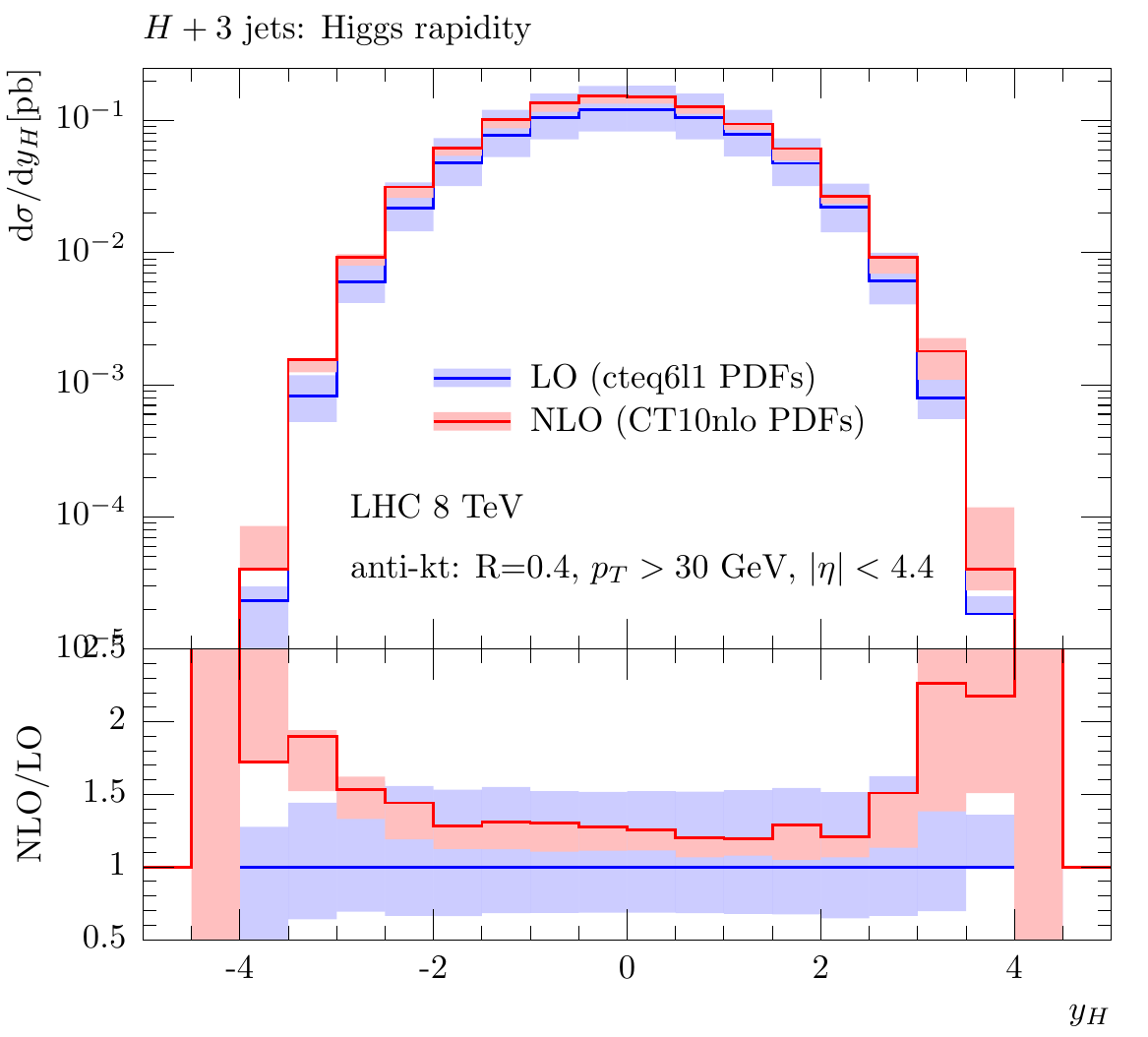}\hskip0mm
  \includegraphics[width=0.33\textwidth]{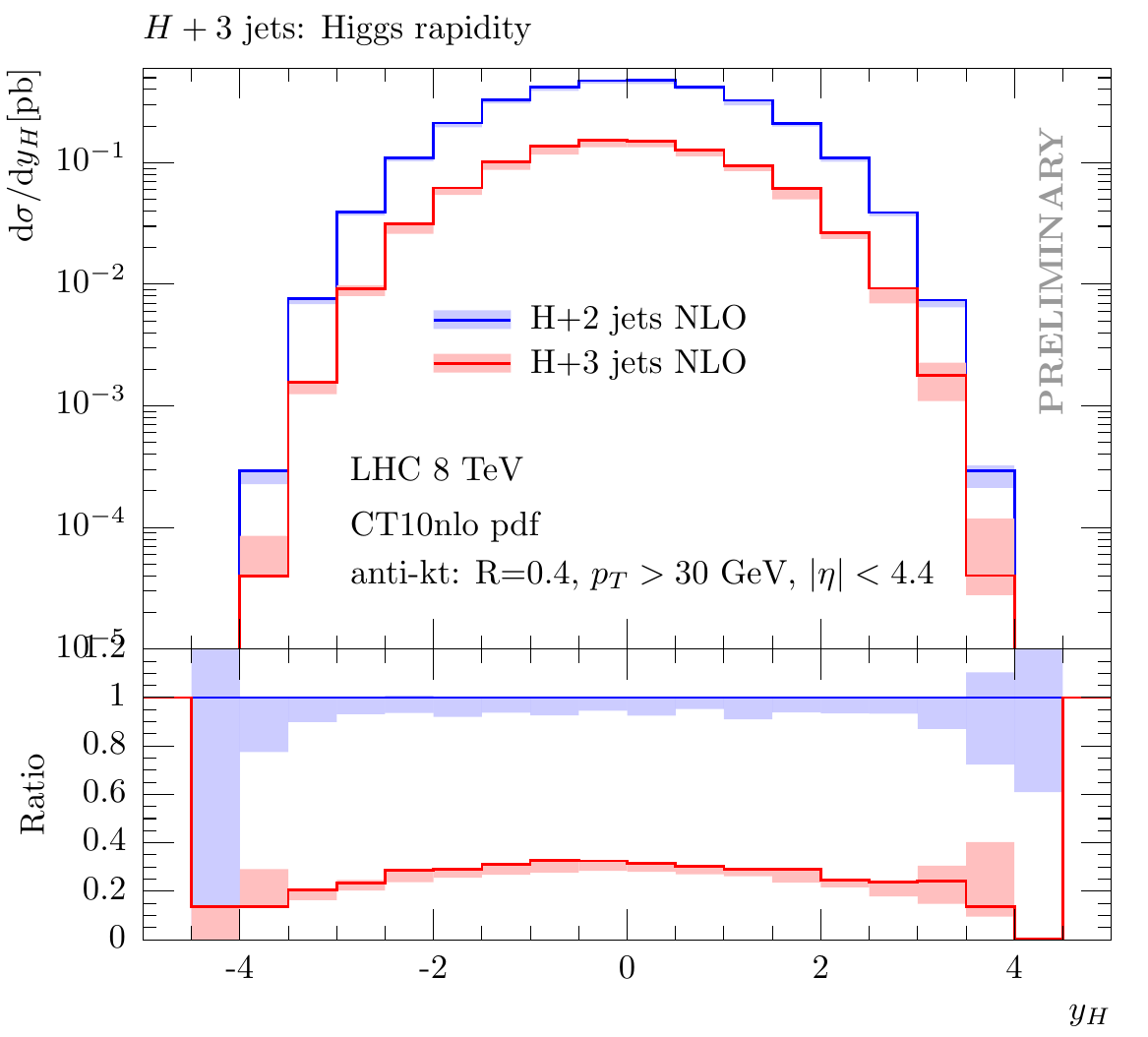}
  \caption{\label{hj3nlo_fig_y}
    Higgs boson rapidity distributions and their $\mu_\mathrm{R,F}$
    scale uncertainties (depicted by pastel-coloured bands) in
    $H$+3-jets production at the $E_\mathrm{cm}=8\mathrm{\:TeV}$ LHC.
    The two plots to the left show a comparison of the NLO (red) to LO
    (blue) predictions where the left panel's LO result has been
    obtained using the same PDF as for the NLO computation. The ratio
    plots at the bottom visualize the $K$-factor variation over the
    Higgs boson's rapidity range. In the rightmost panel, the $y_H$
    distributions of the 3-jets (red) and 2-jets (blue) NLO samples
    (requiring two tagging jets only) are compared to each other, and
    their differential cross section ratio is shown in the lower
    part.}
\end{figure}

We start our discussion by presenting several predictions for the
Higgs boson rapidity distribution in Fig.~\ref{hj3nlo_fig_y}. In all
three panels of this figure, we show the same NLO prediction for
$H$+3-jets production (red lines), and compare it, in the left two
panels, with two different $H$+3-jets predictions of LO accuracy (blue
lines). We observe large, $\mathcal{O}(50\%)$, positive corrections
that are spread almost uniformly over the entire $H$ rapidity range;
they increase in the forward/backward region. In the leftmost panel,
both the LO and NLO predictions were obtained from the same NLO PDF
set, that is \texttt{CT10nlo}. As seen in the middle panel, the
corrections decrease by $\sim20\%$ if one uses an order
$\alpha_\mathrm{s}$ treatment that is consistent between the PDFs and
the parton-level calculations. Half of the effect can then simply be
attributed to the larger $\alpha_\mathrm{s}(M_Z)$ value of the
\texttt{cteq6l1} parametrization that we used to compute the middle
panel's LO result. The scale uncertainties of the central predictions
are shown by the respective envelopes of same colour. Enhancing the
description to NLO accuracy, we find a reduction of these errors from
$\pm50\%$ to less than modulus $30\%$, which also means that the scale
variation bands turn from being fairly symmetric to rather
one-sided. This is a consequence of fixing the central/default scales
right where the NLO cross section plateaus.

\begin{figure}[p!]
  \centering\vskip-12mm
  \includegraphics[width=0.47\textwidth]{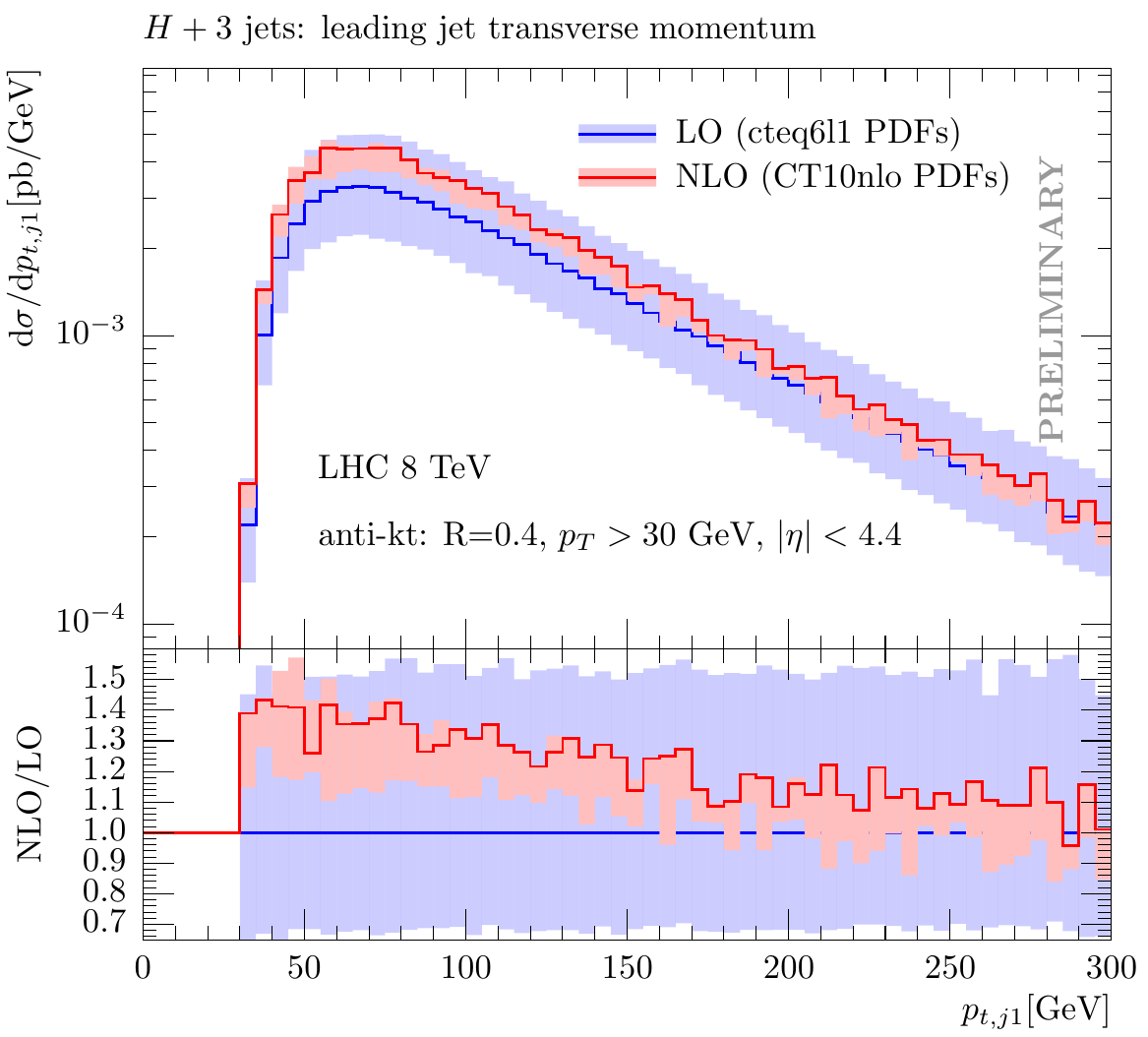}\hskip4mm
  \includegraphics[width=0.47\textwidth]{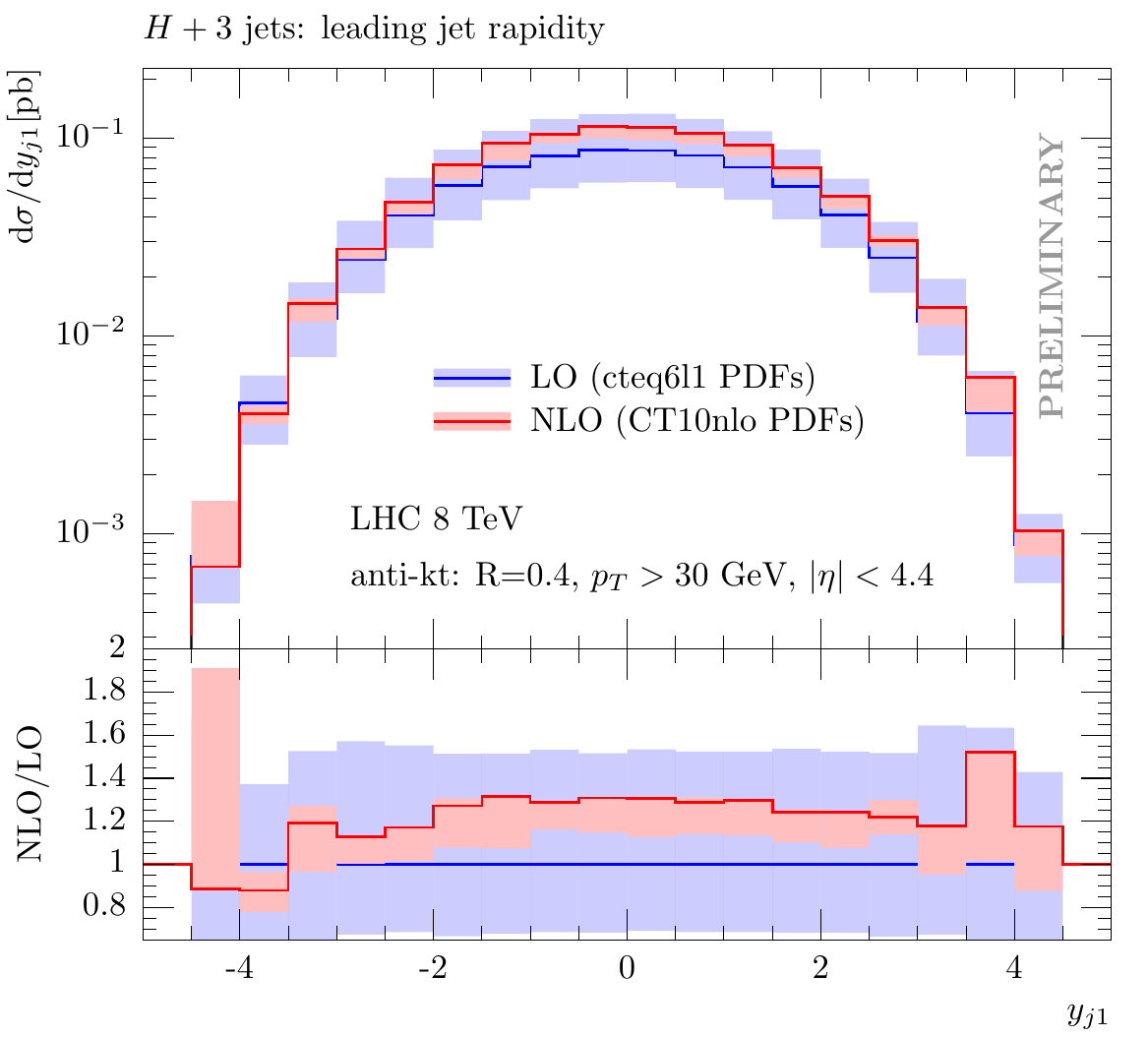}\\[0mm]
  \includegraphics[width=0.47\textwidth]{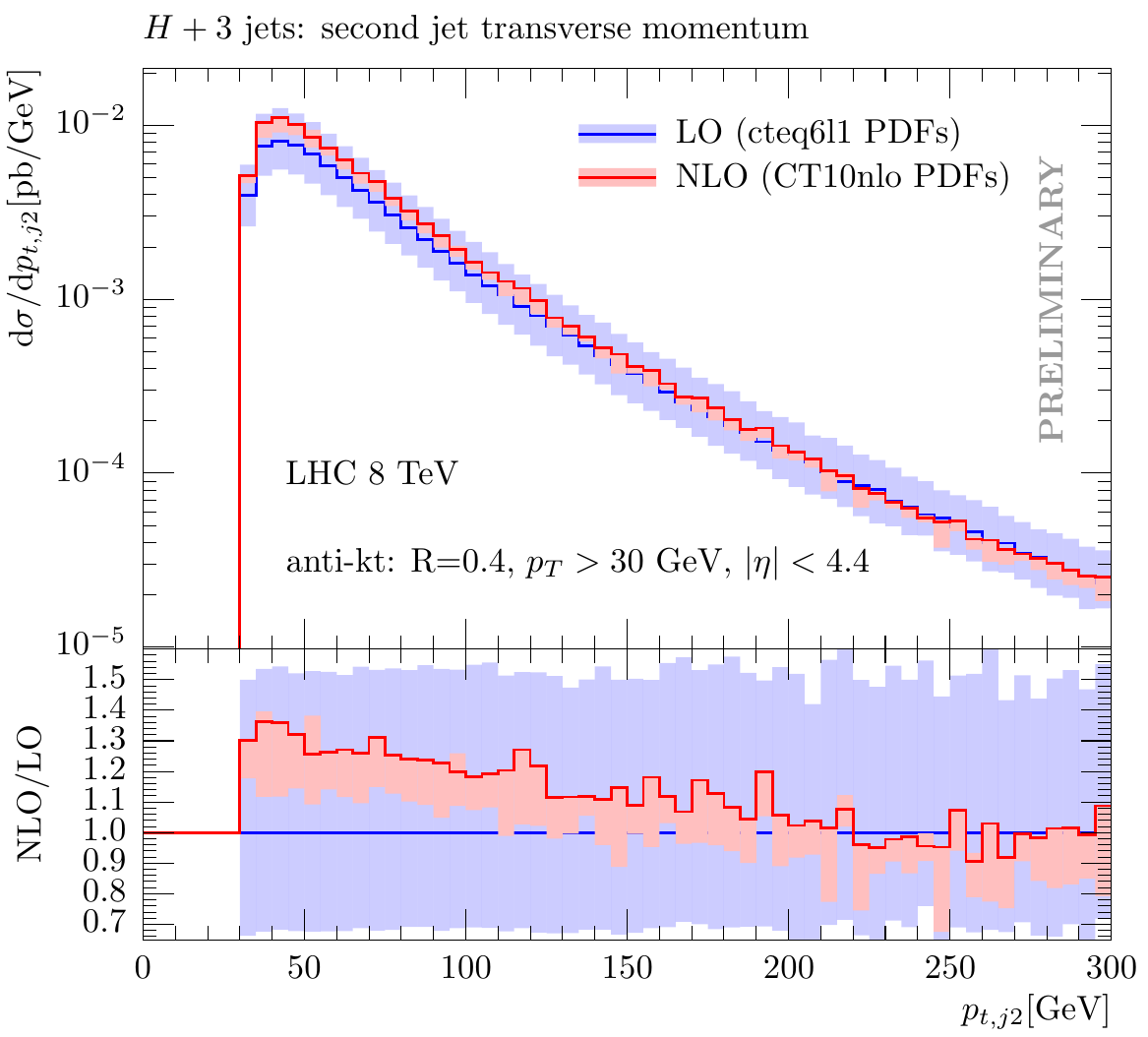}\hskip4mm
  \includegraphics[width=0.47\textwidth]{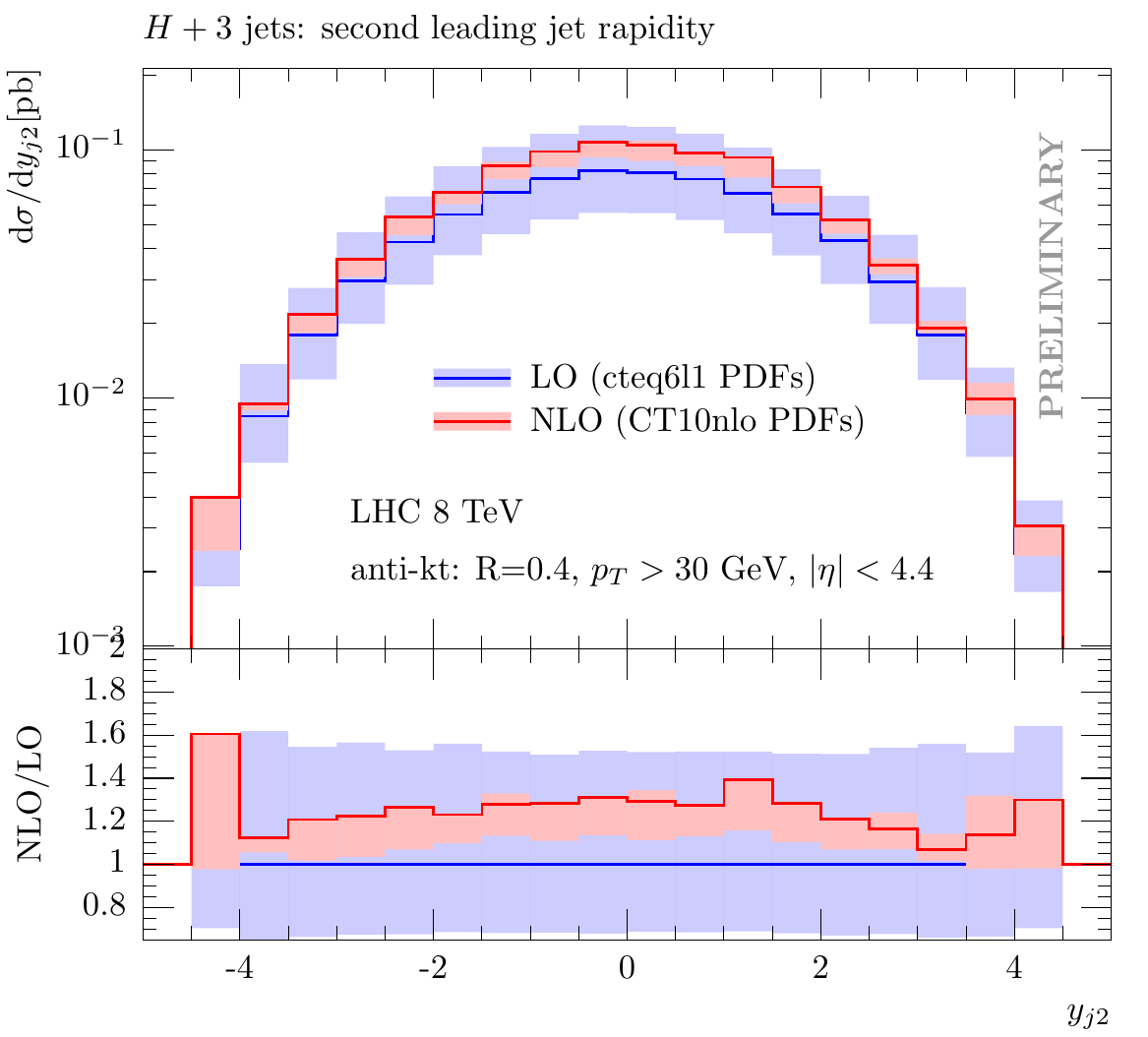}\\[0mm]
  \includegraphics[width=0.47\textwidth]{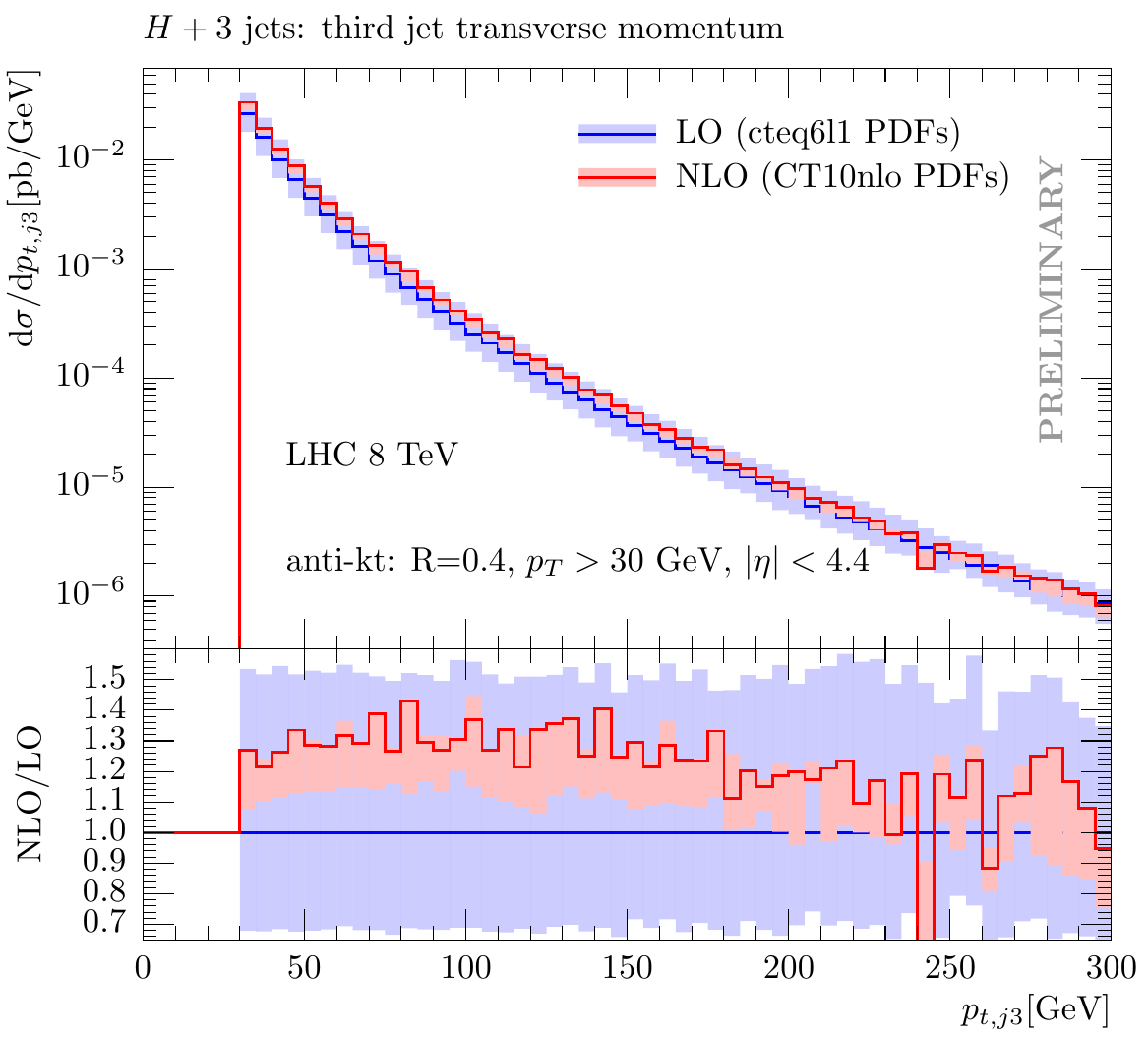}\hskip4mm
  \includegraphics[width=0.47\textwidth]{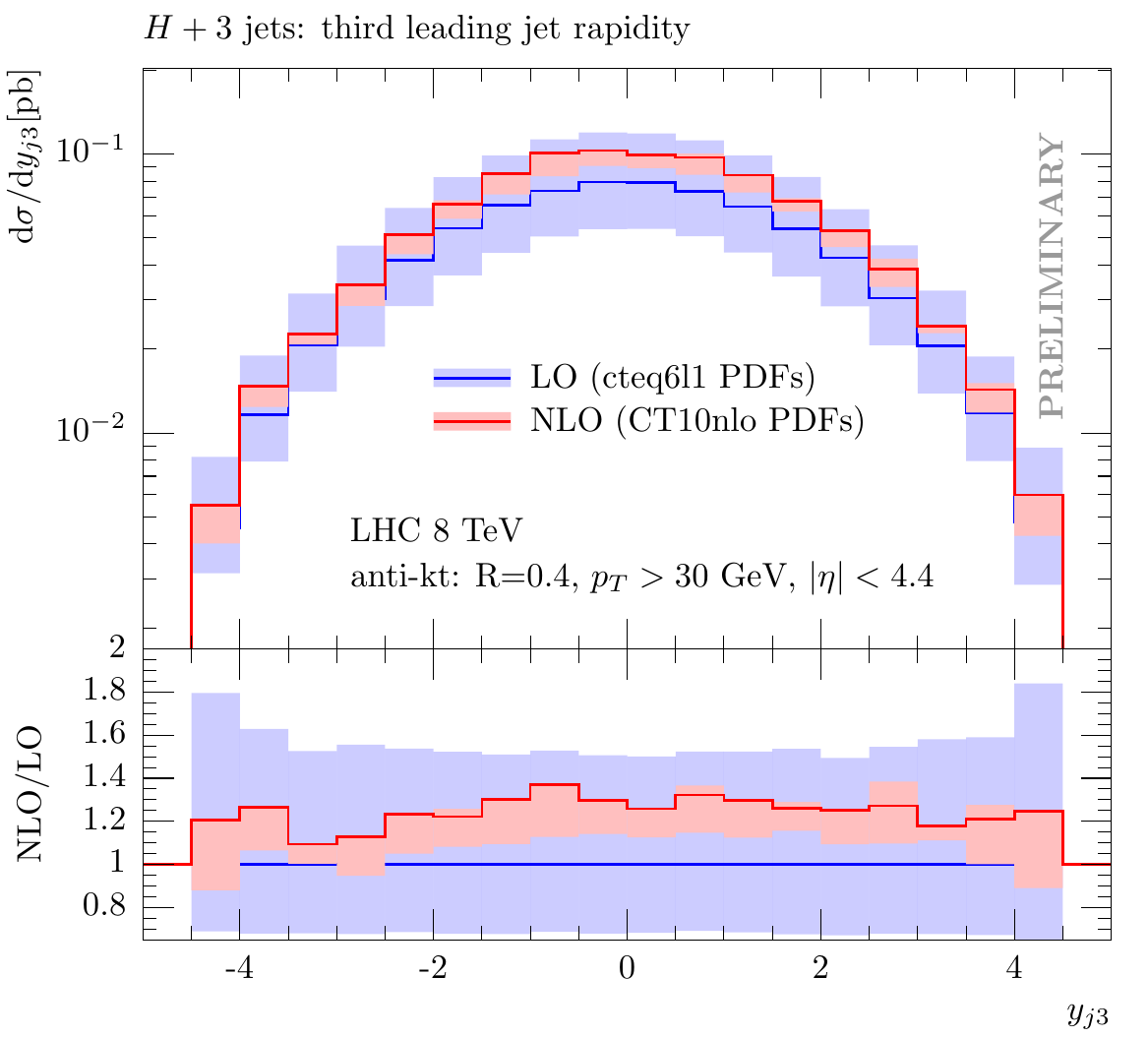}
  \caption{\label{hj3nlo_fig_singlejetKfacs}
    Higgs boson plus three-jet production at leading (blue) and next-to
    leading (blue) order for proton--proton collisions at
    $E_\mathrm{cm}=8\mathrm{\:TeV}$. The transverse momenta and
    rapidities of the three hardest jets are shown in the left and
    right column, respectively. The size of the scale uncertainties
    has been indicated by the lightly coloured bands occurring around
    each central prediction. The lower plots show the $K$-factor
    variations for each of these single jet observables.}
\end{figure}

\begin{figure}[p!]
  \centering\vskip-7mm
  \includegraphics[width=0.47\textwidth]{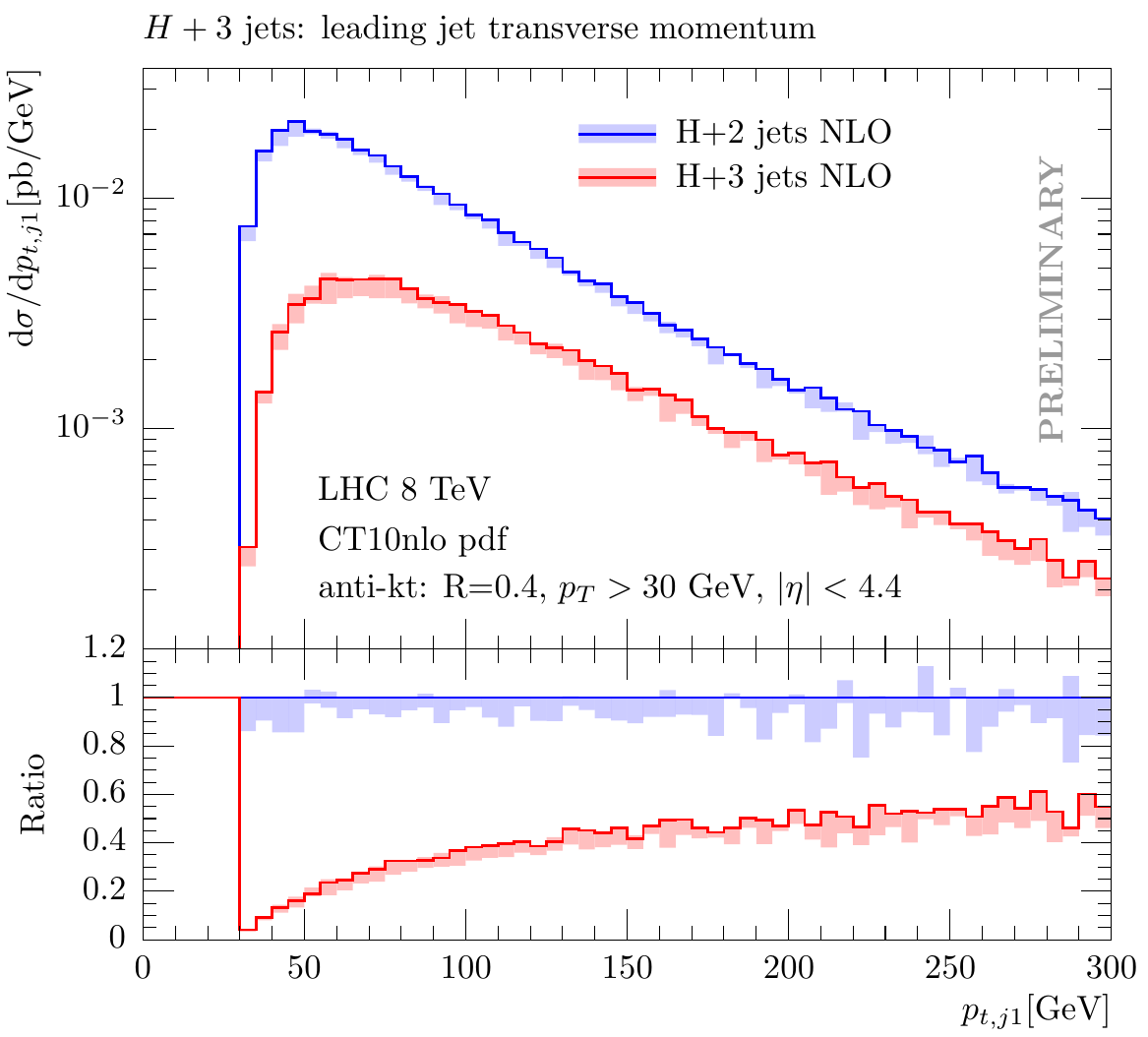}\hskip4mm
  \includegraphics[width=0.47\textwidth]{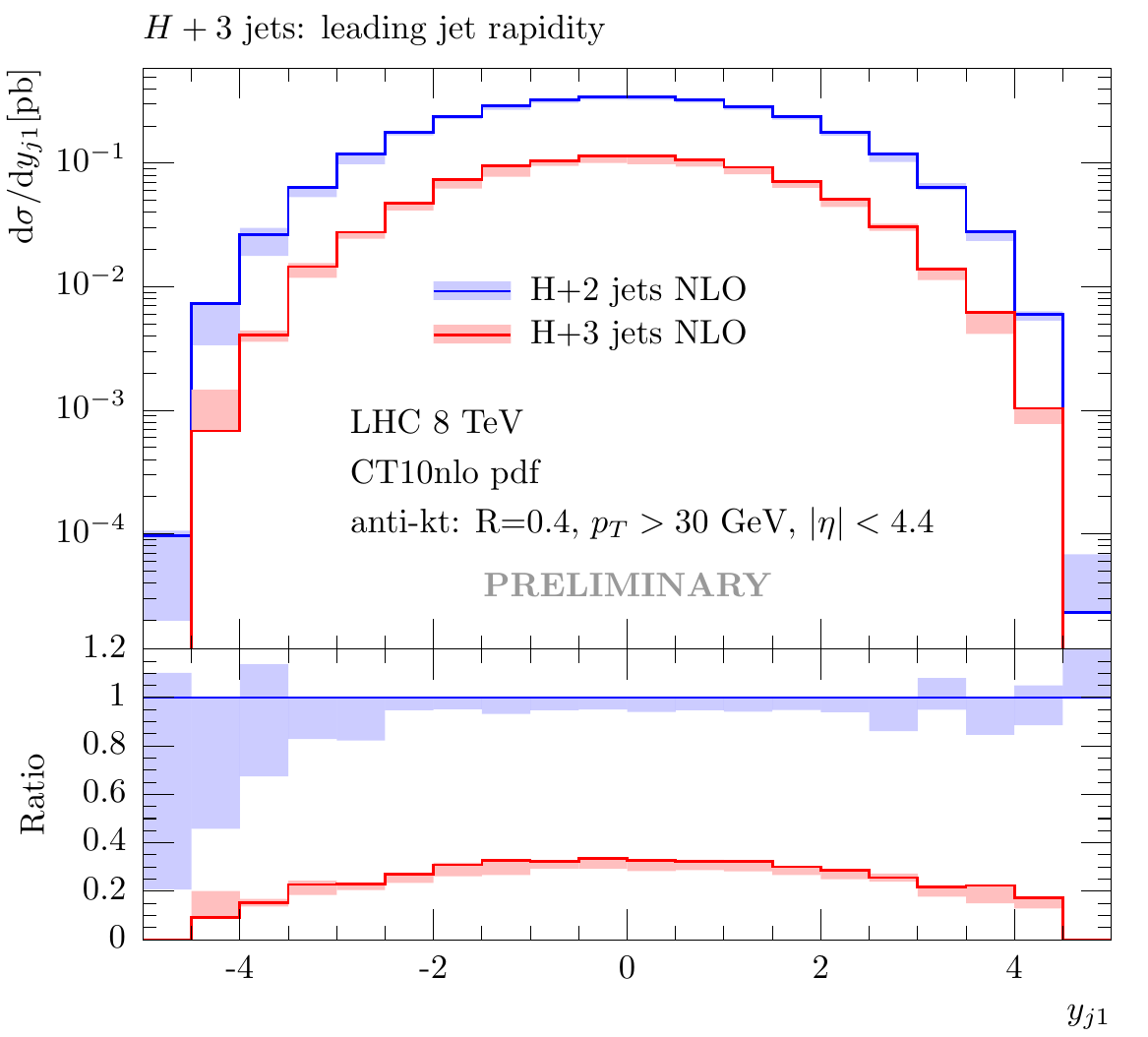}\\[0mm]
  \includegraphics[width=0.47\textwidth]{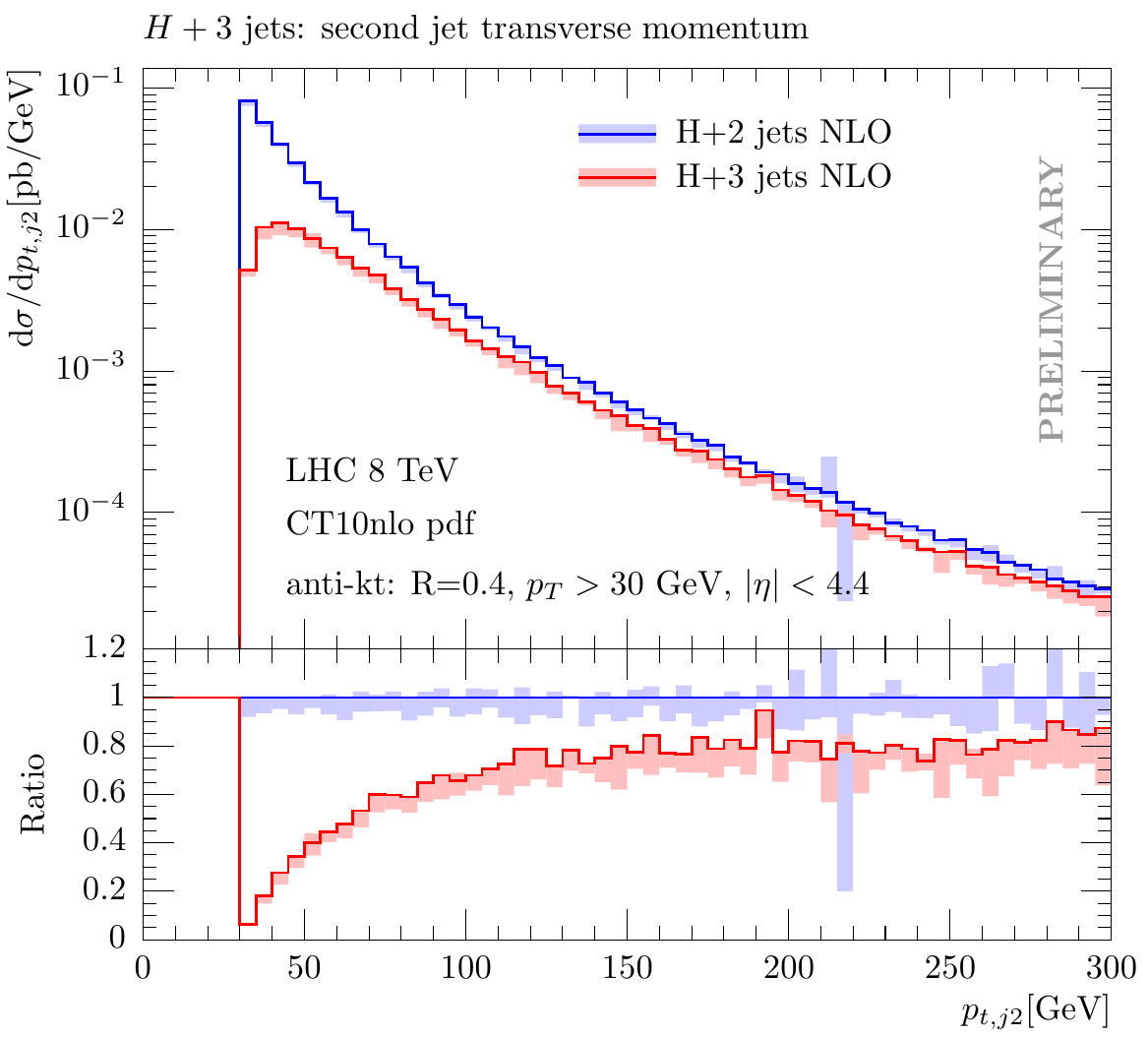}\hskip4mm
  \includegraphics[width=0.47\textwidth]{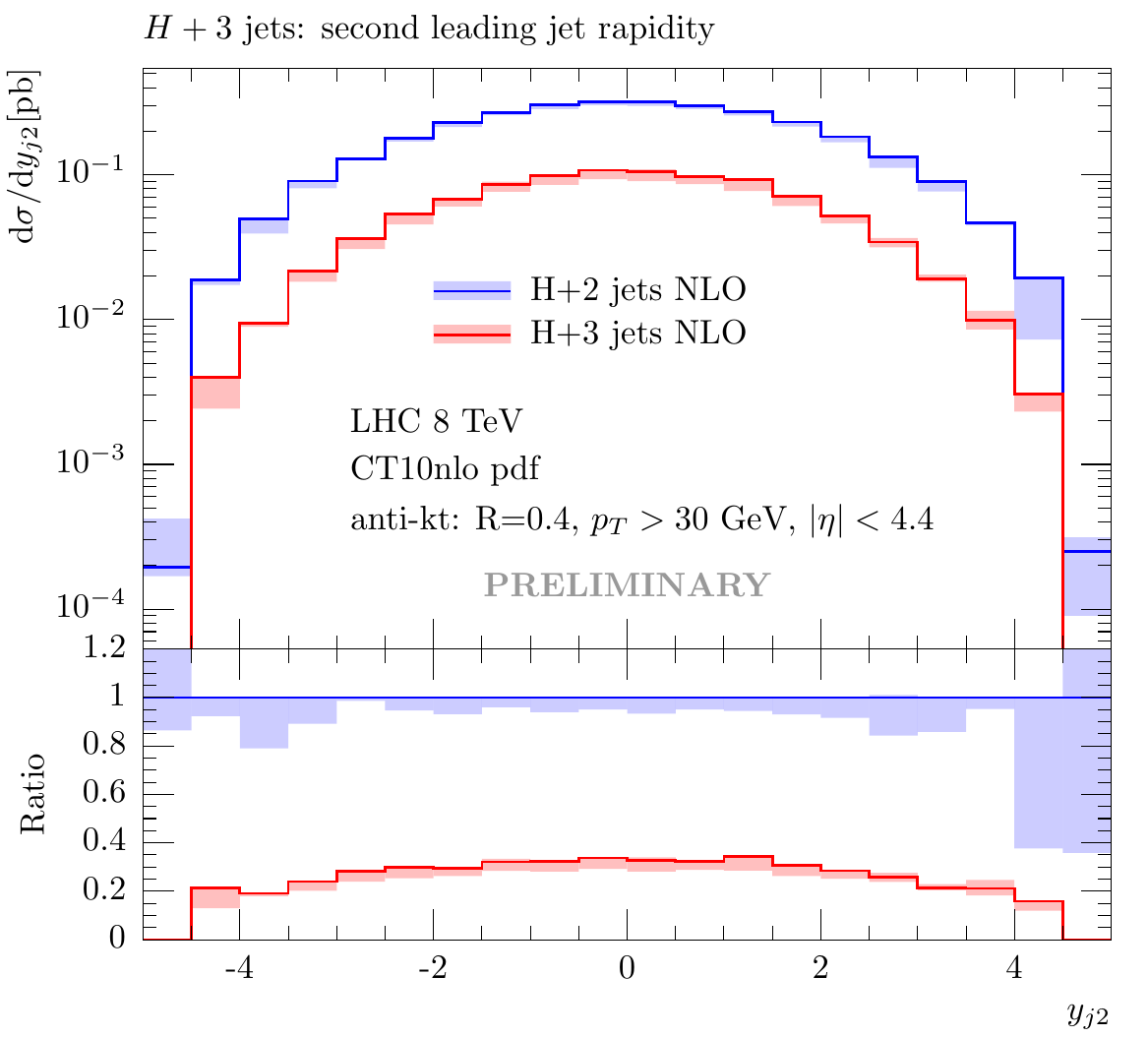}\\[0mm]
  \includegraphics[width=0.47\textwidth]{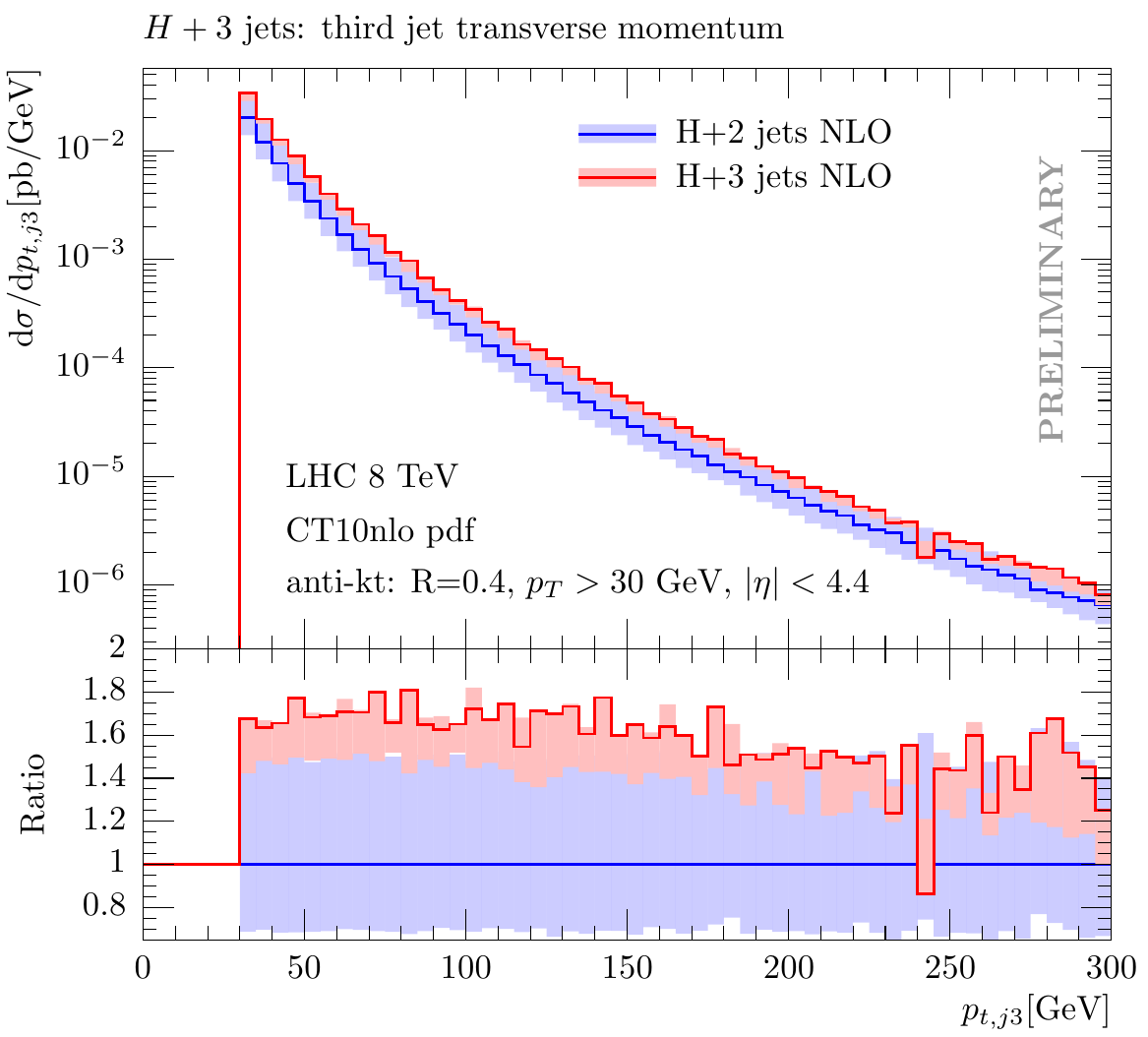}\hskip4mm
  \includegraphics[width=0.47\textwidth]{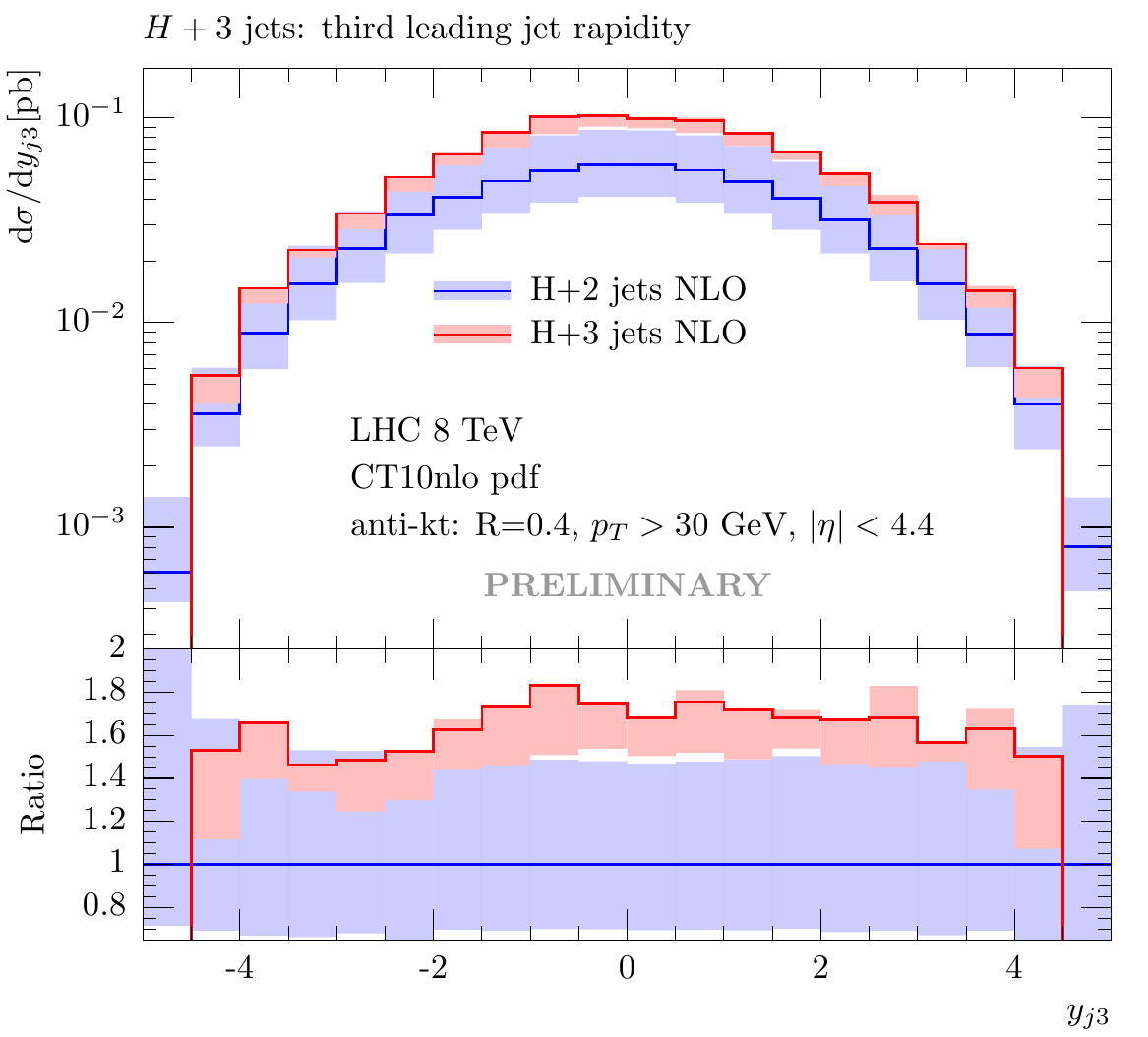}
  \caption{\label{hj3nlo_fig_singlejetcomp}
    Comparison of QCD NLO predictions for inclusive 2-jets (blue) and
    3-jets (red) final states in Higgs boson plus jets production at
    the $E_\mathrm{cm}=8\mathrm{\:TeV}$ LHC. The transverse momenta and
    rapidities of the three hardest jets are shown in the left and
    right column, respectively. The colourful envelops around the
    central predictions depict the size of their scale uncertainties.
    Each lower plot is used to indicate the cross section ratio
    between the three-jet and two-jet samples in dependence on the
    respective single jet observable. Note that in the $H$+2-jet
    calculation, the third-jet spectra are only described with LO
    accuracy.}
\end{figure}

The plot to the right of Fig.~\ref{hj3nlo_fig_y} is a direct
comparison of the $y_H$ distributions given by the 2-jets (blue) and
3-jets (red) NLO samples. Scale variations lead to $\mathcal{O}(20\%)$
uncertainties over a fairly broad $y_H$ range. Although the two NLO
samples differ by one order of $\alpha_\mathrm{s}$, the associated
scale uncertainties are comparable in size, and only somewhat smaller
for the $H$+2-jets case. Since we do not require more than the two
tagging jets, the $H$+3-jets line in the lower plot actually
visualizes the $r_{3/2}$ quantity differentially in dependence on
$y_H$. It varies only mildly around the inclusive, $\sim0.3$ value
given in the table above. The differential ratio also shows that the
production of Higgs bosons in the 3-jets sample is slightly more
central than in the 2-jets sample.

\begin{figure}[t!]
  \centering
  \includegraphics[width=0.47\textwidth]{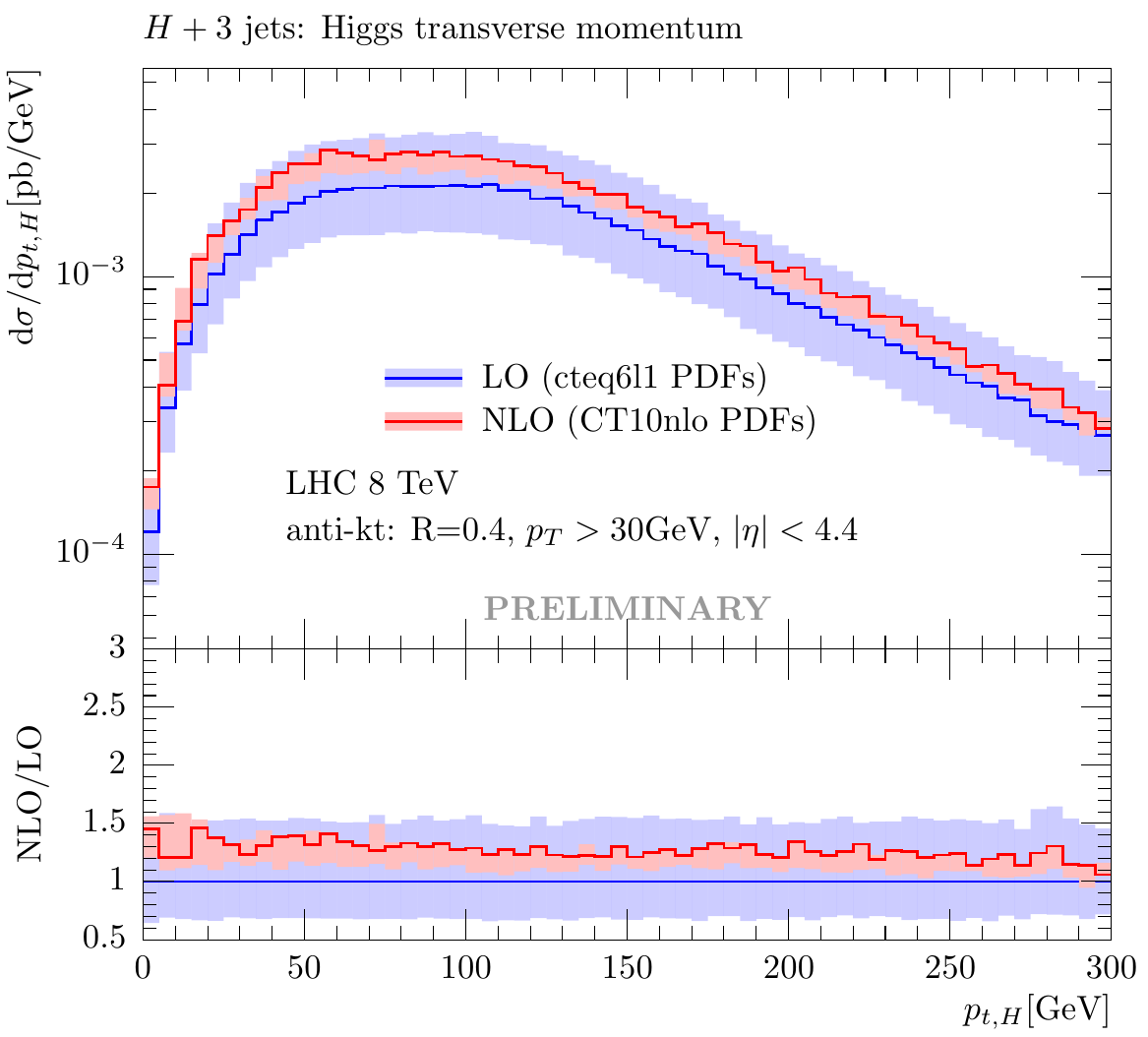}\hskip4mm
  \includegraphics[width=0.47\textwidth]{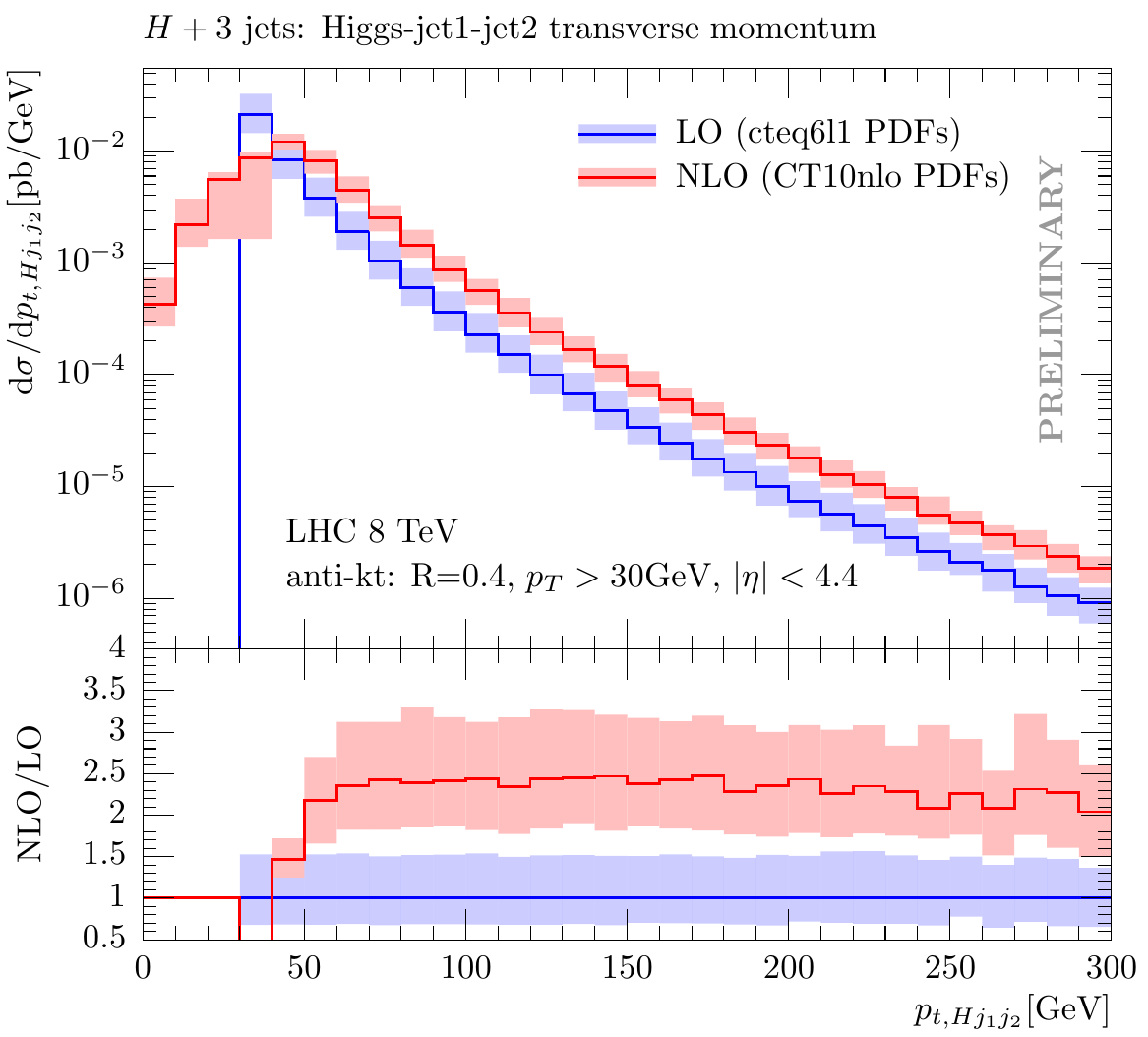}
  \caption{\label{hj3nlo_fig_ptHKfacs}
    Higgs boson plus three-jet production at leading (blue) and next-to
    leading (blue) order for proton--proton collisions at
    $E_\mathrm{cm}=8\mathrm{\:TeV}$. Transverse momentum distributions
    of the Higgs boson (left) and the $H$ plus tagging jet system
    (right) are shown together with their scale uncertainties
    indicated by the lightly coloured bands surrounding each central
    prediction. The lower plots show the $K$-factor variations as
    functions of $p_{T,H}$ and $p_{T,Hjj}$.}
\end{figure}

\begin{figure}[t!]
  \centering
  \includegraphics[width=0.47\textwidth]{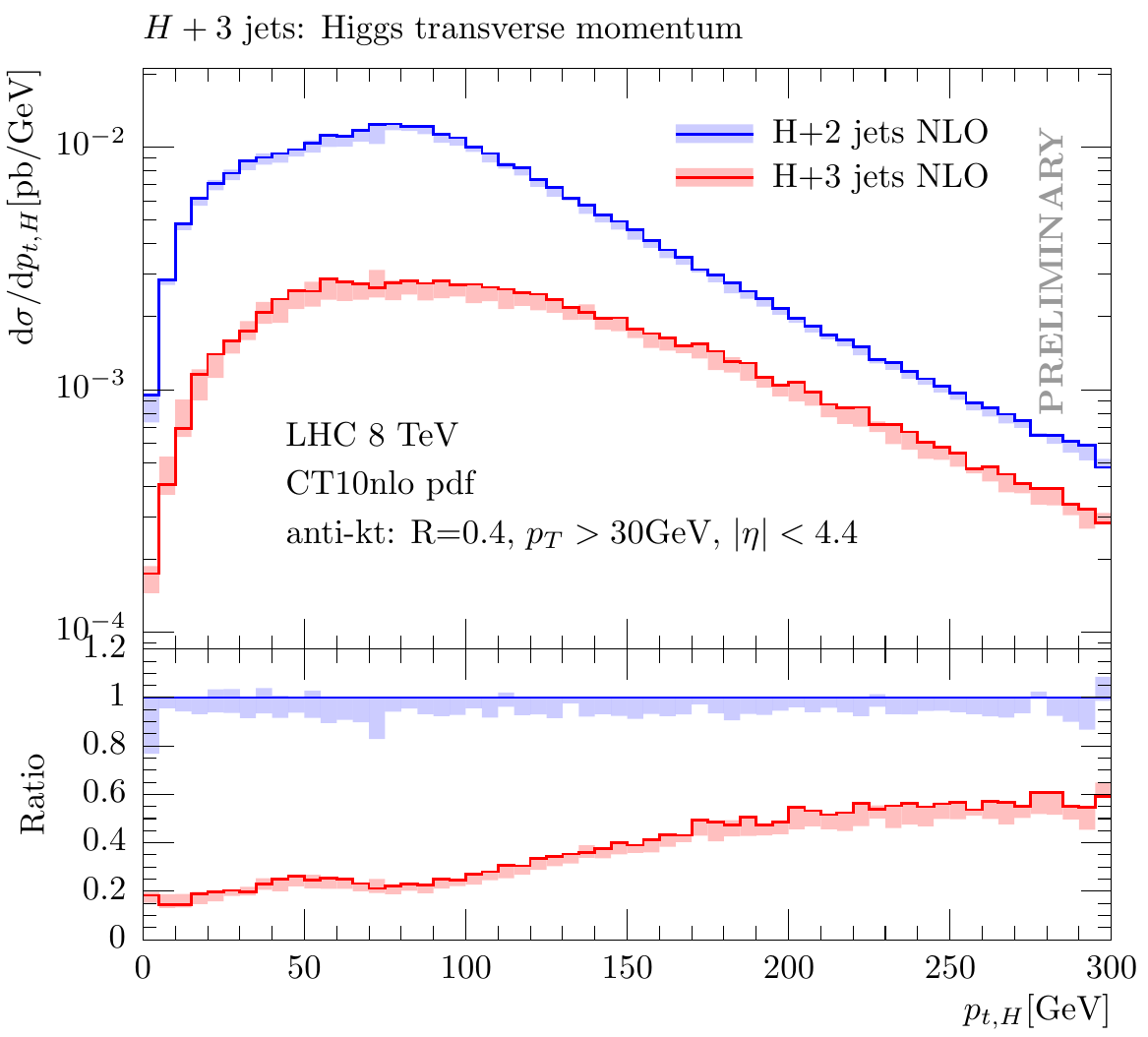}\hskip4mm
  \includegraphics[width=0.47\textwidth]{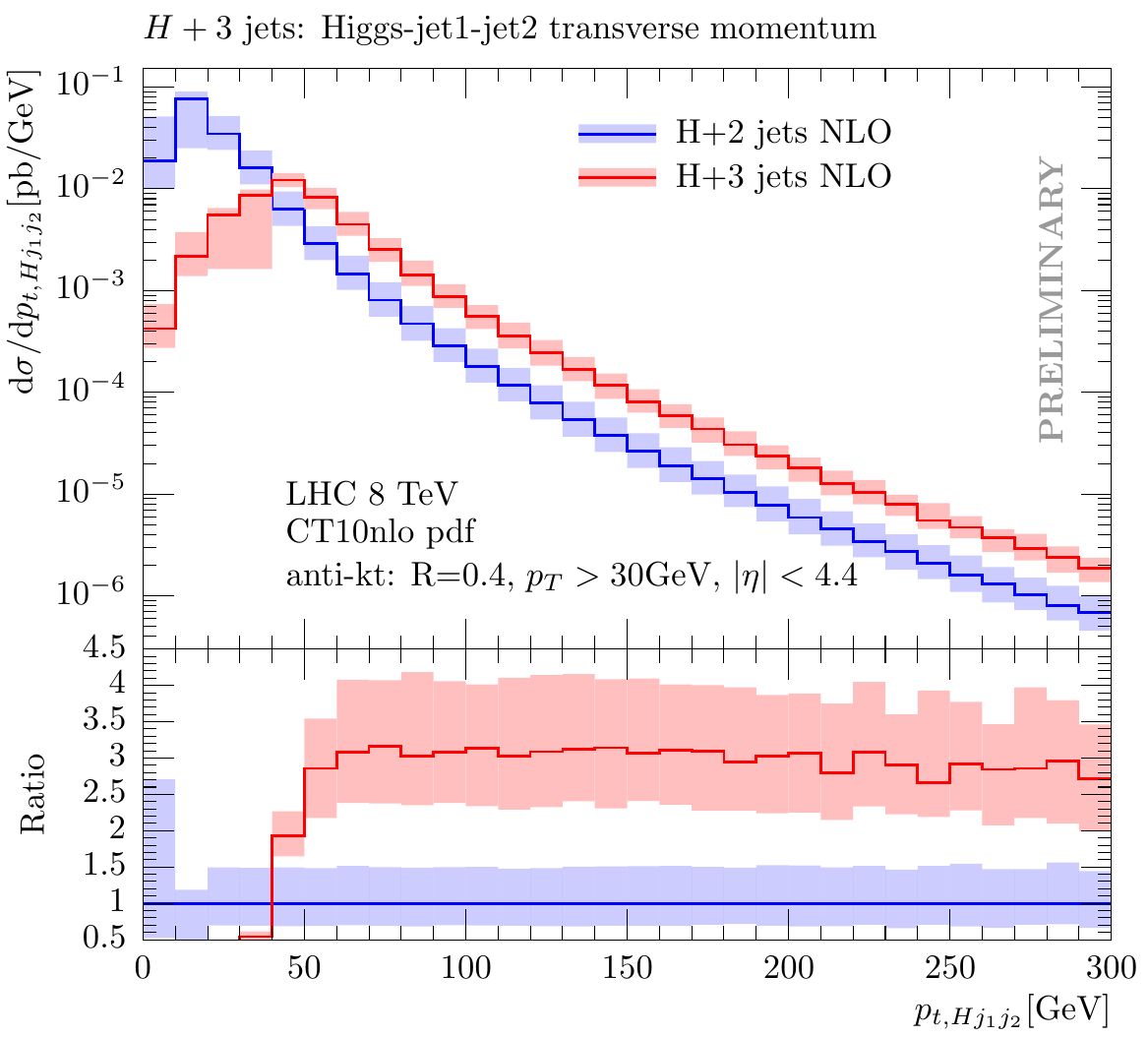}
  \caption{\label{hj3nlo_fig_ptHcomp}
    Comparison of QCD NLO predictions for inclusive 2-jets (blue) and
    3-jets (red) final states in Higgs boson plus jets production at
    the $E_\mathrm{cm}=8\mathrm{\:TeV}$ LHC. The transverse momenta of
    the Higgs boson and Higgs boson plus $j_1j_2$ system are shown in
    the left and right panel, respectively. Colourful envelops are
    used to depict the size of scale uncertainties. The lower plots
    indicate the cross section ratios between the three-jet and
    two-jet samples in dependence on the $p_T$ value. Note that in the
    $H$+2-jets case, the $p_{T,Hjj}$ distribution has LO accuracy only.}
\end{figure}

We now turn to discuss our results, which we obtained for the single
jet observables and the $H$-system $p_t$ spectra.
Figs.~\ref{hj3nlo_fig_singlejetKfacs},~\ref{hj3nlo_fig_ptHKfacs} and
Figs.~\ref{hj3nlo_fig_singlejetcomp},~\ref{hj3nlo_fig_ptHcomp} follow
the same plotting styles/layouts and colour code as chosen for the
center and rightmost subfigures of Fig.~\ref{hj3nlo_fig_y}, respectively.
The former set of figures is therefore used to give a broader comparison
between NLO and LO predictions as generated by the Higgs boson plus
3-jets final state. In the latter set, we then display how the
$H$+3-jets and $H$+2-jets NLO results compare to each other for exactly
the same selection of observables.

Focusing on the jet transverse momenta and rapidities, $p_{t,j_i}$
(left columns) and $y_{j_i}$ (right columns), as depicted for the
three hardest jets in Figs.~\ref{hj3nlo_fig_singlejetKfacs} and
\ref{hj3nlo_fig_singlejetcomp}, we notice that they show very similar
scale variation characteristics and reduction of errors as discussed
for the $y_H$ spectra presented in Fig.~\ref{hj3nlo_fig_y}. Similarly
for the NLO corrections on the jet rapidities, we again find these to
be rather well described by constant positive shifts, which here are
of the order of $20\%$, see Fig.~\ref{hj3nlo_fig_singlejetKfacs}. In
contrast to this, all differential $K$-factors associated with the jet
$p_t$ distributions in this figure show a decline towards larger $p_t$
values. In other words, even that the rate of Higgs associated jet
production gets enhanced at NLO, the $p_t$ tails loose hardness when
taking relative measures. This is due to the extra radiation carried
away from the system by the fourth jet, which shifts the spectra of
all the other jets to lower values. Note that the soft jet $K$-factors
can take values as large as $1.4$.

Turning to Fig.~\ref{hj3nlo_fig_singlejetcomp}, the very same set of
$H$+3-jets NLO predictions is plotted against the corresponding
$H$+2-jets predictions such that the differential $r_{3/2}$ ratios can
be read off easily. As before, they are rather flat for the $y_{j_i}$
predictions, and do not exceed values larger than $35\%$ owing to the
two-tagging-jet requirement. Only if a third jet has to be resolved
(cf.~the last row of Fig.~\ref{hj3nlo_fig_singlejetcomp}), we find
this ratio to be as large as $1.6$, which is no surprise, since it now
denotes a $K$-factor based on a LO third-jet description obtained with
NLO PDFs. For the transverse momenta of the first two jets, the
$r_{3/2}$ variables show a strong dependence on the hardness of the
jets. They rise quickly from very low values of $\mathcal{O}(0.1)$ at
the jet threshold to over $50\%$ for $p_t\sim100\mathrm{\:GeV}$. This
emphasizes the importance of higher jet multiplicity contributions if
one targets a complete description of $H$+jets final states.

Lastly we discuss, in the same manner as before, our findings
regarding the $p_t$ distributions of the Higgs bosons and the
$Hj_1j_2$ three-body objects. The NLO corrections to the Higgs boson
$p_t$ in three-jet final states (Fig.~\ref{hj3nlo_fig_ptHKfacs} to the
left) are of the order of $+30\%$ and decrease very slowly over the $p_t$
range plotted here. The behaviour under scale variations is as before.
Comparing to the $p_{t,H}$ spectrum extracted from the corresponding
2-jets sample, we observe an $r_{3/2}$ ratio that is rising as those
for the tagging jet $p_t$, but on a slower rate; see the left panel in
Fig.~\ref{hj3nlo_fig_ptHcomp}. In particular, for $p_{t,H}$ values
below $100\mathrm{\:GeV}$, it remains locked around $0.2$. For
transverse momenta above $2\,M_H$, the 3-jet contributions are again
as large as $50\%$, with similar implications regarding an inclusive
multi-jet description of $H$+jets final states as pointed out in the
case of the single jet $p_t$.

We round off by discussing our three different predictions for the
transverse momentum distribution of the $Hj_1j_2$ systems. Each of the
right panels in Figs.~\ref{hj3nlo_fig_ptHKfacs} and
\ref{hj3nlo_fig_ptHcomp} shows the $p_t$ spectrum computed from the
$H$+3-jets NLO samples. First we compare it to corresponding LO
predictions (Fig.~\ref{hj3nlo_fig_ptHKfacs}), then to the ones of the
$H$+2-jets NLO samples (Fig.~\ref{hj3nlo_fig_ptHcomp}). The
$p_{t,Hj_1j_2}$ is an interesting variable for our study since it is
sensitive to the description of any additional parton radiation beyond
the two tagging jets. It is out of question that a fixed-order
description of the low $p_t$ region cannot be achieved reliably
without supplementing the necessary resummation contributions.
Nevertheless we can use this variable to point out important features
of the three different calculations as well as discuss the formal
accuracy of the tails of these $p_{t,Hj_1j_2}$ spectra. The prediction
taken from the $H$+3-jets LO sample is just given by the third-jet
$p_t$ distribution of lowest order, with a clear indication of the jet
$p_t$ threshold at $30\mathrm{\:GeV}$, cf.~Fig.~\ref{hj3nlo_fig_ptHKfacs}.
The hardness of the third jet simply determines the recoil of the
$Hj_1j_2$ system. Above threshold, we therefore have a LO accurate
description of the $p_{t,Hj_1j_2}$ spectrum. The situation cannot be
improved by an $H$+2-jets NLO computation (Fig.~\ref{hj3nlo_fig_ptHcomp}):
the difference in the tails is caused by the replacement of LO with
NLO PDFs, again pointing to a $20\%$ effect as discussed earlier. The
region below the jet threshold gets filled by real emission
contributions that are too soft to be resolved as a jet. However, the
resulting low-$p_t$ spectrum is highly unphysical due to the missing
contributions from multiparton emissions.\footnote{Known as the
  Sudakov effect, the very same discussion occurs, for example, for
  the $p_{t,H}$ distribution in Higgs boson production at NLO. The
  reduction seen here for the very first $p_{t,Hj_1j_2}$ bin of the
  $H$+2-jets NLO computation is caused by those events satisfying
  $p_{t,Hj_1j_2}\equiv0$, i.e.~Born, virtual and integrated
  subtraction contributions as well as those from counter events fall
  into the first bin diminishing the large effect from (unresolved)
  real emissions.}
Turning to the NLO accurate evaluation of $H$+3 jets, we finally
improve the precision to which we describe the $p_{t,Hj_1j_2}$ tail.
For transverse momenta larger than $60\mathrm{\:GeV}$, we have
achieved NLO precision. Besides neglecting Sudakov corrections, below
this value we are missing contributions only a full NNLO $H$+2-jet
calculation can provide (contributions such as double unresolved and
virtual plus unresolved emissions). However, contributions with two
more jets beyond the tagged ones or three jets and unresolved extra
emission already lead to $p_t$ balancing as well as $p_t$ enhancing
effects such that we find large, $\mathcal{O}(3)$, fairly constant
corrections for higher $p_{t,Hj_1j_2}$ and a depletion towards
zero $p_{t,Hj_1j_2}$. A last few comments are in order concerning the
scale variation characteristics of the $p_{t,Hj_1j_2}$ predictions:
firstly, a reduction from $\pm50\%$ to $\pm30\%$ is only found for the
prediction from the $H$+3-jets NLO calculation (red envelopes). As
discussed the other evaluations only give LO accurate results.
Secondly, in contrast to other $H$+3-jets NLO predictions, the much
more symmetric band found here is a consequence of the dominance of
the four-jet contributions that feature a LO scale variation
behaviour.\footnote{On a more quantitative level, we note that the
  `BVI' contributions of the $p_{t,Hj_1j_2}$ and $p_{t,j_3}$
  distributions are exactly the same while the `RS' ones of the former
  are considerably harder than those of the latter. This leads to a
  different cancellation pattern when combining the scale varied `BVI'
  and `RS' predictions (they work in opposite directions). For the
  $p_{t,Hj_1j_2}$, we then find the  `RS' uncertainties to be the
  dominating ones for $p_{t,Hj_1j_2}\gtrsim60\mathrm{\:GeV}$.}

\subsection*{Conclusions}
Taking advantage of the recent developments in the automated
computation of NLO predictions we reported on NLO QCD results in an
ATLAS-like analysis of inclusive Higgs boson plus 2-jets and 3-jets
final states.

The loop amplitudes were generated with \textsc{GoSam} and computed
using a new development in integrand reduction techniques, based
on a Laurent expansion, and implemented in the code
\textsc{Ninja}. For tree-level amplitudes and phase space integration
we used \textsc{Sherpa} and the \textsc{MadGraph/Dipole/Event}
framework.

We find that NLO corrections are important and result in a substantial
change of rate and jet hardness. At the level of the total inclusive
cross section we find an increase of almost $30\%$ for both $H$+2-jets
and $H$+3-jets, whereas the scale variation reduces to approximately
$15\%$. Considering differential distributions, we observe that
rapidity distributions for the Higgs boson and the first three hardest
jets obtain a positive shift of about $20\%$, which is rather constant
over the full kinematical range. For the transverse momentum
distributions of the jets instead, we observe a decrease in the
K-factor towards larger $p_t$ values, whereas in the transverse
momentum distribution of the Higgs boson, this decrease is only very
slow. The differential $r_{3/2}$ ratios for the rapidities are rather
flat and never exceed $35\%$, however in the case of the transverse
momentum distributions they reach up to $50\%$ for the leading jet and
the Higgs boson. This shows that $H$+3-jets contributions cannot be
neglected in an inclusive two-jet sensitive analysis. Another
observable, where NLO corrections to $H$+3-jets play an important role
is the $p_t$ distribution of the $Hj_1j_2$ system. In fact, for the
first time, the distribution is described at NLO accuracy for
$p_t>2p^\mathrm{(jet)}_{t,\mathrm{min}}=60$ GeV.

It would be interesting to study the impact of the NLO corrections
presented here when typical VBF cuts are applied. Modern Monte Carlo
tools allow furthermore to study these correction in a matched NLO
plus parton shower framework merged with lower multiplicity
predictions. We postpone these studies to future publication.

\subsection*{Acknowledgements}
We would like to thank the organizers of the Les Houches "Physics at
TeV colliders 2013" workshop for the stimulating environment. 

The work of G.C. was supported by DFG SFB-TR-9 and the EU TMR Network
LHCPHENOnet.  The work of H.v.D., G.L., P.M., T.P., and V.Y. was supported
by the Alexander von Humboldt Foundation, in the framework of the
Sofja Kovaleskaja Award 2010, endowed by the German Federal Ministry
of Education and Research.  G.O. was supported in part by the National
Science Foundation under Grant PHY-1068550. F.T. acknowledges partial
support by MIUR under project 2010YJ2NYW. This research used computing
resources from the Rechenzentrum Garching.



\newcommand{\order}[1]{{\cal O}\left(#1\right)}
\newcommand{\as}{\alpha_s}
\newcommand{\aw}{\alpha_w}
\newcommand{\asCMW}{\alpha_s^{\rm CMW}}
\newcommand{\fb}{\;\mathrm{fb}}
\newcommand{\pb}{\;\mathrm{pb}}
\newcommand{\nb}{\;\mathrm{nb}}
\newcommand{\ptjv}{p_{\rm t,veto}}
\newcommand{\pt}{p_{\rm t}}
\newcommand{\cF}{{\cal F}}
\newcommand{\cO}[1]{{\cal O}\left(#1\right)}
\newcommand{\eff}{\epsilon}

\section{Jet-bin uncertainties estimate through jet efficiencies%
\footnote{P.~F.~Monni (Work in collaboration wit A.~Banfi,  G.P.~Salam, and G.~Zanderighi)}}




\subsection{Introduction}

In order to fully establish whether the new particle recently
discovered at the LHC~\cite{Aad:2012tfa,Chatrchyan:2012ufa} is the SM
Higgs boson, precise measurements of its properties, in particular its
couplings, are essential.  Establishing a significant departure from
the simple SM-like pattern could be a first manifestation of New
Physics.  In fact, the new particle lies in a region of parameter
space where many decay modes have an appreciable branching fraction,
leading to a very rich phenomenology. On the other hand, many decay
channels ({\it e.g.} $H\rightarrow b\bar b$, $H\rightarrow W^+W^-$)
are very difficult to measure because of the large QCD background.
Most Higgs-boson events involve either none or one
jet, since higher multiplicity events are suppressed by powers of the
strong coupling constant, and hence more rare.
It is natural then to impose a veto on additional jets so as to
significantly reduce the dominant background, while reducing only
modestly the signal.
Therefore, it is essential to know how much these tight
kinematical cuts reduce the signal cross section, and what the
uncertainties in the resulting signal are. However, a precise assessment 
of the theory error in the presence of a jet veto is a challenging task.

Tight kinematical cuts applied to QCD radiation give rise to large Sudakov logarithms
of the form $\alpha_s^n \ln ^{2n} (m_H/p_{\rm t,veto})$ in the perturbative series. 
When scale variation
is used to estimate the theory uncertainty, the K-factor effects nearly cancel against the 
large logarithms leading to an underestimate of the actual perturbative error.
Such logarithmic terms spoil the convergence of the perturbative series
in the region where the logarithms are very large, and they must be resummed to all orders
in the strong coupling. Not only does resummation allow for a more precise theory prediction, 
but it also gives a direct handle to estimate the size of higher-order Sudakov logarithms, which
is necessary for a reliable assessment of uncertainties.

Resummed predictions for the Higgs+$0$ jet cross section and
efficiencies have been obtained both in the context of the {\tt
  CAESAR} method~\cite{Banfi:2012yh,Banfi:2012jm}, and in an effective
theory framework~\cite{Becher:2012qa} up to next-to-next-to-leading
logarithmic level. Some terms beyond this accuracy have been recently
obtained in~\cite{Stewart:2013faa,Becher:2013xia}. The impact of heavy
quark masses on the resummed predictions has been also studied for
both the $p_{\rm t,veto}$~\cite{Banfi:2013eda} and $p_{\rm
  t,H}$~\cite{Banfi:2013eda,Grazzini:2013mca} cross sections.  When
more than a tagged jet is present in the process, the infrared
structure of the perturbative series becomes more complex and new
logarithms of the non-global type~\cite{Banfi:2002hw,Dasgupta:2001sh}
appear. For the H+$1$ jet cross section, NLL resummation for the
global class in the regime $p_{\rm t,veto} \ll p_{\rm t,jet}\sim m_H$
was obtained in~\cite{Liu:2013hba}.

In order to optimize the sensitivity, some experimental analyses ({\it
  e.g.} $gg\rightarrow H\rightarrow W^+W^-$, VBF) combine data with
different jet multiplicities. Since resummed predictions are currently
only available for $0$- and $1$-jet bins, it is of primary importance
to have a robust prescription to estimate reliably the uncertainties
associated with a fixed-order calculation. Moreover, a good method
should allow one to incorporate any resummed predictions whenever
available.

A first approach based on the combination of inclusive jet bin
uncertainties was proposed by Stewart and Tackmann
in~\cite{Stewart:2011cf}. Uncertainties in the exclusive jet-bin cross
sections are obtained by treating the errors in the inclusive jet-bins
as uncorrelated.  An alternative prescription was presented
in~\cite{Banfi:2012yh,Banfi:2012jm}, and it is referred to as the
jet-veto efficiency (JVE) method. It is based on the assumption that
the uncertainties in the jet-fractions (efficiencies) with different
multiplicities are uncorrelated with each other and with the
uncertainty in the total cross section. This is physically motivated
by the fact that uncertainties in jet fractions are dominated by
Sudakov suppression effects, whilst K-factor effects (which rule the
uncertainty in the total cross section) largely cancel in the
definition of the jet fractions themselves.  The latter method allows
one to treat resummed predictions for different jet bins on the same
footing as fixed-order ones. For the resummed $0$-jet cross section,
the error obtained with the JVE method is consistent with the direct
estimate obtained from scale variations in the resummed calculation.
More recently, a recipe to combine uncertainties for different
resummed jet-bin cross sections in the context of SCET resummation was
presented in~\cite{Boughezal:2013oha}. When resummed predictions are
not available, the latter method reduces to the standard scale
variation which underestimates the theoretical error.  In the
following sections, we briefly recall the JVE method, and present a
generalisation to any jet multiplicity.

\subsection{The JVE method}
The uncertainties in jet-bin cross sections are due to a combination of K-factor effects which rule the uncertainty in the total cross section, and Sudakov suppression effects due to the presence of severe cuts on the real QCD radiation. The idea behind the JVE approach is to disentangle these two effects and treat them separately. Therefore, we define n-jet efficiencies $\epsilon_n$ as
\begin{equation}
\label{eq:efficiency;qcdsm}
\epsilon_n = \frac{\sigma_{\rm n-jet}}{\sigma_{\rm \ge n-jet}},
\end{equation}
where $\sigma_{\rm n-jet}$ is the exclusive $n-$jet cross section.

To a large extent, K-factor effects cancel in the ratio~(\ref{eq:efficiency;qcdsm}), since the uncertainty in the $n-$jet efficiency is largely ruled by Sudakov suppression.
Therefore, one can treat the uncertainties in the $n-$jet efficiency and in the inclusive $n-$jet cross section as uncorrelated. The uncertainties in the two quantities are computed separately and then combined according to the latter assumption. 

The theory error in the inclusive $n-$jet cross section is obtained by standard scale variation, which is not affected by cancellations in the inclusive case. Threshold resummation effects can be also included~\cite{Heinemeyer:2013tqa}.
To obtain the uncertainty in the $n-$jet efficiency, we start by
noticing that $m+1$ different schemes for the jet fractions can be
defined at N$^m$LO. As an example we consider the $0-$jet efficiency $\epsilon_0$. Using the current state-of-the-art calculations at NNLO for both the $0-$jet and the total cross sections, the following three equivalent schemes can be defined to this perturbative accuracy

  \begin{eqnarray}
    \eff_0^{(a)}(\ptjv) &=& \frac{\Sigma_0(\ptjv)+\Sigma_1(\ptjv)+\Sigma_2(\ptjv)}{\sigma_0+\sigma_1+\sigma_2}\,, 
    \label{eq:NNLOa;qcdsm}\\
    \eff_0^{(b)}(\ptjv) &=&
    \frac{\Sigma_0(\ptjv)+\Sigma_1(\ptjv)+\bar
      \Sigma_2(\ptjv)}{\sigma_0+\sigma_1}\,,
    \label{eq:NNLOb;qcdsm}\\
    \eff_0^{(c)}(\ptjv) &=& 1 +\frac{\bar \Sigma_1(\ptjv)}{\sigma_0}+
    \left(\frac{\bar
        \Sigma_2(\ptjv)}{\sigma_0}-\frac{\sigma_1}{\sigma_0^2}\bar
      \Sigma_1(\ptjv)\right) \,,
    \label{eq:NNLOc;qcdsm}
  \end{eqnarray}
where 
  \begin{eqnarray}
    \Sigma_{\rm 0-jet}(\ptjv) &=& \Sigma_{0}(\ptjv) + \Sigma_{1}(\ptjv) + \Sigma_{2}(\ptjv) + \ldots \\
    \sigma_{\rm\ge 0-jet} &=& \sigma_{\rm tot} =  \sigma_{0} + \sigma_{1} + \sigma_{2} + \ldots ,
  \end{eqnarray}
  and
  \begin{equation}
    \Sigma_{i}(\ptjv) = \bar\Sigma_{i}(\ptjv) + \sigma_i,\qquad 
\bar \Sigma_i(\ptjv) = - \int_{\ptjv}^\infty\, d\pt \frac{d\Sigma_{i}(\pt)}{d\pt}\,.
  \end{equation}
Schemes~(\ref{eq:NNLOb;qcdsm}) and~(\ref{eq:NNLOc;qcdsm}) differ from Eq.~(\ref{eq:NNLOa;qcdsm}) only by terms $\cO{\as^3}$, which are not under control.
The uncertainty in the $0-$jet efficiency is obtained as the envelope of all scale variations for the central scheme~(\ref{eq:NNLOa;qcdsm}) and the central values relative to the remaining two schemes~(\ref{eq:NNLOb;qcdsm}) and~(\ref{eq:NNLOc;qcdsm}). The spread between different schemes is associated with the Sudakov suppression effects and it increases for lower efficiencies, {\it i.e.} when more radiation is suppressed.
When resummation of large Sudakov logarithms is available, it can be included as shown in~\cite{Banfi:2012yh,Banfi:2012jm}, and the variation of the corresponding resummation scale is included in the envelope. Resummed and fixed-order calculations for different jet bins are treated in the same way, and can be straightforwardly combined.
Below we discuss the JVE method for an arbitrary number of exclusive
jet bins, and we explicitly work out the case of only two jet bins.

\subsubsection{Formulation for any jet multiplicity}

In the general case with $n$ exclusive jet bins and one inclusive bin, we define the $n$ exclusive cross sections as
\begin{eqnarray}
\sigma_{\rm 0-jet} =& \epsilon_0 \sigma_{\rm tot},\\
\sigma_{\rm 1-jet} =& \epsilon_1(1-\epsilon_0) \sigma_{\rm tot},\\
\sigma_{\rm 2-jet} =& \epsilon_2(1-\epsilon_1)(1-\epsilon_0) \sigma_{\rm tot},\\
...&\notag\\
\sigma_{\rm n-jet} =& \epsilon_n (1-\epsilon_{n-1})...(1-\epsilon_0) \sigma_{\rm tot},
\end{eqnarray}
and the inclusive one as
\begin{equation}
\sigma_{\rm > n-jet} =  (1-\epsilon_n)(1-\epsilon_{n-1})...(1-\epsilon_0) \sigma_{\rm tot}.
\end{equation}
To express the uncertainty in the jet-bin cross sections, we define the following $n+1$  vectors in the ($n+1$)-dimensional euclidean space 
\begin{equation}
\label{eq:errvec;qcdsm}
\Delta \sigma_{\rm i} = \left(\frac{\partial \sigma_{\rm i}}{\partial \epsilon_n}\delta\epsilon_n,\frac{\partial \sigma_{\rm i}}{\partial \epsilon_{n-1}}\delta\epsilon_{n-1},...,\frac{\partial \sigma_{\rm i}}{\partial \epsilon_0}\delta\epsilon_0,\frac{\partial \sigma_{\rm i}}{\partial \sigma_{\rm tot}}\delta\sigma_{\rm tot}\right),
\end{equation}
 where $\sigma_{\rm i}$ denotes either any of the $n$ exclusive or the
 inclusive jet-bin cross sections, $\delta\epsilon_i$ denotes the
 uncertainty in the $i-$jet efficiency, computed as described above,
 and $\delta\sigma_{\rm tot}$ is the uncertainty in the total cross
 section. They fulfill the following conditions
\begin{eqnarray}
\delta\epsilon_i\cdot\delta\epsilon_j &=& \delta^2\epsilon_i
\delta_{ij},\\
\delta\epsilon_i\cdot \delta\sigma_{\rm tot} &=& 0.
\end{eqnarray}
 The vectors corresponding to the exclusive cross sections read
\begin{eqnarray}
\Delta\sigma_{\rm 0-jet} &=& \left(0\,,...,\,\sigma_{\rm tot}\delta\epsilon_0\,,\,\epsilon_0\delta\sigma_{\rm tot}\right),\\
\Delta\sigma_{\rm 1-jet} &=& \left(0,...,\,(1-\epsilon_0)\sigma_{\rm tot}\delta\epsilon_1\,,\,-\epsilon_1\sigma_{\rm tot}\delta\epsilon_0\,,\,\epsilon_1(1-\epsilon_0)\delta\sigma_{\rm tot}\right),\\
\Delta\sigma_{\rm 2-jet} &=& (0\,,...,\,(1-\epsilon_1)(1-\epsilon_0)\sigma_{\rm tot}\delta\epsilon_2\,,\,-\epsilon_2(1-\epsilon_0)\sigma_{\rm tot}\delta\epsilon_1\,,\notag
\\
&&\,-\epsilon_2(1-\epsilon_1)\sigma_{\rm tot}\delta\epsilon_0\,,\,\epsilon_2(1-\epsilon_1)(1-\epsilon_0)\delta\sigma_{\rm tot}),\\
&...&\notag
\end{eqnarray}
while for the inclusive one we find
\begin{align}
\Delta\sigma_{\rm  > n-jet} = -\bigg(\prod_{i=0; i\ne n}^{i=n}(1-\epsilon_i)\sigma_{\rm tot}\delta\epsilon_n\,&,\,\prod_{i=0; i\ne n-1}^{i=n}(1-\epsilon_i)\sigma_{\rm tot}\delta\epsilon_{n-1}\,,\,...,\notag\\
\,&\prod_{i=0; i\ne 0}^{i=n}(1-\epsilon_i)\sigma_{\rm tot}\delta\epsilon_0\,,\,-\prod_{i=0}^{i=n}(1-\epsilon_i)\delta\sigma_{\rm tot}\bigg),
\end{align}
where we define 
\begin{align}
\prod_{i=0; i\ne 0}^{i=0}(1-\epsilon_i) = 1.
\end{align}
The resulting symmetric covariance matrix reads
\begin{equation}
\label{eq:cov-gen;qcdsm}
{\rm Cov}[\sigma_{\rm 0-jet},\sigma_{\rm 1-jet},...,\sigma_{\rm > n-jet}]=
\left(\begin{array}{cccc} \Delta\sigma_{\rm 0-jet}\cdot\Delta\sigma_{\rm 0-jet} & \Delta\sigma_{\rm 0-jet}\cdot \Delta\sigma_{\rm 1-jet} & ... & \Delta\sigma_{\rm 0-jet}\cdot \Delta\sigma_{\rm > n-jet}\\
 & \Delta\sigma_{\rm 1-jet}\cdot\Delta\sigma_{\rm 1-jet} & ... & \Delta\sigma_{\rm 1-jet}\cdot \Delta\sigma_{\rm > n-jet}\\
 &  &  & \Delta\sigma_{\rm > n-jet}\cdot\Delta\sigma_{\rm > n-jet}\end{array}\right).
\end{equation}

\subsubsection{Case with two jet bins}
As a case of study we analyse the case of a $0-$jet bin and an inclusive $1-$jet bin. Resummed predictions for the $0-$jet efficiency $\epsilon_0$ were obtained in~\cite{Banfi:2012yh,Banfi:2012jm} up to NNLO+NNLL.
The inclusive $1-$jet cross section can be written in terms of $\epsilon_0$ as
\begin{equation}
\sigma_{\rm\ge 1-jet}(\ptjv) = (1-\epsilon_0)\sigma_{\rm tot},
\end{equation}
where $\sigma_{\rm tot}$ is the total cross section for Higgs boson production in gluon fusion, currently known to NNLO. Since we only have two jet bins, we can define the following two 2-dimensional vectors~(\ref{eq:errvec;qcdsm})
\begin{eqnarray}
\Delta\sigma_{\rm 0-jet} &=& \left(\sigma_{\rm tot}\delta\epsilon_0\,,\,\epsilon_0\delta\sigma_{\rm tot}\right),\\
\Delta\sigma_{\rm \ge 1-jet} &=& \left(-\sigma_{\rm tot}\delta\epsilon_0\,,\,(1-\epsilon_0)\delta\sigma_{\rm tot}\right).
\end{eqnarray}
The covariance matrix between the exclusive $0-$jet and the inclusive $1-$jet cross sections can be easily obtained as shown in Eq.~(\ref{eq:cov-gen;qcdsm}) and it can be written as a sum of a totally correlated and a totally anti-correlated terms
\begin{eqnarray}
\label{eq:cov-2bins;qcdsm}
{\rm Cov}[\sigma_{\rm 0-jet},\sigma_{\rm \ge 1-jet}]=
\left(\begin{array}{lr}\epsilon_0^2\delta^2\sigma_{\rm tot} & \epsilon_0(1-\epsilon_0)\delta^2\sigma_{\rm tot}\\
\epsilon_0(1-\epsilon_0)\delta^2\sigma_{\rm tot} & (1-\epsilon_0)^2\delta^2\sigma_{\rm tot}\end{array}\right) +
\left(\begin{array}{lr}\sigma^2_{\rm tot}\delta^2\epsilon_0 & -\sigma^2_{\rm tot}\delta^2\epsilon_0\\
-\sigma^2_{\rm tot}\delta^2\epsilon_0 & \sigma^2_{\rm tot}\delta^2\epsilon_0\end{array}\right).  
\end{eqnarray}
The two matrices in the right hand side of
Eq.~(\ref{eq:cov-2bins;qcdsm}) are uncorrelated. The first term
corresponds to the K-factor contribution and it is proportional to the
uncertainty in the total cross section. This gives rise to a totally
correlated uncertainty between the two jet bins. The second term in
Eq.~(\ref{eq:cov-2bins;qcdsm}) encodes Sudakov effects, and thus it is proportional to the uncertainty in the $0-$jet efficiency, yielding a totally anti-correlated uncertainty between the two jet bins.

\subsubsubsection{Numerical results for LHC}

Below we provide tables with numerical results for cross sections and
efficiencies for the values of veto scales and jet radii, as used in
current LHC analyses. The predictions for the $0-$jet efficiency are computed to NNLO+NNLL and top- and bottom-quark mass effects are included as described in~\cite{Banfi:2013eda}.
All uncertainties have been made symmetric with respect to the central value.

\begin{table}[htp!]
\begin{center}
  \center{\bf{Exact $\bf m_t$ and $\bf m_b$ corrections}}\\ 
  \vspace{0.8em}
  \begin{tabular}{c|c|c|c|c|c}
        R & $\,\ptjv$\,[GeV] & $\sigma_{\rm tot}^{{8\TeV}}$\,[pb] & 
    $\epsilon_0^{(8 \TeV)}$ & \,$\sigma^{(8\TeV)}_{\rm 0-jet}$\,[pb]
    & \,$\sigma^{(8\TeV)}_{\rm\ge 1-jet}$\,[pb]    
    \\[0.2em]\hline
    \phantom{x} & & & & &
    \\[-1em] 
    $0.4$\,& $25$ & \,$19.24\,\pm\,1.78$\, & 
    $0.602\,\pm\,0.070$ & $11.59\,\pm\,1.72$ & $7.66\,\pm\,1.52$
    \\[0.4em] 
    $0.5$\,& $30$ & \,$19.24\,\pm\,1.78$\, &
    \,$0.657\,\pm\,0.070$\, & $12.64\,\pm\,1.79$ & $6.60\,\pm\,1.48$
  \end{tabular}
\end{center}
  \caption{Total inclusive cross section, jet-veto efficiency and
    zero-jet cross section for Higgs production at the 8 TeV LHC for two
    different values of the jet radius $R$ and $\ptjv$. Results
    include exact top and bottom mass dependence. The quoted total
    cross section and the corresponding errors have been computed with
    the {\tt hnnlo} 2.0 code~\cite{Grazzini:2013mca}. }
\label{tab:mtmb;qcdsm}
\end{table}

\begin{table}[htp!]
\begin{center}
  \center{\bf{Large-$\bf m_t$ approximation}}\\ 
  \vspace{0.8em}
  \begin{tabular}{c|c|c|c|c|c}
        R & $\,\ptjv$\,[GeV] & $\sigma_{\rm tot}^{{8\TeV}}$\,[pb] & 
    $\epsilon_0^{(8 \TeV)}$ & \,$\sigma^{(8\TeV)}_{\rm 0-jet}$\,[pb]
    & \,$\sigma^{(8\TeV)}_{\rm\ge 1-jet}$\,[pb]    
    \\[0.2em]\hline
    \phantom{x} & & & & &
    \\[-1em] 
    $0.4$\,& $25$ & \,$19.03\,\pm\,1.76$\, & 
    $0.613\,\pm\,0.064$ & $11.66\,\pm\,1.62$ & $7.36\,\pm\,1.39$
    \\[0.4em] 
    $0.5$\,& $30$ & \,$19.03\,\pm\,1.76$\, &
    \,$0.667\,\pm\,0.058$\, & $12.70\,\pm\,1.61$ & $6.34\,\pm\,1.25$
  \end{tabular}
\end{center}
  \caption{As table~\ref{tab:mtmb;qcdsm} but in the large $m_t$ approximation.}
\label{tab:largemt;qcdsm}
\end{table}

\begin{table}[htp!]
\begin{center}
  \center{\bf{Large-$\bf m_t$ approximation ($\sigma_{\rm tot}^{8\TeV}$ from HXSWG)}}\\ 
  \vspace{0.8em}
  \begin{tabular}{c|c|c|c|c|c}
        R & $\,\ptjv$\,[GeV] & $\sigma_{\rm tot}^{{8\TeV}}$\,[pb] & 
    $\epsilon_0^{(8 \TeV)}$ & \,$\sigma^{(8\TeV)}_{\rm 0-jet}$\,[pb]
    & \,$\sigma^{(8\TeV)}_{\rm\ge 1-jet}$\,[pb]    
    \\[0.2em]\hline
    \phantom{x} & & & & &
    \\[-1em] 
    $0.4$\,& $25$ & \,$19.27\,\pm\,1.45$\, & 
    $0.613\,\pm\,0.064$ & $11.81\,\pm\,1.51$ & $7.46\,\pm\,1.35$
    \\[0.4em] 
    $0.5$\,& $30$ & \,$19.27\,\pm\,1.45$\, &
    \,$0.667\,\pm\,0.058$\, & $12.86\,\pm\,1.47$ & $6.42\,\pm\,1.22$
  \end{tabular}
\end{center}
  \caption{As table~\ref{tab:mtmb;qcdsm} but in the large $m_t$
    approximation. Unlike in table~\ref{tab:largemt;qcdsm} the total
    cross section is taken from the HXSWG~\cite{Heinemeyer:2013tqa}
    and includes finite-width, electro-weak and threshold resummation
    effects.}
\label{tab:largemthxswg;qcdsm}
\end{table}
\vspace{0.7cm}

In the first two tables the total
cross section is computed at NNLO QCD. Quark-mass effects lead
 to a modest decrease of about
half a percent in the zero-jet cross section with respect to the
large-$m_t$ result. The
uncertainties in the cross section increase by about 2\% (and amount to
about 14\%) when top and bottom masses are taken into
account.
In tab.~\ref{tab:largemthxswg;qcdsm} we report the jet-veto efficiency and
cross section using the improved total cross section recommended by
the Higgs Cross Section Working Group
(HXSWG)~\cite{Heinemeyer:2013tqa} instead of the pure NNLO
value. Improvements include the treatment of the Higgs width, NNLL
threshold effects and NLO electro-weak corrections. To figure out only
the size of mass effects one has to compare results including mass
corrections to tab.~\ref{tab:largemt;qcdsm}, rather than
tab.~\ref{tab:largemthxswg;qcdsm}, since the improved predictions for the total
cross section included in tab.~\ref{tab:largemthxswg;qcdsm} are not
available when finite-mass effects are included.

\subsubsection{Case with three jet bins}
Current LHC analyses for $H\rightarrow W^+W^-$ use exclusive $0-$ and
$1-$ jet bins, and an inclusive $2-$jet bin. In addition to the
$0-$jet efficiency $\epsilon_0$ defined in
Eqs.~(\ref{eq:NNLOa;qcdsm}),~(\ref{eq:NNLOb;qcdsm}),~(\ref{eq:NNLOc;qcdsm}),
we introduce the three schemes for the $1-$jet efficiency $\epsilon_1$
available at NNLO
\begin{eqnarray}
\label{eq:NNLOeps1a;qcdsm}
\epsilon_1^{(a)} &=& 1-\frac{\sigma_{\rm\ge 2-jet}^{\rm
    NLO}}{\sigma_{\rm\ge 1-jet}^{\rm NNLO}},\\
\epsilon_1^{(b)} &=& 1-\frac{\sigma_{\rm\ge 2-jet}^{\rm
    NLO}}{\sigma_{\rm\ge 1-jet}^{\rm NLO}},\\
\epsilon_1^{(c)} &=& 1-\frac{\sigma_{\rm\ge 2-jet}^{\rm NLO}}{\sigma_{\rm\ge 1-jet}^{\rm LO}}+\left(\frac{\sigma_{\rm\ge 1-jet}^{\rm NLO}}{\sigma_{\rm\ge 1-jet}^{\rm LO}}-1\right) \frac{\sigma_{\rm\ge 2-jet}^{\rm LO}}{\sigma_{\rm\ge 1-jet}^{\rm LO}}.
\end{eqnarray}
These three schemes differ by NNNLO terms. Notice
that scheme~(\ref{eq:NNLOeps1a;qcdsm}) is currently not available yet,
but the NNLO QCD corrections to $H+1-$jet production is being
computed. The three $3-$dimensional vectors~(\ref{eq:errvec;qcdsm}) are
 \begin{eqnarray}
\Delta\sigma_{\rm 0-jet} &=& \left(0\,,\,\sigma_{\rm tot}\delta\epsilon_0\,,\,\epsilon_0\delta\sigma_{\rm tot}\right),\\
\Delta\sigma_{\rm 1-jet} &=& \left((1-\epsilon_0)\sigma_{\rm tot}\delta\epsilon_1\,,\,-\epsilon_1\sigma_{\rm tot}\delta\epsilon_0\,,\,\epsilon_1(1-\epsilon_0)\delta\sigma_{\rm tot}\right),\\
\Delta\sigma_{\rm\ge 2-jet} &=& (-(1-\epsilon_0)\sigma_{\rm tot}\delta\epsilon_1\,,\,-(1-\epsilon_1)\sigma_{\rm tot}\delta\epsilon_0\,,\,(1-\epsilon_1)(1-\epsilon_0)\delta\sigma_{\rm tot}).
\end{eqnarray}
The covariance matrix reads
\begin{equation}
{\rm Cov}[\sigma_{\rm 0-jet},\sigma_{\rm 1-jet},\sigma_{\rm \ge 2-jet}]=
\left(\begin{array}{ccc} \Delta\sigma_{\rm 0-jet}\cdot\Delta\sigma_{\rm 0-jet} & \Delta\sigma_{\rm 0-jet}\cdot \Delta\sigma_{\rm 1-jet} &  \Delta\sigma_{\rm 0-jet}\cdot \Delta\sigma_{\rm\ge 2-jet}\\
 & \Delta\sigma_{\rm 1-jet}\cdot\Delta\sigma_{\rm 1-jet} &  \Delta\sigma_{\rm 1-jet}\cdot \Delta\sigma_{\rm\ge 2-jet}\\
 &  & \Delta\sigma_{\rm \ge 2-jet}\cdot\Delta\sigma_{\rm \ge 2-jet}\end{array}\right),
\end{equation}
with the following scalar products
\begin{eqnarray}
\Delta\sigma_{\rm 0-jet}\cdot\Delta\sigma_{\rm 0-jet} &=& \sigma_{\rm
  tot}^2\delta^2\epsilon_0+\epsilon_0^2\delta^2\sigma_{\rm tot},\notag\\
\Delta\sigma_{\rm 0-jet}\cdot \Delta\sigma_{\rm 1-jet} &=&
-\epsilon_1\sigma_{\rm 
  tot}^2\delta^2\epsilon_0+\epsilon_0\epsilon_1(1-\epsilon_0)\delta^2\sigma_{\rm 
  tot},\notag\\
\Delta\sigma_{\rm 0-jet}\cdot \Delta\sigma_{\rm\ge 2-jet} &=&
-(1-\epsilon_1)\sigma_{\rm 
  tot}^2\delta^2\epsilon_0+\epsilon_0(1-\epsilon_1)(1-\epsilon_0)\delta^2\sigma_{\rm 
  tot},\notag\\
\Delta\sigma_{\rm 1-jet}\cdot\Delta\sigma_{\rm 1-jet} &=&
(1-\epsilon_0)^2\sigma_{\rm
  tot}^2\delta^2\epsilon_1+\epsilon_1^2\sigma_{\rm
  tot}^2\delta^2\epsilon_0+\epsilon_1^2(1-\epsilon_0)^2\delta^2\sigma_{\rm
  tot},\notag\\
\Delta\sigma_{\rm 1-jet}\cdot\Delta\sigma_{\rm\ge 2-jet} &=&
-(1-\epsilon_0)^2\sigma_{\rm
  tot}^2\delta^2\epsilon_1+\epsilon_1(1-\epsilon_1)\sigma_{\rm
  tot}^2\delta^2\epsilon_0+\epsilon_1(1-\epsilon_1)(1-\epsilon_0)^2\delta^2\sigma_{\rm
  tot},\notag\\
\Delta\sigma_{\rm\ge 2-jet}\cdot\Delta\sigma_{\rm\ge 2-jet} &=&
(1-\epsilon_0)^2\sigma_{\rm
  tot}^2\delta^2\epsilon_1+(1-\epsilon_1)^2\sigma_{\rm
  tot}^2\delta^2\epsilon_0+(1-\epsilon_1)^2(1-\epsilon_0)^2\delta^2\sigma_{\rm
  tot}.
\end{eqnarray}
\subsection*{Conclusions}
The jet-veto-efficiency (JVE) method is generalised to an arbitrary jet multiplicity, and a general parametrization for the covariance matrix is provided. The method presented here can be applied to any exclusive production of a color singlet accompanied by jets.




\newcommand{\lo}       {{\rm LO}}
\newcommand{\nlo}      {{\rm NLO}}
\newcommand{\nnlo}     {{\rm NNLO}}
\newcommand{\lhe}      {{\rm LHE}}
\newcommand{\ckm}      {{\rm CKM}}
\newcommand{\lhc}      {{\rm LHC}}
\newcommand{\pdf}      {{\rm PDF}}
\newcommand{\pdg}      {{\rm PDG}}
\newcommand{\pythia}   {\texttt{PYTHIA}}
\newcommand{\mcfm}     {\texttt{MCFM}}
\newcommand{\fastjet}     {\texttt{Fastjet}}
\newcommand{\py}       {\texttt{PY}}
\newcommand{\hw}       {\texttt{HW}}
\newcommand{\amcatnlo}   {\texttt{aMC@NLO}}
\newcommand{\mcatnlo}   {\texttt{MC@NLO}}
\newcommand{\sherpa}   {\texttt{SHERPA}}
\newcommand{\lauratool}   {\texttt{NLOX}}
\newcommand{\powhel}   {\texttt{PowHel}}
\newcommand{\powheg}   {\texttt{POWHEG}}
\newcommand{\powhegbox}{\texttt{POWHEG-Box}}
\newcommand{\powheldd} {\texttt{PowHel@dd}}
\newcommand{\helacnlo} {\texttt{HELAC-NLO}}
\newcommand{\oneloop} {\texttt{OneLOop}}
\newcommand{\helacnlodd} {\texttt{HELAC-NLO@dd}}
\newcommand{\helaconeloop} {\texttt{HELAC-1LOOP}}
\newcommand{\helaconeloopdd} {\texttt{HELAC-1LOOP@dd}}
\newcommand{\helacdipole} {\texttt{HELAC-Dipole}}
\newcommand{\decayer}   {\texttt{Decayer}}
\newcommand{\helacphegas} {\texttt{HELAC-Phegas}}
\newcommand{\qd}   {\texttt{QD}}
\newcommand{\cuttools}   {\texttt{CUTTOOLS}}
\newcommand{\mint}           {\texttt{MINT}}
\newcommand{\lhapdf}           {\texttt{LHAPDF}}
\newcommand{\ud}       {\mathrm{d}}
\newcommand{\tE}       {\ensuremath{\tilde{E}}}
\newcommand{\ts}       {\ensuremath{\tilde{s}}}
\newcommand{\tx}       {\ensuremath{\tilde{x}}}
\newcommand{\ptmin}    {\ensuremath{p_\bot^{\min}}}
\newcommand{\kt}       {\ensuremath{k_{\bot}}}
\newcommand{\pTmiss}{\ensuremath{\slash\hspace*{-5pt}{p}_{\perp}}}
\newcommand{\W}        {\ensuremath{W}}
\newcommand{\ptt}      {\ensuremath{p_{\bot,\,\mathrm{t}}}}
\newcommand{\ptttbar}  {\ensuremath{p_{\bot,\,\mathrm{t\,\bar{t}}}}}
\newcommand{\ptw}      {\ensuremath{p_{\bot,\,W}}}
\newcommand{\ptj}      {\ensuremath{p_{\bot,\,j}}}
\newcommand{\ptl}      {\ensuremath{p_{\bot,\,\ell}}}
\newcommand{\ptbb}     {\ensuremath{p_{\bot,\,\bq_1\bq_2}}}
\newcommand{\rad}      {\mathrm{rad}}
\newcommand{\tbWp}     {\ensuremath{\mathrm{t\to b\,W^+}}}
\newcommand{\Wpln}     {\ensuremath{\mathrm{W^+\to\bar{\ell}\,}\nu}}
\newcommand{\Wmln}     {\ensuremath{\mathrm{W^-\to\ell\,}\bar{\nu}}}
\newcommand{\WWbB}     {\ensuremath{\mathrm{W^+\,W^-\,b\,\bar{b}}}}
\newcommand{\epmubB}   {\ensuremath{\mathrm{e^+\,\nu_e\,\mu^-\,\bar{\nu}_\mu\,b\,\bar{b}}}}
\newcommand{\qq}       {\ensuremath{\mathrm{q}}}
\newcommand{\qaq}      {\ensuremath{\mathrm{\bar{q}}}}
\newcommand{\qaqp}     {\ensuremath{\mathrm{\bar{q}'}}}
\newcommand{\bq}       {\ensuremath{\mathrm{b}}}
\newcommand{\baq}      {\ensuremath{\mathrm{\bar{b}}}}
\newcommand{\tq}       {\ensuremath{\mathrm{t}}}
\newcommand{\taq}      {\ensuremath{\mathrm{\bar{t}}}}
\newcommand{\bB}       {\ensuremath{\mathrm{b\,\bar{b}}}}
\newcommand{\tT}       {\ensuremath{\mathrm{t\,\bar{t}}}}
\newcommand{\tth}       {\ensuremath{\mathrm{t\,\bar{t}\,H}}}
\newcommand{\pp}       {\ensuremath{p\,p(\bar{p})}}
\newcommand{\tTX}      {\ensuremath{\mathrm{t\,\bar{t}\,X}}}
\newcommand{\ptb}      {\ensuremath{p_{\bot,\,\mathrm{b}}}}
\newcommand{\ptbbar}   {\ensuremath{p_{\bot,\,\mathrm{\bar{b}}}}}
\newcommand{\ptsupp}   {\ensuremath{p_{\bot,\,\mathrm{supp}}}}
\newcommand{\sme}[1]   {\ensuremath{|{\cal M}|^2(#1)}}
\newcommand{\Phiu}     {\ensuremath{\Phi_{\mathrm{u}}}}
\newcommand{\Phid}     {\ensuremath{\Phi_{\mathrm{d}}}}
\newcommand{\mur}      {\ensuremath{\mu_\mathrm{R}}}
\newcommand{\muf}      {\ensuremath{\mu_\mathrm{F}}}
\newcommand{\tot}      {\ensuremath{\mathrm{tot}}}
\newcommand{\cm}       {\ensuremath{\mathrm{CM}}}
\newcommand{\lplus}    {\ensuremath{\ell^+}}
\newcommand{\lminus}   {\ensuremath{\ell^-}}
\newcommand{\tB}       {\ensuremath{\bar{B}}}
\newcommand{\tev}      {\ensuremath{\mathrm{TeV}}}
\newcommand{\gev}      {\ensuremath{\mathrm{GeV}}}
\newcommand{\mev}      {\ensuremath{\mathrm{MeV}}}
\newcommand{\mt}       {\ensuremath{m_{\mathrm{t}}}}
\newcommand{\mz}       {\ensuremath{m_{\mathrm{Z}}}}
\newcommand{\mw}       {\ensuremath{m_{\mathrm{W}}}}
\newcommand{\mh}       {\ensuremath{m_{\mathrm{H}}}}
\newcommand{\mb}       {\ensuremath{m_{\mathrm{b}}}}
\newcommand{\gf}       {\ensuremath{G_{\mathrm{F}}}}






\section{Higgs production in association with top quarks at the LHC: comparison of different approaches for NLO QCD simulations matched with parton shower\footnote{M.V.~Garzelli, H.B.~Hartanto, B.~J\"ager, A.~Kardos, L.~Reina, Z.~Tr\'ocs\'anyi, D.~Wackeroth}}

\subsection*{Research focus}

  We present a comparison of the predictions for \tth\ production at
  the Large Hadron Collider (LHC) including NLO QCD corrections
  matched to parton shower as obtained by different approaches: i)
  \powhel, that uses \helacnlo\ to compute NLO matrix-elements and the
  \powhegbox\ for the matching, ii) the \powhegbox\ combined with a
  routine that provides the analytic virtual matrix elements as
  calculated
  in~\cite{Reina:2001sf,Reina:2001bc,Dawson:2002tg,Dawson:2003zu},
  iii) \sherpa\ combined with the same aforementioned routine.
  Predictions from these different approaches at NLO QCD / LHE level
  show very good agreement among each other, with an uncertainty on
  the inclusive cross section of less than 1\%, whereas the
  dif\-fe\-ren\-tial distributions at the hadron level computed by i)
  and ii) agree within a few percents, i.e. within the uncertainty due
  to the renormalization- and factorization-scale variation.

\subsection{Introduction}

Among the main single-Higgs-boson production mechanisms at the LHC,
the associated production of a Higgs boson with a pair of top quarks,
i.e.  $pp\rightarrow t\bar{t}H$, has the smallest cross section at
both $\sqrt{s}=8$ and $14$ TeV. However, the experimental discovery of
a Higgs boson with a mass around $125-126$~GeV has cast new light on
the role played by \tth\ production. Detailed studies of the
properties of the discovered particle will be used to confirm or
exclude the Higgs mechanism of electroweak symmetry breaking as
minimally implemented in the Standard Model. In this context, the
measurement of the \tth\ production rate can provide a clean direct
access to the top-Higgs Yukawa coupling, probably the most crucial
coupling to fermions, on which the present SM uncertainties are still
very large. Thus, both the ATLAS and CMS experimental collaborations
at LHC have provided analyses trying to isolate \tth\ events,
exploiting the full amount of data available at both $\sqrt{s}$ = 7
and 8
TeV~\cite{ATLAS-CONF-2012-135,Chatrchyan:2013yea,ATLAS-CONF-2013-080,CMS-PAS-HIG-13-019},
and even more dedicated studies are part of the Higgs precision
program for Run II of the LHC.

First analyses of \tth\ events mainly considered the $H\rightarrow
b\bar{b}$ decay channel~\cite{ATLAS-CONF-2012-135,Chatrchyan:2013yea},
since it offers the highest branching ratio for a Higgs boson with a
mass around 125 GeV. However, due to the high uncertainties on the
background for this channel, also related to the difficulties in
reconstructing heavy-flavor jets and distinguishing them from light
jets, more recent analyses have tried to exploit different decay
channels, like $H\rightarrow\tau^+\tau^-$ and
$H\rightarrow\gamma\gamma$~\cite{ATLAS-CONF-2013-080,CMS-PAS-HIG-13-019}.

For most of the analyses published so far, the shapes of the expected
differential distributions used during the experimental studies are
still produced on the basis of LO results, provided by \pythia,
although fully differential theoretical predictions and events
including QCD NLO corrections matched with parton shower are already
publicly available since 2011.  In particular, NLO QCD corrections to
\tth\ have first been computed in
Refs.~\cite{Reina:2001sf,Reina:2001bc,Dawson:2002tg,
  Dawson:2003zu,Beenakker:2001rj,Beenakker:2002nc} and specified to
the case of both the Tevatron and the LHC with
$\sqrt{s}=14$~TeV. These predictions have further been extended to the
case of the LHC with $\sqrt{s} =7$~TeV and 8~TeV in
Ref.~\cite{Dittmaier:2011ti} and Ref.~\cite{Heinemeyer:2013tqa},
respectively, whereas matchings with parton shower have been presented
both in Ref.~\cite{Frederix:2011zi} by means of the \amcatnlo\
approach and in Ref.~\cite{Garzelli:2011vp} by means of the \powhel\
approach. Furthermore, a comparison between the last two approaches,
specified to the case of the LHC with $\sqrt{s}$ = 7 TeV, appeared in
Ref.~\cite{Dittmaier:2012vm} and showed very good agreement at least
at the level of the residual uncertainties due to variations of the
renormalization and factorization scales, of the parton distribution
function (PDF), and of the shower Monte Carlo (SMC)
(\pythia~\cite{Sjostrand:2006za,Sjostrand:2007gs} vs.\
\herwig~\cite{Corcella:2002jc}) used.

In this contribution we present a study of theoretical predictions for
\tth\ at $\sqrt{s}$ = 8 TeV by comparing three different methods: the
already existing \powhel\ approach and two new implementations, both
relying on virtual matrix elements analytically computed on the basis
of Refs.~\cite{Reina:2001sf,Reina:2001bc,Dawson:2002tg,Dawson:2003zu},
interfaced with the \powhegbox~\cite{Alioli:2010xd} and
\sherpa~\cite{Gleisberg:2003xi,Gleisberg:2008ta}, respectively.

\subsection{Short review of the implementations}

\subsubsection{PowHel}
 
The \powhel\ implementation of \tth\ relies on the interface between
the publicly available event generators
\helacnlo~\cite{Bevilacqua:2011xh} and the
\powhegbox~\cite{Alioli:2010xd}. In particular, the \helacnlo\
components (\texttt{HELAC-} \texttt{Phegas}~\cite{Cafarella:2007pc},
\helaconeloop~\cite{vanHameren:2009dr},
\helacdipole~\cite{Czakon:2009ss}) are used to compute all matrix
elements required as input by the \powhegbox, that performs the
matching to parton shower according to the
\powheg~\cite{Nason:2004rx,Frixione:2007vw} formalism. Working in a
general multiparticle framework, 1-loop matrix elements are computed
automatically, by using the OPP method~\cite{Ossola:2006us}
complemented by analytical formulas for calculating the contributions
of the $R_2$ rational terms~\cite{Draggiotis:2009yb}, all implemented
in \helaconeloop~\cite{Bevilacqua:2010mx}.  On the other hand, the
subtraction terms necessary to factor out the infrared NLO
singularities in dimensional re\-gu\-larization are computed
automatically by the \powhegbox, on the basis of the FKS subtraction
formalism~\cite{Frixione:1995ms}.  The \powhel\ implementation of
\tth\ was described in detail and extensively tested in the original
paper~\cite{Garzelli:2011vp} and cross-checked with respect to the
\amcatnlo\ one at both partonic and hadronic level in
Refs.~\cite{Garzelli:2011vp, Dittmaier:2012vm}. The output of \powhel\
are events stored in files in Les Houches format
(LHE)~\cite{Alwall:2006yp}, including up to first radiation emission,
publicly available on the web~\cite{Garzelli:2011iu} at the address:
{\texttt{http://grid.kfki.hu/twiki/bin/view/DbTheory/TthProd}}$\,\,$.
These events can be further showered by means of standard SMC event
generators (so far, both the interfaces with the fortran version of
\pythia~\ and \herwig~\ have been tested and used to produce
theoretical predictions). Se\-ve\-ral millions of \tth\ events
provided by \powhel\ at different energies ($\sqrt{s}$ = 7, 8 and 14
TeV) and in different configurations, have already been made available
and used by the experimental collaborations at LHC for their
preliminary studies aiming at optimizing their \tth\ analysis setup.

\subsubsection{One-loop matrix elements matched with the \powhegbox{} and \sherpa}
In this study we have implemented the one-loop QCD matrix elements for
$q\bar{q}\rightarrow t\bar{t}H$ and $gg\rightarrow t\bar{t}H$
calculated in
Refs.~\cite{Reina:2001sf,Reina:2001bc,Dawson:2002tg,Dawson:2003zu} in
both the \powhegbox~\ and \sherpa. Both implementations have been
tested at the NLO QCD parton level comparing total and differential
cross sections against the original calculation, where the
$O(\alpha_s)$ real-radiation emission is calculated using
phase-space-slicing techniques.  They are in the form of a library of
routines that contains all the analytical building blocks needed to
calculate the $\mathcal{O}(\alpha_s)$ one-loop QCD corrections to the
partonic channels mentioned above and have been thoroughly tested
against the analogous calculation presented in
Refs.~\cite{Beenakker:2001rj,Beenakker:2002nc}. These routines are
already publicly available in \sherpa~\ and will soon be made
available as part of the \powhegbox~\ repository. The tools and
algorithms that have been used in the calculation are thoroughly
explained in the original publications to which we refer for all
details.

\subsection{Checks and predictions}

The NLO predictions discussed in this section were obtained by
considering proton-proton collisions at $\sqrt{s} = 8$~TeV, using the
CT10 NLO set of parton distribution functions with five active
flavours~\cite{Lai:2010vv} as provided by the {\texttt{LHAPDF}}
interface~\cite{Whalley:2005nh} and a running strong coupling computed
accordingly. We fixed masses and couplings using $\mt$ = 172.5 GeV,
$\mh$ = 125 GeV, $\mw$ = 80.385 GeV, $\mz$ = 91.1876 GeV, and the
Fermi constant $\gf$ = 1.16639 $10^{-5}$ GeV$^{-2}$. The
electromagnetic coupling was derived from $\gf$, $\mz$ and $\mw$ in
the $G_\mu$ scheme.  Factorization and renormalization scales were
varied around the central value $\mu_0 = \mur = \muf = \mt + \mh/2$ by
a factor of two, in order to compute scale-uncertainty.  Top quarks,
as well as the Higgs boson, were produced on shell, whereas their
subsequent decays were handled by the SMC code used, according to its
default implementation.  In this study recent fortran versions of \pythia~\
(\pythia~\ 6.4.25 - 6.4.28) were interfaced to both the \powhel~\ and the \powhegbox~\ implementations, whereas \sherpa~\ includes its own parton-shower implementation. 

For each of the implementations described in the previous section, we
generated several millions of events and stored them in the Les Houches
Event (LHE) file format, including up to first radiation emission (LHE
level). This first emission is computed internally according to the
matching scheme of the respective program, i.e. the \powheg~\ scheme
in the \powhegbox~\ and the \mcatnlo~\ scheme in \sherpa. Subsequent
emissions are always computed by the SMC.

\begin{figure}[h!]
\begin{center}
\includegraphics[width=0.495\textwidth]{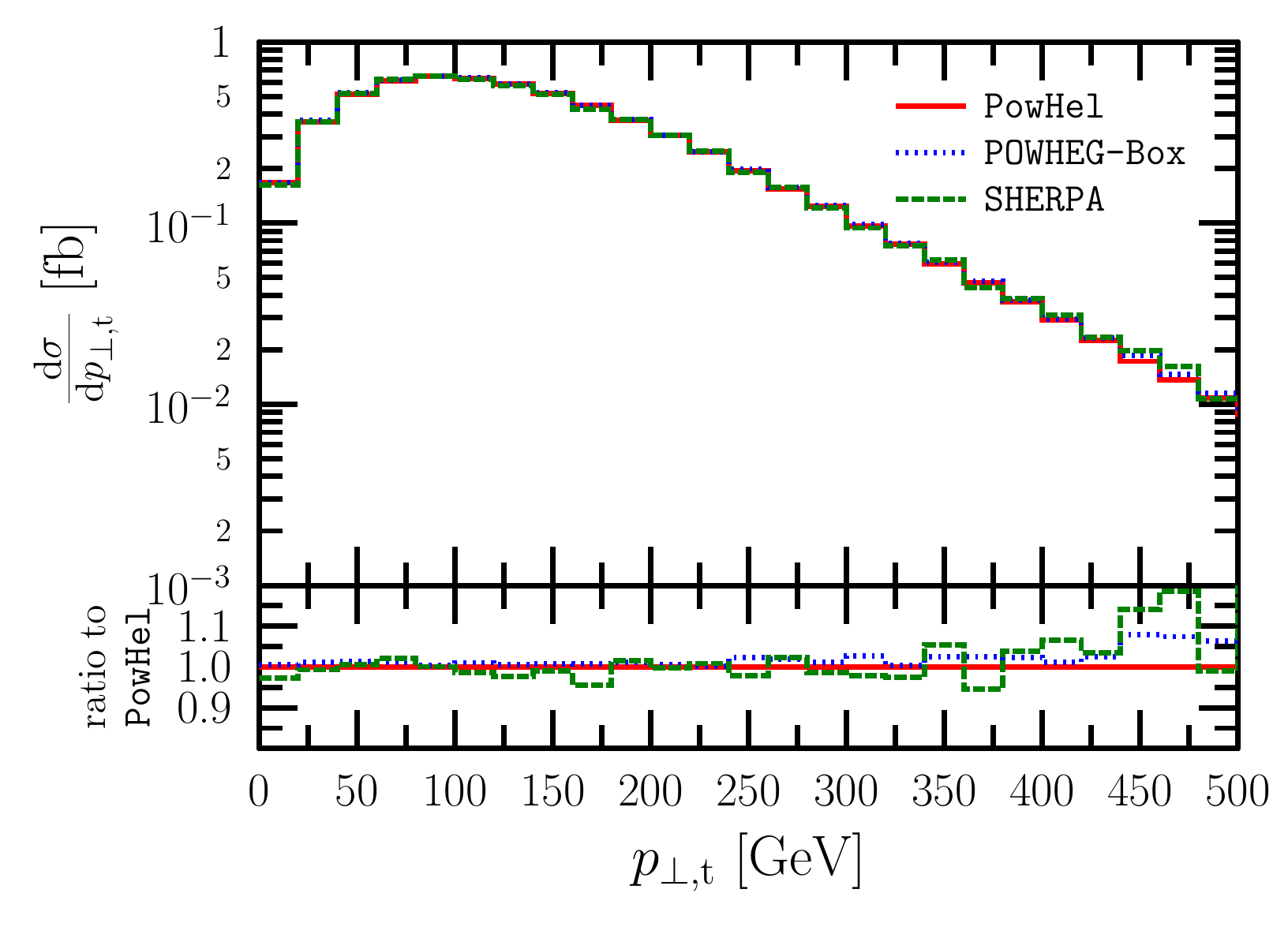}
\includegraphics[width=0.495\textwidth]{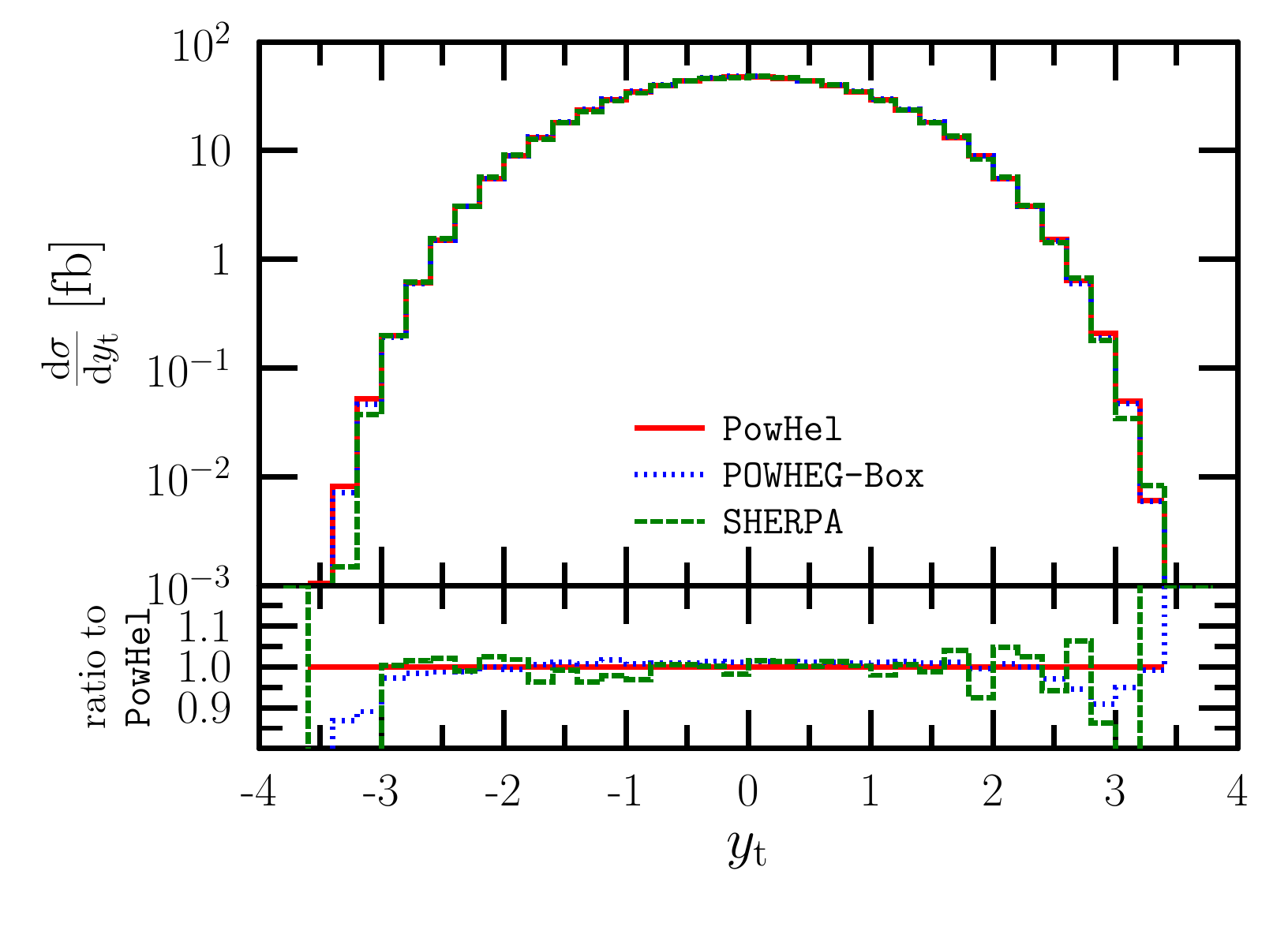}
\includegraphics[width=0.495\textwidth]{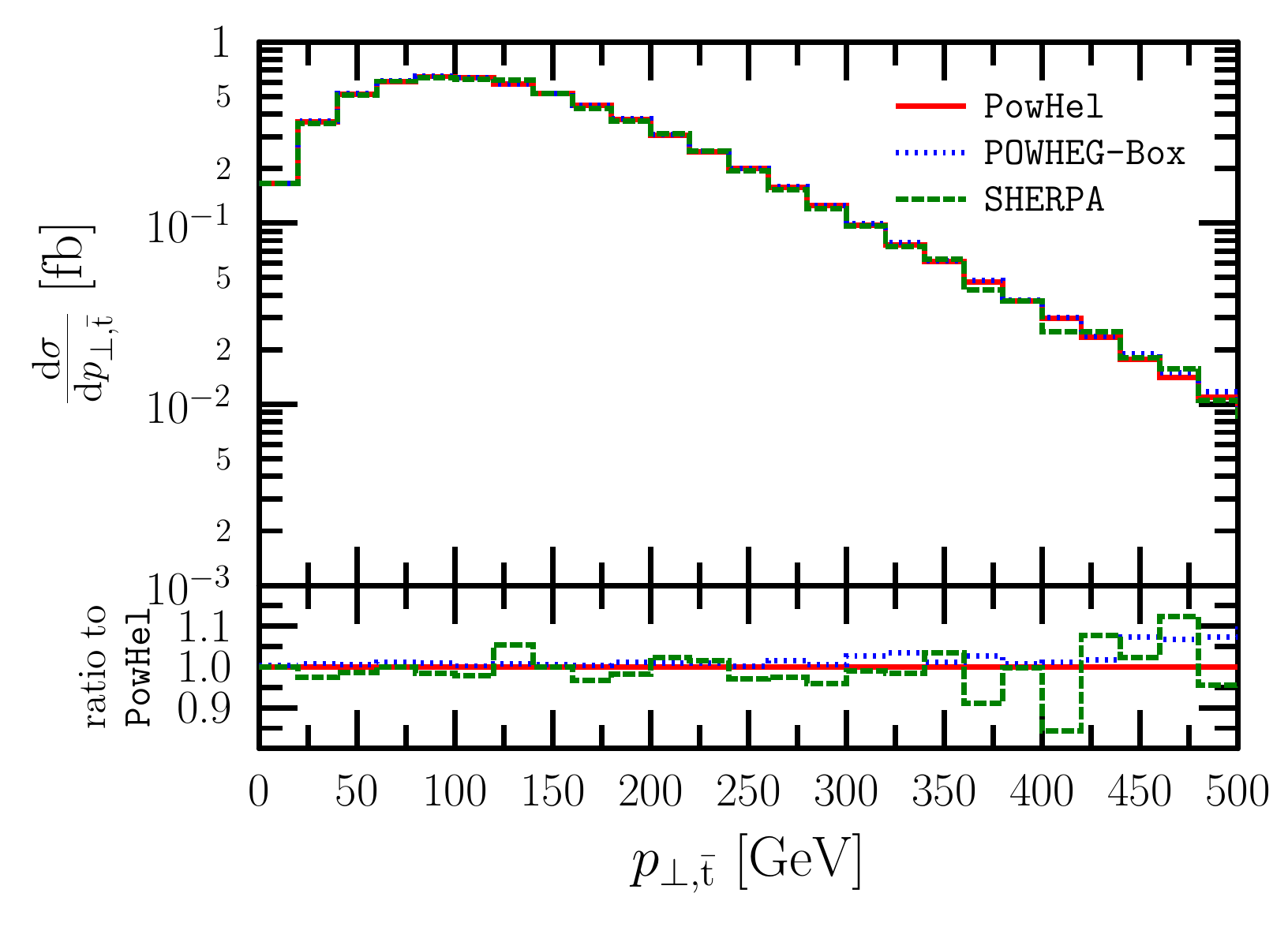}
\includegraphics[width=0.495\textwidth]{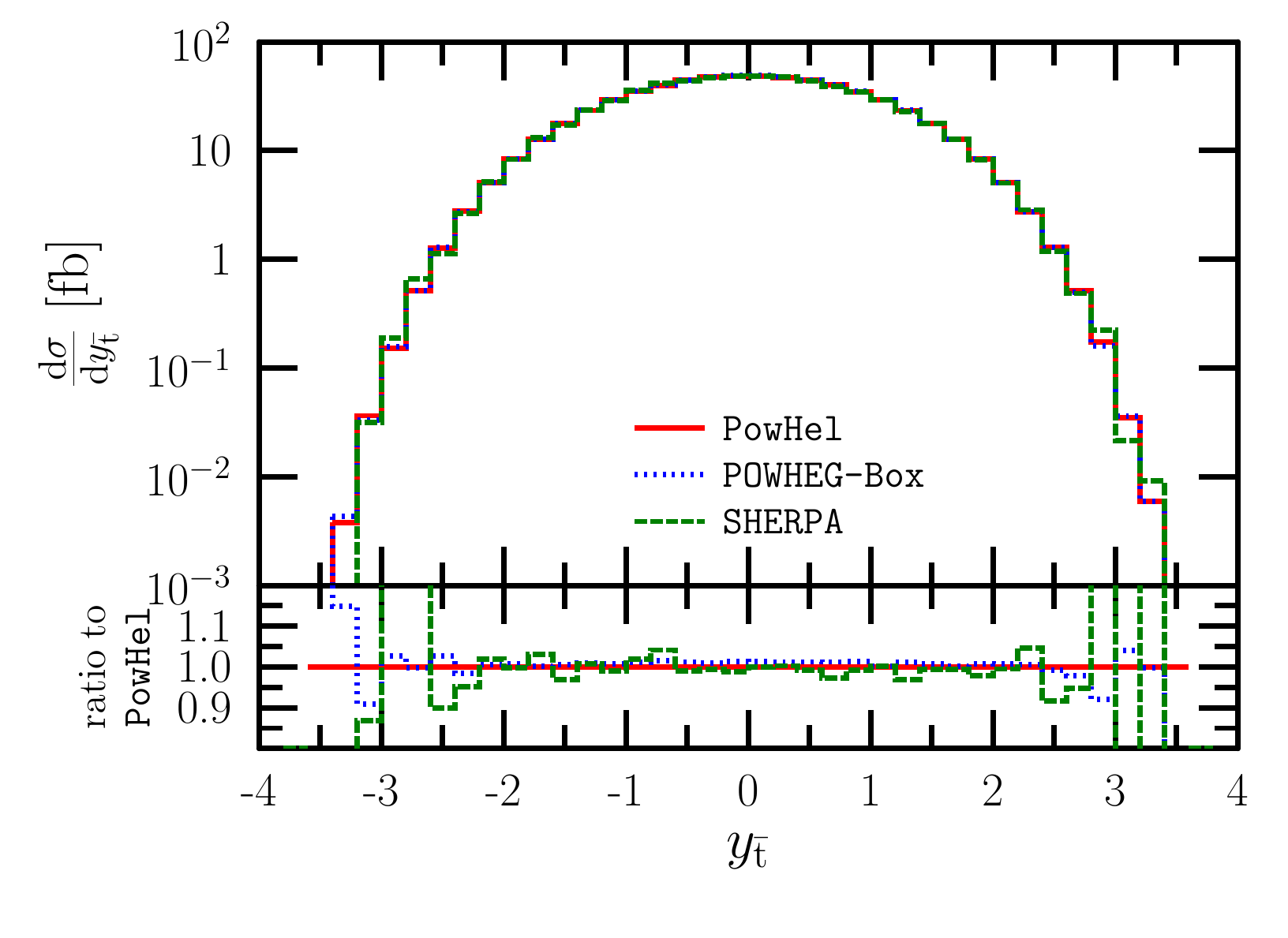}
\includegraphics[width=0.495\textwidth]{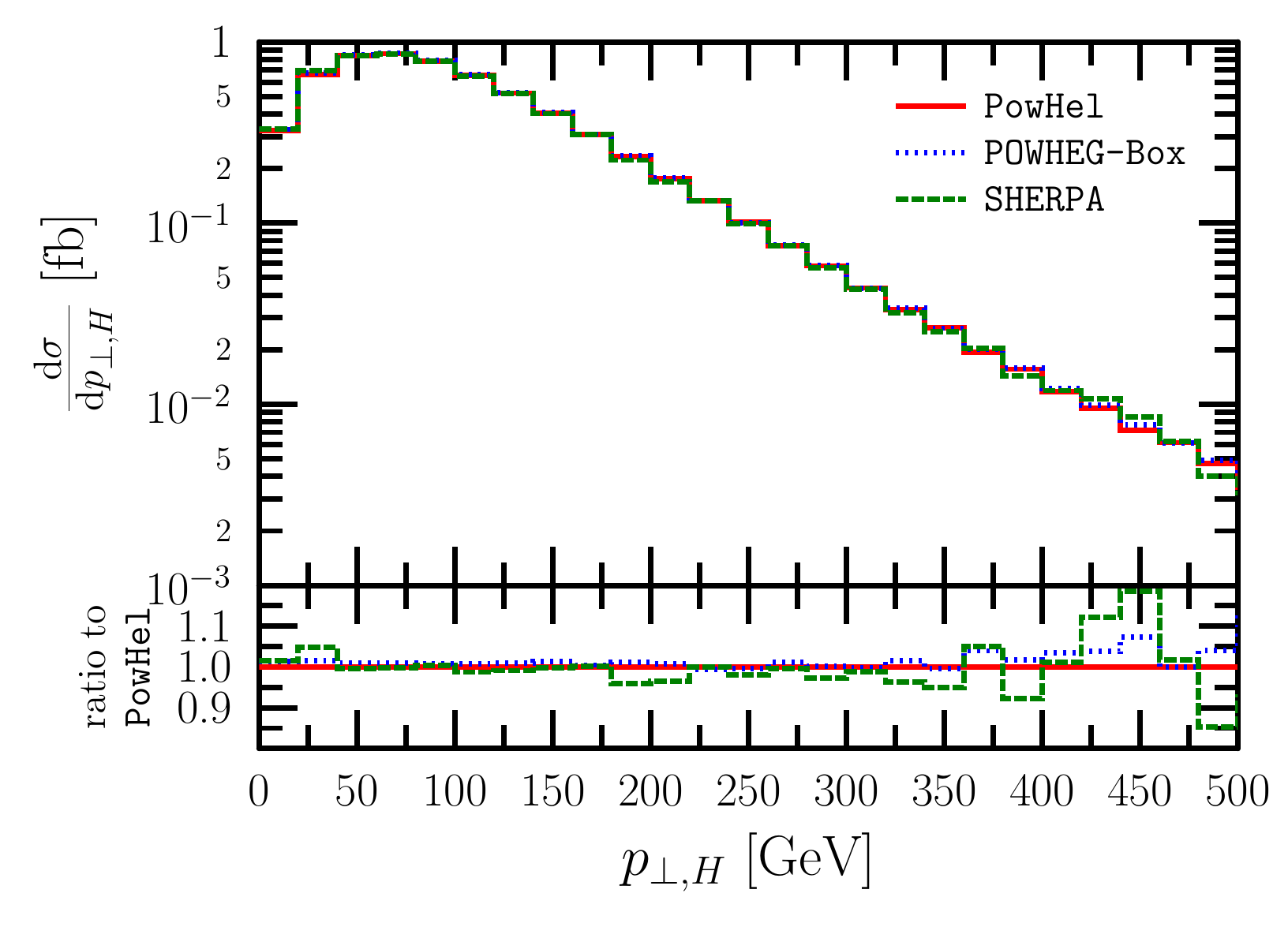}
\includegraphics[width=0.495\textwidth]{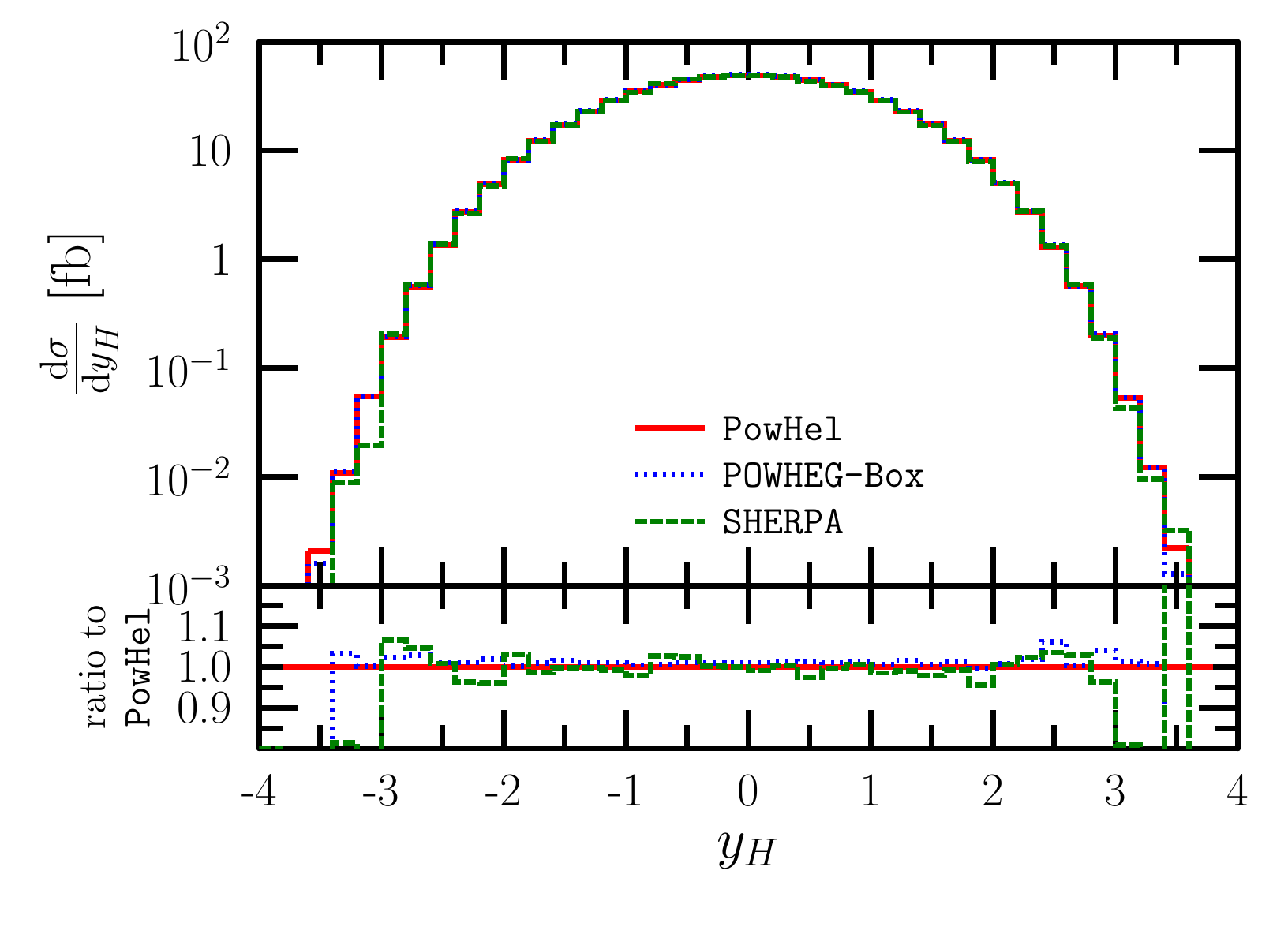}
\caption{\label{fig:tth_fig1;higgs} Transverse momentum and rapidity
  of the top (upper panel), the anti-top (intermediate panel), and the
  Higgs boson (lower panel), as obtained at LHE level by the three NLO
  QCD + PS implementations [\powhel\ (red solid line), the \powhegbox\
  (blue dotted line), and \sherpa\ (green dashed line)]. The ratio of
  the \powhegbox\ and \sherpa\ predictions with respect to those from
  \powhel\ is shown in the lower inset of each panel.}
\end{center}
\end{figure}

\begin{figure}[h!]
\begin{center}
\includegraphics[width=0.495\textwidth]{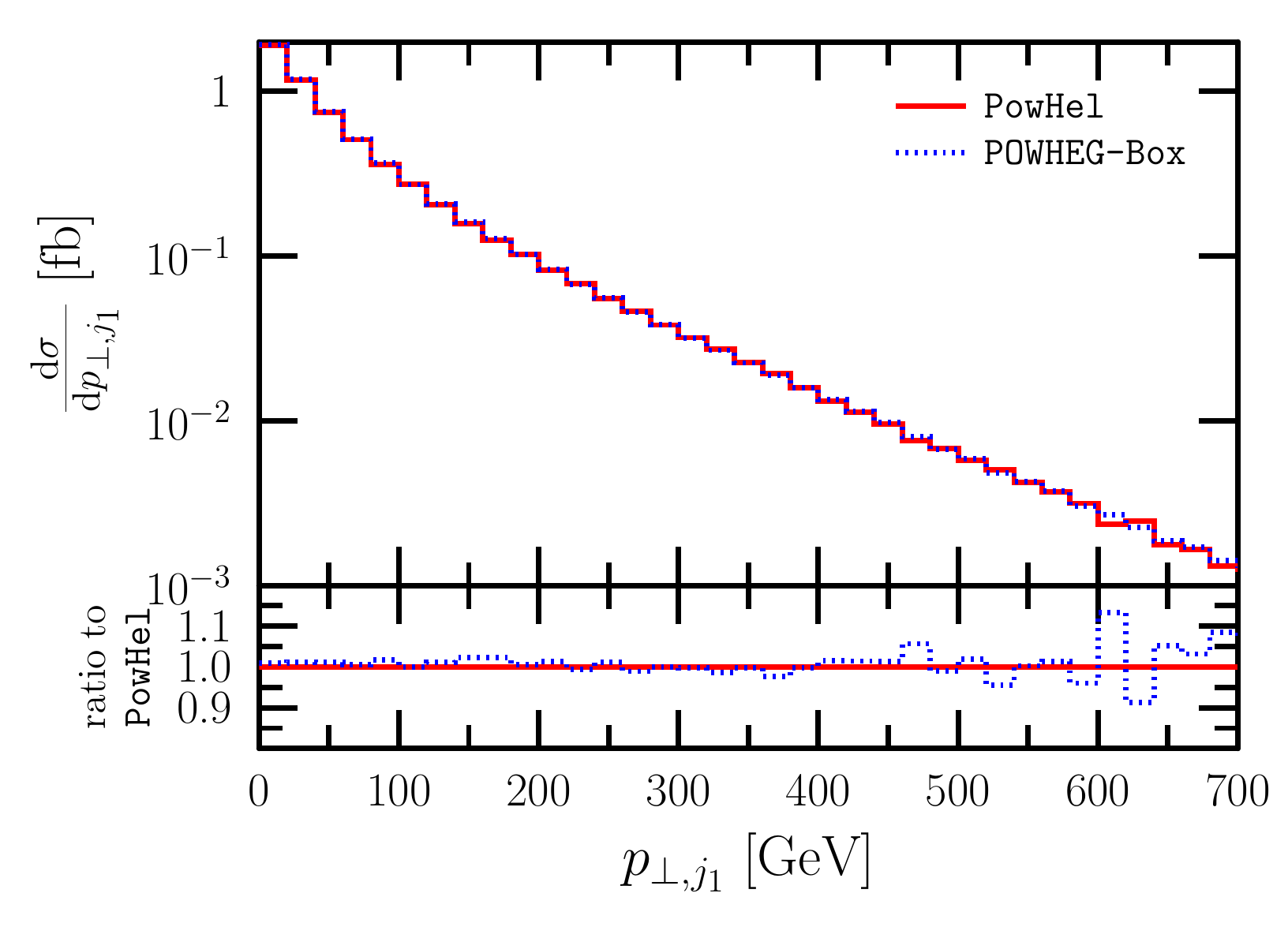}
\includegraphics[width=0.495\textwidth]{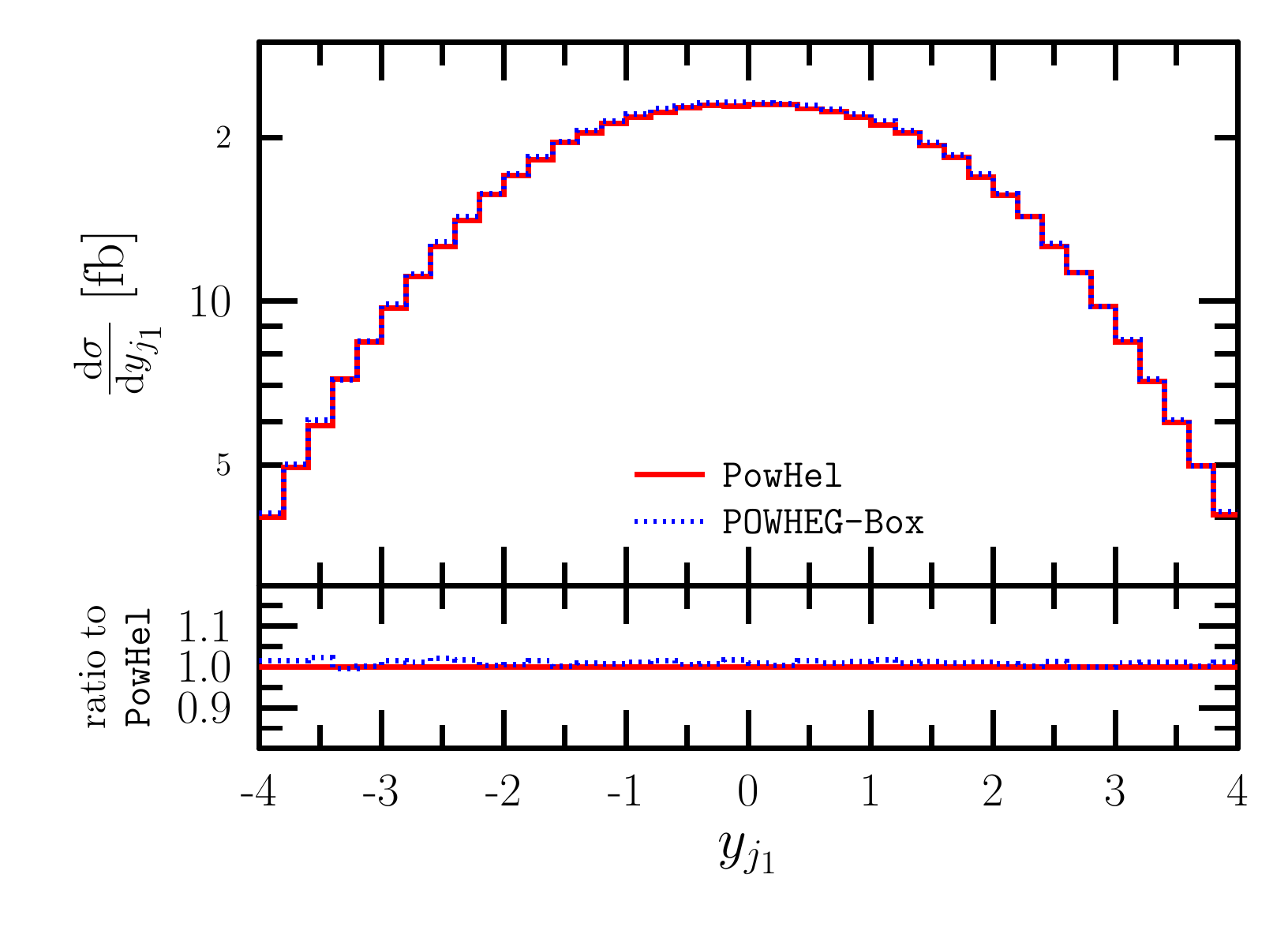}
\caption{\label{fig:tth_fig2;higgs} Transverse momentum and rapidity
  of the hardest resolved jet, as obtained at LHE level by the two
  \powheg-based 
  NLO QCD + PS implementations [\powhel\ (red solid line), and the
  \powhegbox\ (blue dotted line)]. The ratio of the \powhegbox\ and \sherpa\ predictions with
  respect to those from \powhel\ is shown in the lower inset of each
  panel.}
\end{center}
\end{figure}

\begin{figure}[h!]
\begin{center}
\includegraphics[width=0.695\textwidth]{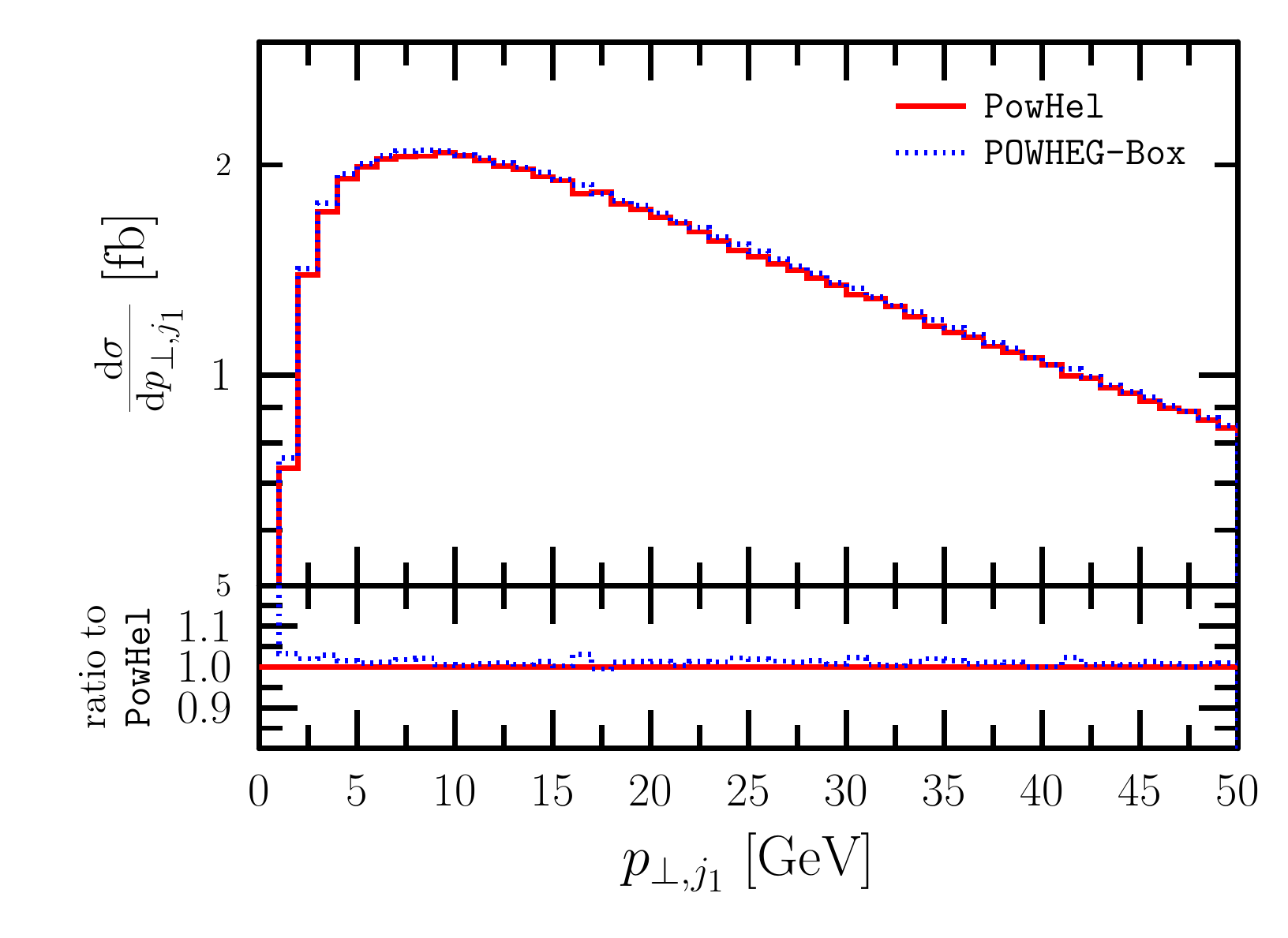}
\caption{\label{fig:tth_fig3;higgs} Zoom of the transverse momentum of
  the hardest resolved jet in the low $p_\bot$ region, as obtained at
  LHE level by the two \powheg-based NLO QCD + PS implementations
  [\powhel\ (red solid line), and the \powhegbox\ (blue dotted line)
  ]. The ratio of the \powhegbox\ 
  predictions with respect to those from \powhel\ is shown in the lower
  inset. 
}
\end{center}
\end{figure}

\begin{figure}[h!]
\begin{center}
\includegraphics[width=0.495\textwidth]{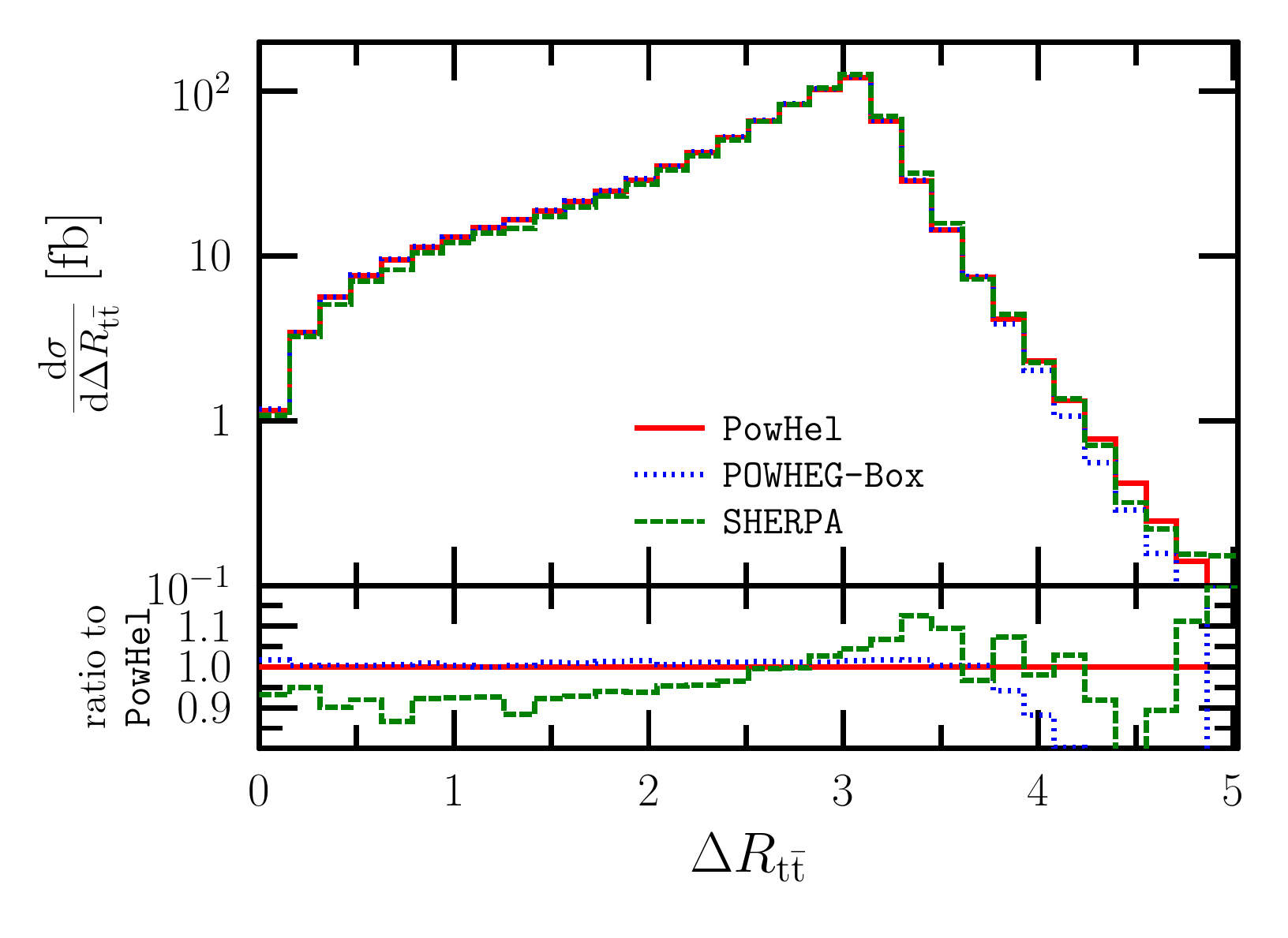}
\includegraphics[width=0.495\textwidth]{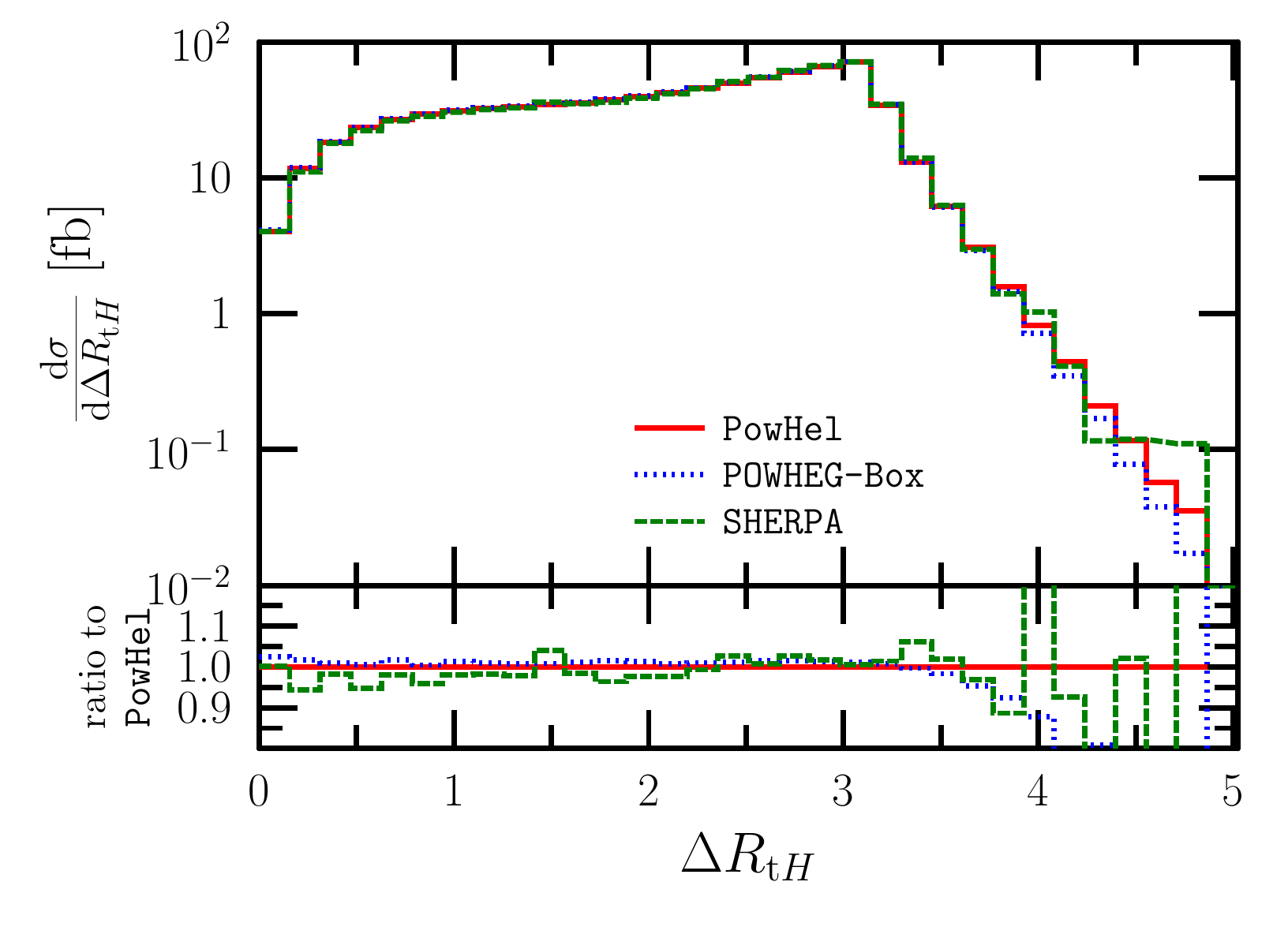}
\includegraphics[width=0.495\textwidth]{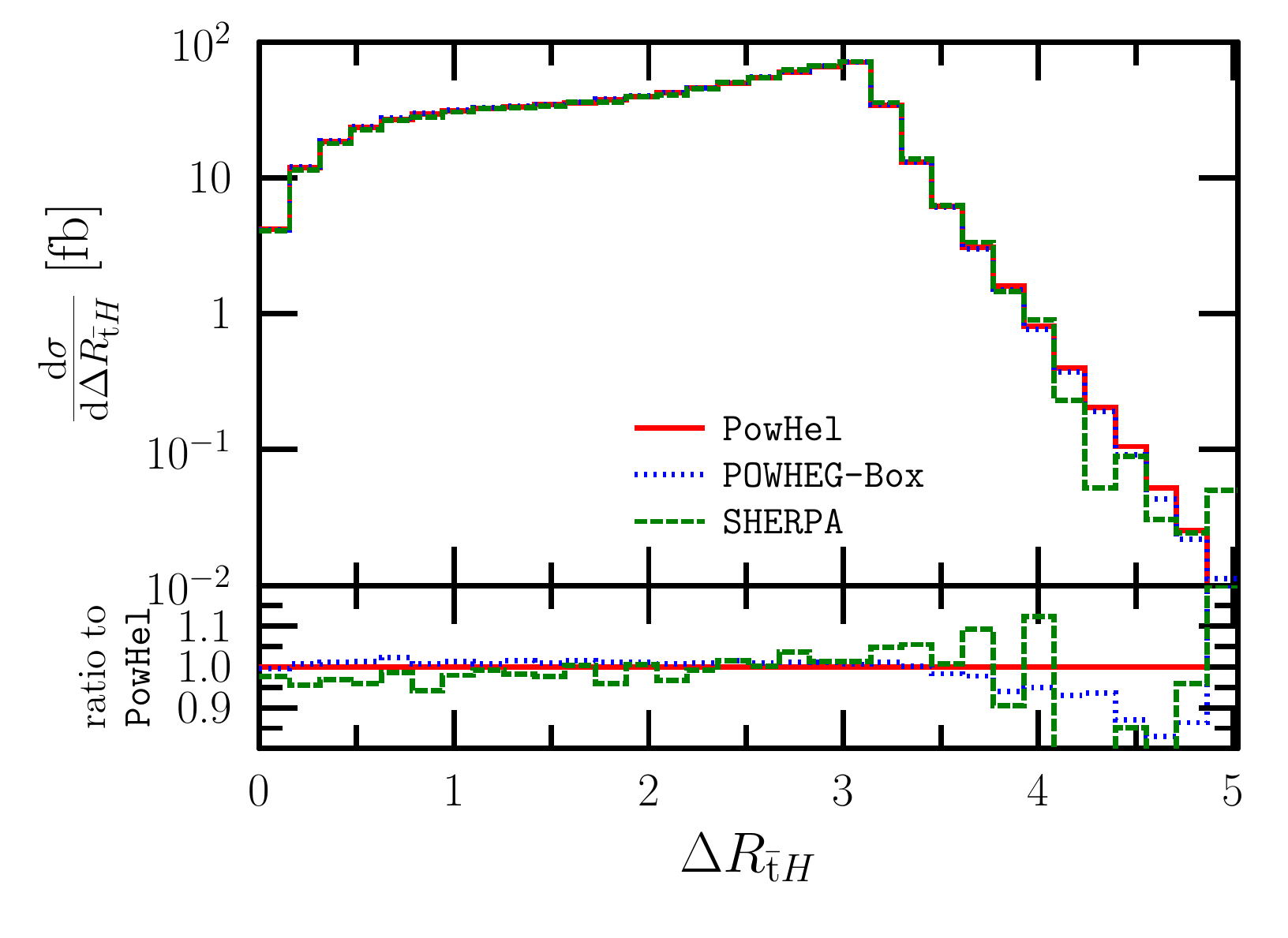}
\includegraphics[bb= 71 239 538 573, width=0.495\textwidth]{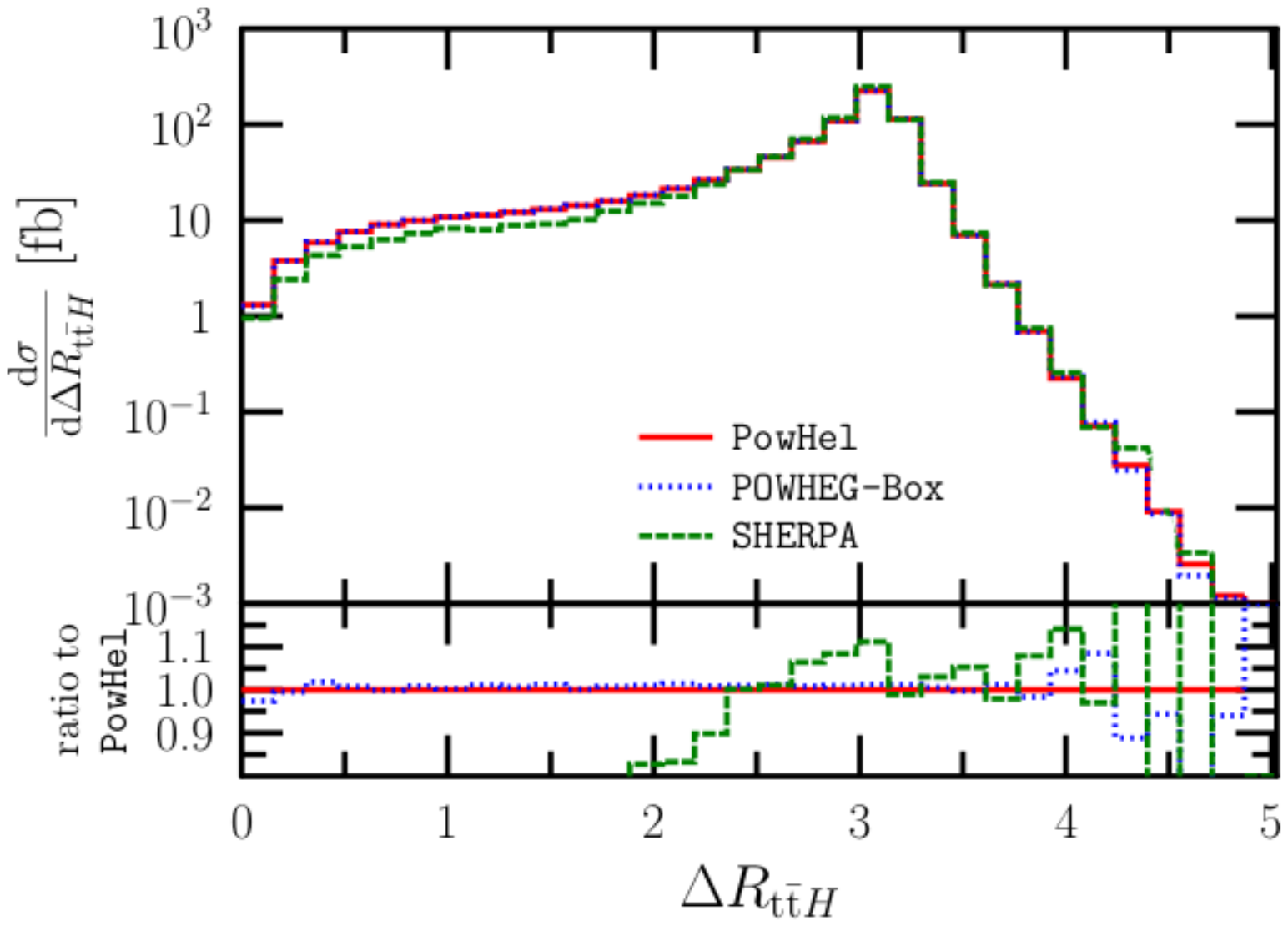}
\caption{\label{fig:tth_fig4;higgs} Separation in the rapidity $-$
  azimuthal-angle plane of different pairs of heavy particles
  [($t$,$\bar{t}$), ($t$,$H$), ($\bar{t}$,$H$), ($t\bar{t}$,$H$)], as
  obtained at LHE level by the three NLO QCD + PS implementations
  [\powhel\ (red solid line), the \powhegbox\ (blue dotted line), and
  \sherpa\ (green dashed line)].  The ratio of the \powhegbox\ and
  \sherpa\ predictions with respect to those from \powhel\ is shown in
  the lower inset of each panel.}
\end{center}
\end{figure}

We first compare inclusive results, without applying any selection
cuts.  The total NLO QCD cross sections from the three implementations
at the central-scale value turned out to be respectively
$\sigma_{\powhel} = 126.8 \pm 0.1$ fb, $\sigma_{\powhegbox} = 126.7
\pm 0.1$ and $\sigma_{\sherpa} = 126.8 \pm 0.1$ fb, showing agreement
within the statistical uncertainty.  Uncertainties related to scale
variation in the interval [$\mu_0/2, 2\mu_0$], as computed by
\powhel\, amount to + 5.2 - 11.8 fb, corresponding to + 4.1\% - 9.3\%
with respect to the central value quoted above.

Differential distributions of the transverse momentum of the top
quark, the antitop quark and the Higgs boson, as well as their
rapidities, are shown in Fig.~\ref{fig:tth_fig1;higgs}. The ratio of
the predictions of the \powhegbox\ and \pythia\ with respect to those
by \powhel\ is also shown in each panel. The slightly larger fluctuations shown
by \sherpa\ predictions with respect to those from \powhel\ and
\powhegbox\ have a statistical origin (the number of events we
generated so far using \sherpa\ is slightly lower than those using the other
two implementations). Agreement within 10\% is found for the three
different implementations considered in this paper, in the $p_\bot$
region below 450 GeV.

In the pre-showered events, a resolved first radiation emission may
emerge, leading to a light jet. The predictions for the transverse
momentum and rapidity of this emission, shown in Fig.~\ref{fig:tth_fig2;higgs}, are very sensitive to the matching. A further zoom in
the Sudakov region is presented in Fig.~\ref{fig:tth_fig3;higgs}.
From these plots we can conclude that the matching implementations in
\powhel\ and the \powhegbox\ are fully consistent among each other. On
the other hand, a consistent comparison with the first radiation
emission from \sherpa\ is more delicate due to the differences in the
matching algorithm, which accounts for the respective different way to
solve the double-counting problem.

In order to further explore the kinematic properties of the produced
particles the separations in the rapidity $-$ azimuthal-angle plane
between different pairs of heavy particles are shown in
Fig.~\ref{fig:tth_fig4;higgs}.  Predictions from \powhel\ and the
\powhegbox\ agree within 3\% in the [0, $\sim$ 3.5] $\Delta R$
interval for all considered pairs, whereas, for larger $\Delta R$,
\powhel\ produces a small excess of events with respect to the
\powhegbox. \sherpa\ predictions for $\Delta R (t,\bar{t})$, $\Delta R (t,\bar{H})$ and $\Delta R (\bar{t},H)$, agree within 10\% with those from the
other two implementations up to $\Delta R$ $\sim$ 3.5, although the
distributions show some shape distortion with a steeper slope with
respect to those from the other two implementations, which are
characterized by a similar stable shape.

After checking the predictions at the NLO accuracy and from
pre-showered events, as a next step we compare predictions after full
SMC simulation, i.e. after parton shower, hadronization, and hadron
decay. We showered the events produced by \powhel\ and
the \powhegbox~\ by means of \pythia, by using the Perugia-0
tune~\cite{Skands:2010ak}, providing a $p_\bot$-ordered shower. Shower emissions beyond the
leading one were forced to occur with a $p_\bot$ smaller than the
first one, already computed by the \powheg~\ matching. In order to
limit jet activity, we forced the Higgs to decay in the $H \rightarrow
\gamma \gamma$ channel. All other particles and hadrons were treated
according to the \pythia~\ default implementation.

\begin{figure}[h!]
\begin{center}
\includegraphics[width=0.495\textwidth]{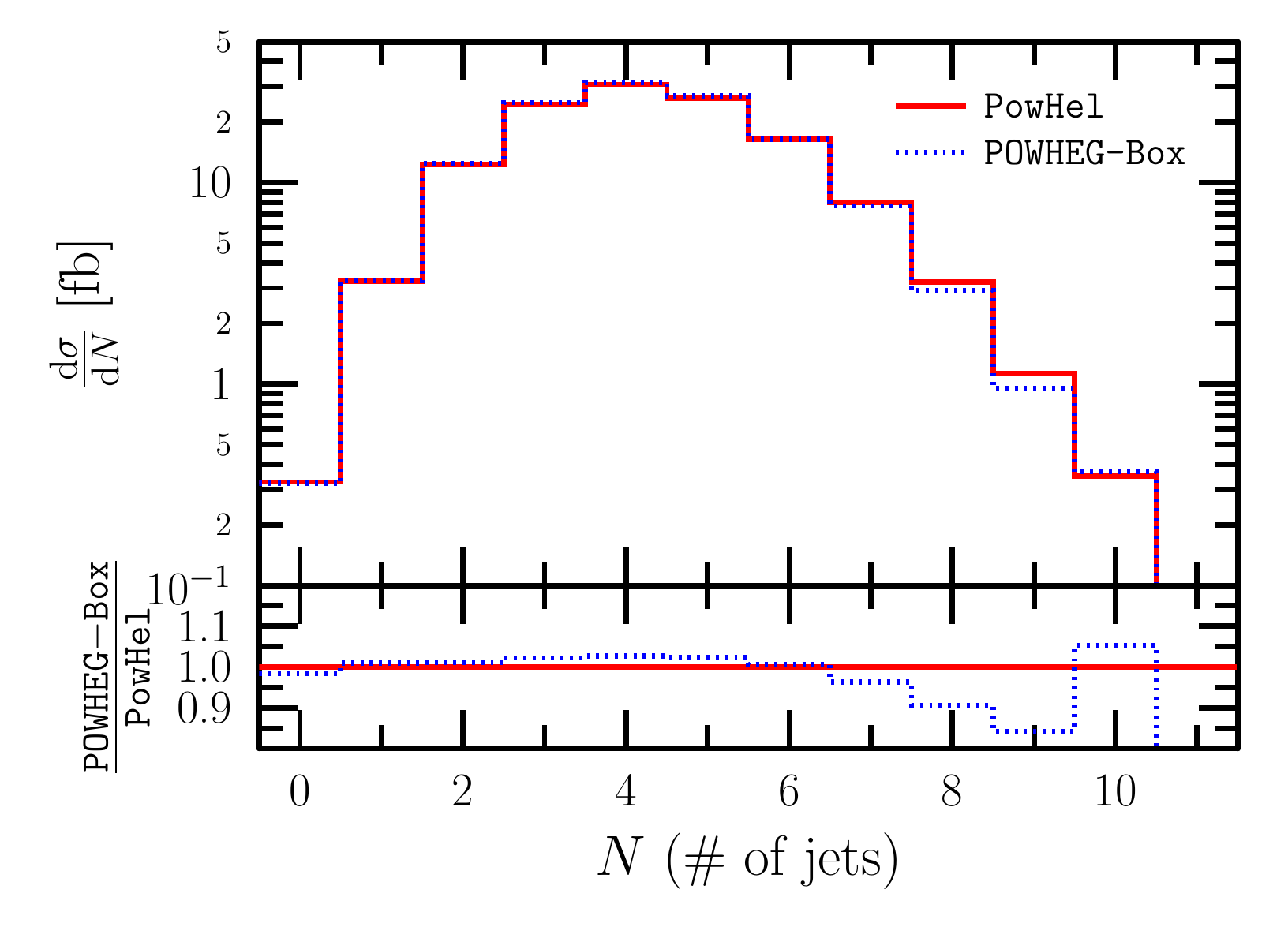}
\includegraphics[width=0.495\textwidth]{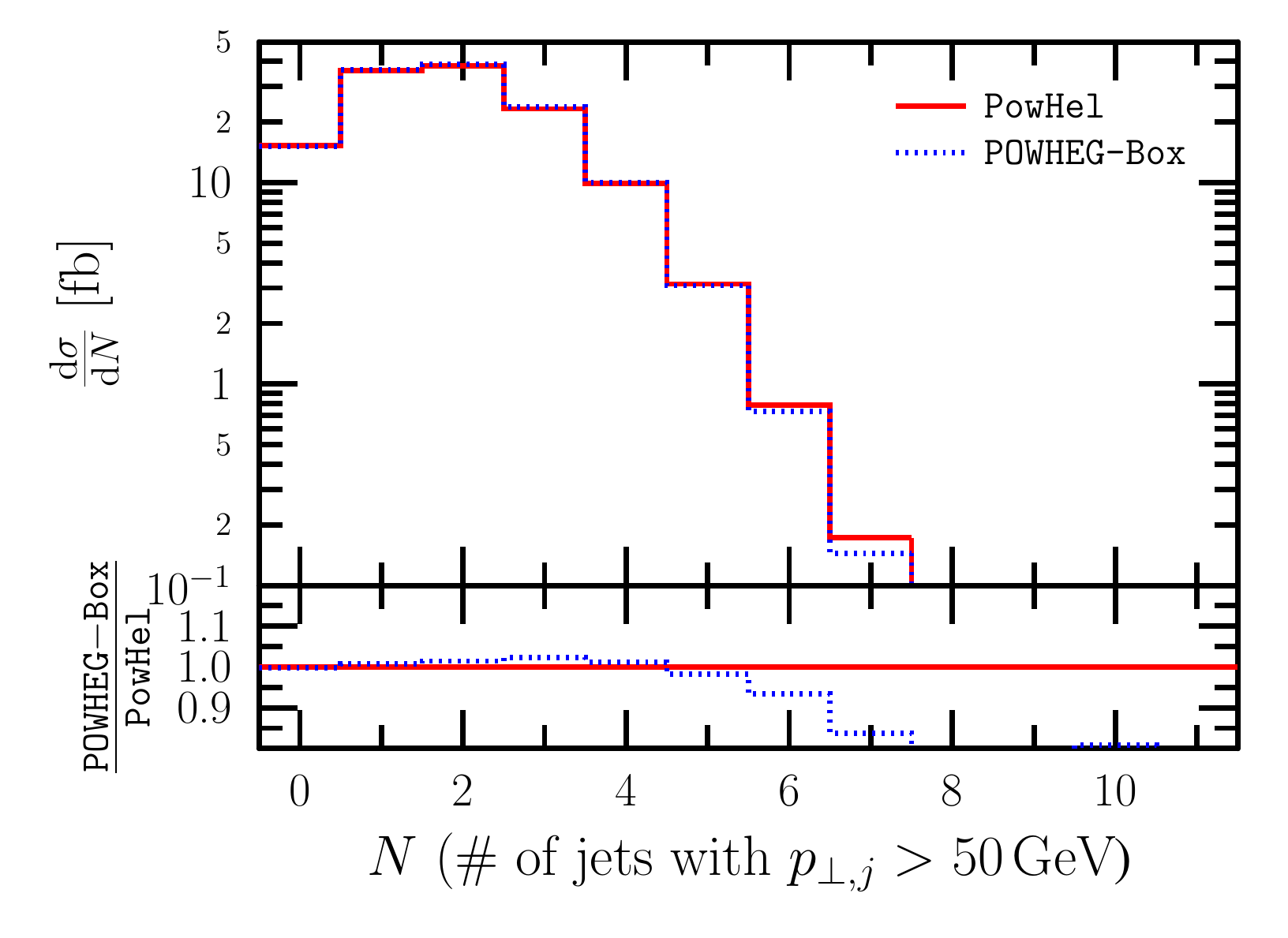}
\caption{\label{fig:tth_fig5;higgs} Number of jets per event and
  number of jets per event with $p_{\bot, j} >$ 50 GeV, as obtained at
  the hadron level by the two NLO QCD + PS implementations [\powhel\
  (red solid line), and the \powhegbox\ (blue dotted line)] interfaced
  to \pythia. The ratio of the \powhegbox\ predictions with respect to
  those from \powhel\ is shown in the lower inset of each panel.}
\end{center}
\end{figure}

\begin{figure}[h!]
\begin{center}
\includegraphics[width=0.495\textwidth]{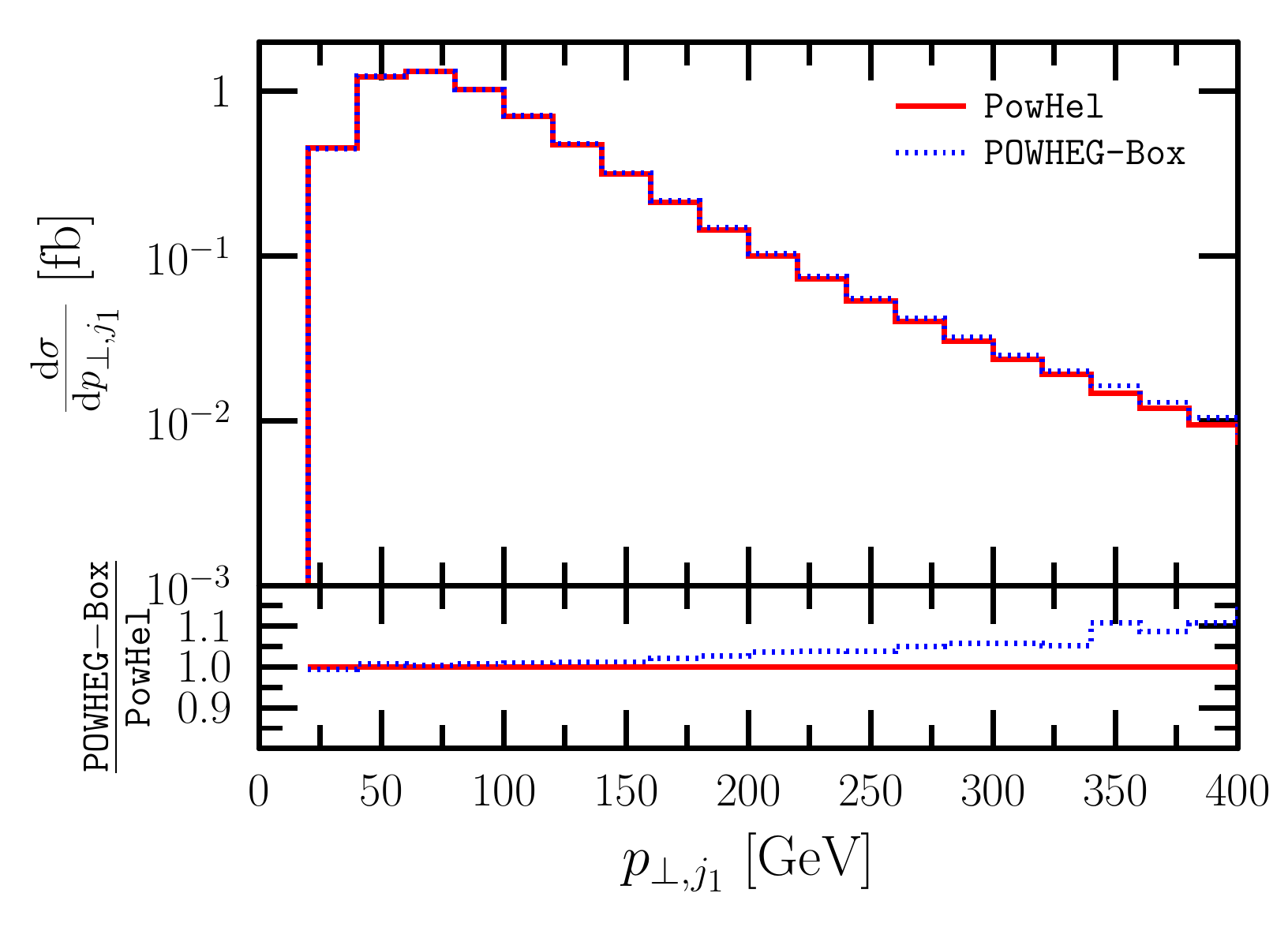}
\includegraphics[width=0.495\textwidth]{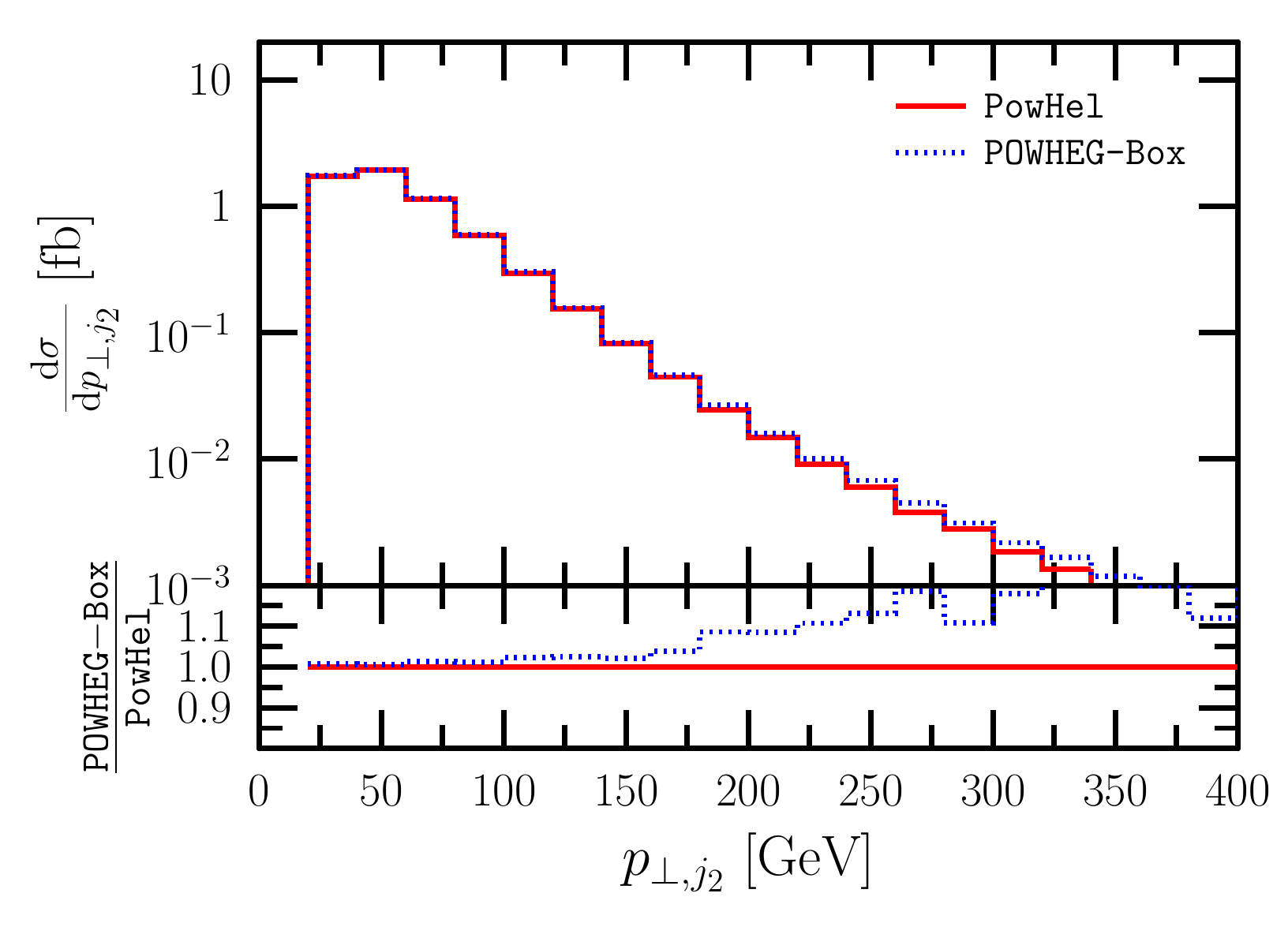}
\caption{\label{fig:tth_fig6;higgs} Transverse momenta of the hardest
  and next-to-hardest jets, as obtained at the hadron level by the two
  NLO~QCD~+~PS implementations [\powhel\ (red solid line), and the
  \powhegbox\ (blue dotted line)] interfaced to \pythia. The ratio of
  the \powhegbox\ predictions with respect to those from \powhel\ is
  shown in the lower inset of each panel.}
\end{center}
\end{figure}

\begin{figure}[h!]
\begin{center}
\includegraphics[width=0.495\textwidth]{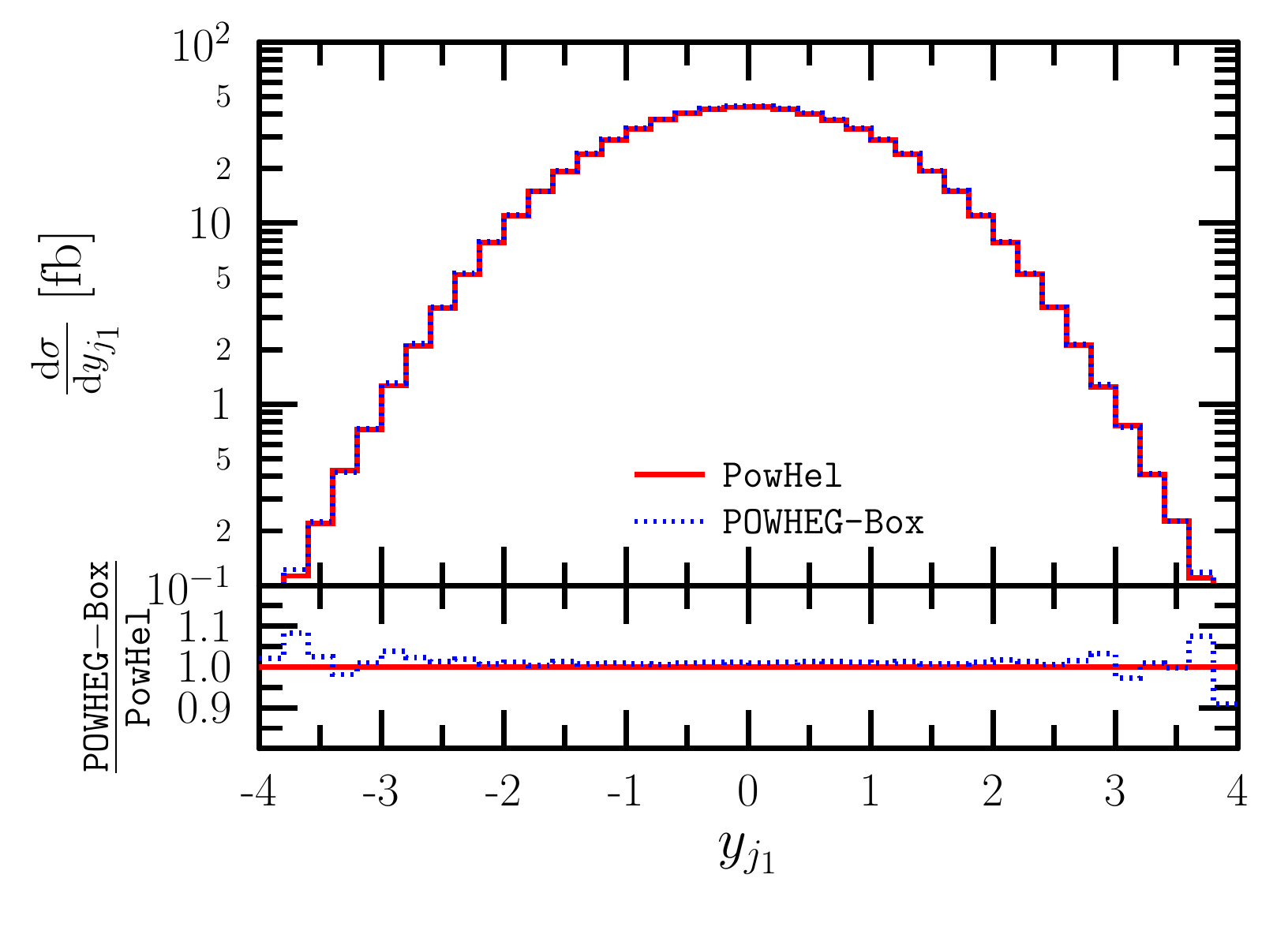}
\includegraphics[width=0.495\textwidth]{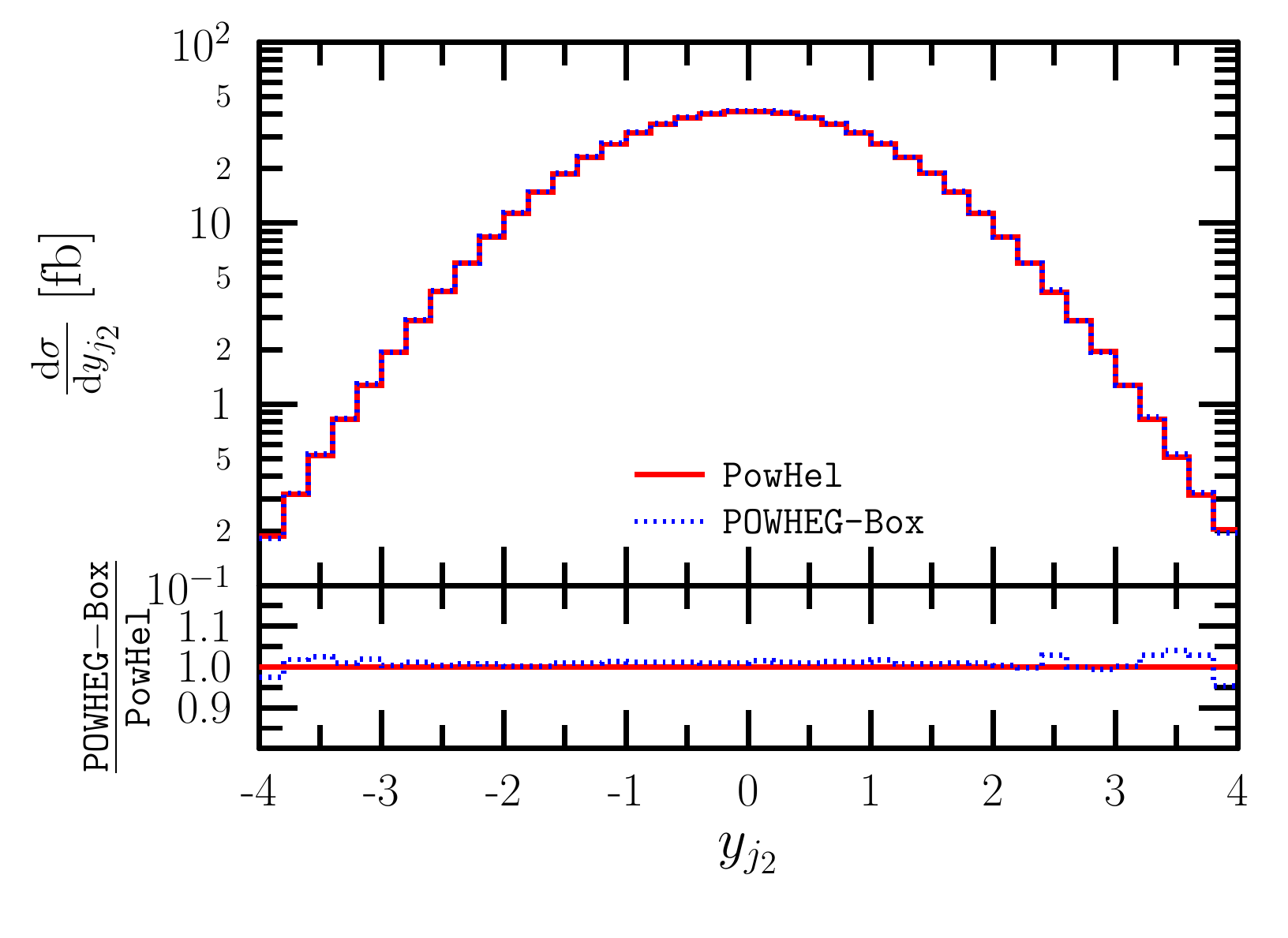}
\caption{\label{fig:tth_fig7;higgs} Rapidities of the hardest and
  next-to-hardest jets, as obtained at the hadron level by the two NLO
  QCD + PS implementations [\powhel\ (red solid line), and the
  \powhegbox\ (blue dotted line)] interfaced to \pythia. The ratio of
  the \powhegbox\ predictions with respect to those from \powhel\ is
  shown in the lower inset of each panel.}
\end{center}
\end{figure}

\begin{figure}[h!]
\begin{center}
\includegraphics[width=0.495\textwidth]{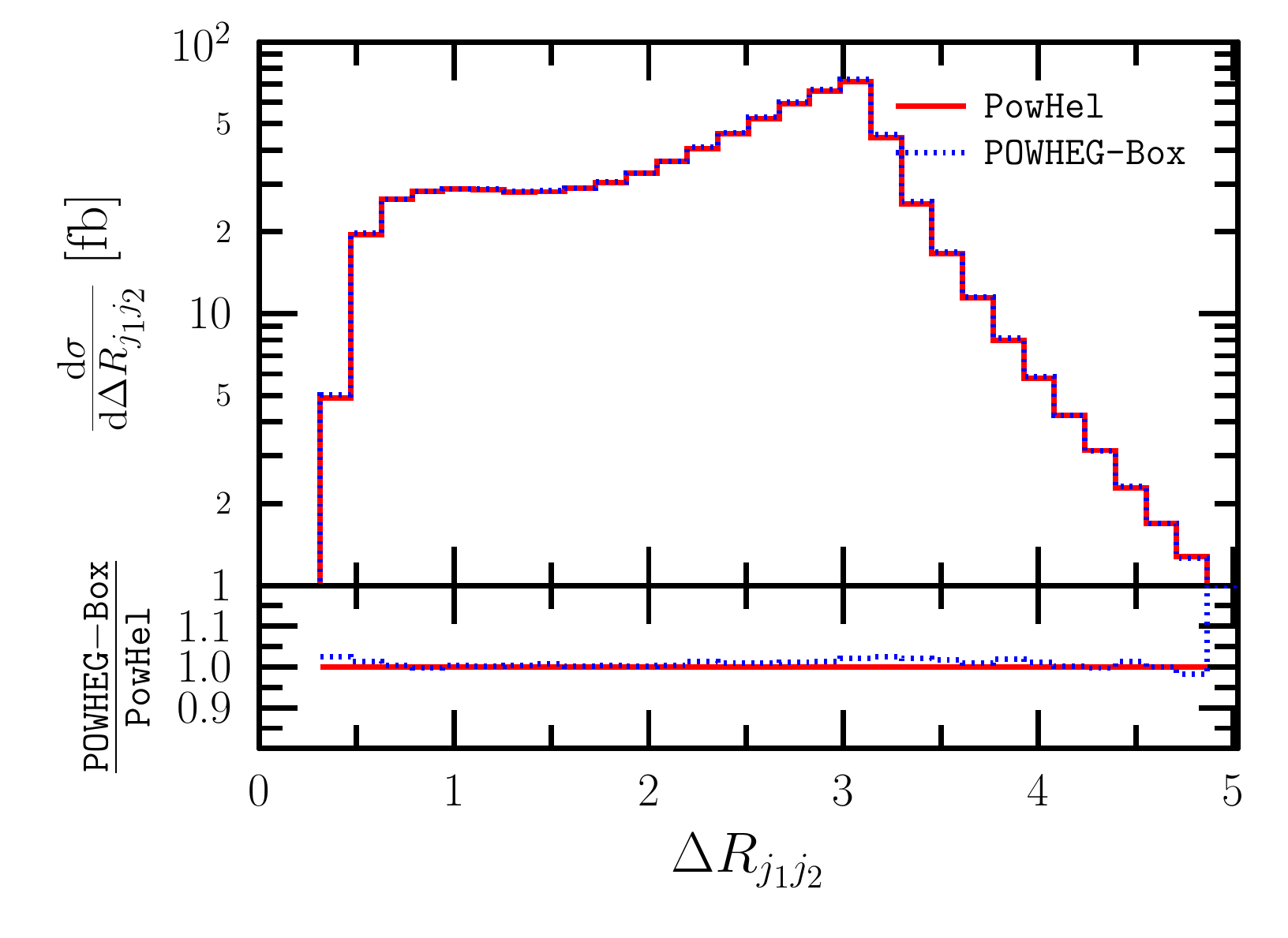}
\includegraphics[width=0.495\textwidth]{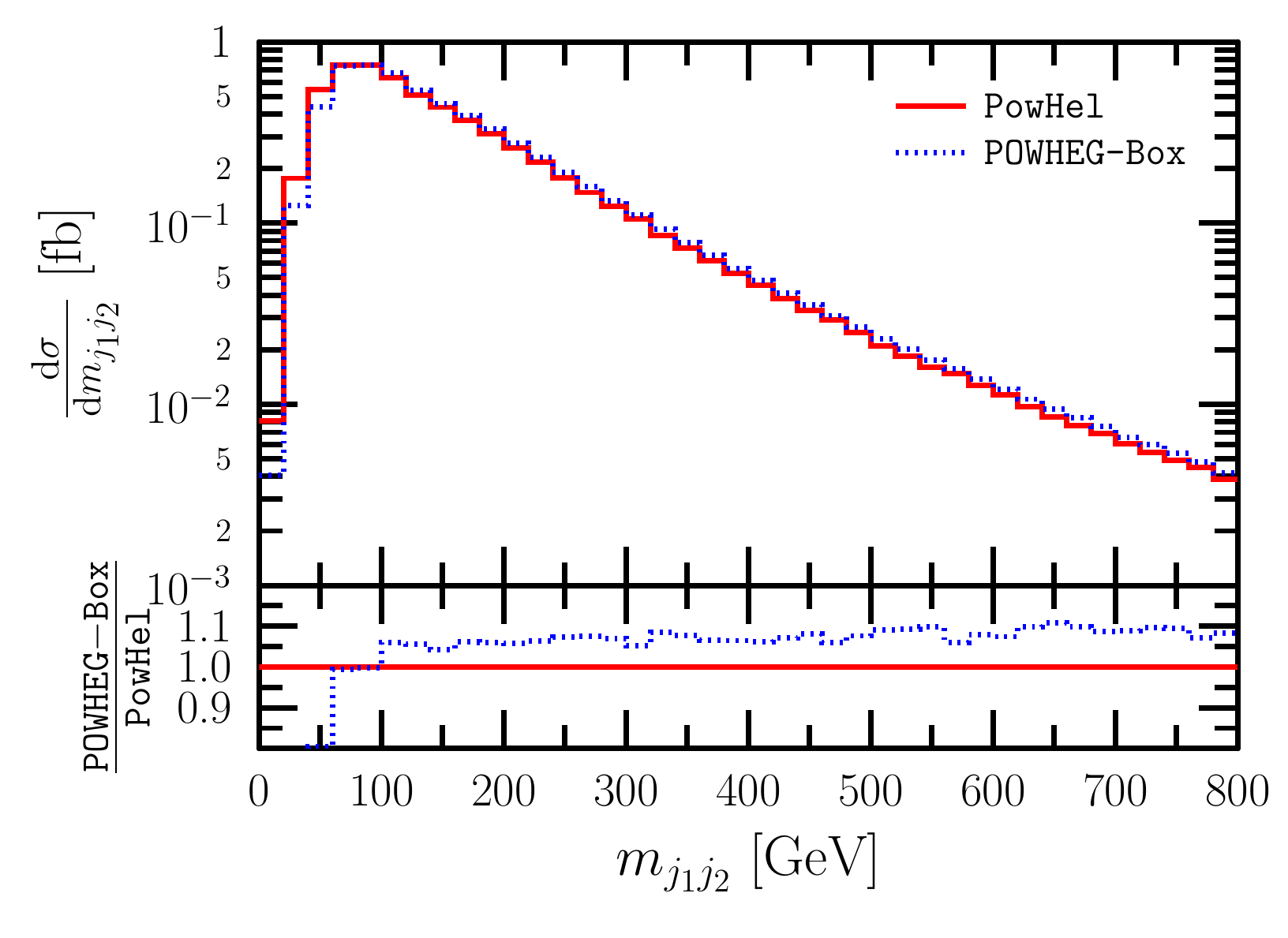}
\caption{\label{fig:tth_fig8;higgs} Separation between the two hardest
  jets in the rapidity $-$ azimuthal-angle plane and their invariant
  mass, as obtained at the hadron level by the two NLO QCD + PS
  implementations [\powhel\ (red solid line),and the \powhegbox\ (blue
  dotted line)] interfaced to \pythia. The ratio of the \powhegbox\
  predictions with respect to those from \powhel\ is shown in the
  lower inset of each panel.}
\end{center}
\end{figure}

In order to investigate the effect of the parton shower on experimentally
accessible distributions, we consider differential cross-sections
involving hard jets and leptons at the SMC level.  We reconstruct jets
out of all hadronic tracks with rapidity $|y| < 5$ by the anti-k$_T$
algorithm~\cite{Cacciari:2008gp} as implemented in the
\fastjet{}~package~\cite{Cacciari:2011ma}, with a resolution parameter
of $R = 0.4$. Jets that do not exhibit a transverse momentum
$p_\bot^\mathrm{jet} > 20$~GeV and a rapidity $|y^\mathrm{jet}| < 4.5$
are discarded.
Similarly, we consider only leptons with a transverse momentum
$p_\bot^\ell > 20$~GeV and rapidity $|y^\ell| < 2.5$ that are
well-separated from all hard jets in the rapidity $-$ azimuthal-angle
plane, i.e. such that  $\Delta R_{\ell,\mathrm{jet}}>0.4$.

Distributions concerning the total number of jets per event and the
number of jets per event with a $p_\bot >$ 50 GeV are shown in
Fig.~\ref{fig:tth_fig5;higgs}, and are peaked around 4 and 2 jets,
respectively, for both implementations. These peaks correspond to the
semileptonic decay of the top quarks, taking into account that in
these simulations the Higgs boson was not allowed to decay
hadronically.  The agreement between \powhel\ and the \powhegbox\ is
within a very few percents from 0 up to 6 jets, whereas, for higher
jet multiplicities, \powhel\ leads to slightly more events with a
larger number of (slightly softer) jets than the \powhegbox.

The distributions of the transverse momentum of the hardest and
next-to-hardest jets are shown in Fig.~\ref{fig:tth_fig6;higgs}, from
where it is clear that \powhel\ and the \powhegbox\ predictions agree
up to about 200 GeV. The small differences at higher transverse
momentum reflect the slight deviation found in the distributions of
$p_{\bot, t}$ and $p_{\bot, \bar{t}}$ in the pre-showered events.

The rapidities of these same jets are presented in
Fig.~\ref{fig:tth_fig7;higgs} showing very good agreement between the
two implementations in the whole phase-space region plotted.

Predictions concerning the $\Delta R$ separation between these two
jets, presented in Fig.~\ref{fig:tth_fig8;higgs}, show again an almost
perfect agreement between \powhel\ and the \powhegbox, whereas those
concerning their invariant mass show agreement within 10\%, with the
\powhegbox\ distribution slightly harder than the \powhel\ one.

\begin{figure}[t!]
\begin{center}
\includegraphics[width=0.495\textwidth]{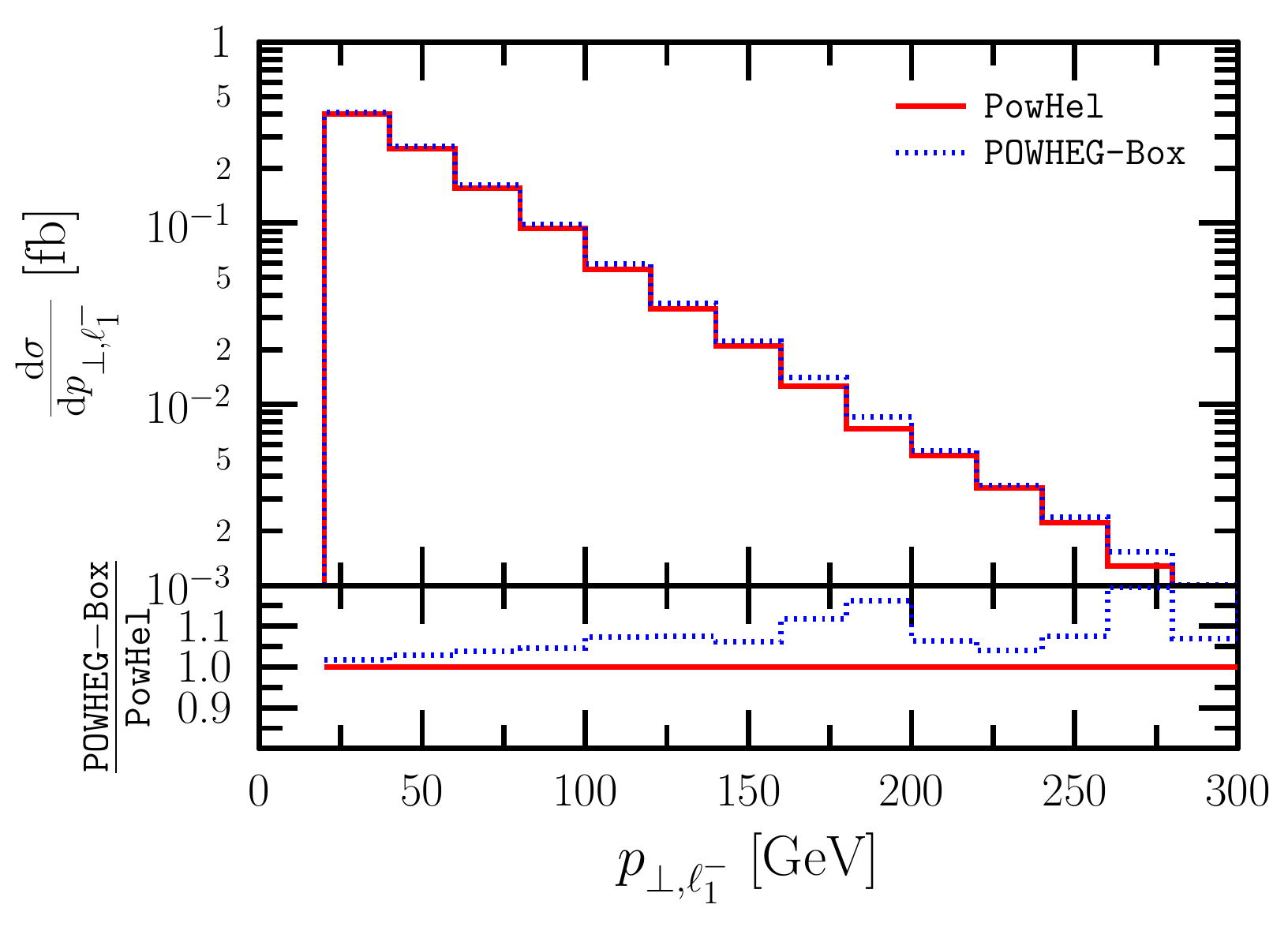}
\includegraphics[width=0.495\textwidth]{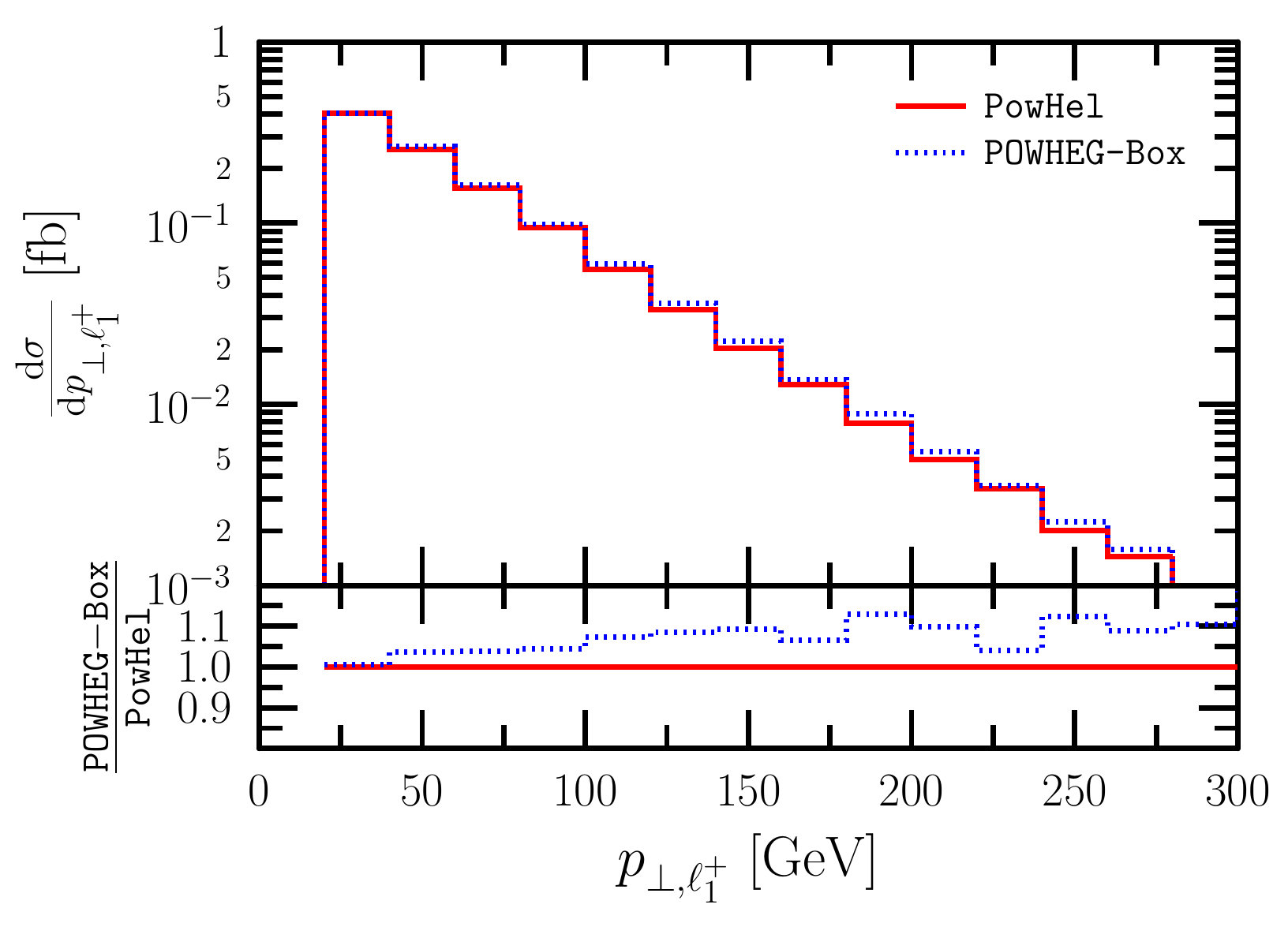}
\caption{\label{fig:tth_fig9;higgs} Transverse momenta of the hardest isolated
  lepton and the hardest isolated antilepton, as obtained at the
  hadron level by the two NLO QCD + PS implementations [\powhel\ (red
  solid line), and the \powhegbox\ (blue dotted line)] interfaced to
  \pythia. The ratio of the \powhegbox\ predictions with respect to
  those from \powhel\ is shown in the lower inset of each panel.}
\end{center}
\end{figure}

\begin{figure}[h!]
\begin{center}
\includegraphics[width=0.495\textwidth]{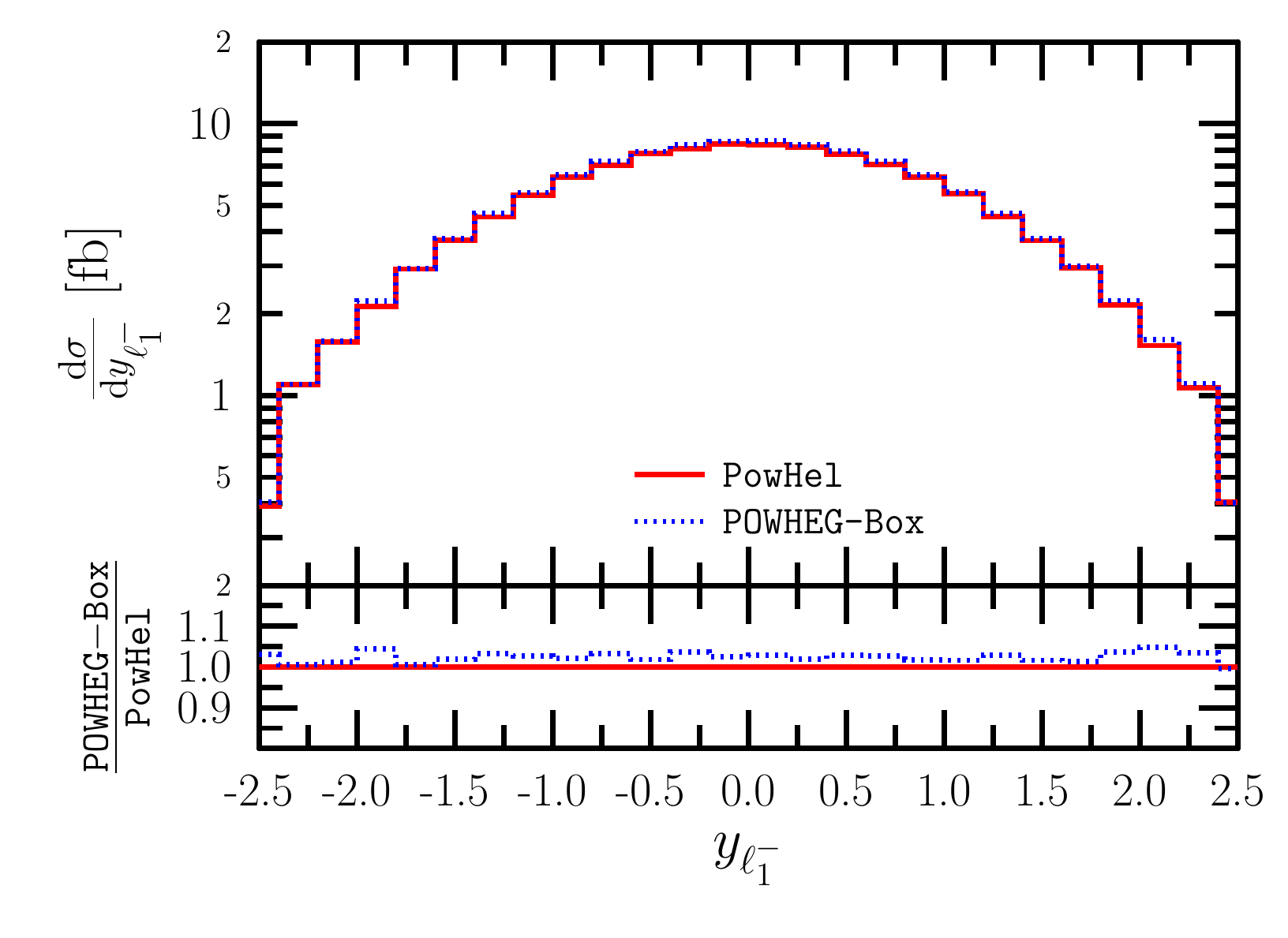}
\includegraphics[width=0.495\textwidth]{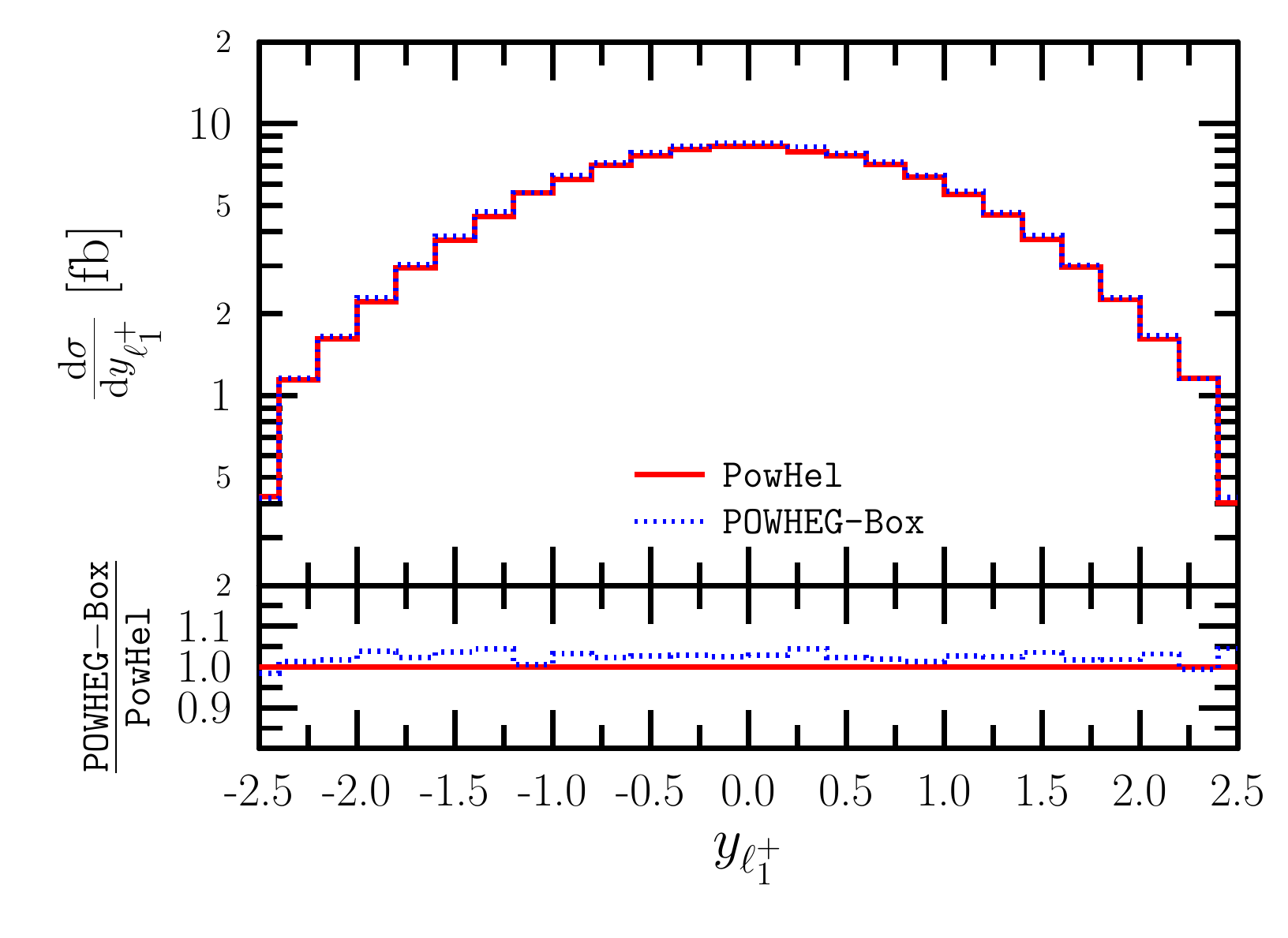}
\caption{\label{fig:tth_fig10;higgs} Rapidities of the hardest
  isolated lepton and the hardest isolated antilepton, as obtained at
  the hadron level by the two NLO QCD + PS implementations [\powhel\
  (red solid line), and the \powhegbox\ (blue dotted line)] interfaced
  to \pythia. The ratio of the \powhegbox\ predictions with respect to
  those from \powhel\ is shown in the lower inset of each panel.}
\end{center}
\end{figure}

Finally, distributions related to the hardest isolated lepton and
antilepton of each event are shown in Fig.~\ref{fig:tth_fig9;higgs}
and~\ref{fig:tth_fig10;higgs}.  As for transverse momentum
distributions, shown in Fig.~\ref{fig:tth_fig9;higgs}, the agreement
between \powhel\ and the \powhegbox\ predictions decreases with
increasing $p_\bot$, staying within 10\% up to $\sim$ 200 GeV, with
the \powhegbox\ predictions increasingly harder than those from
\powhel\ in the tails. As for rapidities, shown in
Fig.~\ref{fig:tth_fig10;higgs}, the agreement between \powhel\ and the
\powhegbox\ predictions is within 5\% over the whole range under study
and no shape distortion occurs.

The comparison of the predictions by \powhel\ + \pythia\ and the
\powhegbox\ + \pythia\ shown in Figs.~\ref{fig:tth_fig5;higgs}
-~\ref{fig:tth_fig10;higgs} with those from \sherpa~\ after its whole
internal SMC evolution is still under way. 

\subsection{Conclusions}

We have compared three different implementations of \tth\ production
at the LHC, including NLO QCD corrections matched to parton
showers. One of these implementations (\powhel) can be considered
mature, and was already extensively checked in the past, whereas the
other two (\powhegbox\ and \sherpa) are new and introduced in this
paper for the first time.  Distributions at the inclusive level are in
agreement within a few percent, i.e. within renormalization and
factorization scale uncertainties, over a large fraction of the
phase-space. The regions where distributions from different
implementations show the largest disagreement turned out to be the
high transverse momentum tails.  Thus, in view of experimental studies
in boosted regions, the comparison of different implementations is
particularly important. The \powhel\ and \powhegbox\ matching
implementations turned out to be fully consistent among each other, as
shown by the very good agreement on the properties of first radiation
emission in the Sudakov region.  A more detailed comparison will be
presented elsewhere.

\subsection*{Acknowledgements}

M.~V.~G., B.~J., A.~K., L.~R., and D.~W. would like to thank the Aspen
Center for Physics where preliminary work on the comparison presented
in this study was initiated, after initial discussions in Les
Houches. Their work in Aspen was supported in part by the National
Science Foundation under Grant No. PHYS-1066293.  B.~J.\ would like to
thank the AHRP alliance for support.  A.~K. and Z.~T. would like to
thank the LHCPhenoNet network PITN-GA-2010-264564 for support, the
Hungarian Scientific Research Fund grant K-101482, and the
Supercomputer, the national virtual lab
TAMOP-4.2.2.C-11/1/KONV-2012-0010 project.  The work of L.R. is
partially supported by the U.S. Department of Energy under grant
DE-FG02-13ER41942, and by the European Research Council under the
European Union's Seventh Framework Programme (FP/2007-2013) ERC Grant
Agreement n. 279972 "NPFlavour".  The work of D.W. is supported in
part by the U.S. National Science Foundation under grant
no. PHY-1118138.

\clearpage



\section{Next-to-next-to-leading order corrections to
Higgs boson pair production at the LHC\footnote{D.~de~Florian and J.~Mazzitelli}}


\subsection{Introduction}

Recently, both ATLAS \cite{Aad:2012tfa} and CMS \cite{Chatrchyan:2012ufa} collaborations discovered a new boson at the Large Hadron Collider (LHC), whose properties are so far compatible with the long sought standard model (SM) Higgs boson \cite{Englert:1964et,Higgs:1964pj,Higgs:1964ia}.
In order to decide whether this particle is indeed responsible for the electroweak symmetry breaking, a precise measurement of its couplings to fermions, gauge bosons and its self-interactions is needed.
In particular, the knowledge of the Higgs self-couplings is the only way to reconstruct the scalar potential.

The possibility of observing Higgs pair production at the LHC have been discussed in Refs. \cite{Baur:2002qd,
Dolan:2012rv,Papaefstathiou:2012qe,
Baglio:2012np,Baur:2003gp,Dolan:2012ac,Goertz:2013kp,
Shao:2013bz,Gouzevitch:2013qca}.
In general, it has been shown that despite the smallness of the signal and the large background its measurement can be achieved at a luminosity-upgraded LHC.


The dominant mechanism for SM Higgs pair production at hadron colliders is gluon fusion, mediated by a heavy-quark loop.
The leading-order (LO) cross section has been calculated in Refs. \cite{Glover:1987nx,Eboli:1987dy,Plehn:1996wb}.
The next-to-leading order (NLO) QCD corrections have been evaluated in Ref. \cite{Dawson:1998py} within the large top-mass approximation and found to be large, with an inclusive $K$ factor close to $2$.
The finite top-mass effects were analysed at this order in Ref. \cite{Grigo:2013rya}, finding that a precision of ${\cal O}(10\%)$ can be achieved if the exact top-mass leading-order cross section is used to normalize the corrections.

Given the size of the NLO corrections, it is necessary to reach higher orders to provide accurate theoretical predictions.
In this proceeding we present the next-to-next-to-leading order (NNLO) corrections for the inclusive Higgs boson pair production cross section\cite{deFlorian:2013jea}.




\subsection{Description of the calculation}
The effective single and double-Higgs coupling to gluons is given, within the large top-mass approximation, by the following Lagrangian
\begin{equation}
{\cal L}_{\text{eff}}=
-\frac{1}{4}G_{\mu\nu}G^{\mu\nu}
\left(C_H\frac{H}{v}-C_{HH}\frac{H^2}{v^2}\right)\,.
\end{equation}
Here $G_{\mu\nu}$ stands for the gluonic field strength tensor and $v\simeq 246\,\text{GeV}$ is the Higgs vacuum expectation value.
While the ${\cal O}(\alpha_{\mathrm{S}}^3)$ of the $C_{H}$ expansion is known \cite{Kramer:1996iq,Chetyrkin:1997iv}, the QCD corrections of $C_{HH}$ are only known up to ${\cal O}(\alpha_{\mathrm{S}}^2)$ \cite{Djouadi:1991tka}. Up to that order, both expansions yield the same result.

The NNLO contributions to the SM Higgs boson pair production squared matrix element can be separated into two different classes: (a) those containing two gluon-gluon-Higgs vertices (either $ggH$ or $ggHH$) and (b) those containing three or four effective vertices.
Given the similarity between $ggH$ and $ggHH$ vertices, the contributions to the class (a) are equal to those of single Higgs production, except for an overall LO normalization (assuming that $C_H=C_{HH}$ up to ${\cal O}\left(\alpha_{\mathrm{S}}^3\right)$). These results can be obtained from Refs. \cite{Harlander:2002wh,Anastasiou:2002yz,Ravindran:2003um}.

\begin{figure}
\begin{center}
\includegraphics[width=0.5\textwidth]{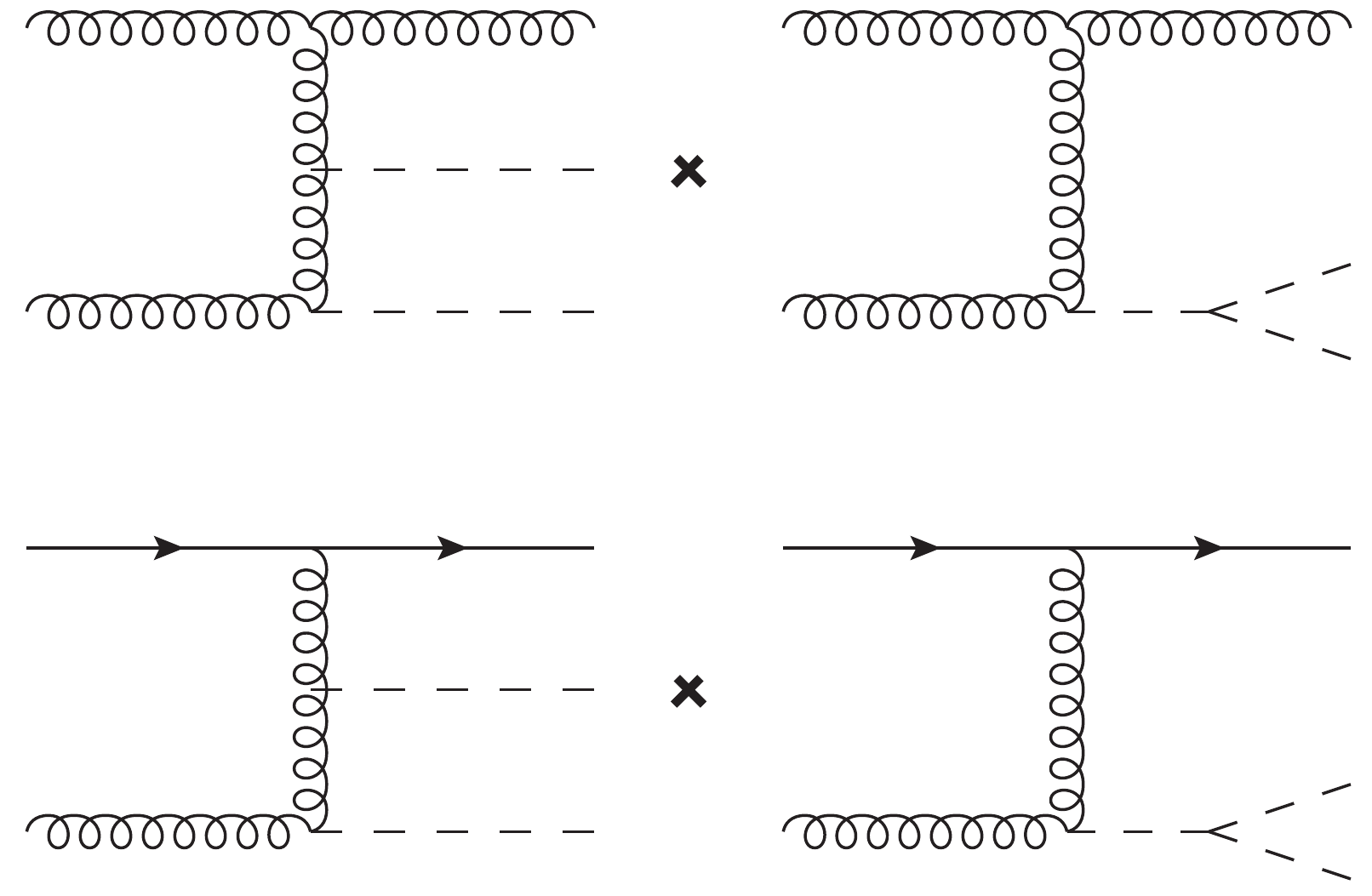}
 \caption{Example of Feynman diagrams needed for the NNLO calculation for $gg\rightarrow HHg$ (top) and $qg\rightarrow HHq$ (bottom) subprocesses.  Other parton subprocesses can be obtained from crossings.}
\label{diagramas}
\end{center}
\end{figure}

Contributions to the class (b) first appear at NLO as a tree-level contribution to the subprocess $gg\rightarrow HH$, given that each $ggH$ and $ggHH$ vertex is proportional to $\alpha_{\mathrm{S}}$.
Then, at NNLO we have one-loop corrections and single real emission corrections.
The former have been calculated in Ref. \cite{deFlorian:2013uza}.
The remaining contributions involve the partonic subprocesses $gg\rightarrow HH+g$ and $qg\rightarrow HH+q$ (with the corresponding crossings). Examples of the Feynman diagrams involved in the calculation are shown in Figure \ref{diagramas}.



To compute this contribution we used the Mathematica packages FeynArts \cite{Hahn:2000kx} and FeynCalc \cite{Mertig:1990an} in order to generate the Feynman diagrams and evaluate the corresponding amplitudes.
The calculation was performed using nonphysical polarizations, which we cancel by including ghosts in the initial and final states.
In order to subtract the soft and collinear divergencies, we used the Frixione, Kunszt, and Signer subtraction method \cite{Frixione:1995ms}. Further details of the calculation, together with the explicit expressions for the NNLO results, can be found in Ref. \cite{deFlorian:2013jea}.


\subsection{Phenomenology}

Here we present the numerical results for the LHC. At each order, we use the corresponding MSTW2008 \cite{Martin:2009iq} set of parton distributions and QCD coupling.
We recall that we always normalize our results using the exact top- and bottom-mass dependence at LO.
For this analysis we use $M_H=126\,\text{GeV}$, $M_t=173.18\,\text{GeV}$ and $M_b=4.75\,\text{GeV}$.
The bands of all the plots are obtained by varying independently the factorization and renormalization scales in the range $0.5\,Q\leq \mu_F,\mu_R \leq 2\,Q$, with the constraint $0.5\leq \mu_F/\mu_R \leq 2$, being $Q$ the invariant mass of the Higgs pair system.


We assume for the phenomenological results that the two-loop corrections to the effective vertex $ggHH$  are the same than those of $ggH$ (that is $C_{HH}^{(2)}=C_{H}^{(2)}$, following the notation of Ref. \cite{deFlorian:2013uza}), as it happens at one-loop order.
We change  its value in the range $0\leq C_{HH}^{(2)}\leq 2C_{H}^{(2)}$ in order to evaluate the impact of this unknown coefficient and find a variation in the total cross section of less than $2.5\%$.


\begin{figure}
\begin{center}
\includegraphics[width=0.5\textwidth]{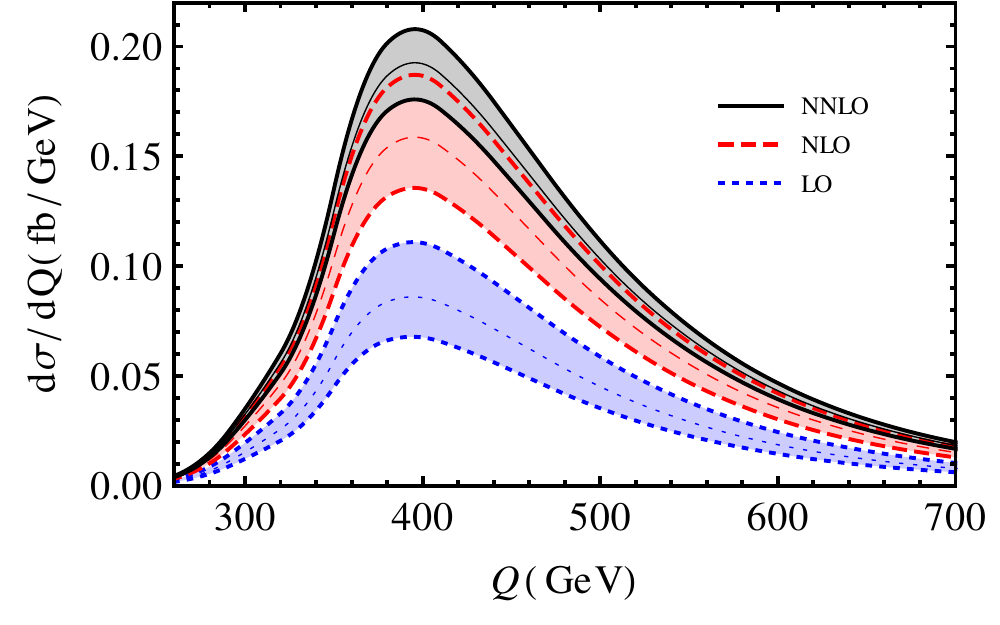}
 \caption{Higgs pair invariant mass distribution at LO (dotted blue), NLO (dashed red) and NNLO (solid black) for the LHC at c.m. energy $E_{cm}=14\,\text{TeV}$. The bands are obtained by varying $\mu_F$ and $\mu_R$ in the range $0.5\,Q\leq \mu_F,\mu_R \leq 2\,Q$ with the constraint $0.5\leq \mu_F/\mu_R \leq 2$.}
\label{q2distr}
\end{center}
\end{figure}

In Figure \ref{q2distr} we present the LO, NLO and NNLO predictions for the hadronic cross section at the LHC as a function of the Higgs pair invariant mass, for a c.m. energy $E_{cm}=14\,\text{TeV}$. As can be noticed from the plot, only at this order the first sign of convergence of the perturbative series appears, finding a nonzero overlap between the NLO and NNLO bands.
Second order corrections are sizeable, this is noticeable already at the level of the total inclusive cross sections,
where the increase with respect to the NLO result is of ${\cal O}(20\%)$, and the $K$ factor with respect to the LO prediction is about $K_{\text{NNLO}}=2.3$.
The scale dependence is clearly reduced at this order, resulting in a variation of about $\pm8\%$ around the central value, compared to a total variation of ${\cal O}(\pm20\%)$ at NLO.

\begin{figure}
\begin{center}
\includegraphics[width=0.5\textwidth]{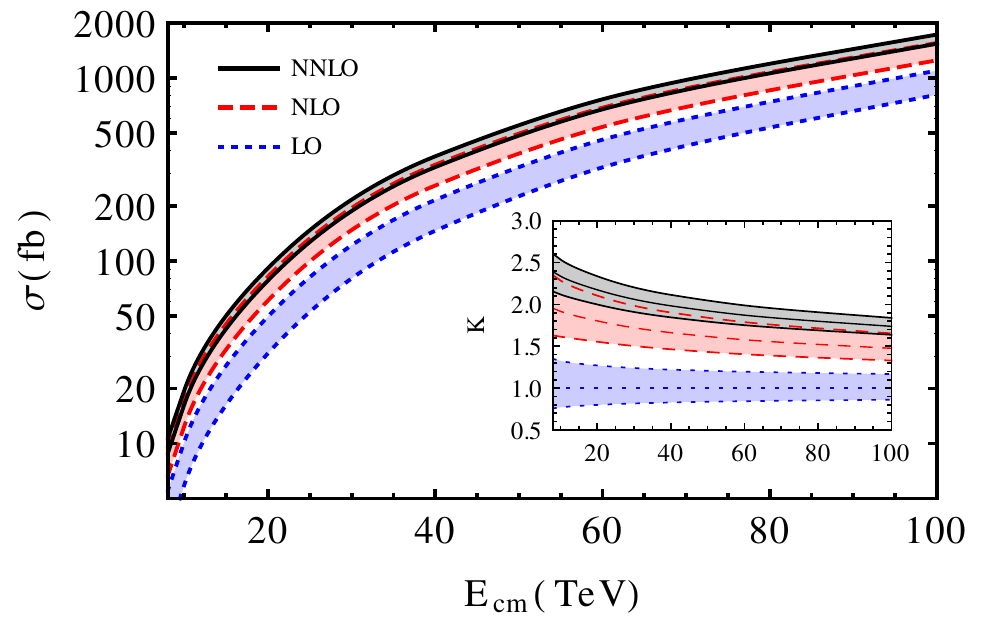}
 \caption{Total cross section as a function of the c.m. energy $E_{cm}$ for the LO (dotted blue), NLO (dashed red) and NNLO (solid black) prediction. The bands are obtained by varying $\mu_F$ and $\mu_R$ as indicated in the main text. The inset plot shows the corresponding $K$ factors.}
\label{sh}
\end{center}
\end{figure}

In Figure \ref{sh} we show the total cross section as a function of the c.m. energy $E_{cm}$, in the range from $8\,\text{TeV}$ to $100\,\text{TeV}$. We can observe that the size of the NLO and NNLO corrections is smaller as the c.m. energy increases. We can also notice that the scale dependence is substantially reduced in the whole range of energies when we include the second order corrections.
The ratio between NNLO and NLO predictions as a function of the c.m. energy is quite flat, running from $1.22$ at $8\,\text{TeV}$ to $1.18$ at $100\,\text{TeV}$.
On the other hand, the ratio between NNLO and LO runs from $2.39$ to $1.74$ in the same range of energies.

Finally, we present in Table \ref{tabla} the value of the NNLO cross section for $E_{cm}=8$, 14, 33 and $100\,\text{TeV}$.
We have taken into account three sources of theoretical uncertainties: missing higher orders in the QCD perturbative expansion, which are estimated by the scale variation, and uncertainties in the determination of the parton flux and strong coupling.
To estimate the parton dinstributions and coupling constant uncertainties we used the MSTW2008 $90\%$ C.L. error PDF sets \cite{Martin:2009bu}, which are known to provide very close results to the PDF4LHC working group recommendation for the envelope prescription \cite{Botje:2011sn}. As we can observe from Table \ref{tabla}, nonperturbative and perturbative uncertainties are of the same order.

\begin{table}
\begin{center}
\begin{tabular}{l  c  c  c  c }
\hline\hline
$E_{cm}$ & $8\text{ TeV}$ & $14\text{ TeV}$ & $33\text{ TeV}$ & $100\text{ TeV}$ \\
\hline
$\sigma_{\text{NNLO}}$ & $9.76\text{ fb}$ & $40.2\text{ fb}$ & $243\text{ fb}$ & $1638\text{ fb}$ \\
Scale $[\%]$ &\footnotesize $+9.0-9.8\,$ &\footnotesize $\,+8.0-8.7\,$ &\footnotesize $\,+7.0-7.4\,$ &\footnotesize $\,+5.9-5.8$ \\ 
PDF $[\%]$ &\footnotesize $+6.0-6.1\,$ &\footnotesize $\,+4.0-4.0\,$ &\footnotesize $\,+2.5-2.6\,$ &\footnotesize $\,+2.3-2.6$ \\ 
PDF$+\alpha_{\mathrm{S}}$  $[\%]$ &\footnotesize $+9.3-8.8\,$ &\footnotesize $\,+7.2-7.1\,$ &\footnotesize $\,+6.0-6.0\,$ &\footnotesize $\,+5.8-6.0$ \\ 
\hline\hline
\end{tabular}
\caption{Total cross section as a function of the c.m. energy at NNLO accuracy. We use the exact LO prediction to normalize our results. The different sources of theoretical uncertainties are discussed in the main text.\label{tabla}}
\end{center}
\end{table}

It is worth noticing that the soft-virtual approximation, which was presented in Ref. \cite{deFlorian:2013uza}, gives an extremely accurate prediction for the NNLO cross section, overestimating for example the $E_{cm}=14\,\text{TeV}$ result by less than $2\%$.
As expected, this approximation works even better than for single Higgs production, due to the larger invariant mass of the final state.

\subsection*{Acknowledgements}
This work was supported in part by UBACYT, CONICET, ANPCyT and the Research Executive Agency (REA) of the European Union under the Grant Agreement number PITN-GA-2010-264564 (LHCPhenoNet).



\section{Photon isolation studies\footnote{L.~Cieri and D.~de~Florian}}

\subsection{Introduction}

Many interesting physics signatures at the LHC involve the presence of single or multiple photons in the final state. These photons may either be produced directly, through the fragmentation of a quark or gluon, or else through the decay of a resonance -- such as e.g. the Higgs boson.

The production of prompt photons is the subject of this analysis. The expression `prompt photons' refers to photons produced at large transverse momenta which do not arise from the decay of hadrons, such as $\pi^{0}$, $\eta$, etc. Prompt photons can be produced according to two possible mechanisms, one of them being fragmentation.

To be precise, the collider experiments at the Tevatron and the LHC do not perform {\it inclusive} photon measurements. The background of secondary photons coming from the decays of $\pi^{0}$, $\eta$, etc., overwhelms the signal by several orders of magnitude. To reject this background, the experimental selection of prompt diphotons requires {\em isolation} cuts. In addition to the rejection of the background of secondary  photons, the isolation cuts (or criteria) also affect the prompt-diphoton cross section itself, in particular by reducing the effect of fragmentation.

The standard cone isolation and the ``smooth'' cone isolation proposed by Frixione \cite{Frixione:1998jh} are two of these criteria. The standard cone isolation is easily implemented in experiments, but it only suppresses a fraction of the fragmentation contribution. The smooth cone isolation (formally) eliminates the entire fragmentation contribution, but its experimental implementation (at least in its original form) is complicated\footnote{There is activity in the experimental implementation~\cite{Binoth:2010ra,Blair:1379880,Wielers:2001suas} of the discretized version of the smooth isolation criterion. An experimental implementation of the smooth isolation criterion was done by the OPAL collaboration~\cite{Abbiendi:2003kf}.} by the finite granularity of the LHC and Tevatron detectors.

On the theoretical side, including fragmentation contributions to photon production can greatly increase the complexity of the calculations, while the application of appropriate isolation cuts can effectively remove those fragmentation contributions. How to apply these isolation criteria to the theoretical tools and how to deal with the fragmentation contributions consistently by the theory side is one of the subjects of this note.

Besides the differences between these two isolation criteria, it is possible (\textit{i.e.} it has physical meaning) to compare theoretical descriptions obtained using the smooth cone isolation criterion and data taken with the standard criterion, because a cross section obtained using the smooth cone isolation criterion provides always a lower bound for a cross section in which the standard criterion was implemented, if one uses the same isolation parameters for both criteria. Furthermore, as we show in this note, this bound turns out to be an excellent approximation to the standard criterion result with an accuracy of the order of the $1 \%$ if tight cuts are imposed.

Given these results, and the fact that in general it is not possible to exactly match the experimental isolation conditions to the theoretical implementation and viceversa,  we propose a pragmatic accord to perform a more precise comparison between the data and the fixed order calculations, that allows to extend the TH computation up to NNLO in some cases.
 That is the major motivation for this note.

The following studies concern Monte Carlo integrators (as {\small \texttt{DIPHOX}}~\cite{hep-ph:9911340}, {\small \texttt{JetFOX}}~\cite{hep-ph:0204023} or {\small \texttt{2$\gamma$NNLO}~\cite{Catani:2011dp}}, etc) for which the fragmentation component is a purely collinear phenomenon. For Monte Carlo generators (parton--shower Monte Carlo), in which the fragmentation photons are emited off quarks at non--zero angle during the showering process, we recomend reference~\cite{AlcarazMaestre:2012vp}.

\subsection{Isolation criteria}
In this section we sumarize the standard and ``smooth'' isolation criteria, including their advantages and problems concerning the theoretical and experimental implementations.

\subsubsection{The standard cone isolation criterion}
The isolation criterion used by collider experiments is schematically as follows. A photon is said to be isolated if, in a cone of radius $R$ in rapidity and azimuthal angle around the photon direction, the amount of deposited hadronic transverse energy $\sum E_{T}^{had}$ is smaller than some value $E_{T \, max}$ chosen by the experiment:  
\begin{equation}\label{standardcriterion}     
\sum E_{T}^{had} \leq E_{T \, max} \;\;\;\; \mbox{inside} \;\;\;\;      
\left( y - y_{\gamma} \right)^{2} +    
\left(  \phi - \phi_{\gamma} \right)^{2}  \leq R^{2}  \;,    
\end{equation} 
where $E_{T \, max}$ can be either, a fixed value\footnote{This requirement was typically used at the Tevatron and was motivated by the fact that most of the energy in the isolation cone results from the underlying event (and pile-up), and so is independent of the photon energy~\cite{Binoth:2010ra}.} or a fraction of the transverse momentum of the photon ($p_T^{\gamma}\epsilon$, where typically $0 < \epsilon \leq 1$). This is the so-called standard cone isolation criterion. In addition to the rejection of the background of secondary  photons, these isolation cuts also affect the prompt-diphoton cross section itself, in particular by reducing the effect of fragmentation, but should be structured so as to have a high efficiency for the retention of real, isolated photons. 

A theoretical description of isolated photons is complicated because of the occurence of collinear singularities between photons and final-state quarks. A finite cross section is only obtained when these singularities are absorbed into the fragmentation functions. As a result the only theoretically well-defined quantity (if we don't use the smooth cone criterion) is the sum of the direct and fragmentation contributions. Once these two contributions are included one can isolate the photon using the cuts of Eq. (\ref{standardcriterion}) in an infrared safe way \cite{Catani:2002ny}.

In addition, a tight isolation cut also has the undesirable effect of making the theoretical prediction unstable~\cite{Catani:2002ny}, due to the restriction of the available phase-space for parton emission.  When the size of the cone used is in the limit of the narrow cone ($R\ll 1$, $R\sim 0.1$) earlier studies reveal potential problems. This leads to a collinear sensitivity in the form of a fairly large dependence on $\ln(1/R)$, which could make the prediction unreliable\footnote{This could even lead to an unphysical result such as an isolated cross section larger than the inclusive one, thereby violating
unitarity.}. In a recent calculation \cite{Catani:2013oma} these large logarithmic terms were ressumed restoring the reliability of the calculation.

\subsubsection{The ``smooth'' isolation criterion}
\label{Sec:Isol_Frix}
There exist an alternative to the standard criterion: the criterion proposed by Frixione in Ref.~\cite{Frixione:1998jh}
(see also Ref.~\cite{Frixione:1999gr,Catani:2000jh}). This criterion modifies Eq.~(\ref{standardcriterion}) in the following way
\begin{equation}\label{Eq:Isol_frixcriterion}     
\sum E_{T}^{had} \leq E_{T \, max}~\chi(r)\;, \;\;\;\; \mbox{inside any} \;\;\;\;      
r^{2}=\left( y - y_{\gamma} \right)^{2} +    
\left(  \phi - \phi_{\gamma} \right)^{2}  \leq R^{2}  \;.    
\end{equation}  
with a suitable choice for the function $\chi(r)$. This function has to vanish smoothly when its argument goes to zero ($\chi(r) \rightarrow 0 \;,\; \mbox{if} \;\; r \rightarrow 0\,$), and  has to verify \mbox{$\; 0<\chi(r)< 1$}, if \mbox{$0<r<R\,\,.$} One possible election is,
\begin{equation}
\label{Eq:Isol_chinormal}
\chi(r) = \left( \frac{1-\cos (r)}{1-\cos R} \right)^{n}\;,
\end{equation}
where $n$ is typically chosen as $n=1$ (unless otherwise stated we rely on this value for the phenomenlogical results presented in this note). This means that, closer to the photon, less hadronic activity is permitted inside the cone. At $r=0$, when the parton and the photon are exactly collinear, the energy deposited inside the cone is required to be exactly equal to zero, and the fragmentation component (which is a purely collinear phenomenon in perturbative QCD) vanishes completely. Since no region of the phase space is forbidden, the cancellation of soft gluon effects takes place as in ordinary infrared-safe cross sections. This is the advantage of this criterion: it eliminates all the fragmentation component in an infrared-safe way.

We can also notice, comparing Eqs.~(\ref{standardcriterion}) and (\ref{Eq:Isol_frixcriterion}), that both criteria coincide at the outer cone ($r=R$, $\chi(R)=1$), and due to the presence of the $\chi(r)$ function which verifies $0\leq\chi(r)\leq 1$, the smooth cone isolation criterion is more restrictive than the standard one. For this reason, we expect smaller cross sections when one uses the smooth cone criterion compared to the case when one implements the standard one if the same parameters are imposed,
\begin{equation}
\sigma_{smooth}\{R,E_{T~max}\}\leq \sigma_{Stand}\{R,E_{T~max}\}\,\,.
\end{equation}
The smooth behaviour of the $\chi(r)$ function is the main obstacle to implement the new isolation criterion into the experimental situation. First, because of the finite size of the calorimeter cells used to measure the electromagnetic shower, the smooth cone criterion must be applied only beyond a minimum distance of approximately $0.1$ (in $\{ \Delta \eta, \Delta \phi \}$ plane). This allows a contribution from fragmentation in the innermost cone and one has to check to which extent the fragmentation component is still suppressed. In addition, the transverse energy in the experimental isolation cone is deposited in discrete cells of finite size. Thus concerning its experimental implementation, the continuity criterion, initially proposed by Frixione has to be replaced by a discretized version consisting of a finite number of nested cones, together with the collection of corresponding maximal values for the transverse energy allowed inside each of these cones.

As was previously noted, another interesting (theoretical) feature of the smooth cone isolation criterion is that it is free of the large logarithmic contributions of the type of $\ln(1/R)$ associated with the narrow cone. 


\subsubsection{Theoretical issues}
As it was previously stated, the inclusion of the fragmentation contributions complicates the calculation. And in some cases, such as when one wants to reach NNLO accuracy in pQCD (like in diphoton production) it would make the calculation very unlikely for at least quite a few years since the machinary for such a computation is not available yet.

For those cases, one can find in the literature theoretical calculations in which the fragmentation component is considered at one pertubative level less than the direct component (\textit{e.g.} the $\gamma \gamma$ and $W/Z \gamma$ production at NLO in pQCD in {\small \texttt{MCFM}} \cite{Campbell:2011bn} or $\gamma \gamma +$ Jet at NLO in pQCD~\cite{Gehrmann:2013aga}, etc.). This procedure, which is a way to approximate the full calculation where the fragmentation component is included at the same perturbative level that the direct component, could introduce unexpected (inconsistent) results in the presence of the standard cone isolation criterion. The following  exercise shows how these problems can easely appear.

Let's consider diphoton production at the LHC ($\sqrt{s}=8$~TeV) computed at NLO, the highest perturbative order at which both direct and fragmentation calculations are available. The acceptance criteria in this case requires: $p_T^{\rm harder}\geq
40$~GeV and $p_T^{\rm softer}\geq 30$~GeV. The rapidity of both photons is restricted to \mbox{$|y_\gamma| \leq 2.5$} and $100$~GeV$<M_{\gamma \gamma}<160$~GeV. The isolation parameters are set to the values $n=1$ (in the case of the smooth cone) and $R=0.4$, and the minimum angular separation between the two photons is $R_{\gamma\gamma}=0.5$. The remain isolation parameter $E_{T~max}$ (or $\epsilon$) is varied in order to understand the cross section dependence on it. All the cross sections are obtained using the {\small \textsf{CTEQ6M}} set of parton distributions functions.
We used for this analysis the {\small \texttt{DIPHOX}} code that includes the full NLO pQCD description. 

\begin{figure}
\begin{center}
\includegraphics[width=0.48\textwidth,height=6.8cm]{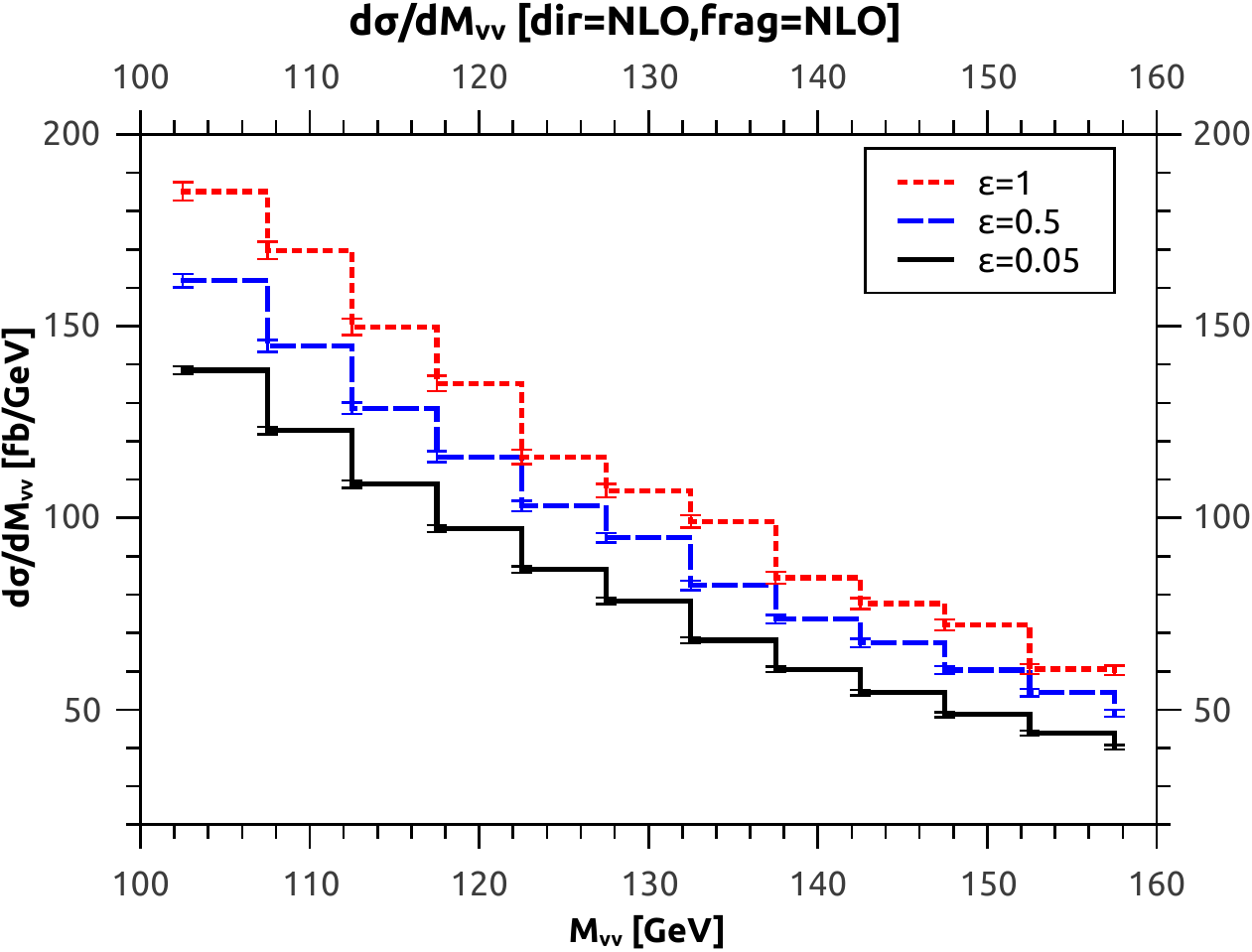}
\includegraphics[width=0.48\textwidth,height=6.8cm]{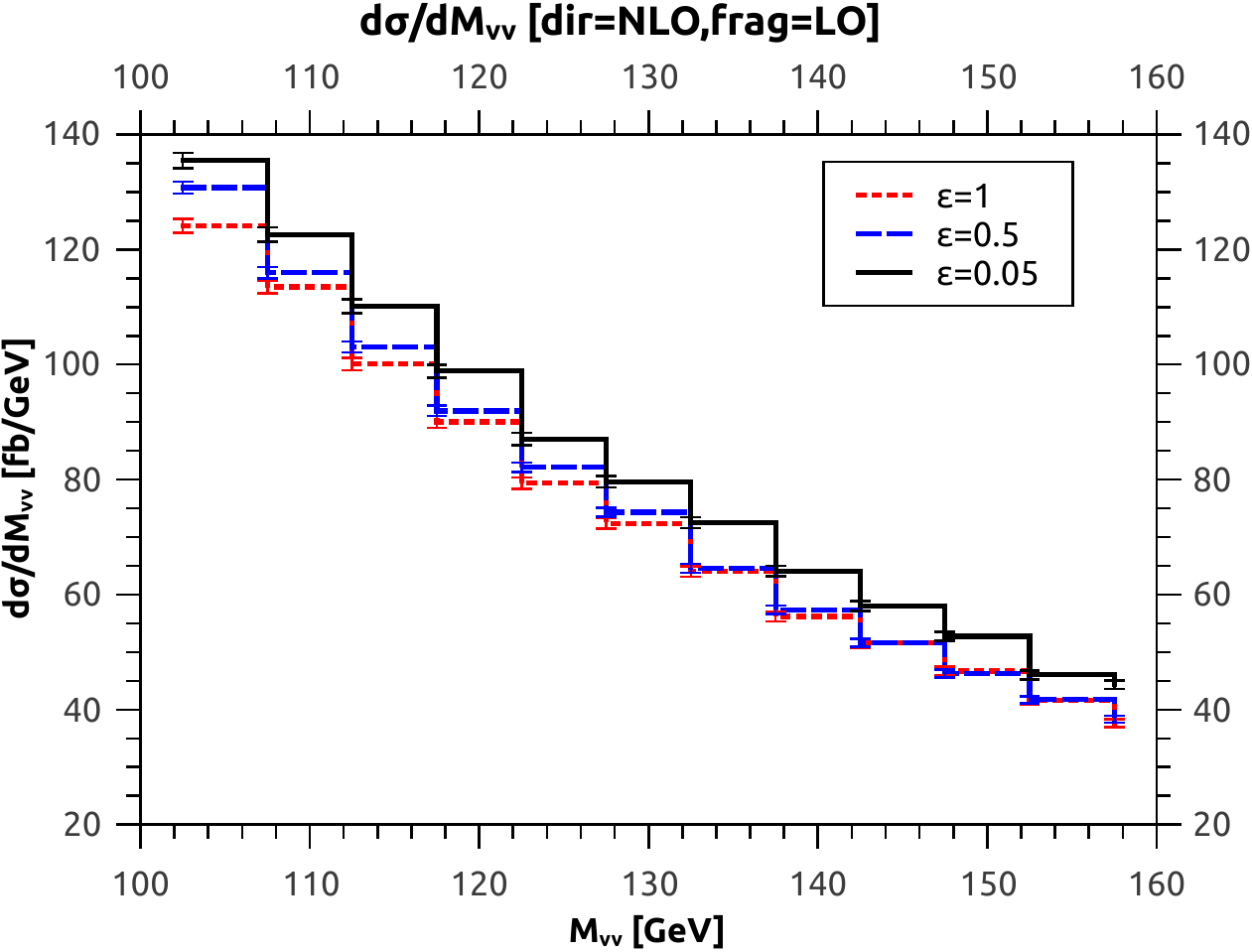}
 \caption{Diphoton cross section as a function of the invariant mass $M_{\gamma\gamma}$. (Left) With NLO fragmentation component. (Right) With LO fragmentation component.}
\label{Fig:Isol_1}
\end{center}
\end{figure}

First,  we compare the calculation using the fragmentation at NLO with the case in which the fragmentation is considered only at LO (one pertubative order less than the direct NLO component). The results in Fig. \ref{Fig:Isol_1} (right) show that for the LO fragmentation one can obtain 
larger cross sections for more severe isolation cuts (smaller $\epsilon$), which is clearly inconsistent. 
The same behaviour was reported in Ref.~\cite{Gehrmann:2013aga} for $\gamma \gamma +$ Jet at NLO, and we obtained similar features for $\gamma \gamma$, $W/Z \gamma$ production at NLO in pQCD with {\small \texttt{MCFM}}. 
On the other hand, the left panel in Fig. \ref{Fig:Isol_1} shows that the  correct behaviour is found when one considers the full result in which the fragmentation contribution is computed at same perturbative level than the direct component (i.e., NLO in this case).

The precedent comparison suggests that one has to be aware that approximating the fragmentation component at one order lower than the direct one can result in unphysical results. The situation can be even more serious when one looks at some extreme kinematical region where the cross section is dominated by higher order contributions. 

In Table \ref{Tab:Isol_CMSNLO} we present the results for the corresponding cross section with different isolation prescriptions and parameters. The values presented there  help to understand the unexpected behaviour in Fig. \ref{Fig:Isol_1} (right). In this case, the results are obtained imposing the cuts used by CMS in a recent measurement of the production cross section for pairs of isolated photons in pp collisions at $\sqrt{s}=7$ TeV~\cite{Chatrchyan:2011qt}. Therefore, we require the harder photon to have a transverse momentum $p_T^{hard}\geq 40$ GeV while for the softer we choose $p_T^{soft}\geq 30$ GeV.
The rapidity of both photons is restricted to $|y_\gamma| \leq 2.5$. Finally, we constrain the invariant mass of the diphotons to lie in the range $100 \,{\rm GeV}\leq M_{\gamma\gamma} \leq 160\,{\rm GeV}$. And to simulate the CMS crack, we reject photons with  $1.442 \leq y_\gamma \leq 1.556$. In the cases in which the standard criterion was applied ($a-f$ in Table \ref{Tab:Isol_CMSNLO}) we observe that as the isolation criterion turns out to be ``loose'' the NLO direct component becomes smaller and the NLO fragmentation (single and double) component larger. The sum of them behaves as expected with respect to the isolation parameters, since the increase in the fragmentation component overcompensates the decrease of the direct  one. We remind the reader that in the standard isolation the theoretical separation between direct and fragmentation components is not physical and the results presented in this note correspond to the conventional $\overline{MS}$ subtraction.
On the other hand, if only the LO calculation is used for the fragmentation contributions, for which the QCD corrections are quite large (with $K-$factors exceding 2), the mismatch in the perturbative order spoils the compensation between the behaviour of the NLO direct and the LO fragmentation terms resulting in the unphysical behaviour observed in Fig. \ref{Fig:Isol_1} (right) as one considers less restringent isolation parameters.

Furthermore, from the cases in which the smooth cone criterion was applied ($h-m$ in Table \ref{Tab:Isol_CMSNLO}) we observed that, as expected, the result in the smooth cone case always provides a lower bound for the one obtained with the standard criterion when the same isolation parameters (energy in this case) are used. In the case of smooth cone isolation the (single and double) fragmentation components are identically null.
Even more interestingly, and with one single exception, the results for the NLO cross sections computed using different isolation precriptions differ by less than $1\%$. This result indicates that using the smooth cone prescription for a theoretical calculation (even when the data is analyzed using the standard one) provides an approximation that it is is far much better than the one consisting in the standard prescription with a lowest order calculation for the fragmentation component. 
The only case where one can observe larger differences (of the order of $10\%$) corresponds to the use a very loose isolation, as  for $\sum E_{T}^{had} \leq 0.5~p_{T}^{\gamma}$, where the fragmentation component in the standard case amounts more than half of the total cross section. 

\begin{table}[ht]
\begin{center}
\resizebox{14cm}{!} {
\begin{tabular}{|c|c|c|c|c|c|c|c|c|}
\hline
&Code & $\sum E_{T}^{had} \leq $& $\sigma_{total}^{NLO}$(fb)& $\sigma_{dir}^{NLO}$(fb) & $\sigma_{onef}^{NLO}$(fb)& $\sigma_{twof}^{NLO}$(fb)&Isolation\\
\hline
\hline
a&\texttt{DIPHOX}& $2$ GeV & $3756$ & $3514$& $239$& $2.6$&Standard\\
\hline
b&\texttt{DIPHOX}& $3$ GeV & $3776$ & $3396$& $374$& $6$&Standard\\
\hline
c&\texttt{DIPHOX}& $4$ GeV & $3796$ & $3296$& $488$& $12$&Standard\\
\hline
d&\texttt{DIPHOX}& $5$ GeV & $3825$ & $3201$& $607$& $17$&Standard\\
\hline
e&\texttt{DIPHOX}& $0.05~p_{T}^{\gamma}$ & $3770$ & $3446$& $320$& $4$&Standard\\
\hline
f&\texttt{DIPHOX}& $0.5~p_{T}^{\gamma}$ & $4474$ & $2144$& $2104$& $226$&Standard\\
\hline
g&\texttt{DIPHOX}& \textit{ incl} & $6584$ & $1186$& $3930$& $1468$&none\\
\hline
h&\texttt{2$\gamma$NNLO}& $0.05~p_{T}^{\gamma}~\chi(r)$ & $3768$ & $3768$& $0$& $0$&Smooth\\
\hline
i&\texttt{2$\gamma$NNLO}& $0.5~p_{T}^{\gamma}~\chi(r)$ & $4074$ & $4074$& $0$& $0$&Smooth\\
\hline
j&\texttt{2$\gamma$NNLO}& $2$ GeV~$\chi(r)$ & $3754$ & $3754$& $0$& $0$&Smooth\\
\hline
k&\texttt{2$\gamma$NNLO}& $3$ GeV~$\chi(r)$ & $3776$ & $3776$& $0$& $0$&Smooth\\
\hline
l&\texttt{2$\gamma$NNLO}& $4$ GeV~$\chi(r)$ & $3795$ & $3795$& $0$& $0$&Smooth\\
\hline
m&\texttt{2$\gamma$NNLO}& $5$ GeV~$\chi(r)$ & $3814$ & $3814$& $0$& $0$&Smooth\\
\hline
\hline

\end{tabular}

}
\end{center}
\caption{{\em Cross sections for the $pp\to \gamma\gamma+X$ process at the LHC at NLO.  All these values are at $1\%$ of statistical accuracy level.}}
\label{Tab:Isol_CMSNLO}
\end{table}

In all cases we have studied, the smooth cone provides an excellent approximation to the standard result as long as the isolation parameters are tight enough, i.e.  $\sum E_{T}^{had} \leq 0.1~p_{T}^{\gamma}$ or $\sum E_{T}^{had} \leq 5$ GeV for the LHC at 7 TeV. Equivalently, one could define the isolation to be tight enough when the contribution from the fragmentation component does not exceeds $\sim 15-20 \%$ of the total cross section.

\begin{figure}
\begin{center}
\includegraphics[width=0.48\textwidth,height=6.8cm]{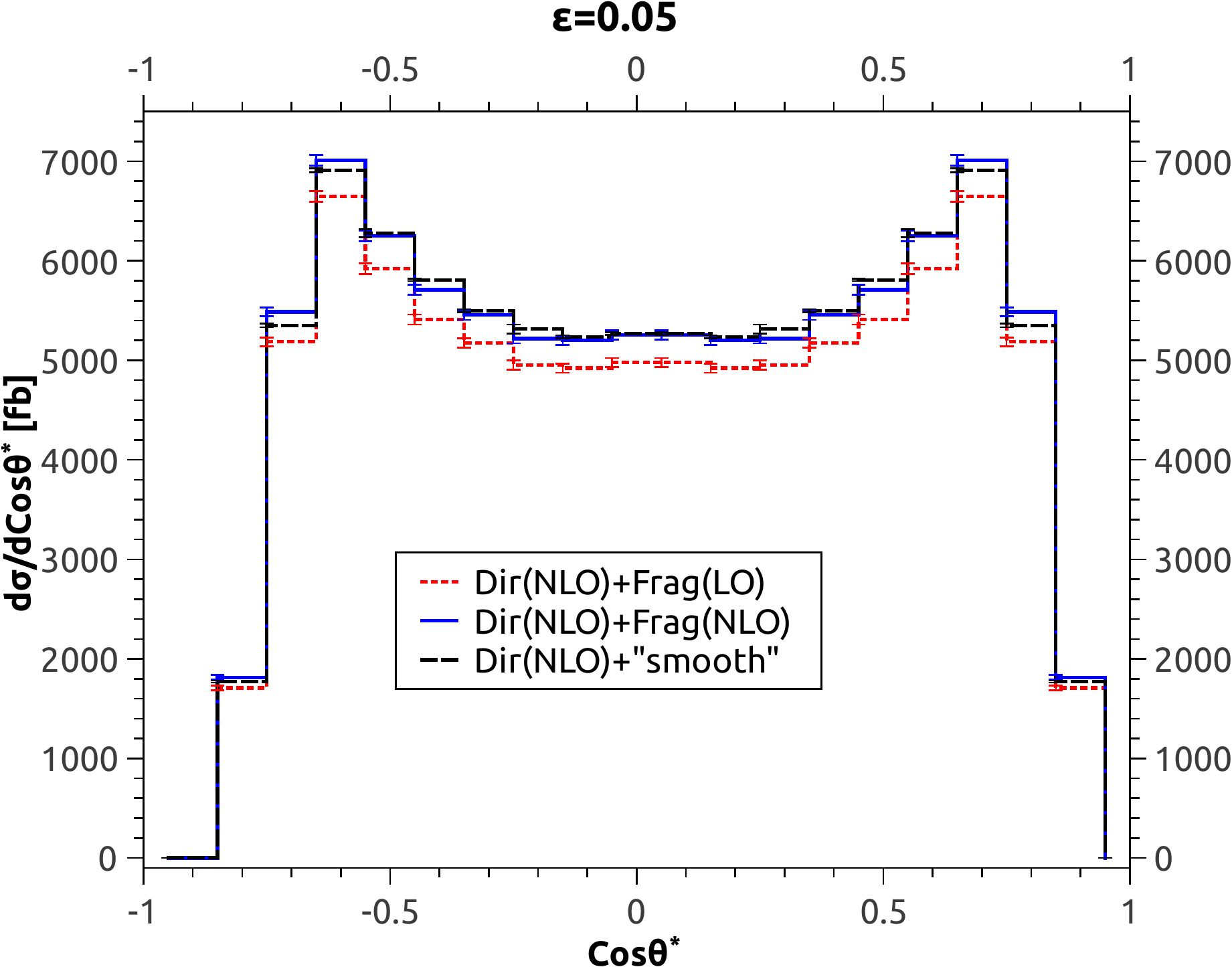}
\includegraphics[width=0.51\textwidth,height=7.0cm]{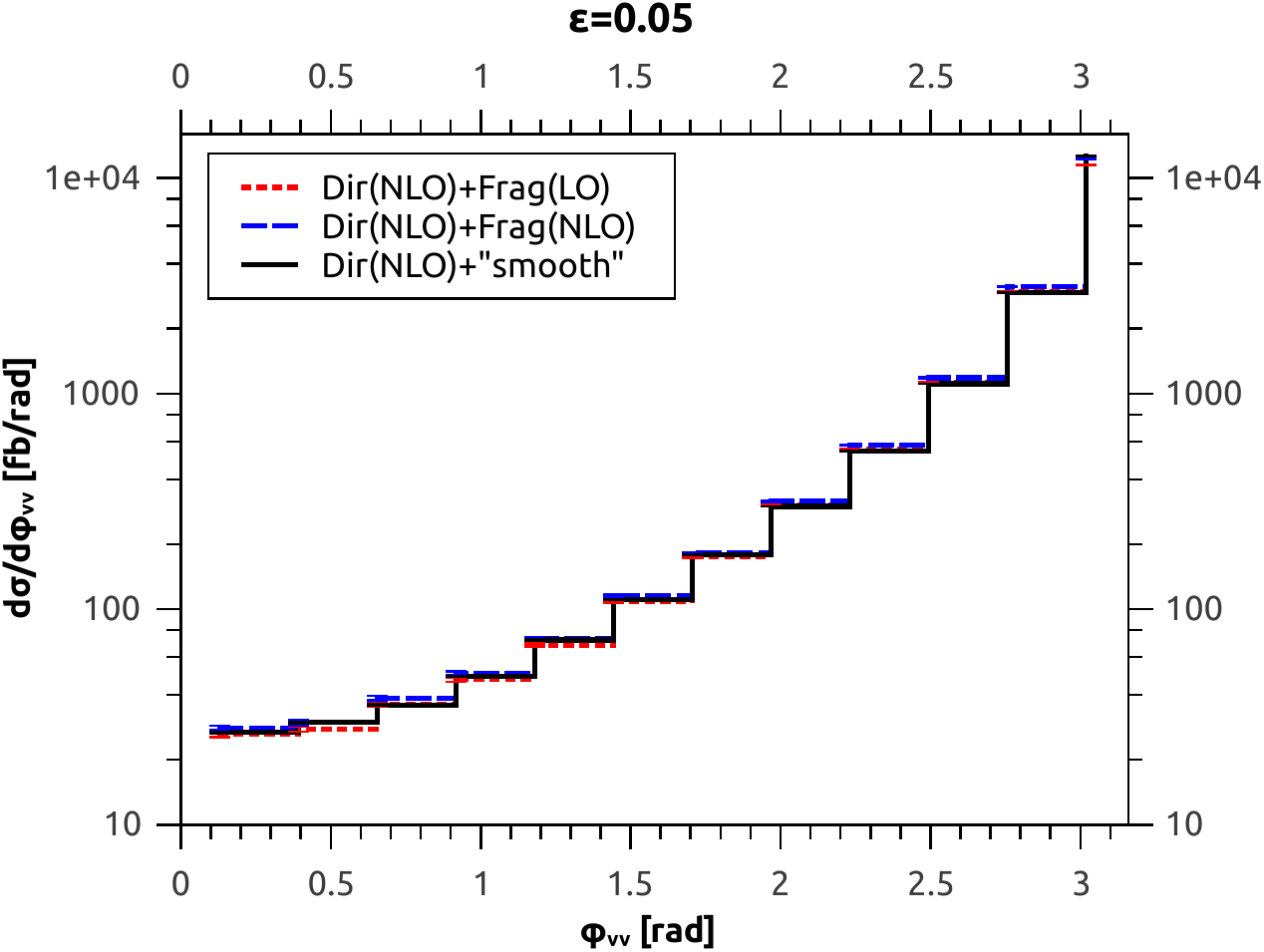}
 \caption{Diphoton cross section as a function of the $\cos \theta^{*}$ (Left), and  the angular separation between the photons $\Delta \phi_{\gamma \gamma}$ (Right). A comparison between the different isolation criteria is showed. The applied cuts are the same as in Fig. \ref{Fig:Isol_1}, and are described in the text.}
\label{Fig:Isol_2}
\end{center}
\end{figure}

\begin{figure}
\begin{center}
\includegraphics[width=0.48\textwidth,height=6.8cm]{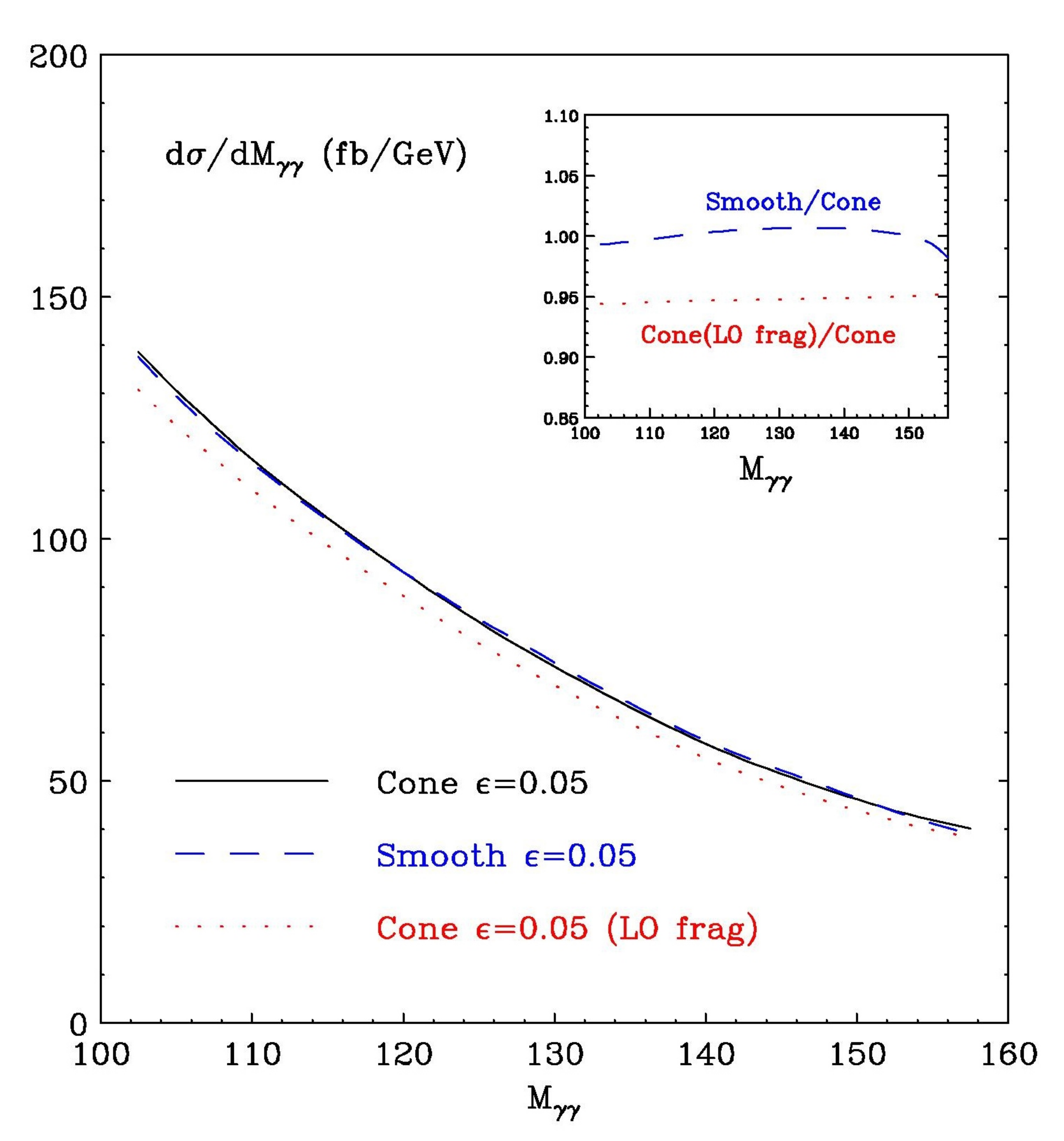}
\includegraphics[width=0.48\textwidth,height=7.0cm]{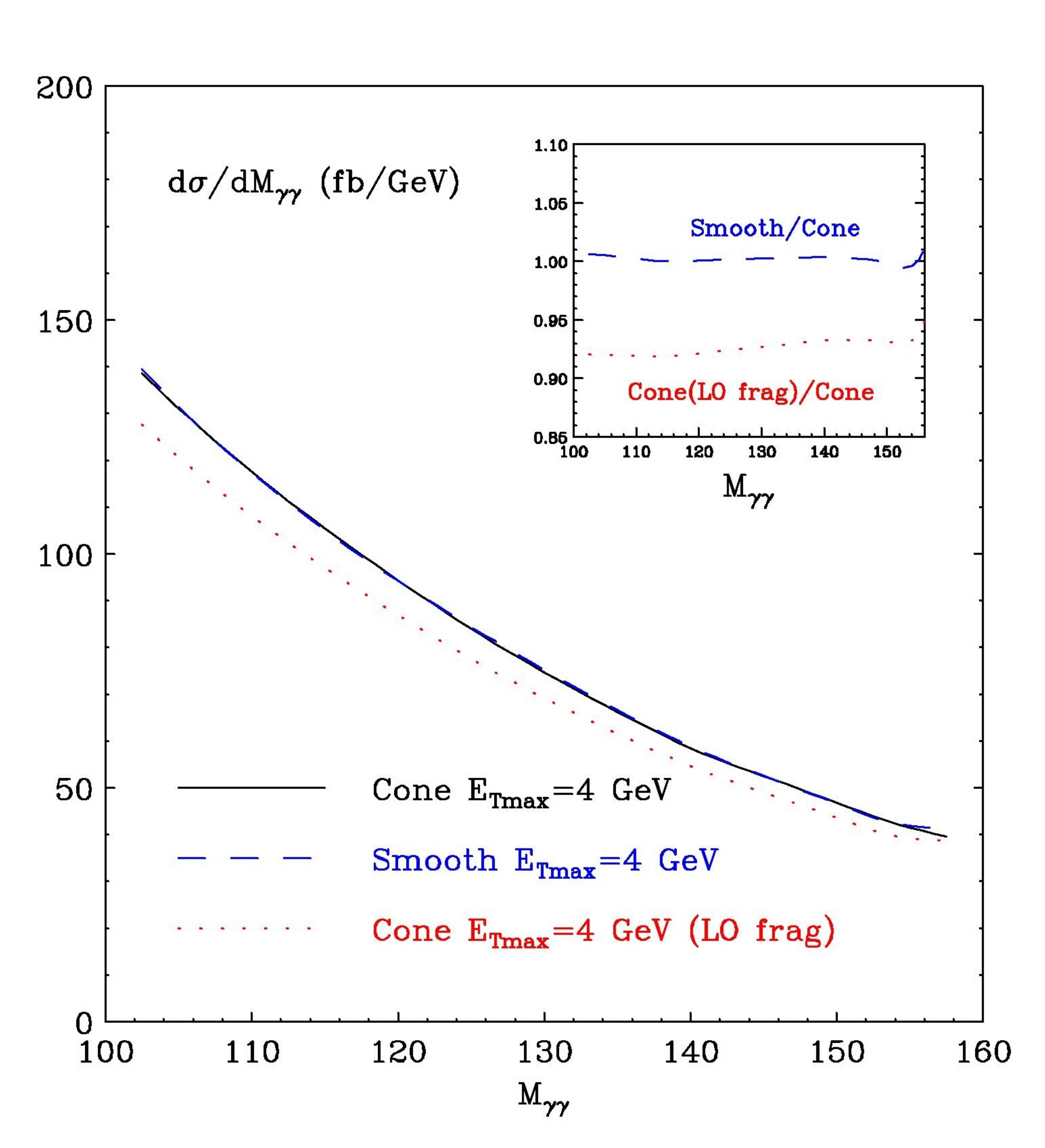}
 \caption{Diphoton cross section as a function of the invariant mass $M_{\gamma\gamma}$.  Cross sections obtained with the standard cone isolation criterion (with LO and NLO fragmentation contributions) are compared with the cross section obtained with the smooth cone criterion using the same isolation parameters.
}
\label{Fig:Isol_compToT}
\end{center}
\end{figure}

While the previous analysis refers only to the fiducial crosss section, it is known that the fragmentation contributions could be larger in kinematical regions far away from the back-to-back configuration\footnote{The low mass region in the invariant mass distribution, the low $\Delta \phi_{\gamma\gamma}$ distribution and the kinematical regions near to $\cos \theta^{*}=\pm 1$ belong to this case.}, and the approximation could in principle become less accurate for those distributions. In order to check that feature, in Fig. \ref{Fig:Isol_compToT} we compare the distributions for the full NLO calculation with the standard prescription, the one obtained using only the LO fragmentation component and the result for the smooth cone with $\sum E_{T}^{had} \leq 0.05~p_{T}^{\gamma}$  for $\cos \theta^{*}$ (left) and $\Delta \phi_{\gamma\gamma}$ (right). In both cases we observe that for all the bins the smooth cone provides the best approximation to the full result, always within a $2.5 \%$ accuracy. A more detailed analysis in presented in Fig. \ref{Fig:Isol_compToT} for the diphoton invariant mass distributions with two different isolation parameters, $\sum E_{T}^{had} \leq 0.05~p_{T}^{\gamma}$ (left) and $\sum E_{T}^{had} \leq 4$ GeV (right). Again, while using the LO fragmentation component fails to reproduce the full NLO result by up to $6\%$, the smooth cone approximation is always better than $1.5\%$ in the same kinematical region.

The discrepancies between the full result and the smooth approach with the LO fragmentation approximation that evidently manifest in the invariant mass distribution (see Fig. \ref{Fig:Isol_compToT}) or in kinematical regions far away from $\cos \theta^{*}=\pm 1$ (see Fig. \ref{Fig:Isol_2} (left)) are ``hidden'' in the $\Delta \phi_{\gamma\gamma}$ distribution, in the bin corresponding to $\Delta \phi_{\gamma\gamma}=\pi$ (the bin containing the back-to-back configurations). Moreover in the low $\Delta \phi_{\gamma\gamma}$ region   we are dealing with events far away from the back-to-back configuration, and the only configuration that survives at NLO (in these kinematical regions) is the real emission at LO which is for the three cases effectively the same contribution under these conditions.

\begin{figure}
\begin{center}
\includegraphics[width=0.50\textwidth,height=6.8cm]{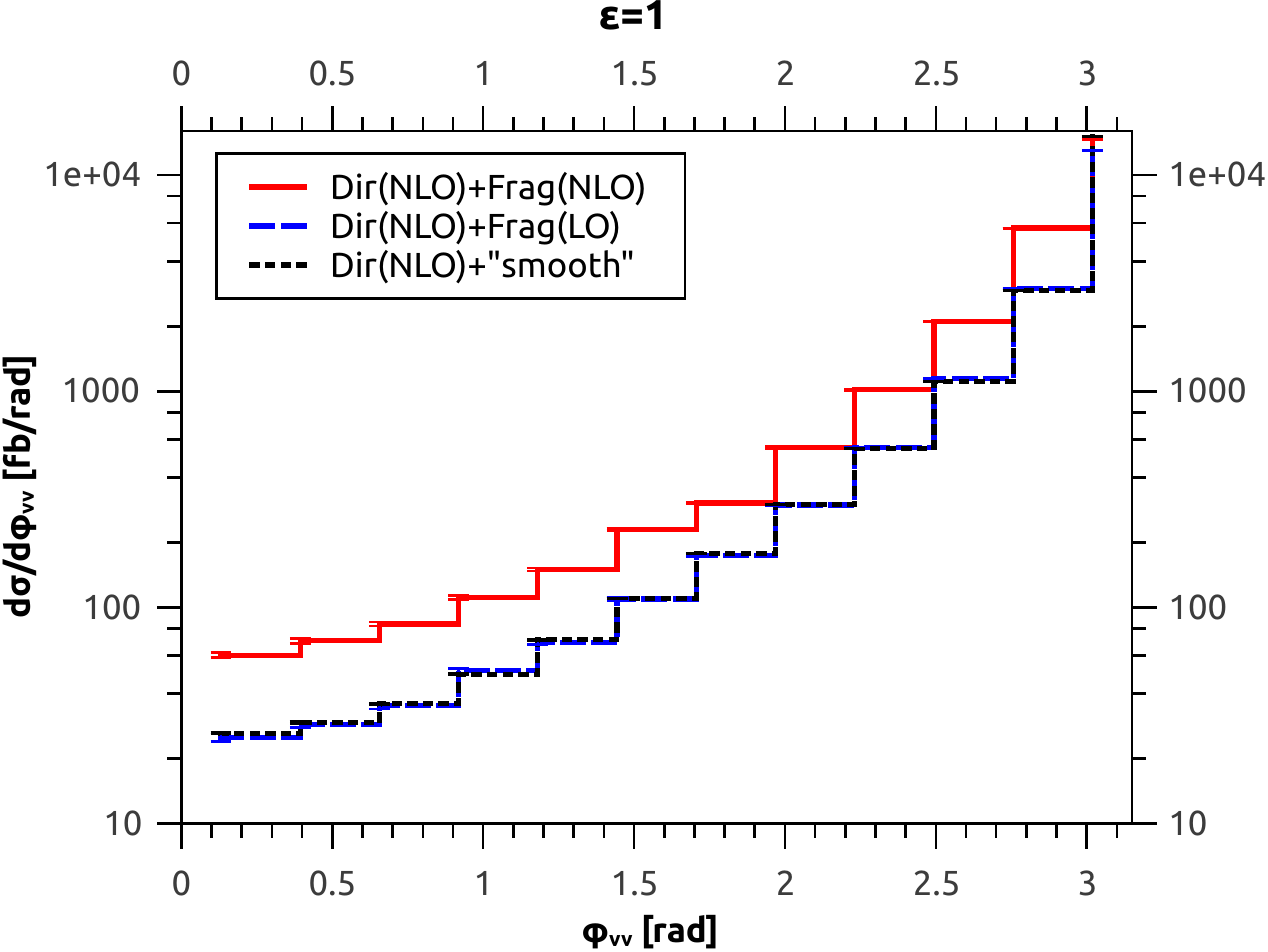}
\includegraphics[width=0.46\textwidth,height=6.8cm]{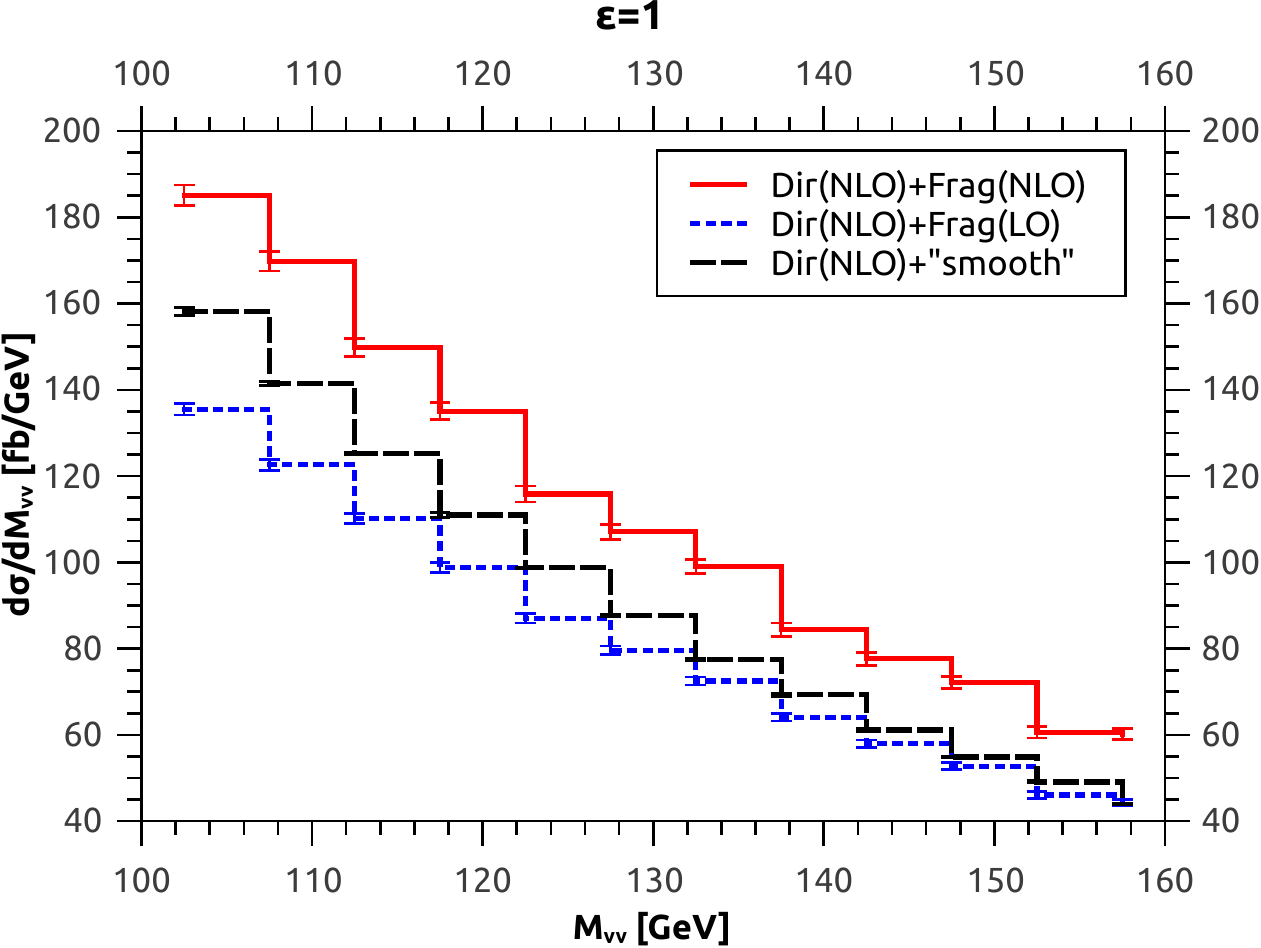}
 \caption{Diphoton cross section as a function of the angular separation between the photons $\Delta \phi_{\gamma\gamma}$ (Left), and  the invariant mass of the diphoton system $M_{\gamma\gamma}$ (Right).}
\label{Fig:Isol_3}
\end{center}
\end{figure}

If one relaxes the isolation and considers $\epsilon =1$ (which is equivalent to $E_{T}^{max}>40$~GeV, allowing for a huge amount of fragmentation contribution), the effects of fragmentation now  strongly manifests at low $\Delta \phi_{\gamma\gamma}$ values, and the full result considerably differs from both the LO fragmentation approximation and the smooth cone criterion as we can observe in Fig. \ref{Fig:Isol_3} (azimuthal distribution in the upper plot and invariant mass in the lower one). Also we can see from Fig. \ref{Fig:Isol_3}, in the bin corresponding to $\Delta \phi_{\gamma\gamma}=\pi$ (the bin containing the back-to-back configurations) the cross section obtained with the smooth cone criterion provides a better approximation  than the LO fragmentation one.

\subsection{Tight isolation accord}
Finally, considering the results presented in the preceeding sections we conclude this note by proposing a pragmatic accord in order to compare experimental data and theoretical calculations obtained at the highest possible perturbative order. Given the fact that matching experimental conditions to theoretical calculations always implies certain degree of approximation, we believe that considering the large QCD corrections to processes involving photons (with NNLO essential to understand diphoton data~\cite{Catani:2011dp}) and the agreement (tipically at the $\%$ level for the diphoton case studied here) between the standard and smooth cone TH calculations, the use of the later for TH purposes is well justified.

We call this approach "pragmatic", in the sense that we do not recommend the experiments to implement the smooth cone isolation, but to proceed to the analysis of the data with the usual standard isolation with cuts tight enough if the interesting observable needs to be an isolated cross section or distribution.
While the definition of "tight enough" might slightly depend on the particular observable (that can always be checked by a lowest order calculation), our analysis shows that at the LHC isolation parameters as $E_{T}^{max}\leq 5$~GeV (or $\epsilon<0.1$), $R\sim 0.4$ and $R_{\gamma\gamma}\sim 0.4$ are safe enough to proceeed.

This procedure would allow to extend available NLO calculations to one order higher (NNLO) for a number of observables, since the direct component is always much simpler to evaluate than the fragmentation part, which identically vanishes under the smooth cone isolation.

We also refer to this approach as pragmatic in a numerical sense: we are certain that the smooth cone isolation applied for the TH calculation is NOT the one used in the experimental data, but considering that NNLO corrections are of the order of $50\%$ for diphoton cross sections~\cite{Catani:2011dp} and a few $100\%$ for some distributions in extreme kinematical configurations, it is far better accepting a few $\%$ error arising from the isolation (less than the size of the expected NNNLO corrections and within any estimate of TH uncertainties!) than neglecting those huge QCD effects towards some "more pure implementation" of the isolation prescription.

We believe that a more detailed analysis of the profile function in the smooth cone isolation can be performed on a case by case basis in order to select, also in a pragmatic way, the most convenient for each observable, even though again, differences are expected to be very small as discussed before.

\subsection*{Acknowledgements}
This work was supported in part by UBACYT, CONICET, ANPCyT, INFN and the Research Executive Agency (REA) of the European Union under the Grant Agree- ment number PITN-GA-2010-264564 (LHCPhenoNet, Initial Training Network).





\section{Diphotons and jets at NLO\footnote{N.~Chanon, T.~Gehrmann, N.~Greiner, G.~Heinrich}}


We study diphoton production in association with up to three jets 
at the LHC. In particular, we compare NLO predictions for up to two jets with 
leading order predictions matched to a parton shower. 
For the one-jet bin, we also include fragmentation contributions 
in the partonic calculation, which enables us to study the impact of 
different isolation criteria.

\subsection{Introduction}

Studies of the Higgs boson decay channel into two photons are of major importance 
in order to scrutinize the Higgs couplings
and to be able to judge whether small deviations 
from the Standard Model predictions are hints of new physics.

Measurements of the inclusive diphoton cross section 
by both ATLAS~\cite{Aad:2012tba} and CMS~\cite{Chatrchyan:2011qt} 
based on the $\sqrt{s}=7$ TeV data set have been published recently. 
The production of photon pairs in association with jets 
are also very interesting processes, despite their smaller cross sections, 
as they constitute the main background to Higgs production in association 
with jets, where the higgs boson decays into two photons. 
In particular, the knowledge of diphoton production in association with two jets 
at next-to-leading order (NLO) accuracy allows to study the impact of 
vector boson fusion (VBF) cuts  and thus helps to disentangle 
Higgs production in gluon fusion from the electroweak production meachnism.

Large-$p_T$ photons in the final state 
can originate either from the hard interaction process 
itself or from the fragmentation of a large-$p_T$ hadron.  
To single out the photons originating from the hard interaction 
from the secondary photons, 
photon isolation criteria need to be applied, 
which are typically formulated in the form of a maximum amount of hadronic energy  
allowed in the vicinity of the photon. 
The photons originating from fragmentation are, in a partonic calculation, described by 
fragmentation functions, which -- in analogy to PDFs in the initial state --
are non-perturbative 
objects that have to be determined from experimental data.
To suppress the dependence of isolated photon cross sections on the
fragmentation functions, a smooth
cone isolation criterion has been proposed~\cite{Frixione:1998hn}, where 
the allowed hadronic energy inside the isolation cone decreases with decreasing 
radial distance from the photon.

While diphoton production without extra jets has been calculated 
at NLO already some time ago, including also the fragmentation contribution at 
NLO in the public code {\sc Diphox}~\cite{Binoth:1999qq}, 
and even the NNLO corrections (without fragmentation contributions) 
have been calculated meanwhile~\cite{Catani:2011qz}, 
diphotons in association with jets at NLO have become available 
only recently. 
Photon pair plus one jet production at NLO has first been calculated 
in \cite{DelDuca:2003uz}, using an isolation criterion~\cite{Frixione:1998hn}
where the fragmentation component is eliminated.
In Ref.~\cite{Gehrmann:2013aga}, the fragmentation component has been included, 
allowing to study the impact of different isolation criteria at NLO, 
and the corresponding code is publicly available~\cite{gosamHepforge}.
The virtual corrections are based on  the automated one-loop program
 \gosam~\cite{Cullen:2011ac}, while the real corrections  are built on 
 the {\sc MadGraph/MadDipole/MadEvent}~\cite{Stelzer:1994ta,Alwall:2007st,Frederix:2008hu,Frederix:2010cj} 
 framework.  

Photon pair plus two jet production at NLO has first been calculated 
in \cite{Gehrmann:2013bga}, also based on \gosam{}+{\sc MadGraph/MadDipole/MadEvent}, 
using a smooth isolation criterion.
NLO results for $\gamma\gamma+2$\,jet production also have been presented in 
\cite{Bern:2013bha, Badger:2013ava,Bern:2014vza}, and Ref.~\cite{Badger:2013ava}
in addition presents results for $\gamma\gamma+3$\,jet production at NLO.

\subsection{Diphoton plus one jet production}

\begin{figure}[htb]
\begin{center}
\begin{minipage}[t]{0.49\textwidth}
\includegraphics[width=0.9\textwidth]{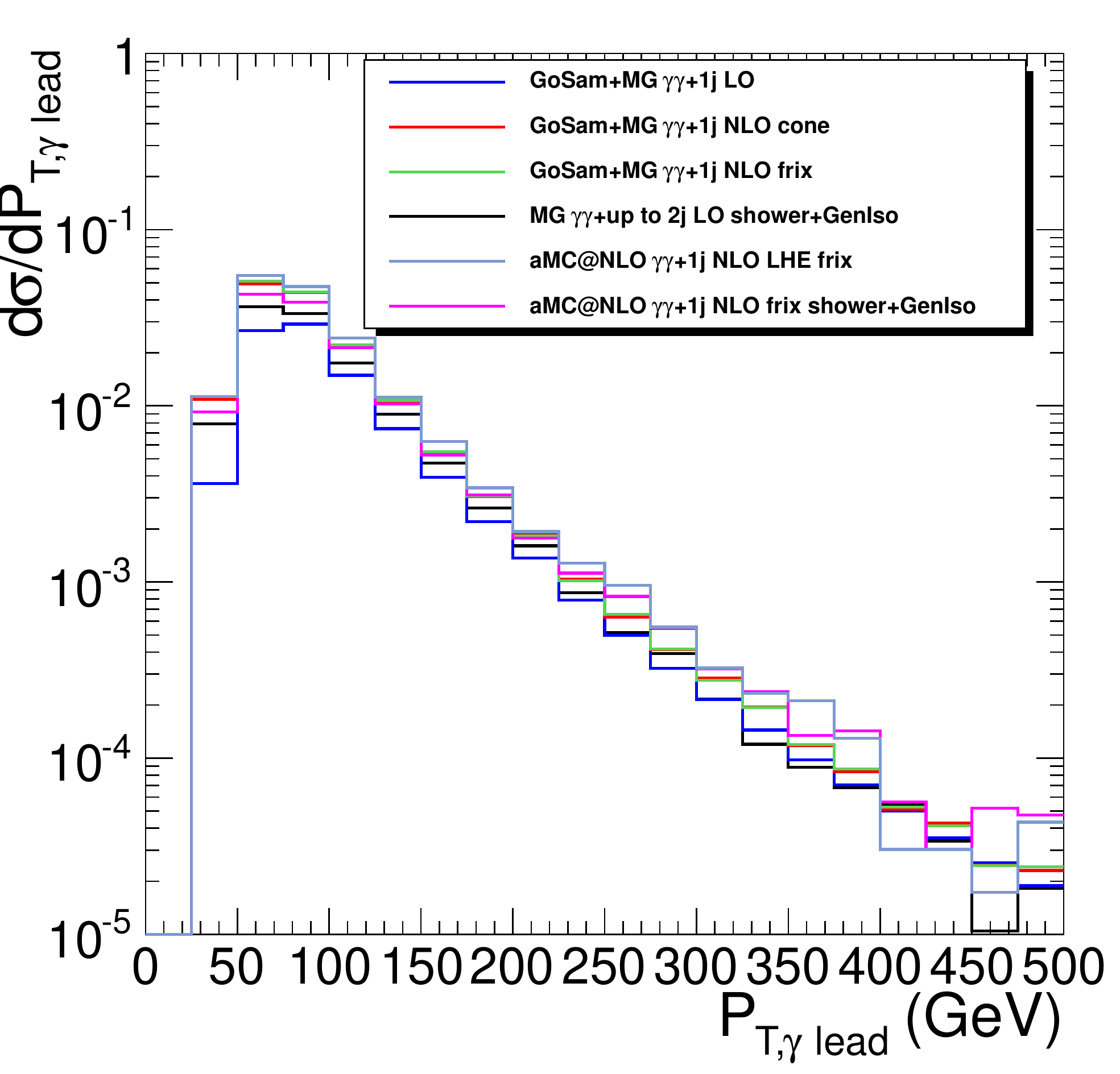}
\end{minipage}
\begin{minipage}[t]{0.49\textwidth}
\includegraphics[width=0.9\textwidth]{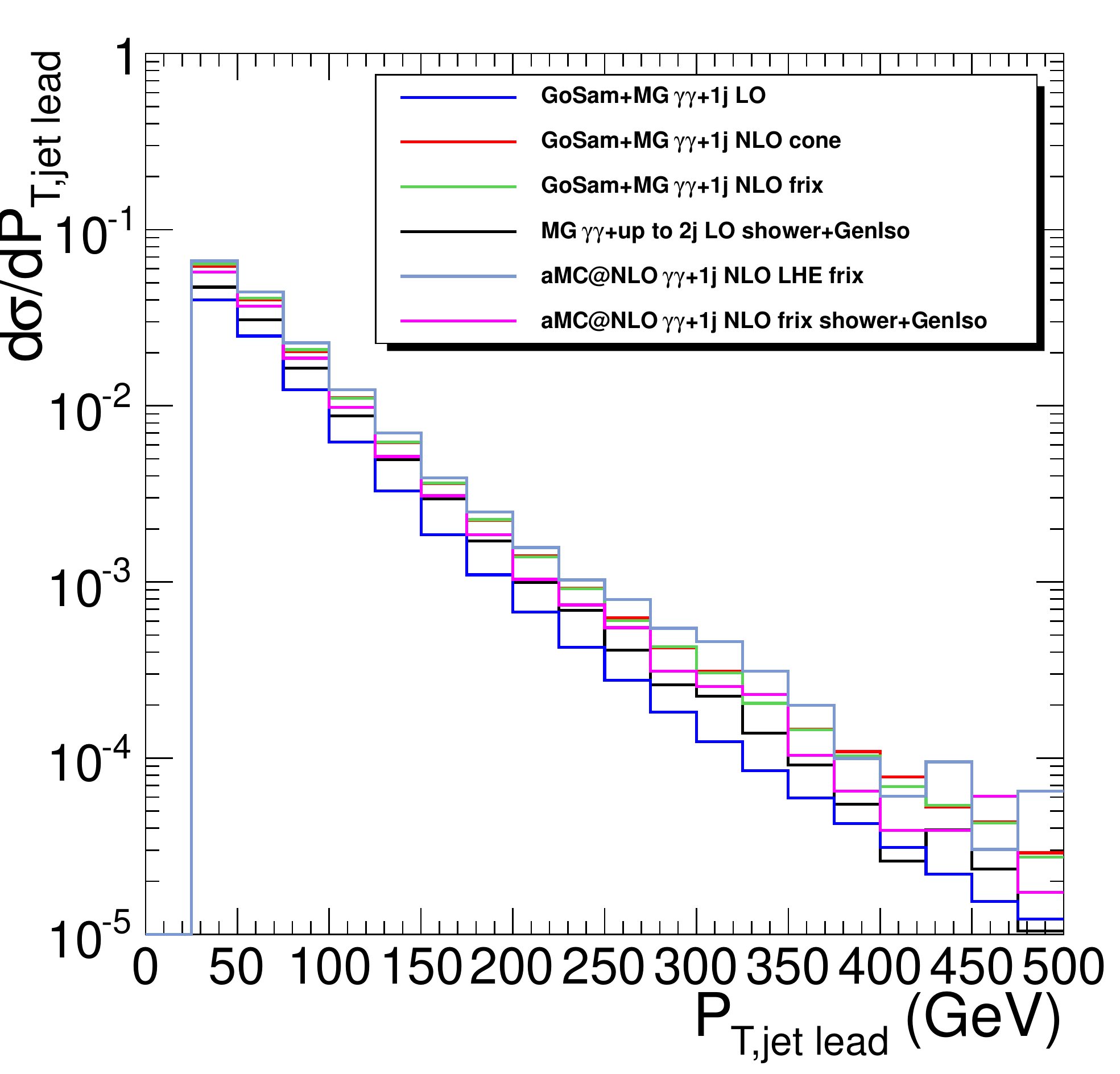}
\end{minipage}
 \caption{Transverse momentum distribution of the leading-$p_T$ photon for the $\gamma\gamma$+1\,jet process.}
\label{fig:ptgamma1j}
\end{center}
\end{figure}

\begin{figure}[htb]
\centering
\begin{minipage}[t]{0.49\textwidth}
\includegraphics[width=0.9\textwidth]{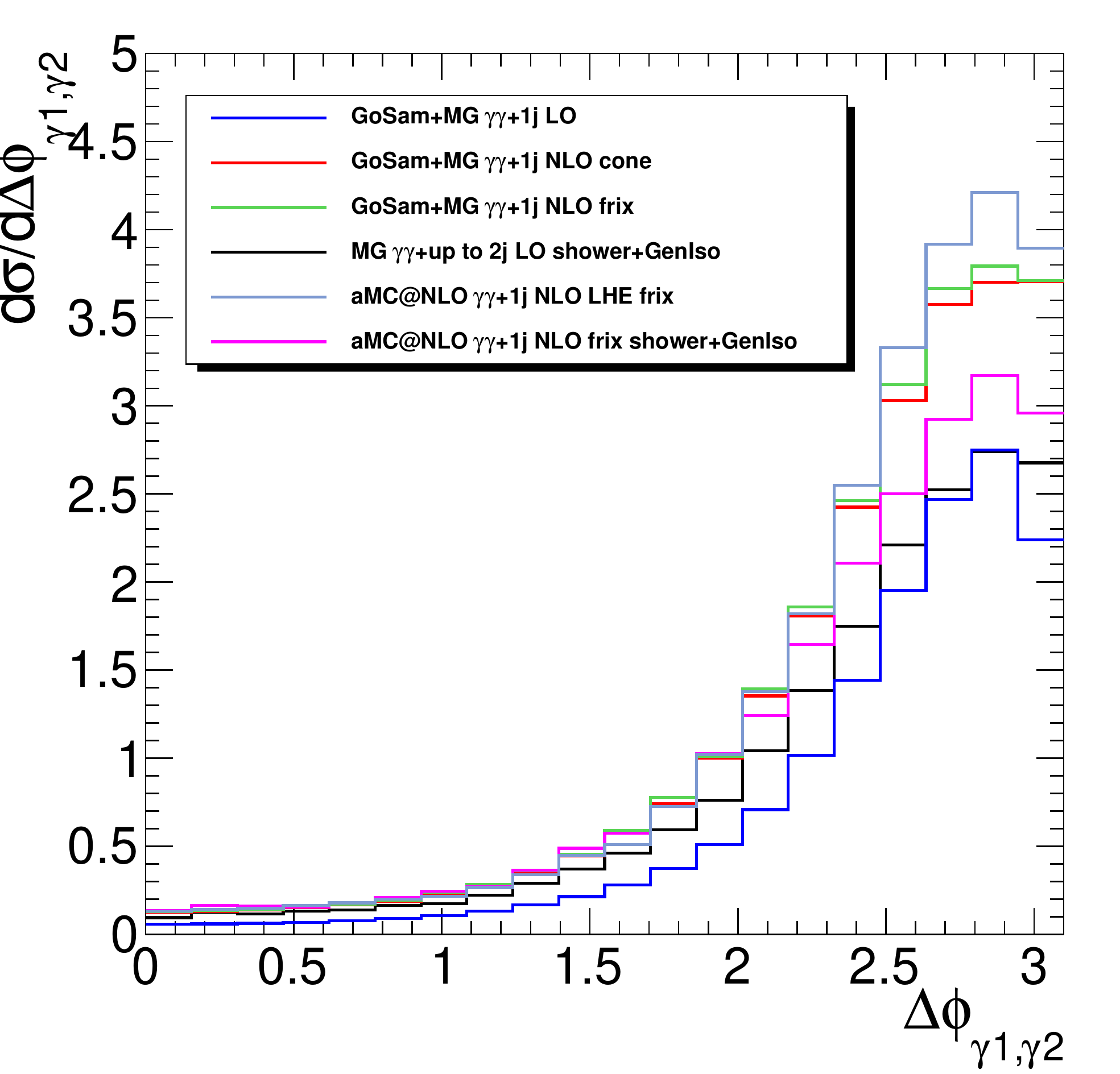}
\end{minipage}
\begin{minipage}[t]{0.49\textwidth}
\includegraphics[width=0.9\textwidth]{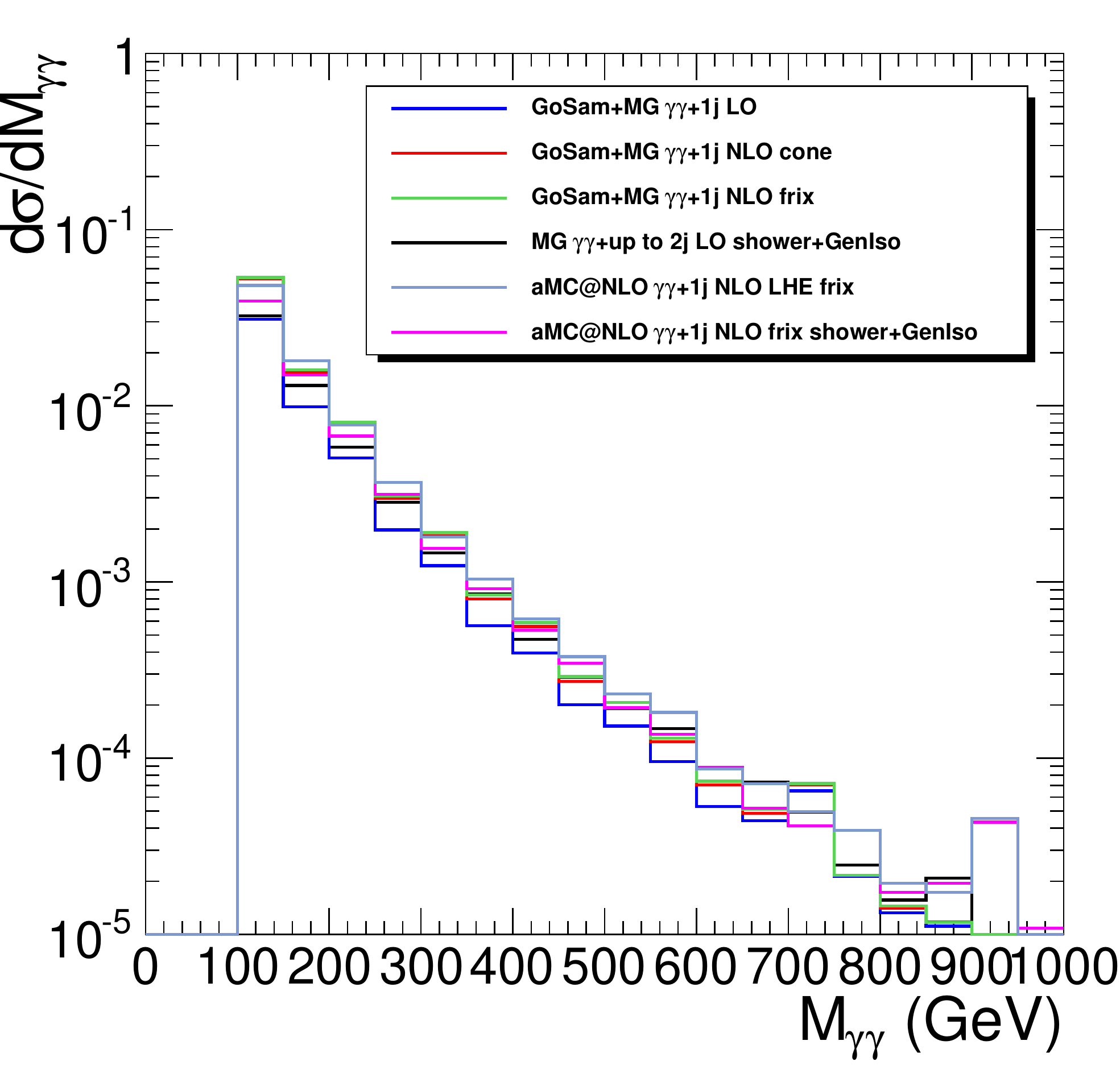}
\end{minipage}
 \caption{$\Delta\phi_{\gamma_1\gamma_2}$ and diphoton invariant mass distributions for the $\gamma\gamma$+1\,jet process.}
\label{fig:deltaphi1j}
\end{figure}

In this section we consider diphoton production in association with 
one tagged jet. 
We compare the following samples:
\begin{enumerate}
\item NLO and LO samples produced by \gosam{}+{\sc MadGraph/MadDipole/MadEvent} (parton level). 
Renormalisation and factorisation scales $\mu$ and $\mu_F$
have been chosen as dynamical scales, 
with the default choice being $\mu_0= \frac{1}{2}\,(m_{\gamma\gamma}+\sum_j p_{T}^{\rm{jet}})$, 
$\mu=\mu_F$. The PDF set used is CT10.
For the NLO sample based on \gosam{}+MG we also compare cone isolation with Frixione\,\cite{Frixione:1998hn} type isolation.
\item NLO samples produced by aMC@NLO~\cite{Hirschi:2011pa} 
interfaced with a {\sc Herwig6} shower~\cite{Corcella:2000bw}, where the 
scale choice is $\mu_0= \frac{1}{2}\,(m_{\gamma\gamma}+\sum p_{T}^{\rm{partons}})$, 
$\mu=\mu_F$. The PDF set used here is CTEQ6M.
\item {\sc MadGraph} LO samples with up to two jets at matrix element level, 
interfaced with a {\sc Pythia6} shower~\cite{Sjostrand:2006za}. 
The scales are set to $\mu_0= p_{T}^{\gamma_1}+p_{T}^{\gamma_2} + \sum p_{T}^{\rm{partons}}$. The PDF set used is CTEQ6L1.
\end{enumerate}
The  LO computations based on {\sc MadGraph} do not need any isolation criterion 
applied at parton level, as we require $R_{\gamma ,j} > 0.5$. 
At NLO, however, 
collinear quark-photon configurations due to extra radiation need to be taken care of.
The singularities related to these configurations are cancelled by absorbing them into the 
``bare" fragmentation functions, analogous to the case of PDFs in the initial state.
Therefore, the fragmentation part needs to be taken into account in an NLO calculation, 
unless it is completely suppressed by an isolation criterion like the one suggested by 
Frixione~\cite{Frixione:1998hn}.
With this criterion, one considers smaller 
cones of radius $r_\gamma$ inside the $R$-cone and calls the photon isolated 
 if the energy in any sub-cone does not exceed
 \begin{displaymath}
 E_{{\rm had, max}} (r_{\gamma}) = \epsilon \,p_{T}^{\gamma} \left( \frac{1-\cos r_\gamma}
 {1-\cos R}\right)^{n}\;.
 \label{eq:frix}
\end{displaymath}
When applying this criterion in our calculations, we use the isolation parameters $R=0.4, n=1$ and $\epsilon=0.05$.\\
However, the latter criterion is difficult to 
implement in a realistic experimental environment, where the photon will always be 
accompanied by some amount of hadronic energy.
Therefore, whenever we produce showered events, 
an isolation is applied at shower level which is a classical cone-type isolation:
$GenIso < 0.05 \times p_{T}^{\gamma}$, where $GenIso$ is computed from the transverse momentum sum 
of all particles falling in a cone $\Delta R<0.4$ around the photon.
In the partonic NLO calculation labelled  ``\gosam{}+MG NLO cone", where the fragmentation component is included, 
we apply the same cone isolation parameters. 
Further we apply the following kinematic requirements on the two photons:
\begin{eqnarray*}
&&
p_T^{\rm{jet}}>30\mbox{ GeV},\quad  p_T^{\gamma_1}>40 \mbox{~GeV}, \quad
p_T^{\gamma_2}>25\mbox{~GeV}, \\
&&m_{\gamma\gamma}>100\mbox{~GeV},\quad |\eta^{\gamma}| \leq 2.5,\quad R_{\gamma, \gamma} >0.45.
\end{eqnarray*}
With \gosam{}+MG, the jets are clustered with the anti-$k_T$ algorithm using a cone of 0.5, applied to the QCD partons only.
For the showered samples, the two hardest $p_{T}$ photons in the event are selected as diphoton candidates, 
and they are removed from the list of particle candidates before the jet clustering algorithm is run. 
We then select  the hardest jet in the event and require:
\begin{eqnarray*}
&&
p_T^{\rm{jet}}>30\mbox{ GeV},\quad
|\eta^{j}| \leq 4.7,\quad R_{\gamma ,j} > 0.5\;.
\end{eqnarray*}
Differential distributions for the samples produced with different generators 
after kinematic and isolation requirements, for $\sqrt{s}=8$\,TeV, are shown
in Fig.~\ref{fig:ptgamma1j} for the leading photon and leading jet transverse momentum distributions.
In particular, we compare, at NLO parton level,  Frixione and cone-type isolation. 
We observe that for the rather tight isolation criteria considered here, 
the differences are quite small.
In addition, we compare to showered samples with the same cone-type isolation. 
As expected, we observe that the effect of the isolation after the shower is considerably 
larger than at parton level.
This can be seen particularly  clearly in 
Fig.~\ref{fig:deltaphi1j}, in distribution of the relative azimuthal angle 
$\Delta\phi_{\gamma_1\gamma_2}$ between the two photons.
Note that the difference between \gosam{} NLO with Frixione isolation 
and aMC@NLO with Frixione isolation at the LHE level 
can be attributed to the different scale choice
(running over the partons for aMC@NLO and over the jets for \gosam{}). 
The difference due to the different PDF set is negligible.
Once the shower with Herwig6 is applied, 
the kinematic acceptance is reduced, giving a lower cross section. 


\subsection{Diphoton plus two jet production}

\begin{figure}[htb]
\begin{center}
\begin{minipage}[t]{0.49\textwidth}
\includegraphics[width=0.9\textwidth]{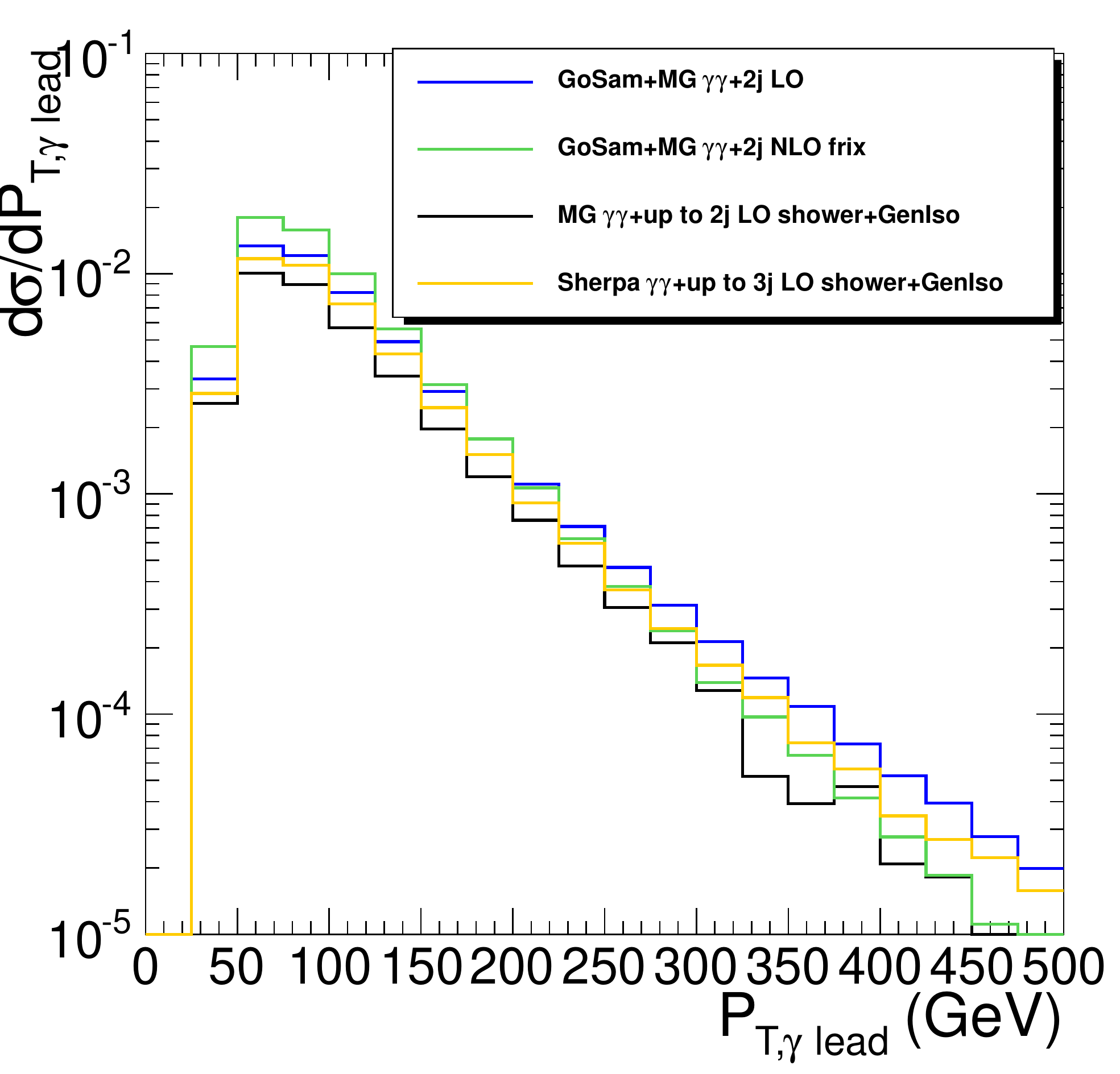}
\end{minipage}
\begin{minipage}[t]{0.49\textwidth}
\includegraphics[width=0.9\textwidth]{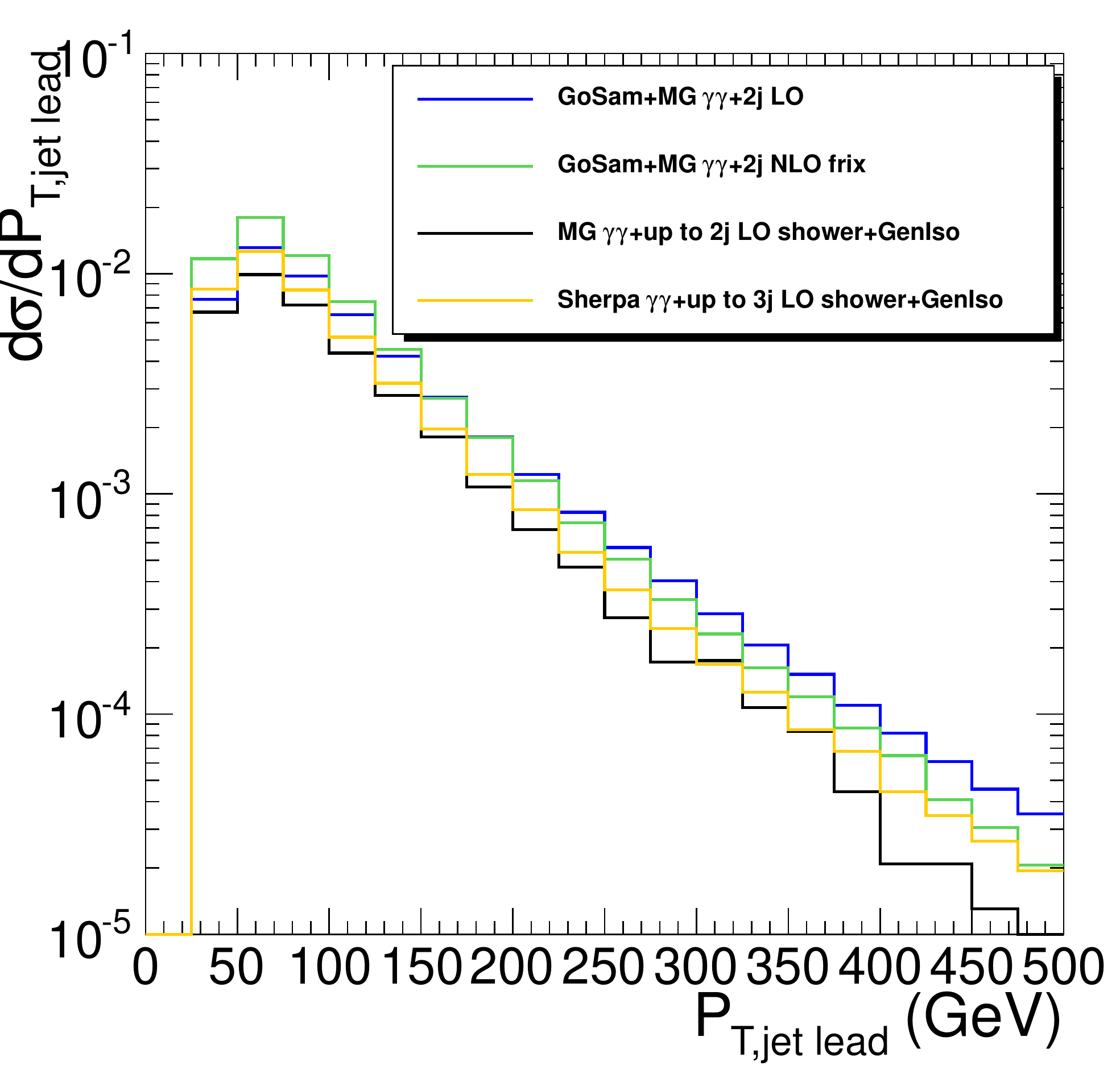}
\end{minipage}
 \caption{Transverse momentum distribution of the leading-$p_T$ photon and jet for the $\gamma\gamma$+2\,jets process.}
\label{fig:ptgamma2j}
\end{center}
\end{figure}

\begin{figure}[htb]
\centering
\begin{minipage}[t]{0.49\textwidth}
\includegraphics[width=0.9\textwidth]{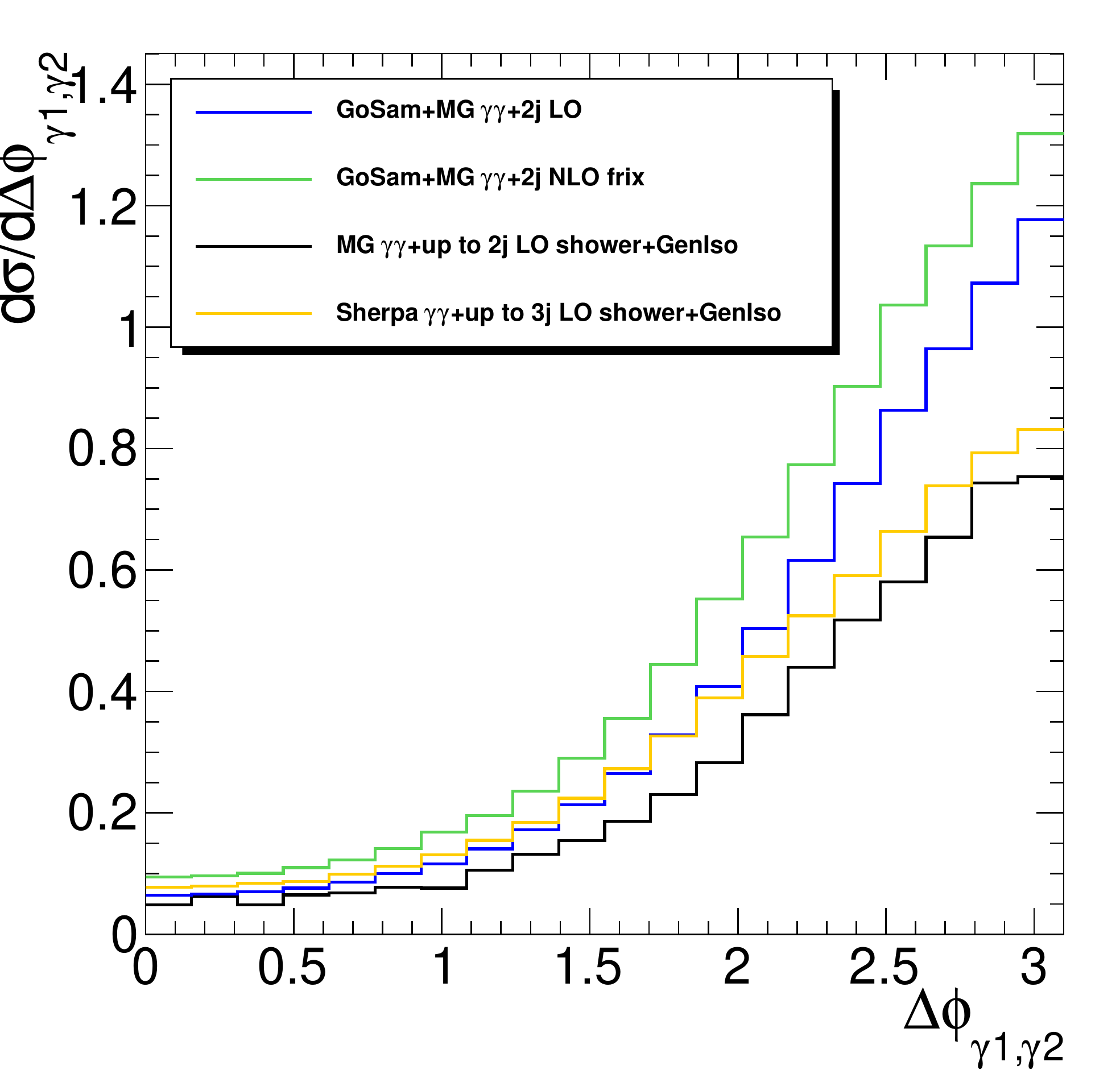}
\end{minipage}
\begin{minipage}[t]{0.49\textwidth}
\includegraphics[width=0.9\textwidth]{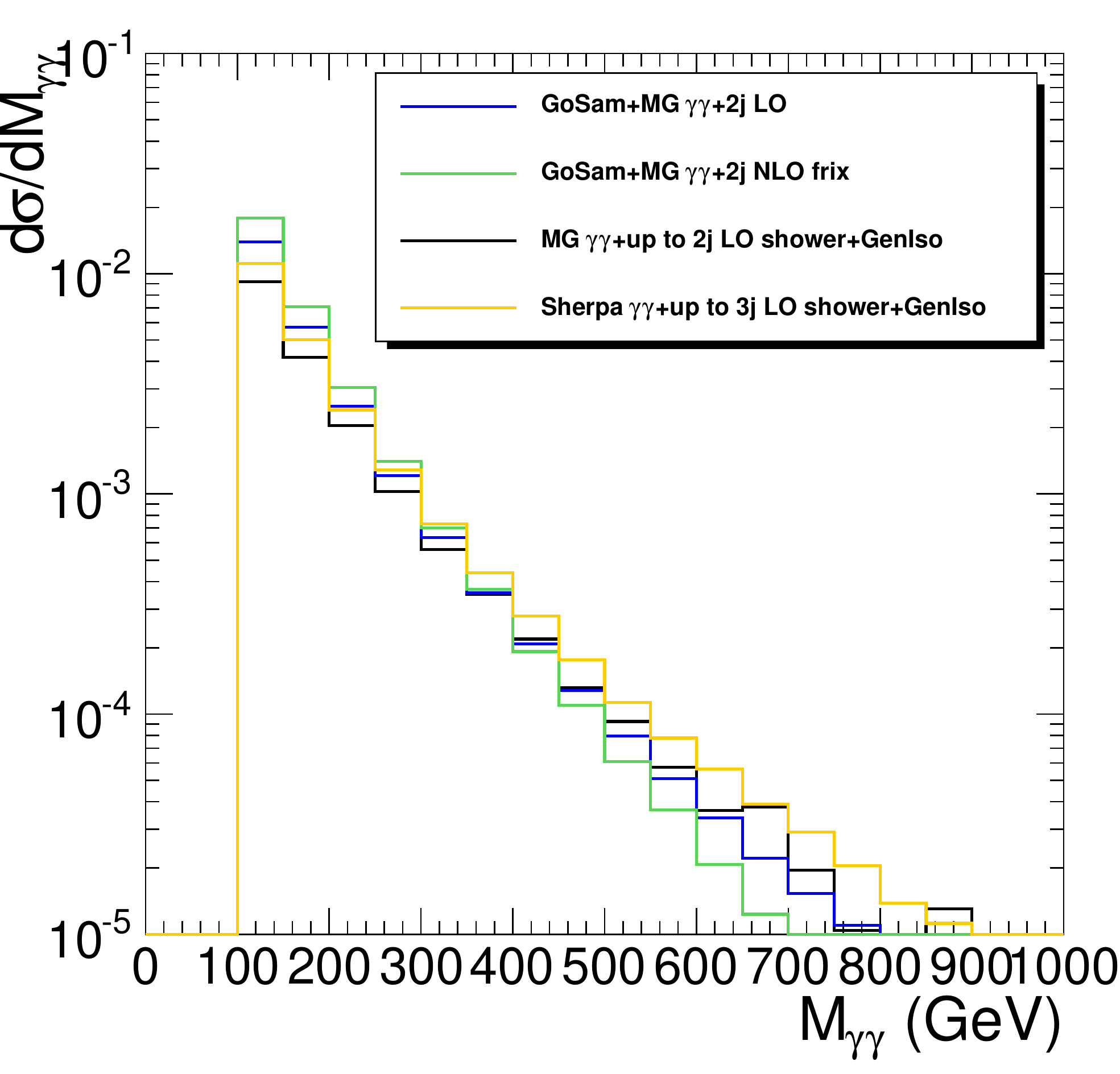}
\end{minipage}
 \caption{$\Delta\phi_{\gamma_1\gamma_2}$ and diphoton invariant mass distribution  for the $\gamma\gamma$+2\,jets process.}
\label{fig:deltaphi2j}
\end{figure}

For the  production of two photons and two jets, we consider similar samples as for the one jet case, 
also adding {\sc Sherpa}~\cite{Gleisberg:2008ta} LO samples with up to three jets at matrix element level.
As the NLO calculation also contains up to three partons accompanying the photons in the final state, 
it makes sense to consider up to three jets at matrix element level in a LO sample.
In the  {\sc Sherpa} sample, the box diagram $gg\to\gamma\gamma$, where the two additional jets are produced by the shower,
is also included, but its numerical impact is negligible.
Factorisation and renormalisation scales are set by the METS scale, 
which uses a modified CKKW matching algorithm. The PDF set used is CT10. 
Differential distributions produced by  the various generators after kinematic and isolation requirements are shown in
Fig.~\ref{fig:ptgamma2j} for the leading photon and leading jet transverse momentum and in 
Fig.~\ref{fig:deltaphi2j} for the difference in azimuthal angle between the two photons
and the diphoton invariant mass. 
One can see that the effect of the parton shower on the LO samples is different in the two-jet case:
the kinematical and isolation cuts reduce the cross-section 
more strongly than in the one-jet case. This is easily understood by the fact that after the shower, 
the probability to identify two hard jets in addition to two isolated photons is much lower than 
the probability to identify just one hard jet.

\subsection{Conclusions}

We have studied the processes of diphoton production in association with one and two jets at the LHC.
For the process $\gamma\gamma+1$\,jet, we have compared partonic LO results, 
LO results matched to a parton shower,  NLO results at parton level with two different isolation criteria, 
and NLO results combined with a parton shower.
As to be expected, we find that the NLO showered results lie between the LO results and the NLO partonic results.
Comparing standard cone isolation with Frixione-type isolation at NLO at parton level, 
we observe that the difference is very small for the tight isolation criteria used in this study.\\
For the process $\gamma\gamma+2$\,jets, we  have compared LO and 
NLO samples at parton level produced with \gosam{}+MadGraph/MadDipole/MadEvent
to two different showered LO samples, produced with MadGraph or Sherpa. 
Comparing to the behaviour in the  $\gamma\gamma+1$\,jet case, we observe that in the presence of the 
additional tagged jet, switching on the parton shower 
leads to a considerably stronger reduction of the (LO) cross section, as the probability 
to identify two hard jets is reduced more strongly after the parton shower than in the one-jet case.

\subsection*{Acknowledgements}
We would like to thank the organisers of  ``Les Houches 2013" 
for the stimulating workshop. We are also grateful to all members of the \gosam{}
project for their contributions to code development.
This work is supported by the Sinergia program of the Swiss National Science
Foundation (SNF) under grant number ~CRSII2-141847.



\clearpage

\clearpage

\chapter{MC tuning and output formats}
\label{cha:mc}






\section{Extensions to the Les Houches Event Format\footnote{A.~Buckley,
    J.~Butterworth, R.~Frederix, B.~Fuks, F.~Krauss, L.~Lonnblad,
    O.~Mattelaer, P.~Nason, C.~Oleari, S.~Padhi, S.~Pl\"atzer,
    S.~Prestel}}

\subsection{Background}

The LHE file format has been around since 2006 and has been very successful. At
the Les Houches workshop in 2009 an update was suggested, mainly to improve the
handling of NLO and matching/merging. The proposal was published in the
proceedings (arXiv:1003.1643), but have never really been properly
used. Nevertheless, some shortcomings have been noticed and some superfluous
features identified, both of which issues we attempt to address in this contribution.

In the discussions at this years workshop some ideas on how to better specify
weights were suggested, and also some other new features were suggested and old
features were suggested to be dropped. Here we summarise the discussion as a set
of agreed-upon proposals to define version 3.0 of the LHE format.

\subsection{Changes in the format of 2009}

Some XML tags that were introduced in 2009 were not adopted by the community,
possibly because their usefulness was very limited. In order not to clutter the
new proposal with unnecessary features, we recommend that the following XML tags
be removed.\\

\noindent
In the \kbd{<init>} block, we recommend that
\begin{description}
\item[\kbd{<mergeinfo>}] Should be removed due to very limited usability.
\end{description}

\noindent
In the \kbd{<event>} block, we recommend that
\begin{description}
\item[\kbd{<weight>}] Should be removed and replaced by \kbd{<weights>} (see below),
   to avoid clashes with the \emph{new} weight tag defined below.
\item[\kbd{<clustering>}] Should be removed due to very limited usability.
\item[\kbd{<pdfinfo>}] Should be removed due to unclear purpose.
\end{description}

\noindent
None of these XML tags seem to have been used. Thus, we do not expect
backward-compatibility problems if they were to be removed, and hence propose
that they be simply ignored by future parsers, as per the normally defined LHE
parsing behaviour for unknown XML tags. Any new LHE-parsing/writing code should
be based on this definition rather than the previous one.

In discussions before and during the Les Houches workshop, it became
obvious that the LHEF standard should be updated. This mainly concerns
the handling of multiple event weights per phase space point. Allowing
for multiple weights can significantly speed up the assessment of
uncertainties as well as performing finely-grained parameter
scans. For a well-defined handling of multiple weights, two proposals
for extending the LHEF format were made, and were later combined into
a single definition agreed on by the authors of the MC generator codes
(both matrix element and parton shower) actively using the format.

\subsection{Additions to the LHEF format for dealing with multiple weights}
The main improvement to the LHEF format is the need to deal with
multiple event weights. We therefore describe the additional XML tags
dealing with them first.

\subsubsection{Multiple weight definition in the event file header}
In the case where additional event weights are computed by the code
that generates the event file, the header of that file should contain
a new block, \kbd{<initrwgt>}, which syntax should follow this example:
\begin{small}
\begin{verbatim}
<initrwgt>
  <weight id='1'> This is the original event weight </weight>
  <weightgroup name='scale variation' combine='envelope'>
     <weight id='2'> muR = 2.0 </weight>
     <weight id='3'> muR = 0.5 </weight>
  </weightgroup>
   <weightgroup name="MRST2008 PDF uncertainty" combine="hessian">
     <weight id='4'> set 01 </weight>
     <weight id='5'> set 02 </weight>
     ...
   </weightgroup>
  <weightgroup name='Qmatch variation' combine='envelope'>
     <weight id='44'> Qmatch=20 </weight>
     <weight id='my_own_id'> Qmatch=40 </weight>
  </weightgroup>
  <weight id='46'> BSM benchmark point number 42B </weight>
</initrwgt>
\end{verbatim}
\end{small}
The information in the \kbd{<weight>} tags should be human-readable
and explain what the weights mean. It can simply contain all the
parameters that were used in to generate this weight; or only the ones
that were changed compared to the original run; or simply a sentence
explaining what this number means. It is up to the user that is doing
the analysis to make sure that this information is correctly used (and
up to the authors of the codes to make sure that the user has enough
information to understand what the weights correspond to). The
\kbd{<weight>} tag has the weight name \kbd{id} as mandatory
argument. Suggested optional arguments are
\begin{description}
\item[\kbd{muf}] The factor multiplying the nominal factorization scale for the event for the given weight
\item[\kbd{mur}] The factor multiplying the nominal renormalization scale for the event for the given weight
\item[\kbd{pdf}] The LHAPDF code corresponding to the given weight
\item[\kbd{pdf2}] The LHAPDF code corresponding to the given weight for the second beam if different from \kbd{pdf}
\end{description}
These arguments can be useful to concatenate a helpful string identifier
when parsing the information through events in HepMC format. In this
context, it is also helpful to choose a descriptive value for \kbd{id}.

The \kbd{<weightgroup>} tag allows to group several weights together and to
have the information about how to combine weights to obtain,
\textit{e.g.}, scale variation or pdf uncertainties. The attribute
\kbd{name} is mandatory and should define the name of the group. The
attribute \kbd{combine} is optional and it indicates how the weights
should be combined to define the associated uncertainties. Possible
arguments are \kbd{none}, \kbd{hessian}, \kbd{envelope} or
\kbd{gaussian}. If not specified, the default choice is
\kbd{combine='none'}, all the curves associated with each weights
being kept independent. For \kbd{combine='hessian'}, the first weight
is the central value and the next weights correspond respectively to
the positive and negative variations along a specific direction of the
parameter space.

\subsubsection{Multiple weights in the \kbd{<event>} block}
As already hinted above, there are two formats for the inclusions of
the weights in each of the events in the \kbd{<event>} block: a
compressed format and a detailed format. The compressed format is
designed to be memory-efficient, by avoiding repetition of identical
strings for each event as much as possible. It amounts to adding a
single \kbd{<weights>} tag to each event. The\kbd{<weights>} tag
contains a space-separated list of weights for the events given in the
order they are defined by the \kbd{<weight>} tags in the
\kbd{<initrwgt>} block. If for some obscure reason a weight source is
absent, its position in the list should be filled with a zero weight
entry. In this case, a typical event will look like:
\begin{small}
\begin{verbatim}
<event>
7 100  0.100E+01  0.200E+00  0.000E+00  0.000E+00
 -2 -1 0 0 0 0 0.126E+01 0.554E+01 0.576E+02 0.579E+02 0.000E+00 0. 0.
  2 -1 0 0 0 0 -.913E+00 0.131E+01 -.349E+02 0.350E+02 0.000E+00 0. 0.
 23  2 1 1 0 0 0.356E+00 0.685E+01 0.226E+02 0.929E+02 0.898E+02 0. 0.
-13  2 3 3 0 0 0.516E+01 0.211E+02 0.539E+02 0.581E+02 0.105E+00 0. 0.
 13  2 3 3 0 0 -.480E+01 -.142E+02 -.312E+02 0.347E+02 0.105E+00 0. 0.
-13  1 0 0 0 0 0.516E+01 0.211E+02 0.539E+02 0.581E+02 0.105E+00 0. 0.
 13  1 0 0 0 0 -.480E+01 -.142E+02 -.312E+02 0.347E+02 0.105E+00 0. 0.
<weights>  1.000e+00  0.204e+00  ...  1.564e+00 </weights>
</event>
\end{verbatim}
\end{small}
In \url{http://home.thep.lu.se/~leif/LHEF} there is a small example class
which is able to read and write this representation of the suggested new
format. Future development and documentation of the format will take place at
\url{http://lhef.hepforge.org}.

For the detailed format the weights are put in \kbd{<wgt>} tags and
are identified by their identifier \kbd{id}; their order is not of
relevance. The list of weights should be put in the \kbd{<rwgt>}
block, and each weight should be on a separate line. In this case a
typical example event looks like:
\begin{small}
\begin{verbatim}
<event>
7 100  0.100E+01  0.200E+00  0.000E+00  0.000E+00
 -2 -1 0 0 0 0 0.126E+01 0.554E+01 0.576E+02 0.579E+02 0.000E+00 0. 0.
  2 -1 0 0 0 0 -.913E+00 0.131E+01 -.349E+02 0.350E+02 0.000E+00 0. 0.
 23  2 1 1 0 0 0.356E+00 0.685E+01 0.226E+02 0.929E+02 0.898E+02 0. 0.
-13  2 3 3 0 0 0.516E+01 0.211E+02 0.539E+02 0.581E+02 0.105E+00 0. 0.
 13  2 3 3 0 0 -.480E+01 -.142E+02 -.312E+02 0.347E+02 0.105E+00 0. 0.
-13  1 0 0 0 0 0.516E+01 0.211E+02 0.539E+02 0.581E+02 0.105E+00 0. 0.
 13  1 0 0 0 0 -.480E+01 -.142E+02 -.312E+02 0.347E+02 0.105E+00 0. 0.
 <rwgt>
  <wgt id='1'> 1.000e+00 </wgt>
  <wgt id='2'> 0.204e+00 </wgt>
  <wgt id='3'> 1.564e+00 </wgt>
  <wgt id='4'> 2.248e+00 </wgt>
  <wgt id='5'> 1.486e+00 </wgt>
  ...
  <wgt id='my_own_id'> -0.839e+00 </wgt>
  <wgt id='46'> -0.899e+00 </wgt>
 </rwgt>
</event>
\end{verbatim}
\end{small}
This representation of the proposed format has been implemented in the
\textsc{MadGraph5\_aMC@NLO}, \textsc{MadAnalysis5} and \textsc{POWHEG BOX V2} packages.

Note that for both formats the weight for the nominal event is still
also given by \kbd{XWGTUP} entry.

\subsection{Additional new tags not related to multiple event weights}
We agreed upon two more tags to become part of the LHEFv3.0 accord. The first one
should be put in the \kbd{<init>} block and helps to transfer the
program version with which the LHEF has been generated:
\begin{description}
\item[\kbd{<generator>}] There should be at least one such tag (potentially
  more) with a \kbd{name} and a \kbd{version} attribute in the \kbd{<init>} section.
\end{description}
A typical example is
\begin{small}
\begin{verbatim}
<generator name='SomeGen' version='1.2.3'> some additional comments </generator>
\end{verbatim}
\end{small}
The additional comments can either identify sub-modules or libraries,
or other information, like an arXiv number.\\

\noindent
The second tag, \kbd{<scales>}, is optional and should be put within
the \kbd{<event>} block. It is defined as:
\begin{description}
\item[\kbd{<scales>}] contains information about different scales used by the
  matrix element generator in the given event. The scales are given as
  attributes, and any attribute name is allowed. However, only the following
  attributes have a pre-defined meaning:
  \begin{description}
  \item[\kbd{muf}] The factorization scale (in \GeV)
  \item[\kbd{mur}] The renormalization scale (in \GeV)
  \item[\kbd{mups}] Suggested shower starting scale (in \GeV)
  \end{description}
  If any of these attributes are missing, then the value in \kbd{SCALUP} is to
  be assumed for the missing attribute.
\end{description}




\subsection*{Acknowledgements}

The authors thank the Les Houches workshop once again for the hospitality and
enjoyable/productive working environment in the Chamonix valley. AB
acknowledges support from a Royal Society University Research Fellowship, a CERN
Scientific Associateship, an IPPP Associateship, and the University of Glasgow's
Leadership Fellow programme.

\clearpage


\section{Proposed updates for HepMC event record\footnote{A.~Buckley, J.~Butterworth, F.~Krauss, L.~Lonnblad, S.~Pl\"atzer, S.~Prestel}}

\subsection{GenRun}

There is a need to store the names and version numbers of all generators used in
producing an event. We propose to define a \kbd{GenRun} object to contain
cross-section (\kbd{GenCrossSection}) and other run-level quantities discussed
below. Each \kbd{GenEvent} should link to the \kbd{GenRun}.

\subsection{Generator names and versions}

Add string array to \kbd{GenRun} with pairs of strings specifying (generator
name, generator version).  All generators used in the production should be
listed, in order of processing, with full version numbers.

\subsection{Vertex ID}

The \kbd{GenVertex.id()} is not currently used/specified. We propose to define six codes:
\begin{description}
\item[0] undefined (no information)
\item[1] matrix element vertex (i.e. the hard process which defines the PDF
  arguments). (Must be the same as the \kbd{GenEvent::signal\_vertex()});
\item[2] secondary partonic scatters (useful for studies of MPI jet production,
  e.g. for suppressing jets from MPI while retaining the low $p_\perp$ tail of
  the underlying event;
\item[3] hard decay (prehadronisation) e.g. of $W$, $H$, $Z$, $t$. Useful
  e.g. for optimisation studies of mass resolution;
\item[4] parton shower (including QED radiation) useful for e.g. identifying
  $g\rightarrow b\bar{b}$ splitting probability; ME/PS matching studies; identifying photons for
  dressing leptons. Note: any parton shower after a hadronisation is ignored,
  and the hadrons from it are flagged as being produced in a hadron decay;
\item[5] primary hadron formation. Useful for e.g re-doing hadronisation,
  applying BEC corrections, studying/tracking decay chains;
\item[6] hadron, tau and muon decays
\end{description}

\noindent We propose that 10--99 be reserved for generator-specific types. Use of this
information must be made carefully, to avoid generator-specific physics
conclusions.

\subsection{Weight names}

Some standardisation of weight names is desirable. We propose several standard
weight name components, to be separated from each other by single underscore
characters:
\begin{description}
\item[MUF0.5, MUF2] factorization scale multiplication. Any float may occur
  between \kbd{MUF} and the underscore divider / end of string;
\item[MUR0.5, MUR1.414] renormalization scale multiplication. Any float may
  occur between \kbd{MUR} and the underscore divider / end of string;
\item[PDF10801, PDF20123] PDF variations, with the number following \kbd{PDF} up
  to the underscore/end being the LHAPDF/PDFLIB unique PDF member ID.
\end{description}

\noindent
The potentially composite names for each weight are proposed to be stored in the
\kbd{GenRun}, linked to the \kbd{WeightContainer} to avoid replication in the
ASCII event format.

\subsection*{Acknowledgements}

All authors thank the Les Houches workshop once again for the hospitality and
enjoyable/productive working environment in the Chamonix valley. AB
acknowledges support from a Royal Society University Research Fellowship, a CERN
Scientific Associateship, an IPPP Associateship, and the University of Glasgow's
Leadership Fellow programme.

\clearpage

\section{Improvements in the Rivet MC analysis toolkit\footnote{A.~Buckley}}

Rivet\,\cite{Buckley:2010ar} is the de facto analysis toolkit used for
phenomenology and experimental MC analysis at the LHC. It comprises a steering
system for efficiently processing events (as HepMC\,\cite{Dobbs:2001ck} code objects)
through a user-specified set of analysis routines, each of which outputs a set
of histograms for comparison to data or to other MC predictions. User-written
analyses can be easily run without rebuilding Rivet, or any of the currently 250
built-in analyses may be used.

Rivet makes a major point of portability between any parton shower MC generators
by reconstructing all quantities from physically reliable final-state (or
decayed hadron) quantities, i.e. those which are at least in principle
measurable and are not subject to the computational ambiguities of perhaps
interfering quantum amplitudes. This has had the side-effect of making Rivet a
useful ``straw man'' system for development of more robust definitions of
physical quantities at ``MC truth'' level.

In this contribution I describe several major updates made to data processing
and physics observable features in the Rivet MC analysis toolkit, which defined
the step from the long-established Rivet version 1 to version 2, and the planned
/ in-development features for the next major release.

\subsection{Histogramming and run merging}

Partially as an historical accident of its genesis, Rivet has never been able to
fully merge generator runs. Primarily this is because it converts all histograms
(i.e. objects which preserve the history of their fills somehow) into simple
``scatter'' plots of $x$ and $y$ points with associated error bars. Histograms
typically compute the visual heights of their bins from stored weighted moments
like $\sum w_i$, $\sum w_i^2$, $\sum w_i x_i$, etc. (where $w_i$ is the weight
of the $i$th histogram fill and $x_i$ the value in that fill), by standard
statistical rules (and of course the width of each bin to preserve the shape
invariance of the observable under rebinning). It is therefore possible to merge
runs by simply adding the summed moments, either directly if the process type in
the different runs is the same, or with a cross-section scaling if they are
different or contain the same process but with complementary kinematic
generation cuts, cf. the classic ``jet slices'' beloved of experiment MC
productions. It is not possible to get the same exact merging behaviour from
``scatter'' graphs, and hence Rivet's run merging has always relied on
asymptotic scaling which break down on unpopulated distribution tails.

In Rivet version 2, a major redevelopment of the histogramming system was made,
replacing the formerly used AIDA system (in a much-hacked implementation called
LWH) with a new histogramming system, YODA\,\cite{yodaweb}. This was written specifically to
support MC analysis idioms and issues but intentionally does not contain any
specific binding to particle physics concepts or usage. Usage of the ubiquitous
ROOT\,\cite{Brun:1997pa} system was also extensively considered, but ROOT's pervasive
global state / threading unfriendliness and issues with object ownership and
histogram behaviour convinced us that it was better to start from scratch and
attempt to ``do it right''. YODA precisely stores the sorts of weighted moments
discussed above, for 1D and 2D histograms and profile histograms, as well as 1D,
2D, and 3D scatters and functions for conversion between these objects and to
their ROOT equivalents.

The YODA statistics implementations are based on those from our LWH code, and
are extremely careful in their treatments of low statistics errors, unfilled
bins, overflow/underflow bins and whole distribution properties, as well as
permitting gaps in binning as required by many historical public datasets in the
HepData system. Extensions to the YODA binning system are intended to allow
\emph{any} type of object to be stored and looked up in an efficient binning
representation. YODA histogram objects can be arbitrarily ``annotated'' with
key--value pairs, which proves extremely useful for e.g. storing information
about scaling operations applied to histograms, as well as style information to
be used by the Rivet plotting system. Unlike in ROOT and AIDA, YODA histograms
do not need to be immediately registered with a unique path or name string: it
is perfectly allowed to handle YODA histograms which are unknown to the system,
and paths are purely used for identification of analysis objects when writing to
or reading from a data file. The main YODA data format is a plain text one
closely related to the simple format previously used by Rivet's plotting system,
and is far simpler and more compact to handle than the verbose AIDA XML
format. Several scripts are provided for merging YODA data files, or for
converting them to or from other formats, including ROOT to the extent that ROOT
can encode all the information stored by YODA's data objects.

Converting Rivet's 200+ built-in analyses to use YODA was an extremely lengthy
and difficult process. The actual conversion of the code was mostly
straightforward, as the majority of analyses do not do anything more complex
than filling and normalisation of histograms and the necessary class renamings
were trivial. Subtractions and additions of histograms were also easily
converted. Some, however, performed quite complex operations and these required
extensive work. This conversion process, and the motivation to keep simple
things simple while making complex ones not unpleasant, was of great help in
designing and iterating the YODA interface. We are confident that the new YODA
mechanisms in every non-trivial case made histogramming in Rivet more intuitive
and expressive than previously.

Simply converting the code was not the whole story, of course: the
Herwig++\,\cite{Bahr:2008pv} Rivet-based internal validation system was used to run sets
of pre-generated events through both the AIDA- and YODA-based Rivet versions,
and numerical agreement of better than one part in $10^5$ was required. Such a
stringent test unavoidably threw up many false positives and a great deal of
time was spent individually going through the discrepant cases and understanding
the source of the disagreement. Again this helped to improve the YODA design and
behaviour, and in several cases bugs were found (and fixed) in the original Rivet
analyses.

\subsection{Lepton dressing}

During the Rivet\,2 development process, discussions arose relating to the
definition of how charged leptons are to be ``dressed'' with photons in
Rivet. The discussion of lepton dressing, with the primary intention of
capturing the physical effects of QED final state radiation (FSR) in $W$ and $Z$
decays, without introducing generator-dependent analysis code, was previously
discussed in the 2009 Les~Houches proceedings\,\cite{Butterworth:2010ym}.

The definition of dressing used in that note noted that ``near-Born''
performance could be achieved by clustering photons in a narrow cone around the
lepton -- the radius of the cone being of order $R = 0.1$ to reflect the typical
granularity of a collider experiment EM calorimeter. This neglects the effect of
wide-angle FSR, which cannot be distinguished from other photons unrelated to
the EW hard process.

The refinement made in Rivet\,2 (and backported to 1.9.0, the final release in
the 1.x series) is to by default restrict the clustered photons to those which
do not have a hadron or tau parent. This classification is made by recursively
walking the HepMC decay chain from final state photons: if any hadrons or tau
leptons with status = 2 are found in a photon's history then it is considered of
hadronic origin and hence emitted on a timescale entirely factorized from that
of the hard-process. This change to the clustering achieves an accuracy for EW
precision observables (e.g. $Z$ $p_\perp$) of order a few per-mille as opposed
to 1\%.

Alternative refinements have also been proposed, particularly those involving a
use of a jet clustering algorithm for photon gathering. Studies so far indicate
no significant advantage to be gained from such an approach. Other suggestions
have included process dependent distinguishing of QED ISR from FSR which, even
if considered desirable, is not sufficiently generic for use in a system like
Rivet.

This change to clustering once again shows the value of Rivet as a concrete
example of a generator-agnostic standard analysis system for driving physically
motivated observable definitions. Isolation of hard leptons is still important,
of course: if two leptons are close enough for their clustering cones to
overlap, then duplicate clusterings and double countings are avoided, but most
likely the photon-to-lepton identifications are not reliable. (This is similar
to the treatment of $b$-tagging with boosted jets, another area where we hope
that Rivet will prove a useful testing ground.)

The charged leptons considered so far just include stable electrons and
muons. Taus are much more troublesome, since they decay (both hadronically and
leptonically) and the decay itself is a source of photon emission, albeit one
unrelated to the hard process of the EW boson decay. Rivet's treatment of taus
will be improved in future, in connection with the evolving experimental
treatment of tau issues.

\subsection{Status, plans and prospects}

Rivet 2.0.0 was finally released in November~2013, using YODA for all
histogramming. It was followed by a catch-up release 1.9.0 to include analyses
provided during the final stages of the v2 development, and shortly thereafter
by an equivalent 2.1.0. No further releases will be made in the Rivet v1
series. All versions of Rivet can be downloaded from the website at
\url{http://rivet.hepforge.org}, along with (updated) bootstrap scripts to help
with installation on systems both with and without CERN AFS available.

Version 2.2 of Rivet is envisaged to contain the following important new
features, all of which have already been significantly developed:
\begin{itemize}
\item \textbf{FastJet integration:} Since Rivet development began concurrently
  with the development of FastJet\,\cite{Cacciari:2011ma}, the original jet clustering
  tools were not based on FastJet but rather on independent implementations. In
  the intervening years FastJet has become the de facto package for jet studies
  (and other IRC-safe clustering applications), and in particular contains many
  powerful features for jet grooming, pile-up/underlying event subtraction, jet
  substructure, etc. While these can be accessed from within Rivet, it does not
  currently integrate well with the Rivet \kbd{Jet} objects. This will be
  improved by basing the Rivet \kbd{Jet} on FastJet's \kbd{PseudoJet} and
  providing conversion operators between them. FastJet v3 is now required by
  Rivet, meaning that Rivet's jet objects will also now automatically hold
  connections to the clustering sequence for jet deconstruction studies. The
  deeper integration with FastJet will also permit more physical $b$, $c$ and
  tau tagging, using new projections developed for finding heavy
  hadrons immediately before their weak decays.

\item \textbf{Kinematic cuts:} The interfaces of many Rivet tools involve
  calling constructors or class methods with arguments expressing the values of
  kinematic cuts, such as $p_\perp$ or pseudorapidity thresholds. All these are
  currently expressed as floating point numbers, meaning that a) it is easy to
  forget the correct orderings and the compiler cannot distinguish between the
  intended meanings of several \kbd{double}s, and b) it is not possible to have
  methods which take arguments of e.g. $(\eta_\text{min}, \eta_\text{max},
  p_\perp^\text{min})$ and $(y_\text{min}, y_\text{max}, E_\perp^\text{min})$
  because again all the compiler sees is three \kbd{double}s. Expressing complex
  cuts such as several disjoint ranges of rapidity is also awkward. This has
  been addressed by creation of an object-based ``cuts'' system, where only one
  argument needs to be passed to represent a collection of kinematic
  selections. Rather than use a pre-approved set of combined cuts, users can
  build them ``inline'' as desired, using logical operators,
  e.g. \kbd{ptMin(10*GeV) \& (etaIn(-2,-1) | etaIn(-0.5,0.5) | etaIn(1,2))}. This
  will make analysis code both more powerful/expressive, and easier to read.

\item \textbf{``Re-entrant'' analysis and general run merging:} While the
  developments of YODA and Rivet\,2 have made ``normal'' merging of MC runs
  possible, they cannot handle all observables. While YODA has the advantage
  that normalized and scaled plots can be combined (which is not normally the
  case, but YODA stores the scaling factor so it can be temporarily undone
  during the combination), it still cannot combine more complex observables such
  as histogram divisions $A/B$ or asymmetries $(A-B)/(A+B)$ constructed in the
  \kbd{finalize} step of the analysis. These objects cease to be histograms in
  the division step -- i.e. it would not make sense to continue calling
  \kbd{fill} on them, and their weighted moments are lost -- hence they cannot
  be combined to get the equivalent result to one large run.

  The solution to this is to write out both the finalized histogram objects
  \emph{and} any temporary objects used to construct them: this latter group
  includes the $A$ and $B$ intermediate histograms above and any weight
  counters, in their state at the start of \kbd{finalize}, after the event loop
  has run. Some machinery is needed to make sure that the necessary information
  is all stored in the YODA data files, and that declarations are passed through
  the processing chain to avoid unwanted plotting of temporary histograms. It is
  envisaged that extra bookkeeping information such as the estimated
  cross-section, number of events, and perhaps a process ID code will also be
  automatically stored to aid automatic run combination.

  The resulting data files, containing a mix of temporary and final analysis
  data objects, will then be combinable using the usual YODA merging mechanisms
  -- although of course only the intermediate object will be correctly
  combinable. The final step in the process will then be to re-run Rivet with
  the merged histogram file as a starting point for the statistical aggregation,
  and usually to jump immediately to the \kbd{finalize} method(s) to create the
  fully merged complex observables. The ability to re-start the Rivet event loop
  based on a previous run obviously has other benefits, such as the ability to
  trivially extend run statistics should they be found to be insufficient, and
  to periodically write out finalized histograms during a run: it is for this
  reason that we call the system ``re-entrant'' analysis.

\item \textbf{NLO and multi-weighted event handling:} The re-entrant system
  described above also provides a clean route for handling awkward event streams
  in which NLO generators place several counter-events in order, requiring that
  each counter-event group be treated as a single statistical unit, or the
  increasing phenomenon of events with very many weights representing systematic
  variations of scales, PDFs, etc. calculated during the generator run. Rivet
  (nor any other analysis system) cannot currently handle such situations
  automatically, and we strongly wish to avoid burdening all authors of analysis
  codes with the need to understand and implement a complex data processing
  procedure which most likely does not apply to their personal usage of the system.

  The planned development is that the histogram objects used in the analysis
  routine are in fact proxies for the permanent histograms (both intermediate
  and final) which will be written out at the end of the run. The user should be
  mainly oblivious to this distinction, but behind the scenes the analysis
  histograms are used automatically for filling an ensemble of histograms for
  each event weight, automatically combined according to the counter-event
  grouping. This places a requirement on the user to register temporary
  histograms with the Rivet histogram booking system, but this requirement
  already exists for the re-entrant histogramming. Conveniently, this
  development will also remove the need for users to explicitly make use of
  event weights when filling histograms, as this will be done automatically by
  the Rivet system.

  A final development for NLO counter-events is the optional use of ``fuzzy''
  histogramming. It is well known that NLO predictions only remain finite when a
  resolution parameter of some form is used, since perfect resolution of soft or
  collinear emissions breaks the necessary cancellation of divergences between
  real and virtual terms in the calculation\,\cite{Kinoshita:1962ur,Lee:1964is}. While
  histogram bins themselves implement this finite resolution, the boundaries
  between them are in a sense perfectly sharp and so can re-introduce
  problems. The classic case of this is when two counter-events fall just on
  opposite sides of a bin boundary, producing a large positive contribution in
  one bin and a large negative entry in its neighbour: the required cancellation
  fails to occur. To handle this situation, we have developed a ``fuzzy
  binning'' procedure, whereby a single fill operation will actually fill two
  bins, identified by and in proportion to the position of the fill within a
  bin: a fill in the middle of a bin will only fill that bin, but one close to a
  bin edge will proportionally contribute both to that bin and to its neighbour
  on the other side of the bin edge. 
  It has not yet been decided whether this feature should be implemented in YODA
  itself, or be performed by Rivet using YODA objects.

\item \textbf{Decay chain tools:} Finally, we mention that Rivet will acquire
  tools to assist with analysis of (mainly hadron) decay chains -- a topic of
  interest for flavour physics analyses in particular but for which Rivet
  provides little assistance, requiring the user to drop into the raw HepMC
  event record. This, as with the need for better tau lepton decay handling
  mentioned above, reflects the increasing trend toward dealing with details of
  event properties and modelling in LHC Run\,2.

\end{itemize}

Rivet continues to be heavily used in the LHC experimental and phenomenological
communities for a variety of analysis types. Limitations of our histogramming
capabilities have long been lamented, and we are pleased to have substantially
improved this area. Work remains to be done, of course, in particular to support
the rise in fully exclusive event simulation with NLO-subtracted matrix elements
and use of internal event reweighting for modelling systematics. Developments
such as incremental improvements in lepton dressing and flavour tagging
algorithms in Rivet are well connected to LHC experiment discussions on the same
topics of truth object definition, which are of key importance as the LHC enters
its precision physics era.

\subsection*{Acknowledgements}

AB thanks the Les Houches workshop once again for the hospitality and
enjoyable/productive working environment in the Chamonix valley, and
acknowledges support from a Royal Society University Research Fellowship, a CERN
Scientific Associateship, an IPPP Associateship, and the University of Glasgow's
Leadership Fellow programme. Much of the key final development and testing of
Rivet\,2, as well as the now-traditional Rivet tutorials, was performed during Les~Houches 2013.

\clearpage

\clearpage



\bibliographystyle{atlasnote}
\bibliography{LH13}

\providecommand{\href}[2]{#2}\begingroup\raggedright\begin{thebibliography}{100}

\bibitem{Badger:2013yda}
S.~Badger, B.~Biedermann, P.~Uwer, and V.~Yundin, {\em {Next-to-leading order
  QCD corrections to five jet production at the LHC}\/},
  \href{http://dx.doi.org/10.1103/PhysRevD.89.034019}{Phys.Rev. {\bf D89}
  (2014)  034019},
\href{http://arxiv.org/abs/1309.6585}{{\tt arXiv:1309.6585 [hep-ph]}}.

\bibitem{Bern:2013gka}
Z.~Bern, L.~Dixon, F.~Febres~Cordero, S.~Hoeche, H.~Ita, et al., {\em
  {Next-to-Leading Order $W + 5$-Jet Production at the LHC}\/},
  \href{http://dx.doi.org/10.1103/PhysRevD.88.014025}{Phys.Rev. {\bf D88}
  (2013)  014025},
\href{http://arxiv.org/abs/1304.1253}{{\tt arXiv:1304.1253 [hep-ph]}}.

\bibitem{Cullen:2013saa}
G.~Cullen, H.~van Deurzen, N.~Greiner, G.~Luisoni, P.~Mastrolia, et al., {\em
  {NLO QCD corrections to Higgs boson production plus three jets in gluon
  fusion}\/},
  \href{http://dx.doi.org/10.1103/PhysRevLett.111.131801}{Phys.Rev.Lett. {\bf
  111} (2013)  131801},
\href{http://arxiv.org/abs/1307.4737}{{\tt arXiv:1307.4737 [hep-ph]}}.

\bibitem{Cullen:2011ac}
G.~Cullen, N.~Greiner, G.~Heinrich, G.~Luisoni, P.~Mastrolia, et al., {\em
  {Automated One-Loop Calculations with GoSam}\/},
  \href{http://dx.doi.org/10.1140/epjc/s10052-012-1889-1}{Eur.Phys.J. {\bf C72}
  (2012)  1889},
\href{http://arxiv.org/abs/1111.2034}{{\tt arXiv:1111.2034 [hep-ph]}}.

\bibitem{Bevilacqua:2011xh}
G.~Bevilacqua, M.~Czakon, M.~Garzelli, A.~van Hameren, A.~Kardos, et al., {\em
  {HELAC-NLO}\/},
  \href{http://dx.doi.org/10.1016/j.cpc.2012.10.033}{Comput.Phys.Commun. {\bf
  184} (2013)  986--997},
\href{http://arxiv.org/abs/1110.1499}{{\tt arXiv:1110.1499 [hep-ph]}}.

\bibitem{Hirschi:2011pa}
V.~Hirschi, R.~Frederix, S.~Frixione, M.~V. Garzelli, F.~Maltoni, et al., {\em
  {Automation of one-loop QCD corrections}\/},
  \href{http://dx.doi.org/10.1007/JHEP05(2011)044}{JHEP {\bf 1105} (2011)
  044},
\href{http://arxiv.org/abs/1103.0621}{{\tt arXiv:1103.0621 [hep-ph]}}.

\bibitem{Cascioli:2011va}
F.~Cascioli, P.~Maierhofer, and S.~Pozzorini, {\em {Scattering Amplitudes with
  Open Loops}\/},
  \href{http://dx.doi.org/10.1103/PhysRevLett.108.111601}{Phys.Rev.Lett. {\bf
  108} (2012)  111601},
\href{http://arxiv.org/abs/1111.5206}{{\tt arXiv:1111.5206 [hep-ph]}}.

\bibitem{Actis:2012qn}
S.~Actis, A.~Denner, L.~Hofer, A.~Scharf, and S.~Uccirati, {\em {Recursive
  generation of one-loop amplitudes in the Standard Model}\/},
  \href{http://dx.doi.org/10.1007/JHEP04(2013)037}{JHEP {\bf 1304} (2013)
  037},
\href{http://arxiv.org/abs/1211.6316}{{\tt arXiv:1211.6316 [hep-ph]}}.

\bibitem{Catani:2011qz}
S.~Catani, L.~Cieri, D.~de~Florian, G.~Ferrera, and M.~Grazzini, {\em {Diphoton
  production at hadron colliders: a fully-differential QCD calculation at
  NNLO}\/},
  \href{http://dx.doi.org/10.1103/PhysRevLett.108.072001}{Phys.Rev.Lett. {\bf
  108} (2012)  072001},
\href{http://arxiv.org/abs/1110.2375}{{\tt arXiv:1110.2375 [hep-ph]}}.

\bibitem{Grazzini:2013bna}
M.~Grazzini, S.~Kallweit, D.~Rathlev, and A.~Torre, {\em {$Z\gamma$ production
  at hadron colliders in NNLO QCD}\/},
\href{http://arxiv.org/abs/1309.7000}{{\tt arXiv:1309.7000 [hep-ph]}}.

\bibitem{Baernreuther:2012ws}
P.~Baernreuther, M.~Czakon, and A.~Mitov, {\em {Percent Level Precision Physics
  at the Tevatron: First Genuine NNLO QCD Corrections to $q \bar{q} \to t
  \bar{t} + X$}\/},
  \href{http://dx.doi.org/10.1103/PhysRevLett.109.132001}{Phys.Rev.Lett. {\bf
  109} (2012)  132001},
\href{http://arxiv.org/abs/1204.5201}{{\tt arXiv:1204.5201 [hep-ph]}}.

\bibitem{Czakon:2012zr}
M.~Czakon and A.~Mitov, {\em {NNLO corrections to top-pair production at hadron
  colliders: the all-fermionic scattering channels}\/},
  \href{http://dx.doi.org/10.1007/JHEP12(2012)054}{JHEP {\bf 1212} (2012)
  054},
\href{http://arxiv.org/abs/1207.0236}{{\tt arXiv:1207.0236 [hep-ph]}}.

\bibitem{Czakon:2012pz}
M.~Czakon and A.~Mitov, {\em {NNLO corrections to top pair production at hadron
  colliders: the quark-gluon reaction}\/},
  \href{http://dx.doi.org/10.1007/JHEP01(2013)080}{JHEP {\bf 1301} (2013)
  080},
\href{http://arxiv.org/abs/1210.6832}{{\tt arXiv:1210.6832 [hep-ph]}}.

\bibitem{Czakon:2013goa}
M.~Czakon, P.~Fiedler, and A.~Mitov, {\em {The total top quark pair production
  cross-section at hadron colliders through $O(\alpha_s^4)$}\/},
\href{http://arxiv.org/abs/1303.6254}{{\tt arXiv:1303.6254 [hep-ph]}}.

\bibitem{Ridder:2013mf}
A.~Gehrmann-De~Ridder, T.~Gehrmann, E.~Glover, and J.~Pires, {\em Second order
  {QCD} corrections to jet production at hadron colliders: the all-gluon
  contribution\/},
  \href{http://dx.doi.org/10.1103/PhysRevLett.110.162003}{Phys. Rev. Lett. {\bf
  110} (2013) no.~16, 162003},
\href{http://arxiv.org/abs/1301.7310}{{\tt arXiv:1301.7310 [hep-ph]}}.

\bibitem{Currie:2013dwa}
J.~Currie, A.~Gehrmann-De~Ridder, E.~Glover, and J.~Pires, {\em NNLO QCD
  corrections to jet production at hadron colliders from gluon scattering\/},
  \href{http://dx.doi.org/10.1007/JHEP01(2014)110}{JHEP {\bf 1401} (2014)
  110},
\href{http://arxiv.org/abs/1310.3993}{{\tt arXiv:1310.3993 [hep-ph]}}.

\bibitem{Boughezal:2013uia}
R.~Boughezal, F.~Caola, K.~Melnikov, F.~Petriello, and M.~Schulze, {\em {Higgs
  boson production in association with a jet at next-to-next-to-leading order
  in perturbative QCD}\/},
  \href{http://dx.doi.org/10.1007/JHEP06(2013)072}{JHEP {\bf 1306} (2013)
  072},
\href{http://arxiv.org/abs/1302.6216}{{\tt arXiv:1302.6216 [hep-ph]}}.

\bibitem{Czakon:2010td}
M.~Czakon, {\em {A novel subtraction scheme for double-real radiation at
  NNLO}\/},  \href{http://dx.doi.org/10.1016/j.physletb.2010.08.036}{Phys.Lett.
  {\bf B693} (2010)  259--268},
\href{http://arxiv.org/abs/1005.0274}{{\tt arXiv:1005.0274 [hep-ph]}}.

\bibitem{GehrmannDeRidder:2005cm}
A.~Gehrmann-De~Ridder, T.~Gehrmann, and E.~N. Glover, {\em {Antenna subtraction
  at NNLO}\/},  \href{http://dx.doi.org/10.1088/1126-6708/2005/09/056}{JHEP
  {\bf 0509} (2005)  056},
\href{http://arxiv.org/abs/hep-ph/0505111}{{\tt arXiv:hep-ph/0505111
  [hep-ph]}}.

\bibitem{Boughezal:2011jf}
R.~Boughezal, K.~Melnikov, and F.~Petriello, {\em {A subtraction scheme for
  NNLO computations}\/},
  \href{http://dx.doi.org/10.1103/PhysRevD.85.034025}{Phys.Rev. {\bf D85}
  (2012)  034025},
\href{http://arxiv.org/abs/1111.7041}{{\tt arXiv:1111.7041 [hep-ph]}}.

\bibitem{Ball:2013bra}
R.~D. Ball, M.~Bonvini, S.~Forte, S.~Marzani, and G.~Ridolfi, {\em {Higgs
  production in gluon fusion beyond NNLO}\/},
  \href{http://dx.doi.org/10.1016/j.nuclphysb.2013.06.012}{Nucl.Phys. {\bf
  B874} (2013)  746--772},
\href{http://arxiv.org/abs/1303.3590}{{\tt arXiv:1303.3590 [hep-ph]}}.

\bibitem{Buehler:2013fha}
S.~Buehler and A.~Lazopoulos, {\em {Scale dependence and collinear subtraction
  terms for Higgs production in gluon fusion at N3LO}\/},
  \href{http://dx.doi.org/10.1007/JHEP10(2013)096}{JHEP {\bf 1310} (2013)
  096},
\href{http://arxiv.org/abs/1306.2223}{{\tt arXiv:1306.2223 [hep-ph]}}.

\bibitem{Anastasiou:2013srw}
C.~Anastasiou, C.~Duhr, F.~Dulat, and B.~Mistlberger, {\em {Soft triple-real
  radiation for Higgs production at N3LO}\/},
  \href{http://dx.doi.org/10.1007/JHEP07(2013)003}{JHEP {\bf 1307} (2013)
  003},
\href{http://arxiv.org/abs/1302.4379}{{\tt arXiv:1302.4379 [hep-ph]}}.

\bibitem{Anastasiou:2013mca}
C.~Anastasiou, C.~Duhr, F.~Dulat, F.~Herzog, and B.~Mistlberger, {\em
  {Real-virtual contributions to the inclusive Higgs cross-section at
  $N^3LO$}\/},  \href{http://dx.doi.org/10.1007/JHEP12(2013)088}{JHEP {\bf
  1312} (2013)  088},
\href{http://arxiv.org/abs/1311.1425}{{\tt arXiv:1311.1425 [hep-ph]}}.

\bibitem{Anastasiou:2014vaa}
C.~Anastasiou, C.~Duhr, F.~Dulat, E.~Furlan, T.~Gehrmann, et al., {\em {Higgs
  boson gluon-fusion production at threshold in N3LO QCD}\/},
\href{http://arxiv.org/abs/1403.4616}{{\tt arXiv:1403.4616 [hep-ph]}}.

\bibitem{Alioli:2010xd}
S.~Alioli, P.~Nason, C.~Oleari, and E.~Re, {\em {A general framework for
  implementing NLO calculations in shower Monte Carlo programs: the POWHEG
  BOX}\/},  \href{http://dx.doi.org/10.1007/JHEP06(2010)043}{JHEP {\bf 1006}
  (2010)  043},
\href{http://arxiv.org/abs/1002.2581}{{\tt arXiv:1002.2581 [hep-ph]}}.

\bibitem{Hoche:2010pf}
S.~Hoche, F.~Krauss, M.~Schonherr, and F.~Siegert, {\em {Automating the POWHEG
  method in Sherpa}\/},  \href{http://dx.doi.org/10.1007/JHEP04(2011)024}{JHEP
  {\bf 1104} (2011)  024},
\href{http://arxiv.org/abs/1008.5399}{{\tt arXiv:1008.5399 [hep-ph]}}.

\bibitem{Platzer:2011bc}
S.~Pl{\"a}tzer and S.~Gieseke, {\em {Dipole Showers and Automated NLO Matching
  in Herwig++}\/},
  \href{http://dx.doi.org/10.1140/epjc/s10052-012-2187-7}{Eur.Phys.J. {\bf C72}
  (2012)  2187},
\href{http://arxiv.org/abs/1109.6256}{{\tt arXiv:1109.6256 [hep-ph]}}.

\bibitem{Hoeche:2011fd}
S.~H{\"o}che, F.~Krauss, M.~Sch{\"o}nherr, and F.~Siegert, {\em {A critical
  appraisal of NLO+PS matching methods}\/},
  \href{http://dx.doi.org/10.1007/JHEP09(2012)049}{JHEP {\bf 1209} (2012)
  049},
\href{http://arxiv.org/abs/1111.1220}{{\tt arXiv:1111.1220 [hep-ph]}}.

\bibitem{Hoeche:2012yf}
S.~H{\"o}che, F.~Krauss, M.~Sch{\"o}nherr, and F.~Siegert, {\em {QCD matrix
  elements + parton showers: The NLO case}\/},
  \href{http://dx.doi.org/10.1007/JHEP04(2013)027}{JHEP {\bf 1304} (2013)
  027},
\href{http://arxiv.org/abs/1207.5030}{{\tt arXiv:1207.5030 [hep-ph]}}.

\bibitem{Lonnblad:2012ng}
L.~L{\"o}nnblad and S.~Prestel, {\em {Unitarising Matrix Element + Parton
  Shower merging}\/},  \href{http://dx.doi.org/10.1007/JHEP02(2013)094}{JHEP
  {\bf 1302} (2013)  094},
\href{http://arxiv.org/abs/1211.4827}{{\tt arXiv:1211.4827 [hep-ph]}}.

\bibitem{Hamilton:2012np}
K.~Hamilton, P.~Nason, and G.~Zanderighi, {\em {MINLO: Multi-Scale Improved
  NLO}\/},  \href{http://dx.doi.org/10.1007/JHEP10(2012)155}{JHEP {\bf 1210}
  (2012)  155},
\href{http://arxiv.org/abs/1206.3572}{{\tt arXiv:1206.3572 [hep-ph]}}.

\bibitem{Hamilton:2012rf}
K.~Hamilton, P.~Nason, C.~Oleari, and G.~Zanderighi, {\em {Merging H/W/Z + 0
  and 1 jet at NLO with no merging scale: a path to parton shower + NNLO
  matching}\/},  \href{http://dx.doi.org/10.1007/JHEP05(2013)082}{JHEP {\bf
  1305} (2013)  082},
\href{http://arxiv.org/abs/1212.4504}{{\tt arXiv:1212.4504}}.

\bibitem{Hamilton:2013fea}
K.~Hamilton, P.~Nason, E.~Re, and G.~Zanderighi, {\em {NNLOPS simulation of
  Higgs boson production}\/},
  \href{http://dx.doi.org/10.1007/JHEP10(2013)222}{JHEP {\bf 1310} (2013)
  222},
\href{http://arxiv.org/abs/1309.0017}{{\tt arXiv:1309.0017 [hep-ph]}}.

\bibitem{Lonnblad:2012ix}
L.~L{\"o}nnblad and S.~Prestel, {\em {Merging Multi-leg NLO Matrix Elements
  with Parton Showers}\/},
  \href{http://dx.doi.org/10.1007/JHEP03(2013)166}{JHEP {\bf 1303} (2013)
  166},
\href{http://arxiv.org/abs/1211.7278}{{\tt arXiv:1211.7278 [hep-ph]}}.

\bibitem{Frixione:2002ik}
S.~Frixione and B.~R. Webber, {\em {Matching NLO QCD computations and parton
  shower simulations}\/},
  \href{http://dx.doi.org/10.1088/1126-6708/2002/06/029}{JHEP {\bf 0206} (2002)
   029},
\href{http://arxiv.org/abs/hep-ph/0204244}{{\tt arXiv:hep-ph/0204244
  [hep-ph]}}.

\bibitem{Frixione:2007vw}
S.~Frixione, P.~Nason, and C.~Oleari, {\em {Matching NLO QCD computations with
  Parton Shower simulations: the POWHEG method}\/},
  \href{http://dx.doi.org/10.1088/1126-6708/2007/11/070}{JHEP {\bf 0711} (2007)
   070},
\href{http://arxiv.org/abs/0709.2092}{{\tt arXiv:0709.2092 [hep-ph]}}.

\bibitem{Dittmaier:2011ti}
{LHC Higgs Cross Section Working Group} Collaboration, S.~Dittmaier et al.,
  {\em {Handbook of LHC Higgs Cross Sections: 1. Inclusive Observables}\/},
\href{http://arxiv.org/abs/1101.0593}{{\tt arXiv:1101.0593 [hep-ph]}}.

\bibitem{Dittmaier:2012vm}
S.~Dittmaier, C.~Mariotti, G.~Passarino, R.~Tanaka, et al., {\em {Handbook of
  LHC Higgs Cross Sections: 2. Differential Distributions}\/},
\href{http://arxiv.org/abs/1201.3084}{{\tt arXiv:1201.3084 [hep-ph]}}.

\bibitem{Heinemeyer:2013tqa}
{LHC Higgs Cross Section Working Group} Collaboration, S.~Heinemeyer,
  C.~Mariotti, G.~Passarino, R.~Tanaka, et al., {\em {Handbook of LHC Higgs
  Cross Sections: 3. Higgs Properties}\/},
\href{http://arxiv.org/abs/1307.1347}{{\tt arXiv:1307.1347 [hep-ph]}}.

\bibitem{Dittmaier:2012nh}
S.~Dittmaier and M.~Schumacher, {\em {The Higgs Boson in the Standard Model -
  From LEP to LHC: Expectations, Searches, and Discovery of a Candidate}\/},
  \href{http://dx.doi.org/10.1016/j.ppnp.2013.02.001}{Prog.Part.Nucl.Phys. {\bf
  70} (2013)  1--54},
\href{http://arxiv.org/abs/1211.4828}{{\tt arXiv:1211.4828 [hep-ph]}}.

\bibitem{Dawson:2013bba}
S.~Dawson, A.~Gritsan, H.~Logan, J.~Qian, C.~Tully, et al., {\em {Higgs Working
  Group Report of the Snowmass 2013 Community Planning Study}\/},
\href{http://arxiv.org/abs/1310.8361}{{\tt arXiv:1310.8361 [hep-ex]}}.

\bibitem{TheATLAScollaboration:2013eia}
T.~A. collaboration,
{\em {Differential cross sections of the Higgs boson measured in the diphoton
  decay channel using 8 TeV pp collisions}\/}, .

\bibitem{Campbell:2010cz}
J.~M. Campbell, R.~K. Ellis, and C.~Williams, {\em {Hadronic production of a
  Higgs boson and two jets at next-to-leading order}\/},
  \href{http://dx.doi.org/10.1103/PhysRevD.81.074023}{Phys.Rev. {\bf D81}
  (2010)  074023},
\href{http://arxiv.org/abs/1001.4495}{{\tt arXiv:1001.4495 [hep-ph]}}.

\bibitem{vanDeurzen:2013rv}
H.~van Deurzen, N.~Greiner, G.~Luisoni, P.~Mastrolia, E.~Mirabella, et al.,
  {\em {NLO QCD corrections to the production of Higgs plus two jets at the
  LHC}\/},  \href{http://dx.doi.org/10.1016/j.physletb.2013.02.051}{Phys.Lett.
  {\bf B721} (2013)  74--81},
\href{http://arxiv.org/abs/1301.0493}{{\tt arXiv:1301.0493 [hep-ph]}}.

\bibitem{Ferrera:2011bk}
G.~Ferrera, M.~Grazzini, and F.~Tramontano, {\em {Associated WH production at
  hadron colliders: a fully exclusive QCD calculation at NNLO}\/},
  \href{http://dx.doi.org/10.1103/PhysRevLett.107.152003}{Phys.Rev.Lett. {\bf
  107} (2011)  152003},
\href{http://arxiv.org/abs/1107.1164}{{\tt arXiv:1107.1164 [hep-ph]}}.

\bibitem{Denner:2011id}
A.~Denner, S.~Dittmaier, S.~Kallweit, and A.~Muck, {\em {Electroweak
  corrections to Higgs-strahlung off W/Z bosons at the Tevatron and the LHC
  with HAWK}\/},  \href{http://dx.doi.org/10.1007/JHEP03(2012)075}{JHEP {\bf
  1203} (2012)  075},
\href{http://arxiv.org/abs/1112.5142}{{\tt arXiv:1112.5142 [hep-ph]}}.

\bibitem{Altenkamp:2012sx}
L.~Altenkamp, S.~Dittmaier, R.~V. Harlander, H.~Rzehak, and T.~J. Zirke, {\em
  {Gluon-induced Higgs-strahlung at next-to-leading order QCD}\/},
  \href{http://dx.doi.org/10.1007/JHEP02(2013)078}{JHEP {\bf 1302} (2013)
  078},
\href{http://arxiv.org/abs/1211.5015}{{\tt arXiv:1211.5015 [hep-ph]}}.

\bibitem{Ferrera:2013yga}
G.~Ferrera, M.~Grazzini, and F.~Tramontano, {\em {Higher-order QCD effects for
  associated WH production and decay at the LHC}\/},
\href{http://arxiv.org/abs/1312.1669}{{\tt arXiv:1312.1669 [hep-ph]}}.

\bibitem{Anastasiou:2011qx}
C.~Anastasiou, F.~Herzog, and A.~Lazopoulos, {\em {The fully differential decay
  rate of a Higgs boson to bottom-quarks at NNLO in QCD}\/},
  \href{http://dx.doi.org/10.1007/JHEP03(2012)035}{JHEP {\bf 1203} (2012)
  035},
\href{http://arxiv.org/abs/1110.2368}{{\tt arXiv:1110.2368 [hep-ph]}}.

\bibitem{Ellis:2013yxa}
J.~Ellis, D.~S. Hwang, K.~Sakurai, and M.~Takeuchi, {\em {Disentangling
  Higgs-Top Couplings in Associated Production}\/},
\href{http://arxiv.org/abs/1312.5736}{{\tt arXiv:1312.5736 [hep-ph]}}.

\bibitem{Dawson:1998py}
S.~Dawson, S.~Dittmaier, and M.~Spira, {\em {Neutral Higgs boson pair
  production at hadron colliders: QCD corrections}\/},
  \href{http://dx.doi.org/10.1103/PhysRevD.58.115012}{Phys.Rev. {\bf D58}
  (1998)  115012},
\href{http://arxiv.org/abs/hep-ph/9805244}{{\tt arXiv:hep-ph/9805244
  [hep-ph]}}.

\bibitem{Grigo:2013rya}
J.~Grigo, J.~Hoff, K.~Melnikov, and M.~Steinhauser, {\em {On the Higgs boson
  pair production at the LHC}\/},
  \href{http://dx.doi.org/10.1016/j.nuclphysb.2013.06.024}{Nucl.Phys. {\bf
  B875} (2013)  1--17},
\href{http://arxiv.org/abs/1305.7340}{{\tt arXiv:1305.7340 [hep-ph]}}.

\bibitem{deFlorian:2013jea}
D.~de~Florian and J.~Mazzitelli, {\em {Higgs Boson Pair Production at
  Next-to-Next-to-Leading Order in QCD}\/},
  \href{http://dx.doi.org/10.1103/PhysRevLett.111.201801}{Phys. Rev. Lett. 111,
  {\bf 201801} (2013)  201801},
\href{http://arxiv.org/abs/1309.6594}{{\tt arXiv:1309.6594 [hep-ph]}}.

\bibitem{Baglio:2012np}
J.~Baglio, A.~Djouadi, R.~Gröber, M.~Mühlleitner, J.~Quevillon, et al., {\em
  {The measurement of the Higgs self-coupling at the LHC: theoretical
  status}\/},  \href{http://dx.doi.org/10.1007/JHEP04(2013)151}{JHEP {\bf 1304}
  (2013)  151},
\href{http://arxiv.org/abs/1212.5581}{{\tt arXiv:1212.5581 [hep-ph]}}.

\bibitem{Frederix:2014hta}
R.~Frederix, S.~Frixione, V.~Hirschi, F.~Maltoni, O.~Mattelaer, et al., {\em
  {Higgs pair production at the LHC with NLO and parton-shower effects}\/},
\href{http://arxiv.org/abs/1401.7340}{{\tt arXiv:1401.7340 [hep-ph]}}.

\bibitem{ATLAS:2012aa}
{ATLAS Collaboration} Collaboration, G.~Aad et al., {\em {Measurement of the
  cross section for top-quark pair production in $pp$ collisions at
  $\sqrt{s}=7$ TeV with the ATLAS detector using final states with two high-pt
  leptons}\/},  \href{http://dx.doi.org/10.1007/JHEP05(2012)059}{JHEP {\bf
  1205} (2012)  059},
\href{http://arxiv.org/abs/1202.4892}{{\tt arXiv:1202.4892 [hep-ex]}}.

\bibitem{Chatrchyan:2012bra}
{CMS} Collaboration, S.~Chatrchyan et al., {\em {Measurement of the $t\bar{t}$
  production cross section in the dilepton channel in $pp$ collisions at
  $\sqrt{s}=7$ TeV}\/},  \href{http://dx.doi.org/10.1007/JHEP11(2012)067}{JHEP
  {\bf 1211} (2012)  067},
\href{http://arxiv.org/abs/1208.2671}{{\tt arXiv:1208.2671 [hep-ex]}}.

\bibitem{Kuhn:2013zoa}
J.~Kühn, A.~Scharf, and P.~Uwer, {\em {Weak Interactions in Top-Quark Pair
  Production at Hadron Colliders: An Update}\/},
\href{http://arxiv.org/abs/1305.5773}{{\tt arXiv:1305.5773 [hep-ph]}}.

\bibitem{Melnikov:2009dn}
K.~Melnikov and M.~Schulze, {\em {NLO QCD corrections to top quark pair
  production and decay at hadron colliders}\/},
  \href{http://dx.doi.org/10.1088/1126-6708/2009/08/049}{JHEP {\bf 0908} (2009)
   049},
\href{http://arxiv.org/abs/0907.3090}{{\tt arXiv:0907.3090 [hep-ph]}}.

\bibitem{Denner:2010jp}
A.~Denner, S.~Dittmaier, S.~Kallweit, and S.~Pozzorini, {\em {NLO QCD
  corrections to WWbb production at hadron colliders}\/},
  \href{http://dx.doi.org/10.1103/PhysRevLett.106.052001}{Phys.Rev.Lett. {\bf
  106} (2011)  052001},
\href{http://arxiv.org/abs/1012.3975}{{\tt arXiv:1012.3975 [hep-ph]}}.

\bibitem{Bevilacqua:2010qb}
G.~Bevilacqua, M.~Czakon, A.~van Hameren, C.~G. Papadopoulos, and M.~Worek,
  {\em {Complete off-shell effects in top quark pair hadroproduction with
  leptonic decay at next-to-leading order}\/},
  \href{http://dx.doi.org/10.1007/JHEP02(2011)083}{JHEP {\bf 1102} (2011)
  083},
\href{http://arxiv.org/abs/1012.4230}{{\tt arXiv:1012.4230 [hep-ph]}}.

\bibitem{Denner:2012yc}
A.~Denner, S.~Dittmaier, S.~Kallweit, and S.~Pozzorini, {\em {NLO QCD
  corrections to off-shell top-antitop production with leptonic decays at
  hadron colliders}\/},  \href{http://dx.doi.org/10.1007/JHEP10(2012)110}{JHEP
  {\bf 1210} (2012)  110},
\href{http://arxiv.org/abs/1207.5018}{{\tt arXiv:1207.5018 [hep-ph]}}.

\bibitem{Cascioli:2013wga}
F.~Cascioli, S.~Kallweit, P.~Maierhöfer, and S.~Pozzorini, {\em {A unified NLO
  description of top-pair and associated Wt production}\/},
\href{http://arxiv.org/abs/1312.0546}{{\tt arXiv:1312.0546 [hep-ph]}}.

\bibitem{Dittmaier:2007wz}
S.~Dittmaier, P.~Uwer, and S.~Weinzierl, {\em {NLO QCD corrections to t anti-t
  + jet production at hadron colliders}\/},
  \href{http://dx.doi.org/10.1103/PhysRevLett.98.262002}{Phys.Rev.Lett. {\bf
  98} (2007)  262002},
\href{http://arxiv.org/abs/hep-ph/0703120}{{\tt arXiv:hep-ph/0703120
  [HEP-PH]}}.

\bibitem{Dittmaier:2008uj}
S.~Dittmaier, P.~Uwer, and S.~Weinzierl, {\em {Hadronic top-quark pair
  production in association with a hard jet at next-to-leading order QCD:
  Phenomenological studies for the Tevatron and the LHC}\/},
  \href{http://dx.doi.org/10.1140/epjc/s10052-008-0816-y}{Eur.Phys.J. {\bf C59}
  (2009)  625--646},
\href{http://arxiv.org/abs/0810.0452}{{\tt arXiv:0810.0452 [hep-ph]}}.

\bibitem{Melnikov:2010iu}
K.~Melnikov and M.~Schulze, {\em {NLO QCD corrections to top quark pair
  production in association with one hard jet at hadron colliders}\/},
  \href{http://dx.doi.org/10.1016/j.nuclphysb.2010.07.003}{Nucl.Phys. {\bf
  B840} (2010)  129--159},
\href{http://arxiv.org/abs/1004.3284}{{\tt arXiv:1004.3284 [hep-ph]}}.

\bibitem{Kardos:2011qa}
A.~Kardos, C.~Papadopoulos, and Z.~Trocsanyi, {\em {Top quark pair production
  in association with a jet with NLO parton showering}\/},
  \href{http://dx.doi.org/10.1016/j.physletb.2011.09.080}{Phys.Lett. {\bf B705}
  (2011)  76--81},
\href{http://arxiv.org/abs/1101.2672}{{\tt arXiv:1101.2672 [hep-ph]}}.

\bibitem{Melnikov:2011qx}
K.~Melnikov, A.~Scharf, and M.~Schulze, {\em {Top quark pair production in
  association with a jet: QCD corrections and jet radiation in top quark
  decays}\/},  \href{http://dx.doi.org/10.1103/PhysRevD.85.054002}{Phys.Rev.
  {\bf D85} (2012)  054002},
\href{http://arxiv.org/abs/1111.4991}{{\tt arXiv:1111.4991 [hep-ph]}}.

\bibitem{Bevilacqua:2010ve}
G.~Bevilacqua, M.~Czakon, C.~Papadopoulos, and M.~Worek, {\em {Dominant QCD
  Backgrounds in Higgs Boson Analyses at the LHC: A Study of $\Pp\Pp \to \Pt
  \bar\Pt + {}$2jets at Next-To-Leading Order}\/},
  \href{http://dx.doi.org/10.1103/PhysRevLett.104.162002}{Phys.Rev.Lett. {\bf
  104} (2010)  162002},
\href{http://arxiv.org/abs/1002.4009}{{\tt arXiv:1002.4009 [hep-ph]}}.

\bibitem{Bevilacqua:2011aa}
G.~Bevilacqua, M.~Czakon, C.~Papadopoulos, and M.~Worek, {\em {Hadronic
  top-quark pair production in association with two jets at Next-to-Leading
  Order QCD}\/},  \href{http://dx.doi.org/10.1103/PhysRevD.84.114017}{Phys.Rev.
  {\bf D84} (2011)  114017},
\href{http://arxiv.org/abs/1108.2851}{{\tt arXiv:1108.2851 [hep-ph]}}.

\bibitem{Hoeche:2014qda}
S.~Hoeche, F.~Krauss, P.~Maierhoefer, S.~Pozzorini, M.~Schonherr, et al., {\em
  {Next-to-leading order QCD predictions for top-quark pair production with up
  to two jets merged with a parton shower}\/},
\href{http://arxiv.org/abs/1402.6293}{{\tt arXiv:1402.6293 [hep-ph]}}.

\bibitem{Kardos:2011na}
A.~Kardos, Z.~Trocsanyi, and C.~Papadopoulos, {\em {Top quark pair production
  in association with a Z-boson at NLO accuracy}\/},
  \href{http://dx.doi.org/10.1103/PhysRevD.85.054015}{Phys.Rev. {\bf D85}
  (2012)  054015},
\href{http://arxiv.org/abs/1111.0610}{{\tt arXiv:1111.0610 [hep-ph]}}.

\bibitem{Garzelli:2011is}
M.~Garzelli, A.~Kardos, C.~Papadopoulos, and Z.~Trocsanyi, {\em {Z0 - boson
  production in association with a top anti-top pair at NLO accuracy with
  parton shower effects}\/},
  \href{http://dx.doi.org/10.1103/PhysRevD.85.074022}{Phys.Rev. {\bf D85}
  (2012)  074022},
\href{http://arxiv.org/abs/1111.1444}{{\tt arXiv:1111.1444 [hep-ph]}}.

\bibitem{Campbell:2012dh}
J.~M. Campbell and R.~K. Ellis, {\em {$t \bar{t} W^{+-}$ production and decay
  at NLO}\/},  \href{http://dx.doi.org/10.1007/JHEP07(2012)052}{JHEP {\bf 1207}
  (2012)  052},
\href{http://arxiv.org/abs/1204.5678}{{\tt arXiv:1204.5678 [hep-ph]}}.

\bibitem{Garzelli:2012bn}
M.~Garzelli, A.~Kardos, C.~Papadopoulos, and Z.~Trocsanyi, {\em {t $\bar{t}$
  $W^{+-}$ and t $\bar{t}$ Z Hadroproduction at NLO accuracy in QCD with Parton
  Shower and Hadronization effects}\/},
  \href{http://dx.doi.org/10.1007/JHEP11(2012)056}{JHEP {\bf 1211} (2012)
  056},
\href{http://arxiv.org/abs/1208.2665}{{\tt arXiv:1208.2665 [hep-ph]}}.

\bibitem{Lazopoulos:2008de}
A.~Lazopoulos, T.~McElmurry, K.~Melnikov, and F.~Petriello, {\em
  {Next-to-leading order QCD corrections to $t \bar{t} Z$ production at the
  LHC}\/},  \href{http://dx.doi.org/10.1016/j.physletb.2008.06.073}{Phys.Lett.
  {\bf B666} (2008)  62--65},
\href{http://arxiv.org/abs/0804.2220}{{\tt arXiv:0804.2220 [hep-ph]}}.

\bibitem{Chatrchyan:2013qca}
{CMS Collaboration} Collaboration, S.~Chatrchyan et al., {\em {Measurement of
  associated production of vector bosons and top quark-antiquark pairs at
  sqrt(s) = 7 TeV}\/},
  \href{http://dx.doi.org/10.1103/PhysRevLett.110.172002}{Phys.Rev.Lett. {\bf
  110} (2013)  172002},
\href{http://arxiv.org/abs/1303.3239}{{\tt arXiv:1303.3239 [hep-ex]}}.

\bibitem{Agashe:2013hma}
{Top Quark Working Group} Collaboration, K.~Agashe et al., {\em {Snowmass 2013
  Top quark working group report}\/},
\href{http://arxiv.org/abs/1311.2028}{{\tt arXiv:1311.2028 [hep-ph]}}.

\bibitem{Aad:2012xca}
{ATLAS Collaboration} Collaboration, G.~Aad et al., {\em {Evidence for the
  associated production of a $W$ boson and a top quark in ATLAS at $\sqrt{s}=7$
  TeV}\/},  \href{http://dx.doi.org/10.1016/j.physletb.2012.08.011}{Phys.Lett.
  {\bf B716} (2012)  142--159},
\href{http://arxiv.org/abs/1205.5764}{{\tt arXiv:1205.5764 [hep-ex]}}.

\bibitem{Chatrchyan:2012zca}
{CMS Collaboration} Collaboration, S.~Chatrchyan et al., {\em {Evidence for
  associated production of a single top quark and W boson in pp collisions at
  $\sqrt{s}$ = 7 TeV}\/},
  \href{http://dx.doi.org/10.1103/PhysRevLett.110.022003}{Phys.Rev.Lett. {\bf
  110} (2013)  022003},
\href{http://arxiv.org/abs/1209.3489}{{\tt arXiv:1209.3489 [hep-ex]}}.

\bibitem{Harris:2002md}
B.~Harris, E.~Laenen, L.~Phaf, Z.~Sullivan, and S.~Weinzierl, {\em {The Fully
  differential single top quark cross-section in next to leading order QCD}\/},
   \href{http://dx.doi.org/10.1103/PhysRevD.66.054024}{Phys.Rev. {\bf D66}
  (2002)  054024},
\href{http://arxiv.org/abs/hep-ph/0207055}{{\tt arXiv:hep-ph/0207055
  [hep-ph]}}.

\bibitem{Campbell:2004ch}
J.~M. Campbell, R.~K. Ellis, and F.~Tramontano, {\em {Single top production and
  decay at next-to-leading order}\/},
  \href{http://dx.doi.org/10.1103/PhysRevD.70.094012}{Phys.Rev. {\bf D70}
  (2004)  094012},
\href{http://arxiv.org/abs/hep-ph/0408158}{{\tt arXiv:hep-ph/0408158
  [hep-ph]}}.

\bibitem{Campbell:2013yla}
J.~Campbell, R.~K. Ellis, and R.~Rontsch, {\em {Single top production in
  association with a Z boson at the LHC}\/},
  \href{http://dx.doi.org/10.1103/PhysRevD.87.114006}{Phys.Rev. {\bf D87}
  (2013)  114006},
\href{http://arxiv.org/abs/1302.3856}{{\tt arXiv:1302.3856 [hep-ph]}}.

\bibitem{Kidonakis:2006bu}
N.~Kidonakis, {\em {Single top production at the Tevatron: Threshold
  resummation and finite-order soft gluon corrections}\/},
  \href{http://dx.doi.org/10.1103/PhysRevD.74.114012}{Phys.Rev. {\bf D74}
  (2006)  114012},
\href{http://arxiv.org/abs/hep-ph/0609287}{{\tt arXiv:hep-ph/0609287
  [hep-ph]}}.

\bibitem{Kidonakis:2011wy}
N.~Kidonakis, {\em {Next-to-next-to-leading-order collinear and soft gluon
  corrections for t-channel single top quark production}\/},
  \href{http://dx.doi.org/10.1103/PhysRevD.83.091503}{Phys.Rev. {\bf D83}
  (2011)  091503},
\href{http://arxiv.org/abs/1103.2792}{{\tt arXiv:1103.2792 [hep-ph]}}.

\bibitem{Dittmaier:2012kx}
S.~Dittmaier, A.~Huss, and C.~Speckner, {\em {Weak radiative corrections to
  dijet production at hadron colliders}\/},
  \href{http://dx.doi.org/10.1007/JHEP11(2012)095}{JHEP {\bf 1211} (2012)
  095},
\href{http://arxiv.org/abs/1210.0438}{{\tt arXiv:1210.0438 [hep-ph]}}.

\bibitem{Aad:2011fc}
{ATLAS} Collaboration, G.~Aad et al., {\em {Measurement of inclusive jet and
  dijet production in $pp$ collisions at $\sqrt{s}=7$ TeV using the ATLAS
  detector}\/},  \href{http://dx.doi.org/10.1103/PhysRevD.86.014022}{Phys.Rev.
  {\bf D86} (2012)  014022},
\href{http://arxiv.org/abs/1112.6297}{{\tt arXiv:1112.6297 [hep-ex]}}.

\bibitem{Aad:2013tea}
{ATLAS Collaboration} Collaboration, G.~Aad et al., {\em {Measurement of dijet
  cross sections in pp collisions at 7 TeV centre-of-mass energy using the
  ATLAS detector}\/},
\href{http://arxiv.org/abs/1312.3524}{{\tt arXiv:1312.3524 [hep-ex]}}.

\bibitem{Chatrchyan:2012bja}
{CMS Collaboration} Collaboration, S.~Chatrchyan et al., {\em {Measurements of
  differential jet cross sections in proton-proton collisions at $\sqrt{s}=7$
  TeV with the CMS detector}\/},
  \href{http://dx.doi.org/10.1103/PhysRevD.87.112002}{Phys.Rev. {\bf D87}
  (2013)  112002},
\href{http://arxiv.org/abs/1212.6660}{{\tt arXiv:1212.6660 [hep-ex]}}.

\bibitem{Chatrchyan:2013txa}
{CMS Collaboration} Collaboration, S.~Chatrchyan et al., {\em {Measurement of
  the ratio of the inclusive 3-jet cross section to the inclusive 2-jet cross
  section in pp collisions at $\sqrt{s}$ = 7 TeV and first determination of the
  strong coupling constant in the TeV range}\/},
  \href{http://dx.doi.org/10.1140/epjc/s10052-013-2604-6}{Eur.Phys.J. {\bf C73}
  (2013)  2604},
\href{http://arxiv.org/abs/1304.7498}{{\tt arXiv:1304.7498 [hep-ex]}}.

\bibitem{Nagy:2001fj}
Z.~Nagy, {\em Three jet cross-sections in hadron hadron collisions at
  next-to-leading order\/},
  \href{http://dx.doi.org/10.1103/PhysRevLett.88.122003}{Phys. Rev. Lett. {\bf
  88} (2002)  122003},
\href{http://arxiv.org/abs/hep-ph/0110315}{{\tt arXiv:hep-ph/0110315}}.

\bibitem{Nagy:2003tz}
Z.~Nagy, {\em Next-to-leading order calculation of three jet observables in
  hadron hadron collision\/},
  \href{http://dx.doi.org/10.1103/PhysRevD.68.094002}{Phys. Rev. D {\bf 68}
  (2003)  094002},
\href{http://arxiv.org/abs/hep-ph/0307268}{{\tt arXiv:hep-ph/0307268}}.

\bibitem{Aad:2011dm}
{ATLAS} Collaboration, G.~Aad et al., {\em {Measurement of the inclusive
  $W^\pm$ and Z/gamma cross sections in the electron and muon decay channels in
  $pp$ collisions at $\sqrt{s}=7$ TeV with the ATLAS detector}\/},
  \href{http://dx.doi.org/10.1103/PhysRevD.85.072004}{Phys.Rev. {\bf D85}
  (2012)  072004},
\href{http://arxiv.org/abs/1109.5141}{{\tt arXiv:1109.5141 [hep-ex]}}.

\bibitem{Chatrchyan:2014mua}
{CMS Collaboration} Collaboration, S.~Chatrchyan et al., {\em {Measurement of
  inclusive W and Z boson production cross sections in pp collisions at sqrt(s)
  = 8 TeV}\/},
\href{http://arxiv.org/abs/1402.0923}{{\tt arXiv:1402.0923 [hep-ex]}}.

\bibitem{Boughezal:2013cwa}
R.~Boughezal, Y.~Li, and F.~Petriello, {\em {Disentangling radiative
  corrections using high-mass Drell-Yan at the LHC}\/},
  \href{http://dx.doi.org/10.1103/PhysRevD.89.034030}{Phys.Rev. {\bf D89}
  (2014)  034030},
\href{http://arxiv.org/abs/1312.3972}{{\tt arXiv:1312.3972 [hep-ph]}}.

\bibitem{Aad:2012sb}
{ATLAS Collaboration} Collaboration, G.~Aad et al., {\em {Determination of the
  strange quark density of the proton from ATLAS measurements of the $W \to
  \ell \nu$ and $Z \to \ell\ell$ cross sections}\/},
  \href{http://dx.doi.org/10.1103/PhysRevLett.109.012001}{Phys.Rev.Lett. {\bf
  109} (2012)  012001},
\href{http://arxiv.org/abs/1203.4051}{{\tt arXiv:1203.4051 [hep-ex]}}.

\bibitem{Denner:2009gj}
A.~Denner, S.~Dittmaier, T.~Kasprzik, and A.~{M\"uck}, {\em {Electroweak
  corrections to W + jet hadroproduction including leptonic W-boson decays}\/},
   \href{http://dx.doi.org/10.1088/1126-6708/2009/08/075}{JHEP {\bf 0908}
  (2009)  075},
\href{http://arxiv.org/abs/0906.1656}{{\tt arXiv:0906.1656 [hep-ph]}}.

\bibitem{Denner:2011vu}
A.~Denner, S.~Dittmaier, T.~Kasprzik, and A.~{M\"uck}, {\em {Electroweak
  corrections to dilepton + jet production at hadron colliders}\/},
  \href{http://dx.doi.org/10.1007/JHEP06(2011)069}{JHEP {\bf 1106} (2011)
  069},
\href{http://arxiv.org/abs/1103.0914}{{\tt arXiv:1103.0914 [hep-ph]}}.

\bibitem{Denner:2012ts}
A.~Denner, S.~Dittmaier, T.~Kasprzik, and A.~{M\"uck}, {\em {Electroweak
  corrections to monojet production at the LHC}\/},
  \href{http://dx.doi.org/10.1140/epjc/s10052-013-2297-x}{Eur.Phys.J. {\bf C73}
  (2013)  2297},
\href{http://arxiv.org/abs/1211.5078}{{\tt arXiv:1211.5078 [hep-ph]}}.

\bibitem{Denner:2013fca}
A.~Denner, L.~Hofer, A.~Scharf, and S.~Uccirati, {\em {Electroweak corrections
  to Z + 2 jets production at the LHC}\/},
\href{http://arxiv.org/abs/1311.5336}{{\tt arXiv:1311.5336 [hep-ph]}}.

\bibitem{Aad:2013ysa}
{ATLAS Collaboration} Collaboration, G.~Aad et al., {\em {Measurement of the
  production cross section of jets in association with a Z boson in pp
  collisions at $\sqrt{s}$ = 7 TeV with the ATLAS detector}\/},
  \href{http://dx.doi.org/10.1007/JHEP07(2013)032}{JHEP {\bf 1307} (2013)
  032},
\href{http://arxiv.org/abs/1304.7098}{{\tt arXiv:1304.7098 [hep-ex]}}.

\bibitem{Aad:2012twa}
{ATLAS Collaboration} Collaboration, G.~Aad et al., {\em {Measurement of $WZ$
  production in proton-proton collisions at $\sqrt{s}=7$ TeV with the ATLAS
  detector}\/},
  \href{http://dx.doi.org/10.1140/epjc/s10052-012-2173-0}{Eur.Phys.J. {\bf C72}
  (2012)  2173},
\href{http://arxiv.org/abs/1208.1390}{{\tt arXiv:1208.1390 [hep-ex]}}.

\bibitem{Chatrchyan:2013yaa}
{CMS Collaboration} Collaboration, S.~Chatrchyan et al., {\em {Measurement of
  the $W^+W^-$ Cross section in $pp$ Collisions at $\sqrt{s} = 7$ TeV and
  Limits on Anomalous $WW\gamma$ and $WWZ$ couplings}\/},
  \href{http://dx.doi.org/10.1140/epjc/s10052-013-2610-8}{Eur.Phys.J. {\bf C73}
  (2013)  2610},
\href{http://arxiv.org/abs/1306.1126}{{\tt arXiv:1306.1126 [hep-ex]}}.

\bibitem{Bierweiler:2012kw}
A.~Bierweiler, T.~Kasprzik, J.~H. {K\"uhn}, and S.~Uccirati, {\em {Electroweak
  corrections to W-boson pair production at the LHC}\/},
  \href{http://dx.doi.org/10.1007/JHEP11(2012)093}{JHEP {\bf 1211} (2012)
  093},
\href{http://arxiv.org/abs/1208.3147}{{\tt arXiv:1208.3147 [hep-ph]}}.

\bibitem{Bierweiler:2013dja}
A.~Bierweiler, T.~Kasprzik, and J.~H. {K\"uhn}, {\em {Vector-boson pair
  production at the LHC to $\mathcal{O}(\alpha^3)$ accuracy}\/},
  \href{http://dx.doi.org/10.1007/JHEP12(2013)071}{JHEP {\bf 1312} (2013)
  071},
\href{http://arxiv.org/abs/1305.5402}{{\tt arXiv:1305.5402 [hep-ph]}}.

\bibitem{Baglio:2013toa}
J.~Baglio, L.~D. Ninh, and M.~M. Weber, {\em {Massive gauge boson pair
  production at the LHC: a next-to-leading order story}\/},
  \href{http://dx.doi.org/10.1103/PhysRevD.88.113005}{Phys.Rev. {\bf D88}
  (2013)  113005},
\href{http://arxiv.org/abs/1307.4331}{{\tt arXiv:1307.4331}}.

\bibitem{Billoni:2013aba}
M.~Billoni, S.~Dittmaier, B.~{J\"ager}, and C.~Speckner, {\em {Next-to-leading
  order electroweak corrections to $\Pp\Pp \to W^+W^-\to4\,$leptons at the LHC
  in double-pole approximation}\/},
  \href{http://dx.doi.org/10.1007/JHEP12(2013)043}{JHEP {\bf 1312} (2013)
  043},
\href{http://arxiv.org/abs/1310.1564}{{\tt arXiv:1310.1564 [hep-ph]}}.

\bibitem{Gieseke:2014gka}
S.~Gieseke, T.~Kasprzik, and J.~H. {K\"uhn}, {\em {Vector-boson pair production
  and electroweak corrections in HERWIG++}\/},
\href{http://arxiv.org/abs/1401.3964}{{\tt arXiv:1401.3964 [hep-ph]}}.

\bibitem{ATLAS:2012mec}
{ATLAS Collaboration} Collaboration, G.~Aad et al., {\em {Measurement of
  $W^+W^-$ production in $pp$ collisions at $\sqrt{s}=7$ TeV with the ATLAS
  detector and limits on anomalous $WWZ$ and $WW\gamma$ couplings}\/},
  \href{http://dx.doi.org/10.1103/PhysRevD.87.112001}{Phys.Rev. {\bf D87}
  (2013)  112001},
\href{http://arxiv.org/abs/1210.2979}{{\tt arXiv:1210.2979 [hep-ex]}}.

\bibitem{Aad:2012awa}
{ATLAS Collaboration} Collaboration, G.~Aad et al., {\em {Measurement of $ZZ$
  production in $pp$ collisions at $\sqrt{s}=7$ TeV and limits on anomalous
  $ZZZ$ and $ZZ\gamma$ couplings with the ATLAS detector}\/},
  \href{http://dx.doi.org/10.1007/JHEP03(2013)128}{JHEP {\bf 1303} (2013)
  128},
\href{http://arxiv.org/abs/1211.6096}{{\tt arXiv:1211.6096 [hep-ex]}}.

\bibitem{Chatrchyan:2012sga}
{CMS Collaboration} Collaboration, S.~Chatrchyan et al., {\em {Measurement of
  the $ZZ$ production cross section and search for anomalous couplings in 2 l2l
  ' final states in $pp$ collisions at $\sqrt{s}=7$ TeV}\/},
  \href{http://dx.doi.org/10.1007/JHEP01(2013)063}{JHEP {\bf 1301} (2013)
  063},
\href{http://arxiv.org/abs/1211.4890}{{\tt arXiv:1211.4890 [hep-ex]}}.

\bibitem{Chatrchyan:2013oev}
{CMS Collaboration} Collaboration, S.~Chatrchyan et al., {\em {Measurement of
  W+W- and ZZ production cross sections in pp collisions at sqrt(s) = 8
  TeV}\/},  \href{http://dx.doi.org/10.1016/j.physletb.2013.03.027}{Phys.Lett.
  {\bf B721} (2013)  190--211},
\href{http://arxiv.org/abs/1301.4698}{{\tt arXiv:1301.4698 [hep-ex]}}.

\bibitem{Bonvini:2013jha}
M.~Bonvini, F.~Caola, S.~Forte, K.~Melnikov, and G.~Ridolfi, {\em
  {Signal-background interference effects for $gg \to H \to W^+ W^-$ beyond
  leading order}\/},
  \href{http://dx.doi.org/10.1103/PhysRevD.88.034032}{Phys.Rev. {\bf D88}
  (2013)  034032},
\href{http://arxiv.org/abs/1304.3053}{{\tt arXiv:1304.3053 [hep-ph]}}.

\bibitem{Aad:2013izg}
{ATLAS Collaboration} Collaboration, G.~Aad et al., {\em {Measurements of
  $W\gamma$ and $Z\gamma$ production in pp collisions at $\sqrt{s}$= 7 TeV with
  the ATLAS detector at the LHC}\/},
  \href{http://dx.doi.org/10.1103/PhysRevD.87.112003}{Phys.Rev. {\bf D87}
  (2013)  112003},
\href{http://arxiv.org/abs/1302.1283}{{\tt arXiv:1302.1283 [hep-ex]}}.

\bibitem{Chatrchyan:2013nda}
{CMS Collaboration} Collaboration, S.~Chatrchyan et al., {\em {Measurement of
  the production cross section for $Z\gamma \to \nu\bar{\nu}\gamma$ in pp
  collisions at $\sqrt{s} =$ 7 TeV and limits on $ZZ\gamma$ and $Z\gamma\gamma$
  triple gauge boson couplings}\/},
  \href{http://dx.doi.org/10.1007/JHEP10(2013)164}{JHEP {\bf 1310} (2013)
  164},
\href{http://arxiv.org/abs/1309.1117}{{\tt arXiv:1309.1117 [hep-ex]}}.

\bibitem{Nhung:2013jta}
D.~T. Nhung, L.~D. Ninh, and M.~M. Weber, {\em {NLO corrections to WWZ
  production at the LHC}\/},
  \href{http://dx.doi.org/10.1007/JHEP12(2013)096}{JHEP {\bf 1312} (2013)
  096},
\href{http://arxiv.org/abs/1307.7403}{{\tt arXiv:1307.7403 [hep-ph]}}.

\bibitem{Aad:2012tba}
{ATLAS} Collaboration, G.~Aad et al., {\em {Measurement of isolated-photon pair
  production in $pp$ collisions at $\sqrt{s}=7$ TeV with the ATLAS
  detector}\/},  \href{http://dx.doi.org/10.1007/JHEP01(2013)086}{JHEP {\bf
  1301} (2013)  086},
\href{http://arxiv.org/abs/1211.1913}{{\tt arXiv:1211.1913 [hep-ex]}}.

\bibitem{Chatrchyan:2011qt}
{CMS} Collaboration, S.~Chatrchyan et al., {\em {Measurement of the Production
  Cross Section for Pairs of Isolated Photons in $pp$ collisions at
  $\sqrt{s}=7$ TeV}\/},  \href{http://dx.doi.org/10.1007/JHEP01(2012)133}{JHEP
  {\bf 1201} (2012)  133},
\href{http://arxiv.org/abs/1110.6461}{{\tt arXiv:1110.6461 [hep-ex]}}.

\bibitem{Denner:1991kt}
A.~Denner, {\em {Techniques for calculation of electroweak radiative
  corrections at the one loop level and results for W physics at LEP-200}\/},
  \href{http://dx.doi.org/10.1002/prop.2190410402}{Fortsch.Phys. {\bf 41}
  (1993)  307--420},
\href{http://arxiv.org/abs/0709.1075}{{\tt arXiv:0709.1075 [hep-ph]}}.

\bibitem{Eidelman:1995ny}
S.~Eidelman and F.~Jegerlehner, {\em {Hadronic contributions to g-2 of the
  leptons and to the effective fine structure constant $\alpha(M_Z^2)$}\/},
  \href{http://dx.doi.org/10.1007/BF01553984}{Z.Phys. {\bf C67} (1995)
  585--602},
\href{http://arxiv.org/abs/hep-ph/9502298}{{\tt arXiv:hep-ph/9502298
  [hep-ph]}}.

\bibitem{Sirlin:1980nh}
A.~Sirlin, {\em {Radiative Corrections in the SU(2)$_L\times$U(1) Theory: A
  Simple Renormalization Framework}\/},
\href{http://dx.doi.org/10.1103/PhysRevD.22.971}{Phys.Rev. {\bf D22} (1980)
  971--981}.

\bibitem{Consoli:1989fg}
M.~Consoli, W.~Hollik, and F.~Jegerlehner, {\em {The Effect of the Top Quark on
  the $M_W-M_Z$ Interdependence and Possible Decoupling of Heavy Fermions from
  Low-Energy Physics}\/},
\href{http://dx.doi.org/10.1016/0370-2693(89)91301-4}{Phys.Lett. {\bf B227}
  (1989)  167}.

\bibitem{Dittmaier:2001ay}
S.~Dittmaier and M.~Kr{\"a}mer, {\em {Electroweak radiative corrections to W
  boson production at hadron colliders}\/},
  \href{http://dx.doi.org/10.1103/PhysRevD.65.073007}{Phys.Rev. {\bf D65}
  (2002)  073007},
\href{http://arxiv.org/abs/hep-ph/0109062}{{\tt arXiv:hep-ph/0109062
  [hep-ph]}}.

\bibitem{Brensing:2007qm}
S.~Brensing, S.~Dittmaier, M.~Kr{\"a}mer, and A.~M{\"u}ck, {\em {Radiative
  corrections to $W^-$ boson hadroproduction: Higher-order electroweak and
  supersymmetric effects}\/},
  \href{http://dx.doi.org/10.1103/PhysRevD.77.073006}{Phys.Rev. {\bf D77}
  (2008)  073006},
\href{http://arxiv.org/abs/0710.3309}{{\tt arXiv:0710.3309 [hep-ph]}}.

\bibitem{Dittmaier:2009cr}
S.~Dittmaier and M.~Huber, {\em {Radiative corrections to the neutral-current
  Drell-Yan process in the Standard Model and its minimal supersymmetric
  extension}\/},  \href{http://dx.doi.org/10.1007/JHEP01(2010)060}{JHEP {\bf
  1001} (2010)  060},
\href{http://arxiv.org/abs/0911.2329}{{\tt arXiv:0911.2329 [hep-ph]}}.

\bibitem{Fadin:1999bq}
V.~S. Fadin, L.~Lipatov, A.~D. Martin, and M.~Melles, {\em {Resummation of
  double logarithms in electroweak high-energy processes}\/},
  \href{http://dx.doi.org/10.1103/PhysRevD.61.094002}{Phys.Rev. {\bf D61}
  (2000)  094002},
\href{http://arxiv.org/abs/hep-ph/9910338}{{\tt arXiv:hep-ph/9910338
  [hep-ph]}}.

\bibitem{Kuhn:1999nn}
J.~H. {K\"uhn}, A.~Penin, and V.~A. Smirnov, {\em {Summing up subleading
  Sudakov logarithms}\/},
  \href{http://dx.doi.org/10.1007/s100520000462}{Eur.Phys.J. {\bf C17} (2000)
  97--105},
\href{http://arxiv.org/abs/hep-ph/9912503}{{\tt arXiv:hep-ph/9912503
  [hep-ph]}}.

\bibitem{Ciafaloni:2000df}
M.~Ciafaloni, P.~Ciafaloni, and D.~Comelli, {\em {Bloch-Nordsieck violating
  electroweak corrections to inclusive TeV scale hard processes}\/},
  \href{http://dx.doi.org/10.1103/PhysRevLett.84.4810}{Phys.Rev.Lett. {\bf 84}
  (2000)  4810--4813},
\href{http://arxiv.org/abs/hep-ph/0001142}{{\tt arXiv:hep-ph/0001142
  [hep-ph]}}.

\bibitem{Hori:2000tm}
M.~Hori, H.~Kawamura, and J.~Kodaira, {\em {Electroweak Sudakov at two loop
  level}\/},  \href{http://dx.doi.org/10.1016/S0370-2693(00)01027-3}{Phys.Lett.
  {\bf B491} (2000)  275--279},
\href{http://arxiv.org/abs/hep-ph/0007329}{{\tt arXiv:hep-ph/0007329
  [hep-ph]}}.

\bibitem{Denner:2001gw}
A.~Denner and S.~Pozzorini, {\em {One loop leading logarithms in electroweak
  radiative corrections. 2. Factorization of collinear singularities}\/},
  \href{http://dx.doi.org/10.1007/s100520100721}{Eur.Phys.J. {\bf C21} (2001)
  63--79},
\href{http://arxiv.org/abs/hep-ph/0104127}{{\tt arXiv:hep-ph/0104127
  [hep-ph]}}.

\bibitem{Melles:2001dh}
M.~Melles, {\em {Resummation of angular dependent corrections in spontaneously
  broken gauge theories}\/},
  \href{http://dx.doi.org/10.1007/s100520200942}{Eur.Phys.J. {\bf C24} (2002)
  193--204},
\href{http://arxiv.org/abs/hep-ph/0108221}{{\tt arXiv:hep-ph/0108221
  [hep-ph]}}.

\bibitem{Beenakker:2001kf}
W.~Beenakker and A.~Werthenbach, {\em {Electroweak two loop Sudakov logarithms
  for on-shell fermions and bosons}\/},
  \href{http://dx.doi.org/10.1016/S0550-3213(02)00171-2}{Nucl.Phys. {\bf B630}
  (2002)  3--54},
\href{http://arxiv.org/abs/hep-ph/0112030}{{\tt arXiv:hep-ph/0112030
  [hep-ph]}}.

\bibitem{Denner:2003wi}
A.~Denner, M.~Melles, and S.~Pozzorini, {\em {Two loop electroweak angular
  dependent logarithms at high-energies}\/},
  \href{http://dx.doi.org/10.1016/S0550-3213(03)00307-9}{Nucl.Phys. {\bf B662}
  (2003)  299--333},
\href{http://arxiv.org/abs/hep-ph/0301241}{{\tt arXiv:hep-ph/0301241
  [hep-ph]}}.

\bibitem{Jantzen:2005xi}
B.~Jantzen, J.~H. {K\"uhn}, A.~A. Penin, and V.~A. Smirnov, {\em {Two-loop
  electroweak logarithms}\/},
  \href{http://dx.doi.org/10.1103/PhysRevD.74.019901,
  10.1103/PhysRevD.72.051301}{Phys.Rev. {\bf D72} (2005)  051301},
\href{http://arxiv.org/abs/hep-ph/0504111}{{\tt arXiv:hep-ph/0504111
  [hep-ph]}}.

\bibitem{Jantzen:2005az}
B.~Jantzen, J.~H. {K\"uhn}, A.~A. Penin, and V.~A. Smirnov, {\em {Two-loop
  electroweak logarithms in four-fermion processes at high energy}\/},
  \href{http://dx.doi.org/10.1016/j.nuclphysb.2005.10.010,
  10.1016/j.nuclphysb.2006.07.004}{Nucl.Phys. {\bf B731} (2005)  188--212},
\href{http://arxiv.org/abs/hep-ph/0509157}{{\tt arXiv:hep-ph/0509157
  [hep-ph]}}.

\bibitem{Denner:2006jr}
A.~Denner, B.~Jantzen, and S.~Pozzorini, {\em {Two-loop electroweak
  next-to-leading logarithmic corrections to massless fermionic processes}\/},
  \href{http://dx.doi.org/10.1016/j.nuclphysb.2006.10.014}{Nucl.Phys. {\bf
  B761} (2007)  1--62},
\href{http://arxiv.org/abs/hep-ph/0608326}{{\tt arXiv:hep-ph/0608326
  [hep-ph]}}.

\bibitem{Christiansen:2014kba}
J.~R. Christiansen and {Sj\"ostrand, Torbj\"orn}, {\em {Weak Gauge Boson
  Radiation in Parton Showers}\/},
\href{http://arxiv.org/abs/1401.5238}{{\tt arXiv:1401.5238 [hep-ph]}}.

\bibitem{Ciafaloni:2006qu}
P.~Ciafaloni and D.~Comelli, {\em {The Importance of weak bosons emission at
  LHC}\/},  \href{http://dx.doi.org/10.1088/1126-6708/2006/09/055}{JHEP {\bf
  0609} (2006)  055},
\href{http://arxiv.org/abs/hep-ph/0604070}{{\tt arXiv:hep-ph/0604070
  [hep-ph]}}.

\bibitem{Baur:2006sn}
U.~Baur, {\em {Weak Boson Emission in Hadron Collider Processes}\/},
  \href{http://dx.doi.org/10.1103/PhysRevD.75.013005}{Phys.Rev. {\bf D75}
  (2007)  013005},
\href{http://arxiv.org/abs/hep-ph/0611241}{{\tt arXiv:hep-ph/0611241
  [hep-ph]}}.

\bibitem{Bell:2010gi}
G.~Bell, J.~{K\"uhn}, and J.~Rittinger, {\em {Electroweak Sudakov Logarithms
  and Real Gauge-Boson Radiation in the TeV Region}\/},
  \href{http://dx.doi.org/10.1140/epjc/s10052-010-1489-x}{Eur.Phys.J. {\bf C70}
  (2010)  659--671},
\href{http://arxiv.org/abs/1004.4117}{{\tt arXiv:1004.4117 [hep-ph]}}.

\bibitem{Chiesa:2013yma}
M.~Chiesa, G.~Montagna, L.~Barzè, M.~Moretti, O.~Nicrosini, et al., {\em
  {Electroweak Sudakov Corrections to New Physics Searches at the LHC}\/},
  \href{http://dx.doi.org/10.1103/PhysRevLett.111.121801}{Phys.Rev.Lett. {\bf
  111} (2013)  121801},
\href{http://arxiv.org/abs/1305.6837}{{\tt arXiv:1305.6837 [hep-ph]}}.

\bibitem{Beenakker:1996kt}
W.~Beenakker et al., {\em {WW cross-sections and distributions}\/},
\href{http://arxiv.org/abs/hep-ph/9602351}{{\tt arXiv:hep-ph/9602351
  [hep-ph]}}.

\bibitem{Arbuzov:1999cq}
A.~Arbuzov, {\em {Nonsinglet splitting functions in QED}\/},
  \href{http://dx.doi.org/10.1016/S0370-2693(99)01290-3}{Phys.Lett. {\bf B470}
  (1999)  252--258},
\href{http://arxiv.org/abs/hep-ph/9908361}{{\tt arXiv:hep-ph/9908361
  [hep-ph]}}.

\bibitem{Placzek:2003zg}
W.~Placzek and S.~Jadach, {\em {Multiphoton radiation in leptonic W boson
  decays}\/},  \href{http://dx.doi.org/10.1140/epjc/s2003-01223-4}{Eur.Phys.J.
  {\bf C29} (2003)  325--339},
\href{http://arxiv.org/abs/hep-ph/0302065}{{\tt arXiv:hep-ph/0302065
  [hep-ph]}}.

\bibitem{CarloniCalame:2003ux}
C.~Carloni~Calame, G.~Montagna, O.~Nicrosini, and M.~Treccani, {\em {Higher
  order QED corrections to W boson mass determination at hadron colliders}\/},
  \href{http://dx.doi.org/10.1103/PhysRevD.69.037301}{Phys.Rev. {\bf D69}
  (2004)  037301},
\href{http://arxiv.org/abs/hep-ph/0303102}{{\tt arXiv:hep-ph/0303102
  [hep-ph]}}.

\bibitem{Golonka:2006tw}
P.~Golonka and Z.~Was, {\em {Next to Leading Logarithms and the PHOTOS Monte
  Carlo}\/},
  \href{http://dx.doi.org/10.1140/epjc/s10052-006-0205-3}{Eur.Phys.J. {\bf C50}
  (2007)  53--62},
\href{http://arxiv.org/abs/hep-ph/0604232}{{\tt arXiv:hep-ph/0604232
  [hep-ph]}}.

\bibitem{Kinoshita:1962ur}
T.~Kinoshita, {\em {Mass singularities of Feynman amplitudes}\/},
\href{http://dx.doi.org/10.1063/1.1724268}{J.Math.Phys. {\bf 3} (1962)
  650--677}.

\bibitem{Lee:1964is}
T.~Lee and M.~Nauenberg, {\em {Degenerate Systems and Mass Singularities}\/},
\href{http://dx.doi.org/10.1103/PhysRev.133.B1549}{Phys.Rev. {\bf 133} (1964)
  B1549--B1562}.

\bibitem{Aad:2011gj}
{ATLAS Collaboration} Collaboration, G.~Aad et al., {\em {Measurement of the
  transverse momentum distribution of $\PZ/\gamma^*$ bosons in proton-proton
  collisions at $\sqrt{s}=7$ TeV with the ATLAS detector}\/},
  \href{http://dx.doi.org/10.1016/j.physletb.2011.10.018}{Phys.Lett. {\bf B705}
  (2011)  415--434},
\href{http://arxiv.org/abs/1107.2381}{{\tt arXiv:1107.2381 [hep-ex]}}.

\bibitem{Martin:2004dh}
A.~Martin, R.~Roberts, W.~Stirling, and R.~Thorne, {\em {Parton distributions
  incorporating QED contributions}\/},
  \href{http://dx.doi.org/10.1140/epjc/s2004-02088-7}{Eur.Phys.J. {\bf C39}
  (2005)  155--161},
\href{http://arxiv.org/abs/hep-ph/0411040}{{\tt arXiv:hep-ph/0411040
  [hep-ph]}}.

\bibitem{Ball:2013hta}
{NNPDF} Collaboration, R.~D. Ball et al., {\em {Parton distributions with QED
  corrections}\/},
  \href{http://dx.doi.org/10.1016/j.nuclphysb.2013.10.010}{Nucl.Phys. {\bf
  B877} (2013) no.~2, 290--320},
\href{http://arxiv.org/abs/1308.0598}{{\tt arXiv:1308.0598 [hep-ph]}}.

\bibitem{Diener:2005me}
K.-P. Diener, S.~Dittmaier, and W.~Hollik, {\em {Electroweak higher-order
  effects and theoretical uncertainties in deep-inelastic neutrino
  scattering}\/},
  \href{http://dx.doi.org/10.1103/PhysRevD.72.093002}{Phys.Rev. {\bf D72}
  (2005)  093002},
\href{http://arxiv.org/abs/hep-ph/0509084}{{\tt arXiv:hep-ph/0509084
  [hep-ph]}}.

\bibitem{CarloniCalame:2007cd}
C.~Carloni~Calame, G.~Montagna, O.~Nicrosini, and A.~Vicini, {\em {Precision
  electroweak calculation of the production of a high transverse-momentum
  lepton pair at hadron colliders}\/},
  \href{http://dx.doi.org/10.1088/1126-6708/2007/10/109}{JHEP {\bf 0710} (2007)
   109},
\href{http://arxiv.org/abs/0710.1722}{{\tt arXiv:0710.1722 [hep-ph]}}.

\bibitem{Denner:2010ia}
A.~Denner, S.~Dittmaier, T.~Gehrmann, and C.~Kurz, {\em {Electroweak
  corrections to hadronic event shapes and jet production in $\Pe^+\Pe^-$
  annihilation}\/},
  \href{http://dx.doi.org/10.1016/j.nuclphysb.2010.04.009}{Nucl.Phys. {\bf
  B836} (2010)  37--90},
\href{http://arxiv.org/abs/1003.0986}{{\tt arXiv:1003.0986 [hep-ph]}}.

\bibitem{Glover:1993xc}
E.~N. Glover and A.~Morgan, {\em {Measuring the photon fragmentation function
  at LEP}\/},
\href{http://dx.doi.org/10.1007/BF01560245}{Z.Phys. {\bf C62} (1994)
  311--322}.

\bibitem{Buskulic:1995au}
{ALEPH Collaboration} Collaboration, D.~Buskulic et al., {\em {First
  measurement of the quark to photon fragmentation function}\/},
\href{http://dx.doi.org/10.1007/s002880050037}{Z.Phys. {\bf C69} (1996)
  365--378}.

\bibitem{Frixione:1998jh}
S.~Frixione, {\em {Isolated photons in perturbative QCD}\/},
  \href{http://dx.doi.org/10.1016/S0370-2693(98)00454-7}{Phys.Lett. {\bf B429}
  (1998)  369--374},
\href{http://arxiv.org/abs/hep-ph/9801442}{{\tt arXiv:hep-ph/9801442
  [hep-ph]}}.

\bibitem{Campbell:2011bn}
J.~M. Campbell, R.~K. Ellis, and C.~Williams, {\em {Vector boson pair
  production at the LHC}\/},
  \href{http://dx.doi.org/10.1007/JHEP07(2011)018}{JHEP {\bf 1107} (2011)
  018},
\href{http://arxiv.org/abs/1105.0020}{{\tt arXiv:1105.0020 [hep-ph]}}.

\bibitem{Binoth:2010ra}
{SM and NLO Multileg Working Group} Collaboration, J.~Andersen et al., {\em
  {The SM and NLO Multileg Working Group: Summary report}\/},
\href{http://arxiv.org/abs/1003.1241}{{\tt arXiv:1003.1241 [hep-ph]}}.

\bibitem{Corcella:2002jc}
G.~Corcella et al., {\em {HERWIG 6.5 release note}\/},
\href{http://arxiv.org/abs/hep-ph/0210213}{{\tt arXiv:hep-ph/0210213
  [hep-ph]}}.

\bibitem{Ciccolini:2007jr}
M.~Ciccolini, A.~Denner, and S.~Dittmaier, {\em {Strong and electroweak
  corrections to the production of Higgs + 2jets via weak interactions at the
  LHC}\/},
  \href{http://dx.doi.org/10.1103/PhysRevLett.99.161803}{Phys.Rev.Lett. {\bf
  99} (2007)  161803},
\href{http://arxiv.org/abs/0707.0381}{{\tt arXiv:0707.0381 [hep-ph]}}.

\bibitem{Ciccolini:2007ec}
M.~Ciccolini, A.~Denner, and S.~Dittmaier, {\em {Electroweak and QCD
  corrections to Higgs production via vector-boson fusion at the LHC}\/},
  \href{http://dx.doi.org/10.1103/PhysRevD.77.013002}{Phys.Rev. {\bf D77}
  (2008)  013002},
\href{http://arxiv.org/abs/0710.4749}{{\tt arXiv:0710.4749 [hep-ph]}}.

\bibitem{Bahr:2008pv}
M.~Bahr, S.~Gieseke, M.~Gigg, D.~Grellscheid, K.~Hamilton, et al., {\em
  {Herwig++ Physics and Manual}\/},
  \href{http://dx.doi.org/10.1140/epjc/s10052-008-0798-9}{Eur.Phys.J. {\bf C58}
  (2008)  639--707},
\href{http://arxiv.org/abs/0803.0883}{{\tt arXiv:0803.0883 [hep-ph]}}.

\bibitem{Arnold:2012fq}
K.~Arnold et al., {\em {Herwig++ 2.6 Release Note}\/},
\href{http://arxiv.org/abs/1205.4902}{{\tt arXiv:1205.4902 [hep-ph]}}.

\bibitem{Gleisberg:2008ta}
T.~Gleisberg, S.~{H\"oche}, F.~Krauss, M.~{Sch\"onherr}, S.~Schumann, et al.,
  {\em {Event generation with SHERPA 1.1}\/},
  \href{http://dx.doi.org/10.1088/1126-6708/2009/02/007}{JHEP {\bf 0902} (2009)
   007},
\href{http://arxiv.org/abs/0811.4622}{{\tt arXiv:0811.4622 [hep-ph]}}.

\bibitem{Cao:2004yy}
Q.-H. Cao and C.~Yuan, {\em {Combined effect of QCD resummation and QED
  radiative correction to $W$ boson observables at the Tevatron}\/},
  \href{http://dx.doi.org/10.1103/PhysRevLett.93.042001}{Phys.Rev.Lett. {\bf
  93} (2004)  042001},
\href{http://arxiv.org/abs/hep-ph/0401026}{{\tt arXiv:hep-ph/0401026
  [hep-ph]}}.

\bibitem{Balossini:2009sa}
G.~Balossini et al., {\em {Combination of electroweak and QCD corrections to
  single W production at the Fermilab Tevatron and the CERN LHC}\/},
  \href{http://dx.doi.org/10.1007/JHEP01(2010)013}{JHEP {\bf 1001} (2010)
  013},
\href{http://arxiv.org/abs/0907.0276}{{\tt arXiv:0907.0276 [hep-ph]}}.

\bibitem{Placzek:2013moa}
W.~Płaczek, S.~Jadach, and M.~Krasny, {\em {Drell-Yan processes with
  WINHAC}\/},  \href{http://dx.doi.org/10.5506/APhysPolB.44.2171}{Acta
  Phys.Polon. {\bf B44} (2013) no.~11, 2171--2178},
\href{http://arxiv.org/abs/1310.5994}{{\tt arXiv:1310.5994 [hep-ph]}}.

\bibitem{Nason:2004rx}
P.~Nason, {\em {A New method for combining NLO QCD with shower Monte Carlo
  algorithms}\/},  \href{http://dx.doi.org/10.1088/1126-6708/2004/11/040}{JHEP
  {\bf 0411} (2004)  040},
\href{http://arxiv.org/abs/hep-ph/0409146}{{\tt arXiv:hep-ph/0409146
  [hep-ph]}}.

\bibitem{Bernaciak:2012hj}
C.~Bernaciak and D.~Wackeroth, {\em {Combining NLO QCD and Electroweak
  Radiative Corrections to W boson Production at Hadron Colliders in the POWHEG
  Framework}\/},  \href{http://dx.doi.org/10.1103/PhysRevD.85.093003}{Phys.Rev.
  {\bf D85} (2012)  093003},
\href{http://arxiv.org/abs/1201.4804}{{\tt arXiv:1201.4804 [hep-ph]}}.

\bibitem{Barze:2012tt}
L.~Barze, G.~Montagna, P.~Nason, O.~Nicrosini, and F.~Piccinini, {\em
  {Implementation of electroweak corrections in the POWHEG BOX: single W
  production}\/},  \href{http://dx.doi.org/10.1007/JHEP04(2012)037}{JHEP {\bf
  1204} (2012)  037},
\href{http://arxiv.org/abs/1202.0465}{{\tt arXiv:1202.0465 [hep-ph]}}.

\bibitem{Barze:2013yca}
L.~Barze, G.~Montagna, P.~Nason, O.~Nicrosini, F.~Piccinini, et al., {\em
  {Neutral current Drell-Yan with combined QCD and electroweak corrections in
  the POWHEG BOX}\/},
  \href{http://dx.doi.org/10.1140/epjc/s10052-013-2474-y}{Eur.Phys.J. {\bf C73}
  (2013)  2474},
\href{http://arxiv.org/abs/1302.4606}{{\tt arXiv:1302.4606 [hep-ph]}}.

\bibitem{Kauer:2007zc}
N.~Kauer, {\em {Narrow-width approximation limitations}\/},
  \href{http://dx.doi.org/10.1016/j.physletb.2007.04.036}{Phys.Lett. {\bf B649}
  (2007)  413--416},
\href{http://arxiv.org/abs/hep-ph/0703077}{{\tt arXiv:hep-ph/0703077
  [hep-ph]}}.

\bibitem{Berdine:2007uv}
D.~Berdine, N.~Kauer, and D.~Rainwater, {\em {Breakdown of the Narrow Width
  Approximation for New Physics}\/},
  \href{http://dx.doi.org/10.1103/PhysRevLett.99.111601}{Phys.Rev.Lett. {\bf
  99} (2007)  111601},
\href{http://arxiv.org/abs/hep-ph/0703058}{{\tt arXiv:hep-ph/0703058
  [hep-ph]}}.

\bibitem{Gambino:1999ai}
P.~Gambino and P.~A. Grassi, {\em {The Nielsen identities of the SM and the
  definition of mass}\/},
  \href{http://dx.doi.org/10.1103/PhysRevD.62.076002}{Phys.Rev. {\bf D62}
  (2000)  076002},
\href{http://arxiv.org/abs/hep-ph/9907254}{{\tt arXiv:hep-ph/9907254
  [hep-ph]}}.

\bibitem{Grassi:2001bz}
P.~A. Grassi, B.~A. Kniehl, and A.~Sirlin, {\em {Width and partial widths of
  unstable particles in the light of the Nielsen identities}\/},
  \href{http://dx.doi.org/10.1103/PhysRevD.65.085001}{Phys.Rev. {\bf D65}
  (2002)  085001},
\href{http://arxiv.org/abs/hep-ph/0109228}{{\tt arXiv:hep-ph/0109228
  [hep-ph]}}.

\bibitem{Bardin:1988xt}
D.~Y. Bardin, A.~Leike, T.~Riemann, and M.~Sachwitz, {\em {Energy Dependent
  Width Effects in e+ e- Annihilation Near the Z Boson Pole}\/},
\href{http://dx.doi.org/10.1016/0370-2693(88)91625-5}{Phys.Lett. {\bf B206}
  (1988)  539--542}.

\bibitem{Beenakker:1996kn}
W.~Beenakker et al., {\em {The Fermion loop scheme for finite width effects in
  $\Pe^{+} \Pe^{-}$ annihilation into four fermions}\/},
  \href{http://dx.doi.org/10.1016/S0550-3213(97)00316-7}{Nucl.Phys. {\bf B500}
  (1997)  255--298},
\href{http://arxiv.org/abs/hep-ph/9612260}{{\tt arXiv:hep-ph/9612260
  [hep-ph]}}.

\bibitem{Denner:1999gp}
A.~Denner, S.~Dittmaier, M.~Roth, and D.~Wackeroth, {\em {Predictions for all
  processes $\Pe^+\Pe^-\to 4\,$fermions${}+\gamma$}\/},
  \href{http://dx.doi.org/10.1016/S0550-3213(99)00437-X}{Nucl.Phys. {\bf B560}
  (1999)  33--65},
\href{http://arxiv.org/abs/hep-ph/9904472}{{\tt arXiv:hep-ph/9904472
  [hep-ph]}}.

\bibitem{Dittmaier:2002ap}
S.~Dittmaier and M.~Roth, {\em {LUSIFER: A LUcid approach to six FERmion
  production}\/},
  \href{http://dx.doi.org/10.1016/S0550-3213(02)00640-5}{Nucl.Phys. {\bf B642}
  (2002)  307--343},
\href{http://arxiv.org/abs/hep-ph/0206070}{{\tt arXiv:hep-ph/0206070
  [hep-ph]}}.

\bibitem{Stuart:1991xk}
R.~G. Stuart, {\em {Gauge invariance, analyticity and physical observables at
  the $\PZ^0$ resonance}\/},
\href{http://dx.doi.org/10.1016/0370-2693(91)90653-8}{Phys.Lett. {\bf B262}
  (1991)  113--119}.

\bibitem{Aeppli:1993rs}
A.~Aeppli, G.~J. van Oldenborgh, and D.~Wyler, {\em {Unstable particles in one
  loop calculations}\/},
  \href{http://dx.doi.org/10.1016/0550-3213(94)90195-3}{Nucl.Phys. {\bf B428}
  (1994)  126--146},
\href{http://arxiv.org/abs/hep-ph/9312212}{{\tt arXiv:hep-ph/9312212
  [hep-ph]}}.

\bibitem{Passarino:2010qk}
G.~Passarino, C.~Sturm, and S.~Uccirati, {\em {Higgs Pseudo-Observables, Second
  Riemann Sheet and All That}\/},
  \href{http://dx.doi.org/10.1016/j.nuclphysb.2010.03.013}{Nucl.Phys. {\bf
  B834} (2010)  77--115},
\href{http://arxiv.org/abs/1001.3360}{{\tt arXiv:1001.3360 [hep-ph]}}.

\bibitem{Goria:2011wa}
S.~Goria, G.~Passarino, and D.~Rosco, {\em {The Higgs Boson Lineshape}\/},
  \href{http://dx.doi.org/10.1016/j.nuclphysb.2012.07.006}{Nucl.Phys. {\bf
  B864} (2012)  530--579},
\href{http://arxiv.org/abs/1112.5517}{{\tt arXiv:1112.5517 [hep-ph]}}.

\bibitem{Grunewald:2000ju}
G.~M. W. et al., {\em {Reports of the Working Groups on Precision Calculations
  for LEP2 Physics: Proceedings. Four fermion production in electron positron
  collisions}\/},
\href{http://arxiv.org/abs/hep-ph/0005309}{{\tt arXiv:hep-ph/0005309
  [hep-ph]}}.

\bibitem{Beneke:2003xh}
M.~Beneke, A.~Chapovsky, A.~Signer, and G.~Zanderighi, {\em {Effective theory
  approach to unstable particle production}\/},
  \href{http://dx.doi.org/10.1103/PhysRevLett.93.011602}{Phys.Rev.Lett. {\bf
  93} (2004)  011602},
\href{http://arxiv.org/abs/hep-ph/0312331}{{\tt arXiv:hep-ph/0312331
  [hep-ph]}}.

\bibitem{Beneke:2004km}
M.~Beneke, A.~Chapovsky, A.~Signer, and G.~Zanderighi, {\em {Effective theory
  calculation of resonant high-energy scattering}\/},
  \href{http://dx.doi.org/10.1016/j.nuclphysb.2004.03.016}{Nucl.Phys. {\bf
  B686} (2004)  205--247},
\href{http://arxiv.org/abs/hep-ph/0401002}{{\tt arXiv:hep-ph/0401002
  [hep-ph]}}.

\bibitem{Hoang:2004tg}
A.~H. Hoang and C.~J. Reisser, {\em {Electroweak absorptive parts in NRQCD
  matching conditions}\/},
  \href{http://dx.doi.org/10.1103/PhysRevD.71.074022}{Phys.Rev. {\bf D71}
  (2005)  074022},
\href{http://arxiv.org/abs/hep-ph/0412258}{{\tt arXiv:hep-ph/0412258
  [hep-ph]}}.

\bibitem{Beneke:2007zg}
M.~Beneke, P.~Falgari, C.~Schwinn, A.~Signer, and G.~Zanderighi, {\em
  {Four-fermion production near the W pair production threshold}\/},
  \href{http://dx.doi.org/10.1016/j.nuclphysb.2007.09.030}{Nucl.Phys. {\bf
  B792} (2008)  89--135},
\href{http://arxiv.org/abs/0707.0773}{{\tt arXiv:0707.0773 [hep-ph]}}.

\bibitem{Falgari:2013gwa}
P.~Falgari, A.~Papanastasiou, and A.~Signer, {\em {Finite-width effects in
  unstable-particle production at hadron colliders}\/},
  \href{http://dx.doi.org/10.1007/JHEP05(2013)156}{JHEP {\bf 1305} (2013)
  156},
\href{http://arxiv.org/abs/1303.5299}{{\tt arXiv:1303.5299 [hep-ph]}}.

\bibitem{Denner:2005fg}
A.~Denner, S.~Dittmaier, M.~Roth, and L.~Wieders, {\em {Electroweak corrections
  to charged-current $\Pe^+\Pe^-\to 4\,$fermion processes: Technical details
  and further results}\/},
  \href{http://dx.doi.org/10.1016/j.nuclphysb.2011.09.001,
  10.1016/j.nuclphysb.2005.06.033}{Nucl.Phys. {\bf B724} (2005)  247--294},
\href{http://arxiv.org/abs/hep-ph/0505042}{{\tt arXiv:hep-ph/0505042
  [hep-ph]}}.

\bibitem{Bredenstein:2006rh}
A.~Bredenstein, A.~Denner, S.~Dittmaier, and M.~Weber, {\em {Precise
  predictions for the Higgs-boson decay $\PH \to \PW\PW/\PZ\PZ \to
  4\,$leptons}\/},
  \href{http://dx.doi.org/10.1103/PhysRevD.74.013004}{Phys.Rev. {\bf D74}
  (2006)  013004},
\href{http://arxiv.org/abs/hep-ph/0604011}{{\tt arXiv:hep-ph/0604011
  [hep-ph]}}.

\bibitem{Bredenstein:2006ha}
A.~Bredenstein, A.~Denner, S.~Dittmaier, and M.~Weber, {\em {Radiative
  corrections to the semileptonic and hadronic Higgs-boson decays $\PH \to
  \PW\PW/\PZ\PZ \to 4\,$fermions}\/},
  \href{http://dx.doi.org/10.1088/1126-6708/2007/02/080}{JHEP {\bf 0702} (2007)
   080},
\href{http://arxiv.org/abs/hep-ph/0611234}{{\tt arXiv:hep-ph/0611234
  [hep-ph]}}.

\bibitem{Actis:2008ug}
S.~Actis, G.~Passarino, C.~Sturm, and S.~Uccirati, {\em {NLO Electroweak
  Corrections to Higgs Boson Production at Hadron Colliders}\/},
  \href{http://dx.doi.org/10.1016/j.physletb.2008.10.018}{Phys.Lett. {\bf B670}
  (2008)  12--17},
\href{http://arxiv.org/abs/0809.1301}{{\tt arXiv:0809.1301 [hep-ph]}}.

\bibitem{Actis:2008uh}
S.~Actis, G.~Passarino, C.~Sturm, and S.~Uccirati, {\em {Two-Loop Threshold
  Singularities, Unstable Particles and Complex Masses}\/},
  \href{http://dx.doi.org/10.1016/j.physletb.2008.09.028}{Phys.Lett. {\bf B669}
  (2008)  62--68},
\href{http://arxiv.org/abs/0809.1302}{{\tt arXiv:0809.1302 [hep-ph]}}.

\bibitem{Passarino:2013bha}
G.~Passarino, {\em {Higgs CAT}\/},
\href{http://arxiv.org/abs/1312.2397}{{\tt arXiv:1312.2397 [hep-ph]}}.

\bibitem{Binoth:2010xt}
T.~Binoth, F.~Boudjema, G.~Dissertori, A.~Lazopoulos, A.~Denner, et al., {\em
  {A Proposal for a standard interface between Monte Carlo tools and one-loop
  programs}\/},
  \href{http://dx.doi.org/10.1016/j.cpc.2010.05.016}{Comput.Phys.Commun. {\bf
  181} (2010)  1612--1622},
\href{http://arxiv.org/abs/1001.1307}{{\tt arXiv:1001.1307 [hep-ph]}}.

\bibitem{Alioli:2013nda}
S.~Alioli, S.~Badger, J.~Bellm, B.~Biedermann, F.~Boudjema, et al., {\em
  {Update of the Binoth Les Houches Accord for a standard interface between
  Monte Carlo tools and one-loop programs}\/},
  \href{http://dx.doi.org/10.1016/j.cpc.2013.10.020}{Comput.Phys.Commun. {\bf
  185} (2014)  560--571},
\href{http://arxiv.org/abs/1308.3462}{{\tt arXiv:1308.3462 [hep-ph]}}.

\bibitem{Catani:1996vz}
S.~Catani and M.~Seymour, {\em {A General algorithm for calculating jet
  cross-sections in NLO QCD}\/},
  \href{http://dx.doi.org/10.1016/S0550-3213(96)00589-5}{Nucl.Phys. {\bf B485}
  (1997)  291--419},
\href{http://arxiv.org/abs/hep-ph/9605323}{{\tt arXiv:hep-ph/9605323
  [hep-ph]}}.

\bibitem{Badger:2012pg}
S.~Badger, B.~Biedermann, P.~Uwer, and V.~Yundin, {\em {Numerical evaluation of
  virtual corrections to multi-jet production in massless QCD}\/},
\href{http://arxiv.org/abs/1209.0100}{{\tt arXiv:1209.0100 [hep-ph]}}.

\bibitem{Kotanski:2011}
J.~Kotanski, J.~Katzy, S.~Pl{\"a}tzer, and Z.~Nagy, {\em {Interfacing nlojet++
  and Herwig++/Matchbox, unpublished}\/}, .

\bibitem{Bellm:2014}
J.~Bellm, S.~Gieseke, N.~Greiner, G.~Heinrich, S.~Pl{\"a}tzer, C.~Reuschle, and
  J.~von Soden-Fraunhofen, {\em {GoSam plus Herwig++/Matchbox, these
  proceedings}\/}, .

\bibitem{Czakon:2009ss}
M.~Czakon, C.~Papadopoulos, and M.~Worek, {\em {Polarizing the Dipoles}\/},
  \href{http://dx.doi.org/10.1088/1126-6708/2009/08/085}{JHEP {\bf 0908} (2009)
   085},
\href{http://arxiv.org/abs/0905.0883}{{\tt arXiv:0905.0883 [hep-ph]}}.

\bibitem{Platzer:2012qg}
S.~Pl{\"a}tzer and M.~Sj{\"o}dahl, {\em {Subleading Nc improved Parton
  Showers}\/},  \href{http://dx.doi.org/10.1007/JHEP07(2012)042}{JHEP {\bf
  1207} (2012)  042},
\href{http://arxiv.org/abs/1201.0260}{{\tt arXiv:1201.0260 [hep-ph]}}.

\bibitem{Platzer:2013fha}
S.~Pl{\"a}tzer, {\em {Summing Large-N Towers in Colour Flow Evolution}\/},
\href{http://arxiv.org/abs/1312.2448}{{\tt arXiv:1312.2448 [hep-ph]}}.

\bibitem{Berger:2008sj}
C.~Berger, Z.~Bern, L.~Dixon, F.~Febres~Cordero, D.~Forde, et al., {\em {An
  Automated Implementation of On-Shell Methods for One-Loop Amplitudes}\/},
  \href{http://dx.doi.org/10.1103/PhysRevD.78.036003}{Phys.Rev. {\bf D78}
  (2008)  036003},
\href{http://arxiv.org/abs/0803.4180}{{\tt arXiv:0803.4180 [hep-ph]}}.

\bibitem{Hahn:2010zi}
T.~Hahn, {\em {Feynman Diagram Calculations with FeynArts, FormCalc, and
  LoopTools}\/},  PoS {\bf ACAT2010} (2010)  078,
\href{http://arxiv.org/abs/1006.2231}{{\tt arXiv:1006.2231 [hep-ph]}}.

\bibitem{Campbell:2010ff}
J.~M. Campbell and R.~Ellis, {\em {MCFM for the Tevatron and the LHC}\/},
  \href{http://dx.doi.org/10.1016/j.nuclphysbps.2010.08.011}{Nucl.Phys.Proc.Suppl.
  {\bf 205-206} (2010)  10--15},
\href{http://arxiv.org/abs/1007.3492}{{\tt arXiv:1007.3492 [hep-ph]}}.

\bibitem{Arnold:2011wj}
K.~Arnold, J.~Bellm, G.~Bozzi, M.~Brieg, F.~Campanario, et al., {\em {VBFNLO: A
  Parton Level Monte Carlo for Processes with Electroweak Bosons -- Manual for
  Version 2.5.0}\/},
\href{http://arxiv.org/abs/1107.4038}{{\tt arXiv:1107.4038 [hep-ph]}}.

\bibitem{Platzer:2012bs}
S.~Pl{\"a}tzer, {\em {Controlling inclusive cross sections in parton shower +
  matrix element merging}\/},
  \href{http://dx.doi.org/10.1007/JHEP08(2013)114}{JHEP {\bf 1308} (2013)
  114},
\href{http://arxiv.org/abs/1211.5467}{{\tt arXiv:1211.5467 [hep-ph]}}.

\bibitem{Bellm:2013lba}
J.~Bellm, S.~Gieseke, D.~Grellscheid, A.~Papaefstathiou, S.~Pl{\"a}tzer, et
  al., {\em {Herwig++ 2.7 Release Note}\/},
\href{http://arxiv.org/abs/1310.6877}{{\tt arXiv:1310.6877 [hep-ph]}}.

\bibitem{Degrande:2011ua}
C.~Degrande, C.~Duhr, B.~Fuks, D.~Grellscheid, O.~Mattelaer, et al., {\em {UFO
  - The Universal FeynRules Output}\/},
  \href{http://dx.doi.org/10.1016/j.cpc.2012.01.022}{Comput.Phys.Commun. {\bf
  183} (2012)  1201--1214},
\href{http://arxiv.org/abs/1108.2040}{{\tt arXiv:1108.2040 [hep-ph]}}.

\bibitem{Skands:2003cj}
P.~Z. Skands, B.~Allanach, H.~Baer, C.~Balazs, G.~Belanger, et al., {\em {SUSY
  Les Houches accord: Interfacing SUSY spectrum calculators, decay packages,
  and event generators}\/},
  \href{http://dx.doi.org/10.1088/1126-6708/2004/07/036}{JHEP {\bf 0407} (2004)
   036},
\href{http://arxiv.org/abs/hep-ph/0311123}{{\tt arXiv:hep-ph/0311123
  [hep-ph]}}.

\bibitem{Platzer:2009jq}
S.~Pl{\"a}tzer and S.~Gieseke, {\em {Coherent Parton Showers with Local
  Recoils}\/},  \href{http://dx.doi.org/10.1007/JHEP01(2011)024}{JHEP {\bf 01}
  (2011)  024},
\href{http://arxiv.org/abs/0909.5593}{{\tt arXiv:0909.5593 [hep-ph]}}.

\bibitem{Badger:2014}
S.~Badger, S.~Pl{\"a}tzer, and V.~Yundin, {\em {The first use case for BLHA2
  extensions: NJet plus Herwig++/Matchbox, these proceedings}\/}, .

\bibitem{Ball:2012wy}
R.~D. Ball, S.~Carrazza, L.~Del~Debbio, S.~Forte, J.~Gao, et al., {\em {Parton
  Distribution Benchmarking with LHC Data}\/},
  \href{http://dx.doi.org/10.1007/JHEP04(2013)125}{JHEP {\bf 1304} (2013)
  125},
\href{http://arxiv.org/abs/1211.5142}{{\tt arXiv:1211.5142 [hep-ph]}}.

\bibitem{Botje:2011sn}
M.~Botje et al., {\em {The PDF4LHC Working Group Interim Recommendations}\/},
\href{http://arxiv.org/abs/1101.0538}{{\tt arXiv:1101.0538 [hep-ph]}}.

\bibitem{Giele:2002hx}
W.~Giele, E.~N. Glover, I.~Hinchliffe, J.~Huston, E.~Laenen, et al., {\em {The
  QCD / SM working group: Summary report}\/},
\href{http://arxiv.org/abs/hep-ph/0204316}{{\tt arXiv:hep-ph/0204316
  [hep-ph]}}.

\bibitem{Salam:2008qg}
G.~P. Salam and J.~Rojo, {\em {A Higher Order Perturbative Parton Evolution
  Toolkit (HOPPET)}\/},
  \href{http://dx.doi.org/10.1016/j.cpc.2008.08.010}{Comput.Phys.Commun. {\bf
  180} (2009)  120--156},
\href{http://arxiv.org/abs/0804.3755}{{\tt arXiv:0804.3755 [hep-ph]}}.

\bibitem{Vogt:2004ns}
A.~Vogt, {\em {Efficient evolution of unpolarized and polarized parton
  distributions with QCD-PEGASUS}\/},
  \href{http://dx.doi.org/10.1016/j.cpc.2005.03.103}{Comput.Phys.Commun. {\bf
  170} (2005)  65--92},
\href{http://arxiv.org/abs/hep-ph/0408244}{{\tt arXiv:hep-ph/0408244
  [hep-ph]}}.

\bibitem{Dittmar:2005ed}
M.~Dittmar, S.~Forte, A.~Glazov, S.~Moch, S.~Alekhin, et al., {\em {Working
  Group I: Parton distributions: Summary report for the HERA LHC Workshop
  Proceedings}\/},
\href{http://arxiv.org/abs/hep-ph/0511119}{{\tt arXiv:hep-ph/0511119
  [hep-ph]}}.

\bibitem{Demartin:2010er}
F.~Demartin, S.~Forte, E.~Mariani, J.~Rojo, and A.~Vicini, {\em {The impact of
  PDF and alphas uncertainties on Higgs Production in gluon fusion at hadron
  colliders}\/},  \href{http://dx.doi.org/10.1103/PhysRevD.82.014002}{Phys.Rev.
  {\bf D82} (2010)  014002},
\href{http://arxiv.org/abs/1004.0962}{{\tt arXiv:1004.0962 [hep-ph]}}.

\bibitem{Watt:2011kp}
G.~Watt, {\em {Parton distribution function dependence of benchmark Standard
  Model total cross sections at the 7 TeV LHC}\/},
  \href{http://dx.doi.org/10.1007/JHEP09(2011)069}{JHEP {\bf 1109} (2011)
  069},
\href{http://arxiv.org/abs/1106.5788}{{\tt arXiv:1106.5788 [hep-ph]}}.

\bibitem{Alekhin:2011sk}
S.~Alekhin, S.~Alioli, R.~D. Ball, V.~Bertone, J.~Blumlein, et al., {\em {The
  PDF4LHC Working Group Interim Report}\/},
\href{http://arxiv.org/abs/1101.0536}{{\tt arXiv:1101.0536 [hep-ph]}}.

\bibitem{Alekhin:2010dd}
S.~Alekhin, J.~Bluemlein, P.~Jimenez-Delgado, S.~Moch, and E.~Reya, {\em {NNLO
  Benchmarks for Gauge and Higgs Boson Production at TeV Hadron Colliders}\/},
  \href{http://dx.doi.org/10.1016/j.physletb.2011.01.034}{Phys.Lett. {\bf B697}
  (2011)  127--135},
\href{http://arxiv.org/abs/1011.6259}{{\tt arXiv:1011.6259 [hep-ph]}}.

\bibitem{Alekhin:2012ig}
S.~Alekhin, J.~Bluemlein, and S.~Moch, {\em {Parton Distribution Functions and
  Benchmark Cross Sections at NNLO}\/},
  \href{http://dx.doi.org/10.1103/PhysRevD.86.054009}{Phys.Rev. {\bf D86}
  (2012)  054009},
\href{http://arxiv.org/abs/1202.2281}{{\tt arXiv:1202.2281 [hep-ph]}}.

\bibitem{CooperSarkar:2007ny}
A.~M. Cooper-Sarkar, {\em {Including heavy flavour production in PDF fits}\/},
\href{http://arxiv.org/abs/0709.0191}{{\tt arXiv:0709.0191 [hep-ph]}}.

\bibitem{Thorne:2012az}
R.~Thorne, {\em {Effect of changes of variable flavor number scheme on parton
  distribution functions and predicted cross sections}\/},
  \href{http://dx.doi.org/10.1103/PhysRevD.86.074017}{Phys.Rev. {\bf D86}
  (2012)  074017},
\href{http://arxiv.org/abs/1201.6180}{{\tt arXiv:1201.6180 [hep-ph]}}.

\bibitem{Ball:2013gsa}
{The NNPDF} Collaboration, R.~D. Ball et al., {\em {Theoretical issues in PDF
  determination and associated uncertainties}\/},
  \href{http://dx.doi.org/10.1016/j.physletb.2013.05.019}{Phys.Lett. {\bf B723}
  (2013)  330--339},
\href{http://arxiv.org/abs/1303.1189}{{\tt arXiv:1303.1189 [hep-ph]}}.

\bibitem{Thorne:2014toa}
R.~Thorne, {\em {The effect on PDFs and $\alpha_s(M_Z^2)$ due to changes in
  flavour scheme and higher twist contributions}\/},
\href{http://arxiv.org/abs/1402.3536}{{\tt arXiv:1402.3536 [hep-ph]}}.

\bibitem{Alekhin:2013nua}
S.~Alekhin, J.~Blümlein, and S.-O. Moch, {\em {ABM news and benchmarks}\/},
  PoS {\bf DIS2013} (2013)  039,
\href{http://arxiv.org/abs/1308.5166}{{\tt arXiv:1308.5166 [hep-ph]}}.

\bibitem{Gao:2013xoa}
J.~Gao, M.~Guzzi, J.~Huston, H.-L. Lai, Z.~Li, et al., {\em {The CT10 NNLO
  Global Analysis of QCD}\/},
\href{http://arxiv.org/abs/1302.6246}{{\tt arXiv:1302.6246 [hep-ph]}}.

\bibitem{CooperSarkar:2011aa}
{H1 and ZEUS} Collaboration, A.~Cooper-Sarkar, {\em {PDF Fits at HERA}\/},  PoS
  {\bf EPS-HEP2011} (2011)  320,
\href{http://arxiv.org/abs/1112.2107}{{\tt arXiv:1112.2107 [hep-ph]}}.

\bibitem{Martin:2009iq}
A.~Martin, W.~Stirling, R.~Thorne, and G.~Watt, {\em {Parton distributions for
  the LHC}\/},
  \href{http://dx.doi.org/10.1140/epjc/s10052-009-1072-5}{Eur.Phys.J. {\bf C63}
  (2009)  189--285},
\href{http://arxiv.org/abs/0901.0002}{{\tt arXiv:0901.0002 [hep-ph]}}.

\bibitem{Ball:2012cx}
R.~D. Ball, V.~Bertone, S.~Carrazza, C.~S. Deans, L.~Del~Debbio, et al., {\em
  {Parton distributions with LHC data}\/},
  \href{http://dx.doi.org/10.1016/j.nuclphysb.2012.10.003}{Nucl.Phys. {\bf
  B867} (2013)  244--289},
\href{http://arxiv.org/abs/1207.1303}{{\tt arXiv:1207.1303 [hep-ph]}}.

\bibitem{Aaron:2009aa}
{H1 and ZEUS} Collaboration, F.~Aaron et al., {\em {Combined Measurement and
  QCD Analysis of the Inclusive e+- p Scattering Cross Sections at HERA}\/},
  \href{http://dx.doi.org/10.1007/JHEP01(2010)109}{JHEP {\bf 1001} (2010)
  109},
\href{http://arxiv.org/abs/0911.0884}{{\tt arXiv:0911.0884 [hep-ex]}}.

\bibitem{Dulat:2013kqa}
S.~Dulat, T.-J. Hou, J.~Gao, J.~Huston, P.~Nadolsky, et al., {\em {Higgs Boson
  Cross Section from CTEQ-TEA Global Analysis}\/},
\href{http://arxiv.org/abs/1310.7601}{{\tt arXiv:1310.7601 [hep-ph]}}.

\bibitem{Abramowicz:1900rp}
{H1, ZEUS} Collaboration, H.~Abramowicz et al., {\em {Combination and QCD
  Analysis of Charm Production Cross Section Measurements in Deep-Inelastic ep
  Scattering at HERA}\/},
  \href{http://dx.doi.org/10.1140/epjc/s10052-013-2311-3}{Eur.Phys.J. {\bf C73}
  (2013)  2311},
\href{http://arxiv.org/abs/1211.1182}{{\tt arXiv:1211.1182 [hep-ex]}}.

\bibitem{Anastasiou:2011pi}
C.~Anastasiou, S.~Buehler, F.~Herzog, and A.~Lazopoulos, {\em {Total
  cross-section for Higgs boson hadroproduction with anomalous Standard Model
  interactions}\/},  \href{http://dx.doi.org/10.1007/JHEP12(2011)058}{JHEP {\bf
  1112} (2011)  058},
\href{http://arxiv.org/abs/1107.0683}{{\tt arXiv:1107.0683 [hep-ph]}}.

\bibitem{Guzzi:2011ew}
M.~Guzzi, P.~M. Nadolsky, H.-L. Lai, and C.-P. Yuan, {\em {General-Mass
  Treatment for Deep Inelastic Scattering at Two-Loop Accuracy}\/},
  \href{http://dx.doi.org/10.1103/PhysRevD.86.053005}{Phys.Rev. {\bf D86}
  (2012)  053005},
\href{http://arxiv.org/abs/1108.5112}{{\tt arXiv:1108.5112 [hep-ph]}}.

\bibitem{Thorne:2006qt}
R.~Thorne, {\em {A Variable-flavor number scheme for NNLO}\/},
  \href{http://dx.doi.org/10.1103/PhysRevD.73.054019}{Phys.Rev. {\bf D73}
  (2006)  054019},
\href{http://arxiv.org/abs/hep-ph/0601245}{{\tt arXiv:hep-ph/0601245
  [hep-ph]}}.

\bibitem{Forte:2010ta}
S.~Forte, E.~Laenen, P.~Nason, and J.~Rojo, {\em {Heavy quarks in
  deep-inelastic scattering}\/},
  \href{http://dx.doi.org/10.1016/j.nuclphysb.2010.03.014}{Nucl.Phys. {\bf
  B834} (2010)  116--162},
\href{http://arxiv.org/abs/1001.2312}{{\tt arXiv:1001.2312 [hep-ph]}}.

\bibitem{Bertone:2013vaa}
V.~Bertone, S.~Carrazza, and J.~Rojo, {\em {APFEL: A PDF Evolution Library with
  QED corrections}\/},
\href{http://arxiv.org/abs/1310.1394}{{\tt arXiv:1310.1394 [hep-ph]}}.

\bibitem{herafitter}
Herafitter program (2012), \href{http://www.herafitter.org}{{\tt
  www.herafitter.org}}.

\bibitem{Botje:2010ay}
M.~Botje, {\em {QCDNUM: Fast QCD Evolution and Convolution}\/},
  \href{http://dx.doi.org/10.1016/j.cpc.2010.10.020}{Comput.Phys.Commun. {\bf
  182} (2011)  490--532},
\href{http://arxiv.org/abs/1005.1481}{{\tt arXiv:1005.1481 [hep-ph]}}.

\bibitem{Aaron:2009kv}
{H1} Collaboration, F.~Aaron et al., {\em {A Precision Measurement of the
  Inclusive ep Scattering Cross Section at HERA}\/},
  \href{http://dx.doi.org/10.1140/epjc/s10052-009-1169-x}{Eur.Phys.J. {\bf C64}
  (2009)  561--587},
\href{http://arxiv.org/abs/0904.3513}{{\tt arXiv:0904.3513 [hep-ex]}}.

\bibitem{Riemersma:1994hv}
S.~Riemersma, J.~Smith, and W.~van Neerven, {\em {Rates for inclusive deep
  inelastic electroproduction of charm quarks at HERA}\/},
  \href{http://dx.doi.org/10.1016/0370-2693(95)00036-K}{Phys.Lett. {\bf B347}
  (1995)  143--151},
\href{http://arxiv.org/abs/hep-ph/9411431}{{\tt arXiv:hep-ph/9411431
  [hep-ph]}}.

\bibitem{Pumplin:2009bb}
J.~Pumplin, {\em {Parametrization dependence and Delta Chi**2 in parton
  distribution fitting}\/},
  \href{http://dx.doi.org/10.1103/PhysRevD.82.114020}{Phys.Rev. {\bf D82}
  (2010)  114020},
\href{http://arxiv.org/abs/0909.5176}{{\tt arXiv:0909.5176 [hep-ph]}}.

\bibitem{Martin:2012da}
A.~Martin, A.~T. Mathijssen, W.~Stirling, R.~Thorne, B.~Watt, et al., {\em
  {Extended Parameterisations for MSTW PDFs and their effect on Lepton Charge
  Asymmetry from W Decays}\/},
  \href{http://dx.doi.org/10.1140/epjc/s10052-013-2318-9}{Eur.Phys.J. {\bf C73}
  (2013)  2318},
\href{http://arxiv.org/abs/1211.1215}{{\tt arXiv:1211.1215 [hep-ph]}}.

\bibitem{Pumplin:2002vw}
J.~Pumplin, D.~Stump, J.~Huston, H.~Lai, P.~M. Nadolsky, et al., {\em {New
  generation of parton distributions with uncertainties from global QCD
  analysis}\/},  \href{http://dx.doi.org/10.1088/1126-6708/2002/07/012}{JHEP
  {\bf 0207} (2002)  012},
\href{http://arxiv.org/abs/hep-ph/0201195}{{\tt arXiv:hep-ph/0201195
  [hep-ph]}}.

\bibitem{Pumplin:2001ct}
J.~Pumplin, D.~Stump, R.~Brock, D.~Casey, J.~Huston, et al., {\em
  {Uncertainties of predictions from parton distribution functions. 2. The
  Hessian method}\/},
  \href{http://dx.doi.org/10.1103/PhysRevD.65.014013}{Phys.Rev. {\bf D65}
  (2001)  014013},
\href{http://arxiv.org/abs/hep-ph/0101032}{{\tt arXiv:hep-ph/0101032
  [hep-ph]}}.

\bibitem{D'Agostini:1993uj}
G.~D'Agostini, {\em {On the use of the covariance matrix to fit correlated
  data}\/},
\href{http://dx.doi.org/10.1016/0168-9002(94)90719-6}{Nucl.Instrum.Meth. {\bf
  A346} (1994)  306--311}.

\bibitem{Ball:2009qv}
{NNPDF} Collaboration, R.~D. Ball et al., {\em {Fitting Parton Distribution
  Data with Multiplicative Normalization Uncertainties}\/},
  \href{http://dx.doi.org/10.1007/JHEP05(2010)075}{JHEP {\bf 1005} (2010)
  075},
\href{http://arxiv.org/abs/0912.2276}{{\tt arXiv:0912.2276 [hep-ph]}}.

\bibitem{Stump:2001gu}
D.~Stump, J.~Pumplin, R.~Brock, D.~Casey, J.~Huston, et al., {\em
  {Uncertainties of predictions from parton distribution functions. 1. The
  Lagrange multiplier method}\/},
  \href{http://dx.doi.org/10.1103/PhysRevD.65.014012}{Phys.Rev. {\bf D65}
  (2001)  014012},
\href{http://arxiv.org/abs/hep-ph/0101051}{{\tt arXiv:hep-ph/0101051
  [hep-ph]}}.

\bibitem{Martin:2010db}
A.~Martin, W.~Stirling, R.~Thorne, and G.~Watt, {\em {Heavy-quark mass
  dependence in global PDF analyses and 3- and 4-flavour parton
  distributions}\/},
  \href{http://dx.doi.org/10.1140/epjc/s10052-010-1462-8}{Eur.Phys.J. {\bf C70}
  (2010)  51--72},
\href{http://arxiv.org/abs/1007.2624}{{\tt arXiv:1007.2624 [hep-ph]}}.

\bibitem{Gao:2013wwa}
J.~Gao, M.~Guzzi, and P.~M. Nadolsky, {\em {Charm quark mass dependence in a
  global QCD analysis}\/},
  \href{http://dx.doi.org/10.1140/epjc/s10052-013-2541-4}{Eur.Phys.J. {\bf C73}
  (2013)  2541},
\href{http://arxiv.org/abs/1304.3494}{{\tt arXiv:1304.3494 [hep-ph]}}.

\bibitem{Nadolsky:2009ge}
P.~M. Nadolsky and W.-K. Tung, {\em {Improved Formulation of Global QCD
  Analysis with Zero-mass Matrix Elements}\/},
  \href{http://dx.doi.org/10.1103/PhysRevD.79.113014}{Phys.Rev. {\bf D79}
  (2009)  113014},
\href{http://arxiv.org/abs/0903.2667}{{\tt arXiv:0903.2667 [hep-ph]}}.

\bibitem{d'Enterria:2012yj}
D.~d'Enterria and J.~Rojo, {\em {Quantitative constraints on the gluon
  distribution function in the proton from collider isolated-photon data}\/},
  \href{http://dx.doi.org/10.1016/j.nuclphysb.2012.03.003}{Nucl.Phys. {\bf
  B860} (2012)  311--338},
\href{http://arxiv.org/abs/1202.1762}{{\tt arXiv:1202.1762 [hep-ph]}}.

\bibitem{Carminati:2012mm}
L.~Carminati, G.~Costa, D.~D'Enterria, I.~Koletsou, G.~Marchiori, et al., {\em
  {Sensitivity of the LHC isolated-gamma+jet data to the parton distribution
  functions of the proton}\/},
  \href{http://dx.doi.org/10.1209/0295-5075/101/61002}{EPL {\bf 101} (2013)
  61002},
\href{http://arxiv.org/abs/1212.5511}{{\tt arXiv:1212.5511}}.

\bibitem{Malik:2013kba}
S.~A. Malik and G.~Watt, {\em {Ratios of $W$ and $Z$ cross sections at large
  boson $p_T$ as a constraint on PDFs and background to new physics}\/},
  \href{http://dx.doi.org/10.1007/JHEP02(2014)025}{JHEP {\bf 1402} (2014)
  025},
\href{http://arxiv.org/abs/1304.2424}{{\tt arXiv:1304.2424 [hep-ph]}}.

\bibitem{Czakon:2013tha}
M.~Czakon, M.~L. Mangano, A.~Mitov, and J.~Rojo, {\em {Constraints on the gluon
  PDF from top quark pair production at hadron colliders}\/},
  \href{http://dx.doi.org/10.1007/JHEP07(2013)167}{JHEP {\bf 1307} (2013)
  167},
\href{http://arxiv.org/abs/1303.7215}{{\tt arXiv:1303.7215 [hep-ph]}}.

\bibitem{Watt:2013oha}
B.~Watt, P.~Motylinski, and R.~Thorne, {\em {The Effect of LHC Jet Data on MSTW
  PDFs}\/},
\href{http://arxiv.org/abs/1311.5703}{{\tt arXiv:1311.5703 [hep-ph]}}.

\bibitem{Beringer:1900zz}
{Particle Data Group} Collaboration, J.~Beringer et al., {\em {Review of
  Particle Physics (RPP)}\/},
\href{http://dx.doi.org/10.1103/PhysRevD.86.010001}{Phys.Rev. {\bf D86} (2012)
  010001}.

\bibitem{Ball:2011uy}
{NNPDF Collaboration} Collaboration, R.~D. Ball et al., {\em {Unbiased global
  determination of parton distributions and their uncertainties at NNLO and at
  LO}\/},  \href{http://dx.doi.org/10.1016/j.nuclphysb.2011.09.024}{Nucl.Phys.
  {\bf B855} (2012)  153--221},
\href{http://arxiv.org/abs/1107.2652}{{\tt arXiv:1107.2652 [hep-ph]}}.

\bibitem{Ball:2011us}
R.~D. Ball, V.~Bertone, L.~Del~Debbio, S.~Forte, A.~Guffanti, et al., {\em
  {Precision NNLO determination of $\alpha_s(M_Z)$ using an unbiased global
  parton set}\/},
  \href{http://dx.doi.org/10.1016/j.physletb.2011.11.053}{Phys.Lett. {\bf B707}
  (2012)  66--71},
\href{http://arxiv.org/abs/1110.2483}{{\tt arXiv:1110.2483 [hep-ph]}}.

\bibitem{Lionetti:2011pw}
S.~Lionetti, R.~D. Ball, V.~Bertone, F.~Cerutti, L.~Del~Debbio, et al., {\em
  {Precision determination of $\alpha_s$ using an unbiased global NLO parton
  set}\/},  \href{http://dx.doi.org/10.1016/j.physletb.2011.05.071}{Phys.Lett.
  {\bf B701} (2011)  346--352},
\href{http://arxiv.org/abs/1103.2369}{{\tt arXiv:1103.2369 [hep-ph]}}.

\bibitem{Anastasiou:2012hx}
C.~Anastasiou, S.~Buehler, F.~Herzog, and A.~Lazopoulos, {\em {Inclusive Higgs
  boson cross-section for the LHC at 8 TeV}\/},  JHEP {\bf 1204} (2012)  004,
\href{http://arxiv.org/abs/1202.3638}{{\tt arXiv:1202.3638 [hep-ph]}}.

\bibitem{Thorne:2011kq}
R.~Thorne and G.~Watt, {\em {PDF dependence of Higgs cross sections at the
  Tevatron and LHC: Response to recent criticism}\/},
  \href{http://dx.doi.org/10.1007/JHEP08(2011)100}{JHEP {\bf 1108} (2011)
  100},
\href{http://arxiv.org/abs/1106.5789}{{\tt arXiv:1106.5789 [hep-ph]}}.

\bibitem{Ball:2010gb}
{NNPDF} Collaboration, R.~D. Ball et al., {\em {Reweighting NNPDFs: the W
  lepton asymmetry}\/},
  \href{http://dx.doi.org/10.1016/j.nuclphysb.2011.03.017,
  10.1016/j.nuclphysb.2011.10.024, 10.1016/j.nuclphysb.2011.09.011}{Nucl.Phys.
  {\bf B849} (2011)  112--143},
\href{http://arxiv.org/abs/1012.0836}{{\tt arXiv:1012.0836 [hep-ph]}}.

\bibitem{Ball:2011gg}
R.~D. Ball, V.~Bertone, F.~Cerutti, L.~Del~Debbio, S.~Forte, et al., {\em
  {Reweighting and Unweighting of Parton Distributions and the LHC W lepton
  asymmetry data}\/},
  \href{http://dx.doi.org/10.1016/j.nuclphysb.2011.10.018}{Nucl.Phys. {\bf
  B855} (2012)  608--638},
\href{http://arxiv.org/abs/1108.1758}{{\tt arXiv:1108.1758 [hep-ph]}}.

\bibitem{Mangano:2012mh}
M.~L. Mangano and J.~Rojo, {\em {Cross Section Ratios between different CM
  energies at the LHC: opportunities for precision measurements and BSM
  sensitivity}\/},  \href{http://dx.doi.org/10.1007/JHEP08(2012)010}{JHEP {\bf
  1208} (2012)  010},
\href{http://arxiv.org/abs/1206.3557}{{\tt arXiv:1206.3557 [hep-ph]}}.

\bibitem{Whalley:2005nh}
M.~Whalley, D.~Bourilkov, and R.~Group, {\em {The Les Houches accord PDFs
  (LHAPDF) and LHAGLUE}\/},
\href{http://arxiv.org/abs/hep-ph/0508110}{{\tt arXiv:hep-ph/0508110
  [hep-ph]}}.

\bibitem{yamlweb}
\url{http://yaml.org/}.

\bibitem{zlibcweb}
\url{https://zlibc.linux.lu/}.

\bibitem{lhapdfweb}
\url{http://lhapdf.hepforge.org/}.

\bibitem{Ellis:1990ek}
S.~D. Ellis, Z.~Kunszt, and D.~E. Soper, {\em One-Jet Inclusive Cross Section
  at Order $\alpha_s^3$: Quarks and Gluons\/},
\href{http://dx.doi.org/10.1103/PhysRevLett.64.2121}{Phys. Rev. Lett. {\bf 64}
  (1990)  2121}.

\bibitem{Ellis:1992en}
S.~D. Ellis, Z.~Kunszt, and D.~E. Soper, {\em Two-Jet Production in Hadron
  Collisions at Order $\alpha_s^3$ in {QCD}\/},
\href{http://dx.doi.org/10.1103/PhysRevLett.69.1496}{Phys. Rev. Lett. {\bf 69}
  (1992)  1496}.

\bibitem{Giele:1994gf}
W.~Giele, E.~N. Glover, and D.~A. Kosower, {\em The Two-Jet Differential Cross
  Section at ${\cal O}(\alpha_s^3)$ in Hadron Collisions\/},
  \href{http://dx.doi.org/10.1103/PhysRevLett.73.2019}{Phys. Rev. Lett. {\bf
  73} (1994)  2019},
\href{http://arxiv.org/abs/hep-ph/9403347}{{\tt arXiv:hep-ph/9403347
  [hep-ph]}}.

\bibitem{Moretti:2006ea}
S.~Moretti, M.~Nolten, and D.~Ross, {\em Weak corrections to four-parton
  processes\/},  \href{http://dx.doi.org/10.1016/j.nuclphysb.2006.09.028}{Nucl.
  Phys. B {\bf 759} (2006)  50},
\href{http://arxiv.org/abs/hep-ph/0606201}{{\tt arXiv:hep-ph/0606201
  [hep-ph]}}.

\bibitem{Scharf:2009sp}
A.~Scharf, {\em Electroweak corrections to b-jet and di-jet production\/},
\href{http://arxiv.org/abs/0910.0223}{{\tt arXiv:0910.0223 [hep-ph]}}.

\bibitem{Chatrchyan:2008aa}
{CMS} Collaboration, S.~Chatrchyan et al., {\em The {CMS} experiment at the
  {CERN} {LHC}\/},
\href{http://dx.doi.org/10.1088/1748-0221/3/08/S08004}{JINST {\bf 3} (2008)
  S08004}.

\bibitem{Cacciari:2011ma}
M.~Cacciari, G.~P. Salam, and G.~Soyez, {\em {FastJet User Manual}\/},
  \href{http://dx.doi.org/10.1140/epjc/s10052-012-1896-2}{Eur.Phys.J. {\bf C72}
  (2012)  1896},
\href{http://arxiv.org/abs/1111.6097}{{\tt arXiv:1111.6097 [hep-ph]}}.

\bibitem{Cacciari:2008gp}
M.~Cacciari, G.~P. Salam, and G.~Soyez, {\em {The Anti-k(t) jet clustering
  algorithm}\/},  \href{http://dx.doi.org/10.1088/1126-6708/2008/04/063}{JHEP
  {\bf 0804} (2008)  063},
\href{http://arxiv.org/abs/0802.1189}{{\tt arXiv:0802.1189 [hep-ph]}}.

\bibitem{Whalley:1989mt}
M.~Whalley, {\em The Durham-{RAL} high-energy physics databases: {HEPDATA}\/},
  \href{http://dx.doi.org/10.1016/0010-4655(89)90282-8}{Comput. Phys. Commun.
  {\bf 57} (1989)  536--537}.
\url{http://durpdg.dur.ac.uk/}.

\bibitem{Buckley:2010jn}
A.~Buckley and M.~Whalley, {\em HepData reloaded: Reinventing the {HEP} data
  archive\/},  PoS {\bf ACAT2010} (2010)  067,
\href{http://arxiv.org/abs/1006.0517}{{\tt arXiv:1006.0517 [hep-ex]}}.

\bibitem{Britzger:2012bs}
{fastNLO} Collaboration, D.~Britzger, K.~Rabbertz, F.~Stober, and M.~Wobisch,
  {\em New features in version 2 of the {fastNLO} project\/},
\href{http://arxiv.org/abs/1208.3641}{{\tt arXiv:1208.3641 [hep-ph]}}.

\bibitem{Sjostrand:2006za}
T.~Sjostrand, S.~Mrenna, and P.~Z. Skands, {\em {PYTHIA 6.4 Physics and
  Manual}\/},  \href{http://dx.doi.org/10.1088/1126-6708/2006/05/026}{JHEP {\bf
  0605} (2006)  026},
\href{http://arxiv.org/abs/hep-ph/0603175}{{\tt arXiv:hep-ph/0603175
  [hep-ph]}}.

\bibitem{Lai:2010vv}
H.-L. Lai, M.~Guzzi, J.~Huston, Z.~Li, P.~M. Nadolsky, et al., {\em {New parton
  distributions for collider physics}\/},
  \href{http://dx.doi.org/10.1103/PhysRevD.82.074024}{Phys.Rev. {\bf D82}
  (2010)  074024},
\href{http://arxiv.org/abs/1007.2241}{{\tt arXiv:1007.2241 [hep-ph]}}.

\bibitem{Ball:2011mu}
R.~D. Ball, V.~Bertone, F.~Cerutti, L.~Del~Debbio, S.~Forte, et al., {\em
  Impact of Heavy Quark Masses on Parton Distributions and LHC
  Phenomenology\/},
  \href{http://dx.doi.org/10.1016/j.nuclphysb.2011.03.021}{Nucl. Phys. B {\bf
  849} (2011)  296},
\href{http://arxiv.org/abs/1101.1300}{{\tt arXiv:1101.1300 [hep-ph]}}.

\bibitem{ATLAS-CONF-2013-017}
{ATLAS} Collaboration, G.~Aad et al., {\em {Search for high-mass dilepton
  resonances in 20/fb of pp collisions at sqrt(s) = 8 TeV with the ATLAS
  experiment}\/},  ATLAS-CONF-2013-017  .

\bibitem{Gavin:2010az}
R.~Gavin, Y.~Li, F.~Petriello, and S.~Quackenbush, {\em {FEWZ 2.0: A code for
  hadronic Z production at next-to-next-to-leading order}\/},
  \href{http://dx.doi.org/10.1016/j.cpc.2011.06.008}{Comput.Phys.Commun. {\bf
  182} (2011)  2388--2403},
\href{http://arxiv.org/abs/1011.3540}{{\tt arXiv:1011.3540 [hep-ph]}}.

\bibitem{Gavin:2012sy}
R.~Gavin, Y.~Li, F.~Petriello, and S.~Quackenbush, {\em {W Physics at the LHC
  with FEWZ 2.1}\/},
\href{http://arxiv.org/abs/1201.5896}{{\tt arXiv:1201.5896 [hep-ph]}}.

\bibitem{Li:2012wn}
Y.~Li and F.~Petriello, {\em {Combining QCD and electroweak corrections to
  dilepton production in FEWZ}\/},
\href{http://arxiv.org/abs/1208.5967}{{\tt arXiv:1208.5967 [hep-ph]}}.

\bibitem{Catani:2007vq}
S.~Catani and M.~Grazzini, {\em {An NNLO subtraction formalism in hadron
  collisions and its application to Higgs boson production at the LHC}\/},
  \href{http://dx.doi.org/10.1103/PhysRevLett.98.222002}{Phys.Rev.Lett. {\bf
  98} (2007)  222002},
\href{http://arxiv.org/abs/hep-ph/0703012}{{\tt arXiv:hep-ph/0703012
  [hep-ph]}}.

\bibitem{Catani:2009sm}
S.~Catani, L.~Cieri, G.~Ferrera, D.~de~Florian, and M.~Grazzini, {\em {Vector
  boson production at hadron colliders: A Fully exclusive QCD calculation at
  NNLO}\/},
  \href{http://dx.doi.org/10.1103/PhysRevLett.103.082001}{Phys.Rev.Lett. {\bf
  103} (2009)  082001},
\href{http://arxiv.org/abs/0903.2120}{{\tt arXiv:0903.2120 [hep-ph]}}.

\bibitem{Hamberg:1991}
R.~Hamberg, W.~L. van Neerven, and T.~Matsuura, {\em A Complete Calculation Of
  The Order Alpha-S**2 Correction To The Drell-Yan K Factor\/},  Nucl. Phys. B
  {\bf 359} (1991)  343.

\bibitem{vrap}
C.~Anastasiou, L.~Dixon, K.~Melnikov, and F.~Petriello, {\em {High precision
  QCD at hadron colliders: Electroweak gauge boson rapidity distributions at
  NNLO}\/},
Phys.~Rev.~D {\bf 69} (2004)  094008.
v0.9 including LHAPDF interface by D. Maitre.

\bibitem{Grazzini:2013_private}
M.~Grazzini, {\em private communications\/},  2013.

\bibitem{Bardin:2012jk}
D.~Bardin, S.~Bondarenko, P.~Christova, L.~Kalinovskaya, L.~Rumyantsev, et al.,
  {\em {SANC integrator in the progress: QCD and EW contributions}\/},
  \href{http://dx.doi.org/10.1134/S002136401217002X}{JETP Lett. {\bf 96} (2012)
   285--289},
\href{http://arxiv.org/abs/1207.4400}{{\tt arXiv:1207.4400 [hep-ph]}}.

\bibitem{PDG}
{Particle Data Group} Collaboration, J.~Beringer et al.,
  \href{http://dx.doi.org/10.1103/PhysRevD.86.010001}{{\em {Review of Particle
  Physics}}}, vol.~86.
\newblock 2012.

\bibitem{Bondarenko:2013nu}
S.~G. Bondarenko and A.~A. Sapronov, {\em {NLO EW and QCD proton-proton cross
  section calculations with mcsanc-v1.01}\/},
  \href{http://dx.doi.org/10.1016/j.cpc.2013.05.010}{Comput.Phys.Commun. {\bf
  184} (2013)  2343--2350},
\href{http://arxiv.org/abs/1301.3687}{{\tt arXiv:1301.3687 [hep-ph]}}.

\bibitem{Golonka:2006photos}
P.~Golonka and Z.~W\c{a}s, {\em {PHOTOS Monte Carlo: a precision tool for QED
  corrections in $Z$ and $W$ decays}\/},  Eur.~Phys.~J. {\bf C45} (2006)
  97--107.

\bibitem{Arbuzov:2012dx}
A.~Arbuzov, R.~Sadykov, and Z.~Was, {\em {QED Bremsstrahlung in decays of
  electroweak bosons}\/},
  \href{http://dx.doi.org/10.1140/epjc/s10052-013-2625-1}{Eur.Phys.J. {\bf C73}
  (2013)  2625},
\href{http://arxiv.org/abs/1212.6783}{{\tt arXiv:1212.6783 [hep-ph]}}.

\bibitem{Li:2012_private}
F.~Petriello and Y.~Li, {\em FEWZ 3.1.a3 EW flags resulting from a private
  communication\/},  July 2012.

\bibitem{Belloni_Klein_2012}
A.~Belloni and U.~Klein, {\em Talk at LPCC EW WG meeting, 27.04.2012\/},
  \url{https://indico.cern.ch/getFile.py/access?contribId=3&resId=0&materialId=slides&confId=188020}.

\bibitem{Klein_2012}
U.~Klein, {\em Talk at LPCC EW WG meeting, 9.10.2012\/},
  \url{https://indico.cern.ch/getFile.py/access?contribId=14&sessionId=2&resId=0&materialId=slides&confId=203748}.

\bibitem{Aad:2013iua}
{ATLAS} Collaboration, G.~Aad et al., {\em {Measurement of the high-mass
  Drell-Yan differential cross-section in pp collisions at $\sqrt{s}$=7 TeV
  with the ATLAS detector}\/},
  \href{http://dx.doi.org/10.1016/j.physletb.2013.07.049}{Phys.Lett. {\bf B725}
  (2013)  223--242},
\href{http://arxiv.org/abs/1305.4192}{{\tt arXiv:1305.4192 [hep-ex]}}.

\bibitem{Martin:2004dh_private}
R.~Roberts and G.~Watt, {\em private communication at PDF4LHC meeting\/},
  December 2012.

\bibitem{Aad:2011ib}
{ATLAS Collaboration} Collaboration, G.~Aad et al., {\em {Search for squarks
  and gluinos using final states with jets and missing transverse momentum with
  the ATLAS detector in $\sqrt{s}=7$ TeV proton-proton collisions}\/},
  \href{http://dx.doi.org/10.1016/j.physletb.2012.02.051}{Phys.Lett. {\bf B710}
  (2012)  67--85},
\href{http://arxiv.org/abs/1109.6572}{{\tt arXiv:1109.6572 [hep-ex]}}.

\bibitem{Collaboration:2011ida}
{CMS Collaboration} Collaboration, S.~Chatrchyan et al., {\em {Search for New
  Physics with Jets and Missing Transverse Momentum in $pp$ collisions at
  $\sqrt{s}=7$ TeV}\/},  \href{http://dx.doi.org/10.1007/JHEP08(2011)155}{JHEP
  {\bf 1108} (2011)  155},
\href{http://arxiv.org/abs/1106.4503}{{\tt arXiv:1106.4503 [hep-ex]}}.

\bibitem{cms:2012mfa}
{CMS Collaboration} Collaboration, S.~Chatrchyan et al., {\em {Search for new
  physics in the multijet and missing transverse momentum final state in
  proton-proton collisions at $\sqrt{s} = 7$ TeV}\/},
  \href{http://dx.doi.org/10.1103/PhysRevLett.109.171803}{Phys.Rev.Lett. {\bf
  109} (2012)  171803},
\href{http://arxiv.org/abs/1207.1898}{{\tt arXiv:1207.1898 [hep-ex]}}.

\bibitem{Ask:2011xf}
S.~Ask, M.~Parker, T.~Sandoval, M.~Shea, and W.~Stirling, {\em {Using
  gamma+jets Production to Calibrate the Standard Model Z(nunu)+jets Background
  to New Physics Processes at the LHC}\/},
  \href{http://dx.doi.org/10.1007/JHEP10(2011)058}{JHEP {\bf 1110} (2011)
  058},
\href{http://arxiv.org/abs/1107.2803}{{\tt arXiv:1107.2803 [hep-ph]}}.

\bibitem{Bern:2011pa}
Z.~Bern, G.~Diana, L.~Dixon, F.~Febres~Cordero, S.~Hoche, et al., {\em {Driving
  Missing Data at Next-to-Leading Order}\/},
  \href{http://dx.doi.org/10.1103/PhysRevD.84.114002}{Phys.Rev. {\bf D84}
  (2011)  114002},
\href{http://arxiv.org/abs/1106.1423}{{\tt arXiv:1106.1423 [hep-ph]}}.

\bibitem{Bern:2012vx}
Z.~Bern, G.~Diana, L.~Dixon, F.~Febres~Cordero, S.~Hoeche, et al., {\em
  {Missing Energy and Jets for Supersymmetry Searches}\/},
  \href{http://dx.doi.org/10.1103/PhysRevD.87.034026}{Phys.Rev. {\bf D87}
  (2013) no.~3, 034026},
\href{http://arxiv.org/abs/1206.6064}{{\tt arXiv:1206.6064 [hep-ph]}}.

\bibitem{Kuhn:2005gv}
J.~H. Kuhn, A.~Kulesza, S.~Pozzorini, and M.~Schulze, {\em {Electroweak
  corrections to hadronic photon production at large transverse momenta}\/},
  \href{http://dx.doi.org/10.1088/1126-6708/2006/03/059}{JHEP {\bf 0603} (2006)
   059},
\href{http://arxiv.org/abs/hep-ph/0508253}{{\tt arXiv:hep-ph/0508253
  [hep-ph]}}.

\bibitem{Maina:2004rb}
E.~Maina, S.~Moretti, and D.~A. Ross, {\em {One loop weak corrections to gamma
  / Z hadroproduction at finite transverse momentum}\/},
  \href{http://dx.doi.org/10.1016/j.physletb.2005.03.064}{Phys.Lett. {\bf B593}
  (2004)  143--150},
\href{http://arxiv.org/abs/hep-ph/0403050}{{\tt arXiv:hep-ph/0403050
  [hep-ph]}}.

\bibitem{Stirling:2012ak}
W.~Stirling and E.~Vryonidou, {\em {Electroweak corrections and Bloch-Nordsieck
  violations in 2-to-2 processes at the LHC}\/},
  \href{http://dx.doi.org/10.1007/JHEP04(2013)155}{JHEP {\bf 1304} (2013)
  155},
\href{http://arxiv.org/abs/1212.6537}{{\tt arXiv:1212.6537 [hep-ph]}}.

\bibitem{Ciafaloni:1998xg}
P.~Ciafaloni and D.~Comelli, {\em {Sudakov enhancement of electroweak
  corrections}\/},
  \href{http://dx.doi.org/10.1016/S0370-2693(98)01541-X}{Phys.Lett. {\bf B446}
  (1999)  278--284},
\href{http://arxiv.org/abs/hep-ph/9809321}{{\tt arXiv:hep-ph/9809321
  [hep-ph]}}.

\bibitem{Beccaria:1999fk}
M.~Beccaria, P.~Ciafaloni, D.~Comelli, F.~Renard, and C.~Verzegnassi, {\em
  {Logarithmic expansion of electroweak corrections to four-fermion processes
  in the TeV region}\/},
  \href{http://dx.doi.org/10.1103/PhysRevD.61.073005}{Phys.Rev. {\bf D61}
  (2000)  073005},
\href{http://arxiv.org/abs/hep-ph/9906319}{{\tt arXiv:hep-ph/9906319
  [hep-ph]}}.

\bibitem{Ciafaloni:2000rp}
M.~Ciafaloni, P.~Ciafaloni, and D.~Comelli, {\em {Electroweak Bloch-Nordsieck
  violation at the TeV scale: 'Strong' weak interactions?}\/},
  \href{http://dx.doi.org/10.1016/S0550-3213(00)00508-3}{Nucl.Phys. {\bf B589}
  (2000)  359--380},
\href{http://arxiv.org/abs/hep-ph/0004071}{{\tt arXiv:hep-ph/0004071
  [hep-ph]}}.

\bibitem{Denner:2000jv}
A.~Denner and S.~Pozzorini, {\em {One loop leading logarithms in electroweak
  radiative corrections. 1. Results}\/},
  \href{http://dx.doi.org/10.1007/s100520100551}{Eur.Phys.J. {\bf C18} (2001)
  461--480},
\href{http://arxiv.org/abs/hep-ph/0010201}{{\tt arXiv:hep-ph/0010201
  [hep-ph]}}.

\bibitem{Ciafaloni:2001vt}
M.~Ciafaloni, P.~Ciafaloni, and D.~Comelli, {\em {Bloch-Nordsieck violation in
  spontaneously broken Abelian theories}\/},
  \href{http://dx.doi.org/10.1103/PhysRevLett.87.211802}{Phys.Rev.Lett. {\bf
  87} (2001)  211802},
\href{http://arxiv.org/abs/hep-ph/0103315}{{\tt arXiv:hep-ph/0103315
  [hep-ph]}}.

\bibitem{Kuhn:2004em}
J.~H. Kuhn, A.~Kulesza, S.~Pozzorini, and M.~Schulze, {\em {Logarithmic
  electroweak corrections to hadronic Z+1 jet production at large transverse
  momentum}\/},
  \href{http://dx.doi.org/10.1016/j.physletb.2005.01.059}{Phys.Lett. {\bf B609}
  (2005)  277--285},
\href{http://arxiv.org/abs/hep-ph/0408308}{{\tt arXiv:hep-ph/0408308
  [hep-ph]}}.

\bibitem{Kuhn:2005az}
J.~H. Kuhn, A.~Kulesza, S.~Pozzorini, and M.~Schulze, {\em {One-loop weak
  corrections to hadronic production of Z bosons at large transverse
  momenta}\/},
  \href{http://dx.doi.org/10.1016/j.nuclphysb.2005.08.019}{Nucl.Phys. {\bf
  B727} (2005)  368--394},
\href{http://arxiv.org/abs/hep-ph/0507178}{{\tt arXiv:hep-ph/0507178
  [hep-ph]}}.

\bibitem{Kuhn:2007qc}
J.~H. Kuhn, A.~Kulesza, S.~Pozzorini, and M.~Schulze, {\em {Electroweak
  corrections to large transverse momentum production of W bosons at the
  LHC}\/},  \href{http://dx.doi.org/10.1016/j.physletb.2007.06.028}{Phys.Lett.
  {\bf B651} (2007)  160--165},
\href{http://arxiv.org/abs/hep-ph/0703283}{{\tt arXiv:hep-ph/0703283
  [hep-ph]}}.

\bibitem{Accomando:2006hq}
E.~Accomando, A.~Denner, and S.~Pozzorini, {\em {Logarithmic electroweak
  corrections to $e^+ e^- \to \nu_e {\bar \nu}_e W^+ W^-$}\/},
  \href{http://dx.doi.org/10.1088/1126-6708/2007/03/078}{JHEP {\bf 0703} (2007)
   078},
\href{http://arxiv.org/abs/hep-ph/0611289}{{\tt arXiv:hep-ph/0611289
  [hep-ph]}}.

\bibitem{Pozzorini:2001rs}
S.~Pozzorini, {\em {Electroweak radiative corrections at high-energies}\/},
\href{http://arxiv.org/abs/hep-ph/0201077}{{\tt arXiv:hep-ph/0201077
  [hep-ph]}}.

\bibitem{Mangano:2002ea}
M.~L. Mangano, M.~Moretti, F.~Piccinini, R.~Pittau, and A.~D. Polosa, {\em
  {ALPGEN, a generator for hard multiparton processes in hadronic
  collisions}\/},  JHEP {\bf 0307} (2003)  001,
\href{http://arxiv.org/abs/hep-ph/0206293}{{\tt arXiv:hep-ph/0206293
  [hep-ph]}}.

\bibitem{Gerwick:2013haa}
E.~Gerwick, {\em {Recursive prescription for logarithmic jet rate
  coefficients}\/},  Phys.Rev. {\bf D88} (2013)  094009,
\href{http://arxiv.org/abs/1305.6319}{{\tt arXiv:1305.6319 [hep-ph]}}.

\bibitem{Gerwick:2012fw}
E.~Gerwick, S.~Schumann, B.~Gripaios, and B.~Webber, {\em {QCD Jet Rates with
  the Inclusive Generalized kt Algorithms}\/},
  \href{http://dx.doi.org/10.1007/JHEP04(2013)089}{JHEP {\bf 1304} (2013)
  089},
\href{http://arxiv.org/abs/1212.5235}{{\tt arXiv:1212.5235}}.

\bibitem{Schumann:2007mg}
S.~Schumann and F.~Krauss, {\em {A Parton shower algorithm based on
  Catani-Seymour dipole factorisation}\/},
  \href{http://dx.doi.org/10.1088/1126-6708/2008/03/038}{JHEP {\bf 0803} (2008)
   038},
\href{http://arxiv.org/abs/0709.1027}{{\tt arXiv:0709.1027 [hep-ph]}}.

\bibitem{Sjostrand:2007gs}
T.~Sj{\"o}strand, S.~Mrenna, and P.~Z. Skands, {\em {A Brief Introduction to
  PYTHIA 8.1}\/},
  \href{http://dx.doi.org/10.1016/j.cpc.2008.01.036}{Comput.Phys.Commun. {\bf
  178} (2008)  852--867},
\href{http://arxiv.org/abs/0710.3820}{{\tt arXiv:0710.3820 [hep-ph]}}.

\bibitem{Bauer:2007ad}
C.~W. Bauer and F.~J. Tackmann, {\em {Gaining analytic control of parton
  showers}\/},  \href{http://dx.doi.org/10.1103/PhysRevD.76.114017}{Phys.Rev.
  {\bf D76} (2007)  114017},
\href{http://arxiv.org/abs/0705.1719}{{\tt arXiv:0705.1719 [hep-ph]}}.

\bibitem{Bauer:2006mk}
C.~W. Bauer and M.~D. Schwartz, {\em {Event Generation from Effective Field
  Theory}\/},  \href{http://dx.doi.org/10.1103/PhysRevD.76.074004}{Phys.Rev.
  {\bf D76} (2007)  074004},
\href{http://arxiv.org/abs/hep-ph/0607296}{{\tt arXiv:hep-ph/0607296
  [hep-ph]}}.

\bibitem{Webber:2010vz}
B.~R. Webber, {\em {QCD Jets and Parton Showers}\/},
\href{http://arxiv.org/abs/1009.5871}{{\tt arXiv:1009.5871 [hep-ph]}}.

\bibitem{Andersen:2009nu}
J.~R. Andersen and J.~M. Smillie, {\em {Constructing All-Order Corrections to
  Multi-Jet Rates}\/},  \href{http://dx.doi.org/10.1007/JHEP01(2010)039}{JHEP
  {\bf 1001} (2010)  039},
\href{http://arxiv.org/abs/0908.2786}{{\tt arXiv:0908.2786 [hep-ph]}}.

\bibitem{Andersen:2009he}
J.~R. Andersen and J.~M. Smillie, {\em {The Factorisation of the t-channel Pole
  in Quark-Gluon Scattering}\/},
  \href{http://dx.doi.org/10.1103/PhysRevD.81.114021}{Phys.Rev. {\bf D81}
  (2010)  114021},
\href{http://arxiv.org/abs/0910.5113}{{\tt arXiv:0910.5113 [hep-ph]}}.

\bibitem{Andersen:2011hs}
J.~R. Andersen and J.~M. Smillie, {\em {Multiple Jets at the LHC with High
  Energy Jets}\/},  \href{http://dx.doi.org/10.1007/JHEP06(2011)010}{JHEP {\bf
  1106} (2011)  010},
\href{http://arxiv.org/abs/1101.5394}{{\tt arXiv:1101.5394 [hep-ph]}}.

\bibitem{BH}
C.~Berger, Z.~Bern, L.~J. Dixon, F.~Febres~Cordero, D.~Forde, H.~Ita, D.~A.
  Kosower, and D.~Ma{\^{\i}}tre, {\em {\it An automated implementation of
  on-shell methods for one-loop amplitudes}\/},
  \href{http://dx.doi.org/10.1103/PhysRevD.78.036003}{Phys. Rev. {\bf D78}
  (2008)  036003},
\href{http://arxiv.org/abs/0803.4180}{{\tt arXiv:0803.4180 [hep-ph]}}.

\bibitem{pureQCD}
Z.~Bern, G.~Diana, L.~J. Dixon, F.~Febres~Cordero, S.~H{\"{o}}che, H.~Ita,
  D.~A. Kosower, D.~Ma{\^{\i}}tre, and K.~Ozeren, {\em {\it Four-jet production
  at the Large Hadron Collider at next-to-leading order in QCD}\/},
  \href{http://dx.doi.org/10.1103/PhysRevLett.109.042001}{Phys.{} Rev.{}
  Lett.{} {\bf 109} (2011)  042001},
\href{http://arxiv.org/abs/1112.3940}{{\tt arXiv:1112.3940 [hep-ph]}}.

\bibitem{sherpa1}
T.~Gleisberg, S.~H{\"{o}}che, F.~Krauss, A.~Sch{\"{a}}licke, S.~Schumann, and
  J.~Winter, {\em {\it SHERPA 1.$\alpha$: A proof of concept version}\/},
  \href{http://dx.doi.org/10.1088/1126-6708/2004/02/056}{JHEP {\bf 0402} (2004)
   056},
\href{http://arxiv.org/abs/hep-ph/0311263}{{\tt arXiv:hep-ph/0311263
  [hep-ph]}}.

\bibitem{sherpa2}
T.~Gleisberg, S.~H{\"{o}}che, F.~Krauss, M.~Sch{\"{o}}nherr, S.~Schumann,
  F.~Siegert, and J.~Winter, {\em {\it Event generation with SHERPA 1.1}\/},
  \href{http://dx.doi.org/10.1088/1126-6708/2009/02/007}{JHEP {\bf 0902} (2009)
   007},
\href{http://arxiv.org/abs/0811.4622}{{\tt arXiv:0811.4622 [hep-ph]}}.

\bibitem{amegicI}
F.~Krauss, R.~Kuhn, and G.~Soff, {\em {\it AMEGIC++ 1.0: A matrix element
  generator in C++}\/},
  \href{http://dx.doi.org/doi:10.1088/1126-6708/2002/02/044}{JHEP {\bf 0202}
  (2002)  044},
\href{http://arxiv.org/abs/hep-ph/0109036}{{\tt arXiv:hep-ph/0109036
  [hep-ph]}}.

\bibitem{amegicII}
T.~Gleisberg and F.~Krauss, {\em {\it Automating dipole subtraction for QCD NLO
  calculations}\/},
  \href{http://dx.doi.org/10.1140/epjc/s10052-007-0495-0}{Eur.{} Phys.{} J.{}
  {\bf C53} (2008)  501--523},
\href{http://arxiv.org/abs/0709.2881}{{\tt arXiv:0709.2881 [hep-ph]}}.

\bibitem{Aad:2011jz}
{ATLAS Collaboration} Collaboration, G.~Aad et al., {\em {Measurement of dijet
  production with a veto on additional central jet activity in $pp$ collisions
  at $\sqrt{s}=7$ TeV using the ATLAS detector}\/},
  \href{http://dx.doi.org/10.1007/JHEP09(2011)053}{JHEP {\bf 1109} (2011)
  053},
\href{http://arxiv.org/abs/1107.1641}{{\tt arXiv:1107.1641 [hep-ex]}}.

\bibitem{Andersen:2007mp}
J.~Andersen, T.~Binoth, G.~Heinrich, and J.~Smillie, {\em {Loop induced
  interference effects in Higgs Boson plus two jet production at the LHC}\/},
  \href{http://dx.doi.org/10.1088/1126-6708/2008/02/057}{JHEP {\bf 0802} (2008)
   057},
\href{http://arxiv.org/abs/0709.3513}{{\tt arXiv:0709.3513 [hep-ph]}}.

\bibitem{Bredenstein:2008tm}
A.~Bredenstein, K.~Hagiwara, and B.~Jager, {\em {Mixed QCD-electroweak
  contributions to Higgs-plus-dijet production at the LHC}\/},
  \href{http://dx.doi.org/10.1103/PhysRevD.77.073004}{Phys.Rev. {\bf D77}
  (2008)  073004},
\href{http://arxiv.org/abs/0801.4231}{{\tt arXiv:0801.4231 [hep-ph]}}.

\bibitem{Dixon:2009uk}
L.~J. Dixon and Y.~Sofianatos, {\em {Analytic one-loop amplitudes for a Higgs
  boson plus four partons}\/},
  \href{http://dx.doi.org/10.1088/1126-6708/2009/08/058}{JHEP {\bf 0908} (2009)
   058},
\href{http://arxiv.org/abs/0906.0008}{{\tt arXiv:0906.0008 [hep-ph]}}.

\bibitem{Plehn:2001nj}
T.~Plehn, D.~L. Rainwater, and D.~Zeppenfeld, {\em {Determining the structure
  of Higgs couplings at the LHC}\/},
  \href{http://dx.doi.org/10.1103/PhysRevLett.88.051801}{Phys.Rev.Lett. {\bf
  88} (2002)  051801},
\href{http://arxiv.org/abs/hep-ph/0105325}{{\tt arXiv:hep-ph/0105325
  [hep-ph]}}.

\bibitem{Klamke:2007cu}
G.~Klamke and D.~Zeppenfeld, {\em {Higgs plus two jet production via gluon
  fusion as a signal at the CERN LHC}\/},
  \href{http://dx.doi.org/10.1088/1126-6708/2007/04/052}{JHEP {\bf 0704} (2007)
   052},
\href{http://arxiv.org/abs/hep-ph/0703202}{{\tt arXiv:hep-ph/0703202
  [HEP-PH]}}.

\bibitem{Dokshitzer:1987nc}
Y.~L. Dokshitzer, S.~Troian, and V.~A. Khoze, {\em {Collective {QCD} Effects in
  the Structure of Final Multi - Hadron States. (In Russian)}\/},
Sov.J.Nucl.Phys. {\bf 46} (1987)  712--719.

\bibitem{Dokshitzer:1991he}
Y.~L. Dokshitzer, V.~A. Khoze, and T.~Sj{\"o}strand, {\em {Rapidity gaps in
  Higgs production}\/},
\href{http://dx.doi.org/10.1016/0370-2693(92)90312-R}{Phys.Lett. {\bf B274}
  (1992)  116--121}.

\bibitem{Andersen:2010zx}
J.~R. Andersen, K.~Arnold, and D.~Zeppenfeld, {\em {Azimuthal Angle
  Correlations for Higgs Boson plus Multi-Jet Events}\/},
  \href{http://dx.doi.org/10.1007/JHEP06(2010)091}{JHEP {\bf 1006} (2010)
  091},
\href{http://arxiv.org/abs/1001.3822}{{\tt arXiv:1001.3822 [hep-ph]}}.

\bibitem{Alwall:2011uj}
J.~Alwall, M.~Herquet, F.~Maltoni, O.~Mattelaer, and T.~Stelzer, {\em {MadGraph
  5 : Going Beyond}\/},  \href{http://dx.doi.org/10.1007/JHEP06(2011)128}{JHEP
  {\bf 1106} (2011)  128},
\href{http://arxiv.org/abs/1106.0522}{{\tt arXiv:1106.0522 [hep-ph]}}.

\bibitem{Campbell:2006xx}
J.~M. Campbell, R.~K. Ellis, and G.~Zanderighi, {\em {Next-to-Leading order
  Higgs + 2 jet production via gluon fusion}\/},
  \href{http://dx.doi.org/10.1088/1126-6708/2006/10/028}{JHEP {\bf 0610} (2006)
   028},
\href{http://arxiv.org/abs/hep-ph/0608194}{{\tt arXiv:hep-ph/0608194
  [hep-ph]}}.

\bibitem{Frederix:2012ps}
R.~Frederix and S.~Frixione, {\em {Merging meets matching in MC@NLO}\/},
  \href{http://dx.doi.org/10.1007/JHEP12(2012)061}{JHEP {\bf 1212} (2012)
  061},
\href{http://arxiv.org/abs/1209.6215}{{\tt arXiv:1209.6215 [hep-ph]}}.

\bibitem{Andersen:2008ue}
J.~R. Andersen and C.~D. White, {\em {A New Framework for Multijet Predictions
  and its application to Higgs Boson production at the LHC}\/},
  \href{http://dx.doi.org/10.1103/PhysRevD.78.051501}{Phys.Rev. {\bf D78}
  (2008)  051501},
\href{http://arxiv.org/abs/0802.2858}{{\tt arXiv:0802.2858 [hep-ph]}}.

\bibitem{Andersen:2008gc}
J.~R. Andersen, V.~Del~Duca, and C.~D. White, {\em {Higgs Boson Production in
  Association with Multiple Hard Jets}\/},
  \href{http://dx.doi.org/10.1088/1126-6708/2009/02/015}{JHEP {\bf 0902} (2009)
   015},
\href{http://arxiv.org/abs/0808.3696}{{\tt arXiv:0808.3696 [hep-ph]}}.

\bibitem{Andersen:2014xx}
J.~R. Andersen, T.~A. Hapola, and J.~M. Smillie To appear  .

\bibitem{ATLAS:2011krm}
{ATLAS} Collaboration, {\em {Further ATLAS tunes of PYTHIA6 and Pythia 8}\/},
ATL-PHYS-PUB-2011-014, ATL-COM-PHYS-2011-1507 (2011)  .

\bibitem{ATLAS:2012uec}
{ATLAS} Collaboration, {\em {Summary of ATLAS Pythia 8 tunes}\/},
ATL-PHYS-PUB-2012-003, ATL-COM-PHYS-2012-738 (2012)  .

\bibitem{Alioli:2008tz}
S.~Alioli, P.~Nason, C.~Oleari, and E.~Re, {\em {NLO Higgs boson production via
  gluon fusion matched with shower in POWHEG}\/},
  \href{http://dx.doi.org/10.1088/1126-6708/2009/04/002}{JHEP {\bf 0904} (2009)
   002},
\href{http://arxiv.org/abs/0812.0578}{{\tt arXiv:0812.0578 [hep-ph]}}.

\bibitem{Campbell:2012am}
J.~M. Campbell, R.~K. Ellis, R.~Frederix, P.~Nason, C.~Oleari, et al., {\em
  {NLO Higgs Boson Production Plus One and Two Jets Using the POWHEG BOX,
  MadGraph4 and MCFM}\/},
  \href{http://dx.doi.org/10.1007/JHEP07(2012)092}{JHEP {\bf 1207} (2012)
  092},
\href{http://arxiv.org/abs/1202.5475}{{\tt arXiv:1202.5475 [hep-ph]}}.

\bibitem{Maltoni:2002qb}
F.~Maltoni and T.~Stelzer, {\em {MadEvent: Automatic event generation with
  MadGraph}\/},  \href{http://dx.doi.org/10.1088/1126-6708/2003/02/027}{JHEP
  {\bf 0302} (2003)  027},
\href{http://arxiv.org/abs/hep-ph/0208156}{{\tt arXiv:hep-ph/0208156
  [hep-ph]}}.

\bibitem{Lonnblad:2001iq}
L.~L{\"o}nnblad, {\em {Correcting the color dipole cascade model with fixed
  order matrix elements}\/},
  \href{http://dx.doi.org/10.1088/1126-6708/2002/05/046}{JHEP {\bf 0205} (2002)
   046},
\href{http://arxiv.org/abs/hep-ph/0112284}{{\tt arXiv:hep-ph/0112284
  [hep-ph]}}.

\bibitem{Lavesson:2005xu}
N.~Lavesson and L.~Lonnblad, {\em {W+jets matrix elements and the dipole
  cascade}\/},  \href{http://dx.doi.org/10.1088/1126-6708/2005/07/054}{JHEP
  {\bf 0507} (2005)  054},
\href{http://arxiv.org/abs/hep-ph/0503293}{{\tt arXiv:hep-ph/0503293
  [hep-ph]}}.

\bibitem{Lonnblad:2011xx}
L.~L{\"o}nnblad and S.~Prestel, {\em {Matching Tree-Level Matrix Elements with
  Interleaved Showers}\/},
  \href{http://dx.doi.org/10.1007/JHEP03(2012)019}{JHEP {\bf 1203} (2012)
  019},
\href{http://arxiv.org/abs/1109.4829}{{\tt arXiv:1109.4829 [hep-ph]}}.

\bibitem{Gehrmann:2012yg}
T.~Gehrmann, S.~H{\"o}che, F.~Krauss, M.~Sch{\"o}nherr, and F.~Siegert, {\em
  {NLO QCD matrix elements + parton showers in $e^+e^-\to$ hadrons}\/},
  \href{http://dx.doi.org/10.1007/JHEP01(2013)144}{JHEP {\bf 1301} (2013)
  144},
\href{http://arxiv.org/abs/1207.5031}{{\tt arXiv:1207.5031 [hep-ph]}}.

\bibitem{Cascioli:2013gfa}
F.~Cascioli, S.~H{\"o}che, F.~Krauss, P.~Maierh{\"o}fer, S.~Pozzorini, et al.,
  {\em {Precise Higgs-background predictions: merging NLO QCD and squared
  quark-loop corrections to four-lepton + 0,1 jet production}\/},
  \href{http://dx.doi.org/10.1007/JHEP01(2014)046}{JHEP {\bf 1401} (2014)
  046},
\href{http://arxiv.org/abs/1309.0500}{{\tt arXiv:1309.0500 [hep-ph]}}.

\bibitem{Cascioli:2013era}
F.~Cascioli, P.~Maierh{\"o}fer, N.~Moretti, S.~Pozzorini, and F.~Siegert, {\em
  {NLO matching for ttbb production with massive b-quarks}\/},
\href{http://arxiv.org/abs/1309.5912}{{\tt arXiv:1309.5912 [hep-ph]}}.

\bibitem{Hoeche:2014lxa}
S.~H{\"o}che, F.~Krauss, and M.~Sch{\"o}nherr, {\em {Uncertainties in MEPS@NLO
  calculations of h+jets}\/},
\href{http://arxiv.org/abs/1401.7971}{{\tt arXiv:1401.7971 [hep-ph]}}.

\bibitem{Dawson:1990zj}
S.~Dawson, {\em {Radiative corrections to Higgs boson production}\/},
\href{http://dx.doi.org/10.1016/0550-3213(91)90061-2}{Nucl.Phys. {\bf B359}
  (1991)  283--300}.

\bibitem{Djouadi:1991tka}
A.~Djouadi, M.~Spira, and P.~Zerwas, {\em {Production of Higgs bosons in proton
  colliders: QCD corrections}\/},
\href{http://dx.doi.org/10.1016/0370-2693(91)90375-Z}{Phys.Lett. {\bf B264}
  (1991)  440--446}.

\bibitem{deFlorian:1999zd}
D.~de~Florian, M.~Grazzini, and Z.~Kunszt, {\em {Higgs production with large
  transverse momentum in hadronic collisions at next-to-leading order}\/},
  \href{http://dx.doi.org/10.1103/PhysRevLett.82.5209}{Phys.Rev.Lett. {\bf 82}
  (1999)  5209--5212},
\href{http://arxiv.org/abs/hep-ph/9902483}{{\tt arXiv:hep-ph/9902483
  [hep-ph]}}.

\bibitem{Ravindran:2002dc}
V.~Ravindran, J.~Smith, and W.~Van~Neerven, {\em {Next-to-leading order QCD
  corrections to differential distributions of Higgs boson production in hadron
  hadron collisions}\/},
  \href{http://dx.doi.org/10.1016/S0550-3213(02)00333-4}{Nucl.Phys. {\bf B634}
  (2002)  247--290},
\href{http://arxiv.org/abs/hep-ph/0201114}{{\tt arXiv:hep-ph/0201114
  [hep-ph]}}.

\bibitem{Hoche:2012wh}
S.~H{\"o}che and M.~Sch{\"o}nherr, {\em {Uncertainties in next-to-leading order
  plus parton shower matched simulations of inclusive jet and dijet
  production}\/},
  \href{http://dx.doi.org/10.1103/PhysRevD.86.094042}{Phys.Rev. {\bf D86}
  (2012)  094042},
\href{http://arxiv.org/abs/1208.2815}{{\tt arXiv:1208.2815 [hep-ph]}}.

\bibitem{AlcarazMaestre:2012vp}
{SM AND NLO MULTILEG and SM MC Working Groups} Collaboration,
  J.~Alcaraz~Maestre et al., {\em {The SM and NLO Multileg and SM MC Working
  Groups: Summary Report}\/},
\href{http://arxiv.org/abs/1203.6803}{{\tt arXiv:1203.6803 [hep-ph]}}.

\bibitem{Buckley:2010ar}
A.~Buckley, J.~Butterworth, L.~L{\"o}nnblad, D.~Grellscheid, H.~Hoeth, et al.,
  {\em {Rivet user manual}\/},
  \href{http://dx.doi.org/10.1016/j.cpc.2013.05.021}{Comput.Phys.Commun. {\bf
  184} (2013)  2803--2819},
\href{http://arxiv.org/abs/1003.0694}{{\tt arXiv:1003.0694 [hep-ph]}}.

\bibitem{RivetAnalysis}
 The analysis can be found at
  \texttt{http://phystev.in2p3.fr/wiki/2013:groups:sm:higgs:hdijets}.

\bibitem{Aad:2014dta}
{ATLAS} Collaboration, G.~Aad et al., {\em {Measurement of the electroweak
  production of dijets in association with a Z-boson and distributions
  sensitive to vector boson fusion in proton-proton collisions at sqrt(s) = 8
  TeV using the ATLAS detector}\/},
\href{http://arxiv.org/abs/1401.7610}{{\tt arXiv:1401.7610 [hep-ex]}}.

\bibitem{DelDuca:2001eu}
V.~Del~Duca, W.~Kilgore, C.~Oleari, C.~Schmidt, and D.~Zeppenfeld, {\em {Higgs
  + 2 jets via gluon fusion}\/},
  \href{http://dx.doi.org/10.1103/PhysRevLett.87.122001}{Phys.Rev.Lett. {\bf
  87} (2001)  122001},
\href{http://arxiv.org/abs/hep-ph/0105129}{{\tt arXiv:hep-ph/0105129
  [hep-ph]}}.

\bibitem{DelDuca:2001fn}
V.~Del~Duca, W.~Kilgore, C.~Oleari, C.~Schmidt, and D.~Zeppenfeld, {\em {Gluon
  fusion contributions to H + 2 jet production}\/},
  \href{http://dx.doi.org/10.1016/S0550-3213(01)00446-1}{Nucl.Phys. {\bf B616}
  (2001)  367--399},
\href{http://arxiv.org/abs/hep-ph/0108030}{{\tt arXiv:hep-ph/0108030
  [hep-ph]}}.

\bibitem{Campanario:2013mga}
F.~Campanario and M.~Kubocz, {\em {Higgs boson production in association with
  three jets via gluon fusion at the LHC: Gluonic contributions}\/},
\href{http://arxiv.org/abs/1306.1830}{{\tt arXiv:1306.1830 [hep-ph]}}.

\bibitem{vanDeurzen:2013aaa}
H.~van Deurzen, N.~Greiner, G.~Luisoni, P.~Mastrolia, E.~Mirabella, et al.,
  {\em {NLO QCD corrections to the production of Higgs plus two jets at the
  LHC}\/},  \href{http://dx.doi.org/10.1016/j.physletb.2013.02.051}{Phys.Lett.
  {\bf B721} (2013)  74--81},
\href{http://arxiv.org/abs/1301.0493}{{\tt arXiv:1301.0493 [hep-ph]}}.

\bibitem{DelDuca:2004wt}
V.~Del~Duca, A.~Frizzo, and F.~Maltoni, {\em {Higgs boson production in
  association with three jets}\/},
  \href{http://dx.doi.org/10.1088/1126-6708/2004/05/064}{JHEP {\bf 0405} (2004)
   064},
\href{http://arxiv.org/abs/hep-ph/0404013}{{\tt arXiv:hep-ph/0404013
  [hep-ph]}}.

\bibitem{Dixon:2004za}
L.~J. Dixon, E.~N. Glover, and V.~V. Khoze, {\em {MHV rules for Higgs plus
  multi-gluon amplitudes}\/},
  \href{http://dx.doi.org/10.1088/1126-6708/2004/12/015}{JHEP {\bf 0412} (2004)
   015},
\href{http://arxiv.org/abs/hep-th/0411092}{{\tt arXiv:hep-th/0411092
  [hep-th]}}.

\bibitem{Badger:2004ty}
S.~Badger, E.~N. Glover, and V.~V. Khoze, {\em {MHV rules for Higgs plus
  multi-parton amplitudes}\/},
  \href{http://dx.doi.org/10.1088/1126-6708/2005/03/023}{JHEP {\bf 0503} (2005)
   023},
\href{http://arxiv.org/abs/hep-th/0412275}{{\tt arXiv:hep-th/0412275
  [hep-th]}}.

\bibitem{Ellis:2005qe}
R.~K. Ellis, W.~Giele, and G.~Zanderighi, {\em {Virtual QCD corrections to
  Higgs boson plus four parton processes}\/},
  \href{http://dx.doi.org/10.1103/PhysRevD.72.054018,
  10.1103/PhysRevD.74.079902}{Phys.Rev. {\bf D72} (2005)  054018},
\href{http://arxiv.org/abs/hep-ph/0506196}{{\tt arXiv:hep-ph/0506196
  [hep-ph]}}.

\bibitem{Ellis:2005zh}
R.~K. Ellis, W.~Giele, and G.~Zanderighi, {\em {Semi-numerical evaluation of
  one-loop corrections}\/},
  \href{http://dx.doi.org/10.1103/PhysRevD.73.014027}{Phys.Rev. {\bf D73}
  (2006)  014027},
\href{http://arxiv.org/abs/hep-ph/0508308}{{\tt arXiv:hep-ph/0508308
  [hep-ph]}}.

\bibitem{Berger:2006sh}
C.~F. Berger, V.~Del~Duca, and L.~J. Dixon, {\em {Recursive Construction of
  Higgs-Plus-Multiparton Loop Amplitudes: The Last of the Phi-nite Loop
  Amplitudes}\/},  \href{http://dx.doi.org/10.1103/PhysRevD.76.099901,
  10.1103/PhysRevD.74.094021}{Phys.Rev. {\bf D74} (2006)  094021},
\href{http://arxiv.org/abs/hep-ph/0608180}{{\tt arXiv:hep-ph/0608180
  [hep-ph]}}.

\bibitem{Badger:2006us}
S.~Badger and E.~N. Glover, {\em {One-loop helicity amplitudes for $H \to$
  gluons: The All-minus configuration}\/},
  \href{http://dx.doi.org/10.1016/j.nuclphysbps.2006.09.030}{Nucl.Phys.Proc.Suppl.
  {\bf 160} (2006)  71--75},
\href{http://arxiv.org/abs/hep-ph/0607139}{{\tt arXiv:hep-ph/0607139
  [hep-ph]}}.

\bibitem{Badger:2007si}
S.~Badger, E.~N. Glover, and K.~Risager, {\em {One-loop phi-MHV amplitudes
  using the unitarity bootstrap}\/},
  \href{http://dx.doi.org/10.1088/1126-6708/2007/07/066}{JHEP {\bf 0707} (2007)
   066},
\href{http://arxiv.org/abs/0704.3914}{{\tt arXiv:0704.3914 [hep-ph]}}.

\bibitem{Glover:2008ffa}
E.~N. Glover, P.~Mastrolia, and C.~Williams, {\em {One-loop phi-MHV amplitudes
  using the unitarity bootstrap: The General helicity case}\/},
  \href{http://dx.doi.org/10.1088/1126-6708/2008/08/017}{JHEP {\bf 0808} (2008)
   017},
\href{http://arxiv.org/abs/0804.4149}{{\tt arXiv:0804.4149 [hep-ph]}}.

\bibitem{Badger:2009hw}
S.~Badger, E.~Nigel~Glover, P.~Mastrolia, and C.~Williams, {\em {One-loop Higgs
  plus four gluon amplitudes: Full analytic results}\/},
  \href{http://dx.doi.org/10.1007/JHEP01(2010)036}{JHEP {\bf 1001} (2010)
  036},
\href{http://arxiv.org/abs/0909.4475}{{\tt arXiv:0909.4475 [hep-ph]}}.

\bibitem{Badger:2009vh}
S.~Badger, J.~M. Campbell, R.~K. Ellis, and C.~Williams, {\em {Analytic results
  for the one-loop NMHV Hqqgg amplitude}\/},
  \href{http://dx.doi.org/10.1088/1126-6708/2009/12/035}{JHEP {\bf 0912} (2009)
   035},
\href{http://arxiv.org/abs/0910.4481}{{\tt arXiv:0910.4481 [hep-ph]}}.

\bibitem{Mastrolia:2012bu}
P.~Mastrolia, E.~Mirabella, and T.~Peraro, {\em {Integrand reduction of
  one-loop scattering amplitudes through Laurent series expansion}\/},
  \href{http://dx.doi.org/10.1007/JHEP06(2012)095}{JHEP {\bf 1206} (2012)
  095},
\href{http://arxiv.org/abs/1203.0291}{{\tt arXiv:1203.0291 [hep-ph]}}.

\bibitem{Nogueira:1991ex}
P.~Nogueira, {\em {Automatic Feynman graph generation}\/},
\href{http://dx.doi.org/10.1006/jcph.1993.1074}{J.Comput.Phys. {\bf 105} (1993)
   279--289}.

\bibitem{Vermaseren:2000nd}
J.~Vermaseren, {\em {New features of FORM}\/},
\href{http://arxiv.org/abs/math-ph/0010025}{{\tt arXiv:math-ph/0010025
  [math-ph]}}.

\bibitem{Reiter:2009ts}
T.~Reiter, {\em {Optimising Code Generation with haggies}\/},
  \href{http://dx.doi.org/10.1016/j.cpc.2010.01.012}{Comput.Phys.Commun. {\bf
  181} (2010)  1301--1331},
\href{http://arxiv.org/abs/0907.3714}{{\tt arXiv:0907.3714 [hep-ph]}}.

\bibitem{Cullen:2010jv}
G.~Cullen, M.~Koch-Janusz, and T.~Reiter, {\em {Spinney: A Form Library for
  Helicity Spinors}\/},
  \href{http://dx.doi.org/10.1016/j.cpc.2011.06.007}{Comput.Phys.Commun. {\bf
  182} (2011)  2368--2387},
\href{http://arxiv.org/abs/1008.0803}{{\tt arXiv:1008.0803 [hep-ph]}}.

\bibitem{Ossola:2006us}
G.~Ossola, C.~G. Papadopoulos, and R.~Pittau, {\em {Reducing full one-loop
  amplitudes to scalar integrals at the integrand level}\/},
  \href{http://dx.doi.org/10.1016/j.nuclphysb.2006.11.012}{Nucl.Phys. {\bf
  B763} (2007)  147--169}, \href{http://arxiv.org/abs/hep-ph/0609007}{{\tt
  arXiv:hep-ph/0609007 [hep-ph]}}.

\bibitem{Ossola:2007bb}
G.~Ossola, C.~G. Papadopoulos, and R.~Pittau, {\em {Numerical evaluation of
  six-photon amplitudes}\/},
  \href{http://dx.doi.org/10.1088/1126-6708/2007/07/085}{JHEP {\bf 0707} (2007)
   085},
\href{http://arxiv.org/abs/0704.1271}{{\tt arXiv:0704.1271 [hep-ph]}}.

\bibitem{Ellis:2007br}
R.~Ellis, W.~Giele, and Z.~Kunszt, {\em {A Numerical Unitarity Formalism for
  Evaluating One-Loop Amplitudes}\/},
  \href{http://dx.doi.org/10.1088/1126-6708/2008/03/003}{JHEP {\bf 0803} (2008)
   003},
\href{http://arxiv.org/abs/0708.2398}{{\tt arXiv:0708.2398 [hep-ph]}}.

\bibitem{Ossola:2008xq}
G.~Ossola, C.~G. Papadopoulos, and R.~Pittau, {\em {On the Rational Terms of
  the one-loop amplitudes}\/},
  \href{http://dx.doi.org/10.1088/1126-6708/2008/05/004}{JHEP {\bf 0805} (2008)
   004},
\href{http://arxiv.org/abs/0802.1876}{{\tt arXiv:0802.1876 [hep-ph]}}.

\bibitem{Mastrolia:2008jb}
P.~Mastrolia, G.~Ossola, C.~Papadopoulos, and R.~Pittau, {\em {Optimizing the
  Reduction of One-Loop Amplitudes}\/},
  \href{http://dx.doi.org/10.1088/1126-6708/2008/06/030}{JHEP {\bf 0806} (2008)
   030},
\href{http://arxiv.org/abs/0803.3964}{{\tt arXiv:0803.3964 [hep-ph]}}.

\bibitem{Mastrolia:2012an}
P.~Mastrolia, E.~Mirabella, G.~Ossola, and T.~Peraro, {\em {Scattering
  Amplitudes from Multivariate Polynomial Division}\/},
  \href{http://dx.doi.org/10.1016/j.physletb.2012.09.053}{Phys.Lett. {\bf B718}
  (2012)  173--177},
\href{http://arxiv.org/abs/1205.7087}{{\tt arXiv:1205.7087 [hep-ph]}}.

\bibitem{Mastrolia:2010nb}
P.~Mastrolia, G.~Ossola, T.~Reiter, and F.~Tramontano, {\em {Scattering
  AMplitudes from Unitarity-based Reduction Algorithm at the
  Integrand-level}\/},  \href{http://dx.doi.org/10.1007/JHEP08(2010)080}{JHEP
  {\bf 1008} (2010)  080}, \href{http://arxiv.org/abs/1006.0710}{{\tt
  arXiv:1006.0710 [hep-ph]}}.

\bibitem{Mastrolia:2012du}
P.~Mastrolia, E.~Mirabella, G.~Ossola, T.~Peraro, and H.~van Deurzen, {\em {The
  Integrand Reduction of One- and Two-Loop Scattering Amplitudes}\/},  PoS {\bf
  LL2012} (2012)  028,
\href{http://arxiv.org/abs/1209.5678}{{\tt arXiv:1209.5678 [hep-ph]}}.

\bibitem{vanDeurzen:2013pja}
H.~van Deurzen, {\em {Associated Higgs Production at NLO with GoSam}\/},
\href{http://dx.doi.org/10.5506/APhysPolB.44.2223}{Acta Phys.Polon. {\bf B44}
  (2013) no.~11, 2223--2230}.

\bibitem{Peraro:2014cba}
T.~Peraro, {\em {Ninja: Automated Integrand Reduction via Laurent Expansion for
  One-Loop Amplitudes}\/},
\href{http://arxiv.org/abs/1403.1229}{{\tt arXiv:1403.1229 [hep-ph]}}.

\bibitem{vanDeurzen:2013saa}
H.~van Deurzen, G.~Luisoni, P.~Mastrolia, E.~Mirabella, G.~Ossola, et al., {\em
  {Multi-leg One-loop Massive Amplitudes from Integrand Reduction via Laurent
  Expansion}\/},
\href{http://arxiv.org/abs/1312.6678}{{\tt arXiv:1312.6678 [hep-ph]}}.

\bibitem{Peraro:2013oja}
T.~Peraro, {\em {Integrand-level Reduction at One and Higher Loops}\/},
\href{http://dx.doi.org/10.5506/APhysPolB.44.2215}{Acta Phys.Polon. {\bf B44}
  (2013)  2215--2221}.

\bibitem{Cullen:2011kv}
G.~Cullen, J.~P. Guillet, G.~Heinrich, T.~Kleinschmidt, E.~Pilon, et al., {\em
  {Golem95C: A library for one-loop integrals with complex masses}\/},
  \href{http://dx.doi.org/10.1016/j.cpc.2011.05.015}{Comput.Phys.Commun. {\bf
  182} (2011)  2276--2284},
\href{http://arxiv.org/abs/1101.5595}{{\tt arXiv:1101.5595 [hep-ph]}}.

\bibitem{Guillet:2013msa}
J.~P. Guillet, G.~Heinrich, and J.~von Soden-Fraunhofen, {\em {Tools for NLO
  automation: extension of the golem95C integral library}\/},
\href{http://arxiv.org/abs/1312.3887}{{\tt arXiv:1312.3887 [hep-ph]}}.

\bibitem{vanHameren:2010cp}
A.~van Hameren, {\em {OneLOop: For the evaluation of one-loop scalar
  functions}\/},
  \href{http://dx.doi.org/10.1016/j.cpc.2011.06.011}{Comput.Phys.Commun. {\bf
  182} (2011)  2427--2438},
\href{http://arxiv.org/abs/1007.4716}{{\tt arXiv:1007.4716 [hep-ph]}}.

\bibitem{Kuipers:2012rf}
J.~Kuipers, T.~Ueda, J.~Vermaseren, and J.~Vollinga, {\em {FORM version
  4.0}\/},
  \href{http://dx.doi.org/10.1016/j.cpc.2012.12.028}{Comput.Phys.Commun. {\bf
  184} (2013)  1453--1467},
\href{http://arxiv.org/abs/1203.6543}{{\tt arXiv:1203.6543 [cs.SC]}}.

\bibitem{Krauss:2001iv}
F.~Krauss, R.~Kuhn, and G.~Soff, {\em {AMEGIC++ 1.0: A Matrix element generator
  in C++}\/},  JHEP {\bf 0202} (2002)  044,
\href{http://arxiv.org/abs/hep-ph/0109036}{{\tt arXiv:hep-ph/0109036
  [hep-ph]}}.

\bibitem{Gleisberg:2007md}
T.~Gleisberg and F.~Krauss, {\em {Automating dipole subtraction for QCD NLO
  calculations}\/},
  \href{http://dx.doi.org/10.1140/epjc/s10052-007-0495-0}{Eur.Phys.J. {\bf C53}
  (2008)  501--523},
\href{http://arxiv.org/abs/0709.2881}{{\tt arXiv:0709.2881 [hep-ph]}}.

\bibitem{Stelzer:1994ta}
T.~Stelzer and W.~Long, {\em {Automatic generation of tree level helicity
  amplitudes}\/},
  \href{http://dx.doi.org/10.1016/0010-4655(94)90084-1}{Comput.Phys.Commun.
  {\bf 81} (1994)  357--371},
\href{http://arxiv.org/abs/hep-ph/9401258}{{\tt arXiv:hep-ph/9401258
  [hep-ph]}}.

\bibitem{Alwall:2007st}
J.~Alwall, P.~Demin, S.~de~Visscher, R.~Frederix, M.~Herquet, et al., {\em
  {MadGraph/MadEvent v4: The New Web Generation}\/},
  \href{http://dx.doi.org/10.1088/1126-6708/2007/09/028}{JHEP {\bf 0709} (2007)
   028}, \href{http://arxiv.org/abs/0706.2334}{{\tt arXiv:0706.2334 [hep-ph]}}.

\bibitem{Frederix:2008hu}
R.~Frederix, T.~Gehrmann, and N.~Greiner, {\em {Automation of the Dipole
  Subtraction Method in MadGraph/MadEvent}\/},
  \href{http://dx.doi.org/10.1088/1126-6708/2008/09/122}{JHEP {\bf 0809} (2008)
   122}, \href{http://arxiv.org/abs/0808.2128}{{\tt arXiv:0808.2128 [hep-ph]}}.

\bibitem{Frederix:2010cj}
R.~Frederix, T.~Gehrmann, and N.~Greiner, {\em {Integrated dipoles with
  MadDipole in the MadGraph framework}\/},
  \href{http://dx.doi.org/10.1007/JHEP06(2010)086}{JHEP {\bf 1006} (2010)
  086}, \href{http://arxiv.org/abs/1004.2905}{{\tt arXiv:1004.2905 [hep-ph]}}.

\bibitem{Cacciari:2005hq}
M.~Cacciari and G.~P. Salam, {\em {Dispelling the $N^{3}$ myth for the $k_t$
  jet-finder}\/},
  \href{http://dx.doi.org/10.1016/j.physletb.2006.08.037}{Phys.Lett. {\bf B641}
  (2006)  57--61},
\href{http://arxiv.org/abs/hep-ph/0512210}{{\tt arXiv:hep-ph/0512210
  [hep-ph]}}.

\bibitem{Aad:2012tfa}
{ATLAS} Collaboration, G.~Aad et al., {\em {Observation of a new particle in
  the search for the Standard Model Higgs boson with the ATLAS detector at the
  LHC}\/},  \href{http://dx.doi.org/10.1016/j.physletb.2012.08.020}{Phys.Lett.
  {\bf B716} (2012)  1--29},
\href{http://arxiv.org/abs/1207.7214}{{\tt arXiv:1207.7214 [hep-ex]}}.

\bibitem{Chatrchyan:2012ufa}
{CMS} Collaboration, S.~Chatrchyan et al., {\em {Observation of a new boson at
  a mass of 125 GeV with the CMS experiment at the LHC}\/},
  \href{http://dx.doi.org/10.1016/j.physletb.2012.08.021}{Phys.Lett. {\bf B716}
  (2012)  30--61},
\href{http://arxiv.org/abs/1207.7235}{{\tt arXiv:1207.7235 [hep-ex]}}.

\bibitem{Banfi:2012yh}
A.~Banfi, G.~P. Salam, and G.~Zanderighi, {\em {NLL+NNLO predictions for
  jet-veto efficiencies in Higgs-boson and Drell-Yan production}\/},
  \href{http://dx.doi.org/10.1007/JHEP06(2012)159}{JHEP {\bf 1206} (2012)
  159},
\href{http://arxiv.org/abs/1203.5773}{{\tt arXiv:1203.5773 [hep-ph]}}.

\bibitem{Banfi:2012jm}
A.~Banfi, P.~F. Monni, G.~P. Salam, and G.~Zanderighi, {\em {Higgs and Z-boson
  production with a jet veto}\/},
  \href{http://dx.doi.org/10.1103/PhysRevLett.109.202001}{Phys.Rev.Lett. {\bf
  109} (2012)  202001},
\href{http://arxiv.org/abs/1206.4998}{{\tt arXiv:1206.4998 [hep-ph]}}.

\bibitem{Becher:2012qa}
T.~Becher and M.~Neubert, {\em {Factorization and NNLL Resummation for Higgs
  Production with a Jet Veto}\/},
  \href{http://dx.doi.org/10.1007/JHEP07(2012)108}{JHEP {\bf 1207} (2012)
  108},
\href{http://arxiv.org/abs/1205.3806}{{\tt arXiv:1205.3806 [hep-ph]}}.

\bibitem{Stewart:2013faa}
I.~W. Stewart, F.~J. Tackmann, J.~R. Walsh, and S.~Zuberi, {\em {Jet pT
  Resummation in Higgs Production at NNLL'+NNLO}\/},
\href{http://arxiv.org/abs/1307.1808}{{\tt arXiv:1307.1808}}.

\bibitem{Becher:2013xia}
T.~Becher, M.~Neubert, and L.~Rothen, {\em {Factorization and
  $N^{3}LL_{p}$+NNLO predictions for the Higgs cross section with a jet
  veto}\/},  \href{http://dx.doi.org/10.1007/JHEP10(2013)125}{JHEP {\bf 1310}
  (2013)  125},
\href{http://arxiv.org/abs/1307.0025}{{\tt arXiv:1307.0025 [hep-ph]}}.

\bibitem{Banfi:2013eda}
A.~Banfi, P.~F. Monni, and G.~Zanderighi, {\em {Quark masses in Higgs
  production with a jet veto}\/},
  \href{http://dx.doi.org/10.1007/JHEP01(2014)097}{JHEP {\bf 1401} (2014)
  097},
\href{http://arxiv.org/abs/1308.4634}{{\tt arXiv:1308.4634 [hep-ph]}}.

\bibitem{Grazzini:2013mca}
M.~Grazzini and H.~Sargsyan, {\em {Heavy-quark mass effects in Higgs boson
  production at the LHC}\/},
  \href{http://dx.doi.org/10.1007/JHEP09(2013)129}{JHEP {\bf 1309} (2013)
  129},
\href{http://arxiv.org/abs/1306.4581}{{\tt arXiv:1306.4581 [hep-ph]}}.

\bibitem{Banfi:2002hw}
A.~Banfi, G.~Marchesini, and G.~Smye, {\em {Away from jet energy flow}\/},
  \href{http://dx.doi.org/10.1088/1126-6708/2002/08/006}{JHEP {\bf 0208} (2002)
   006},
\href{http://arxiv.org/abs/hep-ph/0206076}{{\tt arXiv:hep-ph/0206076
  [hep-ph]}}.

\bibitem{Dasgupta:2001sh}
M.~Dasgupta and G.~Salam, {\em {Resummation of nonglobal QCD observables}\/},
  \href{http://dx.doi.org/10.1016/S0370-2693(01)00725-0}{Phys.Lett. {\bf B512}
  (2001)  323--330},
\href{http://arxiv.org/abs/hep-ph/0104277}{{\tt arXiv:hep-ph/0104277
  [hep-ph]}}.

\bibitem{Liu:2013hba}
X.~Liu and F.~Petriello, {\em {Reducing theoretical uncertainties for exclusive
  Higgs-boson plus one-jet production at the LHC}\/},
  \href{http://dx.doi.org/10.1103/PhysRevD.87.094027}{Phys.Rev. {\bf D87}
  (2013) no.~9, 094027},
\href{http://arxiv.org/abs/1303.4405}{{\tt arXiv:1303.4405 [hep-ph]}}.

\bibitem{Stewart:2011cf}
I.~W. Stewart and F.~J. Tackmann, {\em {Theory Uncertainties for Higgs and
  Other Searches Using Jet Bins}\/},
  \href{http://dx.doi.org/10.1103/PhysRevD.85.034011}{Phys.Rev. {\bf D85}
  (2012)  034011},
\href{http://arxiv.org/abs/1107.2117}{{\tt arXiv:1107.2117 [hep-ph]}}.

\bibitem{Boughezal:2013oha}
R.~Boughezal, X.~Liu, F.~Petriello, F.~J. Tackmann, and J.~R. Walsh, {\em
  {Combining Resummed Higgs Predictions Across Jet Bins}\/},
\href{http://arxiv.org/abs/1312.4535}{{\tt arXiv:1312.4535 [hep-ph]}}.

\bibitem{Reina:2001sf}
L.~Reina and S.~Dawson, {\em {Next-to-leading order results for t anti-t h
  production at the Tevatron}\/},
  \href{http://dx.doi.org/10.1103/PhysRevLett.87.201804}{Phys.Rev.Lett. {\bf
  87} (2001)  201804},
\href{http://arxiv.org/abs/hep-ph/0107101}{{\tt arXiv:hep-ph/0107101
  [hep-ph]}}.

\bibitem{Reina:2001bc}
L.~Reina, S.~Dawson, and D.~Wackeroth, {\em {QCD corrections to associated t
  anti-t h production at the Tevatron}\/},
  \href{http://dx.doi.org/10.1103/PhysRevD.65.053017}{Phys.Rev. {\bf D65}
  (2002)  053017},
\href{http://arxiv.org/abs/hep-ph/0109066}{{\tt arXiv:hep-ph/0109066
  [hep-ph]}}.

\bibitem{Dawson:2002tg}
S.~Dawson, L.~Orr, L.~Reina, and D.~Wackeroth, {\em {Associated top quark Higgs
  boson production at the LHC}\/},
  \href{http://dx.doi.org/10.1103/PhysRevD.67.071503}{Phys.Rev. {\bf D67}
  (2003)  071503},
\href{http://arxiv.org/abs/hep-ph/0211438}{{\tt arXiv:hep-ph/0211438
  [hep-ph]}}.

\bibitem{Dawson:2003zu}
S.~Dawson, C.~Jackson, L.~Orr, L.~Reina, and D.~Wackeroth, {\em {Associated
  Higgs production with top quarks at the large hadron collider: NLO QCD
  corrections}\/},
  \href{http://dx.doi.org/10.1103/PhysRevD.68.034022}{Phys.Rev. {\bf D68}
  (2003)  034022},
\href{http://arxiv.org/abs/hep-ph/0305087}{{\tt arXiv:hep-ph/0305087
  [hep-ph]}}.

\bibitem{ATLAS-CONF-2012-135}
{ATLAS} Collaboration, {\em Search for the Standard Model Higgs boson produced
  in association with top quarks in proton-proton collisions at √s = 7 TeV
  using the ATLAS detector\/},  Tech. Rep. ATLAS-CONF-2012-135,
  ATLAS-COM-CONF-2012-162, CERN, Geneva, Sep, 2012.

\bibitem{Chatrchyan:2013yea}
{CMS} Collaboration, S.~Chatrchyan et al., {\em {Search for the standard model
  Higgs boson produced in association with a top-quark pair in pp collisions at
  the LHC}\/},  \href{http://dx.doi.org/10.1007/JHEP05(2013)145}{JHEP {\bf
  1305} (2013)  145},
\href{http://arxiv.org/abs/1303.0763}{{\tt arXiv:1303.0763 [hep-ex]}}.

\bibitem{ATLAS-CONF-2013-080}
{ATLAS} Collaboration, {\em Search for ttbarH production in the H->gamma gamma
  channel at √s = 8 TeV with the ATLAS detector\/},  Tech. Rep.
  ATLAS-CONF-2013-080, ATLAS-COM-CONF-2013-089, CERN, Geneva, Jul, 2013.

\bibitem{CMS-PAS-HIG-13-019}
{CMS} Collaboration, {\em Search for Higgs Boson Production in Association with
  a Top-Quark Pair and Decaying to Bottom Quarks or Tau Leptons\/},  Tech. Rep.
  CMS-PAS-HIG-13-019, CERN, Geneva, 2013.

\bibitem{Beenakker:2001rj}
W.~Beenakker et al., {\em {Higgs radiation off top quarks at the Tevatron and
  the LHC}\/},  \href{http://dx.doi.org/10.1103/PhysRevLett.87.201805}{Phys.
  Rev. Lett. {\bf 87} (2001)  201805},
\href{http://arxiv.org/abs/hep-ph/0107081}{{\tt arXiv:hep-ph/0107081}}.

\bibitem{Beenakker:2002nc}
W.~Beenakker et al., {\em {NLO QCD corrections to $\PQt\tbar$ H production in
  hadron collisions. ((U))}\/},
  \href{http://dx.doi.org/10.1016/S0550-3213(03)00044-0}{Nucl. Phys. {\bf B653}
  (2003)  151--203},
\href{http://arxiv.org/abs/hep-ph/0211352}{{\tt arXiv:hep-ph/0211352}}.

\bibitem{Frederix:2011zi}
R.~Frederix, S.~Frixione, V.~Hirschi, F.~Maltoni, R.~Pittau, et al., {\em
  {Scalar and pseudoscalar Higgs production in association with a top-antitop
  pair}\/},  \href{http://dx.doi.org/10.1016/j.physletb.2011.06.012}{Phys.Lett.
  {\bf B701} (2011)  427--433},
\href{http://arxiv.org/abs/1104.5613}{{\tt arXiv:1104.5613 [hep-ph]}}.

\bibitem{Garzelli:2011vp}
M.~Garzelli, A.~Kardos, C.~Papadopoulos, and Z.~Trocsanyi, {\em {Standard Model
  Higgs boson production in association with a top anti-top pair at NLO with
  parton showering}\/},
  \href{http://dx.doi.org/10.1209/0295-5075/96/11001}{Europhys.Lett. {\bf 96}
  (2011)  11001},
\href{http://arxiv.org/abs/1108.0387}{{\tt arXiv:1108.0387 [hep-ph]}}.

\bibitem{Gleisberg:2003xi}
T.~Gleisberg, S.~Hoeche, F.~Krauss, A.~Schalicke, S.~Schumann, et al., {\em
  {SHERPA 1. alpha: A Proof of concept version}\/},
  \href{http://dx.doi.org/10.1088/1126-6708/2004/02/056}{JHEP {\bf 0402} (2004)
   056},
\href{http://arxiv.org/abs/hep-ph/0311263}{{\tt arXiv:hep-ph/0311263
  [hep-ph]}}.

\bibitem{Cafarella:2007pc}
A.~Cafarella, C.~G. Papadopoulos, and M.~Worek, {\em {Helac-Phegas: A Generator
  for all parton level processes}\/},
  \href{http://dx.doi.org/10.1016/j.cpc.2009.04.023}{Comput.Phys.Commun. {\bf
  180} (2009)  1941--1955},
\href{http://arxiv.org/abs/0710.2427}{{\tt arXiv:0710.2427 [hep-ph]}}.

\bibitem{vanHameren:2009dr}
A.~van Hameren, C.~Papadopoulos, and R.~Pittau, {\em {Automated one-loop
  calculations: A Proof of concept}\/},
  \href{http://dx.doi.org/10.1088/1126-6708/2009/09/106}{JHEP {\bf 0909} (2009)
   106},
\href{http://arxiv.org/abs/0903.4665}{{\tt arXiv:0903.4665 [hep-ph]}}.

\bibitem{Draggiotis:2009yb}
P.~Draggiotis, M.~Garzelli, C.~Papadopoulos, and R.~Pittau, {\em {Feynman Rules
  for the Rational Part of the QCD 1-loop amplitudes}\/},
  \href{http://dx.doi.org/10.1088/1126-6708/2009/04/072}{JHEP {\bf 0904} (2009)
   072},
\href{http://arxiv.org/abs/0903.0356}{{\tt arXiv:0903.0356 [hep-ph]}}.

\bibitem{Bevilacqua:2010mx}
G.~Bevilacqua, M.~Czakon, M.~Garzelli, A.~van Hameren, Y.~Malamos, et al., {\em
  {NLO QCD calculations with HELAC-NLO}\/},
  \href{http://dx.doi.org/10.1016/j.nuclphysbps.2010.08.045}{Nucl.Phys.Proc.Suppl.
  {\bf 205-206} (2010)  211--217},
\href{http://arxiv.org/abs/1007.4918}{{\tt arXiv:1007.4918 [hep-ph]}}.

\bibitem{Frixione:1995ms}
S.~Frixione, Z.~Kunszt, and A.~Signer, {\em {Three jet cross-sections to
  next-to-leading order}\/},
  \href{http://dx.doi.org/10.1016/0550-3213(96)00110-1}{Nucl.Phys. {\bf B467}
  (1996)  399--442},
\href{http://arxiv.org/abs/hep-ph/9512328}{{\tt arXiv:hep-ph/9512328
  [hep-ph]}}.

\bibitem{Alwall:2006yp}
J.~Alwall, A.~Ballestrero, P.~Bartalini, S.~Belov, E.~Boos, et al., {\em {A
  Standard format for Les Houches event files}\/},
  \href{http://dx.doi.org/10.1016/j.cpc.2006.11.010}{Comput.Phys.Commun. {\bf
  176} (2007)  300--304},
\href{http://arxiv.org/abs/hep-ph/0609017}{{\tt arXiv:hep-ph/0609017
  [hep-ph]}}.

\bibitem{Garzelli:2011iu}
M.~Garzelli, A.~Kardos, and Z.~Trocsanyi, {\em {NLO Event Samples for the
  LHC}\/},  PoS {\bf EPS-HEP2011} (2011)  282,
\href{http://arxiv.org/abs/1111.1446}{{\tt arXiv:1111.1446 [hep-ph]}}.

\bibitem{Skands:2010ak}
P.~Z. Skands, {\em {Tuning Monte Carlo Generators: The Perugia Tunes}\/},
  \href{http://dx.doi.org/10.1103/PhysRevD.82.074018}{Phys.Rev. {\bf D82}
  (2010)  074018},
\href{http://arxiv.org/abs/1005.3457}{{\tt arXiv:1005.3457 [hep-ph]}}.

\bibitem{Englert:1964et}
F.~Englert and R.~Brout, {\em {Broken Symmetry and the Mass of Gauge Vector
  Mesons}\/},
\href{http://dx.doi.org/10.1103/PhysRevLett.13.321}{Phys.Rev.Lett. {\bf 13}
  (1964)  321--323}.

\bibitem{Higgs:1964pj}
P.~W. Higgs, {\em {Broken Symmetries and the Masses of Gauge Bosons}\/},
\href{http://dx.doi.org/10.1103/PhysRevLett.13.508}{Phys.Rev.Lett. {\bf 13}
  (1964)  508--509}.

\bibitem{Higgs:1964ia}
P.~W. Higgs, {\em {Broken symmetries, massless particles and gauge fields}\/},
\href{http://dx.doi.org/10.1016/0031-9163(64)91136-9}{Phys.Lett. {\bf 12}
  (1964)  132--133}.

\bibitem{Baur:2002qd}
U.~Baur, T.~Plehn, and D.~L. Rainwater, {\em {Determining the Higgs boson
  selfcoupling at hadron colliders}\/},
  \href{http://dx.doi.org/10.1103/PhysRevD.67.033003}{Phys.Rev. {\bf D67}
  (2003)  033003},
\href{http://arxiv.org/abs/hep-ph/0211224}{{\tt arXiv:hep-ph/0211224
  [hep-ph]}}.

\bibitem{Dolan:2012rv}
M.~J. Dolan, C.~Englert, and M.~Spannowsky, {\em {Higgs self-coupling
  measurements at the LHC}\/},
  \href{http://dx.doi.org/10.1007/JHEP10(2012)112}{JHEP {\bf 1210} (2012)
  112},
\href{http://arxiv.org/abs/1206.5001}{{\tt arXiv:1206.5001 [hep-ph]}}.

\bibitem{Papaefstathiou:2012qe}
A.~Papaefstathiou, L.~L. Yang, and J.~Zurita, {\em {Higgs boson pair production
  at the LHC in the $b \bar{b} W^+ W^-$ channel}\/},
  \href{http://dx.doi.org/10.1103/PhysRevD.87.011301}{Phys.Rev. {\bf D87}
  (2013)  011301},
\href{http://arxiv.org/abs/1209.1489}{{\tt arXiv:1209.1489 [hep-ph]}}.

\bibitem{Baur:2003gp}
U.~Baur, T.~Plehn, and D.~L. Rainwater, {\em {Probing the Higgs selfcoupling at
  hadron colliders using rare decays}\/},
  \href{http://dx.doi.org/10.1103/PhysRevD.69.053004}{Phys.Rev. {\bf D69}
  (2004)  053004},
\href{http://arxiv.org/abs/hep-ph/0310056}{{\tt arXiv:hep-ph/0310056
  [hep-ph]}}.

\bibitem{Dolan:2012ac}
M.~J. Dolan, C.~Englert, and M.~Spannowsky, {\em {New Physics in LHC Higgs
  boson pair production}\/},
  \href{http://dx.doi.org/10.1103/PhysRevD.87.055002}{Phys.Rev. {\bf D87}
  (2013) no.~5, 055002},
\href{http://arxiv.org/abs/1210.8166}{{\tt arXiv:1210.8166 [hep-ph]}}.

\bibitem{Goertz:2013kp}
F.~Goertz, A.~Papaefstathiou, L.~L. Yang, and J.~Zurita, {\em {Higgs Boson
  self-coupling measurements using ratios of cross sections}\/},
  \href{http://dx.doi.org/10.1007/JHEP06(2013)016}{JHEP {\bf 1306} (2013)
  016},
\href{http://arxiv.org/abs/1301.3492}{{\tt arXiv:1301.3492 [hep-ph]}}.

\bibitem{Shao:2013bz}
D.~Y. Shao, C.~S. Li, H.~T. Li, and J.~Wang, {\em {Threshold resummation
  effects in Higgs boson pair production at the LHC}\/},
\href{http://arxiv.org/abs/1301.1245}{{\tt arXiv:1301.1245 [hep-ph]}}.

\bibitem{Gouzevitch:2013qca}
M.~Gouzevitch, A.~Oliveira, J.~Rojo, R.~Rosenfeld, G.~P. Salam, et al., {\em
  {Scale-invariant resonance tagging in multijet events and new physics in
  Higgs pair production}\/},
  \href{http://dx.doi.org/10.1007/JHEP07(2013)148}{JHEP {\bf 1307} (2013)
  148},
\href{http://arxiv.org/abs/1303.6636}{{\tt arXiv:1303.6636 [hep-ph]}}.

\bibitem{Glover:1987nx}
E.~N. Glover and J.~van~der Bij, {\em {HIGGS BOSON PAIR PRODUCTION VIA GLUON
  FUSION}\/},
\href{http://dx.doi.org/10.1016/0550-3213(88)90083-1}{Nucl.Phys. {\bf B309}
  (1988)  282}.

\bibitem{Eboli:1987dy}
O.~J. Eboli, G.~Marques, S.~Novaes, and A.~Natale, {\em {TWIN HIGGS BOSON
  PRODUCTION}\/},
\href{http://dx.doi.org/10.1016/0370-2693(87)90381-9}{Phys.Lett. {\bf B197}
  (1987)  269}.

\bibitem{Plehn:1996wb}
T.~Plehn, M.~Spira, and P.~Zerwas, {\em {Pair production of neutral Higgs
  particles in gluon-gluon collisions}\/},
  \href{http://dx.doi.org/10.1016/0550-3213(96)00418-X}{Nucl.Phys. {\bf B479}
  (1996)  46--64},
\href{http://arxiv.org/abs/hep-ph/9603205}{{\tt arXiv:hep-ph/9603205
  [hep-ph]}}.

\bibitem{Kramer:1996iq}
M.~Kramer, E.~Laenen, and M.~Spira, {\em {Soft gluon radiation in Higgs boson
  production at the LHC}\/},
  \href{http://dx.doi.org/10.1016/S0550-3213(97)00679-2}{Nucl.Phys. {\bf B511}
  (1998)  523--549},
\href{http://arxiv.org/abs/hep-ph/9611272}{{\tt arXiv:hep-ph/9611272
  [hep-ph]}}.

\bibitem{Chetyrkin:1997iv}
K.~Chetyrkin, B.~A. Kniehl, and M.~Steinhauser, {\em {Hadronic Higgs decay to
  order alpha-s**4}\/},
  \href{http://dx.doi.org/10.1103/PhysRevLett.79.353}{Phys.Rev.Lett. {\bf 79}
  (1997)  353--356},
\href{http://arxiv.org/abs/hep-ph/9705240}{{\tt arXiv:hep-ph/9705240
  [hep-ph]}}.

\bibitem{Harlander:2002wh}
R.~V. Harlander and W.~B. Kilgore, {\em {Next-to-next-to-leading order Higgs
  production at hadron colliders}\/},
  \href{http://dx.doi.org/10.1103/PhysRevLett.88.201801}{Phys.Rev.Lett. {\bf
  88} (2002)  201801},
\href{http://arxiv.org/abs/hep-ph/0201206}{{\tt arXiv:hep-ph/0201206
  [hep-ph]}}.

\bibitem{Anastasiou:2002yz}
C.~Anastasiou and K.~Melnikov, {\em {Higgs boson production at hadron colliders
  in NNLO QCD}\/},
  \href{http://dx.doi.org/10.1016/S0550-3213(02)00837-4}{Nucl.Phys. {\bf B646}
  (2002)  220--256},
\href{http://arxiv.org/abs/hep-ph/0207004}{{\tt arXiv:hep-ph/0207004
  [hep-ph]}}.

\bibitem{Ravindran:2003um}
V.~Ravindran, J.~Smith, and W.~L. van Neerven, {\em {NNLO corrections to the
  total cross-section for Higgs boson production in hadron hadron
  collisions}\/},
  \href{http://dx.doi.org/10.1016/S0550-3213(03)00457-7}{Nucl.Phys. {\bf B665}
  (2003)  325--366},
\href{http://arxiv.org/abs/hep-ph/0302135}{{\tt arXiv:hep-ph/0302135
  [hep-ph]}}.

\bibitem{deFlorian:2013uza}
D.~de~Florian and J.~Mazzitelli, {\em {Two-loop virtual corrections to Higgs
  pair production}\/},
  \href{http://dx.doi.org/10.1016/j.physletb.2013.06.046}{Phys.Lett. {\bf B724}
  (2013)  306--309},
\href{http://arxiv.org/abs/1305.5206}{{\tt arXiv:1305.5206 [hep-ph]}}.

\bibitem{Hahn:2000kx}
T.~Hahn, {\em {Generating Feynman diagrams and amplitudes with FeynArts 3}\/},
  \href{http://dx.doi.org/10.1016/S0010-4655(01)00290-9}{Comput.Phys.Commun.
  {\bf 140} (2001)  418--431},
\href{http://arxiv.org/abs/hep-ph/0012260}{{\tt arXiv:hep-ph/0012260
  [hep-ph]}}.

\bibitem{Mertig:1990an}
R.~Mertig, M.~Bohm, and A.~Denner, {\em {FEYN CALC: Computer algebraic
  calculation of Feynman amplitudes}\/},
\href{http://dx.doi.org/10.1016/0010-4655(91)90130-D}{Comput.Phys.Commun. {\bf
  64} (1991)  345--359}.

\bibitem{Martin:2009bu}
A.~Martin, W.~Stirling, R.~Thorne, and G.~Watt, {\em {Uncertainties on alpha(S)
  in global PDF analyses and implications for predicted hadronic cross
  sections}\/},
  \href{http://dx.doi.org/10.1140/epjc/s10052-009-1164-2}{Eur.Phys.J. {\bf C64}
  (2009)  653--680},
\href{http://arxiv.org/abs/0905.3531}{{\tt arXiv:0905.3531 [hep-ph]}}.

\bibitem{Blair:1379880}
R.~Blair, B.~Brelier, F.~Bucci, S.~Chekanov, M.~Stockton, and M.~Tripiana, {\em
  {NLO Theoretical Predictions for Photon Measurements Using the PHOX
  Generators}\/},  Tech. Rep. CERN-OPEN-2011-041, CERN, Geneva, Sep, 2011.

\bibitem{Wielers:2001suas}
M.~Wielers, {\em {Isolation of Photons}\/},  Tech. Rep. ATL-PHYS-2002-004,
  ATL-COM-PHYS-2001-024, CERN-ATL-PHYS-2002-004, 2001.

\bibitem{Abbiendi:2003kf}
{OPAL Collaboration} Collaboration, G.~Abbiendi et al., {\em {Measurement of
  isolated prompt photon production in photon photon collisions at s(ee)**(1/2)
  = 183 GeV - 209-GeV}\/},
  \href{http://dx.doi.org/10.1140/epjc/s2003-01359-1}{Eur.Phys.J. {\bf C31}
  (2003)  491--502},
\href{http://arxiv.org/abs/hep-ex/0305075}{{\tt arXiv:hep-ex/0305075
  [hep-ex]}}.

\bibitem{hep-ph:9911340}
T.~Binoth, J.~Guillet, E.~Pilon, and M.~Werlen, {\em {A Full next-to-leading
  order study of direct photon pair production in hadronic collisions}\/},
  \href{http://dx.doi.org/10.1007/s100520050024}{Eur.Phys.J. {\bf C16} (2000)
  311--330},
\href{http://arxiv.org/abs/hep-ph/9911340}{{\tt arXiv:hep-ph/9911340
  [hep-ph]}}.

\bibitem{hep-ph:0204023}
S.~Catani, M.~Fontannaz, J.~Guillet, and E.~Pilon, {\em {Cross-section of
  isolated prompt photons in hadron hadron collisions}\/},
  \href{http://dx.doi.org/10.1088/1126-6708/2002/05/028}{JHEP {\bf 0205} (2002)
   028},
\href{http://arxiv.org/abs/hep-ph/0204023}{{\tt arXiv:hep-ph/0204023
  [hep-ph]}}.

\bibitem{Catani:2011dp}
S.~Catani, L.~Cieri, D.~de~Florian, G.~Ferrera, and M.~Grazzini, {\em {Diphoton
  production at hadron colliders: a fully-differential QCD calculation at
  NNLO}\/},
  \href{http://dx.doi.org/10.1103/PhysRevLett.108.072001}{Phys.Rev.Lett. {\bf
  108} (2012)  072001},
\href{http://arxiv.org/abs/1110.2375}{{\tt arXiv:1110.2375 [hep-ph]}}.

\bibitem{Catani:2002ny}
S.~Catani, M.~Fontannaz, J.~Guillet, and E.~Pilon, {\em {Cross-section of
  isolated prompt photons in hadron hadron collisions}\/},
  \href{http://dx.doi.org/10.1088/1126-6708/2002/05/028}{JHEP {\bf 0205} (2002)
   028},
\href{http://arxiv.org/abs/hep-ph/0204023}{{\tt arXiv:hep-ph/0204023
  [hep-ph]}}.

\bibitem{Catani:2013oma}
S.~Catani, M.~Fontannaz, J.~P. Guillet, and E.~Pilon, {\em {Isolating Prompt
  Photons with Narrow Cones}\/},
  \href{http://dx.doi.org/10.1007/JHEP09(2013)007}{JHEP {\bf 1309} (2013)
  007},
\href{http://arxiv.org/abs/1306.6498}{{\tt arXiv:1306.6498 [hep-ph]}}.

\bibitem{Frixione:1999gr}
S.~Frixione and W.~Vogelsang, {\em {Isolated photon production in polarized $p
  p$ collisions}\/},
  \href{http://dx.doi.org/10.1016/S0550-3213(99)00575-1}{Nucl.Phys. {\bf B568}
  (2000)  60--92},
\href{http://arxiv.org/abs/hep-ph/9908387}{{\tt arXiv:hep-ph/9908387
  [hep-ph]}}.

\bibitem{Catani:2000jh}
S.~Catani, M.~Dittmar, D.~Soper, W.~J. Stirling, S.~Tapprogge, et al., {\em
  {QCD}\/},
\href{http://arxiv.org/abs/hep-ph/0005025}{{\tt arXiv:hep-ph/0005025
  [hep-ph]}}.

\bibitem{Gehrmann:2013aga}
T.~Gehrmann, N.~Greiner, and G.~Heinrich, {\em {Photon isolation effects at NLO
  in gamma gamma + jet final states in hadronic collisions}\/},
  \href{http://dx.doi.org/10.1007/JHEP06(2013)058}{JHEP {\bf 1306} (2013)
  058},
\href{http://arxiv.org/abs/1303.0824}{{\tt arXiv:1303.0824 [hep-ph]}}.

\bibitem{Frixione:1998hn}
S.~Frixione, {\em {Isolated photons in perturbative QCD}\/},
  \href{http://dx.doi.org/10.1016/S0370-2693(98)00454-7}{Phys.Lett. {\bf B429}
  (1998)  369--374},
\href{http://arxiv.org/abs/arXiv hep-ph/9801442}{{\tt arXiv hep-ph/9801442}}.

\bibitem{Binoth:1999qq}
T.~Binoth, J.~Guillet, E.~Pilon, and M.~Werlen, {\em {A Full next-to-leading
  order study of direct photon pair production in hadronic collisions}\/},
  \href{http://dx.doi.org/10.1007/s100520050024}{Eur.Phys.J. {\bf C16} (2000)
  311--330},
\href{http://arxiv.org/abs/hep-ph/9911340}{{\tt arXiv:hep-ph/9911340
  [hep-ph]}}.

\bibitem{DelDuca:2003uz}
V.~Del~Duca, F.~Maltoni, Z.~Nagy, and Z.~Trocsanyi, {\em {QCD radiative
  corrections to prompt diphoton production in association with a jet at hadron
  colliders}\/},  JHEP {\bf 0304} (2003)  059,
\href{http://arxiv.org/abs/hep-ph/0303012}{{\tt arXiv:hep-ph/0303012
  [hep-ph]}}.

\bibitem{gosamHepforge}
{http://www.gosam.hepforge.org/diphoton}.

\bibitem{Gehrmann:2013bga}
T.~Gehrmann, N.~Greiner, and G.~Heinrich, {\em Precise QCD Predictions for the
  Production of a Photon Pair in Association with Two Jets\/},
  \href{http://dx.doi.org/10.1103/PhysRevLett.111.222002}{Phys. Rev. Lett. {\bf
  111} (2013)  222002}, \href{http://arxiv.org/abs/1308.3660}{{\tt
  arXiv:1308.3660 [hep-ph]}}.
\url{http://link.aps.org/doi/10.1103/PhysRevLett.111.222002}.

\bibitem{Bern:2013bha}
Z.~Bern, L.~Dixon, F.~Febres~Cordero, S.~Hoeche, H.~Ita, et al., {\em
  {Next-to-leading order diphoton+2-jet production at the LHC}\/},
\href{http://arxiv.org/abs/1312.0592}{{\tt arXiv:1312.0592 [hep-ph]}}.

\bibitem{Badger:2013ava}
S.~Badger, A.~Guffanti, and V.~Yundin, {\em {Next-to-leading order QCD
  corrections to di-photon production in association with up to three jets at
  the Large Hadron Collider}\/},
\href{http://arxiv.org/abs/1312.5927}{{\tt arXiv:1312.5927 [hep-ph]}}.

\bibitem{Bern:2014vza}
Z.~Bern, L.~Dixon, F.~Febres~Cordero, S.~Hoeche, H.~Ita, et al., {\em
  {Next-to-Leading Order Gamma Gamma + 2-Jet Production at the LHC}\/},
\href{http://arxiv.org/abs/1402.4127}{{\tt arXiv:1402.4127 [hep-ph]}}.

\bibitem{Corcella:2000bw}
G.~Corcella, I.~Knowles, G.~Marchesini, S.~Moretti, K.~Odagiri, et al., {\em
  {HERWIG 6: An Event generator for hadron emission reactions with interfering
  gluons (including supersymmetric processes)}\/},  JHEP {\bf 0101} (2001)
  010,
\href{http://arxiv.org/abs/hep-ph/0011363}{{\tt arXiv:hep-ph/0011363
  [hep-ph]}}.

\bibitem{Dobbs:2001ck}
M.~Dobbs and J.~B. Hansen, {\em {The HepMC C++ Monte Carlo event record for
  High Energy Physics}\/},
\href{http://dx.doi.org/10.1016/S0010-4655(00)00189-2}{Comput.Phys.Commun. {\bf
  134} (2001)  41--46}.

\bibitem{yodaweb}
\url{http://yoda.hepforge.org/}.

\bibitem{Brun:1997pa}
R.~Brun and F.~Rademakers, {\em {ROOT: An object oriented data analysis
  framework}\/},
\href{http://dx.doi.org/10.1016/S0168-9002(97)00048-X}{Nucl.Instrum.Meth. {\bf
  A389} (1997)  81--86}.

\bibitem{Butterworth:2010ym}
J.~Butterworth, A.~Arbey, L.~Basso, S.~Belov, A.~Bharucha, et al., {\em {The
  Tools and Monte Carlo working group Summary Report}\/},
\href{http://arxiv.org/abs/1003.1643}{{\tt arXiv:1003.1643 [hep-ph]}}.

\end{thebibliography}\endgroup


\providecommand{\href}[2]{#2}\begingroup\raggedright\begin{thebibliography}{10}

\bibitem{Frixione:1998jh}
S.~Frixione, {\em Phys.Lett.} {\bf B429} (1998) 369--374,
  [\href{http://xxx.lanl.gov/abs/hep-ph/9801442}{{\tt hep-ph/9801442}}].

\bibitem{Binoth:2010ra}
J.~Andersen {\em et.~al.},, {\bf SM and NLO Multileg Working Group}
  Collaboration \href{http://xxx.lanl.gov/abs/1003.1241}{{\tt 1003.1241}}.

\bibitem{Blair:1379880}
R.~Blair, B.~Brelier, F.~Bucci, S.~Chekanov, M.~Stockton, and M.~Tripiana,
  Tech. Rep. CERN-OPEN-2011-041, CERN, Geneva, Sep, 2011.

\bibitem{Wielers:2001suas}
M.~Wielers, Tech. Rep. ATL-PHYS-2002-004, ATL-COM-PHYS-2001-024,
  CERN-ATL-PHYS-2002-004, 2001.

\bibitem{Abbiendi:2003kf}
G.~Abbiendi {\em et.~al.},, {\bf OPAL Collaboration} Collaboration {\em
  Eur.Phys.J.} {\bf C31} (2003) 491--502,
  [\href{http://xxx.lanl.gov/abs/hep-ex/0305075}{{\tt hep-ex/0305075}}].

\bibitem{hep-ph:9911340}
T.~Binoth, J.~Guillet, E.~Pilon, and M.~Werlen, {\em Eur.Phys.J.} {\bf C16}
  (2000) 311--330, [\href{http://xxx.lanl.gov/abs/hep-ph/9911340}{{\tt
  hep-ph/9911340}}].

\bibitem{hep-ph:0204023}
S.~Catani, M.~Fontannaz, J.~Guillet, and E.~Pilon, {\em JHEP} {\bf 0205} (2002)
  028, [\href{http://xxx.lanl.gov/abs/hep-ph/0204023}{{\tt hep-ph/0204023}}].

\bibitem{Catani:2011dp}
S.~Catani, L.~Cieri, D.~de~Florian, G.~Ferrera, and M.~Grazzini, {\em
  Phys.Rev.Lett.} {\bf 108} (2012) 072001,
  [\href{http://xxx.lanl.gov/abs/1110.2375}{{\tt 1110.2375}}].

\bibitem{AlcarazMaestre:2012vp}
J.~Alcaraz~Maestre {\em et.~al.},, {\bf SM AND NLO MULTILEG and SM MC Working
  Groups} Collaboration \href{http://xxx.lanl.gov/abs/1203.6803}{{\tt
  1203.6803}}.

\bibitem{Catani:2002ny}
S.~Catani, M.~Fontannaz, J.~Guillet, and E.~Pilon, {\em JHEP} {\bf 0205} (2002)
  028, [\href{http://xxx.lanl.gov/abs/hep-ph/0204023}{{\tt hep-ph/0204023}}].

\bibitem{Catani:2013oma}
S.~Catani, M.~Fontannaz, J.~P. Guillet, and E.~Pilon, {\em JHEP} {\bf 1309}
  (2013) 007, [\href{http://xxx.lanl.gov/abs/1306.6498}{{\tt 1306.6498}}].

\bibitem{Frixione:1999gr}
S.~Frixione and W.~Vogelsang, {\em Nucl.Phys.} {\bf B568} (2000) 60--92,
  [\href{http://xxx.lanl.gov/abs/hep-ph/9908387}{{\tt hep-ph/9908387}}].

\bibitem{Catani:2000jh}
S.~Catani, M.~Dittmar, D.~Soper, W.~J. Stirling, S.~Tapprogge, {\em et.~al.},
  \href{http://xxx.lanl.gov/abs/hep-ph/0005025}{{\tt hep-ph/0005025}}.

\bibitem{Campbell:2011bn}
J.~M. Campbell, R.~K. Ellis, and C.~Williams, {\em JHEP} {\bf 1107} (2011) 018,
  [\href{http://xxx.lanl.gov/abs/1105.0020}{{\tt 1105.0020}}].

\bibitem{Gehrmann:2013aga}
T.~Gehrmann, N.~Greiner, and G.~Heinrich, {\em JHEP} {\bf 1306} (2013) 058,
  [\href{http://xxx.lanl.gov/abs/1303.0824}{{\tt 1303.0824}}].

\bibitem{Chatrchyan:2011qt}
S.~Chatrchyan {\em et.~al.},, {\bf CMS Collaboration} Collaboration {\em JHEP}
  {\bf 1201} (2012) 133, [\href{http://xxx.lanl.gov/abs/1110.6461}{{\tt
  1110.6461}}].

\end{thebibliography}\endgroup



\end{document}